\documentclass[11pt]{report}
\usepackage{indentfirst}
\usepackage{makeidx}

\makeindex

\newcommand{\Adami}{Adami\index{Adami, Chris}}
\newcommand{\Adler}{Adler\index{Adler, Steven}}
\newcommand{\Aharonov}{Aharonov\index{Aharonov, Yakir}}
\newcommand{\AlbertD}{Albert\index{Albert, David Z.}}
\newcommand{\Alley}{Alley\index{Alley, Carol}}
\newcommand{\Archimedes}{Archimedes\index{Archimedes}}
\newcommand{\Aristotle}{Aristotle\index{Aristotle}}
\newcommand{\BakerD}{Baker\index{Baker, David B. L.}}
\newcommand{\BakerH}{Baker\index{Baker, Howard}}
\newcommand{\Ballentine}{Ballentine\index{Ballentine, Leslie E.}}
\newcommand{\Banville}{Banville\index{Banville, John}}
\newcommand{\Barad}{Barad\index{Barad, Karen}}
\newcommand{\Barnum}{Barnum\index{Barnum, Howard}}
\newcommand{\Baudrillard}{Baudrillard\index{Baudrillard, Jean}}
\newcommand{\Baytch}{Baytch\index{Baytch, Nurit E.}}
\newcommand{\Bell}{Bell\index{Bell, John S.}}
\newcommand{\Beller}{Beller\index{Beller, Mara}}
\newcommand{\Benioff}{Benioff\index{Benioff, Paul}}
\newcommand{\Benka}{Benka\index{Benka, Stephen G.}}
\newcommand{\Bennett}{Bennett\index{Bennett, Charles H.}}
   \newcommand{\Charlie}{Charlie\index{Bennett, Charles H.}}
\newcommand{\Theo}{Theo\index{Bennett, Theodora}}
\newcommand{\TonyBennett}{Tony~Bennett\index{Bennett, Tony}}
\newcommand{\Bernardo}{Bernardo\index{Bernardo, Jos{\'e} M.}}
\newcommand{\Bernoulli}{Bernoulli\index{Bernoulli, James}}
\newcommand{\Bernstein}{Bernstein\index{Bernstein, Herbert J.}}
   \newcommand{\Herb}{Herb\index{Bernstein, Herbert J.}}
\newcommand{\Berthiaume}{Berthiaume\index{Berthiaume, Andr\'e}}
\newcommand{\Beyler}{Beyler\index{Beyler, Richard H.}}
\newcommand{\Bhattacharya}{Bhattacharya\index{Bhattacharya, Tanmoy}}
\newcommand{\Bilodeau}{Bilodeau\index{Bilodeau, Doug}}
\newcommand{\Bohm}{Bohm\index{Bohm, David}}
\newcommand{\Bohr}{Bohr\index{Bohr, Niels}}
\newcommand{\Boole}{Boole\index{Boole, George}}
\newcommand{\Bopp}{Bopp\index{Bopp, Friedrich (Fritz)}}
\newcommand{\Born}{Born\index{Born, Max}}
\newcommand{\Brassard}{Brassard\index{Brassard, Gilles}}
   \newcommand{\Gilles}{Gilles\index{Brassard, Gilles}}
\newcommand{\Braunstein}{Braunstein\index{Braunstein, Samuel L.}}
\newcommand{\Bretthorst}{Bretthorst\index{Bretthorst, G. Larry}}
\newcommand{\Broglie}{Boglie\index{Broglie, Louis}}
\newcommand{\Brukner}{Brukner\index{Brukner, \v{C}aslav}}
\newcommand{\Brun}{Brun\index{Brun, Todd A.}}
   \newcommand{\Todd}{Todd\index{Brun, Todd A.}}
\newcommand{\Bub}{Bub\index{Bub, Jeffrey}}
\newcommand{\Buck}{Buck\index{Buck, Joseph R.}}
\newcommand{\Buhrman}{Buhrman\index{Buhrman, Harry}}
\newcommand{\Busch}{Busch\index{Busch, Paul}}
\newcommand{\Bush}{Bush\index{Bush, George W.}}
\newcommand{\Butterfield}{Butterfield\index{Butterfield, Jeremy}}
\newcommand{\Buzek}{Bu\v{z}ek\index{Bu\v{z}ek, Vladimir}}
\newcommand{\Cabello}{Cabello\index{Cabello, Adan}}
\newcommand{\Cardano}{Cardano\index{Cardano, Girolamo}}
\newcommand{\Carvalho}{Carvalho\index{Carvalho, Alvaro}}
\newcommand{\Caves}{Caves\index{Caves, Carlton M.}}
   \newcommand{\Carl}{Carl\index{Caves, Carlton M.}}
\newcommand{\Cerf}{Cerf\index{Cerf, Nicolas J.}}
\newcommand{\Chaitin}{Chaitin\index{Chaitin, Gregory J.}}
\newcommand{\Chau}{Chau\index{Chau, H. F.}}
\newcommand{\Chuang}{Chuang\index{Chuang, Isaac L.}}
\newcommand{\Church}{Church\index{Church, Alonzo}}
\newcommand{\Cirac}{Cirac\index{Cirac, Juan I.}}
\newcommand{\Cleve}{Cleve\index{Cleve, Richard}}
\newcommand{\Clifton}{Clifton\index{Clifton, Rob}}
\newcommand{\Coleman}{Coleman\index{Coleman, Sidney R.}}
\newcommand{\Colgate}{Colgate\index{Colgate, Stirling A.}}
\newcommand{\Comer}{Comer\index{Comer, Gregory L.}}
   \newcommand{\Greg}{Greg\index{Comer, Gregory L.}}
\newcommand{\Crepeau}{Cr{\'e}peau\index{Cr{\'e}peau, Claude}}
\newcommand{\DAriano}{D'Ariano\index{D'Ariano, G. Mauro}}
\newcommand{\Finetti}{de Finetti\index{de Finetti, Bruno}}
\newcommand{\DeMartini}{De~Martini\index{De~Martini, Francesco}}
\newcommand{\Democritus}{Democritus\index{Democritus}}
\newcommand{\Derrida}{Derrida\index{Derrida, Jacques}}
\newcommand{\Deutsch}{Deutsch\index{Deutsch, David}}
\newcommand{\Dewdney}{Dewdney\index{Dewdney, Chris}}
\newcommand{\DeWitt}{DeWitt\index{DeWitt, Bryce S.}}
\newcommand{\Dirac}{Dirac\index{Dirac, Paul A. M.}}
\newcommand{\DiVincenzo}{DiVincenzo\index{DiVincenzo, David P.}}
\newcommand{\Duerr}{D\"urr\index{D\"urr, Detlef}}
\newcommand{\Dyson}{Dyson\index{Dyson, Freeman J.}}
\newcommand{\Ehrenfest}{Ehrenfest\index{Ehrenfest, Paul}}
\newcommand{\Einstein}{Einstein\index{Einstein, Albert}}
\newcommand{\Ekert}{Ekert\index{Ekert, Artur}}
\newcommand{\Enk}{Enk\index{Enk, Steven J. van}}
\newcommand{\Enz}{Enz\index{Enz, Charles P.}}
\newcommand{\Epicurus}{Epicurus\index{Epicurus}}
\newcommand{\Everett}{Everett\index{Everett, Hugh}}
\newcommand{\Euclid}{Euclid\index{Euclid}}
\newcommand{\Exner}{Exner\index{Exner, Franz}}
\newcommand{\Ezekiel}{Ezekiel\index{Ezekiel}}
\newcommand{\Farhi}{Farhi\index{Farhi, Edward}}
\newcommand{\Faye}{Faye\index{Faye, Jan}}
\newcommand{\Feinstein}{Feinstein\index{Feinstein, Amiel}}
\newcommand{\Fermi}{Fermi\index{Fermi, Enrico}}
\newcommand{\Feuer}{Feuer\index{Feuer, Lewis S.}}
\newcommand{\Fichte}{Fichte\index{Fichte, J. G.}}
\newcommand{\Fierz}{Fierz\index{Fierz, Markus}}
\newcommand{\Fine}{Fine\index{Fine, Arthur}}
\newcommand{\Flatt}{Flatt\index{Flatt, Lester}}
\newcommand{\Fleck}{Fleck\index{Fleck, Ludwik}}
\newcommand{\Folse}{Folse\index{Folse, Henry J.}}
\newcommand{\Forman}{Forman\index{Forman, Paul}}
\newcommand{\Forney}{Forney\index{Forney, G. D.}}
\newcommand{\Fortun}{Fortun\index{Fortun, Mike}}
\newcommand{\Fowler}{Fowler\index{Fowler, Alan}}
\newcommand{\Frost}{Frost\index{Frost, David}}
\newcommand{\Albert}{Albert\index{Fuchs, Albert Winston}}
\newcommand{\Emma}{Emma\index{Fuchs, Emma Jane}}
\newcommand{\Kiki}{Kiki\index{Fuchs, Kristen M. (Kiki)}}
\newcommand{\Wizzy}{Wizzy\index{Fuchs, Wizzy}}
\newcommand{\Furusawa}{Furusawa\index{Furusawa, Akira}}
\newcommand{\Galavotti}{Galavotti\index{Galavotti, Maria Carla}}
\newcommand{\GarciaAlcaine}{Garcia-Alcaine\index{Garcia-Alcaine, Guillermo}}
\newcommand{\Gardner}{Gardner\index{Gardner, Martin}}
\newcommand{\Garfunkel}{Garfunkel\index{Garfunkel, Art}}
\newcommand{\Garisto}{Garisto\index{Garisto, Robert}}
\newcommand{\Garrett}{Garrett\index{Garrett, Anthony J. M.}}
\newcommand{\GellMann}{Gell-Mann\index{Gell-Mann, Murray}}
\newcommand{\Gershenfeld}{Gershenfeld\index{Gershefeld, Neil A.}}
\newcommand{\Ghirardi}{Ghirardi\index{Ghirardi, Giancarlo}}
\newcommand{\Ghose}{Ghose\index{Ghose, Partha}}
\newcommand{\Giacometti}{Giacometti\index{Giacometti, Alberto}}
\newcommand{\Giere}{Giere\index{Giere, Ronald N.}}
\newcommand{\Gingrich}{Gingrich\index{Gingrich, Robert M.}}
\newcommand{\Ginsberg}{Ginsberg\index{Ginsberg, Allen}}
\newcommand{\Gisin}{Gisin\index{Gisin, Nicolas}}
\newcommand{\Giuntini}{Giuntini\index{Giuntini, Roberto}}
\newcommand{\Gleason}{Gleason\index{Gleason, Andrew M.}}
\newcommand{\Goedel}{G\"odel\index{G\"odel, Kurt}}
\newcommand{\Goldstein}{Goldstein\index{Goldstein, Sheldon}}
\newcommand{\Goldstone}{Goldstone\index{Goldstone, Jeffrey}}
\newcommand{\Good}{Good\index{Good, Irving John}}
\newcommand{\Gordon}{Gordon\index{Gordon, J. P.}}
\newcommand{\Gottesman}{Gottesman\index{Gottesman, Daniel}}
\newcommand{\Gottfried}{Gottfried\index{Gottfried, Kurt}}
\newcommand{\Graham}{Graham\index{Graham, Neill}}
\newcommand{\Griffiths}{Griffiths\index{Griffiths, Robert B.}}
\newcommand{\Gutmann}{Gutmann\index{Gutmann, Sam}}

\newcommand{\Hacking}{Hacking\index{Hacking, Ian}}
\newcommand{\Haering}{Haering\index{Haering, Theodor}}
\newcommand{\Hagar}{Hagar\index{Hagar, Sammy}}
\newcommand{\Hains}{Hains\index{Hains, Chris}}
\newcommand{\Hall}{Hall\index{Hall, Michael J. W.}}
\newcommand{\Hanle}{Hanle\index{Hanle, Paul A.}}
\newcommand{\Hanna}{Hanna\index{Hanna, Marty}}
\newcommand{\Hardy}{Hardy\index{Hardy, Lucien}}
\newcommand{\Harris}{Harris\index{Harris, Paul}}

\newcommand{\Hartle}{Hartle\index{Hartle, James B.}}
\newcommand{\Hausladen}{Hausladen\index{Hausladen, Paul}}
\newcommand{\Hawking}{Hawking\index{Hawking, Stephen W.}}
\newcommand{\Hayden}{Hayden\index{Hayden, Patrick}}
\newcommand{\Hegel}{Hegel\index{Hegel, G. W. F.}}
\newcommand{\Heilbron}{Heilbron\index{Heilbron, J. L.}}
\newcommand{\Heisenberg}{Heisenberg\index{Heisenberg, Werner}}
\newcommand{\Heitler}{Heitler\index{Heitler, Walter}}
\newcommand{\Hellman}{Hellman\index{Hellman, Geoffrey}}
\newcommand{\Hepburn}{Hepburn\index{Hepburn, Audrey}}
\newcommand{\Herling}{Herling\index{Herling, Gary}}
\newcommand{\Hermann}{Hermann\index{Hermann, Grete}}
\newcommand{\Herzfeld}{Herzfeld\index{Herzfeld, Karl F.}}
\newcommand{\Hillery}{Hillery\index{Hillery, Mark}}
\newcommand{\Hirota}{Hirota\index{Hirota, Osamu}}
\newcommand{\Hitler}{Hitler\index{Hitler, Adolf}}
\newcommand{\Holevo}{Holevo\index{Holevo, Alexander S.}}
\newcommand{\Holladay}{Holladay\index{Holladay, Wendall}}
\newcommand{\Holton}{Holton\index{Holton, Gerald}}
\newcommand{\Honner}{Honner\index{Honner, John}}
\newcommand{\Hood}{Hood\index{Hood, Christina J.}}
\newcommand{\HorodeckiM}{Horodecki\index{Horodecki, Micha{\l}}}
\newcommand{\HorodeckiP}{Horodecki\index{Horodecki, Pawe{\l}}}
\newcommand{\HorodeckiR}{Horodecki\index{Horodecki, Ryszard}}
\newcommand{\Hughes}{Hughes\index{Hughes, Richard J.}}
\newcommand{\Hughston}{Hughston\index{Hughston, Lane P.}}
\newcommand{\Hume}{Hume\index{Hume, David}}
\newcommand{\Isham}{Isham\index{Isham, Chris J.}}
\newcommand{\Jacobs}{Jacobs\index{Jacobs, Kurt}}
\newcommand{\JamesS}{James\index{James, Henry Sr.}}
\newcommand{\JamesH}{James\index{James, Henry}}
\newcommand{\James}{James\index{James, William}}
\newcommand{\Jammer}{Jammer\index{Jammer, Max}}
\newcommand{\Janzing}{Janzing\index{Janzing, Dominik}}
\newcommand{\Jaynes}{Jaynes\index{Jaynes, Edwin T.}}
\newcommand{\Jeffrey}{Jeffrey\index{Jeffrey, Richard C.}}
\newcommand{\Jolly}{Jolly\index{Jolly, Phillip von}}
\newcommand{\Jordan}{Jordan\index{Jordan, Pascual}}
\newcommand{\Jozsa}{Jozsa\index{Jozsa, Richard}}
\newcommand{\Jung}{Jung\index{Jung, Carl G.}}
\newcommand{\Kahn}{Kahn\index{Kahn, Karen L.}}
\newcommand{\Kalckar}{Kalckar\index{Kalckar, J{\o}rgen}}
\newcommand{\Kampen}{Kampen\index{Kampen, N. G. van}}
\newcommand{\Kant}{Kant\index{Kant, Immanuel}}
\newcommand{\Kennedy}{Kennedy\index{Kennedy, John B.}}
\newcommand{\KennedyJF}{Kennedy\index{Kennedy, John F.}}
\newcommand{\Kent}{Kent\index{Kent, Adrian}}
\newcommand{\Kepler}{Kepler\index{Kepler, Johannes}}
\newcommand{\Kermode}{Kermode\index{Kermode, Frank}}
\newcommand{\Kevles}{Kevles\index{Kevles, Daniel J.}}
\newcommand{\Kierkegaard}{Kierkegaard\index{Kierkegaard, S{\o}ren}}
\newcommand{\Khrennikov}{Khrennikov\index{Khrennikov, Andrei}}
\newcommand{\Kimble}{Kimble\index{Kimble, H. Jeffrey}}
\newcommand{\Kochen}{Kochen\index{Kochen, Simon}}
\newcommand{\Kolmogorov}{Kolmogorov\index{Kolmogorov, Andrei N.}}
\newcommand{\Koresh}{Koresh\index{Koresh, David}}
\newcommand{\Kraus}{Kraus\index{Kraus, Karl}}
\newcommand{\Kuhn}{Kuhn\index{Kuhn, Thomas S.}}
\newcommand{\Kyburg}{Kyburg\index{Kyburg, Henry E.}}
\newcommand{\Laflamme}{Laflamme\index{Laflamme, Raymond}}
\newcommand{\Lange}{Lange\index{Lange, Wolfgang}}
\newcommand{\Lahti}{Lahti\index{Lahti, Pekka J.}}
\newcommand{\Lambalgen}{Lambalgen\index{Lambalgen, Michiel van}}
\newcommand{\Landau}{Landau\index{Landau, Lev D.}}
\newcommand{\Landauer}{Landauer\index{Landauer, Rolf}}
\newcommand{\Laplace}{Laplace\index{Laplace, Pierre Simon}}
\newcommand{\Laurikainen}{Laurikainen\index{Laurikainen, Kalervo V.}}
\newcommand{\Leibniz}{Leibniz\index{Leibniz, Gottfried W.}}
\newcommand{\Levin}{Levin\index{Levin, Leonid}}
\newcommand{\Lieb}{Lieb\index{Lieb, Elliott H.}}
\newcommand{\Lifshitz}{Lifshitz\index{Lifshitz, Evgenii M.}}
\newcommand{\Lindblad}{Lindblad\index{Lindblad, G\"oran}}
\newcommand{\Lloyd}{Lloyd\index{Lloyd, Seth}}
\newcommand{\Lo}{Lo\index{Lo, Hoi-Kwong}}
\newcommand{\Mabuchi}{Mabuchi\index{Mabuchi, Hideo}}
\newcommand{\Macchiavello}{Macchiavello\index{Macchiavello, Chiara}}
   \newcommand{\Hideo}{Hideo\index{Mabuchi, Hideo}}
\newcommand{\MacKinnon}{MacKinnon\index{MacKinnon, Edward}}
\newcommand{\Mann}{Mann\index{Mann, Ady}}
\newcommand{\MartinLof}{Martin-L\"of\index{Martin-L\"of, Per}}
\newcommand{\Mayers}{Mayers\index{Mayers, Dominic}}
\newcommand{\McCartney}{McCartney\index{McCartney, Paul}}
\newcommand{\Mermin}{Mermin\index{Mermin, N. David}}
\newcommand{\Merzbacher}{Merzbacher\index{Merzbacher, Eugen}}
\newcommand{\Meyer}{Meyer\index{Meyer, David A.}}
\newcommand{\Minnelli}{Minnelli\index{Minnelli, Liza}}
\newcommand{\Mises}{Mises\index{Mises, Richard von}}
\newcommand{\Mittelstaedt}{Mittelstaedt\index{Mittelstaedt, Peter}}
\newcommand{\Mor}{Mor\index{Mor, Tal}}
\newcommand{\Muller}{M\"uller\index{M\"uller, Hermann}}
\newcommand{\Murdoch}{Murdoch\index{Murdoch, Dugald}}
\newcommand{\Neapolitan}{Neapolitan\index{Neapolitan, R. E.}}
\newcommand{\Newton}{Newton\index{Newton, Isaac}}
\newcommand{\Nicholson}{Nicholson\index{Nicholson, Jeffrey W.}}
\newcommand{\Nielsen}{Nielsen\index{Nielsen, Michael A.}}
\newcommand{\Niewood}{Niewood\index{Niewood, Gerry}}
\newcommand{\Niu}{Niu\index{Niu, Chi-Sheng}}
\newcommand{\Ochs}{Ochs\index{Ochs, Wolfgang}}
\newcommand{\Omnes}{Omn\`es\index{Omn\`es, Roland}}
\newcommand{\Oppenheimer}{Oppenheimer\index{Oppenheimer, J. Robert}}
\newcommand{\Orwell}{Orwell\index{Orwell, George}}
\newcommand{\Pagels}{Pagels\index{Pagels, Heinz R.}}
\newcommand{\Pais}{Pais\index{Pais, Abraham}}
\newcommand{\Pauli}{Pauli\index{Pauli, Wolfgang}}
\newcommand{\Pearle}{Pearle\index{Pearle, Philip}}
\newcommand{\Peierls}{Peierls\index{Peierls, Rudolf}}
\newcommand{\Peirce}{Peirce\index{Peirce, Charles Sanders}}
\newcommand{\Penrose}{Penrose\index{Penrose, Roger}}
\newcommand{\Peres}{Peres\index{Peres, Asher}}
   \newcommand{\Asher}{Asher\index{Peres, Asher}}
\newcommand{\Petersen}{Petersen\index{Petersen, Aage}}
\newcommand{\Piaget}{Piaget\index{Piaget, Jean}}
\newcommand{\Pierce}{Pierce\index{Pierce, John R.}}
\newcommand{\Pike}{Pike\index{Pike, Rob}}
\newcommand{\Pitowsky}{Pitowsky\index{Pitowsky, Itamar}}
\newcommand{\Plaga}{Plaga\index{Plaga, Rainer}}
\newcommand{\Planck}{Planck\index{Planck, Max}}
\newcommand{\Plath}{Plath\index{Plath, Sylvia}}
\newcommand{\Plato}{Plato\index{Plato}}
\newcommand{\Plenio}{Plenio\index{Plenio, Martin B.}}
\newcommand{\Plotnitsky}{Plotnitsky\index{Plotnitsky, Arkady}}
\newcommand{\Podolsky}{Podolsky\index{Podolsky, Boris}}
\newcommand{\Poincare}{Poincar\'e\index{Poincar\'e, Henri}}
\newcommand{\Polzik}{Polzik\index{Polzik, Eugene S.}}
\newcommand{\Popescu}{Popescu\index{Popescu, Sandu}}
\newcommand{\Popper}{Popper\index{Popper, Karl R.}}
\newcommand{\Preskill}{Preskill\index{Preskill, John}}
\newcommand{\Ptolemy}{Ptolemy\index{Ptolemy}}
\newcommand{\Quine}{Quine\index{Quine, Willard V.}}
\newcommand{\Ralph}{Ralph\index{Ralph, Timothy C.}}
\newcommand{\Rasco}{Rasco\index{Rasco, B. Charles}}
\newcommand{\Reichenbach}{Reichenbach\index{Reichenbach, Hans}}
\newcommand{\Renes}{Renes\index{Renes, Joseph}}
\newcommand{\Revzen}{Revzen\index{Revzen, M.}}
\newcommand{\Rimini}{Rimini\index{Rimini, Alberto}}
\newcommand{\Robb}{Robb\index{Robb, Daniel}}
\newcommand{\Rorty}{Rorty\index{Rorty, Richard}}
\newcommand{\Rosen}{Rosen\index{Rosen, Nathan}}
\newcommand{\Rosenfeld}{Rosenfeld\index{Rosenfeld, L\'eon}}
\newcommand{\Rovelli}{Rovelli\index{Rovelli, Carlo}}
\newcommand{\Ruskai}{Ruskai\index{Ruskai, Mary Beth}}
\newcommand{\Savage}{Savage\index{Savage, Leonard J.}}
\newcommand{\SchackD}{Dorothee\index{Schack, Dorothee}}
\newcommand{\Schack}{Schack\index{Schack, R\"udiger}}
   \newcommand{\Ruediger}{R\"udiger\index{Schack, R\"udiger}}
\newcommand{\Schilpp}{Schilpp\index{Schilpp, P. A.}}
\newcommand{\Schopenhauer}{Schopenhauer\index{Schopenhauer, Arthur}}
\newcommand{\Schroedinger}{Schr\"odinger\index{Schr\"odinger, Erwin}}
\newcommand{\Schumacher}{Schumacher\index{Schumacher, Benjamin W.}}
\newcommand{\Schumann}{Schumann\index{Schumann, Robert H.}}
\newcommand{\Sciama}{Sciama\index{Sciama, Dennis W.}}
\newcommand{\Scruggs}{Scruggs\index{Scruggs, Earl}}
\newcommand{\Petra}{Petra\index{Scudo, Petra F.}}
\newcommand{\Schweber}{Schweber\index{Schweber, Sylvan S.}}
\newcommand{\Seife}{Seife\index{Seife, Charles}}
\newcommand{\Shannon}{Shannon\index{Shannon, Claude E.}}
\newcommand{\Shenker}{Shenker\index{Shenker, Orly R.}}
\newcommand{\Shimony}{Shimony\index{Shimony, Abner}}
\newcommand{\Shor}{Shor\index{Shor, Peter W.}}
\newcommand{\SimonB}{Simon\index{Simon, Barry}}
\newcommand{\Simon}{Simon\index{Simon, Paul}}
\newcommand{\Simone}{Simone\index{Simone, Nina}}
\newcommand{\Sipe}{Sipe\index{Sipe, John E.}}
\newcommand{\Smith}{Smith\index{Smith, Adrian F. M.}}
\newcommand{\Smolin}{Smolin\index{Smolin, John A.}}
\newcommand{\SmolinL}{Smolin\index{Smolin, Lee}}
\newcommand{\Sobottka}{Sobottka\index{Sobottka, Stanley}}
\newcommand{\Socrates}{Socrates\index{Socrates}}

\newcommand{\Sommerfeld}{Sommerfeld\index{Sommerfeld, Karl}}
\newcommand{\Sorensen}{S{\o}rensen\index{S{\o}rensen, Jens L.}}
\newcommand{\Specker}{Specker\index{Specker, Ernst}}
\newcommand{\Spekkens}{Spekkens\index{Spekkens, Robert W.}}
\newcommand{\Spinoza}{Spinoza\index{Spinoza, Baruch}}

\newcommand{\Starkman}{Starkman\index{Starkman, Glenn D.}}
\newcommand{\Stapp}{Stapp\index{Stapp, Henry P.}}
\newcommand{\Steane}{Steane\index{Steane, Andrew M.}}
\newcommand{\Steiner}{Steiner\index{Steiner, Rudolf}}
\newcommand{\Stevens}{Stevens\index{Stevens, Wallace}}
\newcommand{\Styer}{Styer\index{Styer, Daniel}}
\newcommand{\Suppes}{Suppes\index{Suppes, Patrick}}
\newcommand{\Tegmark}{Tegmark\index{Tegmark, Max}}
\newcommand{\Terhal}{Terhal\index{Terhal, Barbara M.}}

\newcommand{\Thacker}{Thacker\index{Thacker, William D.}}
\newcommand{\Thapliyal}{Thapliyal\index{Thapliyal, Ashish V.}}
\newcommand{\Turchette}{Turchette\index{Turchette, Quentin A.}}
\newcommand{\Turing}{Turing\index{Turing, Alan}}
\newcommand{\Uffink}{Uffink\index{Uffink, Jos}}
\newcommand{\Uhlhorn}{Uhlhorn\index{Uhlhorn, Ulrich}}
\newcommand{\Uhlmann}{Uhlmann\index{Uhlmann, Armin}}
\newcommand{\Vaidman}{Vaidman\index{Vaidman, Lev}}
\newcommand{\Vidal}{Vidal\index{Vidal, Guifre}}
\newcommand{\Vitanyi}{Vitanyi\index{Vitanyi, Paul}}
\newcommand{\Wallace}{Wallace\index{Wallace, David}}
\newcommand{\Wallach}{Wallach\index{Wallach, Nolan}}
\newcommand{\Walther}{Walther\index{Walther, Herbert}}
\newcommand{\Washington}{Washington\index{Washington, Dinah}}
\newcommand{\Waskan}{Waskan\index{Waskan, Jonathan A.}}
   \newcommand{\Jon}{Jon\index{Waskan, Jonathan A.}}
\newcommand{\Weaver}{Weaver\index{Weaver, Leslie}}
\newcommand{\Weber}{Weber\index{Weber, Tullio}}
\newcommand{\Weinfurter}{Weinfurter\index{Weinfurter, Harald}}
\newcommand{\Westmoreland}{Westmoreland\index{Westmoreland, Michael D.}}
\newcommand{\Weyl}{Weyl\index{Weyl, Hermann}}
\newcommand{\Wheeler}{Wheeler\index{Wheeler, John Archibald}}
\newcommand{\Whitaker}{Whitaker\index{Whitaker, Andrew}}
\newcommand{\Wiesner}{Wiesner\index{Wiesner, Stephen J.}}
\newcommand{\Wigner}{Wigner\index{Wigner, Eugene P.}}
\newcommand{\Wocjan}{Wocjan\index{Wocjan, Pawel}}
\newcommand{\Wolf}{Wolf\index{Wolf, Stefan}}
\newcommand{\Wolpert}{Wolpert\index{Wolpert, David H.}}
\newcommand{\Wootters}{Wootters\index{Wootters, William K.}}
   \newcommand{\Bill}{Bill\index{Wootters, William K.}}
\newcommand{\Yablonovitch}{Yablonovitch\index{Yablonovitch, Eli}}
\newcommand{\Zajonc}{Zajonc\index{Zajonc, Arthur}}
\newcommand{\Zbinden}{Zbinden\index{Zbinden, Hugo}}
\newcommand{\Zeilinger}{Zeilinger\index{Zeilinger, Anton}}
   \newcommand{\Anton}{Anton\index{Zeilinger, Anton}}
\newcommand{\Zurek}{Zurek\index{Zurek, Wojciech H.}}

\newtheorem{mermin}{Merminition}
\newtheorem{asher}{Asherism}
\newtheorem{schack}{Schackcosm}
\newtheorem{wootters}{Woottersism}
\newtheorem{herb}{Herbal Treatment}
\newtheorem{kent}{Kentism}
\newtheorem{baker}{Bakercise}
\newtheorem{jeff}{El Jeffy}
\newtheorem{greg}{Comerism}
\newtheorem{charlie}{Bennettism}
\newtheorem{waskan}{Waskanism}
\newtheorem{brun}{Brunism}
\newtheorem{caves}{Cavesism}

\newtheorem{barnum}{Barnumism}
\newtheorem{preskill}{Preskillism}
\newtheorem{bub}{Bubism}
\newtheorem{benioff}{Benioffer}
\newtheorem{conj}{Conjecture}
\newtheorem{schumann}{Schumannism}
\newtheorem{shimony}{Shimonyism}
\newtheorem{renes}{Renesism}
\newtheorem{ruskai}{Ruskai-ism}
\newtheorem{folse}{Folse-ism}
\newtheorem{bilodeau}{Bilodeau-ism}

\newcommand{\bv}{\begin{verse}}
\newcommand{\ev}{\end{verse}}
\newcommand{\be}{\begin{equation}}
\newcommand{\ee}{\end{equation}}
\newcommand{\bea}{\begin{eqnarray}}
\newcommand{\eea}{\end{eqnarray}}
\newcommand{\bq}{\begin{quotation}}
\newcommand{\eq}{\end{quotation}}
\newcommand{\bdm}{\begin{mermin}\protect$\!\!${\em\bf :}$\;\;$}
\newcommand{\edm}{\end{mermin}}
\newcommand{\bap}{\begin{asher}\protect$\!\!${\em\bf :}$\;\;$}
\newcommand{\eap}{\end{asher}}
\newcommand{\brs}{\begin{schack}\protect$\!\!${\em\bf :}$\;\;$}
\newcommand{\ers}{\end{schack}}
\newcommand{\bbw}{\begin{wootters}\protect$\!\!${\em\bf :}$\;\;$}
\newcommand{\ebw}{\end{wootters}}
\newcommand{\bhbe}{\begin{herb}\protect$\!\!${\em\bf :}$\;\;$}
\newcommand{\ehbe}{\end{herb}}
\newcommand{\bak}{\begin{kent}\protect$\!\!${\em\bf :}$\;\;$}
\newcommand{\eak}{\end{kent}}
\newcommand{\bhba}{\begin{baker}\protect$\!\!${\em\bf :}$\;\;$}
\newcommand{\ehba}{\end{baker}}
\newcommand{\bjn}{\begin{jeff}\protect$\!\!${\em\bf :}$\;\;$}
\newcommand{\ejn}{\end{jeff}}
\newcommand{\bgc}{\begin{greg}\protect$\!\!${\em\bf :}$\;\;$}
\newcommand{\egc}{\end{greg}}
\newcommand{\bcb}{\begin{charlie}\protect$\!\!${\em\bf :}$\;\;$}
\newcommand{\ecb}{\end{charlie}}
\newcommand{\bjw}{\begin{waskan}\protect$\!\!${\em\bf :}$\;\;$}
\newcommand{\ejw}{\end{waskan}}
\newcommand{\btb}{\begin{brun}\protect$\!\!${\em\bf :}$\;\;$}
\newcommand{\etb}{\end{brun}}
\newcommand{\bcc}{\begin{caves}\protect$\!\!${\em\bf :}$\;\;$}
\newcommand{\ecc}{\end{caves}}
\newcommand{\bhb}{\begin{barnum}\protect$\!\!${\em\bf :}$\;\;$}
\newcommand{\ehb}{\end{barnum}}
\newcommand{\bjp}{\begin{preskill}\protect$\!\!${\em\bf :}$\;\;$}
\newcommand{\ejp}{\end{preskill}}
\newcommand{\bjb}{\begin{bub}\protect$\!\!${\em\bf :}$\;\;$}
\newcommand{\ejb}{\end{bub}}
\newcommand{\bpb}{\begin{benioff}\protect$\!\!${\em\bf :}$\;\;$}
\newcommand{\epb}{\end{benioff}}
\newcommand{\bconj}{\begin{conj}\protect$\!\!${\em\bf :}$\;\;$}
\newcommand{\econj}{\end{conj}}
\newcommand{\brschu}{\begin{schumann}\protect$\!\!${\em\bf :}$\;\;$}
\newcommand{\erschu}{\end{schumann}}
\newcommand{\bas}{\begin{shimony}\protect$\!\!${\em\bf :}$\;\;$}
\newcommand{\eas}{\end{shimony}}
\newcommand{\bjr}{\begin{renes}\protect$\!\!${\em\bf :}$\;\;$}
\newcommand{\ejr}{\end{renes}}
\newcommand{\bbr}{\begin{ruskai}\protect$\!\!${\em\bf :}$\;\;$}
\newcommand{\ebr}{\end{ruskai}}
\newcommand{\bhf}{\begin{folse}\protect$\!\!${\em\bf :}$\;\;$}
\newcommand{\ehf}{\end{folse}}
\newcommand{\bdb}{\begin{bilodeau}\protect$\!\!${\em\bf :}$\;\;$}
\newcommand{\edb}{\end{bilodeau}}

\title{\huge Notes on a Paulian Idea\\
\bigskip \Large\it Foundational, Historical, Anecdotal and Forward-Looking\\
Thoughts on the Quantum \bigskip\bigskip\bigskip\bigskip \\
\normalsize \rm Selected Correspondence, 1995--2001\\
\bigskip 
\bigskip\bigskip \bigskip\bigskip}

\author{Christopher A. Fuchs\medskip\\
\footnotesize Computing Science Research Center
\\
\footnotesize Bell Labs, Lucent Technologies
\\
\footnotesize Room 2C-420, 600--700 Mountain Ave.
\\
\footnotesize Murray Hill, New Jersey 07974, U.S.A.\medskip\\
\bigskip\bigskip\bigskip\bigskip\bigskip
\\
\vspace{1in}}

\date{10 May 2001}

\begin{document}

\pagenumbering{roman}

\maketitle

\thispagestyle{empty} \vspace*{3.7in}

\begin{center}
\copyright~2001, Christopher A. Fuchs
\end{center}

\pagebreak

\vspace*{1.5in}

\begin{center}
\Large \bf Introduction
\end{center}

\bq
This document is the first installment of three in the {\sl Cerro
Grande Fire Series}.  The Cerro Grande Fire left many in the Los
Alamos community acutely aware of the importance of backing up the
hard drive. I could think of no better instrument for the process
than LANL itself. This is a collection of letters written to various
friends and colleagues (several of whom regularly circuit this
archive), including Howard {\Barnum}, Paul {\Benioff}, Charles
{\Bennett}, Herbert {\Bernstein}, Doug {\Bilodeau}, {\Gilles}
{\Brassard}, Jeffrey {\Bub}, {\Carl}ton {\Caves}, {\Greg}ory
{\Comer}, Robert {\Griffiths}, Adrian {\Kent}, Rolf {\Landauer},
{\Hideo} {\Mabuchi}, David {\Mermin}, David {\Meyer}, Michael
{\Nielsen}, {\Asher} {\Peres}, John {\Preskill}, Joseph {\Renes},
Mary Beth {\Ruskai}, {\Ruediger} {\Schack}, Robert {\Schumann}, Abner
{\Shimony}, William {\Wootters}, {\Anton} {\Zeilinger}, and many
others.

In a way, I hope this book evokes images of the kind of dusty,
pipe-smoke infused gems one sometimes finds in the far corner of a
used bookstore---something not unlike the copy of William {\James}'s
collected letters I once owned.  It takes a funny person to read such
a book: one who is willing to dig in the far corner. But I would not
want that imagery taken too far. For despite its nod to the
happenstance that brought it about, this is not a book about the
past, but about our open future. Its singular theme is the quantum.
Without exception, every letter in the book is devoted to coming to
grips with and extolling the virtues of our quantum world. The
content ranges from the foundational, to the historical, to the
anecdotal, but every piece sings or whispers of the quantum. For
myself, I see some of the letters as my best efforts to date at
defining a vague thought that keeps creeping into my mind---{\it the
Paulian idea}. To the extent I have communicated its faint shadow to
my correspondents and seen a head turn, it seemed worthwhile to try
to give it more life. The idea pleads to be made precise. But for
this, it must possess the souls of more than me.
\eq

\pagebreak

\begin{center}
\bf \LARGE Foreword
\end{center}
\medskip

\bq
\noindent On September 18, 1996 I received email that began
\begin{center}
\parbox{4.85in}{\small \tt \noindent Dear Dr.\ Mermin,\smallskip \\
I encountered your ``Ithaca Interpretation'' paper this morning on \\
the quant-ph archive \ldots\ and, I must say, I've been walking
around \\ with a nice feeling since.  There are some things in it
that I \\ like very much!}
\end{center}

Although I didn't really know my correspondent, it was clear from
what followed that he had thought hard --- probably harder than I
had --- about many of the matters I was trying to address.  His
scholarship and intelligence were so evident that I started walking
around with a nice feeling myself at having received unsolicited
praise from so thoughtful a source.

Little did I know that those heartwarming words of appreciation were
just bait to lure me into a long critical exchange which, to my
pleasure and enlightenment, has been going on ever since.  A part of
my subsequent education (there have also been several invariably
instructive and delightful real-life meetings) can be found in
Chapter 18 below.

If Chris Fuchs (rhymes with ``books'') did not exist then God would
have been remiss in not inventing him.  Foundations of quantum
mechanics is a unique blend of poetry and analysis.  Without the
poetic vision the analysis tends to chase its own tail into more and
more convoluted realms of intricate triviality.  Without the analysis
the poetry easily degenerates into self-indulgent doggerel.  To
achieve an adequate understanding of how quantum mechanics captures
our efforts to express the ``relations between the manifold aspects
of our experience'' [Bohr] requires both the poetry and the analysis.
Analysis is needed to pin down the structure of the relations; poetry
is required to characterize how they reflect the experience.  Nobody
today writing about quantum mechanics combines poetry and analysis to
better effect than Chris Fuchs.  By deliberate authorial choice, the
500 pages that follow are long on poetry and short on analysis, but a
search on ``Fuchs'' will uncover in these Archives many examples
where the balance is more even.

The past decade has seen the growth of intense interest in
applications of quantum mechanics to information processing, brought
about by its deep intellectual richness, in fortuitous but
sociologically significant resonance with our cultural obsession with
keeping secrets.  Chris Fuchs is the conscience of the field.  He
never loses sight of the real aim of these pursuits, and if you
yourself thought it had to do with secure data transmission, RSA code
cracking, fast searches, fighting decoherence, concocting ever more
ingenious tricks, and such, you should look up from your beautiful
algorithms or candidate qubits for a few hours every now and then to
browse through these pages.

The real issue is nothing less than how you and I can each construct
a representation of the manifold aspects of our individual
experiences (loosely known as a world), and the constraints that my
representation imposes on yours, and vice-versa.  By focusing
explicitly on the strange information-processing capabilities
inherent in the quantum mechanical description of physical reality,
the new discipline of quantum information offers an opportunity to
put on a sound foundation what was only hinted at in the convoluted
prose of Bohr, the facile sensationalism of Heisenberg, the
aphorisms of Pauli, and the poetic mysticism of Schr\"odinger.  If
it hasn't occurred to you that this is the real justification for
your quantum information-theoretic pursuits, then you owe it to
yourself to pause and peruse these pages.

Those who do not come at the subject from the perspective of quantum
information theory may well be irritated or even outraged by some of
the views expressed herein about their fondest notions.  But you
would have to be an exceptionally conservative Bohmian, a dry-as-dust
consistent historian, a stubbornly literal-minded dynamical
collapsian, a supremely surrealistic many-worldsian, or an
irresponsibly post-modern correlations-without-correlatan not to be
at least entertained and amused and even, every now and then,
instructed by this highly readable commingling of scholarship,
intellectual passion, philosophical vision, and biographical
glimpses into the daily life of a young scientist.

Indeed, one of the difficulties I have encountered at a busy time in
writing this Foreword with dispatch, is the distracting presence, in
another window on my screen, of the text itself, constantly luring me
away from my own painful efforts at expression, into its charming,
and by no means fully explored, byways.

The death of letters as a high literary form brought about by the
telephone turns out to have been only a lengthy coma --- a 20th
century aberration.  Clearly the rise of email had, from the
beginning, the potential to resuscitate the patient.  Now we have an
existence proof. The thought-provoking pages that follow, which can
either be read like a Nabokov novel, or dipped into from time to
time, like a collection of poems or short stories, gloriously
provide an early 21st century demonstration that the art form is
once again alive and well --- and also, of course, that there remain
profound questions to ask and to strive to answer about the real
meaning of quantum mechanics.\bigskip

\begin{flushright}
\parbox{2in}{N. David Mermin\\
Ithaca, New York\\
May 8, 2001}
\end{flushright}
\eq

\pagebreak


\begin{center}
\baselineskip=12pt
\parbox{3.0in}{\baselineskip=12pt
DISCLAIMERS:  \medskip\\
{\bf I.} This document represents a unique, and hopefully
entertaining, method to communicate some happy thoughts on the
quantum. For precisely this reason, however, it carries a great
danger to my friends.  It is after all a collection of
correspondence. There are two things that should not be mistaken: 1)
The potential of my memory to be faulty when reporting the views of
others, and 2) that the quotes taken from my correspondents were
composed in anything other than a casual manner for {\it private\/}
use only. With regard to the latter, I assert the right of my
correspondents to deny---without apologies!---that their quotes
represent accurate accounts of their thoughts. I have tried to guard
against misrepresentation by keeping the number of quotes and
correspondent-replies to a minimum:  The ones that are used, are used
mainly as springboards for {\it my\/} tendentiousness.
\medskip\\
{\bf II.} Various deletions of text have been made to the original
letters. The purpose of the vast majority of these is to spare the
reader of the ``merely personal'' in my life. A smaller fraction are
for the sake of protecting the innocent, protecting the
correspondents, and protecting the illusion that I am good-natured.
The same holds as well for a small number of explicit changes of
phrase (in my own writings, {\it never\/} the correspondents). In
most cases, I have tried to make the process look as seamless as
possible, with no evidence that the text may have been otherwise. In
my own writings, bare ellipses should be interpreted as punctuation;
bracketed ellipses indicate true editorial changes.
\medskip\\
{\bf III.} There is no claim that all the ideas presented here are
coherent. The hope is instead that the incoherent ones will earn
their keep by their entertainment value.}
\end{center}

\medskip

\begin{center}
\baselineskip=12pt
\parbox{3.0in}{\baselineskip=12pt
ACKNOWLEDGEMENTS: \medskip\\
I thank {\Todd} {\Brun}, Jeff {\Bub}, {\Carl} {\Caves}, Steven van
{\Enk}, and David {\Mermin} for the subtle influence that gave this
project life.  I thank my correspondents for the use of their
quotes. Most importantly, I thank my wife {\Kiki} for allowing me to
type in the middle of the night.}
\end{center}

\pagebreak

\bq
\noindent \small I want you to frame a question, as sharp and clear as
possible---one to which you do not yet know the answer, but
desperately want to know, and expect someday to know. Pretend to be
David Hilbert.  The Millennium is approaching.  Issue a challenge to
the quantum theorists of the 21st century.  List the key questions
they should seek to answer.  Hard questions, but not hopelessly
hard, questions whose answers could transform our understanding of
how the physical world works. {\it I need to know what the question
is. Then, perhaps, I can be more engaged in the search for the
answer.}
%
\hspace{0.2in}
--- {\rm John {\Preskill}}
\eq
\vspace{0.10in}

\noindent {\large I give it my best shot:}
\begin{center}
\begin{tabular}{|ll||ll|}
\hline
\multicolumn{4}{|c|}{} \\
\multicolumn{4}{|c|}{\bf Quantum Mechanics:} \\
\multicolumn{4}{|c|}{\it The Axioms and Our Imperative!} \\
\multicolumn{4}{|c|}{} \\
\hline\hline
& & & \\
\hspace{.1in} {\bf 1.} & States correspond to density & &
   {\it Give an information theoretic reason} \hspace{.05in} \\
   & operators $\rho$ over a Hilbert space $\cal H$. &  &
   \hspace{.1in} {\it if possible!} \\
& & & \\
\hspace{.1in} {\bf 2.} & Measurements correspond to positive & & \\
   & operator-valued measures (POVMs) & &
   {\it Give an information theoretic reason}\\
   & $\{E_b\}$ on $\cal H$. &  &
   \hspace{.1in} {\it if possible!} \\
& & & \\
\hspace{.1in} {\bf 3.} & $\cal H$ is a complex vector space, & & \\
   & not a real vector space, not a & &
   {\it Give an information theoretic reason}\\
   & quaternionic module. &  &
   \hspace{.1in} {\it if possible!} \\
& & & \\
\hspace{.1in} {\bf 4.} & Systems combine according to the tensor & & \\
   & product of their separate vector & &
   {\it Give an information theoretic reason}\\
   & spaces, ${\cal H}_{\rm\scriptscriptstyle AB}=
   {\cal H}_{\rm\scriptscriptstyle A}\otimes
   {\cal H}_{\rm\scriptscriptstyle B}$. &  &
   \hspace{.1in} {\it if possible!} \\
& & & \\
\hspace{.1in} {\bf 5.} & Between measurements, states evolve & & \\
   & according to trace-preserving completely & &
   {\it Give an information theoretic reason}\\
   & positive linear maps. &  &
   \hspace{.1in} {\it if possible!} \\
& & & \\
\hspace{.1in} {\bf 6.} & By way of measurement, states evolve & & \\
   & (up to normalization) via outcome- & &
   {\it Give an information theoretic reason}\\
   & dependent completely positive linear maps. \hspace{.01in} & &
   \hspace{.1in} {\it if possible!} \\
& & & \\
\hspace{.1in} {\bf 7.} & Probabilities for the outcomes & & \\
   & of a measurement obey the {\Born} rule & &
   {\it Give an information theoretic reason}\\
   & for POVMs ${\rm tr}(\rho E_b)$. & &
   \hspace{.1in} {\it if possible!} \\
& & & \\
\hline
\end{tabular}
\end{center}
\vspace{0.1in}
\bq
\noindent \small The distillate that remains---the bare piece of quantum
theory with no information theoretic significance---will be our {\it
first\/} unadorned glimpse of ``quantum reality.''  Far from being
the end of the journey, I well believe, placing this new conception
of nature in open view will be the start of the greatest adventure
yet in physics.
\eq

\pagebreak

\vspace*{1.75in}

\begin{center}
\baselineskip=12pt
\parbox{2.79in}{\baselineskip=12pt
``In the new pattern of thought we do not assume any longer the {\it
detached observer} \ldots\ but an observer who by his indeterminable
effects creates a new situation, theoretically described as a new
state of the observed system.  In this way every observation is a
singling out of a particular factual result, here and now, from the
theoretical possibilities, thereby making obvious
the discontinuous aspect of the physical phenomena.''
\\
\hspace*{\fill} --- {\it Wolfgang {\Pauli}}\\}
\end{center}

\vspace{.75in}

\begin{center}
\baselineskip=12pt
\parbox{2.77in}{\baselineskip=12pt
``Like an ultimate fact without any cause, the {\it individual\/}
outcome of a measurement is, however, in general not comprehended by
laws.  This must necessarily be the case \ldots''
\\
\hspace*{\fill} --- {\it Wolfgang {\Pauli}}\\}
\end{center}

\pagebreak

\tableofcontents

\pagebreak

\pagenumbering{arabic}

\chapter{Letters to David {\BakerD}}

\section{22 July 1996, \ ``Noodles of Nothing''}

How's your academic life?  Have you been reading any?  Do you still
subscribe to {\sl Texas Monthly}?  Or was it {\sl The Smithsonian}, I
forget which?  Right now, in my spare time, I'm reading {\sl Robert
{\Oppenheimer}:\ Letters and Recollections}.  Mostly it's a
collection of letters from when he was in college and graduate
school.  The fellow was an amazing writer: it would be hard to tell
he was a scientist if you didn't know otherwise.  Of course, my real
interest is in peering into his thoughts as he was learning quantum
mechanics; that's why I'm putting time into the book.  But it is
always good to absorb something of someone else's style.

Have you ever read anything by William {\James} (the psychologist and
philosopher---not to be confused with Henry {\JamesH})?  For some
reason I'm just fascinated with his writing style; I would love to
be able to pull the same tricks, and have it acceptable to do so. If
you get a chance, take a look at the first few pages of his essay
``The Dilemma of Determinism'' just for the style.  (You'll be able
to find it in the library in any collection of his essays, probably
with titles like {\sl The Will to Believe}; the one I have at home
is a Dover edition.)

There's a new phenomenon going on in my field: we've apparently
reached the level where it's time for some books on the subject.
While in Torino, I was approached by four people with some mention
of the book they're writing on quantum computers or quantum
information in general.  That makes a total of five books on the
drawing board that I know about personally.  I wish I were in the
group, but I'm lucky enough to be able to write a paper every now
and then.  Probably, if I'd just write a little less e-mail \ldots\
!!!

\section{27 July 1996, \ ``Life on Long Island''}

I asked whether you had read anything by William {\James}, and then
said that he should not be confused with (his brother) Henry
{\JamesH}. That's OK; I'm sure I can take your answer to mean that
you haven't read either of the two.  William was one of the great
American philosophers from the end of the last century.  He's also
known for his work in psychology.  I suppose his biggest
contribution to thought was in the founding (along with Charles
Sanders {\Peirce}) of a philosophical system known as
``pragmatism''---something I like very much, actually.  Henry was a
novelist and literary critic (and probably quite a bit more famous
in general circles than his brother).  I've never read anything by
Henry.

I got to thinking about Mr.\ {\James} again because of my last note
to you. So, I was very pleased to find a book at my favorite used
bookstore today titled {\sl The Philosophy of Henry {\JamesS}, Sr}.
I had never heard of this member of the family.  As the ``Sr.''
implies, he was William and Henry's father.  Apparently he had
something of a philosophical system of his own \ldots\ which, the
author of this book claims, is of interest in and of itself and not
chiefly because of its relation to William's ideas.  Anyway, I
snatched the book, ten (Canadian) bucks.  Also finally bought
{\Kierkegaard}'s {\sl Either/Or\/} (two volumes), and three books by
{\Piaget}: {\sl The Construction of Reality in the Child}, {\sl The
Child's Conception of the World}, and {\sl The Child's Conception of
Physical Causality}. (The philosophical mysteries of quantum theory
plague me just as always; I'm not proud, I look for clues wherever I
can!)

The thing that intrigues me about {\James} and {\Peirce} is that they
both rejected the mechanical view of the world that was the rage of
their Victorian time.  {\James}, in particular, was lead to believe
something that had quite the flavor of the ``many-worlds
interpretation of quantum mechanics''---something some of the
foolhardies of today would say is uniquely implied by the quantum
mechanical formalism. (Many-worlds quantum mechanics was introduced
in 1957 by one of {\Wheeler}'s Ph.D. students Hugh {\Everett};
{\James} died in 1910.)
\bq
\noindent
To that view, actualities seem to float in a wider sea of
possibilities from out of which they are chosen; and, somewhere,
indeterminism says, such possibilities exist, and form a part of the
truth. \\
\hspace*{\fill} --- {\James}, 1884 (address to the Harvard Divinity
Students)
\eq

I'm sure I spent too much money on books again, but it's a nasty
itch. The store I went to today is really excellent, even though
it's much smaller than our apartment here.  It's a wonder what an
owner with taste can do.  The bookstore is also especially good for
me because the owner's husband is a physicist.  He does much of the
work in purchasing their science and philosophy collections.  I got
to meet the fella today for the first time; very nice---he was
curious to know what quantum cryptography is.  One day I was talking
to his wife, and she was telling me how she had the good fortune to
see Feynman lecture once.  I said that I had never seen him, but
that I knew his graduate advisor.  She said, ``My God, he must be
old.''  I said, ``well yes,'' and then went on to explain that it
was John {\Wheeler}.  She said, ``I know of {\Wheeler}; around our
house {\Wheeler} is God!''  That's how I found out that her husband
is a physicist (particle physicist in particular).

I just poured a beer, a Sleeman's Lager.  I like it.  Actually I
like a lot of the beers up here: different ones for different
moods.  When you come up, I'll introduce you to the whole entourage
of Qu\'ebecois beers. About your coming up \ldots\ let's see, what
can I say?

I think you'll have no problem finding things to do while I'm tied
up.  The neighborhood is full of surprises (and bars), and every
museum is a metro (subway) stop away.  I do have two ``must see''
bars for you: The Mad Hatter's Library (near McGill U) and The Yoda
Den (near a smoked-meat joint).  Oh yeah, there's also The City Pub
\ldots\ it's got a nice {\Kant}ian/``quantum mechanics is a law of
thought'' feel that I like (and good cheap food to go with the beer).

\section{03 August 1996, \ ``Lonely Jazz''}

Ms Holiday is playing in the background and I've got my first cup of
the morning.  I'm still pretty much exhausted.  I ended up staying
at IBM an extra day, and the whole affair was pretty intense from
start to finish. I think we found something very, very nice \ldots\
which, of course, was expected---that's why we got together in the
first place.  The idea is simple and it's this.  Suppose Alice needs
to communicate to Bob one bit of information, either a 0 or a 1. Also
let us suppose the resources available to her for carrying out this
task are two (noisy) fiber optic cables and two photons, one for
each cable.  The bit will, in some sense, be carried in the
polarization of the photons; we allow two photons in the game to
give Alice a little redundancy to help her get past the noise. Note
that I've said nothing about how the photons are produced; I didn't
say anything about whether they came from a single source (localized
in some small region of space) or whether they are generated by two
coordinated but independent sources.  Now the question is this: can
the transmission's fidelity be helped by allowing the photons to be
generated at a single source AND combining them back together before
before Bob performs his measurement (for gathering the bit)?  That
is to say, can we increase Bob's chances of guessing the bit
correctly by first generating the photon at a single source---so the
two are ``entangled'' in a strange quantum mechanical way---and then
allowing Bob to make a measurement on the two together, a
measurement on the whole being greater than a sum of its parts?  The
answer is yes, and I find that so wonderful.

Why, you ask?  Because, it's wonderful \ldots\ period.  And a bit
unexpected too actually.  I guess the thing I really like the most
about so many of the questions we've been asking lately is that they
really put ``entanglement'' to use.  I just ran across a wonderful
quote by {\Oppenheimer} the other day, from a letter he wrote to
Fowler soon after Hahn and Strassmann discovered Uranium fission:
\bq
\noindent
The U business is unbelievable. \ldots\  What do you think?  It is I
think exciting, not in the rare way of positrons and mesotrons, but
in a good honest practical way.
\eq

The idea of entanglement has it's origin in {\Einstein}, perhaps as
early as 1930 \ldots\ though I can't recall for sure right now.  In
any case, a pretty clear statement of it and what he didn't like
about it came out in 1935 in the paper of {\Einstein}, {\Podolsky},
and {\Rosen}. (It's through Nathan {\Rosen} that I have my
``{\Einstein} Number'' 3: {\Rosen} having written many papers with
{\Einstein}, {\Peres} having written many papers with {\Rosen}, and
I having written a paper with {\Peres}.) It's existence was
something EPR considered clear-cut evidence that quantum theory
could not be a ``complete theory.'' (I.e., by this time {\Einstein}
had already given up the idea that the theory was wrong, he just
didn't think that it could be the whole story.) Since then, the
notion just pretty much stayed an oddity, only thought about for the
most part by the philosophically minded \ldots\ that is, until its
practical resurrection in quantum information theory.

By the way, I should point out that EPR were wrong in the sense that
it was finally shown in 1964 that quantum theory could not be
``completed'' in their sense.  Any such completion (that took
entanglement away) would contradict experiment---and very fine
experiments on this phenomena have been done.  What we have now is
that, not only is entanglement required for consistency with
observation, but also that it can be exploited for something
interesting (and perhaps practical).

Enough physics, right?  Ms {\Simone} is on by now, I should say.
{\Kiki} and I are about to step out for Ethiopian.  It's a place
with a lunchtime buffet; we haven't tried it before.  I had
Ethiopian in DC once and was really taken with it, but I haven't
been able to recapture the experience since then.  Perhaps the
fourth time's a charm.  Then we're gonna do a little CD shopping.

Here I'd thought that I was gonna answer all your questions about
Canadian beer and such this morning, but instead I just got carried
away with science.  Sorry about that.  I'll be back later in the day
with something more on the mundane (not the derogatory meaning of
the word, but rather ``earthly as opposed to heavenly'').

\section{17 August 1996, \ ``More Meaning''}

A lazy Saturday afternoon.  I'm listening to a new CD, {\sl Liza
{\Minnelli} (From Radio City Music Hall)}.  Having an afternoon
coffee, after an afternoon beer---a strange combination.  New York,
New York.  {\Kiki}'s lying on the bed asleep, and I'm dreaming.
Today was a day stranger than most. {\Albert} apparently had a
stroke this morning.  He sort of lost control of much of his left
side for a while, maybe 5--7 minutes.  As the day has gone he's
regained more and more control. Now everything is pretty much normal
except he's still a bit wobbly on his left hind leg and I find
myself more sentimental than usual. (Excuse me, now we have Mel
Torme.)

The things I live and breath for. {\sl Twilight Zone}, {\sl Star
Trek}, Quantum Mechanics, and an open future.  Maybe that summarizes
it all.  (It's been a long time since I've lapsed into Gertrude
Steinisms.)  The Mars rock has really set me off lately \ldots\ so
much so that I was even willing to spend \$4.00 for a {\sl Time
Magazine}. I didn't find out much more about Mars that I didn't
already know, but at least I learned that Audrey {\Hepburn} was a
notch on Jack {\KennedyJF}'s bedpost. Actually, that really
depressed me; I have quite a crush on her \ldots\ the 1961--64
version of her, that is.

Night Thoughts of a Quantum Theorist.  Tonight it's my turn to
cook.  (A chicken dish I suspect.)  The problem's not in the
seasoning, but in finding the main ingredient.  Those with little
imagination restrict themselves to games with seasonings.
Unfortunately, that's where I stand now.  It's a good thing {\Kiki}
makes me cook every Saturday; maybe I'll get the hang of it
eventually.

\section{19 August 1996, \ ``Self Promotion''}

What's with the ``cheers'' thing?  Corporate lingo, I suppose.  I
picked it up from the Brits, who always sign their letters to me that
way.  However, it's becoming a relatively general salutation in my
clique.  Depending upon my mood, I rotate it with, ``best regards,''
``best wishes,'' ``kind regards,'' ``very best wishes,'' etc.  On
special occasions, and, in particular, if the addressee is an old
German professor, I might end with something like, ``with warm good
wishes.''  But, I tell ya, that's nothing compared to some of the
curlicues I saw in the {\Oppenheimer} letter collection!

\section{01 September 1997, \ ``Mo' Investigative Work''}

Allow me to reach into the depths of your knowledge of history.  A
couple of days ago, I came across the following question in Trivial
Pursuit:  Who said, `` The victor will never be asked if he told the
truth''?  The answer was ``Adolf {\Hitler}.''  Now I would like to
pin down the actual source of that quote (i.e., from what speech or
what private conversation of {\Hitler}'s, to whom and when did he say
it). Do you think you might be able to tackle this task?  Of course
my motivation for getting this straight has to do with some silly
thoughts about quantum mechanics \ldots\ as you could have guessed.

\section{21 February 1998, \ ``Home a Short While''}

Sorry to hear about your painful taste of dharma.  Try to remember
instead that, in the end, Brahman $=$ Atman.  That which is without
is within.  I've always tended to find more solace in that anyway.

\section{02 March 1998, \ ``Long Finicky Flight''}

Well here I am again.  A long flight in front of me \ldots\ they say
10 hours, 58 minutes.  You may not get this note for quite a while,
March 10 in particular:  I doubt that I'll be connecting much if any
while in Japan (due to the difference in network protocols, etc.).
My guess is that this will turn into a long note, as it usually does
when I'm on a flight and I don't feel much like working.  But right
now I don't have a clue as to what the subject is going to be.

What can I tell you?  To a large extent I've been pretty damned
brainless the last few weeks.  I'm not completely sure why that is,
but it is.  A lot of it is ``shut down'' I'm sure.  I've got a
zillion and one things I need to be doing, more than I can possibly
handle.  So instead of tackling what I can, I tackle none.  A
corollary to that is that I watch a lot of TV now.

Only 1.5 hours into this flight, and already I have one heck of a
back ache.  Do you really want to travel so much as you say you do?
I'm still not completely sure how this summer's travels are going to
turn out.  The dates for the Torino conference are now set at June
29 -- July 19.  But it looks like I'll also be invited to a
conference in Benasque, Spain (a small village in the Pyrenees) July
5--25.  I'm not yet sure how I want to apportion the time.  Almost
certainly I will go to Spain for some time of that.  However, I
really can't say whether I'll be going to Torino.  The travel has
just gotten to be too much:  the more I travel, the more shallow my
thoughts become \ldots\ and I've got to stop that.  Last week I found
out that I had the opportunity to go to Mexico City (expense free for
the American taxpayer) in August:  I think it's the first time I've
done this, but I turned it down.

OK, maybe it's time for a little personal philosophy.  Though my
tolerance for equations is becoming less and less, I am finding that
my view of the world is becoming firmer and firmer.  Despite my own
depression, I must say that I am finding myself believing that the
world is more vibrant and {\it alive\/} than I ever have before.
When was the last time I sent you a compilation of my philosophical
ramblings?  This much I have really started to share with my friend
{\Herb} {\Bernstein}: the notion that there is a ``reality'' above
and beyond man, everlasting and eternal, is simply outdated.  It
comes from a time when science could only make progress by
extricating the human element from things; it comes from a
back-reaction to religion.  But now, with some hindsight from the
quantum revolution, it seems clear to me that the world is so much
more.  It's far more surprising than Baconian science would have us
believe, and it's far more participatory than any of the western
religions (or eastern, for that matter) ever dreamt.  The world, its
description, and the laws that govern it, are not simply there
independent of our actions.  There was a time when they were, before
complex organic molecules, but now that's not the case.  The world
and its laws seem to me to be every bit as evolutionary as life
itself.  And just as the idea of radical Darwinism becomes outdated
when one realizes that random natural selection fails to hold the
second one being can say to another, ``I love you,'' so it is with
the universe.  The world is a big pushme-pullyou.  If I could talk
to the animals \ldots\ .\@  Is that what I've been doing?

But I know you want to hear a little more philosophy.  {\Herb}
describes our explorations as trying to get at a new category.  He
calls it ``realit{\it t\/}y'' \ldots\ that is to say, reality with a
little something extra thrown in.  It describes the fact that the
world pushes back in an unpredictable way when you push on it.  And
the way it pushes depends on what you do to it.  And, finally, that
that push is not inconsequential in the least bit.  Maybe you
remember John {\Wheeler}'s ``game of twenty questions (surprise
version)''; I guess I subscribe to it more than ever.

What is this thing called language, and how does it fit into the
whole of everything I wrote you above?  I always think of Linda
Henderson and the introduction to postmodernism that she tried to
give us way back when I ask a question like this.  I don't think I
would have ever believed in 1983 that I would be thinking of her
words 15 years down the road.  Looking back on it, I have to wonder
whether she had just been making a bedtime reading of Foucault or
{\Derrida}, and had been trying to share it with us.  You probably
don't remember this, but John Simpson and I fought some of the
things she said as silliness, tooth and nail. Linda once said,
``Without language there can be no thought.''  We said, ``That's
simply ridiculous.''  Now I find myself saying to myself, ``Without
language and the collective action that it leads to (through the
demagogues, the communicated scientific ethic, the body politic, the
media, etc.), the world we see would scarcely be the same.''

But maybe that's enough for now.  I'll get back to you after
something of a nap.

\section{14 April 1998, and to {\Herb} {\Bernstein}, \
``{\Hawking} on Evolution''}

I know that you have listened to me patiently in my speculation about
Darwinism becoming more and more outdated as we are ever more able to
inject planning and purpose into evolution.  However, in case you
haven't had enough, you may be interested in listening to Stephen
{\Hawking}'s side of the same story:  it's a pretty good talk on the
whole I would say.  (I certainly don't agree with some of his
Everettista views nor his ``end of physics'' sermons, but much of
the rest of it is really good stuff.)  You can find a webcast of it
at the following site, {\tt
http://www.sun.com/newmedia/whitehouse/%
stephen\underline{\phantom{a}}hawking.html}.

\section{17 May 1998, \ ``The Electronic Itch''}

To be honest, I don't have much to say, but still I have the itch,
the need to write something down.  This is my essence as far as it
goes.

I think I am in the last two hours of my flight to Chicago.  Then an
hour and a half in the airport, and finally three more hours back to
LA.  I arrive just at the peak of rush-hour traffic; so more than
likely it'll still take me one hour from there to get home.  What a
life.

I come back from my second trip to the fatherland still impressed.
The food in this part of Germany, very near Luxembourg, was quite
different from the Bavarian style that I was getting used to.  But
still it was very, very good.  The landscape was beautiful and full
of life.  Yesterday and last night, I stayed in Mainz---an old town,
just on the river \underline{\phantom{Danube?}} (let's see if you
remember your geography)---feeling lonely and searching through the
archetypal archive in my soul.  Weird things like this make me take
a little stock in {\Jung}'s thought.  The country seems to do
something to me.

It's funny contemplating this future of ours.  I keep thinking more
and more about it.  Some of that is spurred by my now constant watch
of the exponential growth in computing power.  And I am sure some of
it comes about by my predisposition to think that we've yet to
discover any real laws of physics---the last four hundred years being
an elaborate phenomenology to fill in the gaps.  I keep feeling that
something really, really big is about to happen.  Sixteen months ago,
I bought this 133 MHz Pentium I machine with 16 megabytes of memory,
a ~1 gigabyte hard drive, $800 \times 600$ resolution screen, and a
6x CD-ROM, all for \$3000.  Right now I could buy a 266 MHz Pentium
II machine, with 64 megabytes of memory, a 4 gigabyte hard drive,
$1024 \times 768$ resolution screen, and a 24x CD-ROM, all for
\$3000. (This is much more like a factor of 4 difference, not the
factor 2 that everyone talks about in ``Moore's law.'')  Eighteen
months from now, laptop computers will be a least twice as powerful
as this one. Where is this going to lead?  Where can it lead?  It
makes me wonder; it makes me religious in a rather strange sense.  I
don't think it unreasonable to expect that it won't be long before
we'll see a full-scale reconstruction of our species.  Certainly
less than two hundred years from now.  But then what?  This question
eats at me sometimes.

\section{14 August 1999, \ ``Rampant Repine''}

And the boredom overtook us \ldots

Guess what I'm doing again?  (Buzz like a bee.)  I'm scheduled for
an early arrival in Los Angeles in a little over an hour.  What I
hate about that is that I always get my hopes up only to have them
dashed:  inevitably an early arrival leads to sitting on the runway
for extra time.  It never fails that there's another plane still
parked in the scheduled spot.  Two seats in front of me is Hugh
Grant, the actor.  (For real, escorted on and everything.)

Have you ever read Borges?  I bought a collection of his short
stories while I was living in Canada, but I hadn't really sat down
to read it until now.  I read ``The Garden of Forking Paths'' the
other night and really enjoyed it; it was quite eerie.

Lately too I've been turning my psychotic side to thinking about
acid.  Namely, how it must induce certain kinds of connections in
the brain that aren't normally there.  And similarly how it must
suspend other ones that we normally rely upon.  Is it a priori
obvious that that sort of rearrangement of the brain would be a bad
thing?  We've always been told that it is surely so, that it
suspends our function in society.  But I've been wondering what
might happen if we took a large community (whose transportation is
based on the bicycle rather than the car!) like Amsterdam, and
surreptitiously gave the residents a small dose of acid in their
water supply for something like five or ten years.  Would anything
interesting and permanent crop up.  The acid-eaters would in this
case not be isolated in society (as they always have been in the
past) but would be the complete community.  What form would that
community evolve into?  What form would their art and literature
take?  And most importantly for me---i.e., the real reason I'm
thinking about this---what form would there scientific
investigations and insights start to take?  What would they be able
to see in systematic ways that we cannot see at all?  Could they
capture in a scientific way, whole aspects of the world that we are
just blind to?  Would they in their discussions ask (fruitful)
questions about nature that would never have occurred to us?

There's a lot of reasons I've been thinking about this.  But one
certainly takes its roots in a slide that I use in some of my talks.
It's a chart of the raw genetic differences percentage-wise between
various species of animals.  I have both man and dog marked with a
yellow marker.  The wonderful thing is that there is only an 11\%
difference between the two species!!  We're so accustomed to
thinking that mankind is the pinnacle of creation---and surely we
are---but how this hints at a wonderful new slant on the story. This
difference in intelligence and understanding that we (subjectively)
suppose as almost infinite, might not be infinite at all.  In
reality it might itself only be 11\%.  How wonderful that would be!

\section{28 July 2000, \ ``The Role of Registered Phenomena''}

I have been thinking a lot about John {\Wheeler}'s registered
phenomena lately.  And, I have decided that on this count John was
just wrong. There simply is no real world ``out there'' completely
independent of our interventions into it.  Now that our kind is
here, we are an integral component of that which constitutes nature
as a whole. John had hoped that there was still something of a
bedrock to the world---most likely a trapping from his training in
classical physics---and it was to that role he assigned the
``registered phenomenon.''  I am forced instead to ask myself over
and over which aspects of nature can we at least treat as
effectively real, if not real in any absolute sense.  For certainly
we see an independent world around us:  We stub our toes on rocks
when we least expect it.

My opinion is lately this.  We tend to call something real when it is
beyond our control to change it.  But that is only a subjective
state of affairs, one controlled largely by our lack of information
or technology, and sometimes, sadly, by our lack of genuine will.  As
{\Archimedes} told the king, ``Give me firm support and I shall move
the earth.''  When we know all that we can know about a physical
system, modern quantum mechanics tells us that we can mold it to our
purposes. And, as such, it can no longer retain an independent
reality.  But when we have less than maximal knowledge of it, degree
by degree, it becomes every bit as real as the rocks and trees about
us.  By this account, the independent reality of our world comes
about solely from the mystery it holds for us.

Congratulations on the birth of your daughter.  She was at the
moment of her birth, and will be from that time forward, a mystery
with which you must reckon.  She will likely be the most real thing
to ever arise in your life, for you will never know her completely.
Every day she will bring you a new surprise.  And when, years from
now, she is on her own and away from your protection, you can look
back and know that you partook in creation in the most absolute of
senses.

\chapter{Letters to Howard {\BakerH}}

\section{29 September 1997, \ ``Quantum Information
Questions''}

I think I'll answer your questions somewhat out of the order in which
you asked them.  I hope you don't mind: it may fit what I'm thinking
more closely that way.

\bhba
How important do you think these papers are to Quantum Information
and the quantum technologies (computing, communication, cryptography
and teleportation)? Are they a major step forward? And if so, why?
\ehba

In my opinion, the {\Holevo}-{\Schumacher}-{\Westmoreland} (HSW)
result was the most technically difficult one in Quantum Information
Theory to be proved last year.  In the long run, the papers are
certain to be classics.  As far as importance goes, it was certainly
one of the top two results (the other being a consortium of papers
to do with fault-tolerant computation on quantum computers).

The HSW papers solve a long-standing problem that was on {\Holevo}'s
mind as early as 1978.  {\Holevo} actually had a paper conjecturing
this result in 1979, but somehow it escaped all of our attention
until Richard {\Jozsa} and I met him last September in Japan.  He
told us of the old paper then.  (I've done a citation search since
then, and, believe it or not, this paper had only been cited five
times in its life.  Stranger still, three of those citations were
for the wrong reason---the result they were citing was in his 1973
paper!)  Once {\Holevo} knew of the earlier {\Hausladen}, et.\ al.,
result---which he learned of at the Japan conference, 25--30
September 1996---things must have fallen into place pretty quickly
\ldots\ because his paper appeared on the Los Alamos e-print archive
14 November 1996.

As best I can piece it together {\Schumacher} and {\Westmoreland}
must have found the result almost simultaneously with {\Holevo}.
They saw his paper on the archive as they were in the course of
writing theirs. They were disheartened a bit, but felt that the
methods of their proof were sufficiently different that they would
go ahead and finish their writing and submit the paper anyway.
(They related this to me when we met at the PhysComp conference in
Boston around Nov 20.)

Is this a major step forward?  Yes, because I think this is the first
paper that shows an effective way to deal with asymptotic problems to
do with quantum communication channels.  In a way, this result does
for classical communication on quantum channels what {\Shannon} did
for purely classical channels in 1948.  (Though it should be noted
that there was no rigorous proof of {\Shannon}'s ``theorem'' until
1954---the first was due to {\Feinstein}.)  It asks, given that I am
building messages out of some finite alphabet of signals, what is
the best use I can make of a noisy communication channel?  And it
answers it, though this time we had to wait 17 years for a rigorous
proof. (Actually that last remark may not be all that fair:
{\Holevo}'s 1979 conjecture was in no way near the stage of
completion that {\Shannon}'s 1948 stuff was \ldots\ and
certainly---for whatever reason---it didn't draw the same amount of
attention. Also, without the whole apparatus of classical
information theory, the proof of this new theorem wouldn't have been
imagined, nor would it have been conjectured.)

\bhba
What information do you think Quantum Information and these papers
give us about the nature of the quantum world.
\ehba

It's pretty clear that Quantum Information Theory is giving us a load
of new conceptual tools with which to explore the essence of quantum
phenomena.  In fact, I think a fairly major attitude change has come
along with Quantum Information Theory (i.e., in the last ten years or
so).  Previous to that, certain basic facts of quantum theory---such
as quantum indeterminism and {\Bell} inequality violations---were
viewed as mostly negative principles.  The {\Heisenberg} uncertainty
relations were seen as these darned things that just got in the way
of our precision measurements.  The {\Bell} inequality violations
were similarly these darned things that got in the way of our
constructing a comfortable-feeling hidden-variable theory to go
underneath quantum mechanics.  Now these things are turned on their
head:  quantum indeterminism is something really good (quantum
cryptography), quantum entanglement---i.e., the thing giving rise to
{\Bell} inequality violations---is something really good (quantum
teleportation), and so on.  This new attitude (and the tools it
leads us to search for) gives us a way of looking at quantum theory
in a way that wasn't available before \ldots\ and that can only lead
to a deeper understanding of nature.  Maybe, in some ways, it's
teaching us to broaden or diversify our notions of the elementary
essences of the world:  now along with energy, etc., perhaps we
should include ``indeterminism'' and ``entanglement.'' (I tried to
say some of these things---with respect to entanglement in
particular---in a slightly poetic way to a friend the other day;
perhaps I'll attach that letter to the bottom of this file \ldots\
for whatever it's worth.)

John {\Wheeler} (a long-time professor at Princeton and later at the
University of Texas) was a great advocate that information theory had
something deep to say about quantum mechanics.  He always exuded this
air of urgency about him:  ``It is imperative for us to understand
the meaning of quantum mechanics in the grand scheme of things! We
must make as many mistakes as we can, as fast as we can, so that we
can hope to obtain an understanding within our lifetime!'' The last
time I saw him was in 1994 at a little conference in Santa Fe---the
one, in fact, where Peter {\Shor}'s factoring algorithm was
announced. (Actually, come to think of it, I saw him one time later
that same year \ldots\ at his 83rd birthday festschrift.)  Anyway,
at Santa Fe, {\Wheeler} gave a talk (probably titled ``How Come the
Quantum?'')\ that he closed with a slide depicting {\Planck}'s head
(maybe etched on a coin or something).  I remember, he said
(roughly), ``In 1900 {\Planck} discovered the quantum.  The end of
the century is drawing near.  We only have six years left to
understand why it is that it's here. Wouldn't that be a tribute?!''

I think there is something of a feeling in the Quantum Information
community that this may not be complete hogwash.  In ways, there
really is an exciting, sort of revolutionary feel to it all.  I wish
I could put my finger on it more carefully for you.  In the end, of
course, it's still just quantum mechanics \ldots\ so there's nothing
revolutionary in that sense.  Instead, it's more like waking up one
morning and realizing that ``maybe it's not coincidental or an
accident that gravitational and inertial mass are numerically the
same in Newtonian mechanics.''  This is surely the sort of thing that
hit {\Einstein} at some point:  it was then just a question of
counting the time until something truly wonderful came out of it.

With all that in mind, let me focus on the specifics of the HSW
result.  There is one thing that is truly intriguing about the effect
that powers it.  When separate particles are ``entangled'' as they
are in the {\Einstein}-{\Podolsky}-{\Rosen} pairs, there is a real
sense in which the ``whole is not reducible to a sum of its
parts.''  A more precise way of putting this is really something of
its negative: ``maximal knowledge of the whole does not imply
maximal knowledge of the parts for entangled entities.''  The idea
that I'm trying to convey with this is that, for entangled
particles, even though the whole of the system may be in a
completely specified quantum mechanical state, the separate parts
will not have a precise state of their own.  This sort of thing is
useful in its own right, as I've already pointed out.  The effect
that powers the HSW result is something of a dual to this, though it
again boils down to a statement that the ``whole is not reducible to
the sum of its parts.'' Let me try to explain.

Here one uses nonentangled particle states in a such a way that the
whole is ``more than the sum of its parts.''  Specifically, even
classically correlated quantum states of separate entities are more
than the sum of their parts if one is willing to allow a little
entanglement to assist in the measurement.  This physical fact was
found by {\Holevo} in 1979 and independently by {\Peres} and
{\Wootters} in 1991.  The HSW result carries that effect to its most
extreme form in the context of a specific communication problem.

Is this result deep?  I think so, specifically because of the way it
stands in a dual relation with more standard entanglement.  In some
ways it is just as shocking as entanglement itself.  The full
implications of this new sort of ``non-locality'' have yet to be
fleshed out.

Another way in which the HSW result gives rise to surprising things
has to do with I something I wrote about in Physical Review Letters
last month or so.  Because of the HSW theorem, it turns out that
there are clear-cut situations where it is better to use a signaling
alphabet composed of nonorthogonal quantum states.  This in itself is
rather shocking, because from the ``classical perspective'' (i.e.,
the bad intuition we get stuck with by our everyday experiences)
nonorthogonal signals correspond to {\it noisy\/} signals.  That is
to say, it's very much like the telephone operator asking you to
mumble so that you have a better chance of being understood at the
receiver's end!  Clearly that is nonsense within the world of
classical physics, but quantum mechanics, on the other hand, has this
way of making life more exciting!

\bhba
Is it possible to give some figures on the amount of information that
could be carried and read from a photon or other quantum carrier?
\ehba

Let me first answer your question in generality---I'm not sure
whether you're addressing specifics of the HSW theorem or whether
you're asking something of a broader nature.  There is a certain
sense in which an arbitrarily large amount of classical information
can be loaded onto a single qubit.  This is because there are a
continuous infinity of ways in which it can be prepared.  However, it
is impossible to obtain back that much information from it.  For a
qubit, the maximum amount of information that can be retrieved is a
single bit's worth.  You can find a fairly elementary exposition of
these points in a paper {\Carl} {\Caves} and I wrote:  ``Quantum
Information: How Much Information in a State Vector?'', LANL e-print
archive {\tt quant-ph/9601025}.

So the real answer to the general question is that it is not the raw
information that can be loaded onto a quantum object that counts.
Rather it is in the novel ways that that information can be used.

Concerning the HSW result specifically, yes I can supply you with
some numbers.  Consider a communication channel that consists of
individual two-state atoms as the qubits; that is to say, we shall
use the two-dimensional Hilbert space due to an atom's ground and
excited states as the spot where we place our signals.  The channel's
noise is due to the possibility that the atom's excited state can
decay to the ground by dumping a photon into the electromagnetic
field.  Suppose the coupling between the atom and the field is such
that the probability of this decay is 30\% (as the atom makes its way
from sender to receiver).  Then, before the HSW result, the best rate
at which one could hope to transmit classical information on this
channel was 0.229 bits per transmission.  With the HSW result, we
know that there are ways to use this physical system to transmit
0.300 bits/transmission if we insist on using
orthogonal (i.e., noiseless) signaling alphabets, and 0.311 bits/%
transmission if we allow nonorthogonal (i.e., noisy) signals into the
game.  This represents squeezing 36\% more performance out of the
fundamental physical laws than had been expected. (There are likely
examples that make this difference much more dramatic, i.e., on the
order of 100\% or more, but this is the one I know off the top of my
head.)

\bhba
Do you see us using quantum information and the quantum technologies
in the near future?
\ehba

How do you define ``near future''?  I think it is inevitable that
quantum information technologies will one day make an effect on the
common man's life.  Technology always seems to have a way of coming
into existence if it's not excluded by physical law.  (If it doesn't
come into existence, it means we will have found something better.)
It's just a question of when.

About the HSW result in particular, {\Hideo} {\Mabuchi} (here at
Caltech and soon to be on its faculty) tells me that with a slight
modification of the {\Turchette}, et.\ al., experiment for building
quantum gates in ``cavity quantum electrodynamics'' the rudiments of
the HSW result could be seen pretty easily.  (What I mean by
``rudiments'' here is that with a little entanglement in the
measurement one could realize transmission rates that one could not
without that entanglement.)  However, one should not lose sight of
the fact that the HSW result is an asymptotic one.  Thus it
establishes a fundamental limit on the rate at which a channel can be
used, but to obtain that rate one needs ever larger amounts of
entanglement in the measurement.  So its real role is really in
establishing how far an engineer can hope to push a system.

\section{02 October 1997, \ ``Next Round''}

Below I try to answer all your questions.

\bhba
What's the next big thing to be done do you think?
\ehba

I wish I knew!  The big things are usually surprises.  In
general---to to give you something of a workaday guide---we need a
deeper understanding of the similarities and differences between
quantum and classical information.  But from that general statement,
there are several ways to go.

For instance, it would be nice to have a result similar to the HSW
one but for the ``(one-way) quantum capacity of a quantum channel.''
That would aid in the comparison.  But I haven't seen anything too
satisfactory in that direction yet.    This is something that'll
just have to build up slowly.

Another direction is to just get a better understanding of how to
quantify entanglement.  Maybe the best efforts in that direction so
far have come from the IBM group (Charlie {\Bennett} and company) and
{\Bill} {\Wootters}.  Once that is done, then one can hope to
separate cleanly the amount of classical vs.\ quantum correlation
between two systems.

I hope that gives you some of the flavor of things in the quantum
information theory world.  In our sister wolrd, the world of quantum
computation, things are about the same.  Of course it would be good
to find {\it other\/} new algorithms that can do wonderful things on
quantum computers.  But maybe the most (systematically) important
question is to pinpoint in a very precise way just where it is that a
quantum computer gets its power from.  That is to say, in a way that
is deeper than using some pat phrases like ``quantum parallelism'' or
``quantum entanglement, of course.''  Then, one may have a hope of
understanding the general class of problems that, in principle, can
be sped up on a quantum computer.

\bhba
In what way was he {\em [{\Holevo}]} thinking about these problems in
1978. Surely not in the present terms of Quantum Information?
\ehba

{\Holevo} is an impressive soul!  He posed exactly the HSW problem in
\ldots\ doh(!) \ldots\ actually it was 1977 in a paper titled
``Problems in the Mathematical Theory of Quantum Communication
Channels'' ({\it Reports on Mathematical Physics}, {\bf 12}(2), pp.
273--278, 1977). The paper was received December 10, 1976.  In a
``note added in proof'' he writes something that essentially says
that he has an {\it example\/} that shows that the classical capacity
(of a quantum channel) can be increased from what had been previously
thought by allowing the receiver to perform entangled measurements.
(In the body of the paper, he had already said it was an open
problem.)  By the 1979 paper, he had {\it conjectured\/} what is now
the HSW formula.

Now let me try to clarify what sorts of things were in vogue in the
early to mid 1970s (especially in Russia) along these lines. You are
correct in guessing that things were not stated in terms of the
present lingo of Quantum Information Theory.  However, the question
of how best to use a quantum mechanical system to transmit {\it
classical information\/} goes back at least to a 1962 Masters Thesis
at MIT by G. D. {\Forney}; the earliest published paper on the
question may be one due J.~P. {\Gordon} in 1964.  That's as far as
I've been able to trace it back.  Now you do understand what is
meant by this: using a quantum channel to send ``classical
information'' is to use a quantum mechanical system to carry
distinguishable symbols that can be decoded at a receiver.  Examples
of sending classical information are putting ink on a newspaper (the
quantum carrier is the paper), talking on the telephone (the quantum
carriers are the electrical currents in the wire), or broadcasting a
television show (the quantum carrier is the electromagnetic field).
This is to be held distinct from the uses of a quantum channel that
came up in the 1980s and 1990s: that is in using quantum mechanical
systems to carry incompletely distinguishable (i.e., nonorthogonal)
states and/or entanglement from the sender to receiver.  These
applications are new:  {\Holevo} certainly never thought of
them---he's told me this, in fact.  Nevertheless the concept of
entanglement has been around for a long time, at least since
{\Schroedinger} coined (a German version of) the term in 1935.  What
{\Holevo} proposed was, essentially, to use entanglement in the aid
of the measurement or decoding step in the classical information
problem.

{\it Aside\/}: The English version of the word ``entanglement'' is
also due to {\Schroedinger}: it too came out in 1935.  Maybe it'll
also be useful for you to know that, in a rudimentary form,
{\Einstein} was aware of the concept as early as 9 July 1931.  This
can be seen from a letter of that date from Paul {\Ehrenfest} to
Niels {\Bohr} that describes a conversation with {\Einstein}.
{\Einstein} gave a rudimentary version of the
{\Einstein}-{\Podolsky}-{\Rosen} argument. Max {\Jammer} wrote a
nice historical article on this; if you'd like I can try to dig up
the exact reference.

\bhba
\bq
\noindent
{\em ``I had been working on the problem myself somewhat, but was
never able to pinpoint a measurement good enough for the decoding.
That's where the real genius in their proof lay.''}
\eq
Can you explain this?
\ehba

Yeah, sure.  There was a paper by {\Hausladen}, {\Jozsa},
{\Schumacher}, {\Westmoreland}, and {\Wootters} that proved a very
limited case of the HSW result.  Some versions of the paper were
floating around in 1995. The paper was a little muddled as to the
purpose or goal of the problem, but it was all essentially there.
Anyway, at the end of the paper, they conjectured the full HSW
result.  But then upon meeting and talking to {\Holevo} and reading
his 1979 paper, I said, ``Aha, that's what it's all about.''  Plus
{\Holevo}, in that paper, had also proven that the conjecture was at
least an upper bound on the true answer. So I talked to Richard
{\Jozsa} about working on the thing, and he said, ``Sure, jump in''
(or something to that effect).

So here's where it stood.  To use a (noisy) quantum channel to send
classical information, here's what you have to do.  The sender and
receiver have to decide on the set of possible messages that may be
sent.  (For instance, you and I have tacitly agreed to speak in
English rather than in French.)  Then they have to decide on a good
coding of the messages so that they will become resilient to the
noise.  (As an example here, if we knew that there was an evil
``third letter thief'' listening in to our conversation we might
start douubling thee thiird lettter of eveery worrd thaat we wriite
to eacch othher.)  Finally, the sender has to decide how she will
actually implement this code in terms of preparations of a quantum
system (the ones to be sent down the channel), and the receiver has
to decide upon which measurements he will perform to help reveal the
message.  (You see, when Alice and Bob communicate via written
letters, it is clear what Bob has to do to retrieve Alice's messages
in the most efficient manner:  he has to ``look'' at the words on the
paper.  However, in general, there are other things he could have
done.  For instance---I know this is silly---he could have felt out
the indentations on the paper instead of looking.)  With regards to
quantum systems, it is generally not obvious as to which of the many
possibilities the optimal measurement will turn out to be.  ``Should
I look at spin in the $x$-direction, or spin in the $z$-direction?
Quantum mechanics won't let me do both, so I'd better be careful in
picking my protocol.''

Of course all these steps in building a good use of the channel are
intertwined.  Luckily the coding part of the scenario above can be
taken out pretty easily by a minor extension of classical information
theory---this {\Holevo} did in his 1979 paper.  After that, one can
simplify the problem by holding fixed everything but the
decoding/measurement part at the receiver's end.  Even with that,
though, the problem becomes a bear.  And that's what I was talking
about.  {\Holevo} and {\Schumacher}/{\Westmoreland} both finally
stumbled into a measurement that was good enough to achieve this
upper bound that {\Holevo} had already proven.  That's where all the
genius in these last articles lay.

\bhba
\bq
\noindent
{\em ``Is this result deep?  I think so, specifically because of the
way it stands in a dual relation with more standard entanglement.  In
some ways it is just as shocking as entanglement itself.  The full
implications of this new sort of ``non-locality'' have yet to be
fleshed out.''}
\eq
Are you suggesting there is a sort of residual entanglement still
available to classically correlated states?
\ehba

Not really.  I think putting it the way you did in this question is
probably dangerous---you should steer clear of that.  The problem is
that I don't understand the effect so much yet that I have a good
everyday-language way of saying it.  (You see, the reason we use so
much mathematics in physics is that we don't have to think too hard
then!)  Maybe the best way to put it is simply that these classical
correlations between quantum states have more visibility than meets
the classical eye.  That is, if you allow entanglement to be used in
the measuring process, then more about those correlations can be
ferreted out than would have been the case otherwise.  However, there
is no residual entanglement in the states themselves: the acid test
for this is that there is no way to make these classically correlated
states violate any {\Bell} inequalities.

{\Peres} and {\Wootters} put it like this in their 1991 article ({\it
Phys.\ Rev.\ Lett.} {\bf 66} p.~1119, 1991):
\bq
It is well known that composite quantum systems, consisting of
noninteracting parts, can possess nonlocal properties.  In
particular, a composite system can exhibit correlations which cannot
be reproduced by any theoretical model that involves only variables
belonging to each subsystem separately.  \ldots\  In this Letter, we
consider a different kind of composite system.  Its parts never
interacted in the past.  They may have been prepared in different
laboratories.  However, they are prepared according to the same set
of instructions.  Therefore, these subsystems are in the same quantum
state---insofar as their internal variables are concerned.  \ldots\
Our work suggests that one can indeed obtain more information by
measuring the two particles together \ldots\
\eq

I hope that helps.

\bhba
\bq
\noindent
{\em ``There is a certain sense in which an arbitrarily large amount
of classical information can be loaded onto a single qubit.  \ldots\
For a qubit, the maximum amount of information that can be retrieved
is a single bit's worth.''}
\eq
I need to make this clear before I start writing the feature so
forgive me for repeating what might be the very obvious.

Is this statement correct?
\ehba

Yes it is.  Though I think you will be able to understand the flavor
of it better if you read the first four pages of the thing I wrote
with {\Caves}.

The remark you quote above, as I say, is true.  But it generally
addresses a different sort of scene than the one addressed in the HSW
problem.

\bhba
If you have classical information on a quantum channel (the HSW
system). You can load a large amount of information on but you can
only take one bit off if you use orthogonal states. If you use
non-orthogonal states you are limited to the von Neumann entropy
which will be less than one bit.

Is this statement correct?
\ehba

Somewhat.  Regardless of whether your encoding takes (classical) bits
to orthogonal or non-orthogonal states, the most information an
observer can hope to retrieve about the encoding is the von Neumann
entropy of the density operator associated with the states in that
encoding.  The von Neumann entropy of any density operator is bounded
above by the logarithm of the dimension of the Hilbert space.  For a
single qubit, the von Neumann entropy is bounded above by $\log
2=1\mbox{ bit}$.

But that issue is not so important for the HSW question.  Their
question is more to do with:  what is the best way for me to stuff
information into a qubit so that I have the greatest chance of
retrieving it after the qubit has encountered noise.  In particular,
it can be shown that it is never a good idea to try to stuff more
than 4 classical bits into a qubit if your ultimate goal is to
retrieve the most information at the (noisy) channel's output.  (If
there is no noise at all, it is never a good idea to try to stuff
more than a single classical bit into it.)

\bhba
If you have quantum information on a quantum channel the main problem
is to keep the quantum system intact (quantum fidelity) until it
reaches the other end. There it will be used by a system needing
quantum information input e.g.\ a quantum computer.
\ehba

Yes this is correct.  There are at least two tasks to which one might
submit a quantum communication channel. The first, as we've already
discussed at great length, is to reliably send classical bits down
it.  The second, which we haven't touched on too much, is to use it
to transmit quantum information.  This breaks into two categories: A)
to use it to send arbitrary quantum states drawn from some Hilbert
space, and B) to transmit entanglement over it, so that the sender
and receiver become entangled to their hearts' content.  In category
A, the criterion of a good transmission is that the outputs be as
much like the inputs as possible.  I.e., the fidelity between input
and output be good.  In category B, it's that the entanglement be as
high as possible.  Now the way to ensure that these things happen is
to spread the real thing you're trying to send over many
transmissions through the channel.  That's what Peter {\Shor}'s
insight into quantum ``error-correction'' is about.  Once you know
that such a thing will help, then the next question is: how many
transmissions do I need per qubit to make the signal fidelity or
final entanglement good?  That's the question addressed by the issue
of the ``quantum capacity of a quantum channel.''  (But we are quite
some distance from that answer.)

\bhba
Can you answer these questions?

Is there a way this quantum information can be used by a classical
system, e.g., a television, normal computer? And if so, how much
information can be taken off? From what I have read so far it should
still be one bit, but there are hints it might be more, much more.
What is the correct answer?
\ehba

Nope, sorry, it doesn't really look like that's the case.  What I can
tell you is what I've already said.  I guess the thing I'd like you
to take away from this conversation is that the thing we're learning
from Quantum Information is a lot of new ways to think: we're
learning to view quantum mechanics in a really positive light now.
Entanglement, for instance, is a new resource that should be taken
into account when approaching a problem.  Sometimes it's helpful and
sometimes it's even a gold mine (like when used for factoring large
numbers, {\Shor}'s factoring algorithm).  But it's not a panacea.
Learning all the things we can about this stuff, including the
limitations, is our great challenge.  Good and even amazing things
will come---in my opinion---but they'll likely be concerned with
things we don't have the imagination to think of today.  Quantum
Information just establishes a good direction in which to think.

\section{06 October 1997, \ ``49th Parallel''}

\bhba
I have another question prompted by your remarks on entanglement and
energy. How close is the parallel?  Do you think there is an $E =
mc^2$ for entanglement or a $q=DU-w$ or $G=H-TS$?
\ehba

Wouldn't that be great?!  But, unfortunately, I don't think anyone
can say presently how close the concepts might be.  Some of the
questions raised concerning entanglement certainly have the flavor of
thermodynamics, but I think it's just too premature to draw any
rigorous parallel.  Sorry.

\chapter{Letters to Howard {\Barnum}}

\section{02 June 1999, \ ``Please Do''}

You know what I'm really trying to get at here:  I want the structure
of quantum mechanics to be extremal for some kind of more abstract
(i.e., QM-independent) information-disturbance problem.  But the
first thing that needs to be done is to pin down what quantum
mechanics itself has to say.

Another thing that keeps coming to mind is that the assumptions
behind {\Gleason}'s theorem might have something to do with the
allegory I presented you with at {\Charlie} and {\Theo}'s house. God
said, ``Ok I will make information gathering invasive, but not so
invasive that you will be forever science-less.''  The connection is
this.  I think one can split the assumption behind {\Gleason} up
into at least two parts:  (1)  The questions that can be asked of a
quantum system {\it only\/} correspond to orthonormal bases over
some vector space, with there being no good notion of measuring two
distinct questions simultaneously (there being no good notion of an
AND operation).  And (2) It is the task of physical theory to give
probabilities for the outcomes of those questions, but we can say at
least this much about the probabilities.  They are
context-independent in the sense that, for a given outcome, it does
not matter which physical question (which basis) we've associated it
with. I.e., the probabilities can be assumed to be of the form of a
frame function.

It seems to me that the first assumption captures to some extent the
idea that information gathering is invasive.  If you gather some
information and I gather other information, there is no guarantee
that we can put the two pieces of information into a consistent
picture.  The second assumption, though, appears to be more of the
flavor that, nevertheless, that information gathering is not {\it
too\/} very invasive. For otherwise I might imagine the probabilities
to depend upon the full context.

\section{17 June 1999, \ ``So Slow''}

{\Kochen}-{\Specker}, no it's not too important.  It just immediately
throws out a certain naive class of hidden variable theories. That's
good for illustrative value, but not much more.  The thing that
intrigues me about David {\Meyer}'s result is that if we work over
the rational field, then we can no longer get that illustrative
example.  Because of that, I keep wondering whether there is
something fishy here \ldots\ it's just a vague feeling.

{\Wallach}'s result has to do with trimming down the assumptions
behind {\Gleason}'s theorem on tensor-product spaces.  It's not
related at all to the stuff about {\Meyer}.  About extra assumptions
for getting {\Gleason}'s result on rational Hilbert spaces, I don't
know.  I was just speculating:  for instance, linearity would
probably give it. But then that already throws away all of the
beauty of it.

All this does cause me to wonder about on which kinds of discrete
structure we might hope to prove an analogue of {\Gleason}'s theorem.
For instance, let us not fixate on the field over which the Hilbert
space is defined.  But perhaps instead let us consider as valid
quantum mechanical ``questions'' (i.e., measurements, but I no longer
like that term), any orthonormal basis that is rotated away from a
standard basis by angles which are rational fractions of $\pi$.  Can
we prove {\Gleason}'s theorem on such a structure?  I suppose I
should ask either {\Wallach} or {\Pitowsky} this question.

\section{04 July 1999, \ ``No Microphysical Reality?''}

Oh the accusations you make about my poor philosophy!  It's the
middle of the night, and there's a wonderful thunderstorm going on
right now.  I've been up the last hour or so rereading a lot of your
emails from last week.

It seems you think I believe there is NO reality?  Anyway, I think
the notes below best express my present views.  Have patience with
them:  read all the way to the bottom, and then come back and read
the top.

[NOTE:  In place of the notes I orginally recommended to Howard, see
this whole book!]

\section{18 August 1999, \ ``Modalities''}

It dawned on me on my flight home that we've never discussed the
so-called ``modal interpretations'' of quantum mechanics.  Take for
instance the {\Spekkens} and {\Sipe} poster we saw in Baltimore as a
good example.  The essence of these interpretations seems to be the
following.  One starts with a Hilbert space for the universe and
assumes that it has a fixed tensor-product structure.  The state
vector of the universe is not identified with reality immediately
but instead gives a catalog of what is {\it possible}.  The
properties that are {\it real\/} in this theory are the state vectors
associated with the factors of the big Hilbert space---but only one
of those product-states ``actually'' obtains.  For instance, in the
{\Spekkens}-{\Sipe} proposal one takes the universal wavefunction,
decomposes it into a superposition of orthogonal product states
(with a certain extra entropy-minimizing criterion) and that defines
what is possible.  The possible physical states of the universe are
those products states.  Their (objective) probability for being real
is given by the absolute-square of the associated coefficient in the
decomposition.  But the real world only actually has the properties
associated with one of those components.  At every moment in time
one has this, and thus the world has its properties stochastically
and nonlocally determined.

It seems that the main difference between this point of view and the
many-worlds view is that in this set-up only one world actually
attains.  Also I suppose that it is crucial that there is a fixed
tensor-product structure and at least some adherents of the
many-worlds view would say that that's not so crucial.

What's your opinion of all this?

\section{30 August 1999, \ ``It's All About Schmoz''}

I read your paper [``Dieks' Realistic Interpretation of Quantum
Mechanics: A Comment''] on the plane the other day: essentially I
agree with you that the modal interpretation is just a hidden
variable theory. Jeffrey {\Bub} in his new book even classifies
Bohmian mechanics as a modal interpretation. For some reason the
modal guys just don't seem to like the label ``hidden-variableists.''

\bhb
the only thing that can tell us the other branches are there is a
``disentangling'' interaction \ldots\ something like the reversal of
a measurement.  This is what I think is usually meant when e.g.\
relative state'er's tell you that the reason for having the other
branches around is ``interference'':  we might, {\em in principle},
be able to set up an experiment involving us {\em and} the rest of
the universe (or whatever portion of it is entangled with us in
respect of the measurement we're trying to reverse), and arrange to
interfere them in such a way as to disentangle the state, erasing our
observation record.   Sure this is incredibly difficult.  But, do we
have a rigorous theory of its {\em impossibility}?
\ehb

This passage lays bare the very core of our difference!  In the
{\Everett} interpretation, we can never ARRANGE to ``disentangle the
state.'' There is no sense in which this phrase or the gist of your
passage is even meaningful in that interpretation.  The {\Everett}
world just IS\@. Experimenters have NO choice to arrange anything in
an {\Everett} multiverse; their actions are at best a grand
conspiracy set by the initial condition of the universe.  They have
no will, no possibility to do anything that wasn't preordained.  If
an experimenter in the {\Everett} theory perceives himself reversing
the entanglement of two systems, it can be but ILLUSION that he had
any choice in the matter.

\bhb
no-one [\ldots] has ever reversed a measurement whose result they
were conscious of [\ldots].  So, as a
state-vector-isn't-real-but-is-just-%
a-representation-of-our-knowledgeist, you might say these
interference effects haven't been observed, and you needn't worry
about them as you blithely reduce the wavefunction, and say you just
got new knowledge, changing what you can say about the probability of
future experimental outcomes.  Fine, but now your theory differs from
no-collapse theories on the outcomes of at least some {\em
in-principle} possible experiments.
\ehb

You just don't get it!  It is precisely for the
state-vector-is-knowledgeist, that there is any substantial sense to
being able to reverse the evolution of a quantum system.  There is
nothing extra to standard UNITARY quantum mechanics in this picture:
one simply must be sober enough in thought to recognize that the
state vector is NOT reality and that is that.

Suppose I know everything that I can validly know about an electron,
and so represent that knowledge of it by some pure state
$|\psi\rangle$. Suppose too that you believe the same thing of the
electron that I believe, and hence you yourself use $|\psi\rangle$.
Finally, suppose that just as with the electron, I know everything I
can validly know of you including your intent to measure the
electron's spin.  I describe you via the quantum state
$|\eta\rangle$.  Time passes and true to form you measure the
electron.  You learn something about the electron and thus update
your state of knowledge (perhaps discontinuously as we usually
presume); you change the electron's state assignment to something
new, perhaps $|\tau\rangle$.  I, not interacting with you or the
electron any further, describe the state of affairs via some unitary
evolution that entangles you and the electron.  My state of knowledge
about you AND the electron is now some entangled state.  The
difference in our two DESCRIPTIONS is one due to an outside point of
view as compared to an inside point of view.  In that sense, it is
the same disparity as one encounters in the old Maxwell demon
problem.

Now, what about reversibility?  It is precisely this:  my state of
knowledge of you and the electron is still maximal.  The fact that
from the inside point of view you have learned something and thus
made some discontinuous change in your description of the electron
makes not one iota of difference.  My state of maximal knowledge
allows me certain privileges.  For instance, since I know both you
and the electron so well, I can in principle construct some
complicated apparatus to control you in any way I wish.  In
particular, I can, IF I SO CHOOSE, reverse your measurement, wiping
out all memory in you of ever having gone through the motions of the
measurement.  There is no inconsistency here; there is nothing
external to unitary quantum dynamics.  The only thing that one must
do to see the logic is HUMBLE oneself to the point of view that
quantum mechanics is about what we know of systems and how we can
manipulate them in the light of that knowledge.  In this point of
view, the ontological content of quantum mechanics is not in the
state vector \ldots\ it lies somewhere else in the theory.

You used the phrase, ``no one has ever reversed a measurement whose
result they were conscious of.''  This I agree with wholly.  When one
learns information, that process is by its very nature irreversible:
within the {\Shannon} theory, information gain cannot even be defined
without the notion of prior ignorance.  Probabilities must change
irreversibly for information to have been gained.  But REALIZE that
is from the inside point of view.  IF there is an outside point of
view (as there was in the example above), then there is nothing
sacred about irreversibility \ldots\ but that is purely with respect
to that outside point of view.

In general what I dislike about the {\Everett} extravagance is that
it makes a concerted attempt to ignore the agent's---be it a man, a
scientist, a chimpanzee, or a HAL-9000's---place in science and place
in the universe.  It is a mysticism not unlike {\Kepler}'s theory of
planetary motion.

If you do one favor for me this year, please read and think about all
that I said above.  And please read and think thoroughly about the
old {\Fierz} paper that I'll place below.  Please try to put
yourself in my mindset when reading all these words \ldots\ and I
think you'll find it more reasonable than you ever thought you could.

\section{05 September 1999, \ ``New Schmoz Cola''}

You know, I probably shouldn't even be writing you, if for no other
reason out of punishment: I still haven't seen your crypto paper on
the archive (or in my mailbox)!!!

\bhb
Ah!  I have come to understand (or remember, perhaps) the importance
of leaving a place for free will in your views on the foundations of
qm.  I'm sure you realize that a Newtonian, or perhaps better,
Laplacian universe would have exactly the same problems?  This
didn't cause physicists to abandon Newtonian mechanics, though
\ldots.
\ehb
I agree with your second sentence pretty much, though I think
{\Everett} is still worse for other reasons.  As I told {\Herb} a
few weeks ago:
\bq
\noindent
I think many-worlds empties the world of content in a way that's even
worse than classical determinism.  Let me explain.  In my mind, both
completely deterministic ontologies and completely indeterministic
ones are equally unpalatable.  This is because, in both, all our
consciousnesses, all our great works of literature, everything that
we know, even the coffee maker in my kitchen, are but dangling
appendages, illusions.  In the first case, the only truth is the
Great Initial Condition.  In the second, it is the great ``I Am That
I Am.''  But many-worlds compounds that trouble in a far worse
fashion by stripping away even those small corners of mystery.  It is
a world in which anything goes, and everything does.  What could be
more empty than that?
\eq
The Laplacian universe did indeed have the problem you mentioned, and
NO that didn't cause physicists to abandon Newtonian mechanics.  But,
as you saw from my last note, I have no intention of abandoning
unitary quantum mechanics:  your reprimand slides off me like water.
The solution is in taking the {\Fierz} article to heart.  Physical
theories are here to help us make our way through the world; they
should, for the most part, be viewed as predictional meat grinders
and not ontological statements.  In the classical case, the input is
the initial Liouville distribution, the output is the final Liouville
distribution. Both are states of knowledge:  the inputs and outputs
should not be viewed as states of the world.

\bhb
{}From the inside, the experimenter has a choice; from the outside,
no.  Here's a caricature, so feel free to object:  {\Bell}'s worry
about the foundations of QM has been:  that we have ``measurement''
as an ``unanalyzed primitive'' of the theory.  {\Everett} shows us
how to get around that.  You don't like {\Everett}'s resolution
because you {\em want\/} to have an unanalyzed primitive around so
it can be the locus of free will.
\ehb
Not exactly.  From the inside, the experimenter has a choice; from
the outside, the experimenter has a choice.  Quantum mechanical
statements are statements about the extent to which things can be
predicted.  Get out of the ontological mind-set!  Concerning
``{\Everett} showing us how to get around that,'' as much effort as I
have given it, I have never been able to see it as more than a
statement of faith.  It, like most religions in this world---it seems
to me---leads to not much more than comfort.  About my ``wanting a
locus for free will,'' there is little doubt about that.  The thing
about quantum mechanics is that it helps me believe we're closer to
that situation than I once thought possible.

I find it hard to gulp that my whole life is an illusion.  I find it
immensely improbable that the design of this Dell computer was
written in some partial way in an ontological initial condition for
the universe.  It strikes me as almost laughable that though we are
essentially automata---even if splitting into minor variations
higgledy-piggledy---we are nevertheless automata lucky enough to have
the built-in illusion that we are understanding and discovering
things anew with each day.

This will be hard for you to believe, but if you ask me, I think my
view is far more Copernican than yours.  You're willing to risk the
guess---when you're playing devil's advocate for {\Everett}, that
is---that the Everettista model can be placed in an encompassing
correspondence with all that is.  ``This little thing that OUR minds
dreamed up captures reality in its very essence.''  That is an
extreme anti-Copernican chutzpah, it seems to me.  In contrast, in
order to get my view off the ground all I have to do is acknowledge
that our knowledge is limited, that I am perhaps only 11\% better off
in my understanding of the world than my dogs.  I do acknowledge that
our actions can change the world in a very limited way (at present at
least),  but I do NOT acknowledge that they can change things
``before we were here.''  It is a humble point of view, but a point
of view with hope (for in it there are aspects of the world that are
not set in immutable stone).

\bhb
Your discussion of the experimenter (Dr.\ Science) who does the
reversal, is fascinating.  I am glad you (and also {\Herb}) agree
that Dr.\ Science could wipe out your memory of the measurement.  The
inside/outside business (also emphasized by {\Herb} at lunch the
other day) is nice.   But, if I (who will [undertake] the
experiment), know what Dr.\ Science is going to do, won't {\em I\/}
want to make predictions that agree with Dr.\ Science's?
\ehb

But there is no sense in which the two descriptions disagree, if you
take the appropriate view of what it is the state vector symbolizes.
When one has a pure-state description of a system, one has {\em
maximal\/} knowledge, but maximal knowledge is not complete knowledge
in quantum mechanics---there is nothing in quantum mechanics that
plays the role of complete knowledge in classical physics, i.e., the
phase space point.  I assert these sentences with great confidence:
it would however be an excellent research project for my point of
view to find a formal way to show there is enough leeway in the
alternative quantum descriptions that one never gets into trouble.

What is it that the knowledge is of?  I'll bet your gut reaction is
to think I'm thinking the knowledge is of some kind of underlying
``reality.''  Well, I'm not.  Lately I've been thinking the best
language to describe the situation is the same as in L.~J.
{\Savage}'s treatment of classical decision theory.  Quantum
``measurement'' is more rightly viewed as being about decisions,
acts, and consequences. When confronted with a quantum system, we
can {\it decide\/} to act upon it in one of any of a number of
nonequivalent ways. The set of acts available to us is the set of
positive operator valued measures: Each POVM corresponds to an {\it
act\/} we might perform on the system. The set of possible {\it
consequences\/} associated with an act corresponds to the separate
operators within the act's affiliated POVM.  The quantum system is
the conduit from the acts to the consequences; it takes the place of
the ``states of the world'' in the {\Savage} system, but it is a
black box that cannot be cracked.  The quantum state is only an
``if-then.'' It stands for a compendium of probabilities that
summarize our beliefs about the consequences of our acts.  That is
its sole content.  [As an aside, you might notice that I associated
the POVM elements rather than the associated quantum operations with
the consequences of an act.  That was intentional; in my view the
operations have to do solely with the updating of subjective beliefs
in analogy to Bayes rule.]

\bhb
{\Herb} at lunch kept saying (I paraphrase severely) ``sure these
other terms exist, but only in the wavefunction from the outside
point of view \ldots\ not in the wavefunction you can use.''  But,
that's exactly what's being argued about:  if an experiment is going
to happen, and I, while ``inside'' the experiment, can understand
what's going to happen by taking the viewpoint you label
``outside'', who are you to tell me I can't take that viewpoint? One
could then see the {\Everett} interpretation as a systematic attempt
to be as ``outside'' as one can in one's model of the world (viewed
as fundamentally quantum mechanical).
\ehb

I've said it before, I'll say it again:  Can a dog collapse a state
vector?  Dogs don't use state vectors.  I myself didn't collapse a
state vector until I was 20 years old.  More to the point,
\begin{center}
QUANTUM STATES DO NOT EXIST.
\end{center}
If you take the {\Savage} language presented above seriously, then
it is an immediate consequence that you on the inside cannot see
things as if you were on the outside.  (Knowing of an outside
viewpoint's POSSIBLE existence is a different thing than taking
it.)  You cannot take a view that includes yourself within it
because you can act upon what you can act upon and nothing more.
How can you get outside yourself to act upon yourself?

It does appear to be true that I cannot stop you from attempting to
take the most outside view (naively) imaginable, the one that
{\Everett} wishes. But I would only be a broken record if I were to
say that I don't see how that point of view can be but stale and
counterproductive.  When you ontologize the state vector, you force
the world to be an empty shell---that's the way I see it.

\bhb
Not ignore, but to allow the possibility of representing the agent as
made of the same stuff as the rest of the world \ldots\ a very
monistic kind of theory \ldots\ and to allow the possibility of
taking a point of view ``outside'' the agent as well as the
laboratory apparatus and quantum systems under study, or whatever.
I'm just afraid you dislike this approach so intensely that you want
to claim it's impossible, illogical, etc \ldots\ whereas my view of
the {\Everett} interpretation is that it shows there's nothing
special about quantum mechanics that makes taking the outside view
impossible, although perhaps it makes it weirder, and less appealing,
to take that view.
\ehb

In my view, I explicitly do NOT see the agent as being made of
different stuff than the rest of the universe.  If I did, I would not
be able to trust modern medical science to develop new drugs to cure
old diseases; I would not be able to imagine a Dr.\ Science who might
reverse my measurements.  That is just the point.  There is the
world, and then there is what we can say about how it reacts to our
experiments, our punches and our pushes.  Quantum mechanics is about
what we know of the right-hand part of that sentence.

I do dislike the Everettista movement fairly intensely, but I don't
feel that I have gone so far as to claim that it is ``impossible,
illogical, etc.''  My main points are these.  1) I have not been
convinced that Everettism is less of a religion than my own personal
point of view.  To me, in fact, it seems the opposite:  I see almost
no possibility of extracting the world of our common experience from
an imagined universal wave function.   This could be a weakness on my
part; I am not all knowing. But if that is true, I still have: 2)
Everettism simply seems to be a dead end for the furtherance of our
understanding this wonderful world.  I could be following a false
prophet, but if my view succeeds then we will know that there is room
for something new under the sun.  How could we not want that?

\chapter{Letters to Paul {\Benioff}}

\section{09 April 1996, \ ``Randomness''}

I don't know if you remember me, but we met at the Santa Fe Institute
Workshop on the Physics of Information two summers ago.  I spoke to
you about your old papers on randomness and quantum mechanics.

In any case, I am very much interested in them again.  (Presently,
in particular, the ones titled ``Finite and Infinite Measurement
Sequences in Quantum Mechanics and Randomness: The {\Everett}
Interpretation'' JMP {\bf 18} (1977) 2289, and ``A Note on the
{\Everett} Interpretation of Quantum Mechanics'' Found.\ Phys.\ {\bf
8} (1978) 709.) {\Carl} {\Caves} and I are writing a book on Quantum
Information and are trying to get a feel for how far these sorts of
considerations really go toward deriving the standard probability
expression of quantum theory.

Last night I reread the latter of the two papers mentioned above.
The main question on my mind now is about Section 4 ``The Asymptotic
Description.''  At the end you state, ``Thus it is an open question
whether or not there is any possible state description of the
asymptotic situation that includes component states each
corresponding to a possible universe as perceived by an observer
with memory trace $\nu$.''  Has there been any progress in this
direction since then??  If there has, could you please send me a
reference or outline the solution?

\subsection{Paul's Reply}

\bq
I think I do remember you from the Santa Fe conference, which I
enjoyed very much.  I am sorry for the delay in responding to your
letter.  I dug out my paper and reread it as I had forgotten about
it.  Yes the problem still exists, as one cannot discuss in quantum
mechanics within Hilbert space infinite sequences without using some
restrictive tail conditions to limit consideration to a countable
infinity of sequences described in a separable Hilbert space. From a
measure theoretic viewpoint, any countable set is a set of measure
zero.

One way around this may be the use of quantum field theoretic
techniques such as lattice gauge theory.  In this manner systems
with an infinite number of degrees of freedom can be treated, but I
have not pursued this approach.

I have been much interested in randomness especially in relation to
quantum mechanics.  As you may know, there are an infinite number of
different definitions of randomness each depending on the definition
of statistical tests for randomness.  In 1976 [JMP vol.~17, pp.~618,
629 (1976)] I showed that for sufficiently strong definitions,
physics had something to say about the foundations of mathematics in
that there were mathematical universes as models of ZF Set Theory
which had no random sequences and thus could not be mathematical
universes for physics and especially quantum mechanics.  The problem
was that I could not prove and still cannot prove that these very
strong definitions are necessary.  Weaker ones, e.g.\ those of
{\Chaitin} and others, seem sufficient.  It appears that this
approach is not productive.

In any case I support your efforts in this direction.  Please keep
me posted on your progress and give my greetings to Professor
{\Caves}. I hope we meet again at some conference.
\eq

\section{15 April 1996, \ ``More Randomness''}

Thanks for clarifying things for me:  I was mainly interested in
knowing whether there are other mathematical tools lying around for
the asymptotic situation that might make my life easier.  If you
don't know of any, that's OK; in the end, I haven't been convinced
that the full asymptotic situation is important anyway.

Mostly I'm interested in recasting these old results by you and by
{\Hartle} (AJP, vol.\ 36, p.\ 704, 1968) in a Bayesian (or
subjectivistic) light anyway.  For that I don't think I need to get
a handle on the full asymptotic case.

Thanks also for your comments on your 1976 JMP papers.  These are
two that I was not aware of!  (Though I should have been because they
are referenced in your others.)  Let me ask you one question right
off.  Is your result true for finite dimensional quantum systems? Or
rather must you consider repeated measurements on identical
preparations of {\it infinite\/} dimensional systems to make it go?
If the first option is the case, then I'll certainly try to dig them
up at the library here.  (Unfortunately I have to make a special
request to get hold of older papers around here.)

There was a time when I was a strong believer that quantum mechanical
probabilities are more closely allied with the frequentist notion of
probability than with the subjective or Bayesian versions. However I
had hoped that that would somehow make itself known through the
already-given structure of standard quantum mechanics \ldots\ that is
to say, rather than introducing ``randomness'' explicitly as you did.
But, of course, I had no clear idea about how to check that such a
thing might actually be true.

Presently I'm more taken with Bayesian notions.  (Evidence that
{\Carl} {\Caves} has had some influence on me!)  In case you're
interested in what I mean by this, I'll attach a letter I wrote to
Michiel van {\Lambalgen} a few days ago.  van {\Lambalgen} is also
interested in ``randomness,'' though he does not believe that it is
properly defined within the framework of classical, Platonistic
mathematical logic.  (He's told me that he's corresponded with you
before.)

\subsection{Paul's Reply}

\bq
First to answer your question re my 1976 papers in JMP. The results
are true for finite dimensional systems and their limits. That is
one infinite repetition of a measurement process in which one
prepares a system makes a measurement and discards the system is
considered.  One does not have to consider finite or infinite
repetitions of infinite repetitions of measurements.

My 76 papers are based on the existence of many different
definitions of randomness.  (This is briefly discussed in the last
section of the paper.)  In fact there are an infinite number of
possible definitions.  The basic problem is that I know of no way to
use physics or mathematics to place a floor under the definitions
other than the relatively weak floor, namely that each statistical
test must be at least effectively enumerable ({\MartinLof} or
{\Chaitin}).  The problem is very intriguing though because
randomness is so much more an essential part of quantum mechanics
than of classical mechanics.  This is what motivated my work in the
area.

I remember a brief conversation I had with Leonid {\Levin} many years
ago on a bus at the end of a meeting.  To him it was obvious that
the proper definitions of randomness should be very strong so that
the results of my 76 paper would apply.  However I did not
understand his arguments at the time and did not pursue the matter
further.  Dr.\ {\Levin} was at Boston University the last I knew.
Keep me posted if you pursue this further.
\eq

\section{30 April 1998, \ ``Randomness Again''}

Once again I am thinking about randomness and quantum mechanical
indeterminism in a relatively serious way.  In particular, I just
read your article:
\bq
\noindent P.~{\Benioff}, ``On the Correct Definition of Randomness,'' in
{\sl PSA 1978: Proceedings of the 1978 Biennial Meeting of the
Philosophy of Science Association, Vol.~2}, edited by P.~D. Asquith
and I.~{\Hacking} (Philosophy of Science Association, East Lansing,
Michigan, 1981), pp.~63--78.
\eq
and the follow-up article by {\Hellman}:
\bq
\noindent G.~{\Hellman}, ``Randomness and Reality,'' in {\sl PSA 1978:
Proceedings of the 1978 Biennial Meeting of the Philosophy of Science
Association, Vol.~2}, edited by P.~D. Asquith and I.~{\Hacking}
(Philosophy of Science Association, East Lansing, Michigan, 1981),
pp.~79--97.
\eq
I must say, I find many of {\Hellman}'s criticisms pretty convincing.
In particular, his remark at the top of page 86 and his Note \#4
resonates well with me.  Did you ever write anything countering his
arguments?  Or do you have any thoughts on the matter that you never
wrote down?  If you're interested in recording them, I'd love to
hear!

\subsection{Paul's Reply}

\bq
Randomness is a fascinating subject that won't go away.  I had
forgotten about the papers you referred to but found them.  I have
not written a critique or rebuttal.  But the following expresses my
views.

I have much criticism with {\Hellman}'s critique.  One main one is
that he does not distinguish carefully between classical mechanics
and quantum mechanics. In CM randomness is epistemic as it is due to
an observers lack of knowledge about the properties of a system.  In
QM it is present in principle and has nothing to do with what an
observer happens to know or not know or can know. about a system.

I do not understand what {\Hellman} is driving at in his P.\ 86
comments.  I do not use the terms as nondeterminstic and acausal.  A
process such as an infinite repetition of a preparation of a QM
system and measurement of an observable that does not commute with
the prepared state of the system generates a random sequence of
outcomes. (For simplicity assume a new system is prepared for each
measurement.)  A product probability measure describes this ordered
ensemble of repeated measurements.  The randomness of this outcome
sequence is a physical property of the ensemble just as position or
momentum is a property of the individual systems in the ensemble.

For me the main question is if the outcome sequence is random, how
random is it or random according to what definition?  I do not know
the answer to this question.  The main point about my J.\ Math Phys.\
papers was to show that if all such sequences are random according
to sufficiently strong definitions, e.g.\ Solovay random, then
physics has something to say about  the foundations of mathematics
in that not all ZF models can be used as the carriers of the
mathematics of physics.  {\Hellman} does not find this of interest
because the usual model of ZF is much bigger than the minimal
model.  I do not agree. Thus for {\Hellman} to emphasize, P.\ 81
bottom, ``if we believe in ZF there is nothing for physics to say''
is not right.  If a sufficiently strong definition applies, then
physics would have shown that we must go outside ZF at least for a
mathematical description of randomnes of QM outcome sequences.

The requirement I proposed in the paper that weakest definition of
randomness be such that no contradiction can be derived from an
empirical outcome sequence seems reasonable.  However I have to take
my own medicine. This may or may not be relevant to the strength of
randomness associated with a QM outcome sequence.  In this sense I
agree with {\Hellman}'s criticisms.

However the last part of {\Hellman}'s note 4, on P.\ 91 is simply
wrong in my view.  The invariance of limit means under subsequence
selections is intuitively assumed by every physics experimenter.  If
the even numbered elements of a sequence showed evidence of
converging to a different limit mean than the odd elements the
experimenter would conclude something was wrong with the
experimental setup.  This holds also for the sequential tests of
randomness even in the widely accepted and weak definitions of
randomness, e.g.\ martin lof or chaitin randomness.    This also
applies to every outcome sequence generated by experiment, not
``almost every one'' whatever that means physically.

I do not know what definition of randomness applies to the sequences
of outcomes in QM.  So far no one has shown that there are any
problems with the weak definitions, however I do not know how hard
people have looked.  I stopped working on the problem because the
approach did not seem fruitful. But the problem has not gone away
for me.

As you might guess I am extremely interested in the connections
between the foundations of physics and mathematics.  This is a main
motivation of my work on quantum computers and recent work on
quantum robots and environments.

I am quite interested in your thoughts on this.  Keep me posted on
what you come up with.
\eq

\section{10 June 1999, \ to Paul {\Benioff}, \ ``Definitive List?''}

I just finished compiling what I think may be a definitive list of
all your writings to do with quantum mechanical randomness.  It's
pasted below.  Would you mind having a look over it to see if I've
missed anything?  If I have missed something, could you email me the
appropriate references?  Thanks a million.

Just recently I read your paper ``Some Foundational Aspects of
Quantum Computers and Quantum Robots'' and enjoyed it very much!
Indeed, you point out an issue that is in the greatest need of
exploration:  how do the theory-making entities in this quantum world
bootstrap their way into a theory of it?  In turn how do they
validate the theory they've come across?  I think if we understood
that, we would have a much better grasp on the ``why'' of quantum
mechanics than we do today.  (Today, I think we have essentially {\it
no\/} grasp on why the theory takes the precise structure that it
does. The contrast of this situation to the one in special
relativity, for instance, is stark.  There one can see in almost an
immediate way the physical assumptions behind the apparatus of the
theory.  It is almost a crying shame that we are nowhere close to
that with quantum mechanics, given that it is over 70 years old now.)

I expressed my own ideas in this direction to an old friend in a
rather poetic way a couple of months ago.  I'll attach that text too,
in case it might pique your interest.  [See note to {\Greg} {\Comer}
dated 22 April 1999.] (Don't take the religious imagery as a serious
reflection of my views; I used it only to help drive a point home.)
Too bad we couldn't get together at the Gordon conference to discuss
these things.  Will you be going to the Newton Institute meeting this
summer?  I'll be there June 16 through July 17.

\begin{enumerate}
\item
P.~A. {\Benioff}, ``Some Properties of the Contact between Theory and
Experiment,'' J. Math.\ Phys.\ {\bf 9}, 514--524 (1968).

\item
P.~A. {\Benioff}, ``The Imposition of Empirical Acceptability
Conditions on Sequences of Single Measurements,'' Z. Naturforsch.\
{\bf 24a}, 86--96 (1969).

\item
P.~A. {\Benioff}, ``Some Aspects of the Relationship between
Mathematical Logic and Physics.~I''  J. Math.\ Phys.\ {\bf 11},
2553--2569 (1970).

\item
P.~A. {\Benioff}, ``Some Aspects of the Relationship between
Mathematical Logic and Physics.~II''  J. Math.\ Phys.\ {\bf 12},
360--376 (1971).

\item
P.~{\Benioff}, ``A Theorem on Undefinability,'' J. Symb.\ Logic {\bf
36}, 377 (1971).

\item
P.~A. {\Benioff}, ``Decision Procedures in Quantum Mechanics'' J.
Math.\ Phys.\ {\bf 13}, 908--915 (1972).

\item
P.~{\Benioff}, ``Possible Strengthening of the Interpretative Rules
of Quantum Mechanics,'' Phys.\ Rev.\ D {\bf 7}, 3603--3609 (1973).

\item
P.~{\Benioff}, ``On the Relationship Between Mathematical Logic and
Quantum Mechanics,'' J. Symb.\ Logic {\bf 38}, 547--548 (1973).

\item
P.~{\Benioff}, ``On Definitions of Validity Applied to Quantum
Theories,'' Found.\ Phys.\ {\bf 3}, 359--379 (1973).

\item
P.~{\Benioff}, ``Some Consequences of the Strengthened Interpretative
Rules of Quantum Mechanics,'' J. Math.\ Phys.\ {\bf 15}, 552--559
(1974).

\item
P.~{\Benioff}, ``On Procedures for the Measurement of Questions in
Quantum Mechanics,'' Found.\ Phys.\ {\bf 5}, 251--255 (1975).

\item
P.~A. {\Benioff}, ``Models of Zermelo Frankel Set Theory as Carriers
for the Mathematics of Physics.~I'' J. Math.\ Phys.\ {\bf 17},
618--628 (1976).

\item
P.~A. {\Benioff}, ``Models of Zermelo Frankel Set Theory as Carriers
for the Mathematics of Physics.~II'' J. Math.\ Phys.\ {\bf 17},
629--640 (1976).

\item
P.~A. {\Benioff} and H.~Ekstein, ``States and State-Preparing
Procedures in Quantum Mechanics,'' Il Nuov.\ Cim.\ {\bf 40B}, 9--26
(1977).

\item
P.~A. {\Benioff}, ``Finite and Infinite Measurement Sequences in
Quantum Mechanics and Randomness: The {\Everett} Interpretation,''
J.\ Math.\ Phys.\ {\bf 18}, 2289--2295 (1977).

\item
P.~{\Benioff}, ``A Note on the {\Everett} Interpretation of Quantum
Mechanics,'' Found. Phys. {\bf 8}, 709--720 (1978).

\item
P.~{\Benioff}, ``On the Correct Definition of Randomness,'' in {\sl
PSA 1978: Proceedings of the 1978 Biennial Meeting of the Philosophy
of Science Association}, Vol.~2, edited by P.~D. Asquith and
I.~{\Hacking} (Philosophy of Science Association, East Lansing,
Michigan, 1981), pp.~63--78.

\item
P.~{\Benioff}, ``Randomness Yet Again,'' personal communication to
C.~A. Fuchs, 2 May 1998.
\end{enumerate}

\subsection{Paul's Reply}

\bq
It is very pleasing to see your research program direction as your
views have many points in common with mine.  In essence I believe in
the need for a coherent theory of mathematics and physics (as quantum
mechanics) that (1) refers to its own validity and completeness to
the maximum extent possible (TMEP) and (2) is maximally complete and
valid.  Such a theory is also a theory of everything. The  theory
will also include a description of quantum systems (e.g.\ us)
constructing and validating the theory. I have made these points in a
recent paper {\tt quant-ph/9811055} which will appear in Phys.\ Rev A
June 99.

Probably the most radical aspect of this viewpoint is that one does
not start with the basic aspects of physical reality as given. Rather
they are consequences of the requirement that a coherent theory of
math and physics refer to its own validity TMEP and be maximally
valid and complete.  (An equivalent way to state this requirement is
that the theory be maximally internally self consistent.) In this
view the basic aspects of physical reality, such as why there are 3
space and 1 time dimension and why the basic forces and particles
have the properties they do, is a consequence of these requirements.
Physical reality in its basic aspects is thus an emergent property.
It emerges from the more basic requirement of maximal internal self
consistency that a coherent theory of physics and mathematics must
satisfy.

More details on this viewpoint are in the material appended to this
letter. It represents a section I wrote for inclusion in the paper
you just read on ``Some Foundational Aspects--'' but had to delete
because of length restrictions imposed by the editor. It expresses
with minor differences my views today.  As always comments are
appreciated.
\eq

\section{19 March 2000, \ ``Probability Validation''}

Thanks for your encouraging remarks about the {\sl Physics Today\/}
article  {\Asher} and I wrote.

You write,
\bpb
Regarding your example of the weather prediction of 35\% rain, it is
true that there is only one unique time series of tomorrows.  But
this series is used by the observer and the weather predictors to
determine if the percentage predicted is valid or not.  I won't take
an umbrella if the predictor has no credibility with his prediction.
You did not discuss in your article how one determines validity for
probability predictions.
\epb
In an earlier draft, I wrote the following passage:
\bq
Here it is essential to understand that the statistical nature of
quantum theory does not restrict its validity to situations where
there is a large number of similar systems. Statistical predictions
do apply to single events. When we are told that the probability of
precipitation tomorrow is 35\%, there is only one tomorrow. This
tells us that it is advisable to carry an umbrella.  Probability
theory is the formal quantification of how to make rational decisions
in the face of uncertainty; this carries over as much to quantum
phenomena as it does to anything else.  When one makes a probability
statement concerning a quantum measurement outcome, one is
essentially making a {\it bet\/} about what will be seen.  As long as
the probability calculus is used, an adversarial gambler can never
force the bettor to a sure loss.
\eq

Of course this won't clarify too much for you, but it does push a
little further the direction in which I am thinking.  For the
Bayesian, there is no such thing as a valid or invalid probability
prediction \ldots\ as far as a particular numerical value is
concerned. The only empirically enforced criterion is that the
probability {\it calculus\/} be obeyed.  What I am talking about
above is something called the Dutch book argument.  You can find it
all nicely laid out in the following two books:
\begin{enumerate}
\item
H.~E. {\Kyburg}, Jr.\ and H.~E. Smokler, eds., {\sl Studies in
Subjective Probability}, Second Edition, (Robert~E. Krieger
Publishing, Huntington, NY, 1980).

\item
J.~M. {\Bernardo} and A.~F.~M. {\Smith}, {\sl Bayesian Theory},
(Wiley, New York, 1994).
\end{enumerate}

Below I'll place a piece of an exchange I had with  {\Asher} when I
was reading through a draft of one of his other papers.  They make a
further more detailed connection to your comment.  [See note to
 {\Asher} {\Peres}, dated 1 December 1998, titled ``Here Comes the
Judge.'']

\section{20 March 2000, \ ``Small Addendum''}

I was thinking about something you wrote me a couple of years ago,
and I thought I should say something about it before you readdress
the point in the context of yesterday's email.

\bpb
I have much criticism with {\Hellman}'s critique.  One main one is
that he does not distinguish carefully between classical mechanics
and quantum mechanics.  In CM randomness is epistemic as it is due
to an observer's lack of knowledge about the properties of a
system.  In QM it is present in principle and has nothing to do with
what an observer happens to know or not know or can know about a
system.
\epb

I agree with you wholeheartedly that in quantum mechanics the
``randomness'' of measurement outcomes is NOT epistemic.  However
that does not preclude my view that all probabilities (including
quantum mechanical ones) are epistemic in nature.  They quantify how
much we can say about a phenomena based upon what we know.  It so
happens in the quantum world that we cannot tighten up our knowledge
to the point of removing all ignorance (about the consequences of
our interventions), and in that sense the randomness is
ontological---it is a property of the world that was here long
before we ever showed up on the scene.  But it takes epistemic tools
to describe that property.  That's the direction I'm coming from.

\section{20 October 2000, \ ``Not Instructions, but Information''}

Let me try to answer your questions, which I think are good ones.
(I'm on a flight now heading for Texas and should have plenty of
time to answer you.)

\bpb
To me that is an interpretation of QM.  Interpretations are what
give otherwise empty theories their meaning.
\epb

You're quite right about that.  What  {\Asher} and I wrote about is
indeed a kind of interpretation of the quantum mechanical
formalism.  The title and the ending words of the article were more
for attention-getting than anything else.  Also, though, the words
were meant to be a small slap in the face to some of the extremes
people have gone to (like {\Everett} worlds, {\Bohm} trajectories,
and {\Ghirardi}-{\Rimini}-{\Weber} stochastic collapses) just to
hold on to a philosophic view that came around long before quantum
mechanics was ever heard of.  (Talk about people being set in their
ways!)

\bpb
If I recall correctly you wrote a paper with {\Peres} on QM without
interpretation but stated that a quantum state corresponds to an
algorithm for preparing it.
\epb

In this, though, you're confusing two things.   {\Asher} wrote a
paper in the early 1980s for AJP titled ``What is a State Vector?''
and in that paper he took the point of view you mention.  That is,
that a quantum state corresponds to nothing more than the
(equivalence class of) instructions for preparing a system one way
or the other. I've never felt completely comfortable with that point
of view, so in the paper that  {\Asher} and I co-wrote for {\sl
Physics Today\/} last year we worked around that.  I think also that
I may have even swayed  {\Asher}'s opinion on this issue, but you
would have to check with him directly.  (Perhaps I'll just carbon
copy this letter to  {\Asher}.) Here's how I put it in a note to
 {\Asher} 15 November 1999: [\ldots]

I believe this point of view is adequately expressed in the article
 {\Asher} and I wrote for Physics Today:  a quantum state is nothing
more and nothing less than one's best (probabilistic) information on
how a system will react to our experimental interactions with it.
How we may have come by that information---be it through a
preparation, through a sheer guess based on all available evidence,
or the principle of maximum entropy---is something I view as largely
outside quantum theory proper.  The structure of quantum theory
instead codifies how we should manipulate our information (this is
what time evolution and the collapse rule is about) and enumerates
the varied ways with which we may gather new information (this is
what the structure of observables or POVMs is about).

In this sense, I would call what we are talking about an
``information interpretation'' or ``Bayesian interpretation'' of
quantum theory, rather than an ``algorithmic interpretation.''  For
the most part, however, I would like to avoid the wording of an
interpretation.

I will go further---and this is one point where I may diverge from
 {\Asher}---and say that I do suspect that we will one day be able to
point to some ontological content within quantum theory.  But that
ontological statement will have more to do with our interface with
the world---namely that in learning about it, we change it---than
with the world itself (whatever that might mean).

\bpb
To my main question, which is related to your algorithmic belief.  In
essence exactly what is meant by saying a state is physically
preparable, an observable is physically measureable, or a Hamiltonian
physically realizable?  If one accepts the belief that a state is
preparable if and only if there exists a set of instructions for
preparing it, then, by cardinality arguments, most states are not
preparable.  The same holds for observable measurements and
Hamiltonian realizability.
\epb

There is a lovely line in John {\Wheeler}'s paper
\begin{itemize}
\item
J.~A. {\Wheeler}, ``World as System Self-Synthesized by Quantum
Networking,'' IBM J. Res.\ Develop.\ {\bf 32}, 4--15 (1988).
\end{itemize}
that he took from W.~V. {\Quine} about the real numbers being a sort
of convenient fiction.  I wish I could repeat that line now, because
I think it summarizes my point of view about this question.  In fact,
it's only the tip of an iceberg for me.

Let me give an example, suppose the {\it only\/} thing I know about a
quantum system (beside the dimensionality of its Hilbert space) is
the expectation value of some observable.  Suppose even that that
expectation value happens to be a rational number.  The only
question that's relevant to me is how do I use the information I
have to make an estimate of the probabilities of outcomes for any
other measurement I might make.  Someone who believed in an
ontological content for the quantum state itself might just stop and
give up at this point:  for to him the probabilities I am speaking
of are actual physical properties pertaining within the system.  But
for me that makes no never-mind:  Bayesian probability theory gives
a host of things I can do in situations of incomplete knowledge.  In
particular, {\Jaynes}' principle of maximum entropy is perfect for
this situation (where an expectation value is absolutely the only
information I possess).  Through the principle of maximum entropy I
am able to make a quantum state assignment, and when I have done
that, I have done the most I can do.

Now, that quantum state assignment allows me to say things about all
kinds of observables.  For some of these observables, it will
predict irrational even uncomputable probabilities.  In fact, it
will even let me predict probabilities for the outcomes of
uncomputable observables (I'll presume you understand what I mean by
this phrase).  Does that bother me?  No.  That is because these
numbers are derived from a quantum state which is in itself, as far
as I am concerned, a kind of convenient fiction.  These numbers and
that state simply stand for the best I can say about any potential
measurement (whether that measurement has a finite description
itself or not).

\bpb
The main problem with this view, which I think is probably right, is
what is the exact meaning of a set of instructions?  In essence the
problem is that, unlike the case for computable functions, there is
no {\Church}-{\Turing} thesis equivalent for preparable states,
measureable observables, or physically realizable Hamiltonians.
There does not seem to be some precise concept which is generally
accepted as equivalent to the informal notions of preparable states,
etc.

What are your views on this?  Can you shed any light on this problem?
\epb

Above, I said that seeing the real numbers as a convenient fiction
is just the tip of an iceberg for me.  This is because I suspect the
rational numbers themselves share no less in that sin.  Indeed I
sometimes think the that the mathematization of reality (in any form
or fashion) is nothing but a convenient fiction, one that helps us
coordinate our activities and communicate with each other more
easily.

Should there be some kind of kind of {\Church}-{\Turing} thesis for
quantum observables?  For all the reasons above, I suspect I'm less
moved by this question than you are.  But still I think one can make
some sense of it, if one insists.  If you are interested in my
detailed views on this point, I will try to articulate them in a
later note.

\chapter{Letters to {\Charlie} {\Bennett}}
\setcounter{section}{0}

\section{25 December 1997, \ ``{\Bennett} Festschrift''}
\medskip

\begin{center}
The Day I Was Worth 30 Pounds
\end{center}

Charles {\Bennett} is so much more than a great scientist:  his
personality is the source of great science.  In some ways, I never
feel more comfortable or secure than when I'm in his presence. Last
year, while visiting England, I decided on a whim to attend a small
quantum computing conference with a friend.  The friend had no
official travel funds, so we were trying to spend as little as
possible.  It turned out that {\Charlie} was at the same conference.

When time rolled around for the conference banquet, {\Charlie}
happened to catch me in the hotel lobby and asked, ``Where is this
banquet? You're going, aren't you?''

``No! It's too expensive for my blood; the price is 30 pounds.''  You
see, I thought I would sound a little noble that way.  I really did
want {\Charlie}'s company, but there was my absent friend to think
of, and he couldn't afford it.

{\Charlie} said, ``Well look, I make a lot of money now; I'll pay
your way.''

I was in a pinch!  Bad choice for an excuse: everyone knows what a
sponge I can be!  But, just then, my friend walked in, and I had an
out.  ``Actually we're going to try to find a pub or something.  He
feels the same as I do about this ridiculous price.''

``Oh come to the banquet.  I'll contribute 15 pounds toward each of
you \ldots\ for good conversation.''

``No, we can't really.  That's too much of an imposition.  A pub
would be so much better than a stodgy old banquet anyway.''
{\Charlie} yielded:  he would join us at the pub!  (But you must
realize:  there was no arm twisting.  It was his decision \ldots\
though with a little reluctance.  After all, he {\it was\/} an
invited speaker. The conference organizers {\it had\/} flown him all
the way to England. But no one had told him that he was expected at
the banquet \ldots\ or so he reasoned.)

We turned to the concierge to get directions and, in the process,
picked up another companion for the trek.  This one was a stray; none
of us had ever met him before.  But he was a nice fellow.  And he was
enthusiastic, wanting to learn everything he could about entanglement
and the analogy between it and a shared secret key. Off we went.

One block.  Two blocks.  Almost across the campus, when we run into a
nicely dressed man---jacket, tie, shiny shoes---who seemed to know
{\Charlie} like a friend.  ``Where are you going,'' he asked.
``Taxis for the banquet are this way.''

{\Charlie}'s ``little reluctance'' got the better of him.  He looked
to the slick fellow, ``Do you think I'm expected to be there?''

``Oh, why yes.  Of course.''  And he threw in a couple of proper
English ``hurrummphs'' for good measure.

{\Charlie}'s face turned a little red with embarrassment.  ``Look
guys we have to huddle.  I really should go to this banquet.  I think
that's probably expected of me; these guys did spring for my airfare
and all.  I tell you what, I know this banquet is expensive; I'll
contribute 10 pounds toward each of your meals.  Come on, we'll have
a good time.''

But there was really no changing the plan now:  we held fast, thanked
him for his kindness.  {\Charlie} went his way; we three went ours.

What does this have to do with science?  Nothing really.  But it does
have everything to do with the atmosphere of warm comfort that
{\Charlie} has built around his group at IBM.  I think it goes some
way toward explaining how so many wonderful results have come from
that little area of New York in the last few years. At the end of the
evening, after a little contemplation, I knew I was worth 30 pounds.
And that may be one of the nicest compliments I have ever received.

\section{05 May 1998, \ and to many others, \ ``Probability Does Not
Exist''}

To all of my friends who may have been influenced by Ed {\Jaynes} as
much as I was, I thought I should spread the word.  Professor
{\Jaynes} left this world last Thursday, April 30.  Beside the human
loss, we as scientists will lose so much from the fact that he was
not able to finish the magnificent book he was writing {\sl
Probability Theory:\ The Logic of Science}.  His guidance will be
missed.

\section{07 December 1998, \ and to the other teleporters, ``Swedish Bikini
Team''}

The Swedish television crew came and went yesterday with little
incident.  No questions at all about the soul this time:  I was
spared (and maybe so were you).  But I didn't pass up the chance to
try to get an unsuspecting audience to think deeply about the wave
function.  (This audience, in fact, really will be unsuspecting: the
footage was for a Swedish children's show.)

The question came, ``What material is teleported?  What is it that is
transported instantaneously from Alice to Bob?''

I said,  ``The only thing that is teleported in quantum teleportation
is what the preparer HAS THE RIGHT TO SAY about Bob's system.  It is
his description, his predictions, that jump instantaneously from one
system to the other.''

A blank stare, ``But what MATERIAL is teleported?''

``That is the material.''

He seemed pleased and left it at that.  When he left, I had a look at
my atlas to refresh the old memory:  Sweden was indeed not so very
far from Copenhagen.

\section{10 December 1998, \ ``More Swedish Bikini Team''}

\bcb
Why bikini? That is even harder to understand than meat slicer.
\ecb

Don't you remember the old beer commercials (I think it was Miller
Lite), that flashed their viewers with poses of the ``Swedish Bikini
Team''?  The team consisted of about 12 buxom, snow-white haired,
darkly tanned women in blue bikinis.  Why did I title my note after
them? Because I wanted you guys to read it; I figured the SBT was
just as relevant to my note as it was to drinking beer.

\bcb
By ``material'' I think they [meant] physical matter, with mass,
etc., not material as the word is sometimes used by wordsmiths to
mean information, as in the background material behind a newspaper
story.
\ecb

Oh, I'm sure of that.  That was meant to be implied by what I wrote;
did you not get it?  But the point I was slyly trying to make---I'm
sure it made no impression on them---is that not only is information
physical, but it's just about all there is.

\bcb
If they asked me what material is transported I would say none.  Only
information is transported, indeed a subtle kind of information that
would be spoiled by observing it.
\ecb

Maybe there's hope for you yet, Charles {\Bennett}.  Maybe one day
you actually will be induced to leave the tranquilizing comfort of
that silly church you attend (the one of the larger Hilbert space).
I wonder, and I wonder all the time, if the structure of quantum
mechanics isn't just a result of a more fundamental fact:  that
information gathering and disturbance (to others' information) go
hand in hand in our world.  In such a world, we construct the best
scientific theory we can subject to that constraint and \ldots\
(poof!) it's quantum mechanics.

By the way, since it's after 1:00 AM and I'm sleeping and dreaming
anyway, tell me this.  What could the phrase ``information is
physical'' possibly mean in an inanimate, mechanistic world?  Imagine
a world so unlucky that information processing units like ourselves
or my two dogs or Deep Blue never arose.  Indeed what would
information alone mean in such a world?  My only knowledge of
information---through the {\Shannon} theory---always requires a
probability distribution around in order to define information.  One
needs ``ignorance'' before one can have ``information.''  Do you
really feel comfortable with ignorance as a concept defined within
the church?  If you start talking about partial traces and von
Neumann entropies right off the bat, I probably won't be satisfied.
What gives the right to associate such things with ignorance?  Our
main reason now is that it matches the {\Shannon} theory in all the
appropriate cases.  And if you start talking about algorithmic
information right off the bat, I probably won't be satisfied either.
(For other religious reasons.)  But still, if you have time, you
should give it your best shot.

\section{10 December 1998, \ ``More Bikinis''}

By the way, let me ask you a more serious question.  You said:

\bcb
If they asked me what material is transported I would say none.  Only
information is transported, indeed a subtle kind of information that
would be spoiled by observing it.
\ecb

On the other hand, I had said:

\bq
I said,  ``The only thing that is teleported in quantum teleportation
is what the preparer HAS THE RIGHT TO SAY about Bob's system.  It is
his description, his predictions, that jump instantaneously from one
system to the other.''
\eq

Do you see our answers as different?  From my point of view, they
look about the same (except I didn't mention the disturbance business
here). If you see a difference, what is it?

\subsection{{\Charlie}'s Reply}

\bq
I have a couple of serious (for me) answers.  My first is that my
idea feels right, while yours is hard to get my head around.  The
capital letters make me think that even you may be using them to
overcompensate for an otherwise shaky idea. Also rights are the
province of lawyers and ethicists, a very ugly province indeed
compared to physics.  But I am probably wrong about both these
things.  Probably to one uncorrupted by many worlds, your idea seems
perfectly sensible and beautiful.  My second thought is the
realization is that you encountered {\Jaynes} in your formative
years, at around the mental age I was when I encountered whatever
makes me loyal to my Church.  In other words, in both cases, we
encountered what might be described as a pedophilic idea, an idea so
seductive as to be dangerous to our youthful selves and make us
dangerous to others when we grew up, or were thought by others to
have grown up.
\eq

\section{10 December 1998, \ ``500 Swedish Irregular Verbs''}

You made my Thursday morning!  Now I'll always keep our pedophilic
tendencies in mind when discussing foundational issues with you!  So
you're trying to make me feel perverse:  Is that because I shunned a
date with your photos?  Thanks for the commentary: it's quite useful
to see what phraseology causes trouble.  Discarding the ``right,''
however, I still don't see the difference between the words
``information'' and the ``sum total of what one can say.''  They're
both just euphemisms for the quantum state, no?

By the way, it's interesting that you wrote the title you did.  Just
yesterday I wrote two notes, one titled ``177 words of curly hair,''
and the other ``175 words that touch upon truth.''  The first
contained an abstract that Wojciech {\Zurek} wrote for a paper about
his existential interpretation of QM (whatever that is).  The second
was as an abstract I wrote for a paper that should capture the
essence of my pedophilia.  I used this trick to try help convince my
coauthors that my abstract wasn't too long --- probably won't work,
I use too many tricks.

\section{13 January 1999, \ and to many, many others, \ ``{\Emma}
Jane Fuchs''}

\noindent Dear Family, Friends, and Colleagues, \medskip

{\Emma} Jane Fuchs was born into the world healthy (and by all
indications happy) yesterday 12 January 1999 at 9:50 PM.  She is
beautiful like her mother and already striving hard to understand
her place in the world like her father.  The labor near the end was
a tough one, though looking back on it, it is no wonder:  8 lbs, 8.8
ounces is a lot of little girl to get through a birth canal.  {\Kiki}
is doing just wonderfully now, very proud of her daughter and the
cleft chins the two of them share.

On {\Emma}'s behalf, I send greetings to all. \medskip

\section{10 February 1999, \ ``Computational Power of N w/o E??''}

I'm going to be traveling and out of email contact for a few days,
but I wanted to record the following vague thought before it slips
out of my head \ldots\ and see if it elicits any reaction.

Have any of you thought about the ``computational power'' of
nonlocality without entanglement?   By this I mean the following.
What tasks can be performed efficiently by a quantum computer or
{\Turing} machine whose discrete time steps are always restricted to
take it from one separable state to another?  This of course means
that the state space available to such a computer is a
Cartesian-product space and not a tensor-product space---{\Ekert} and
{\Jozsa} make a big to do about the difference of the two products in
{\tt quant-ph/9803072} where they argue that entanglement is the
essential feature of quantum computation---but it's a heck of a lot
of a larger space than the Cartesian-product space of classical
states (all of which would be orthogonal presumably).  Don't you
think that the 9-state example hints that there just might be
something here? The key would be in making use of all the
nonorthogonal possibilities in the separate spaces.  Another hint
comes from the original {\Deutsch} problem (or was it D-J?), where
one need not entangle the target and the register to get something
resembling a true blue quantum computation.

Could we invent a {\it computational\/} example where something good
(something distinctly nonclassical) comes out of nonlocality without
entanglement?

\section{05 June 1999, \ ``{\Emma} Jane''}

\bcb
PS What is {\Emma} Jane up to these days?
\ecb

\begin{itemize}

\item
smiling at people when they smile at her

\item
smiling when one parent or the other rescues her from her lonely bed
first thing in the morning (sometimes there is a race to get this
lucky prize)

\item
laughing when her arm pits are tickled

\item
laughing when her dad mimics the dogs barking outside the window

\item
sleeping through most nights

\item
getting cranky most evenings just before bedtime

\item
bouncing with delight in her bouncing contraption that hangs from
the ceiling

\item
struggling to crawl (great sounds of exasperation when her legs and
arms move, but that belly on the floor just has too much friction)

\item
rolling over rather easily, but clockwise only (this is her best
form of locomotion presently --- on one occasion she's moved at
least four feet like this)

\item
taking in all the sights in her daily walks hanging from mom or
dad's belly --- I think this is her most content part of the day;
she never complains and stays awake as long as she can

\item
teething:  gnawing on her mom's fingers, gnawing on her little green
beany frog, gnawing on her teething blanket, gnawing on the edge of
her high chair, gnawing, gnawing, gnawing

\item
learning to voice her opinion when she wants to be played with
(``Damn it, what am {\it I\/} getting out of your cooking dinner?
Come play with me!'')

\item
eating solid---what a euphemism!---foods surprisingly well (``If any
falls on the edge of the table, not to worry:  I can suck it right
up.  You may need a spoon; I don't!'')

\item
getting familiar with the numbers 1 through 10:  one kiss (smooch),
two kisses (smooch, smoooch), four kisses (smooch, smoooch,
smooooch, smooch), three fingers, ...

\item
having the attention span for about one book per sitting

\item
loving her bath time, playing in the water, acting like an Olympic
swimmer

\item
inspiring her parents to great happiness

\item
giving her dad more reason to suspect that the world is so much more
than a mechanical contraption clinking along
\end{itemize}

\section{27 November 2000, \ ``An Emoticon''}

I like the title of your talk:
\bq
\noindent
Charles {\Bennett} is Technion's guest of honor in the program
``Israel Pollak Distinguished Lecture Series''. He will give two more
lectures: \medskip \\
Wednesday 6 December, 15:00, Physics Building, Auditorium 323
Lecture of general interest: \medskip \\
QUANTUM INFORMATION PROCESSING: Uses for a kind of information so
fragile that it cannot be observed without disturbing it
\eq

Now that sounds like a talk that even I would want to attend!  Are
you going to mention as one of your ``uses'' that it also gives us a
clue as to what quantum mechanics is about in the first place?
(Grin.)

\chapter{Letters to {\Herb} {\Bernstein}}
\setcounter{section}{0}

\section{25 December 1996, \ ``Reality Steaks''}

Holiday cheers!  I'm sitting in Geneva, connected to my machine at
Caltech, thinking about an old fart in Massachusetts.  I was just
reading an article (in the {\sl New Yorker}) about Woody Allen and
came across the most wonderful quote:  ``I hate reality, but, you
know, where else can you get a good steak dinner?''  Like it? Quantum
mechanics everywhere you turn.

\section{17 February 1997, \ ``Prophetic Herberts''}

This is the second time you've intrigued me with a phrase or two: I'm
not letting you off the hook this time.  Please explain in more
detail what you mean by the following.  The first quote comes from 25
December 1996:
\bhbe
Actually that Austrian reaction to enhancing classical communication
was probably part of a reality-loving or at least a
quantum-preferring inclination which is something beneficial to the
reality-seekers amongst us---if everyone were so crazed about
measurement and what-all, {\Charlie} would never have figured out
that it wasn't the knowing of an answer which introduced the
irreversibility; it was the erasure of the ``garbage'' produced in
the calculation.  And without all the fuss over reversible classical
computation, we wouldn't have had so much fun with quantum comp.
\ehbe
The second quote comes from 12 February 1997:
\bhbe
I was impressed similarly by {\Charlie}'s hint that thinking beyond
the ``big deal'' everyone since Szilard seemed to make of
measurements was crucial to his realization that the entropy
generated by Maxwell's demon for a 1-molecule gas came from
forgetting which side of the door it was on. The entropy generation
didn't come in measuring which side of the barrier contained the
molecule, and memory of the side would make the process reversible,
etc.\ etc.\ etc.
\ehbe
Are you saying that {\Charlie}'s solution of the ``Maxwell Demon
Paradox'' might have something to say about the solution to the
``Quantum Measurement Paradox''?  If so, let me know what you're
thinking---I'm intrigued.

\subsection{{\Herb}'s Reply}

\bq
I am saying that I suspect {\Charlie} was UNimpressed by the notion
that something very special occurs in the act of measurement {\it
per se} -- and I do mean UN.  He was able to see past where Szilard
went unlike many other people, because he didn't believe the
measurement automatically meant irreversibility.  And he analyzed
what it would take to make the ``measurement'' i.e.\ determination
of which side of the barrier the molecule was on recoverable; indeed
he pushed the analysis until he could say for sure the state was
unrecoverable, and discovered that the issue was getting rid of the
garbage that accumulates in a series of measurements, like the
garbage that accumulates during a long calculation.

So the moral of this long-winded story is that we who are impressed
with the quantum requirement to make a decision somewhere about what
it means to do certain interactions, i.e.\ those of us who remain
Bohrian enough to believe a phenomenon is not a piece of reality
until it is a registered phenomenon, must be cautious.  We admit it
is an assumption that this implies reality creation IS a part of
scientific inquiry and we examine this---and others---of our
assumptions always; by doing so we will also clarify when they are
wrong or misleading and when they just slightly misdirect our gaze.

In short there is a rather complex and muddled connection between C.\
{\Bennett}'s ``Maxwell's Demon'' work and our continuing
investigation of the need and privilege to CREATE the reality we
study in quantum mechanical situations of all kinds.
\eq

\section{23 February 1997, \ ``Herbertic Realities''}

Thanks for your comments on {\Charlie} and demons; I understand you
better now.  I especially liked your comment: ``\ldots\ indeed he
pushed the analysis until he could say for sure the state was
unrecoverable, and discovered that the issue was getting rid of the
garbage \ldots'' I'm not sure what I want to say about that, but I
want to say something---I just love the intent there.  What is the
difference between getting rid of information by dumping it into the
unknown and getting rid of it by spreading the information to an
uncontrollable group of free-willed communicators.  Give me a few
days to try to formalize it all.

\section{27 March 1997, \ ``Consistent Herb''}

In the next two mailings, I'll send you the correspondence I had with
{\Griffiths} on ``consistent histories.''  (The first is my writing,
the second is his.)  I would still like to construct a detailed reply
concerning this analogy he likes to draw between ``frames of
reference'' in special relativity and ``Boolean frameworks'' within
his version of quantum mechanics.  I don't like the analogy and I
think it is misleading precisely because quantum mechanical
probabilities are not about real existent things that are or are not
the case---as best I can tell, they are solely about contingencies.
Alas, though, my life and time are not infinite.  (I'll get to it
eventually.)

I had a very nice conversation with {\Charlie} on the drive back to
Wendell from your house, that, I think, has allowed me to sharpen
what I think about things in quantum mechanics.  In particular, I
would like to work on a point of view that substitutes in place of
the ``Church of the Larger Hilbert Space,'' something along the lines
of ``Church of the Big-Enough Hilbert Space.''  However, it'll
require some writing for me to get it down coherently.  I'll try to
combine it with my thoughts on {\Mermin}.

Anyway, I had a really great time last week at your place.  This
quantum mechanics business is so wonderful, it's almost no surprise
we can entertain ourselves with that, a few fried potatoes, and not
much more!  Greetings to Mary and the family.

\section{29 March 1997, \ ``More Reality''}

Did you, by chance, read the article titled ``Get Unreal'' in the
March 17th {\sl New Yorker}?  In case you didn't, I'll share with you
my favorite quotes from it (that I want to tabulate anyway \ldots\ I
love to use friends as good excuses).

\bq
``People experimenting with high-definition television, or HDTV
\ldots\ report that the resolution of the image is such that details
invisible on ordinary television screens, like scuff marks on the
anchorperson's desk \ldots\ leap out at you on HDTV.  The Times
predicted the other day that television production designers,
formerly accustomed to tacking a studio set together out of plywood
and duct tape, will have to start building the real thing.  On
standard television, mahogany stain looks like mahogany.  On HDTV, it
looks like stain.  To maintain a realistic image, in other words,
television will have to construct a better reality.  The simulation
is getting ahead of the simulated.''
\eq
\bq
``In the case of the music industry, the rationale for persuading
people to ditch their phonographs and LPs and replace them with CD
equipment was that digital CD sound was more lifelike than the analog
sound of the old LPs.  Actually there was nothing wrong with analog
sound.  The ear learned to block out the occasional hiss and pop, and
found the result adequately lifelike.  You do not, after all, listen
to live music in a soundproof chamber.  \ldots\  Once CDs became the
standard, though, {\it they\/} became `lifelike,' and analog
recordings started sounding wobbly and staticky.''
\eq
\bq
``Human kind cannot bear very much reality,'' says the little bird in
T. S. Eliot's ``Four Quartets.''
\eq
\bq
``The craving for more and more realistic representations of reality
is at bottom an aesthetic craving, and one that people are perfectly
capable of indulging purely for its own sake---an indulgence
beautifully illustrated by the tremendous expense and effort being
invested in getting a computer image to look and function exactly
like a piece of paper.  But the effort to satisfy such cravings kind
of misses the point of what the true pleasure of representation is.
The true pleasure of a representation does not come from its
indistinguishability from the real thing.  It comes from its
distinguishability.  An Elvis impersonator gives pleasure precisely
because he's not Elvis, and it is crucial to the effect that we never
forget he's not.''
\eq

\section{07 April 1997, \ ``Howling Quanta''}
\medskip

\begin{flushleft}
\parbox{2.9in}{
``The universe is a thing of dream\\ substance naught \& Keystone
void\\ vibrations of symmetry Yes No\\ Foundation of Gold Element
Atom\\ all the way down to the first Wave\\ making opposite Nothing a
mirror\\ which begat a wave of Ladies marrying\\ waves of Gentlemen
till I was born in 1926\\ in Newark, New Jersey under the sign of\\
sweet Gemini''\\
\hspace*{\fill} --- {\it Allen {\Ginsberg}}\\}
\end{flushleft}
I guess you've heard that Mr.\ {\Ginsberg} passed on?  Maybe that
makes today the appropriate day for writing this note.  You ask, why
are we (you and I) so in sync on this issue of reality creation?  I
suspect it's most likely a chance fluctuation \ldots\ or a
``selection effect''---why else would {\it we\/} end up working on
quantum information.

Of course, the 60's did play a part, apparently for both of us.  But
for me, I suspect, in a very different way than for you.  I was a kid
in south Texas in the late sixties, very far from hippiedom.  My
sixties were filled with conservative democrats, cowboys, oil fields,
and an occasional trip to the big city of Houston---there was no
great connection to a world of ideals.  I think surreality slipped
into me from a much smaller world of turmoil.    Somehow in the
early years I gained the habit of making things look magical to
myself.  In my head, the most mundane objects became spaceships.  For
years I imagined that I (and an imaginary friend) were in search of
the ``end of space.''  The fantasy mostly consisted of us traveling
outward, and building ever faster vehicles as we were doing so. There
was no time to waste!

Did this shape my thoughts on quantum mechanics and, in particular,
make me more receptive to it as an almost mystical structure?  I
think so.  I'll place a couple of passages below that build some
imagery about it.  The first I wrote sometime in 1993; the second is
a little poem that came out in the DFW airport about a month or two
ago.

You asked me if I have ever thought about the immense responsibility
entailed by the possibility of reality creation.  Yes I have, but
never in any very organized fashion.  In relation to this, my
thoughts keep coming back mostly to the Holocaust, but also to
instances of self-devastation like with the Jim Jones gang in Guiana.
I think it was just this sort of thing that {\Einstein} feared most
about quantum mechanics:  that reality \ldots\ and consequently right
and wrong(!)\ \ldots\ might even be creations of the participants.
(By the way, you can find an expression of this in the bad ``free
form'' poetry below.)  I understand the fear of E \ldots\ but, also,
I don't know how to get around it.

\bq
\noindent Everyday Albuquerque reveals.  Hollywood knew the quantum;
giant scorpions in a desert.  Fluid reality.  The shape of
entertainment. Notorious youth making out in cars; paying no
attention to surreality itself.  Paying no attention to the mystery
and depth.  There is a faint image that lives within me; sometimes
it speaks.  And sometimes I know it contains the room across the
hall. Sometimes I know the room across the hall was foreseen
somewhere. Something is built in. Somehow it registers; it picks out
a unique reality. The tome of this world.  {\Einstein} must have
foreseen the evil in our dear quantum.  I don't envy the pain; the
blood is so very frightening.  No evil in $A$-bomb, but the evil of
no image; the evil of no substrate; the burning people in Waco; my
father tied to a bed.  The poor roaming wisdom of E. On a movie set;
money to be gained.  Sexuality in images for money. This I see in
THE Friday night.  And this gnaws; frightens; feeds; asks for
stability.  An image that asks for expression.  But with mysticism
that can't be controlled.  In paint delicately placed on a manifold.
Spacetime is there somewhere in all the projects asking for flight.
I see jeeps with water canisters on the back.  I see a snake near a
water faucet.  I see a child giving up.  Emptiness in a world
nonexistent. Emptiness as definition; and emptiness for all its
sake.  There was an old stump with a few nails of differing size
driven into it; it was a console and control.  There were mysteries
at the grass and at the curb.  Tall grass; dreams of soldiers; boys
dying across the world. The fence near the control protected; gave
barrier.  The nurturer of surreality was somewhere there too.
{\Orwell} came to visit somewhere in those dreams, at the control
and at the creek. He gave me a gun from afar, and a yellow bicycle.
And with the gun I learned to count.  And with the bicycle I learned
fear and dread. Dreams of reason in this reasoned state.  {\Orwell}
saw my youth from afar; and then gave his mind away.  His novel is my
world; the scaffold on which to build what can be built.  {\Orwell}
shaved his head and came to visit when I was sick. The tears came
when they said ``love.''  Albuquerque made its presence known
somewhere in Texas long ago.  The Germanic blood in my veins saw it
then.  The
world was portended. \\
\hspace*{\fill} --- {\it CAF in a weird mood, 1993}
\eq

\section{05 June 1997, \ ``Dreams of a More Ethereal Quantum''}

I thought I'd just say hi and let you know that I've been visiting
``reality town'' for the last couple days.  I've finally (after all
this time!) really started thinking about {\Mermin}'s Ithaca
Interpretation.  (He's now jokingly calling it the ``75 Hickory Road
Interpretation'' since not everyone else in Ithaca is so enthused
about it.  {\Charlie} B. asked him if everyone in his family was
actually in agreement!)  Much of this was spurred by my meeting him
last week in Montr\'eal.

I'm not sure what to make of it.  I guess I think it's all too vague
for me to have any strong opinion one way or the other yet.  This has
been helped out by my impression that {\Mermin} himself doesn't
really know what he's trying to get after yet.  The key ideas seem
to be:

1) The density operator for a ``system'' is a convenient
icon/shorthand for all possible correlations between all possible
``subsystems.'' However, the quotation marks in the preceding
sentence are absolutely necessary; for it is the correlations
themselves and not the individual properties of the systems that form
the ``elements of reality.''

2) The real interpretational problems in quantum theory are really in
understanding a proper notion of ``objective probability.''  The only
hint we have for understanding such a notion presently is the quantum
theory itself---the very thing we were working hard to interpret in
the first place.  [This attitude, by the way, traces back to at least
his 1983 essay ``The Great Quantum Muddle'' critiquing {\Popper}'s
books, where he writes, ``\ldots\ what is most marvelously intricate
and subtle in the behavior is just a mystery and a horror to be
dispelled by some clear thinking about [[objective]] probability.'']

You can see from Appendix B in the paper (on the {\Hardy} Paradox)
that he wants to tie these two ideas together very closely.  Namely
that the flavor of objective probability is that some conditional
probabilities cannot be properly formed.  However, I don't see really
how that is so distinct from Copenhagen: ``unperformed measurements
have no outcome.'' What is it that's really new here? (\ldots\ beside
the over-use of the phrase ``objective probability''?)  What am I
missing?  I guess the idea is to use the standard interpretation to
figure out which (conditional) probability statements are meaningless
and then abstract/bootstrap these properties into the new
interpretation.

I just looked back at the last note you wrote me on this subject. I'm
not sure I can answer your questions.  Have you thought about them
more?  The best I can say is that I don't think he intended the
density matrix as the end-all and be-all of everything \ldots\ only
of the correlations between the subsystems which it encompasses.  The
density operator itself says nothing of the how the overall system is
correlated to the rest of the world (which I think is what your
example was getting at).  Apparently the only thing that one can
glean \ldots\ or, at least this is what I think he'd say \ldots\ is
that if the state is not pure then the system of interest is
definitely correlated to something without (i.e., external to it).

I'll keep thinking about these things.  Within a day or two I'll be
composing a note to {\Mermin} on ``objective probabilities''; I'll
{\tt cc:}\ you something when it's written.

In the mean time, I place below---for your amusement---a note I just
wrote  {\Asher}.  I think it contains the most succinct statement
I've written yet concerning the program of ``Mechanica quantica ex
mente orta lex est.''

\section{04 August 1997, \ ``The Nielsbohricon''}

Anyway, I was sitting here feeling low about not having yet finished
the long note on quantum probabilities that I am in the process of
writing you.  And I was thinking about your Niels {\Bohr}
``roundtable'' plans.  You guys at the liberal artsy places get to do
such fun things!  In case it's of any use to you, below I give a
small offering of a ``starting point'' for your fun and games:  these
are some of the things that I myself have perused and found useful.
(The list has no pretension of being complete.)

I think these things are useful to read \ldots\ not because they'll
paint a consistent picture of what {\Bohr} thought (I doubt there's
one to be painted) \ldots\ but because they give a good hint of all
the issues that our own project must come up against.  The reality
project, that is.  I agree with {\Beller} in that,  ``My aim is not
to cure this `schizophrenia' by eliminating the inconsistencies, but
to analyze the sources, uses, and aims of such shifting philosophical
positions \ldots''

I think the best things in the list are {\Honner}, {\Holton}, and
{\Faye}'s three books.  {\Kalckar} is good because it has a lot of
previously unpublished personal letters to and from {\Bohr} on meaty
reality issues. If you're looking for the most on M{\o}ller's book,
take a look at {\Feuer}, {\Faye}, {\MacKinnon}, and {\Holton}.  The
collection edited by {\Faye} and {\Folse} is really good too.  I'd
especially recommend ``Description and Deconstruction: Niels {\Bohr}
and Modern Philosophy,'' by John {\Honner}, in it.  The book by
{\Plotnitsky} was strongly recommended by David {\Mermin}, but I
haven't made much headway into it---the subject is
deconstructionism, {\Derrida}, and the connection of all that to
{\Bohr} and {\Goedel}.

I have still several more things to put in the list for you, but
you'll have to give me some time to dig them up.
\begin{enumerate}

\item
Mara {\Beller}, ``The Rhetoric of Antirealism and the Copenhagen
Spirit,'' {\it Philosophy of Science\/} {\bf 63}, 183--204 (1996).

\item
Jan {\Faye} and Henry J.~{\Folse}, eds., {\it Niels {\Bohr} and
Contemporary Philosophy}, (Kluwer, Dordrecht, 1994).

\item
Jan {\Faye}, {\it Niels {\Bohr}: His Heritage and Legacy (An
Anti-Realist View of Quantum Mechanics)}, (Kluwer, Dordrecht, 1991).

\item
Lewis S.~{\Feuer}, {\it {\Einstein} and the Generations of Science},
(Basic Books, New York, 1974).

\item
Henry J.~{\Folse}, {\it The Philosophy of Niels {\Bohr}: The
Framework of Complementarity}, (North Holland, Amsterdam, 1985).

\item
A.~P. French and P.~J. Kennedy, {\it Niels {\Bohr}: A Centenary
Volume}, (Harvard University Press, Cambridge, MA, 1985).

\item
Gerald {\Holton}, {\it Thematic Origins of Scientific Thought:
{\Kepler} to {\Einstein}}, Revised Edition, (Harvard University
Press, Cambridge, MA, 1988).

\item
John {\Honner}, {\it The Description of Nature: Niels {\Bohr} and the
Philosophy of Quantum Physics}, (Oxford University Press, Oxford,
1987).

\item
J{\o}rgen {\Kalckar}, {\it Niels {\Bohr}, Collected Works: Volume 7,
Foundations of Quantum Physics II (1938-1958)}, (Elsevier, Amsterdam,
1996).

\item
Edward M.~{\MacKinnon}, {\it Scientific Explanation and Atomic
Physics}, (University of Chicago Press, Chicago, 1982).

\item
Dugald {\Murdoch}, {\it Niels {\Bohr}'s Philosophy of Physics},
(Cambridge University Press, Cambridge, 1987).

\item
Abraham {\Pais}, {\it Niels {\Bohr}'s Times, in Physics, Philosophy,
and Polity}, (Clarendon Press, Oxford, 1991).

\item
Arkady {\Plotnitsky}, {\it Complementarity: Anti-Epistemology after
{\Bohr} and {\Derrida}}, (Duke University Press, Durham, NC, 1994).
\end{enumerate}

\section{12 October 1997, \ ``Reality as Realty''}

I'm just thinking of you, as I often do on my leisurely Sundays.
Today's treat comes from the LA Times Book Review: an article by
Martin {\Gardner}, titled ``Mathematical Realism and Its
Discontents.''
\bq
Consider $2^{1,398,269} - 1$.  Not until 1996 was this giant number
of 420,921 digits proved to be prime \ldots\ \@.  A realist does not
hesitate to say that this number was prime before humans were around
to call it prime and that it will continue to be prime if human
culture vanishes.  It would be found prime by any extraterrestrial
culture with sufficiently powerful computers.

Social constructivists prefer a different language.  Primality has no
meaning apart from minds.  Not until humans invented counting
numbers, based on how units in the external world behave was it
possible for them to assert that all integers are either prime or
composite \ldots\ \@.  In a sense, therefore, a computer did discover
that $2^{1,398,269} - 1$ is prime, even though it is a number that
wasn't ``real'' until it was socially constructed.  All this is true,
of course, but how much simpler to say it in the language of realism!
\eq

The last sentence just pegs it for me.  Take reality when you can, I
say!  But when it's simpler to let it go, you've got to do that too.
As I see it, that's the place quantum mechanics leaves us.  For
instance, {\Bohm}'s realistic version of QM seems to be perfectly
consistent and to give the same observable results.  All this is
true, but how much simpler to say it in the language of anti-realism!

Oh well, enough of that.

\section{13 November 1997, \ ``Tria Juncta In Uno''}

I'm starting to think that writing proposals might be more useful
than I could have imagined.  Look at what I just found in my
electronic thesaurus for the word ``REAL''.  In particular notice
``well-documented''!
\bq
\noindent REAL\@: real, essential, substantive, substantial,
not imagined, uninvented, actual, positive, factual, genuine,
well-documented, historical, grounded, well-grounded, true, natural,
of nature, physical, flesh and blood, material, concrete, solid,
tangible, dense
\eq

\section{23 November 1997, \ ``Sunday {\Baudrillard}''}

I'm reading Jean {\Baudrillard} and eating {\Kiki}'s famous sourdough
raisin-pecan biscuits this morning.  Ever heard of the guy?
{\Baudrillard}, that is.  Anyway, I ran across the following little
quote:
\bq
\noindent The transition from signs which dissimulate something to
signs which dissimulate that there is nothing, marks the decisive
turning point.
\eq
When do you think that was?  1926-27, maybe?

\section{28 November 1997, \ ``Muddling Through Reality''}

I look forward to reading your book:  just send it when you can.  (If
it arrives here before December 20, I can take it to Texas with me
\ldots\ for reading on my vacation, Dec.\ 20 through Jan.\ 2.)

Yes, {\Baudrillard} is contemporary.  The reference to 1926--1927 in
my first mention of him was a joke!  It was meant to signify the
birth of quantum mechanics.

\section{10 December 1997, \ ``Reality Alert I''}

I've got a tip for you!  Yesterday, I ran across a wonderful article
that I know will please you immensely.  It's by S. S. {\Schweber}:
``The Metaphysics of Science at the End of a Heroic Age,''  in {\em
Experimental Metaphysics}, edited by R.~S. Cohen, M.~Horne, and
J.~Stachel (Kluwer, Dordrecht, 1997), pp.\ 171--198.

I think one of the S's in {\Schweber}'s name stands for Sam \ldots\
is that true?  Anyway, the subject is roughly speaking ``law without
law'' \ldots\ the mutability of physical laws \ldots\ the creation of
reality \ldots\ and an evolutionary universe.  Roughly all the stuff
of which {\Wheeler} used to speak.  The twist here is that he spends
a little time exploring ethics and morals in such a kind of world.
Let me give you an excerpt:
\bq
\noindent Since \ldots\ the scientific enterprise is presently
involved in the creation of novelty---in the design of objects
that never existed before in the universe and in the creation of
conceptual frameworks to understand the complexity and novelty
that can emerge from the {\em known\/} foundations and
ontologies---they must assume moral responsibility for these
objects and representations.  My emphasis on the act of creation
\ldots\
\eq
and, quoting C.~S. {\Peirce},
\bq
\noindent Under this conception, the ideal of conduct will be to
execute our little function in the operation of the creation by
giving a little hand toward rendering the world more reasonable
whenever, as the slang is, it is `up to us' to do so.
\eq

Besides, the article also has a lot of wonderful references to stuff
on ``law without law'' that I hadn't known about.  You should take a
look. (There is, however, an annoying thing in it:  (1) he
consistently misspells {\Peirce}'s middle name---it is Sanders not
Saunders.)

David {\Mermin} spent four days here last week, and we had a lot of
time to discuss the ``Ithaca Interpretation of QM.''  I would say
that, to a large extent, he's on our side of the game.  I.e., the
anti-realism side.  That, somehow, though hasn't come out in this
particular twist of his thinking.  I showed him something you once
wrote me complaining that he's slipping away from our side, and he
had a good laugh.  He's a very good guy; I like him a lot.

\section{10 December 1997, \ ``List of Sins''}

\noindent {\Herb}ert, {\Herb}ert, {\Herb}ert, \medskip

I am so disappointed in you.  What's it take for a guy to be treated
with respect by you \ldots\ perhaps a {\it faculty\/} appointment at
Caltech??? Let me list your sins:

\bv
{\bf 1)} Not only did you send {\Hideo}'s copy of your book by
priority mail (\$4.00 postage), while only sending mine by ``special
standard mail,'' \medskip
\\
but \medskip
\\
{\bf 2)} You placed \$1.28 of postage on it when it apparently
required \$2.24 worth.  Caltech had to pay 96 cents for me to have
the privilege of your words!
\ev

I just hope you know that I'm hurt.  Please return the e-mail I wrote
you this morning about {\Schweber}'s article:  I can no longer bear
knowing that I'm sharing my thoughts with you.  (This time though,
please use the correct postage!)\medskip

\noindent Christopher \medskip\\
\noindent PS.  The zip code at Caltech is 91125, not 91109.

\section{27 December 1997, \ ``Reality Alert II''}

Thanks for your comments about {\Schweber}, etc.  This ``reality''
issue is becoming more acute, isn't it?  I've very enthusiastic
about it all again lately.  I read the preamble and the amble of
your book the other day on the plane.  So far at least, I've been
enjoying it immensely.

{\Mermin} really is a good guy.  His take on reality is, I think,
somewhere between the {\Bernstein}-Fuchses of the world and the
hardcore realists.  In general, he gives a thumb's down to reality
\ldots\ so I guess that makes him much closer to our side of the
mark.  He just doesn't seem to think that quantum mechanics itself
is about reality creation.  That takes place---according to him,
that is \ldots\ or at least I think so---at a somewhat higher level.

I wrote him an extensive critique of the Ithaca Interpretation (in
its latest incarnation).  After he puts his long paper on the
archive, I may make my note available for public perusal (but it's
not something I would put on the net).  In general, there is a lot
that I like about his way of viewing things---but I think it points
much more strongly than he believes to the idea that quantum
mechanics is about {\it information\/} and nothing more.  By studying
him, I've been able to come to much better grips concerning what I
think quantum mechanics is about.  I've been summarizing my latest
understanding with the pithy phrase, ``{\Bohr} was Bayesian!''

\section{29 December 1999, \ ``Cradle of Reality''}

Guess where {\Kiki} and I are heading off to this morning?  Austin,
Texas: for me, the cradle of reality-creation.  It was there that I
first became acquainted with John {\Wheeler}'s ideas, and somehow the
town has a mystical hold on me.  Wish me luck for a little more
insight.

\section{04 January 1998, \ ``{\Bohr}, Bayesians, and {\Bernstein}s''}

I'm back in California now.  On the ride home, I managed to get
another chapter read from your book with Mike {\Fortun}.  (Yes, I
would like to meet him.)  I'm sorry that my reading is going slower
than I had imagined it would.  I just found that I was in hardly any
mood to be intellectual over the holidays.  So I still can't give
you much feedback on it other than that I am wholly in agreement
with the point of view espoused in Chapter 1.  Most important to me,
in particular, is the issue of articulating what exactly this new
category ``realit{\it t\/}y'' consists of.  What concept or category
can we take to replace ``reality'' in a world where our actions
actually make a difference?  I think that is a lovely question!!  I
don't think I had ever seen it posed so pointedly before reading this
Chapter.  So thanks for giving me the opportunity.  (If I can dig
them up, I'll send you some of my old meager attempts at {\it even\/}
formulating this question---I didn't come quite so close to the mark
as you guys have!)

Thanks for commenting on my phrase ``{\Bohr} was a Bayesian'' and the
idea that quantum mechanics is about information.  Actually I didn't
mean by it anything so detailed as {\Bohr}'s particular take on the
``frequency interpretation of probability,'' etc.  I think what I
really mean by the phrase is encapsulated in the following
two-sentence explanation.  {\Bohr} was willing to own up to the idea
that quantum theory is a theory of ``what we have the right to say''
in a world where the observer cannot be detached from what he
observes.  He was willing to own up to the idea that it is that and
nothing more.  (Making these two sentences rigorous and useful is
roughly the big picture of my present research program.  Did you, by
chance, read the research proposal that I sent you?  If you have any
feedback, I {\it would\/} like to hear it.)  My most up-to-date
attempt to say this clearly is in a couple of notes---one to {\Comer}
and one to {\Mermin}---I'll go ahead and forward them on to you in
case you're interested.  (Read the one to {\Comer} first, for the
overview.)

Oh yeah, and on a related note, I did want to comment on something
you said,
\bq
\noindent I got the idea that ``Ithaca'' was about the possibility
that q-mech somehow was only connected to information. To my mind
this is a bit close to the folks in our field who think the universe
is only information. That always seemed to trivialize what a
marvelous theory quantum mechanics is -- both philosophically and
practically. Just misses the whole point, like the related stance of
many-worldism.
\eq
To say something like that is to ``ontologize'' information.  When I
say, ``Quantum mechanics is almost totally about information,'' that
is not at all the sort of thing that I am trying to imply.  Instead,
I am trying to relate that almost all the formal structure of quantum
theory (i.e., Hilbert spaces, state vectors, unitary evolution,
etc.)\ is not really about {\it physics\/} at all.  There is almost
no physics in that formal structure:  it is rather something of a
more Bayesian-like character, it is about the formal tools for
describing what we know.  The {\it physics\/} behind quantum theory
is that: ``reality'' {\it must\/} be replaced by ``realit{\it t\/}y''
(to borrow a little from your terminology).

\section{14 January 1998, \ ``Quote-A-Day''}

New quote.  (My computer gives me one at random every day.)  I'm not
sure why I like it, but somehow it strikes me as having something to
do with the ``muddled middle'' you speak of in your book.
\bq
\noindent The very purpose of existence is to reconcile the glowing
opinion we have of ourselves with the appalling things that other
people think about us.
\eq
It's by Quentin Crisp (b.\ 1908), British author, from {\sl How to
Become a Virgin}, ch.\ 2 (1981).

\section{25 January 1998, \ ``Shame Dance''}

I'm ashamed to say, but only now I've just finished reading Chapter 2
(``Articulating Experiments'') of your book with {\Fortun}.  Time
(and much of my sanity) has just slipped away from me this semester:
true to your analysis, my life is very much a kludge job.  All I can
say still is that I very much like this idea of ``realit{\it t}y.''
It shines through to me as the most important of ideas.  But with
all the silly travel that I have to do this coming month (and the
preparation of talks for it), it'll probably be some time before I
can really get into the book again.  Let me apologize for that right
now:  I know that I'm not being of very much use to you in your
publishing process.  But there certainly is enough in the book to
keep me wanting to come back to it.

\section{02 March 1998, \ ``Early Morning Japan''}

It's early morning in Japan right now, and I'm having trouble
sleeping. My clock's still running on California time.  I was
thinking of you, wondering how you're doing.  I've spent the last
few listless hours in bed trying to compose a little realit{\it
t\/}y, trying to give the world a little form in exchange for the
form it gave me.  You should see this joint I'm staying at!  It's
the Tamagawa University guest house, and I'm sure by Japanese
standards it is a palace.  Very comfortable. Alexander {\Holevo} and
I are sharing it for seven days.  Then I move on to visit the
Communications Research Laboratory of the Ministry of Posts \&
Telecommunications for three days.  I've been warned that my living
arrangements there will not be so nice.

\section{08 March 1998, \ ``The Soul of Daibutsu''}

This is just a short note to wish you well before I leave Japan.  A
few days ago I saw the most moving sight: The Great Buddha of
Kamakura, Daibutsu.  It's a 44 foot tall bronze buddha; I was even
able to walk inside its belly.  Dreams of philosophy made my heart
pound.  At a nearby shrine, I placed a 500 yen wish on their ``wish
board'' (with about 10,000 previous wishes):  Wish for Realit{\it
t\/}y.  I hope you won't mind my pirating your phrase, but there was
just nothing else to better describe what I was truly wishing for at
just that moment. \medskip

\noindent Most honorable regards.  Your faithful bhikku, \medskip\\
\noindent Christopher

\section{29 March 1998, \ ``Philosophical Purgation''}

I'm doing my usual thing of being philosophical on Sunday.  I've come
across a little piece of an interview with David {\Bohm} that I sort
of like the flavor of \ldots\ so I'm entering it into my usual Sunday
database: my friends.  I'll pick on you because I know that your wife
(and maybe you) resonates with some of Mr.\ {\Bohm}'s ideas.

\bq
Q:  You said that there was difficulty in understanding quantum
mechanics. \medskip

A:  Yes.  I think that the difficulty is that we have no way of
understanding what is actually happening, or what I call the {\it
actual fact}.  If I may paraphrase {\Bohr}, we have only the
phenomena, i.e., the observed phenomena, which are essentially
classical in their description.  Ordinary classical phenomena---the
observation of a dot or a click---were previously understood to
signify information about particles, and the particles were
independent of these phenomena.  Now, if you analyse the
{\Heisenberg} microscope experiment, you come to the conclusion that
the experiment cannot give you unambiguous information about the
structures you are supposed to be observing.  Therefore, there is no
clear way of considering the unknown reality which is responsible
for the experimental results.
\medskip

Q:  Wouldn't {\Bohr} have said that this is a fundamental property of
the world? \medskip

A:  In effect he did say that.  I don't think he ever said it
directly, but it was implied.  But if he said that it is fundamental,
then I ask: how does he know it's fundamental?  It's only fundamental
as long as the present theory works, and there are many ways in which
it doesn't work, as we know.  We certainly just can't accept it on
authority that it is fundamental.  \ldots\
\eq

I think the reason I like this little passage is that {\Bohm} shows
that he did indeed have a good understanding of the Copenhagen
attitude. He simply chose to reject it.  There's nothing in my mind
better than an honest assessment of one's motives.  I myself keep
playing with the idea that all of quantum mechanics is about one
fundamental ontological statement:  {\Wheeler}'s game of twenty
questions. Almost all of the detailed structure of the theory is
about reasoning, betting, assessing, and dealing with a world
possessing exactly that fundamental property.  Once we clear up that
that is what quantum mechanics is really about, then we will really
be in a position to find something truly new and wonderful.  So the
point that {\Bohm} wishes to question is just the point that I want
to take as basic, the one I wish to see if it can be built upon.

That's all.  Happy philosophy.

\section{02 January 1999, and to {\Greg} {\Comer}, \ ``A New Year's
Toast''}

Happy New Year!  Please allow me to toast it in with a small gift,
one that I think concerns all three of us.  It's a beautiful little
article from {\sl Science\/} magazine by John {\Banville} (a
reporter at the {\it Irish Times\/} in Dublin).  I scanned the
article in in its entirety because the whole thing seemed so
relevant.  It contains a truth that I think we three in particular
know all too well \ldots\ a truth that we three in particular want
to see become a {\it productive\/} truth.

\bq
\begin{center}
Beauty, Charm, and Strangeness: Science as Metaphor\\ (From Science
Magazine Online)\\ by John {\Banville}
\end{center}
\medskip

\small I wish to advance a thesis which, were they to take note of
it, the academies would decry as scandalous.  My thesis is that
modern science, particularly physics, is being forced, under
pressure of its own advances, to acknowledge that the truths it
offers are true not in an absolute but in a poetic sense, that its
laws are contingent, that its facts are a kind of metaphor.  Of
course, art and science are fundamentally different in their
methods, and in their ends.  The doing of science involves a level
of rigor unattainable to art.  A scientific hypothesis can be
proven---or, perhaps more importantly, {\it disproven}---but a poem,
a picture, or a piece of music, cannot. Yet in their {\it origins\/}
art and science are remarkably similar. It was a scientist, Niels
{\Bohr}, who declared that a great truth is a statement whose
opposite is also a great truth.  Oscar Wilde would have agreed.

Since the Enlightenment, the chasm between art and science has yawned
ever wider with each new stage in the campaign to subdue nature to
man's will.  The human race cannot abide nature's indifference, and
uses the physical sciences to attempt to wring from it a word of
acknowledgment.  Yet what we today think of as science is for the
most part not science at all, but {\it applied\/} science, that is,
technology. The machinery of modern science is so elaborate, and the
building of it requires so much ingenuity---requires, indeed, so much
{\it science}---that we naturally confuse the thinking with the
doing. The great particle accelerator at CERN, for example, is for us
the very image of modern science:  a vast and inconceivably expensive
machine built to perform minute and unimaginably complex operations
whose results can be interpreted only by a handful of physicists. But
we are willing to pay the cost of building these machines, are
willing to allow the physicists their arcane rules and specialized
language, because we believe that they are getting their hands into
the very bowels, or, rather, the very synapses, of nature.  And at
some point, we believe, they will bring forth news of another
advance, another boiled-down version of the world's variousness,
another $E = mc^2$, only bigger and better.  Perhaps this time they
may even discover the final equation, the Grand Theory of Everything.
Then, as Stephen {\Hawking} puts it, ``we shall all, philosophers,
scientists, and just ordinary people [I am struck by that
distinction, by the way], be able to take part in the discussion of
the question of why it is that we and the universe exist.  If we find
the answer to that, it would be the ultimate triumph of human
reason---for then we would know the mind of God.''

Such foolhardy talk, from such an eminent source, misleads us into
the notion that the aim of science is to find the ``meaning'' of the
world.  That there must be a meaning seems certain, otherwise how is
it that there is such a thing as progress?  Science keeps uncovering
more and more secrets, keeps getting closer and closer to \ldots\
well, to {\it something}, in the same way that computations in the
infinitesimal calculus keep approaching nearer and nearer to infinity
without ever getting there.  Progress must be progress {\it toward\/}
something, surely, some final end to the quest for knowledge?  But to
my mind the world has no meaning.  It simply {\it is}.  {\Leibniz}'s
thrilling question, ``Why is there something rather than
nothing?''~is significant not because an answer to it is possible,
but because out of the blind, boiling chaos that is the world, a
species should have emerged that is capable of posing such a
question.

Science and art are different ways of looking at the same thing,
namely, the world.  Let us take the case of Goethe.  In his role as
amateur scientist, he was vehemently opposed to {\Newton}'s
mechanistic model of reality.  He was mistaken---that is to say, his
science was bad science, although his scientific writings are not
bad philosophy, and still less are they bad poetry.  Goethe demanded
that science should always hold to the human scale.  He opposed the
use of the microscope, since he believed that what cannot be seen
with the naked eye should not be seen, and that what is hidden from
us is hidden for a purpose.  In this, Goethe was a scandal among
scientists, whose first, firm, and necessary principle is that if
something {\it can\/} be done, then it {\it should\/} be done.  Yet
his furious denial of {\Newton} was more than merely the bloodshot
jealousy of one great mind drawing a bead on another.  Goethe's
theory of light is wrong insofar as the science of optics is
concerned, yet in the expression of his theory Goethe achieves a
pitch of poetic intensity that is as persuasive, in its way, as
anything {\Newton} did.  But persuasive at what level?

There is a world beyond politics, says the poet Wallace {\Stevens},
and we might adapt that to say that there is a world beyond science,
or, at least, there is a world beyond {\it the current state\/} of
science.  At the end of the 19th century professors of physics in the
great European universities were steering students away from the
discipline because they believed that there was very little of
interest left to be discovered about the nature of physical reality.
Then came {\Einstein}.  As we approach the end of the 20th century,
we are still guilty of hubris, as evidenced by Stephen {\Hawking}'s
statements quoted above. Probably a Unified Field Theory will be
achieved, and will seem for a time, perhaps even as long as the
period between {\Newton}'s {\it Principia\/} and {\Einstein}'s first
paper on the theory of relativity, to explain everything; then a
{\Heisenberg} or a {\Goedel} will come forward and point to a loose
end which, when pulled, will unravel the entire structure.

This is a truth that both clearsighted artists and scientists---that
is, those not blinded by hubris, or a cramped imagination, or
both---have always acknowledged:  There is no end to the venture. The
difference between the two, however, is that while the artist
acknowledges that in art there is nothing new to be said, only new
ways of saying the old things, new combinations of old materials---a
process, paradoxically, that {\it makes\/} a new thing, namely, the
work of art---science seems always to be pressing on into hitherto
uncharted territory.  Yet the fact is, science is not {\it making\/}
this new landscape, but {\it discovering\/} it.  {\Einstein} remarked
more than once how strange it is that reality, as we know it, keeps
proving itself amenable to the rules of man-made science.  It
certainly is strange; indeed, so strange, that perhaps it should make
us a little suspicious.  More than one philosopher has conjectured
that our thought extends only as far as our capacity to express it.
So too it is possible that what we consider reality is only that
stratum of the world that we have the faculties to comprehend.  For
instance, I am convinced that quantum theory flouts commonsense logic
only because commonsense logic has not yet been sufficiently
expanded.

I am not arguing that art is greater than science, more universal in
its concerns, and wiser in its sad recognition of the limits of human
knowledge.  What I am proposing is that despite the profound
differences between them, at an essential level art and science are
so nearly alike as to be indistinguishable.  The only meaningful
distinction I can see between the two is that science has a practical
extension into technology, and art does not.  But this is a
distinction only in terms of utility.  At the level that concerns me,
the level of {\it metaphor}, art and science are both blithely
inutile---at this level, for instance, the theory of relativity has
nothing to do with the atomic bomb.

The critic Frank {\Kermode} has argued, persuasively, I believe, that
one of art's greatest attractions is that it offers ``the sense of an
ending.''  The sense of completeness that is projected by the work of
art is to be found nowhere else in our lives.  We cannot remember our
birth, and we shall not know our death; in between is the ramshackle
circus of our days and doings.  But in a poem, a picture, or a
sonata, the curve is completed.  This is the triumph of form.  It is
a deception, but one that we desire, and require.

The trick that art performs is to transform the ordinary into the
extraordinary and back again in the twinkling of a metaphor.  Here is
Wallace {\Stevens} again, in lines from his poem {\it Notes Toward a
Supreme Fiction\/} (1942):
\begin{verse}
You must become an ignorant man again\\ And see the sun again with an
ignorant eye\\ And see it clearly in the idea of it.
\end{verse}
This is the project that all artists are embarked upon:  to subject
mundane reality to such intense, passionate, and unblinking scrutiny
that it becomes transformed into something rich and strange while yet
remaining solidly, stolidly, itself.  Is the project of pure science
any different?  When Johannes {\Kepler} recognized that the planets
move in elliptical orbits and not in perfect circles, as received
wisdom had for millennia held they must do, he added infinitely to
the richness of man's life and thought.  When Copernicus posited the
horrifying notion that not the Earth but the sun is the center of our
world, he literally put man in his place, and he did it for the sake
of neither good nor ill, but for the sake of demonstrating {\it how
things are}.  When, probably early in the next millennium, quantum
theory gives up its secrets, we shall see the world again with a new,
ignorant eye---not as a blizzard of atoms, not as a speck whirling in
an unimaginable immensity of darkness, not even as that
blue-and-white marble photographed by the first moon travelers, the
beauty of which took our breath away---as our ordinary and always
known home, which is the world that art and science alike, even in
their most seemingly transcendental modes, are concerned with.  In
the 1970s, when quantum theory began employing such terms as
``beauty,'' ``charm,'' and ``strangeness'' to signify the various
properties of quarks, a friend turned to me and said: ``You know,
they're waiting for you to give them the words.''  I saw what he
meant, but he was not quite right:  Science does not need art to
supply its metaphors.  Art and science are alike in their quest to
reveal the world.  Rainer Maria Rilke spoke for both the artist and
the scientist when he said:
\bq
\noindent Are we, perhaps, {\it here\/} just for saying: House, Bridge,
Fountain, Gate, Jug, Fruit tree, Window,---possibly: Pillar, Tower?\
\ldots\ but for {\it saying}, remember, oh, for such saying as never
the things themselves hoped so intensely to be.
\eq
\eq

\section{03 January 1999, \ ``Mary and {\Banville}''}

I forgot to tell you:  Also please share the {\Banville} article with
Mary. I would be interested in knowing her opinion on it.  Does the
artist think this is as much hogwash as the average physicist
probably does? (You and I aren't so average.)

The point I find most eloquently stated in that article is where
{\Banville} says:
\bq
\noindent
{\Einstein} remarked more than once how strange it is that reality,
as we know it, keeps proving itself amenable to the rules of man-made
science.  It certainly is strange; indeed, so strange, that perhaps
it should make us a little suspicious.  More than one philosopher has
conjectured that our thought extends only as far as our capacity to
express it.  So too it is possible that what we consider reality is
only that stratum of the world that we have the faculties to
comprehend.
\eq

To bring a similar point home in a talk I give on the ``real meaning
of quantum information,'' I use three slides that I'm particularly
proud of.  This first is one I stole from Seth {\Lloyd} from his
talk at the QUIC Kickoff meeting.  It shows a tall mountain labeled
with ``novel quantum states'' at the bottom and ``factoring'' at the
top, and some clouds in between.  I say, ``It has become popular to
show slides like this \ldots\ to depict graphically the mountain we
must climb in quantum computing.  It gives the idea that the thing we
really want is that peak.  But if you ask me, the real reason to
climb from the base to the peak, is to gauge the distance between
the two.'' Then I change slides to one of iceberg that {\Kiki} drew
for me. (There's a little boat near the iceberg, and a penguin
sitting on its peak.)  ``If we can gauge that distance, then we'll
have a better feel for the other 8/9 that's still below the
surface!'' Then I say that the source of that belief has no logical
justification, but it seems that physicists have lately gotten in a
habit of ignoring a pretty basic fact of the world.  Then I show a
slide copied from a biology book, ``Amino Acid Difference Matrix for
26 Species of Cytochrome c.''  I've got {\it man\/} and {\it dog\/}
highlighted in yellow on it; the difference is only 11 percent. Then
I say, ``My dog isn't even close to a Grand Unified Theory of the
universe. There are some things he just can't see, no matter how
hard I try to train him.  Why should we be much more than 11\% ahead
of him in the game?''

\section{24 April 1999, \ ``Genetic Genesis''}

OK, I still haven't read through {\Fleck} again.  But I keep thinking
I'm going to.  Does that count for anything?

Anyway I really would like you to write up a report on that
wonderful conversation we had with {\Charlie} the other night.  It
would be sad if it faded from our collective memory.  Of the three
of us, you seemed to have the deepest understanding of all that was
being said and, on top of that, to have a glimpse of how to
conciliate of all three points of view.  So please, please do it for
me.

As a bit of a peace offering in the mean time, let me send you two
things.  The first is a small essay I wrote a couple of days ago to
firm up how I might present my research program in a sort of
evocative way.  It's based on the Adam and Eve slide I showed you. I
think you'll enjoy it; it's placed below.  [See note to {\Greg}
{\Comer}, titled ``Fuchsian Genesis,'' dated 22 April 1999.]  The
other thing, coming in the next email, is a passage from a paper by
Doug {\Bilodeau}.  [See letter to  {\Asher} {\Peres}, titled ``The
Deep Intervention,'' dated 26 December 1998.]  It expresses very
clearly what I was trying to convey to {\Charlie} about the
scientist as the setter of initial conditions. He thinks that is
absolutely incidental and unimportant for science (an
epiphenomenon?), but I just don't think so.  I guess the essay below
and {\Bilodeau}'s passage both express this.  I'd love to hear your
opinions (along with the stuff I begged for above).

\section{08 May 1999, \ ``A {\Fleck} of Quantization''}

The last time I wrote you about Ludwik {\Fleck}'s book {\sl Genesis
and Development of a Scientific Fact}---1 October 1995, can you
believe it!?---I said the following: \medskip

\noindent ------------------------------
\bq
Another piece of news is that I've finally finished reading
{\Fleck}'s book.  It was indeed worth taking a look at, though I'm
not yet sure what to make of it. In this ``reality generation''
business, I see the quantum as a crucial ingredient.  But {\Kuhn} and
{\Fleck} seem to argue otherwise.  It's amazing how deeply ingrained
I find the ``realist'' tendency.  Because quantum theory apparently
forces me to relinquish some of my intuitions, I'm willing to give
these spookier ideas a shot \ldots\ but always some resistance
remains. In any case, I was intrigued that {\Fleck} and I have come
to similar ideas about the ``stability of reality.'' (Cf.\ one of my
old notes titled ``Insights from Cuero'' or something like
that---it's the one that contains stuff about the burnings in Waco.)

These are the little notes I compiled; maybe you'll find them useful
too.
\begin{enumerate}
\item
crossing thought style -- p. 2
\item
reality -- p. 10, 28 (bottom), 127 ``Our reality did not exist for
them'', 156
\item
resistance -- p. 27
\item
individual cognition dependent upon community -- p. 38
\item
contribution of the individual -- p. 40
\item
slacker reference -- p. 44
\item
stability of reality -- p. 47, 99, 102
\item
thought collective being cultural and not across the board
  for humankind -- p. 174 (note p. 49)
\item
French bread -- p. 5
\item
existence is lawless -- p. 51
\item
truth -- p. 100, 116, 125
\item
fact as thought-collective resistance -- p. 101
\item
objects created by thought -- p. 181
\item
how to use philosophical principles -- p. 181
\item
symbols -- p. 125
\item
name as property -- p. 136
\item
meaning as a property of the object -- p. 137
\end{enumerate}
\eq
------------------------------

Apparently this made little impression on you:  you told me you
didn't even remember my sending it!  So let me bless you with
something of an expanded version.

In what follows I revisit those marked sections and copy down what it
was that I found interesting.  In a small number of cases, I couldn't
figure out what I found interesting last time; so those citations got
skipped this time around.  Also, a few of the subjects got renamed.

Reading over it all again, I guess I find that I'm significantly more
impressed than I was last time.  Perhaps the years have made me
wiser.  Or perhaps they've caused me to return to my youth:  you'll
see what I mean shortly.  I think your forcing me to look through
{\Fleck} again was particularly timely.  This is because the little
tale ``Fuchsian Genesis'' I sent you the other day.  I continue to
think quantum mechanics supplies a crucial piece that Mr.\ {\Fleck}
couldn't have foreseen.

Throughout I'll make commentary here and there, and even annotate
with some of my old writings.  At the end, I'll sum up my present
thoughts on the {\it Denkkollektiv}. And then you must fulfill your
promise!

\subsection{A Few of {\Fleck}'s Thoughts (with Annotation)}

\noindent {\it Resistance -- p. 27:}
\bq
Once a structurally complete and closed system of opinions consisting
of many details and relations has been formed, it offers enduring
resistance to anything that contradicts it.

A striking example of this tendency is given by our history of the
concept of ``carnal scourge'' in its prolonged endurance against
every new notion.  What we are faced with here is not so much simple
passivity or mistrust of new ideas as an active approach which can be
divided into several stages.  (1) A contradiction to the system
appears unthinkable.  (2) What does not fit into the system remains
unseen; (3) alternatively, if it is noticed, either it is kept
secret, or (4) laborious efforts are made to explain an exception in
terms that do not contradict the system.  (5) Despite the legitimate
claims of contradictory views, one tends to see, describe, or even
illustrate those circumstances which corroborate current views and
thereby give them substance.
\eq
{\it Individual cognition dependent upon the community -- p. 38-39:}
\bq
In comparative epistemology, cognition must not be construed as only
a dual relationship between the knowing subject and the object to be
known.  The existing fund of knowledge must be a third partner in
this relation as a basic factor of all new knowledge.  It would
otherwise remain beyond our understanding how a closed and
style-permeated system of opinions could arise, and why we find, in
the past, rudiments of current knowledge which at the time could not
be legitimized by any ``objective'' reasons and which remained only
pre-ideas.

Such historical and stylized relations within knowledge show that an
interaction exists between that which is known and the act of
cognition.  What is already known influences the particular method of
cognition; and cognition, in turn, enlarges, renews, and gives fresh
meaning to what is already known.

Cognition is therefore not in individual process of any theoretical
``particular consciousness.''  Rather it is the result of a social
activity, since the existing stock of knowledge exceeds the range
available to any one individual.

The statement, ``Someone recognizes something,'' whether it be a
relation, a fact, or an object, is therefore incomplete.  It is no
more meaningful as it stands than the statements, ``This book is
larger,'' or ``Town A is situated to the left of town B.''  Something
is still missing, namely the addition, ``than that book,'' to the
second statement, and either, ``to someone standing on the road
between towns A and B while facing north,'' or ``to someone walking
on the road from town C to town B,'' to the third statement.  The
relative terms ``larger'' and ``left'' acquire a definite meaning
only in conjunction with their appropriate components.

Analogously, the statement, ``Someone recognizes something,'' demands
some such supplement as, ``on the basis of a certain fund of
knowledge,'' or, better, ``as a member of a certain cultural
environment,'' and, best, ``in a particular thought style, in a
particular thought collective.''

If we define ``thought collective'' as {\it a community of persons
mutually exchanging ideas or maintaining intellectual interaction, we
will find by implication that it also provides the special
``carrier'' for the historical development of any field of thought,
as well as for the given stock of knowledge and level of culture.
This we have designated thought style.}  The thought collective thus
supplies the missing component.
\eq
{\it Contribution of the individual -- p. 40-41:}
\bq
Cognition therefore means, primarily, to ascertain those results
which must follow, given certain preconditions.  The preconditions
correspond to active linkages and constitute that portion of
cognition belonging to the collective.  The constrained results
correspond to passive linkages and constitute that which is
experienced as objective reality. The act of ascertaining is the
contribution of the individual.

The three factors involved in cognition---the individual, the
collective, and objective reality (that which is to be known)---do
not signify metaphysical entities; they too can be investigated, for
they have further relations with respect to one another.

These further relations consist in the facts that, on the one hand,
the collective is composed of individuals and that, on the other,
objective reality can be resolved into historical sequences of ideas
belonging to the collective.  It is therefore possible from the
viewpoint of comparative epistemology to eliminate one or perhaps
even two factors.

Although the thought collective consists of individuals, it is not
simply the aggregate sum of them. The individual within the
collective is never, or hardly ever, conscious of the prevailing
thought style, which almost always exerts an absolutely compulsive
force upon his thinking and with which it is not possible to be at
variance.
\eq
{\it Minimum thought collective -- p. 43-44:}
\bq
A kind of superstitious fear prevents us from attributing that which
is the most intimate part of human personality, namely the thought
process, also to a collective.  A thought collective exists wherever
two or more people are actually exchanging thoughts.  He is a poor
observer who does not notice that a stimulating conversation between
two persons soon creates a condition in which each utters thoughts he
would not have been able to produce either by himself or in different
company.  A special mood arises, which would not otherwise affect
either partner of the conversation but almost always returns whenever
these persons meet again. Prolonged duration of this state produces,
from common understanding and mutual misunderstanding, a thought
structure that belongs to neither of them alone but nevertheless is
not at all without meaning. Who is its carrier and who its
originator?  It is neither more nor less than the small collective of
two persons.  If a third person joins in, a new collective arises.
The previous mood will dissolve and with it the special creative
force of the former small collective.

We could agree with anybody who calls the thought collective
fictitious and the personification of a common result produced by
interaction.  But what is any personality if not the personification
of many different momentary personalities and their common
psychological Gestalt?  A thought collective, by analogy, is composed
of different individuals and also has its special rules of behavior
and its special psychological form. As an entity it is even more
stable and consistent than the so-called individual, who always
consists of contradictory drives.
\eq

A moral?  Even as little as two students at Columbine High School in
Littleton, Colorado can form a thought collective.  And indeed they
did: look at the drastic reality they were able to create (destroy).
\medskip

\noindent {\it Stability of reality -- p. 46-47:}
\bq
Gumplowicz expressed himself very poignantly on the importance of the
collective.  ``The greatest error of individualistic psychology is
the assumption that a {\it person\/} thinks.  This leads to a
continual search for the source of thought within the individual
himself and for the reasons why he thinks in a particular way and not
in any other.  Theologians and philosophers contemplate this problem,
even offer advice on how one ought to think.  But this is a chain of
errors.  What actually thinks within a person is not the individual
himself but his social community.  The source of his thinking is not
within himself but is to be found in his social environment and in
the very social atmosphere he `breathes.'  His mind is structured,
and necessarily so, under the influence of this ever-present social
environment, and {\it he cannot think in any other way}.''

Jerusalem dealt with this problem in a number of essays, the last of
them bearing the apposite title ``Social Conditioning of Thinking and
of Thought Patterns.''  ``{\Kant}'s firm belief in a timeless,
completely immutable logical structure of our reason, a belief that
has since become the common heritage of all who adopt an a priori
point of view and is maintained with great tenacity also by the
latest representatives of this direction of thinking, has not only
failed to be confirmed by the results of modern ethnology but proved
to be definitely erroneous.''  ``The primitive individual feels
himself only a member of his tribe and clings to its traditional way
of interpreting sensory perceptions with absolutely incredible
tenacity.''  ``I have no doubt, and it is confirmed through the
diverse institutions found in primitive societies, that tribesmen
reinforce each other's belief in the ubiquity of spirits and demons,
which is already sufficient to give these figments of the imagination
some degree of reality and stability.  This process of mutual
corroboration is by no means confined exclusively to primitive
societies.  It is rather prevalent today, fully effective in our
everyday lives.  I wish to designate this process and any structure
of belief formed and fortified by it {\it social consolidation}.''
``Even particular and objective observations \ldots\ require
confirmation by the observation of others.  Only then will they
become common property and thus suitable for practical utilization.
Social consolidation functions actively even in science.  This is
seen particularly clearly in the resistance which as a rule is
encountered by new directions of thought.''
\eq
{\it Stability of reality -- p. 99:}
\bq
Because it belongs to a community, the thought style of the
collective undergoes social reinforcement, as will shortly be
discussed.  Such reinforcement is a feature of all social structures.
The thought style is subject to independent development for
generations.  It constrains the individual by determining ``what can
be thought in no other way.''  Whole eras will then be ruled by this
thought constraint. Heretics who do not share this collective mood
and are rated as criminals by the collective will be burned at the
stake until a different mood creates a different thought style and
different valuation.
\eq
\pagebreak {\it Stability of reality -- p. 102:}
\bq
The fact thus defined as a ``signal of resistance by the thought
collective'' contains the entire scale of possible kinds of
ascertainment, from a child's cry of pain after he has bumped into
something hard, to a sick person's hallucinations, to the complex
system of science.

Facts are never completely independent of each other. They occur
either as more or less connected mixtures of separate signals, or as
a system of knowledge obeying its own laws.  As a result, every fact
reacts upon many others.  Every change and every discovery has an
effect on a terrain that is virtually limitless.  It is
characteristic of advanced knowledge, matured into a coherent system,
that each new fact harmoniously---though ever so slightly---changes
all earlier facts.  Here every discovery is actually a re-creation of
the whole world as construed by a thought collective.

A universally interconnected system of facts is thus formed,
maintaining its balance through continuous interaction.  This
interwoven texture bestows solidity and tenacity upon the ``world of
facts'' and creates a feeling both of fixed reality and of the
independent existence of the universe.  The less interconnected the
system of knowledge, the more magical it appears and the less stable
and more miracle-prone is its reality, always in accordance with the
thought style of the collective.
\eq

This issue about community-created realities---from whence comes
their stability---used to be on my mind a lot.  Let me throw in at
this point a little essay I wrote my friend {\Greg} {\Comer} way back
(28 March 1993).  I think it is quite in line with much of what
{\Fleck} is saying:

\begin{center}
Insights from Cuero
\end{center}
\bq
I login to the computer only to find an ``empty tray'' representing
the status of my new electronic mail.  Woe is me.  You know it is the
mail that makes me live and thrive.

I wonder how close I really am to getting this idea of the quantum
straight.  Sometimes I think I'm very close and then at others (like
today) I feel completely lost. And it's you that gets to hear about
all this over and over.

I wish I could fill your ears with the technical details of how to
``derive'' the quantum from a few simple desiderata for LWL.  But I
can't and that's nothing new.  So to fill my time and make my fingers
tired I plan to fill this note with pure unadulterated philosophy
\ldots\ Insights from Cuero.  There'll be no hint of real science
here; I make no apologies.

{}From where and of what utility comes this notion of an ``objective
world'' independent of man, woman, animal, and plant? My opinion is
that (in the end) ``objective reality'' is posited for nothing more
than to have a device for coordinating the various experiences common
to all those who communicate.  Mark Twain once wrote, ``If you tell
the truth, you don't have to remember anything.'' It seems to me that
that pretty much sums it up.  The objective truth saves us from
having to make sure that the stories we tell are consistent; it saves
us from having to remember all aspects of the past.  It gives us a
means for determining whether someone's behavior is insane.  It gives
us a means by which to determine the guilt or innocence of an accused
murderer.  Is there any other real motivation for positing an
objective reality?  I can think of none \ldots\ but why should I, the
reason listed above should be powerful enough argumentation for
anyone involved in the sciences.

Nevertheless, just how essential is this notion of an objective
reality?  The ``idealists'' have been trying to do away with it since
the days of Bishop George Berkeley. ``Esse est percipi.''  Why be
fixated on a ``material reality'' that can never be confirmed; all
that is truly available to our discerning is our thoughts and our
sense impressions.  As far as thinking beings are concerned,
material/objective reality is a mental construct \ldots\ whether it
is indeed ultimately ``out there'' or not.  So just do away with it;
superfluousness.  All that is really needed to make sense of this
world is ``mind.''

Cute idea, I say.  But what a strain when the so much simpler picture
of ``objective reality'' will do.  This is the side of the razor I'd
put Ockham on if I could.  Mais, c'est un monde que je ne vois pas.

The problem we face in modern times is that the existence of quantum
phenomena seems to cry out for the relinquishing of this objective
reality notion.  How else can we consolidate the experimentally
confirmed violation of the {\Bell} inequalities with the equally well
experimentally confirmed Lorentz character of spacetime? (Well, of
course, there are ways \ldots\ if one is of such a mind. {\Bohm}'s
1952 nonlocal hidden variable theory, for instance, is a step in that
direction.  Or at worst, one could say that Allah wills each and
every quantum mechanical measurement outcome.  But what a price these
both are to pay.  In the first case, one relinquishes simplicity in
the equations.  In the second, one relinquishes science and reasoned
philosophy.)  The case in the end seems to be that quantum mechanical
systems have properties only insofar as those properties are {\it
created\/} by (freely-chosen) human ``measurements.''

So what are we left with? Either we decide it's simpler to work on
some muck like the {\Bohm} hidden variable theory to save objective
reality \ldots\ risking the simplicity of the best physical theory
we've yet constructed, not to mention the fact that so far it has
seemed to lead to no new physics. Or we decide it's simpler to work
out how our experiences can be coordinated without an objective
reality.  You know which way I lean.

But how can we easily get by with simply ``consistency'' and
``coordination'' for our ``community of communicators'' instead of
``reality''?  From where now can we pull a man-independent notion of
consistency?  The answer must be ``nowhere.''  We are left with
pulling it from the community itself.  And it is of course this very
fact that worries us---the individual and community opinions being
demonstrably (through standard interpersonal relations and history)
so {\it damned\/} fluid.  You asked a few days ago:
\bq
\noindent
People are different, so surely the rules must constantly change,
even if slightly.  This is what troubles me these days:  the rules
must constantly change.  \ldots\  If there are no laws, then how can
we count on the permanence of anything, especially the requirement of
``consistent''?  \ldots\  But if people are truly different how can
nature be consistent?
\eq
No answers today, of course, but maybe some hints:  force and
diplomacy.

Sometime around the year 1985 or 1986, I was walking across the
University of Texas campus \ldots\ from the Student Union (I think)
to the Perry-Castenada Library.  On the South Mall between the UT
Tower and the TX State Capital, I approached a man talking to himself
fairly audibly walking in the same direction as myself.  I slid up
behind him as closely as I could without being detected and tried to
listen in on the conversation.  (He did, by the way, later notice me
\ldots\ and I became scared stiff as he subsequently followed me all
the way into the library!)  Sure enough, the conversation seemed, as
far as I could tell, like any other conversation one might hear on a
Sunday dusk \ldots\ only one could not hear or see the second
participant.  (The day, by the way, was indeed Sunday.)  This caused
pause for a lot of reflection in that short time \ldots\ and
consequently over the years.

The first thought, of course, was that the man was crazy.  There was
no one else there for him to talk to. But then I started thinking
about what might happen if I were to confront him with this
observation.  One possibility that occurred to me was that he would
look at me, tell me I was ``crazy'', and introduce his friend Joe
(for instance).  At that point, given exactly that situation, how on
earth would I really know that it was he and not I that was crazy? Of
course, by nature, I would have to believe that it was he that was
insane, but how could I really prove to myself that that was the
case?  It dawned on me that I needed only to find a passerby to
confirm my opinion.  But what if our man of interest said, ``You're
both crazy.  Can't you see my friend?  This is the strangest thing
I've ever encountered; two crazies at one time.''??? How could we
know that we weren't both crazy?  Well, in the same vein, we could
just ask for confirmation of the man's insanity from a third person;
then we would feel quite confident, wouldn't we? Yeah, we probably
would, but this is now a question of principle. What if the man said,
``All three of you are crazy as birds.  My friend Joe is here as
plain as day.  Joe, we need to call the authorities about these
people. You heard what I heard didn't you? I thought you did. Yeah,
that's a good idea; you run off to the library.  There's a phone
there.''  Just what if?  Then we would have to pull in a fourth
person for that extra bit of confirmation to assure ourselves of our
own sanity.  And so on the story could go.

What is the point of all this? Without an objective reality
underlying our experiences, even one bad seed can destroy the
possibility of a comfortable background for their coordination.  We
can never be sure that that one ``crazy'' man is not in fact sane and
we, on the other hand, are all insane. This tells us that questions
of principle---to some extent---go down the tubes when the notion of
objective reality is wiped from our repertoire of physical concepts.
That said, though, what would happen {\it in practice\/} in a
scenario like that described above?  Easy. The singular man would be
locked away in some asylum in San Antonio, TX. Majority rules \ldots\
or those with power rule \ldots\ or the ruling class rules \ldots\ or
\ldots. The point is some faction makes the rules, and those rules
are enforced. It is this enforcing that gives a certain uniformity
to our existence.  Any unbearable non-conformity is either isolated
or destroyed.

Before we run with this idea a little further, let's take a look at a
slightly more difficult case. Consider the events that have been
taking place in Waco, TX the last few weeks. There a cult leader,
David {\Koresh}, has declared himself Jesus Christ. He and his
following (originally over 100 strong) have blockaded themselves
along with quite a bit of firepower within some compound either near
or in the city.  At least six or so people have been killed in the
ensuing struggle.  Can we consider these hundred people sane?  I
would say, ``no!''  One would think they would surely sense the
consequences of their actions \ldots\ if they were sane.  They will
eventually lose; they will all eventually be arrested or killed.
There is no doubt about this \ldots\ {\it to all of us who observe
from the outside}. From the ``inside'', though, it must be a
completely different story: David {\Koresh} {\it is\/} Jesus Christ;
God {\it will\/} save them; killing others {\it is\/} justified.  By
any standard criterion of insanity, this must be a ``mass
insanity.''  (In perhaps different words, many of the convicted will
probably plead this when trial comes.)  No longer here are we
isolated to a {\it single\/} man doing crazy things; now it is a
mass of 100 people.  [Of special interest, I find it particularly
enlightening that {\Koresh}'s right-hand man is a ``Harvard educated
lawyer.'' !!! This is just a hint of what we know must eventually be
the case in this observer created reality we are trying to
investigate. ``Rational powers of thought'' (as exhibited by the
LSAT score required for Harvard) must have nothing to do with
anything in the end. Reality is defined by the community.  Period.]
The point now:  be it the insanity of one or a hundred, the faction
making the rules will snuff that insanity out in one of the ways
listed above.

Now let's carry this to the extreme.  What of Germany in 1941,
perhaps at the peak of its Nazi prowess?  It very well could have
easily become the ``faction making the rules.''  And then it would
have enforced those rules. Surely an insane faction by present-day
standards!  But if it had become the ruling faction, it is our
present-day notions that would be insane. This is what we have to
gulp in our picture of a reality created by the community of
communicators---not only for social interactions but for the
``physical world.''

Just in summary let me reiterate this small insight or missing link:
Not only is reality created by the community, but also {\it
enforced\/} by the community.  It is this enforcing that gives a
certain uniformity to our existence and ultimately circumscribes any
good notion of ``consistency.''

How formal this notion can be made, I'm not sure.  I don't even want
to contemplate it now.  What I'd like to do in the remainder of this
note is briefly sketch the importance of {\Wheeler}'s phrase:  ``The
past exists only insofar as it is recorded in the present.''  Lately
here in Albuquerque we have been a little worried about the problem
of the arrow of time and its relation to the Second Law of
Thermodynamics. I've told you this before.  David {\Wolpert} asks,
``Why can we only remember the past and not the future given that all
physical laws are time symmetric?''  If we take {\Wheeler}'s
standpoint, this is not a problem at all but a tautology, a
triviality.  For what is in memory is, by definition, ``the past.''
The past has no existence otherwise. It too is not simply something
``out there'' independent of man, woman, animal and plant. This
point we can use to tie together a loose string in our discussion.

If a ruling faction enforces a certain uniformity and consistency in
the present, then by {\Wheeler}'s maxim, it also enforces that same
uniformity and consistency in the past. You ask how can
``consistency'' be stable? Well here's the direction of an answer
\ldots\ if we take all this seriously. The present makes the past and
thus the continuity of all its policies with those of the past
\ldots\ including the continuity of physical law. Perhaps there is a
bit more than social commentary to be learned from George {\Orwell}'s
novel {\sl 1984}! Recall how history was constantly being rewritten
in that novel by an official government agency.  Any dissenters from
the official account were snuffed out. No less can we expect in our
account of physical law.

This ends our tale so far. Please, please recall though that I warned
you that this note would contain only pure unadulterated philosophy.
Not a bit of science. So I expect no less than a million loopholes in
these thoughts. Nevertheless I hope there is at least something
worthwhile in here.
\eq
Does this tell you that I have been inclined to the idea of the {\it
Denkkollektiv\/} for a long, long time?  Notice again, though, that
my motivation came largely from quantum mechanics.
\medskip

\noindent {\it Requirement that there be a community -- p. 174 (note
from p. 49):}
\bq
But we soon read: ``Not every observation by an individual must in
itself be valued as an experience.  Only after a stock of general and
well-confirmed knowledge has formed as a result of mutual agreement
and reinforcement in the course of continued cooperation of the
intellects involved should we speak of experience.  General and
well-confirmed experience, however, must be considered the sole
criterion of truth.''  Confrontation of these contradictions does not
constitute a criticism of Jerusalem.  It merely exemplifies that when
new thought styles are evolving, contradictions sets in as an
expression of the intellectual ``contest of the fields of view.''
\eq
{\it French bread -- p. 50:}
\bq
What is the reason for this special position of current scientific
statements as required by the philosophers just quoted?

They believe that our present-day scientific opinions are in complete
contrast with all other ways of thinking.  As if we had become wise
and our eyes had been opened, they believe that we have simply
discarded the naive self-consciousness of thought processes which are
primitive or archaic.  We are supposedly in possession of ``correct
thinking'' and ``correct observation,'' and therefore what we declare
to be true {\it is\/} ipso facto {\it true}.  What those others such
as the primitives, the old people, the mentally ill, or the children
declare to be true {\it seems to be true only to them}. This
arch-naive view, which prevents the building up of a scientific
epistemology, reminds us very much of the theory of a French
philologist of the eighteenth century who claimed that {\it pain,
sitos, bread, Brot, panis\/} were arbitrary, different descriptions
of the same thing.  The difference between French and other
languages, according to this theory, consisted in the fact that what
is called bread in French really was bread.
\eq
{\it Connection to {\Wheeler}'s ``Law without Law'' -- p. 51:}
\bq
The views outlined here should not be construed as skepticism.  We
are certainly capable of knowing a great deal.  If we cannot know
``everything,'' according to the traditional position, it is simply
because we cannot do much with the term ``everything,'' for every new
finding raises at least {\it one\/} new problem: namely an
investigation of what has just been found.  The number of problems to
be solved thus becomes infinite and the term ``everything''
meaningless.

An ``ultimate'' or set of fundamental first principles from which
such findings could be logically constructed is just as nonexistent
as this ``everything.''  Knowledge, after all, does not repose upon
some substratum.  Only through continual movement and interaction can
that drive be maintained which yields ideas and truth.
\eq

I find this passage greatly reminiscent of David {\Deutsch}'s
description of John {\Wheeler}'s idea of ``law without law''
[D.~{\Deutsch}, ``On {\Wheeler}'s notion of `Law without Law' in
Physics,'' Found.\ Phys.\ {\bf 16}, 565--572 (1986)].  {\Deutsch}
wrote:
\bq
Is it possible that there is an ultimate law of physics, a principle
$\cal P$ from which follows everything that is knowable about the
material world? If so, what can we already infer or postulate about
the form that this principle takes?  {\Wheeler}'s idea of ``law
without law'' is an attempt to begin to answer these questions.

If there were no all-explanatory physical principle $\cal P$
approachable by the methods of science, this would presumably mean
that there exist aspects of the natural world that are fundamentally
inaccessible to science. This would run directly counter to
rationalism and to our view of physics as the universal science,
which have hitherto been the driving forces behind progress in the
subject and which we should be extremely reluctant to abandon.

But if $\cal P$ were itself a law of physics, then the problem of
$\cal P$'s own origin---why that particular principle holds in nature
rather than some other---would be forever insoluble. And hence $\cal
P$ would not be all-explanatory within physics. So, paradoxically
$\cal P$, the ultimate principle of physics, cannot be a ``law'' (of
physics). Hence the expression ``law without law.''
\eq

Honestly, though, I'm not sure to what extent {\Wheeler} really saw
this argument as his own motivation: it might just be a case of
{\Deutsch} looking at {\Wheeler} through Oxford-colored glasses. I
don't recall ever seeing {\Wheeler} write the argument himself.
Instead, it seems that his motivation for ``law without law'' had
been that every law he had ever seen formulated had ultimately been
``transcended.'' This itself built a bit of scepticism.  But then on
top of that was quantum mechanics with its random measurement
outcomes, ``each individual one shunning all law.''
\medskip

\noindent {\it What is truth? -- p. 100:}
\bq
A historical connection thus arises between thought styles.  In the
development of ideas, primitive pre-ideas often lead continuously to
modern scientific concepts.  Because such ideational developments
form multiple ties with one another and are always related to the
entire fund of knowledge of the thought collective, their actual
expression in each particular case receives the imprint of uniqueness
characteristic of a historic event.  It is, for instance, possible to
trace the development of the idea of an infectious disease from a
primitive belief in demons, through the idea of a disease miasma, to
the theory of the pathogenic agent.  As we have already hinted, even
this latter theory is already close to extinction.  But while it
lasted, only one solution to any given problem conformed to that
style. \ldots\ {\it Such a stylized solution, and there is always
only one, is called truth}.  Truth is not ``relative'' and certainly
not ``subjective'' in the popular sense of the word.  It is always,
or almost always, completely determined within a thought style.  One
can never say that the same thought is true for A and false for B.
If A and B belong to the same thought collective, the thought will be
either true or false for both.  But if they belong to different
thought collectives, it will just {\it not\/} be {\it the same\/}
thought!  It must either be unclear to, or be understood differently
by, one of them.  Truth is not a convention, {\it but rather\/} (1)
{\it in historical perspective, an event in the history of thought},
(2) {\it in its contemporary context, stylized thought constraint}.
\eq

This is a wild idea, but notice how the sum total of thought
collectives almost hints of a ``partial Boolean algebra'' structure.
That is to say, the set of thought collectives looks a little like a
Hilbert space: the set of propositions concerning a quantum system
can be viewed as a collection of ``local'' Boolean algebras with a
certain ``pasting'' condition for connecting them together---in total
one gets a so-called Boolean manifold. So too, it might be the case
with thought collectives.
\medskip

\noindent {\it Fact as thought-collective resistance -- p. 101-102:}
\bq
In the field of cognition, {\it the signal of resistance\/} opposing
free, arbitrary thinking is called a {\it fact}.  This notice of
resistance merits the adjective ``thought collective,'' because every
fact bears three different relations to a thought collective: (1)
{\it Every fact must be in line with the intellectual interests of
its thought collective}, since resistance is possible only where
there is striving toward a goal.  Facts in aesthetics or
jurisprudence are thus rarely facts for science.  (2) {\it The
resistance must be effective within the thought collective.  It must
be brought home to each member as both a thought constraint and a
form to be directly experienced.}  In cognition this appears as the
connection between phenomena which can never be severed within the
collective.  This linkage seems to be truth and conditioned only by
logic and content.  Only an investigation in comparative
epistemology, or a simple comparison after a change has occurred in
the thought style, can make these inevitable connections accessible
to scientific treatment.  The principle of immutability of species
characteristics was valid for classical bacteriology, according to
the interpretation of the time.  If a scientist of that time had been
asked why the principle was accepted or why the characteristics of
species were conceived in this way, he could only have answered,
``Because it is true.''  Only after a change in thought style did we
learned that the opinion was constrained mainly by the methods
applied.  The passive linkage between these principles was
transformed into an active one.  (3) {\it The fact must be expressed
in the style of the thought collective.}
\eq
{\it Objects created by thought -- p. 181:}
\bq
The boundary line between that which is thought and that which is
taken to exist is too narrowly drawn.  Thinking must be accorded a
certain power to create objects, and objects must be construed as
originating in thinking; but, of course, only if it is the
style-permeated thinking of a collective.
\eq
{\it How to use philosophical principles -- p. 181:}
\bq
\noindent
\ldots\ philosophical principles are like money.  They are very good
servants but very bad masters.  Principles should be made use of, but
not blindly accepted as guides.
\eq
{\it Symbols -- p. 125:}
\bq
This is how chemistry was described before it entered the modern age.
Such mystical allegories and comparisons and the strongly emotional
images exhale an atmosphere that is completely alien to our
scientific thinking.  The comparison of gold with the sun and of
silver with the moon survives only in popular imagination.
Associating lead with Saturn and tin with the devil has lost all
meaning even in popular thinking.  It is a special, self-contained
style, consistent from its point of view.  Those people thought and
saw differently than we do.  They accepted certain symbols that to us
appear fanciful and contrived.  What if we can present our
symbols---the potential, or physical constants, or the gene of
heredity, etc.---to thinkers of the Middle Ages?  Could we expect
them to be delighted with the ``correctness'' of the symbols and
instantly listen to reason?  Or, conversely, would they find our
symbolism just as fanciful, contrived, and arbitrarily devised as we
find theirs?
\eq
{\it Name as a property -- p. 136:}
\bq
A name here has a completely different significance from what it has
today.  It is not an arbitrary, conventional designation or one that
arose by historical accident.  The meaning is inherent in the name,
and its investigation constitutes an integral part of acquiring
knowledge about what it names.  The name ranks as a property of its
object of reference.
\eq
{\it Meaning as a property of the object -- p. 137:}
\bq
We are thus confronted with ideograms, or graphic representations of
certain ideas and certain meanings.  It involves a kind of
comprehending where the meaning is represented as a property of the
object illustrated.
\eq

\subsection{The Denkkollektiv and the Quantum}

The Denkkollektiv, what do I make of it?  Something about it sounds
so nice and right, but still I worry.  Let me try to make much
clearer what I already tried to express to you in the old letter of
October 1995.

The essential point of {\Fleck}, as I see it, is that science never
grasps the {\it actual\/} ``thing in itself.''  The most science can
do is toy with the ``things in themselves'' that it itself
constructs---that is, as part of a community-oriented project to
grasp and codify the world.  With this, I am inclined to believe. But
still one can ask, is there an actual thing in itself, the measly
idea of which can at least be abstracted from the phenomena we
observe? Could it be that science, despite its socially constructed
character, is steadily (or even jerkily) moving toward being a better
and better reflection of {\it what is\/} (i.e., that which was there
prior to our attempts to understand it)?  As far as I can recall,
{\Fleck} gave no sound argument for why that {\it could not\/} be the
case.  As far as I can tell, nothing about {\Fleck}'s system would
crumble if underneath it all there really were a ``real world'' that
science might or might not grasp.  Of course, the scientist may never
know that he is there, but that doesn't preclude the existence of
such a substrate.

This contrasts with the world indicated by quantum mechanics.  It is
a world so sensitive to the touch that the most natural understanding
of it may {\bf only} come about via Fleckian lines of thought.  This
is my present feeling.  And this is what I was trying to express in
the note I forwarded to you titled ``Fuchsian Genesis'' (it was
attached to the bottom of a note titled ``Genetic Genesis.'')

{\Charlie} {\Bennett} might say, ``Of course, there's a reality out
there. That's what gives the scientist a reason to be.'' But maybe
there isn't: the quantum points in just that direction.  Quantum
mechanics teaches us that the questions we can ask of the world have
this wonderful non-Boolean property: I can ask a question $X$ or a
question $Y$, but there's no good sense in which I can ask the
question $X\!\wedge Y$.  Thus my free will (and your free will) play
a fundamental role in the evidence we amass for our world picture,
and necessarily so. This in turn also makes the Denkkollektiv more
fundamental than one might have thought.  It is not the case that if
we as scientists cleaned up our act sufficiently well, we might find
ourselves sitting atop that hallowed edifice called reality.

The reality we have is in nonnegligible part the one we create.  And,
it seems to me, there is no reasonable way of getting around that
when---but only when---quantum phenomena are recognized as part of
the world.

\section{15 May 1999, \ ``Tsk, Tsk''}

Almost like I was 28 years younger, I bounded out of bed this
morning with this greatest of glee!  But this time it wasn't for the
Saturday morning cartoons, no.  Somehow I was sure that today you
would surprise me with some meaty philosophical speculation about
the realities we create---a report on the connections between your
thought, my thought, and Mr.\ {\Bennett}'s.  I opened my mailbox in
childlike anticipation \ldots\ \ldots\ and then \ldots\ slowly
\ldots\ \ldots\ sulked back to bed.  There was no more comfort than
the protection of the sheets over my head.  The world was left better
unfaced.

\section{22 May 1999, \ ``Reality in Lottery''}

{\Kiki} told me you called.  Sorry I haven't gotten back in touch
with you; lots of things have been going on this week, and I've
hardly had a chance to get any email done.  If you'd still like to
talk, give me a call Saturday.  Or if you want, I can call you Sunday
(when rates are significantly cheaper for us).  Let me know (when I
should be listening for the phone or commanding its dial).

In the mean time, have a look at what I swiped from the New York
Times this morning.  It's from an article titled ``Living Off the
Daily Dream of Winning a Lottery Prize.''  Would you call that
reality creation? Would you say Newsome lives in one Denkkollektiv,
while the writer of the article lives in another?
\bq
Newsome, who works at a nearby powder-coating plant, applying
protective coatings to metal products, has developed his own system
of tracking past winning numbers to determine patterns. In a ritual
that is performed in the family's house each morning, Newsome rises
at 4:30 a.m., puts on his robe and walks downstairs to the kitchen.

After turning on a pot of water for his instant coffee, he sits at
the table and in front of him he places several memo pads and worn
file cards that have hundreds of three-digit numbers written on them.

Painstakingly, Newsome creates charts based on factors like which
lottery hostess on television picks the numbers on which day of the
week. (A lottery drawing is held every day except Christmas in New
Jersey.) Although the drawing is totally random, Newsome is convinced
that there is a pattern.

``I see here that 994 has come out three times on Sunday,'' he said
as he sat down to make his selections one recent Sunday morning.
Lighting up a Marlboro 100 and scooping four heaping tablespoons of
sugar into a large mug of coffee, he predicted, ``I think that 994 is
going to come back today.'' (It did not.) Based on the information
Newsome compiles each morning, he writes down his 20 favorite numbers
for the day, and leaves a copy on the table for his wife. She decides
which ones she will play.

It is hard to tally how much the Newsomes spend or win, because in
addition to the game itself, there is a little gamesmanship between
the two. A couple of months ago, Mrs. Newsome played one of the
numbers her husband had suggested and she won more than \$300. When
Newsome asked her whether she had played it, she told him she had
not. ``I know he's hitting a lot of times and ain't saying nothing to
me,'' she said.

Last June, according to Newsome's day planner, the couple won on four
consecutive days, for a total of more than \$1,000. ``That was a good
week,'' Newsome said. But throughout most of two months early this
year, the Newsomes were playing every day and were not having any
luck.

By the middle of one month, Newsome was growing anxious. ``We could
sure use the money,'' he said one evening as he walked up to the
counter at Home Dairy and played 172, 578, 198, and 574. (They were
all losers; the winning three-digit number that evening was 416.)
\eq

\section{02 August 1999, \ ``Epiphenomena Chez Dyer''}

Thanks so much for the long note about {\Charlie}, Howard, and the
other many-worlders.  I enjoyed it very much and think we're finally
getting somewhere:  the conflicting desires for many-worlds and
Copenhagen are rooted in different ideas about the goals of science.
That does have a ring of truth, doesn't it?

I'm sorry to be writing back at such a late date over this.  My time
in Europe turned out to be much more hectic than I imagined it could
be. Actually I started to write you a note on the conversation Chez
Dyer over two months ago, just after I got your first long note about
it. But for one reason or other I just never finished it.  I've
decided to paste below what {\it was\/} written and simply not bother
with finishing it. I think the language I was using in that note was
a little overblown anyway.  Included in that note is a quote of Henry
{\Stapp} that I never got to the point of explaining; it's from his
article ``Attention, Intention and Will in Quantum Physics'' {\tt
quant-ph/9905054}.    It is from that article that I snagged the
phrase ``dangling appendage'' as you'll see momentarily.  In general
I disagree with almost everything {\Stapp} says, but I thought this
quote was noteworthy and quite relevant to our conversation Chez
Dyer.

Essentially the thing that struck me at Chez Dyer was the uncanny
feeling that all three of us were saying precisely the same thing.
But for some reason, {\Charlie} felt that he was saying something
different. That confused me greatly and, in fact, still confuses me.
In {\Charlie}'s mind, somehow you and I have crossed the bounds of
what is real science; somehow he views his language as more neutral
(less anthropocentric) than ours \ldots\ and therefore better.

The greatest danger I see in the many-worlds/one-Hilbert-space point
of view (beside the ridiculous silliness of it all) is the degree to
which it is a dead end.  The degree to which it is morally bankrupt.
{\Charlie}, by thinking that he has taken some of the
anthropocentrism out of the picture, has actually emptied the world
of all content.

Beyond that though, I think, many-worlds empties the world of content
in a way that's even worse than classical determinism.  Let me
explain.  In my mind, both completely deterministic ontologies and
completely indeterministic ones are equally unpalatable.  This is
because, in both, all our consciousnesses, all our great works of
literature, everything that we know, even the coffee maker in my
kitchen, are but dangling appendages, illusions.  In the first case,
the only truth is the Great Initial Condition.  In the second, it is
the great ``I Am That I Am.'' But many-worlds compounds that trouble
in a far worse fashion by stripping away even those small corners of
mystery.  It is a world in which anything goes, and everything does.
What could be more empty than that?

My most technical criticism though, is that I don't see anything in
the quantum formalism that compels the many-world point of view.  One
could have constructed such a free-for-all world in 1884.  William
{\James} came close:
\bq
\indent Indeterminism, on the contrary, says that the parts have a
certain amount of loose play on one another, so that the laying down
of one of them does not necessarily determine what the others shall
be.  It admits that possibilities may be in excess of actualities,
and that things not yet revealed to our knowledge may really in
themselves be ambiguous.  Of two alternative futures which we
conceive, both may now be really possible; and the one become
impossible only at the very moment when the other excludes it by
becoming real itself.  Indeterminism thus denies the world to be one
unbending unit of fact.  It says there is a certain ultimate
pluralism in it; and, so saying, it corroborates our ordinary
unsophisticated view of things.  To that view, actualities seem to
float in a wider sea of possibilities from out of which they are
chosen; and, somewhere, indeterminism says, such possibilities exist,
and form a part of the truth.
\eq
The only difference between {\James} and many-worlds is that in
many-worlds actualities do not float in a WIDER sea of possibilities;
the seas are of the same size for they are identified with each
other.  One could, for instance, take a classical phase space and
declare that all initial conditions are equally real and refuse to
discriminate between any of the them.  Each initial condition is a
world, and that is that.  Who is to say that one is more real than
the other?  [[The standard many-worlder usually objects to me at this
point by saying something like, ``But there is no notion of
superposition there!''  So I ask, ``What does that mean?'' and ``What
role does it play?'':  they always fall flat.  The reason is they
don't know what role it plays; they likely would have been
predisposed to many-worlds even if they had known no quantum
mechanics.  This is one reason I believe there is a lot of truth in
the assessment you gave in the last note.]]  If you want to see how
ridiculous one can take this point of view---even to the point of
completely forgetting about physics when one speaks of the many
worlds---have a look at Max {\Tegmark}'s article, ``Is `the theory of
everything' merely the ultimate ensemble theory?'' Ann. Phys. {\bf
270}, 1--51.  You can also find the article on the net at {\tt
http://www.sns.ias.edu/$\,\tilde{\;\;}\!$max/toe.html}.  I don't
think there is any better technical argument against many worlds than
to read this paper (though certainly Mr.\ {\Tegmark} wouldn't see it
that way!!).  Who needs quantum mechanics to have many worlds?

But that's a little bit of an aside; let me get back to {\Charlie}.
The other day he said to me---and this is a direct quote---``It is
the fabric of possibilities that is real.''  That's his way of
describing the many-worlds point of view.  He sees it that the task
of science is to delimit what is possible and not go beyond that.
Anything beyond that is a kind of religion, or chauvinism as he
calls it. There is a way in which I am mildly in agreement with
this, but I don't see any way of grounding the word ``possibility''
in a way that does not take into account what is ``known'' \ldots\
and for that one needs a knower (always). The real extent of my mild
agreement is that I don't believe the {\it actual\/} outcomes of
quantum mechanical experiments, or the mechanism by which they
arise, are contained within the structure of the theory:  it is a
theory of what is possible, nay, probable, but based on what one
knows. Wave functions don't collapse because of any mechanism in the
world that changes them abruptly, they collapse when and only when
our knowledge changes.  [[Can a dog collapse a wave function?  Dogs
don't use wave functions.  Heck I didn't collapse a wave function
until I was at least 20 years old.  We should never confuse the
world with what little bit we know of it.]]

It seems to me that there must be a sense in which the world just
is---i.e., the sort of thing that {\Charlie} and Howard are striving
for---but that way of describing the world is not open to the methods
of science.  How can one or even a community of observers bootstrap
themselves to such a point of view?  And indeed even if it could be
done, what would it mean for the world that ``is'' to be nothing more
than a list of possibilities without actualities (or, to be more
accurate to {\Charlie}'s point of view, a list of possibilities that
are all equally actual)?  All that is just too high falutin' for me
\ldots\ and as I tried to explain above, simply too empty to be a
useful or interesting guide to the world.

The beauty I see in quantum mechanics is precisely the beauty of
Copenhagen.  As I tried to say it crisply to {\Bill} {\Wootters} the
other day:  the great lesson of quantum mechanics is that the world
can be moved.  There is a reason we are stuck with a physics that is
``the ability to win a bet'' instead a physics that is a static
portrait of ``what is.''  If the world can be moved, we simply can't
ask for more; it would be self contradictory.  There has to be some
room left within the physical theory; it can't be a closed book if
the book itself isn't closed.

But I ramble.  Let me get back to some of the questions you asked.
What should we do with all these ideas?  I'm certainly not averse to
our writing something together.  If we can make the points clearly,
it could be a good service to the community.  Any more concrete
ideas as to how we could go about it?  Do you have an outline in
mind?

OK, that's enough for now.  Ping me your thoughts on all that I said
here and I'll pong you back.  (Don't forget to look at the further
stuff below.)

Below follows the never completed note including the {\Stapp}
quote:\medskip

\noindent
-----------------------------------------------------------------
\medskip

Thanks again for the long note putting some thoughts in order about
the nice conversation we had with {\Charlie} April 4 (Easter).

I think I do agree with you that,

\bhbe
[I]t seems that there is a great crossroads in the next century,
whether we continue as a society to ignore the creation of reality in
\& by science or we choose to use that creativity for good purposes.
\ehbe

Do you have anything concrete worked out along these lines?  Have you
written anything expanding upon this sentence?  Is there a particular
part of Muddling Through that I should take a look at?

Anyway, I thought in this note I'd try to add my two cents to your
valiant summary of our discussion chez Dyer.  The most important
thing I could see, from my perspective, was the unnerving similarity
between the following two thoughts (one {\Charlie}'s and one mine).

1) Chris:  The quantum mechanical formalism does not address a
process or mechanism by which the singular, particular measurement
outcomes in experiments come about.  That aspect of our world is just
not contained within the formalism and cannot be, but that is no
blemish to the theory.

2) {\Charlie}:  The quantum mechanical formalism does not address a
process or mechanism by which the singular, particular measurement
outcomes in experiments come about.  That aspect of our world is just
not contained within the formalism and cannot be, but that is no
blemish to the theory.

See the difference?  If not, then read one more time.

The only great distinction as far as I could tell was in the
intonation with which they were presented!  Actually, that's an
exaggeration, but these sentences did have two vastly different
thoughts behind them.

The reason I would say such a sentence (in my intonation) is because
it seems clear to me that quantum theory is about and only about our
knowledge.  It is the best we can say---the most we can
predict---about the outcomes of our prodding of the world.  It is the
best estimate we can give of the world's reaction to our
interventions.  I don't condemn the weatherman when he knows no
physics or chaos theory, when he knows not the details of how rain is
actually made; his predictions are useful nonetheless and it is worth
my while to pay him for his services.  So to, quantum mechanics
should not be blamed for not providing the omniscience we have come
to expect from classical physics:  it cannot predict the outcomes of
my prods with complete certainty because the world itself does not
know how it will react (nor can it know in advance if or how I will
prod).  That it gives no mechanism is because it is about our
knowledge; that it cannot give a mechanism is because there is none
there to be found.  ``But what about the weatherman?  We can imagine
one that is infinitely good, can't we?  One that is on the mark each
and every time?  This says your assessment of quantum theory need not
be correct.''  Well, we can imagine unicorns too.  Just because a
theory is about knowledge, it does not automatically mean that it can
be improved.  (And that statement need not be a renunciation of the
goals of science.)

\bq
A controversy is raging today about the power of our minds.
Intuitively we know that our conscious thoughts can guide our
actions.  Yet the chief philosophies of our time proclaim, in the
name of science, that we are mechanical systems governed,
fundamentally, entirely by impersonal laws that operate at the level
of our  microscopic constituents.

The question of the nature of the relationship between conscious
thoughts and physical actions is called the mind-body problem.  Old
as philosophy itself it was brought to its present form by the rise,
during the seventeenth century, of what is called `modern science'.
[\ldots]  The central idea is that the physical universe is composed
of ``material'' parts that are localizable in tiny regions, and that
all motion of matter is completely determined by matter alone, via
local universal laws.  This {\it local\/} character of the laws is
crucial. It means that each tiny localized part responds only to the
states of its immediate neighbors:  each local part ``feels'' or
``knows about'' nothing outside its immediate microscopic
neighborhood.  Thus the evolution of the physical universe, and of
every system within the physical universe, is governed by a vast
collection of local processes, each of which is `myopic' in the
sense that it `sees' only its immediate neighbors.

The problem is that if this causal structure indeed holds then there
is no need for our human feelings and knowings.  These experiential
qualities clearly correspond to large-scale properties of our brains.
But if the entire causal process is already completely determined by
the `myopic' process postulated by classical physical theory, then
there is nothing for any unified graspings of large-scale properties
to do. Indeed, there is nothing that they {\it can\/} do that is not
already done by the myopic processes.  Our conscious thoughts thus
become prisoners of impersonal microscopic processes:  we are,
according to this ``scientific'' view, mechanical robots, with a
mysterious dangling appendage, a stream of conscious thoughts that
can grasp large-scale properties as wholes, but exert, as a
consequence of these graspings, nothing not done already by the
microscopic constituents.
\eq

\section{17 August 1999, \ ``A {\Bernstein} Off the Earth?''}

Where've you been?  Did you fall off the world?  Most importantly,
did you get the long note I wrote you a couple of weeks ago
concerning THE conversation chez Dyer?  I think it's my best attempt
yet to pin down what I hate about many-worlds.  (It also pins down
what I hate about modal interpretations of quantum mechanics.)  I
would love to hear your comments.  Also, I'd be interested in
hearing further ideas about our pursuing this issue in the public
eye, i.e., by writing some of it down in a paper.  (Remember the
only rule I have in writing papers is that the author lists be
alphabetically ordered.)

I wish you had been at the foundations conference in Maryland last
week.  In my talk (which I think went quite well this time), I took
a strong stance on the interpretation of QM.  Namely I stated very
explicitly that the most one can hope for in interpreting the state
vector is that it corresponds to a state of knowledge.  If you use
state vectors, you can collapse them; if you don't, then you can't.
They're just not physical entities; the don't exist in and of
themselves.  In fact I started off the whole talk with a slide of
that wonderful quote by Bruno de {\Finetti} (suitably modified to the
quantum context):
\begin{quote}
\small My thesis, paradoxically, and a little provocatively, but
nonetheless genuinely, is simply this:
\begin{center}
QUANTUM STATES DO NOT EXIST.
\end{center}
The abandonment of superstitious beliefs about the existence of
Phlogiston, the Cosmic Ether, Absolute Space and Time, ..., or
Fairies and Witches, was an essential step along the road to
scientific thinking. The quantum state, too, if regarded as something
endowed with some kind of objective existence, is no less a
misleading conception, an illusory attempt to exteriorize or
materialize the information we possess. \hspace*{\fill} --- {\it the
ghost of Bruno de {\Finetti}}
\end{quote}

One thing that was really funny was how in conversations, these
ideas were referred to variously as ``the knowledge interpretation
of quantum mechanics'' or ``Fuchs's interpretation.''  Finally, at
the last lunch, just as everyone was parting, I turned to Lucien
{\Hardy} and Robert {\Garisto} (an editor at PRL) and said, ``You
know, I never said this because I knew that I would be able to
convey the ideas more effectively if I didn't \ldots\ but the point
of view I've been advocating is nothing other than the Copenhagen
interpretation.''

I had some really superbly productive conversations with Lucien
{\Hardy}.  If you ever get the chance, get to know him well; he is a
scholar in the best sense.  I think he's one of the few already
poised to help us stretch and tone our point of view.  (He has even
admitted that he finds it somewhat attractive.)

Write me some time (soon)!

\section{14 September 1999, \ ``A {\Fleck} of {\Fleck}''}

This morning while reading Markus {\Fierz}'s book {\sl Girolamo
{\Cardano}:~1501--1576, Physician, Natural Philosopher,
Mathematician, Astrologer, and Interpreter of Dreams}, I came across
a passage of {\Cardano}'s that reminded me of something I had sent
you about {\Fleck}. In particular, recall that I was struck by the
following two passages in {\Fleck}:

\bq
\noindent
{\it Name as a property -- p.\ 136:} \ A name here has a completely
different significance from what it has today.  It is not an
arbitrary, conventional designation or one that arose by historical
accident.  The meaning is inherent in the name, and its investigation
constitutes an integral part of acquiring knowledge about what it
names.  The name ranks as a property of its object of reference.
\eq

and

\bq
\noindent
{\it Meaning as a property of the object -- p.\ 137:} \ We are thus
confronted with ideograms, or graphic representations of certain
ideas and certain meanings.  It involves a kind of comprehending
where the meaning is represented as a property of the object
illustrated.
\eq

{\Fleck} was spurred to these passages by a reading of Fontanus's
epitome of Vesalius's {\sl Anatomy\/}.  I think it was published in
1642.

Anyway it appears that {\Cardano} was already fighting this
thought-style when he wrote {\sl De Libris propriis\/}, ca.\ 1562. In
a passage offering some basic suggestions about how books should be
written, {\Cardano} writes:

\bq
\noindent
Pay attention to the things Galen was criticized for, and don't think
that they will bring you praise.  Always remember:  Words are there
to describe things, not things to illustrate words.
\eq

I wonder how widespread this practice---the one of Fontanus, i.e.,
that of thinking of the symbol as a property---was by the time
{\Cardano} wrote this?  I can tie these thoughts to one other thread
by telling you something I read in a paper by Charles {\Enz} [C. P.
{\Enz}, ``The Wavefunction of Correlated Quantum Systems as Objects
of Reality,'' in {\sl Vastakohtien todellisuus: Juhlakirja
professori K. V. {\Laurikainen} 80-vuotisp\"aiv\"an\"a}, edited by U.
Ketvel, et al. (Helsinki U. Press, 1996), pp.~61--76.]  It has to do
with a concept that {\Pauli} called the ``idea of the reality of the
symbol.''  {\Enz} writes,
\bq
For a Chinese or a Japanese the reality of symbols is exemplified by
the Chinese character representing his name.  Indeed for a Chinese
{\it a calligraphic character thus is, after all, like the cipher of
his identity}.  In the same way as the Chinese characters have to be
appreciated not as static pictures but by recreating in one's mind
the dynamics of the calligrapher's mind, body and brush drawing the
successive strokes, the understanding of a specific quantum
phenomenon---which, as I tried to show, is most accurately described
in terms of the wavefunction $\psi$---is a {\bf dynamical process}.
This suggests that the quest for quantum reality may be best
characterized as {\bf active realism}.  Indeed, in order to
understand the {\bf symbolism of $\psi$} one has to recreate in one's
mind the {\bf properties expressed by $\psi$} much in the way Chinese
characters have to be read.
\eq

I think this is perhaps closely connected to something much more
technical I wrote Howard the other day \ldots\ but you can discuss
that with Howard.

\section{01 September 1999, \ ``The Allure of Texas''}

To help answer a question you asked a long time ago.  From J.~L.
Casti, {\sl Paradigms Lost:\ Images of Man in the Mirror of
Science}, (William Morrow and Co., New York, 1989):

\bq
Texas may call itself the Lone Star State but Texans have always
done things in a big way, so when the agenda item is reality
generation no one will be surprised to find that the ``lone star''
is magically transformed into an entire universe of glowing objects,
the centerpiece being nothing less than the meaning of meaning
itself.  The chief architect of this Texas-sized version of reality
is John A. {\Wheeler}, director of the Center for Theoretical Physics
at the University of Texas at Austin.

The heart of the Austin Interpretation championed by {\Wheeler} is
the idea of a reality created by the observer through exercise of the
measurement option.  The Austin school believes that we are wrong to
think of the past as having a definite existence ``out there.'' The
past exists only insofar as it is present in the records we have
today.  And the very nature of those records is dictated by the
measurement choices we exercised in generating them.  Thus, if we
chose to measure an electron's position yesterday in the lab and
recorded the resulting observation, then that electron's position
from yesterday exists but its velocity doesn't.  Why not?  Simply
because we chose to measure the position and not the velocity.

Because this very act of {\it choosing\/} is always involved in what
we measure, {\Wheeler} feels that the act of observation is ``an
elementary act of creation.''  \ldots

We should hasten to note that the Austin Interpretation champions an
{\it observer}-creat\-ed reality, not a consciousness-created one.
The Austin view, while differing from Copenhagen in significant ways,
still accepts some of the crucial aspects of {\Bohr}'s position.
Most important, the two schools agree that scientists can communicate
unambiguously only about the final results of measurement.  For
{\Wheeler}, the essence of existence (reality) is meaning, and the
essence of meaning is communication defined as the joint product of
all the evidence available to those who communicate.  In this view
meaning rests on action, which means decisions, which in turn force
the choice between complementary questions and the distinguishing of
answers.  Putting all these links together, out pops the Austin
Interpretation of reality generation by exercise of the quantum
measurement option.
\eq

\section{09 September 2000, \ ``More Fleckulation''}

The ``{\Fleck} materials'' book is quite nice.  There are several
articles by {\Fleck} himself, one of them previously unpublished:

\bv
1) Some Specific Features of the Medical Way of Thinking [1927] \\
2) On the Crisis of `Reality' [1929] \\
3) Scientific Observation and Perception in General [1935] \\
4) The Problem of Epistemology [1936] \\
5) Problems of the Science of Science [1946] \\
6) To Look, To See, To Know [1947] \\
7) Crisis in Science [unpublished, 1960]
\ev

It also includes 14 articles about {\Fleck}'s ideas and their context
in Polish philosophy.  Finally, it contains a full bibliography of
{\Fleck}'s writings.

\bq
\noindent R.~S. Cohen and T.~Schnelle, {\sl Cognition and Fact:\
Materials on Ludwik {\Fleck}}, (D. Reidel, Dordrecht, 1986).
\eq

You should have a look at it if you get a chance.

\section{26 October 2000, \ ``Activating or Catalyzing?''}

Thanks a million for the notes!  I'll incorporate them soon.

I've had a tremendously difficult time trying to come up with a name
for this document.  Here are the ones I've tried below.  Any votes?
And please explain why you think what you think.

\bv
The Undetached Observer: \\
The Activating Observer: \\
The Catalyzing Observer: \\
The Malleable Reality: \\
The Malleable Substrate: \medskip\\
\it Resource Material for a {\Pauli}an--{\Wheeler}ish Conception of Nature
\ev

\section{29 November 2000, \ ``Anecdote''}

Too bad you're not coming to Vienna.  I'm letting the soul (but not
the libido) of {\Schroedinger} slowly seep into me.

Let me tell you a quick anecdote while I have your ear.  It's one
you'll appreciate given your great respect for Ludwik {\Fleck}.  I
was talking to Experimentalist X the other night, asking him to
evaluate his various students so I could tumble over in my head
whether any should be approached about employment at Bell Labs.
Sadly, one of his students didn't get the highest of
recommendations.  The main point was that the student didn't seem to
appreciate the difference between making an experiment happen and
keeping himself/herself busy with work.  ``If a part doesn't work,
don't waste time trying to fix it---kill it, and get a new one,'' he
said. ``It can be tough in the lab.  It's almost as if you have to
look at your equipment and say, `{\it I will you to work}'. You have
to command nature.  And Student Y just doesn't seem to have that.''

Have you ever heard anything more marvelous come directly from the
mouth of an experimentalist?

\chapter{Letters to Doug {\Bilodeau}}

\section{22 August 1999, \ ``Your Article''}

I am writing to let you know that I enjoyed your article {\tt
quant-ph/9812050}, ``Why Quantum Mechanics is Hard to Understand,''
very much.  I was in particular very pleased with your Section IV
``Dynamics vs.\ History.''  It may interest you to know that Markus
{\Fierz} expressed a similar point of view in his article: ``Does a
physical theory comprehend an `objective, real, single process'?''
in {\sl Observation and Interpretation in the Philosophy of Physics},
edited by S. K\"orner (Dover, NY, 1957), pp.\ 93--96.  I will append
the full text below for your enjoyment.  [See note to {\Ruediger}
{\Schack}, dated 29 August 1999.]

Most importantly, I would like to ask you for the final coordinates
of your paper.  Was it published somewhere?  I would like to cite it
in an upcoming paper.

\section{28 August 1999, \ ``Not Much''}

I don't have a good idea about where to send your paper.  I think
chances are that AJP will be tougher than FP.  In any case, it had
an impact on me.  My old advisor used to say of our joint papers,
``If three people read this paper, we'll be doing OK!''  If you know
of two others with yours, you're probably doing OK.

I like this focus on the concept of ``object.''  I look forward to
anything you might have to say about it.  I have some slides that I
use as a rallying cry to the idea that quantum information theory
may have some impact on quantum foundations.  The first is a list of
axioms for quantum mechanics.  It starts off with Axiom 0:  Systems
exist.  Then I point out the stark contrast between this long list
of rather abstract looking axioms and the ones of special
relativity:  for overemphasis, I have a slide that only lists,
\begin{center}
$c$ is constant\\
physics is constant
\end{center}
Then I joke that until we can reduce quantum mechanics to such a
simple crisp statement, we will do things like have opinion polls
about people's various interpretations.  Then I put up a slide
titled ``The Jim {\Hartle} 1968 (section IV) Interpretation of
Quantum Mechanics (suitably modified)'' and say that I usually can't
even vote because I'm never represented.  Then I read a very
Copenhagenish sounding passage from Jim's old paper, changing a few
words here and there. The upshot is that a quantum state should be
understood as a state of knowledge not a state of nature.  And so I
finally return to the slide with the axioms, and I put an overlay
over it.  Beside each axiom EXCEPT Axiom 0, it says, ``Give an
information theoretic justification.'' This, I say, is the deepest
duty of quantum information theory.

So further elucidation of the concept of object would be great.  I
do think that {\Kant} still has a lot of useful stuff hidden in his
system.  If you can wait about a month I'll send you a large
compilation of things that I've had to say about this; you might
enjoy it.

The paper I alluded to is one by {\Carl} {\Caves}, {\Ruediger}
{\Schack} and myself titled ``Bayesian Probability in Quantum
Mechanics.''  It will have a companion, more technical paper, titled
``On Unknown Quantum States.''  We'll be citing you in the first.

What is your situation?  Are you a student?  Postdoc?  Professor?
(None of the above?)

Anyway, again, thank you for the long note.

\subsection{Doug's Reply}

{\it Object\/} is a practical, functional concept, not ontological or
constitutive in a mechanical sense.  On the other hand,
paradoxically, our whole concept of ``objective'' reality is based on
``objects'' (naturally) and all we can know or understand of the
``real'' structure of the physical world must come by way of them.
Hence the need for {\Kant}ian subtlety.  I know that's not very
clear, but I'm working on it.  For now, I just want a general quantum
mechanical method applicable to all cases which spells out the
relationship between objects and systems.  So I'm in the process of
going through a number of simple but real-world applications of QM,
trying to find an optimal general system for describing what's going
on.  What people usually do is fall back on the concept of particle,
and treat the particles as a classical objects whenever possible,
sliding back and forth between quantum and classical properties as
needed.  But that is neither clear nor consistent nor applicable to
all cases.

I think you're right that information theory is tied to quantum
foundations.  But not in the sense that {\Wheeler} meant with his
``{\it It from Bit}'' slogan, if I understand him.  Information (like
objects themselves) is a feature of the way we experience and
interact with the world. I like your simple summary of relativity.
We do have to do the same with QM -- the opposite of complex
formalizations like quantum logic.  But even Axiom 0 is not without
difficulties. ``Exist'' is a loaded word.  There are phenomena
correctly described by quantum systems, but people tend to confuse a
system with an object or a component of a mechanical world.

\section{06 September 1999, \ ``Labor Day Lounging''}

\bdb
One of the positive responses was from Tom Siegfried, a science
writer for the {\sl Dallas Morning News}.  He wrote a column about
the paper which appeared Jan.\ 4 of this year.
\edb

Tom is a nice guy; I've met him a couple of times at conferences. He
does seem to take a real interest in the foundations of quantum
mechanics.  That has been both good and bad.  One of the bad times
was when he wrote a couple of articles saying that Chris {\Adami} and
Nicolas {\Cerf} had ``solved the measurement problem in quantum
mechanics.'' Take a look at {\tt quant-ph/9806047} for instance and
some of the connected papers and tell me if you don't think that was
a misjudgment.

I would enjoy having a look at this article of his.  If you have a
copy of the words in your machine, please forward them to me.

\bdb
But even Axiom 0 is not without difficulties.  ``Exist'' is a loaded
word.
\edb

Agreed.  The point mostly was to make a point:  to show that there
must be some solid background for the remainder of the theory (just
as {\Rosenfeld} expresses in the passage below).

\bdb
BTW, the {\Rosenfeld} paper says some very good things.  As I read
over it again, I realized how much it influenced me.
\edb

I know of that paper too; it influenced me in the same way.  In fact
I thought it expressed itself so clearly that I scanned a passage in
and sent it to David {\Mermin}.  The result was quite positive; it
turned his head a little too.  I'll attach that below.  [See note to
David {\Mermin}, dated 17 April 1998, titled ``How Do I Sleep?'']

\section{06 September 1999, \ ``Another Request''}

I just had a look at your webpage.  If you have your paper
``Physics, Machines, and the Hard Problem'' written in \TeX\ or
\LaTeX\ or MS Word, I would appreciate getting an electronic copy to
have a read through.

Also I had forgotten to tell you \ldots\ another {\sl large\/} source
of lucid writings by {\Rosenfeld} can be found in:
\bq
\noindent
L.~{\Rosenfeld}, {\sl Selected Papers of L\'eon {\Rosenfeld}},
edited by R.~S. Cohen and J.~J. Stachel, Boston Studies in the
Philosophy of Science, Vol.\ 21, (D.~Reidel, Dordrecht, 1979).
\eq

\section{05 November 1999, \ ``Setting the Initial Condition''}

First let me apologize for not replying to your nice letters before
now.  This move to Los Alamos has just been so very hectic that I've
hardly had time for anything enjoyable lately.  But fate imposed a
little break upon me the other day:  I became infected with
salmonella Monday, and that put me out of active commission for
three days \ldots\ enough so that I had to cancel giving an invited
plenary talk (!) at the New England Section APS Fall Meeting this
week.  (I can still see my career crumbling before me because of my
misadventures with Albuquerque restaurants!)

But every cloud has a silver lining---so they say---and mine was the
opportunity to read or reread everything that you've sent me.  I
remain quite impressed by your collection of thoughts about quantum
mechanics.  I think there is quite some correlation between our
points of view (to the extent that mine is yet firmly defined).  Do
you come across the same impression, or do you see great places
where we diverge?  I would very much like to hear your thoughts on
my collection of ramblings, dreams, and desires.  As you are (to use
your words) ``intensely dedicated to this program of
clarification,'' I am intensely dedicated to have some completely
new physics come out of finally taking completely seriously the
proper point of view about QM.  So your thoughts will be immensely
welcome.

One question right away though.  In Footnote 3 of your JCS paper,
you write:
\bq
\noindent
When I write of the physicist's `freedom' of action in setting up
experiments and controlling parameters, I am not taking a position on
the philosophical question of `freedom of the will'.  I mean here
only a pragmatic freedom which is independent of the physical
entities being observed.
\eq
Can you explain in more detail what you meant by this?  Can you
think of any other references that I might read that influenced you
on this point?  In that regard, it may be helpful if you could
comment on my notes to Howard {\Barnum} (starting with the one titled
``It's All About Schmoz''), to John {\Preskill} (starting with the
one titled ``Two Rabbis in a Bar,'' but only right after Preskillism
3).  (Somewhat tangentially, but also connected to this query, you
might also enjoy the note to {\Herb} {\Bernstein} titled
``Epiphenomena Chez Dyer.'')

Have I answered all the questions that you've asked me?  Most likely
not, but let me now try \ldots\ or at least add a little meat to some
of my previous answers.

\bdb
I am curious what is the subject of the paper for which you wanted
the citation, if you don't mind describing it.  I am encouraged by
what appears to be a recent convergence of experimental and
theoretical focus on phenomena which will help bring about a genuine
and substantial clarification of fundamental concepts.
\edb

Actually, as it stands I now have three projects going on that have
some connection to the view we share about quantum mechanics.  The
first is an ``opinion piece'' for {\sl Physics Today\/} that I'm
writing with  {\Asher} {\Peres}.  Its title will be ``Quantum
Mechanics Needs No Interpretation.''  We're presenting that as
something of a rebuttal to two articles that PT published within the
last year:  one endorsed Bohmianism, and one endorsed Consistent
Histories.  This may be the ``project from hell,'' I don't know---
{\Asher} and I certainly diverge on some of our opinions about
quantum mechanics. It will be nice to see what part of our opinions
can be put into a coherent whole.  Perhaps the greatest divergence
between us is attitude:  I believe I am much more inclined than
 {\Asher} to see the clarification of quantum mechanics as recognizing
that we are on the tip of a great iceberg.  I think he's more
inclined to see it as the closing of a book.  This article should be
finished by the end of November.

The second project (with {\Caves} and {\Schack}) is one that must be
finished before I go to the Naples meeting on ``Chance in Physics''
just after Thanksgiving.  I've been titling my talks about it ``On
Unknown Quantum States,'' but I don't yet know what we'll be calling
the paper.  A recent talk abstract for the idea goes as follows:
\bq
There is hardly a paper in the field of quantum information theory
that does not make use of the idea of an ``unknown quantum state.''
Unknown quantum states can be protected with quantum error
correcting codes.  They can be teleported.  They can be used to
check whether an eavesdropper is listening in on a communication
channel.  But what does the term ``unknown state'' mean?

In this talk, I will make sense of the term in a way that breaks
with the vernacular:  an unknown quantum state can always be viewed
as a known state---albeit a mixed state---on a larger
``multi-trial'' Hilbert space.  The technical result is a quantum
mechanical version of the de {\Finetti} representation theorem for
exchangeable sequences in probability theory:  a density operator on
an infinite tensor product of complex Hilbert spaces has complete
exchange symmetry if and only if it can be expressed as a convex
combination of identical product states.  Interestingly, this
theorem fails for real Hilbert spaces.  The implications of this
theorem for the point of view that quantum states represent nothing
over and above one's knowledge of a quantum system will be discussed.
\eq
In particular, one result is that quantum states {\sl only\/} have a
good interpretation as ``states of knowledge, not states of nature''
if the Hilbert spaces of QM are over a complex number field instead
of the reals.  This seems to say that {\sl complex\/} quantum
mechanics is crucial for the understanding you and I have.  This is
one example of the thing I call for in the samizdat I sent you: pick
an axiom of quantum mechanics and ``give an information theoretic
justification for it.''

The final project (again with {\Caves} and {\Schack}) is a much
larger one and it probably won't be finished until next spring.  The
tentative title is ``Bayesian Probability in Quantum Mechanics (The
Unexpurgated Version).''  Here's a tentative abstract:
\bq
We show that---despite their being specified by fundamental physical
law---quantum probabilities are best understood within the Bayesian
approach to probability theory. In that approach, probability always
quantifies a state of knowledge, obtaining an operational definition
only through a subject's consistent betting behavior.  The
distinction between classical and quantum probabilities lies not in
their definition but in the nature of the information they encode.
In the classical world, {\it maximal\/} information about a physical
system is complete in the sense of providing definite predictions for
all possible questions that can be asked of the system.  In the
quantum world, maximal information is {\it not\/} complete and cannot
be completed.  This distinction provides a novel way to define the
meaning of quantum indeterminism and randomness.  Through this we
find a stronger connection between probability and frequency than can
be justified classically.  Finally we reconsider the notion of an
``unknown quantum state''---an oxymoron within the Bayesian approach.
The solution to this conundrum is found in a quantum version of de
{\Finetti}'s representation theorem for exchangeable sequences.
\eq

There, I hope that gives you something of a feeling about what I'm
up to.

\bdb
Also, I've started to look at your papers and a few others in
quantum computing/information theory.  Any suggestions for recent
general reading in that area?
\edb

I'm very glad to hear that.  As I see it, quantum mechanics has
always been about information \ldots\ but we're only now looking for
the proper tools with which to express it correctly.  (But be aware,
that is a minority opinion within my subfield; most people in the
field seem to be followers of `many worlds'.)  Anyway, a good place
to start learning quantum information theory is John {\Preskill}'s
lecture notes for his course at Caltech.  You can obtain them at his
website {\tt http://www.theory.caltech.edu/%
people/preskill/index.html}. Also there you can find a paper of his
``The Future of Quantum Information'' where you can see how much his
opinion about all this diverges from mine.  And you can find some
nice public lectures and other links.

OK, it's getting late and I should get back to bed:  complete
recovery is not upon me yet.  I look forward to hearing from you
when you get a chance.

\subsection{Doug's Reply}

\bq
I do think that our views are very close; at least we are motivated
by many of the same insights.  The main difference is that I am
still leery about ``information'' as a foundational concept.  But I
have an open mind about it.  I think most of the conceptual
confusion in QM today is related to the inadequacy of the idea of
``particle'', which carries with it too much of the old Cartesian
notion of the geometrical basis of physical existence.  Feynman has
greatly advanced our understanding of quantum physics, but his
attachment to the particle idea has persisted and made it more
difficult to advance further conceptually.  Perhaps the concept of
information can help to clarify the dynamical rather than
geometrical/ontological nature of the quantum.  [\ldots]

To answer your first question:  When I wrote the comment in the JCS
paper about freedom of will, I was saying that to establish causal
connections or verify patterns in phenomena, it is necessary that
the actions of the experimenter not be correlated with the
contingent details of the phenomena under investigation.  To know
what ``free will'' in an absolute sense means is a problem more
subtle and difficult even than quantum mechanics.  It is possible
that a lack of correlation or the idea of measurement in general
will turn out to imply some kind of property of ontological
independence in the observer, but for the purpose of doing physics,
I think it is sufficient to assume that the behavior of the
observer/experimenter is not determined by some external agency
which also controls the phenomena being observed -- e.g.\ that the
brightness of a star does not change just because I decide to look
at it or vice versa. In any case, the observer can be that way and
still be free or deterministic, I think.  The whole question of free
will is perhaps not well posed.  I have read some of the writings
you mention, but will say more in a couple of days when I have had a
chance to digest it better.

Re: your prospective article for {\it Physics Today}.  I have long
thought it would be possible to blow Bohmian mechanics and
many-worlds out of the water with a simple analysis of what physics
does and why those viewpoints were introduced and how they fail to
accomplish what they set out to do. I think they fail pretty
drastically and cover up their failures with obviously feeble
rationalizations. I'll try to say more about that in a couple of
days, too. Consistent histories I'm not sure that I really
understand at all. Accounts I've seen in several papers seem really
opaque and poorly motivated to me, and I don't see that it means
anything except in simple cases where it reduces to ordinary
calculation of Feynman amplitudes.  So I don't know how to critique
it.
\eq

\section{21 March 2000, \ ``You Might Enjoy''}

I discovered that Tom Siegfried has a book now.  I even discovered
that he makes a brief mention of me in it.  He writes, ``Chris Fuchs
(rhymes with books) applied {\Landauer}'s principle \ldots''!  I
wrote everyone in my family, ``Now that's a reporter!  He even cared
that the reader get the pronunciation of our name!''  You might enjoy
it.  The opening sentence is, ``John {\Wheeler} likes to flip
coins.'' Unfortunately, he didn't seem to report your work:  it
looks like he may have written most of it before he discovered your
paper.

Here's the reference,
\bq
\noindent
T.~Siegfried, {\sl The Bit and the Pendulum: From Quantum Computing
to M Theory---The New Physics of Information}, (Wiley, New York,
2000).
\eq

\section{22 March 2000, \ ``{\Wheeler}''}

Yes I do know John.  I first met him in his undergraduate course
``Great Men, Great Minds of Science.''  He paid attention to me a
little because I made a 105 on his final exam.  The five bonus
points came from completing the sentence ``No elementary quantum
phenomenon is a phenomenon until \ldots''  Strangely no one else in
the class (of 50 or so) got the bonus!  Anyway, then I did a
semester-long research course with him, mostly under the guidance of
one of his students.  The project wasn't so interesting (it was in
computational general relativity):  much better was getting to hang
around the rest of his research group who were doing things with
quantum mechanics.

Also John is my academic great-grandfather.  The lineage goes back
like this:  Fuchs--{\Caves}--Thorne--{\Wheeler}--{\Herzfeld}.  Look
at the letter I wrote to  {\Asher} {\Peres}, 05 April 1998 titled
``Other Things.''

I like to accumulate things like this.  For instance, I have an
``{\Einstein} number'' of three.  This is because I have written
papers with {\Peres} who has written papers with {\Rosen} who has
written papers with {\Einstein}.  I have a {\Bohr} number of five or
better: Fuchs--{\Caves}--Thorne--{\Wheeler}--{\Bohr}.  And I have a
{\Pauli} number of four, I believe.  That also gives me a
{\Heisenberg} number of five or better.

\section{15 April 2000, \ ``{\Pierce} Cringes''}

Thank you for the wonderful, wonderful note.  You have indeed given
me a lot of food for thought with this one.  I will comment in depth
on it once it has all sunk in better.

But in the mean time, let me make a quick comment on one of your
points that brought a smile to my face.
\bdb
5. Redevelop the foundations of information theory and its
connections to physics in light of the above.
\edb

There is a slide I've made to open some of my talks on quantum
mechanical channels.  It's a quote from a paper by J.~R. {\Pierce}
commemorating the first 25 years of information theory (IEEE Trans.\
Inf.\ Theory, vol IT-19, 3--8 (1973)).  After reading it to the
audience, I say, ``The amazing thing is that even after over 25 more
years, we still don't know how to answer {\Pierce}'s question.''  But
with you, I should emphasize another point!  Here's the quote:
\bq
\noindent
I think that I have never met a physicist who understood information
theory.  I wish that physicists would stop talking about
reformulating information theory and would give us a general
expression for the capacity of a channel with quantum effects taken
into account rather than a number of special cases.
\eq

\section{08 June 2000, \ ``Resend?''}

Could you do me a favor and send me another copy of
\bq
\noindent
D.~J. {\Bilodeau}, ``Physics, Machines, and the Hard Problem,'' J.
Consc.\ Stud.\ {\bf 3}, 386--401 (1996).
\eq

My last version was all marked up with notes about what parts I had
wanted to enter into my computer.  Now, I'll just have to read it
again!

\bdb
I don't know that I would agree that the past is as malleable as the
future.  But I do believe strongly that the {\it meaning\/} of the
past is in part dependent on what we do in the future.
\edb

I didn't say I believed it; I said I wanted to believe it.  But, in
any case, I am intrigued by your last remark.  Could you expand on
it?

Like so many times when I've written you, I'm flying across the
waters again.

Despite the fire, I went on with my ``Quantum Foundations in the
Light of Quantum Information'' meeting.  It went quite well, I think.
People really got into the spirit, and I think something tangible
will come of it.  I'll forward on to you the pre-meeting problem set
I had sent to all the attendants.  One of the questions was answered
fairly easily already, but there's still work to be done on the
others.  Also, it turns out that everyone contributed problems just
as I had hoped.  So the total list is substantially longer now.
Perhaps the best thing that came out of the meeting for me is that
{\Schumacher}, {\Schack} and I developed a novel argument for
linearity based purely on inference issues (related to the quantum
de {\Finetti} representation).

\subsection{Doug's Reply}

\bq
\noindent [Referring to:  ``I didn't say I believed it; I said I wanted
to believe it.  But, in any case, I am intrigued by your last remark.
Could you expand on it?'']\medskip

It's a little like when I'm trying to draw something, and my hand
slips and makes an unintended mark on the paper, but then afterwards
the shape of the mark looks suggestive and may end up as part of a
newly-imagined picture.  Events are far more complex than how we
conceptualize them. Correlations are more extensive and subtle than
we generally expect (e.g.\ ``karma''?).  Looking back, I think we can
find ways in which accidents become opportunities, and in which our
own apparently random, mindless actions take on an effectively
intentional nature if we have the insight and imagination to
incorporate them into a larger ``life-pattern''.  Maybe what we
thought was accidental wasn't really.  So part of the trick in life
is to keep our minds open to the unexpected, and in particular be
ready to turn the seemingly random into positive and constructive
additions to our world-building.  Seems to me to work that way,
anyway. There is a kind of new-ageish formula in meditation or
prayer I've heard which says, ``This or something better.''  I.e., we
should be careful not to let our wishing for a certain outcome
exclude an even better possibility. (There's another good formula --
``for the highest good of all concerned'' -- there may exist win-win
solutions we would never see if our imaginations were locked into a
zero-sum state space.  We reshape the world with imagination,
intention, and expectation.  Much of the process is unconscious.  We
have already done much work of which we might not even be aware. But
the meaning of a seed depends on what we do with it.
\eq

\chapter{Letters to {\Gilles} {\Brassard}}

\section{18 November 1997, \ ``Bolt from the Blue''}

I saw Adrian {\Kent}'s four papers on the server early this morning,
and I was reminded of L\'eon {\Rosenfeld}'s description of {\Bohr}'s
reaction to the EPR paper:
\bq
\noindent This onslaught came down upon us as a bolt from the blue.
Its effect on {\Bohr} was remarkable.  We were then in the midst of
groping attempts at exploring the implications of the fluctuations of
charge and current distributions\ldots.  A new worry could not come
at a less propitious time.  Yet, as soon as {\Bohr} had heard my
report of {\Einstein}'s argument, everything else was abandoned
\ldots\
\eq

I suspect the air in Montr\'eal will be just as this when you all
awake today.  How I wish I were there to see the excitement!

I hope for the sake of our field, and a generally more exciting
world, that {\Kent} is right!  As you all know, this skeptical
physicist (i.e., me) was always a little wary of the
{\Mayers}-{\Lo}-{\Chau} strong claim of having considered all
possible quantum protocols.  I would like to hear the Montr\'eal
verdict once it is in.

A good day to all!

\section{31 August 1999, \ ``Subliminal Messages''}

Quantum Foundations in the Light of Quantum Cryptography

Quantum Foundations in the Light of Quantum Cryptography

Quantum Foundations in the Light of Quantum Cryptography

Quantum Foundations in the Light of Quantum Cryptography

Quantum Foundations in the Light of Quantum Cryptography

Quantum Foundations in the Light of Quantum Cryptography

Quantum Foundations in the Light of Quantum Cryptography

Quantum Foundations in the Light of Quantum Cryptography

Quantum Foundations in the Light of Quantum Cryptography

Quantum Foundations in the Light of Quantum Cryptography

Quantum Foundations in the Light of Quantum Cryptography

Quantum Foundations in the Light of Quantum Cryptography

Quantum Foundations in the Light of Quantum Cryptography

Quantum Foundations in the Light of Quantum Cryptography



\section{23 January 2000, \ ``I See Why Bit Commitment!''}

Now on a more positive note.  (This note was started just after I
wrote you the negative one about \ldots\@.  Now, however, I am in the
comfort of my office at home \ldots\ having coffee, thinking Sunday
thoughts.)  Let me just say that I had a bit of an epiphany in the
shuttle bus at Dulles Airport:  for the first time I have understand
why you want to take {\it both\/} the EXISTENCE of secure key
distribution and the NONEXISTENCE of bit commitment as pillars in
your sought-after derivation of QM.  You have been thinking more
deeply than me since the beginning!

Let me place at the end of this note a little piece from my
samizdat.  [See letter to {\Greg} {\Comer}, 22 April 1999, titled
``Fuchsian Genesis.'']  It sort of presents what I've been trying to
get at in a dramatic way:  it may be my best presentation of the
idea of why quantum key distribution has something to do with the
foundations of quantum mechanics.  But more to the present point,
let me tell you about a second way I use to get the point across.
I've used this slide in a few talks.  (I'll place a PostScript file
of it in the next mail.)  It consists of five frames with the
following little story.

\bq
In the first frame God starts to speak to Adam at a time just before
Genesis, ``Adam, I am going to build you a world.  Do you have any
suggestions?''
\medskip\\
{\bf Adam}:  Mostly I don't want to be alone.  I want to have friends
\ldots\ and enemies to spice things up \ldots\ and generally just
plenty of people to talk to.
\medskip\\
{\bf God}:  Done.  I'll give you a world populated with loads of
other people.  But you ask for a bit of an engineering feat when you
ask to be able to talk to them.  If you want to communicate, the
world can't be too rigid; it has to be a sort of malleable thing.
It has to have enough looseness so that you can write the messages
of your choice into its properties.  It will make the world a little
more unpredictable than it might have been for me---I may not be
able to warn you about impending dangers like droughts and hurricanes
anymore---but I can do that if you want.
\medskip\\
{\bf Adam}:  Also God, I would like there to be at least one special
someone---someone I can share all my innermost thoughts with, the
ones I'd like to keep secret from the rest of the world.
\medskip\\
{\bf God}:  Now you ask for a tall order!  You want to be able to
communicate with one person, and make sure that no one else is
listening?  How could I possibly do that without having you two
bifurcate into a world of your own, one with no contact whatsoever
with the original?  How about we cut a compromise?  Since I'm
already making the world malleable so that you can write your
messages into it, I'll also make it sensitive to unwanted
eavesdropping.  I'll give you a means for checking whether someone
is listening in on your conversations:  whenever information is
gathered from your communication carriers, there'll be a reciprocal
loss in what you could have said about them otherwise.  There'll be
a disturbance.  Good enough?  You should be able to do something
clever enough with that to get by.
\medskip\\
{\bf Adam}:  Good enough!
\medskip\\
{\bf God}:  Then now I'll put you in a deep sleep, and when you awake
you'll have your world.
\medskip\\
{\bf Adam}:  Wait, wait!  I overlooked something!  I don't want an
unmanageable world, one that I'll never be able to get a scientific
theory of.  If whenever I gather information about some piece of the
world, my colleagues lose some of their information about it, how
will we ever come to agreement about what we see?  Maybe we'll never
be able to see eye to eye on anything.  What is science if it's not
seeing eye to eye after a sufficient amount of effort?  Have I
doomed myself to a world that is little more than chaos as far as my
description of it goes?
\medskip\\
{\bf God}:  No, actually you haven't.  I can do this for you:  I'll
turn the information-disturbance tradeoff knob just to the point
where you'll still be able to do science.  What could be better?  You
have both privacy and science.
\medskip\\
So Adam fell into a deep sleep, and God set about making a world
consistent with his desires.  And, poof(!), there was QUANTUM
MECHANICS.
\eq

That's the tale.  But now I see the crucial spot of outlawing bit
commitment within it.  God could have supplied Adam with a set of
impenetrable boxes (and keys to open them) where he could place his
information whenever he wanted some secrecy.  A bit commitment
protocol could certainly be used in that secondary fashion.  But God
chose to make all information open for all the world to see:  he
just left the possibility of an imprint whenever someone has a look.

Now for the old piece from the samizdat.  Have fun.

\section{22 March 2000, \ ``It from Bit''}

I was just lying in bed thinking about our conference and planning
the sorts of things that I will say in my ``Setting the Flavor of
the Meeting'' talk.  One thing that occurred to me is that I will
most surely start off with a slide of a quote from one of John
{\Wheeler}'s letters (written to Carol {\Alley}).  I'll place it
below; you ought to read it.  I don't think there's anything more
appropriate to start the meeting with.

In that connection, I was thinking that it is precisely because of
him and his influence that at least four participants would even
dream that quantum information would have something to say about the
foundations.  ({\Bernstein}, Fuchs, {\Schumacher}, {\Wootters}, that
is.) Also in that connection I remembered one of John's talks in
1994.  I wrote up a little story about it once; let me insert that
here:

\bq
John {\Wheeler} (a long-time professor at Princeton and later at the
University of Texas) was a great advocate that information theory had
something deep to say about quantum mechanics.  He always exuded this
air of urgency about him:  ``It is imperative for us to understand
the meaning of quantum mechanics in the grand scheme of things! We
must make as many mistakes as we can, as fast as we can, so that we
can hope to obtain an understanding within our lifetime!'' The last
time I saw him was in 1994 at a little conference in Santa Fe---the
one, in fact, where Peter {\Shor}'s factoring algorithm was
announced. (Actually, come to think of it, I saw him one time later
that same year \ldots\ at his 83rd birthday festschrift.)  Anyway,
at Santa Fe, {\Wheeler} gave a talk (probably titled ``How Come the
Quantum?'')\ that he closed with a slide depicting {\Planck}'s head
(maybe etched on a coin or something).  I remember, he said
(roughly), ``In 1900 {\Planck} discovered the quantum.  The end of
the century is drawing near.  We only have six years left to
understand why it is that it's here. Wouldn't that be a tribute?!''
\eq

Well, this year is the 100th anniversary of {\Planck} discovering the
quantum of action.  Here's the punchline.  What would you think
about inviting John to the meeting as a token participant?  Just a
thought, really.

\section{15 May 2000, to the attendants of the Montr\'eal Meeting, \
``Problem Set Coming''}

In the next email, you will find a longer letter from that I started
a few days ago and brought to something of a closing point this
morning.  It is a problem set I wrote up to indicate the sort of
playfulness I hope you will all join in on in drawing up your set of
``concrete problems'' for our meeting.

I apologize that I was not able to write things in as much detail as
I would have liked.  However, I still hope you can catch on to the
drift of most questions.  As some of you know, my house and
everything in it burnt up in the Los Alamos fires last week.  Life
has become more hectic than I ever imagined it could be.

Still I look forward to a fun and truly productive meeting starting
Wednesday.  I think we're all finally in a position to really take
quantum mechanics by the tail.  There are untold treasures out there
if we'll all just take the trouble to look for them.

\section{15 May 2000, \ ``Problem Set Based on
Information-Disturbance Foundation Quest''}

I hope you've had a chance to think about the request {\Gilles} and I
made in our invitation letter:  namely, to compile a list of
concrete problems whose solutions might shed some light on the
foundations of quantum theory.  What we were thinking in particular
is that no point of view about quantum foundations is worth its salt
if, at this stage, it doesn't raise as many questions as it answers.
Why should we buy into a point of view if it doesn't lead to more
fun or, at the very least, something more concrete than a stale
philosophical satisfaction?

With that in mind, I've decided to grease the gears a bit by giving
you a preview of some of the problems motivated by my particular
ish-ism. If you haven't yet created a set of your own problems
(based on your ish-ism of course), I hope this will give you a
flavor of what we were thinking when we made our request. Certainly
the more varied the sets of problems everyone brings, the greater
the chance we have for making some real progress!

The point of view I'm likely to represent at our meeting is, I
think, best captured (though perhaps a little flamboyantly) by a
manifesto I wrote a couple of years ago.  Let me reproduce that here
as an introduction and motivation to the problems that follow.

\bq
\small
\begin{center}
\bf Genesis and the Quantum
\end{center}
\bq \noindent In the beginning God created the heaven and the
earth.  And the earth was without form, and void; and darkness was
upon the face of the deep. And the Spirit of God moved upon the face
of the waters. And God said, Let there be light: and there was
light.  And God saw the light, that it was good; and God divided the
light from the darkness. And God called the light Day and the
darkness he called Night.  And the evening and the morning were the
first day. \ldots\ [And so on through the next five days until
finally \ldots] And God saw everything that he had made, and behold,
it was very good. And there was evening and there was morning, a
sixth day.  Thus the heavens and the earth were finished, and all
the host of them. \eq

But in all the host of them, there was no science.  The scientific
world could not help but {\it still\/} be without form, and void.
For science is a creation of man, a project not yet finished (and
perhaps never finishable)---it is the expression of man's attempt to
be less surprised by this God-given world with each succeeding day.

So, upon creation, the society of man set out to discover and form
physical laws.  Eventually an undeniable fact came to light:
information gathering about the world is not without a cost.  Our
experimentation on the world is not without consequence.  When {\it
I\/} learn something about an object, {\it you\/} are forced to
revise (toward the direction of more ignorance) what you could have
said of it.  It is a world so ``sensitive to the touch'' that---with
that knowledge---one might have been tempted to turn the tables, to
suspect a priori that there could be no science at all.  Yet
undeniably, distilled from the process of our comparing our notes
with those of the larger community---each expressing a give and take
of someone's information gain and someone else's consequent
loss---we have been able to construct a scientific theory of much
that we see. The world is volatile to our information gathering, but
not so volatile that we have not been able to construct a successful
theory of it.  How else could we, ``Be fruitful, and multiply, and
replenish the earth, and subdue it?''  The most basic, low-level
piece of that understanding is quantum theory.

The {\sl speculation\/} is that quantum theory is the unique
expression of this happy circumstance:  it is the best we can say in
a world where {\it my\/} information gathering and {\it your\/}
information loss go hand in hand.\footnote{Why is that a happy
circumstance?  Because it implies in part that the book of Nature
may not yet be a written product.  ``The world can be moved.''} It
is an expression of the ``laws of thought'' best molded to our lot
in life.  What we cannot do anymore is suppose a physical theory
that is a direct reflection of the mechanism underneath it all: that
mechanism is hidden to the point of our not even being able to
speculate about it (in a scientific way). We must instead find
comfort in a physical theory that gives us the means for describing
what we can {\it know\/} and how that {\it knowledge\/} can change
(quantum states and unitary evolution). The task of physics has
changed from aspiring to be a static portrait of ``what is'' to
being ``the ability to win a bet.''

This speculation defines the larger part of my present research
program.
\eq

\subsection{A. Some Concrete Problems}

\subsubsection{\protect\hspace*{.2in} Problem {\#}1:~~Pre-{\Gleason}, or Why Orthogonality?}

Andrew {\Gleason}'s 1957 theorem is an extremely powerful result for
the foundations of quantum theory.  This is because it indicates the
extent to which the {\Born} probability rule and even the state-space
structure of density operators are {\it dependent\/} upon the
theory's other postulates.  Quantum mechanics is a tighter package
than one might have first thought.

The formal statement of the theorem runs as follows.  Let ${\cal
H}_d$ be a (complex or real) Hilbert space of dimension $d\ge3$, and
let ${\cal S}({\cal H}_d)$ denote the set of one-dimensional
projectors onto ${\cal H}_d$.  We shall suppose that whatever a
``quantum measurement'' is, it always corresponds to some complete
orthogonal subset of ${\cal S}({\cal H}_d)$. Particularly, within
each such orthogonal set, the individual projectors are the
theoretical expressions for the possible outcomes of the measurement
associated with it.

Assume now that it is the task of the theory to assign probabilities
to the outcomes of all conceivable measurements. Suppose all that we
know of the way it does this is the following: There exists a
function
\be
p: {\cal S}({\cal H}_d)\longrightarrow[0,1]
\label{Heimlich}
\ee
such that
\be
\sum_{i=1}^d p(\Pi_i)=1
\label{Maneuver}
\ee
whenever the projectors $\Pi_i$ form a complete orthonormal set. It
might seem a priori that there should be loads of functions $p$
satisfying such a minimal set of properties.  But there isn't.
{\Gleason}'s result is that for any such $p$, there exists a density
operator $\rho$ such that
\be
p(\Pi)={\rm tr}(\rho\Pi)\;.
\ee
In words, {\Gleason}'s theorem derives the standard {\Born}
probability rule {\it and}, in the process, identifies the quantum
state-space structure to be the density operators over ${\cal H}_d$.
Moreover, he gets this from assumptions that are ostensibly much
weaker than either of the end results. This theorem is quite
remarkable in that it requires no further conditions on the class of
allowed functions $p$ beyond those already stated. In particular,
there is not even an assumption of continuity on the functions $p$.

A question on my mind is to what extent, if any, does the structure
of this theorem support an information-disturbance foundation for
quantum mechanics?  I think this might be fruitfully explored by
thinking in the following way.  The assumptions behind {\Gleason}'s
theorem naturally split into two pieces. (A)~The questions that can
be asked of a quantum system {\it only\/} correspond to orthogonal
projectors onto ${\cal H}_d$. A consequence of this is that there is
no good notion of measuring two distinct questions
simultaneously---that is, there is no good notion of an AND
operation for two measurements. And (B), it is the task of physical
theory to give probabilities for the outcomes of these questions,
and we can say at least this much about the probabilities:  They are
{\it noncontextual\/} in the sense that, for a given outcome, it
does not matter which physical question (i.e., which orthogonal set)
we've associated it with. This is the content of the assumption that
the probability rule is of the form of a ``frame function'' (a
function satisfying Eqs.~(\ref{Heimlich}) and (\ref{Maneuver})).

It seems to me that the first assumption to some extent captures the
idea that information gathering is invasive.  If you gather some
information and I gather some other information, there is no
guarantee that we can put the two pieces of information into a
consistent picture: my information gathering has disturbed the
relevance of the information you've already gathered. The second
assumption, however, appears to be more of the flavor that
nevertheless such information gathering is not {\it too\/} invasive.
For otherwise one might imagine the probabilities for a
measurement's outcomes to depend upon the full specification of the
orthogonal set used in its definition. The {\Born} probability rule
clearly has a much weaker dependence on the measurement than it
might have had.

A question whose answer could bolster (or discourage) this point of
view is the following.  Why is the invasiveness of quantum
measurement specifically captured by identifying measurement
outcomes with orthogonal sets of projectors?  Hilbert space has a
lot of structure; why single out precisely the orthogonal projectors
for defining the notion of measurement? To get a handle on this, we
could try to see how it might have been otherwise.

As a wild example, consider an imaginary world where quantum
measurements are not only associated with orthogonal projectors, but
with the projectors onto {\it any\/} complete linearly independent
set of vectors.  This would be a notion of measurement that made use
solely of the linear structure of ${\cal H}_d$, eschewing any
concern for its inner product.  What kinds of probability rules can
arise for such a notion of measurement?  In particular, can one have
an interesting ``noncontextual'' probability rule in the spirit of
{\Gleason}'s theorem?  More precisely, what kind of functions $p$ can
satisfy Eqs.~(\ref{Heimlich}) and (\ref{Maneuver}) but with the
summation in the latter equation satisfied for any linearly
independent set?

Well, it's not hard to see that the only noncontextual probability
rule that works for all ``measurements'' of this kind would have to
be the trivial probability assignment of $1/d$ for each outcome, no
matter what the measurement.  To give an example of how to see this,
visualize three linearly independent unit vectors $v_1$, $v_2$, and
$v_3$ in $R^3$ and imagine assigning them probabilities $p_1$,
$p_2$, and $p_3$.  Hold $v_1$ and $v_2$ fixed and rotate the third
vector whichever way you wish. As long as it doesn't fall on the two
lines spanned by $v_1$ and $v_2$, then the projector associated with
it must always be assigned the same probability, namely $p_3$.  Now
do the same thing with vector $v_2$, holding $v_1$ and $v_3$ fixed.
This will make almost all vectors on the unit sphere associated with
projectors of probability $p_2$, proving that $p_2=p_3$.  Finally,
one does the same trick with $v_1$, proving that $p_1=p_2=p_3=1/3$.

The lesson is simple:  if every linearly independent complete set of
vectors in ${\cal H}_d$ constituted a measurement, one could not
hope to retain a noncontextual probability assignment for
measurement outcomes without making the world an awfully dull place!

But maybe this version of the game is just too dumb.  So, let's try
to spice it up a bit by explicitly using the inner product structure
of ${\cal H}_d$, but in a nonstandard way.  Again consider $R^3$,
the smallest Hilbert space on which {\Gleason}'s standard theorem can
be proved. Suppose now that a ``measurement'' corresponds to any
three vectors with a fixed angle relation between themselves. What
I'm thinking of here is to start with three vectors $v_1$, $v_2$,
and $v_3$ whose angles (moving around them cyclically) are $\alpha$,
$\beta$, and $\gamma$.  Now rigidly rotate that structure in all
possible ways to generate all possible measurements.  Are there any
interesting {\it noncontextual\/} probability rules---again in the
spirit of {\Gleason}---that one can associate with this notion of
``measurement?''

Here I don't know the answer.  But I do know of some special cases
where one again gets only the trivial assignment $p_1=p_2=p_3=1/3$.
For instance, take the cases where $\alpha=\beta=\gamma$ and
$\alpha$ is such that if we rotate around $v_1$, $v_2$ will fall
back upon itself after an odd number of ``clicks''---what I mean by
a click here is rotating $v_2$ into $v_3$ and so on \ldots\ click,
click, click.  What happens if we run through such a ``clicking''
process?  Well, $v_1$ must be constantly associated with the same
probability value $p_1$ by the assumption of noncontextuality.  But
then by that same assumption, as $v_2$ rotates into the old $v_3$, it
must pick up the probability $p_3$.  And so on it will fluctuate up
and down: $p_2$, $p_3$, $p_2$, $p_3$ \ldots\ until it finally falls
upon its original position.  If this happens in an odd number of
clicks, then it will have to be the case that $p_2=p_3$ or the
assumption of noncontextuality would be broken.  Similarly we can
see that the whole circle generated by rotating $v_2$ and $v_3$
around $v_1$ must be ``colored'' with the same probability value.
Finally, run through the same process but by rotating about the
vector $v_2$.  This will generate a second circle that intersects
with the first.  From that and the assumption of noncontextuality,
it follows that $p_1=p_2=p_3=1/3$ and this will be true of any triad
by the very same argumentation.

Another case where one can see the same effect is in the single
qubit Hilbert space $C^2$.  There the game would be that any two
vectors with a fixed angle $\alpha$ between them would constitute a
measurement.  Thinking about the Bloch-sphere representation of
$C^2$, one can use the argument similar to the one above to see that
whenever $\alpha\ne 90^\circ$ the only possible noncontextual
probability assignment is $p_1=p_2=1/2$ for all possible
measurements.

\bconj
In the case of $R^3$, whenever one of the angles $\alpha$, $\beta$,
or $\gamma$ is not identically $90^\circ$, then the only possible
noncontextual probability assignment for measurements outcomes will
be the trivial one $p_1=p_2=p_3=1/3$.
\econj

How to tackle such a problem?  I think it may not be too hard
actually, especially if one assumes that the noncontextual
probability assignments, whatever they are, must be continuous
functions. The starting point would be to try to trace through
{\Asher} {\Peres}'s derivation of the standard {\Gleason} theorem in
his textbook. There, something will surely fail when one looks at the
expansion of the proposed ``frame functions'' in terms of spherical
harmonics. (I hope someone will bring  {\Asher}'s book with them to
the meeting:  I would bring mine, but I don't have it anymore.)

What's to be learned from this problem?  I'm not quite sure, but I
think mostly it will help reinforce the idea that our standard
notion of quantum measurement is not simply an arbitrary structure.
It's there for a reason, a reason we still need to ferret out.

\subsubsection{\protect\hspace*{.2in} Problem {\#}2:~~{\Wootters} Revamped with POVMs.}

{\Bill} {\Wootters} in his Ph.D. thesis explored an alternative
derivation of the quantum probability rule.  His work was based on
the hope that it could be obtained via an extremization principle
much in the spirit of the principle of least action in classical
mechanics.  (I believe he may talk about this very problem at the
meeting.)  The quantity extremized was the {\Shannon} information a
measurement reveals about a system's preparation, under the
assumption that one has many copies of the system all with identical
measurements.

Specifically the scenario was this.  Consider a {\it real\/} Hilbert
spaced of dimension $d$ and a fixed orthogonal basis within that
space.  One imagines that one has possession of $N$ copies of a
quantum system with that Hilbert space, all with precisely the same
quantum state $|\psi\rangle$.  Which quantum state?  One drawn
randomly with respect to the unique unitarily invariant measure on
the rays of ${\cal H}_d$, only one doesn't know which.  The fixed
orthogonal basis represents a measurement that one will perform on
the separate copies in an attempt to ascertain the unknown
preparation.\bigskip

\noindent [{\bf NOTE:}
{\it Here, through the remainder of this problem set,
the words were thrown together hurriedly after the fire.}]\bigskip

Anyway, {\Bill}'s attempt of a derivation didn't really work so
nicely for complex Hilbert spaces.  The question here is, can we
make it work after all, if we start thinking of POVMs as a primitive
notion of measurement in its own right.

Specifically, the first thing we must ask is does there always exist
an informationally complete set of rank-one POVM elements all of
equal weighting on ${\cal H}_d$.  And it might be even nicer if that
set could taken to have precisely $d^2$ elements. That is, for each
$d$, does there exist a set of $d^2$ projectors $|b\rangle\langle
b|$ and a positive number $g$ such that
\be
g\sum_{b=1}^{d^2}|b\rangle\langle b|=I\;?
\ee
We know that there does exist such a set if $d=2$ or $d=3$. When
$d=2$, just take any four states corresponding to the vertices of a
regular tetrahedron on the Bloch sphere.  For the case $d=3$, {\Bill}
has explicitly worked out an example that perhaps he can remind us
of. Also I remember some vague murmurings by Armin {\Uhlmann} that
``of course'' they exist in all dimensions. But still, we should
treat the existence in all $d$ as an open question---I'm not sure
how much of Armin's talk was statement of known fact, how much was
conjecture, and how much was faith.

If such a set exists always, then we can ask of it precisely the
same question that {\Bill} did in his thesis.  Assume we don't yet
know the quantum probability law: we only know that there is some
function $f$ for which the probability $p_b$ is given by
\be
p_b=f(|\langle\psi|b\rangle|^2)
\ee
when the system's ``unknown'' preparation is $|\psi\rangle$. What
function $f$ extremizes the information we gain about $|\psi\rangle$
when we have only one copy of the system available? What function
$f$ extremizes the information when we have a very large number of
copies available?

What does this have to do with my manifesto?  Perhaps nothing. But I
have always felt that {\Bill}'s attempt at derivation was missing
something in that nowhere in it did it make use of the idea that
quantum measurements are invasive beasts:  it talked about
information, but it didn't talk about disturbance. If {\Bill}'s
derivation does turn out to work nicely by the addition of POVMs,
then maybe that will be some motivation for me to rethink my ish-ism.

\subsubsection{\protect\hspace*{.2in} Problem {\#}3:~~Post-{\Gleason}, or Should I Think von
Neumann Is Special?}

For a long time, I have disliked the tyranny of thinking of von
Neumann measurements as more fundamental than other POVMs. Here's a
question that might break some of that orthodoxy.

{\bf Suppose} such a set of informationally complete POVMs as
described in the last problem exists.  Let us think of the class of
all ``primitive'' measurements on ${\cal H}_d$ as those that can be
gotten from acting the unitary group on that set.  For instance, for
a single qubit, the primitive measurements would correspond to all
possible regular tetrahedra draw on the Bloch sphere.

Let us now imagine a notion of ``frame function'' as in Problem
{\#}1 for these kinds of measurements.

Question 1:  Is there a {\Gleason}-like theorem for these structures?
And in, particular does the extra freedom of having $d^2$ outcomes
to play with simply the proof of {\Gleason}'s result?

Question 2:  By use of the Church of the Larger Hilbert space, can
we construct an arbitrary POVM with this notion of primitive
measurement.  That is, is there a kind of Neumark extension theorem
for this notion of measurement?

\subsubsection{\protect\hspace*{.2in} Problem {\#}4:~~Where Did Bayes Go?}

From my point of view, quantum states are best interpreted as states
of knowledge, not states of nature.  Quantum mechanics is {\bf
mostly} a ``law of thought'' in that it provides a firm method of
reasoning and making probabilistic estimates in light of the
fundamental {\bf physical} situation that the world is ``sensitive
to our touch.''

With that in mind, I have to ask myself why doesn't wavefunction
collapse look more like Bayes rule for updating probabilities under
the acquisition of new information.  Recall Bayes rule for when we
acquire some data $D$ about a hypothesis $H$:
\be
P(H|D)=\frac{P(H)P(D|H)}{P(D)}\;.
\ee
On the other hand, when we perform an efficient POVM $\{E_b\}$ and
find outcome $b$, we should update our quantum state $\rho$
according to
\be
\rho\; \longrightarrow\; \tilde\rho_b= \frac{1}{p_b}U_b
E_b^{1/2}\rho E_b^{1/2} U_b^\dagger\;,
\ee
where $U_b$ is some unitary operator and $p_b={\rm tr}\rho E_b$.

Forgetting about the unitary $U_b$ for the present discussion,
notice the difference in expression of these two ``collapse''
rules.  Bayes rule involves states of knowledge alone:  it is
constructed solely of probabilities.  Quantum collapse, on the other
hand, appears to involve two distinct kinds of entities: density
operators and POVMs.  Can we put it into a form more reminiscent of
Bayes rule and perhaps learn something in the process.

Here's one way that I think might be fruitful.  For each density
operator $\rho$ and each POVM $\{E_b\}$, we can construct a
canonical decomposition or refinement of $\rho$:  just multiply the
equation $I=\sum E_b$ from the left and the right by $\rho^{1/2}$.
We get,
\be
\rho=\sum_b p_b \rho_b\;,
\ee
where
\be
\rho_b= \frac{1}{p_b} E_b^{1/2}\rho E_b^{1/2}\;.
\ee
Note that with this, and just a little bit of algebra, we can
rewrite the collapse rule (again forgetting about the $U_b$) to be
\be
\tilde\rho_b=\rho^{-1/2}\left(\rho^{1/2}\sqrt{\rho^{-1/2}
\rho_b\rho^{-1/2}}\rho^{1/2} \right)^2 \rho^{-1/2}
\ee

This expression is, I think, quite intriguing.  This is because it
turns that the quantity in the large parentheses above,
\be
G(\rho_b,\rho)\equiv \rho^{1/2}\sqrt{\rho^{-1/2}
\rho_b\rho^{-1/2}}\rho^{1/2}
\ee
has been characterized independently in the mathematical literature
before.  It appears to be the most natural generalization of the
notion of ``geometric mean'' from positive numbers to positive
operators.  Here are some references:
\begin{enumerate}
\item W.~Pusz and S.~L. Woronowicz, ``Functional Calculus for
Sesquilinear Forms and the Purification Map,'' Rep.\ Math.\ Phys.\
{\bf 8}, 159--170 (1975).

\item T.~Ando, ``Concavity of Certain Maps on Positive Definite
Matrices and Applications to Hadamard Products,'' Lin.\ Alg.\ App.\
{\bf 26}, 203--241 (1979).

\item F.~Kubo and T.~Ando, ``Means of Positive Linear Operators,''
Mathematische Annalen {\bf 246}, 205--224 (1980).

\item M.~Fiedler and V. Pt\'ak, ``A New Positive Definite
Geometric Mean of Two Positive Definite Matrices,'' Lin.\ Alg.\
App.\ {\bf 251}, 1--20 (1997).
\end{enumerate}

Can someone bring these references?  As you know, my copies don't
exist anymore.  (If you only have time to copy one, perhaps the last
one is best \ldots\ as it gives a large summary of all things
known.  {\Jozsa} may be most interested in the first reference by
Pusz and Woronowicz.  Kubo and Ando would be the third most useful.

Something fun to do is to rewrite the classical Bayes rule in a
similar form as this revamped quantum rule.  Indeed the standard
geometric mean crops up in precisely this way.  Now try to further
the analogy if you can.

One possible fact I seem to recall is that the operator geometric
mean can be characterized in the following way.  Start with $\rho_b$
and $\rho$ and consider the following matrix, where $X$ is also a
positive semidefinite matrix:
\be
\left(\begin{array}{cc}
\rho &  X\\
X & \rho_b\end{array}\right)
\ee
Then $X=G(\rho_b,\rho)$ is the matrix that maximizes the above
matrix in the sense of making it as large as it can be in the
standard matrix ordering sense.

Can we learn something about why the collapse rule takes the form it
does from this exercise.  (Perhaps Jeff {\Bub} can give us an
introduction to his other characterization of the L\"uders collapse
rule.)

\subsection{B. Some Not-So Concrete Problems}

\subsubsection{\protect\hspace*{.2in} Proto-Problem {\#}1:~~Computing Power vs.\ Error
Correctability.}

We know (suspect) that quantum computing gives us a speed up over
classical computing.  But we also know that we have to strain
slightly harder to get error correction and fault tolerance for it.
Imagine now the set of all computational models (whatever that might
mean).  Within that set will be both classical computing and quantum
computing, but also a lot of other things. Could it be the case that
quantum computing hits some kind of happy medium, the one where the
ratio of speed-up to error correction resources is best?  (You can
see that this is directly motivated by my parable:  could the
speed-up of quantum computing be due to this world's wonderful
sensitivity to our touch?)

\subsubsection{\protect\hspace*{.2in} Proto-Problem {\#}2:~~Entanglement Monogamy and
{\Schroedinger}'s Insight.}

Can we think of a way of viewing quantum entanglement as a secondary
effect?  The primary effect being information disturbance tradeoff.
Let me just cut and paste an old email here.  [See the note to
{\Todd} {\Brun}, titled ``Information Theoretic Entanglement,''
dated 8 June 1999, and the note to Howard {\Barnum}, titled ``It's
All About Schmoz,'' dated 30 August 1999.]

\subsubsection{\protect\hspace*{.2in} Proto-Problem {\#}3:~~Down with Beables, Up with
Dingables.}

Quantum logicians (and presumably Jeff {\Bub}), like to think that
quantum mechanics is about ``what is,'' i.e., beables not
observables (a phrase coined by {\Bell}).  The only way they can do
that is by introducing a kind of logic about the ``facts of the
world'' that is different from our usual logic of AND, OR, and NOT.
I, on the other hand, like to think that that kind of change of
logic is a strong indicator that we just can't get a notion of a
``free-standing reality'' within physics.

So the question on my mind is what kind of algebraic structure (if
any) does my parable indicate/motivate for a purely algebraic
approach.  I've already said, I don't think AND makes any sense at
all in such a world.  Do the other notions from standard logic still
make sense though?

I realize, I've left too much vague in this question.  Talk to me at
the meeting.

\subsubsection{\protect\hspace*{.2in} Proto-Problem {\#}4:~~Intrinsic
Characterization of Complete Positivity.}

I really don't like the Church of the Larger Hilbert space. Let's
see how far we can get toward CPMs and POVMs without ever having to
assume it.  Is there a physically motivated criterion for CPMs that
makes no reference to the Church.

\subsubsection{\protect\hspace*{.2in} Proto-Problem {\#}5:~~Doing
{\Gleason} with Algebraic Numbers.}

Problem motivated by {\Bennett} remark.  Recently {\Pitowsky} and
{\Meyer} have raised some interesting questions about {\Gleason}'s
theorem when the number field of quantum mechanics is restricted to
the rationals.  It might be useful for our understanding of QM to
ferret out what is essential and what is not in its formulation.
{\Cabello} and {\Peres} completely discount {\Pitowsky} and {\Meyer}
because their world has no superposition principle.  But who cares?
Well, maybe if {\Gleason} still works in the minimal world with a
superposition principle (i.e., algebraic number fields), we should
think harder about the meaning of the ``superposition principle.''

\subsubsection{\protect\hspace*{.2in} Proto-Problem {\#}6:~~Challenge to Everettistas.}

\begin{itemize}
\item
Would your interpretation still work if the number field of ${\cal
H}_d$ were the reals instead of the complexes?  If it were the
rationals instead?
\item
Would your interpretation still work if the time evolution of the
universe as a whole were nonlinear instead of unitary?
\item
Would your interpretation still work if the collapse rule of QM were
anything different from the standard one?
\item
Would your interpretation still work if \underline{\hspace{.5in}}?
[You fill in the blank, challenge yourself.]
\end{itemize}

I understand that I'm being belligerent, but I {\it suspect\/} your
answers to each of these will be ``yes.''  Cf.\ Any of David Lewis
philosophical works on the modal logics and the plurality of worlds,
or Max {\Tegmark}'s paper {\tt gr-qc/9704009}, ``Is the `theory of
everything' merely the ultimate ensemble theory?''  The tentative
conclusion I draw from this is that the Everettista has a contentless
interpretation.

\section{17 June 2000, \ and to Montr\'eal Meeting participants, \
``Memories of Montr\'eal''}

I hope you all had a nice and thought-filled time at our little
meeting in Montr\'eal last month.  I know that I did:  In a way it
was a dream come true for me.  I was especially pleased with all the
interaction I got a chance to see---{\Bill} and Ben working out a set
of foil theories with which to compare quantum mechanics \ldots\
David and {\Ruediger} finding some common ground between {\Peierls}'
remarks and the Dutch book argument \ldots\ {\Ruediger}, Ben, and I
concocting a Bayesian argument for the linearity of time evolutions
\ldots\ {\Charlie} cautiously admitting a tear in his Everettista
fatigues because of Ben's talk \ldots\ and so on.  (I'll place the
full list of talks below.)

Perhaps I'm just being mystical, but I feel strongly that we'll find
the greatest things (and technologies) to come out of quantum
mechanics when we finally grasp the parts of it that make us feel
the most uncomfortable.  The theory is begging us to ask something
new and profound of Nature.

As most of you know---I think everyone but {\Charlie} and {\Herb},
that is---there was a general consensus at the end of the meeting
that all the participants should write up their questions and
distribute them to the rest of the crowd.  This was considered to be
a first step toward the writing of a set of articles for a special
issue of SHPMP.  (SHPMP stands for the journal {\sl Studies in
History and Philosophy of Modern Physics\/} which Jeremy
{\Butterfield} has kindly offered as a home to our articles.  Also
it turns out that Jeff and David are on its editorial board.)  I
hope everyone who hasn't started that project will do so soon,
before the magic of the meeting starts to fade from your soul.  If
we could churn some articles out by the end of the summer that would
be great.

So mainly I send this note out as a reminder to all. \medskip

\underline{\hspace{1in}}

\bv
{\bf Workshop on Quantum Foundations in the Light of Quantum
Information}
\smallskip
\\
Centre de Recherches Math\'ematiques Universit\'e de Montr\'al
\smallskip
\\
16 -- 19 May 2000 \smallskip
\\
{\Gilles} {\Brassard} and Christopher A. Fuchs, organizers
\bigskip \\

\underline{17 May}\medskip\\

{\Gilles} {\Brassard}, Universit\'e de Montr\'eal\\
\hspace*{.4in} {\it Quantum Foundations in the Light of Quantum
Cryptography} \smallskip\\

Christopher A. Fuchs, Los Alamos National Laboratory\\
\hspace*{.4in} {\it Quantum Foundations in the Light of Quantum
Information} \smallskip\\

William K. {\Wootters}, Williams College\\
\hspace*{.4in} {\it Quantum Mechanics from Distinguishability?}\smallskip\\

Benjamin W. {\Schumacher}, Kenyon College\\
\hspace*{.4in} {\it Doubting {\Everett}} \bigskip\\

\pagebreak

\underline{18 May}\medskip\\

N. David {\Mermin}, Cornell University\\
\hspace*{.4in} {\it Pre-{\Gleason}, Post-{\Peierls} \& Compumentarity}\smallskip\\

{\Herb}ert J. {\Bernstein}, Hampshire College\\
\hspace*{.4in} {\it Why Quantum Mechanics?}\smallskip\\

Richard {\Jozsa}, University of Bristol\\
\hspace*{.4in} {\it Foundations of an Interpretation of Quantum Mechanics\\
\hspace*{.4in} in the Light of Quantum Computing} \smallskip\\

Charles H. {\Bennett}, IBM Research at Yorktown Heights\\
\hspace*{.4in} {\it Entanglement-Assisted Remote State Preparation} \bigskip\\

\underline{19 May}\medskip\\

{\Ruediger} {\Schack}, University of London -- Royal Holloway \\
\hspace*{.4in} {\it Quantum Gambling and Bayesian Probability in
Quantum Mechanics}\smallskip\\

Patrick M. {\Hayden}, Oxford University\\
\hspace*{.4in} {\it Two Lessons from the {\Heisenberg} Representation, or\\
\hspace*{.4in} How I Learned to Stop Worrying \& Love Non-Commutativity}\smallskip\\

Jeffrey {\Bub}, University of Maryland\\
\hspace*{.4in} {\it Some Reflections on Quantum Logic}\\
\ev

\section{26 July 2000, \ ``Interaction without Interactoids?''}

Remember my excitement in Mykonos about how the question ``What is
real about a quantum system?'' should be answered with the reply
``The ZING!\ that gives rise to quantum key distribution.  That is,
what is real is nothing more than the full set of
information--disturbance curves.''  This is probably wacky, but now
I'm having dreams of connecting up that idea with gravity.  That the
quantum ``sensitivity to touch,'' the ZING!, might be a kind of
energy in its own right, and therefore couple to gravity.  Like I
said, kind of wacky, but it has me extremely excited.

\chapter{Letters to Jeffrey {\Bub}} \label{BubChapter}

\section{03 April 2000, \ ``Foundations in the Light of
Information''}

Below is a slightly modified version of the invitation letter I sent
everyone.  (The only change is an update on the attendance.) Looking
forward to seeing you again.  (BTW, I hope you'll take the problems
list seriously.)\medskip

\noindent --------------- \medskip

\noindent Dear friends, \medskip

{\Gilles} {\Brassard} and I are organizing a little meeting in
Montr\'eal from Wednesday May 17 through Friday May 19 on the
subject of ``Quantum Foundations in the Light of Quantum Information
and Cryptography.'' We hope you can come.  Mainly the purpose of the
meeting is to gather a group of our friends who think that some
aspects of quantum foundations are a little more mysterious than
they ought to be and, importantly, are intrigued by the idea of
applying quantum information to the task of cleaning things up.  The
atmosphere of the meeting should be quite relaxed with plenty of
time for discussion and/or private brooding: we only ask that you be
prepared to give a 20 to 30 minute talk (though longer talks are
welcome too) expressing your point of view and any avenues for
progress you foresee.  Attendants are also encouraged to compile and
share a list of concrete problems whose solution would tell us
something novel about the quantum foundations.

A lucky few attendants will stay in the Rockledge Apartments near the
University, where one living room will be made available for some of
the more informal discussions.  (This venue has been quite
productive in the past:  it helped give rise to quantum
teleportation, for instance.)  The remainder of the attendants will
stay at Hotel Chateau Versailles in the downtown area (a short bus
ride from the university).  Attendants may also consider a Saturday
night stay-over May 20 for saving on airfare and perhaps talking
some more.  Living expenses will be provided.  Travel assistance may
be available for those in need.

The list of potential attendants is presently this:

\begin{tabbing}
\hspace{.2in} \= \underline{Invitee} \hspace{1in} \=
\underline{Status} \hspace{1in} \= \underline{Known Orientation}\\
1) \> Charles {\Bennett}      \> (confirmed)   \>    Many-Worlds-ish
\\
2) \> {\Herb} {\Bernstein}   \> (confirmed)   \>    Copenhagen-ish
\\
3) \> {\Gilles} {\Brassard} \>  (confirmed) \>      -ish
\\
4) \> Jeffrey {\Bub} \> (confirmed) \>  Quantum-Logic-ish
\\
   \>             \>             \>
\\
\> \underline{Invitee}  \> \underline{Status} \> \underline{Known
Orientation}
\\
5) \> Chris Fuchs \> (confirmed) \>  Quantum-Mechanics-is-a-Law- \\
   \>             \>             \>  of-Thought-in-a-World- \\
   \>             \>             \>  Sensitive-to-the-Touch-ish
\\
6) \> Lucien {\Hardy} \>  (confirmed) \> Bohm-ish
\\
7) \> Patrick {\Hayden} \> (confirmed) \>  Local-Action-ish
\\
8) \> Richard {\Jozsa}  \> (confirmed) \>  Nonlocal-Action-ish
\\
9) \> David {\Mermin}  \>  (confirmed) \>
Correlation-without-Correlata-ish
\\
10) \> Ben {\Schumacher} \> (confirmed) \> Terribly-Unhappy-ish
\\
11) \> Umesh Vazirani \>   ??          \> ??
\\
12) \> {\Bill} {\Wootters} \> (confirmed) \> Wavefunction-Straddles-the\\
    \>               \>             \> Fence-between-Objective-and-\\
    \>               \>             \> Subjective-ish
\\
13) \> Arthur {\Zajonc} \>  ??           \>
Light-of-Nature/Light-of-Mind-ish
\end{tabbing}
Please let us know at your earliest convenience whether you can
attend. \medskip

\noindent Best regards,\medskip

\noindent Chris and {\Gilles}

\section{30 September 2000, \ ``Commit the Bit!''}

\bjb
I downloaded your mammoth `Notes on a {\Pauli}an Idea' and have been
dipping into it. Really fascinating stuff. I resonate to your idea
that we live in a world where there are some funny constraints on
gathering and communicating information -- and that this ultimately
involves an implicit ontological claim. The world is like THIS and
not like that (say, a classical world). This raises the question of
what kind of science we can have in such a world. We presumably have
the science -- quantum mechanics -- but the issue is what attitude
we should have to this science, how we should interpret it, given
the constraints on information. This puts the interpretation problem
in a new light. I might even change my mind about some things.

Have you had a chance to think about the emails I sent you at
{\Gilles}' urging?
\ejb

\bjb
{\em [From a letter originally to {\Gilles} {\Brassard}.]} I'm sorry
if I've been instrumental in publicizing this conjecture [\ldots],
but I do think that this is too intriguing a conjecture to keep under
wraps. I think there's a good chance the conjecture is correct (just
a gut feeling again), and if not then it will likely turn out that
what you want is {\tt key distribution $+$ no bit commitment $+$ X},
where this {\tt X} is something well-defined and interesting. To me,
this is what is so fascinating about the work on quantum information
and quantum cryptography: the foundational significance of the
results. In fact, your conjecture -- whether or not it's correct, or
just close to correct -- is the reason people interested in
foundations of physics should be interested in bit commitment, and
it provides the rationale for submitting the paper to Foundations of
Physics. So I think if I cut this out of the paper, it's a really
big cut. I can easily replace the paper with a new version, but I
can't force myself to stop thinking about key distribution and bit
commitment in the light of this conjecture. Of course, you are not
asking me to do that, but it feels rather odd for me to censor the
paper in this way and not mention something that I think is crucial.
I think the idea should be out there for people to think about --
and realistically, there's no way to keep the idea within a small
closed circle. So I do hope you'll change your mind. [\ldots]

I should also mention that I have been invited to give a number of
talks this semester: in Spain the beginning of October (a talk in a
`magisterial session,' whatever that means, that is a tribute to John
{\Bell}), at Johns Hopkins and at Yale in November. I have also
organized a  symposium on quantum information and quantum
computation at the biennial Philosophy of Science Association
meeting in Montr\'eal the first week of November. In all of these
talks I will be talking about quantum entanglement, nonlocality, and
the significance of the work on quantum cryptography, and I had
intended bringing in your conjecture. So I would have to censor
these talks as well, and it will feel something like telling a joke
without the punchline. This is what makes the subject interesting
philosophically.
\ejb

I've finally had a chance to read all the notes you sent me.  I was
especially pleased to learn how seriously you take the idea {\Gilles}
and I wish to see pursued.  I really don't have a problem with you
mentioning our speculation in your {\sl Foundations of Physics\/}
paper or in your upcoming talks.  The idea needs a significant
amount of refinement before anything can come of it, and probably
the only way that will ever happen is to have more of our community
thinking about it and participating in an open semi-random walk to
the truth.  So, with {\Gilles}' countenance, I would encourage you
to do just what you had planned.

Now, how to cite it properly?  I've been spouting the half about
``the existence of quantum key distribution being part of the
foundations of quantum mechanics'' since I was working for {\Gilles}
as a postdoc in 1996.  I've been able to find at least two relevant
passages in the Paulian-idea samizdat:  10--16 December 1997 in a
letter to David {\Mermin} and 6 July 1998 in a letter to Rolf
{\Landauer}.  You can cite that document and list my website as the
source:  I'll have a Bell Labs website by the end of the week if you
can wait that long.   [By the way, prompted by the fire, I've been
toying with the idea of posting an extended version of the samizdat
on {\tt quant-ph} itself.  Especially, if I can get someone
respectable like David {\Mermin} to legitimate it with Foreword.  Do
you have any opinion on this silly idea?]  Also though, I've got
three papers that allude to the idea. Here they are:
\bq
\noindent
C.~A. Fuchs, ``Information Gain vs.\ State Disturbance in Quantum
Theory,'' Fortschritte der Physik {\bf 46}(4,5), 535--565 (1998).
[Reprinted in {\sl Quantum Computation: Where Do We Want to Go
Tomorrow?}, edited by S.~L. {\Braunstein} (Wiley--VCH Verlag,
Weinheim, 1999), pages 229--259.]
\eq
See the last section in particular.  (That was actually placed on
{\tt quant-ph} in 1996; it took a long time to appear properly.)
Also,
\bq
\noindent
C.~A. Fuchs, ``Just {\it Two\/} Nonorthogonal Quantum States,'' in
{\sl Quantum Communication, Computing, and Measurement 2}, edited by
P.~Kumar, G.~M. {\DAriano}, and O.~{\Hirota} (Kluwer, Dordrecht,
2000), pages~11--16.
\eq
And, finally, for the most thorough discussion:
\bq
\noindent
C.~A. Fuchs and K.~{\Jacobs}, ``An Information Tradeoff Relation for
Finite-Strength Quantum Measurements,'' submitted to Physical Review
A.  See {\tt quant-ph/0009101}.
\eq

To the best of my recollection, {\Gilles} first mentioned the half
about bit commitment to me at the {\Newton} Institute workshop in
Cambridge last summer as we were discussing these earlier ideas.
Some time after that I had a fairly stirring epiphany when I
realized that I {\it really\/} did want to incorporate {\Gilles}'
idea into my program.  I'll dig up the note explaining that and
forward it to you in a minute.

It is a pretty speculation, isn't it?  I don't think {\Gilles} really
has anything to worry about:  even if you do manage to get other
people thinking about it, I think it'll be a slow process to settle
the issue.  We'll have plenty of chance to make a technical
contribution if we so desire.

\subsection{Jeff's Reply}

\bq
Just a quick note to say thanks for your input on this, and for the
various references. I've just returned from Spain and have a backlog
of things to take care of for tomorrow.

Re your query about posting the samizdat on {\tt quant-ph}, I think
that's an excellent idea. I must say your question: `What if God
made the world so that information gathering and exchange is
characterized by a specific sort of limited privacy ($=$ the quantum
world)?' is really resonating with me.

I now see my book as showing that if you want to interpret quantum
mechanics as giving you an ontology with a `detached observer,' to
use {\Pauli}'s phrase, then that ontology will be like {\Bohm}'s
theory, in the sense that it will (i) violate Lorentz invariance,
and (ii) involve a mechanism that will prevent you from seeing the
violation. But this would violate {\Einstein}'s dictum: `God is
subtle, but not malicious.' There would be physically different
situations associated with alternative space-like separated
measurements that would be in principle empirically inaccessible.
(In {\Bohm}'s theory this would happen because you could never
localize a Bohmian particle more precisely than the localization
given by a quantum state. So you could never control the particle
position precisely enough to instantiate the two different
situations.)

So it seems to me that you have two ways to go. You could say that
the way to understand quantum mechanics is `like {\Bohm}'s theory'
(in a sense that would include modal interpretations and a bunch of
other things, but not `collapse' theories), or you could say that the
ontology is `veiled' (d'Espagnat's term) because of something about
the way in which information works.

Now one might say that God is just as malicious this way, in
`veiling' the ontology. But my sense is that the `information' view
is really different. To continue with the theological metaphor,
there would seem to be no point to the Bohmian sort of `veiling,'
which then becomes malicious. Or putting it differently, the
question `why should it be like that?' has no interesting answer. The
difference between the `Bohmian' view and the `information' view is
something like the difference between the Lorentz-Fitzgerald theory
(in which you have real spatial contraction because of motion
through the ether, but you can never measure this) and relativity.

So the interesting question (philosophically) now becomes: What sort
of science can we have if the world is like this, i.e., if the world
has the feature that information gathering and exchange is
constrained in a certain way? This makes the Paulian view you push
in the samizdat seem more sophisticated that the Bohmian or
Einsteinian alternative.

My gut feeling was always with {\Einstein} and against {\Pauli} as
far as the `detached observer' issue is concerned. That is, it always
seemed to me that you should look for a physics in which the
observer is irrelevant. So if you end up with a physics in which
there appears to be an essential reference to the observer, then you
are on the wrong track and have to think again. But what if the way
the world is put together means that the kind of science possible is
restricted in some way because of constraints on information? That
strikes me as a perfectly good ontological ($=$ observer-free)
possibility -- it has a sort of {\Kant}ian flavour, but with a new
twist.
\eq

\section{30 October 2000, \ ``Curiosity''}

I was in a bit of a conversation with Richard {\Jozsa} and I
forwarded to him your email from October 9 where you said that the
Paulian idea is ``resonating'' with you.  This was his reply:

\bq
Those remarks by {\Bub} are very interesting. I think I would
actually go along with much of what he says in spirit, but I cannot
accept your view (if I am representing you not incorrectly) that
physics is about the state of knowledge of an observer, rather than
about an ``objective'' reality. I would interpret {\Bub}'s use of the
term ``information'' as corresponding to something objective too, in
contrast to just the state of knowledge of some observer.
\eq

Now I'm just curious to know how you react to this, especially his
interpretation of your use of the term ``information.''  My own
(snippy) reaction to Richard was this:

\bq
Somehow you always miss the point.  I thought Jeff expressed what I'm
seeking quite adequately:  the point is that deep within quantum
theory we can hope to find an ontological statement.  But that direct
ontological statement will refer to our interface with the world:
the world is wired in such and such a way that the surface terms in
the theory (the density operators) can only refer to our subjective
states of knowledge.  That does not negate that there is a real world
out there and that we strive to say useful things about it.
\eq

\subsection{Jeff's Reply}

\bq
Here's what I would say. The position I'm moving to (following
reading your Paulian manifesto and re-thinking some things) is this:

It may be that our universe (i.e., objective reality) is structured
in such a way that there are certain constraints on gathering and
communicating information. These constraints seem to be expressed as
a limitation on privacy (the possibility of secure key distribution
and the impossibility of secure bit commitment, if you and {\Gilles}
are right). I take this as a perfectly objective statement about the
ontology of our universe: ontologically, the situation is such that
information acquisition and communication is constrained in a
certain way.

Now, if this is the case, then we have to ask the question about
what sort of science we can have in such a world. That is, there are
constraints on epistemology, given that the ontology is of such and
such a sort. (There's a very {\Kant}ian flavour to this, by the
way.) I would now say that the science we can have in such a world
is a science about the sort of information we can have, and how this
information can be moved around.

Classical physics can be interpreted `descriptively,' as saying what
there is in the world, what sorts of physical systems there are,
what their properties are, and how these change as the systems
interact. This is descriptive in an observer-free sense -- the whole
story can be told without mentioning an observer.

Now, I always thought that it was important to keep to the same
observe-free format in formulating quantum mechanics. In my book I
showed that if you want to do this with quantum mechanics, then you
are going to end up with something like Bohmian mechanics. That
means that the theory will not be Lorentz invariant, but it will
incorporate a mechanism for hiding this: empirically, you won't be
able to see the violation of relativity. I think this is
unsatisfactory because it makes God malicious and not subtle. (You
recall {\Einstein}'s quote: `God is subtle, but not malicious.')
Bohmian quantum mechanics is to standard quantum mechanics like the
Lorentz-Fitzgerald theory is to special relativity.

But it may be that the way quantum mechanics is formulated, in terms
of a notion of state that refers to preparation and measurement --
our manipulations of the world and the information we obtain from
such manipulations -- rather than in terms of a notion of state that
specifies a list of properties at a particular time as in classical
mechanics, is just the right way to formulate a physical theory
because of the way the world is wired, as you put it. It may be that
an observer-free physics is not an option for our world because of
the way the world is wired. So we have to re-think what a
fundamental, objective theory of mechanics should look like if the
world is wired in this way, how it should be formulated and what it
should be about. I would say that it should be about the sort of
information we can have, and how this information can be moved
around.

I think all this is perfectly `objective.' It's just that the nature
of objective science has to be re-thought if the ontology has this
sort of structure (involving constraints on information).
\eq

\section{10 December 2000, \ ``The Oyster and the Quantum''}

I have been terribly rude to you and I hope you will accept my
apology!  When someone else takes one's ideas seriously (as you have
done with mine), the last thing one should do is return only
silence!!  More seriously, there was a time when I welcomed and even
craved letters like the last one you sent me.  But since arriving at
Bell Labs, the duties and the travels have been falling upon me so
quickly that I am afraid I am getting buried in an avalanche.  I
haven't been able to reply to {\it any\/} of the more thoughtful
emails in my mailbox.

But, all that said, thank you for your email!  Let me try to make up
a little for my silence now.  I am in the last few days of a visit
to Vienna, and I think I may have finally wrangled my delinquent
email account back to a manageable level.  So, if you reply to the
present note, I will do my best to return a prompt response.  (It is
most helpful that the Christmas holidays are coming up soon.)

Where do I begin?  First off, I don't know that I can say what I'm
trying to get at any more cleanly than in the little blurb ``Genesis
and the Quantum'' I placed in my problem set for the Montr\'eal
conference.  The only thing that competes with that might be the
Introduction I wrote for {\tt quant-ph/0009101} ``Information
Tradeoff Relations for Finite-Strength Quantum Measurements'' and the
important footnote 27 therein.  If you haven't read that
Introduction, please try to have a look at it.

Maybe one further stab at it comes from the silly abstract I wrote
for the Caltech meeting last month.

\bq
\noindent
{\bf Title:} The Oyster and the Quantum \medskip
\\
{\bf Abstract:} I say no interpretation of quantum mechanics is worth
its salt unless it raises as many technical questions as it answers
philosophical ones.  In this talk, I hope to convey the essence of a
salty, if not downright briny, point of view about quantum theory:
The deepest truth of quantum information and computing is that our
world is a world wildly sensitive to the touch.  When we irritate it
in the right way, the result is a pearl.  The speculation is that
this sensitivity alone gives rise to the whole show, with the
quantum calculus portraying the best agreement we can come to in
such a world.  True to form, I will ask more questions than I know
how to answer.  However, along the way, I will give a variant of
{\Gleason}'s theorem that works even for rational Hilbert
spaces---even two-dimensional ones---and prove some constraints on
how quantum states change under measurement.
\eq

But let me try to clarify that further by explicitly addressing what
you wrote me.

\bjb
I have a question about your conjecture that quantum mechanics is
equivalent to (can be derived from?) the possibility of secure key
distribution and the impossibility of secure bit commitment.

Secure key distribution makes it possible for two people to
communicate privately, in the sense that they can discover any
attempt to eavesdrop on their private communication. As I understand
you, you see the impossibility of secure bit commitment as a
limitation on this privacy -- a limitation that makes science
possible.

What I don't get is how this makes science possible. What is the
connection?
\ejb

Most importantly, I would warn that you not take the statement
``{\tt QKD + NQBC = QM}'' too literally.  The ingredients I see as
crucial to the endeavor are instead:  (1) That none of us can ever
completely hide the effects of our interactions with the world.
There are no lock boxes for information.  Or, in the conception of
{\tt quant-ph/0009101}, every agent in the world has it within his
power to partially align his state of knowledge about a system with
any other agent's (even if only ever so slightly).  And (2) that
there are ways to write information onto a system so that any
surreptitious gathering of the information will necessarily decrease
the writer's predictability about how the system will react to his
further interactions with it.

Item (1) sounds a lot like the nonexistence of bit commitment.  Item
(2) sounds a lot like the existence of QKD.  But I don't know that I
would really carry it any more literally than that (though I have
certainly contemplated it).  The main point is that in order for
science to exist we have to be able to come into alignment---at
least to some extent---concerning our predictions on how systems
will react to our further interactions with them.  On the other
hand, the existence of the information-disturbance tradeoff
(especially in quantum key distribution protocols) expresses that
there is a limit to the extent with which we can come into
alignment.  The latter effect has to be generally weaker than the
former effect or we would be sunk in the game of science.

That is very roughly the idea, and I just don't know how to make it
more precise than that yet.  For whatever reason, however, I do have
a tremendous faith that this is at least the right direction with
which to chip away at the problem of understanding the quantum world.

What I want to know, for instance, is WHY the
information-disturbance curve in my paper with  {\Asher} [PRA {\bf
53}, 2038 (1996)] doesn't rise more steeply than it does in quantum
mechanics.  If it did, would we not be able to have science?  Or,
look at the property I proved in Appendix A of {\tt
quant-ph/0009101}. This says that for each quantum measurement
(i.e., positive operator valued measure), there is a way of
performing the measurement so that on average I will be left with
more predictability about how the system will react to further
interventions upon it.  It need not have been so.  One could imagine
other worlds.  What goes wrong in those worlds?

\bjb
It has just occurred to me that I may have misunderstood what you
have in mind by a 'limited privacy.'
\ejb

I don't know that I ever used the term `limited privacy.'  I think
you were the one who has always emphasized that phrase (and probably
also invented it).

\bjb
The way in which private communication is secured in our quantum
world is not this way. Rather, Alice and Bob can communicate
privately by exploiting the fact that an eavesdropper can always be
detected, to any level of security. Now this feature of the world
depends on such things as 'no cloning' and 'no information gain
without disturbance.' But these features essentially (?) involve
entanglement.  If you have entanglement, then you can't have bit
commitment.
\ejb

I don't know how much these features really require entanglement
{\it per se}.  Somehow I imagine that one need only deal with
strictly weaker features of the tensor product structure in order to
see them arise.  See for instance our paper ``Nonlocality without
Entanglement'' [PRA {\bf 59}, 1070 (1999)].  Here is a set of states
that have a novel information-disturbance tradeoff in them with
respect to local observers even though they are orthogonal and have
no entanglement.

Moreover, going back to the scenario that  {\Asher} and I explored.
Imagine that we equip quantum mechanics with a kind of discrete time
evolution that never entangles systems; it always leaves them in
product states (but is otherwise inner-product preserving at those
discrete time steps).  With it one would have a no-cloning theorem
and an information-disturbance principle to boot.  But never any
entanglement.

Overall, I'm inclined to think that the key ingredient is the
noncommutivity of states.  It seems to me that entanglement is only
a derivative concept.  (I can actually make this notion
precise---especially within this context---but you're going to have
to wait a few weeks for the new paper I'm writing up.)

\bjb
You want to take quantum mechanics as a theory of the way information
is represented and the limitations on the communication of
information, and not a description of the behavior of particles, as
in classical mechanics. Granted, the way the world is hard-wired
might impose limitations on the gathering of information and the
exchange of information -- limitations expressed precisely by the
'limited sort of privacy' we have (i.e., secure kd, but no secure
bc), hence by science necessarily taking the form of quantum
mechanics. But how does it follow from this that we must interpret
quantum mechanics as a theory of information, and not as a
descriptive theory in the sense of classical mechanics?
\ejb

It doesn't.  You're completely on track there.  Do you remember my
slide where I listed the axioms of quantum mechanics in Montr\'eal?
In my presentation I said how I'm always struck by the stark
contrast between that list of axioms and the ones we take for our
other cornerstone theory of the world (referring to special
relativity):  (1) the speed of light is constant, and (2) physics is
the same in all frames.  The debate over the foundations of quantum
mechanics will not end until we can reduce the theory to such a set
of crisp physical statements---I believe that with all my heart.
However, just as special relativity will always be interpretable in
Lorentz's way, quantum mechanics will likely always be interpretable
in {\Bohm}'s way.  There's nothing we can do about that.

What I'm really searching for is just a polite way to say, ``Ahh,
blow it out your butt.  You can believe that Lorentzian way of
looking at things if you want to, but why when have this absolutely
simple alternative conceptual structure?''

What I'm looking for is just a couple of crisp physical
statements---contradictory appearing even, just as
{\Einstein}'s---that can characterize what quantum mechanics is all
about.  Something like (but more precise than):
\begin{enumerate}
\item
The effects of our interventions into the world are nondiminishable.
And
\item
But still we have science; all the world is not simply a dream of
our own concoction.
\end{enumerate}

Once we get that cleared up, we'll finally be ready to move to the
next stage of physics, much like {\Einstein} was ready to move on to
general relativity once he had reduced the Lorentz contractions to
the two statements above.

\subsection{Jeff's Reply}

\bq
Thanks for your long letter. I've printed out several of your papers
that I want to study. I'm particularly intrigued by your `Nonlocality
Without Entanglement' paper. I'll get back to you shortly with some
observations -- I've been snowed under with end of term duties.

I understand what you say about QKD and information trade-offs, and
I think there is definitely something deep and important there. But
I'm still a bit puzzled by what you say about bit commitment.

You'll probably be surprised to hear me saying that I now agree
entirely with your `blow it out your butt' comment on {\Bohm} and
modal interpretations, and your reference to Lorentz here. I don't
think these approaches shed any light at all on what's going on in
quantum cryptography and quantum computation -- in fact, they obscure
things, as far as I can see.
\eq

\section{11 January 2001, \ ``Thanks''}

Thanks a million for [\ldots]

So, let me reward your kindness with the pleasure of a question.  I
wonder whether you've thought about this before.  Suppose I have two
noncolorable sets of rays (in the sense of {\Kochen} and {\Specker}).
One set lives in one vector space and the other lives in another.
The dimensions of these two spaces can be anything you like, say $d$
and $f$.  Suppose also there are $m$ rays in one set and $n$ rays in
the other.  Now imagine forming all possible tensor product rays from
the two sets.  What I mean by this, really, is to form the set of
all tensor products of projection operators and then to turn those
into rays.  Anyway, there will be $mn$ such rays, and they will live
in a $df$-dimensional vector space.

The question is this.  Can we say anything about the colorability of
such a set?  That is, more prosaically, is the tensor product of two
noncolorable sets also noncolorable?  Or, are there
counterexamples?  Conversely, one can ask is the tensor product of
two colorable sets always colorable?  I haven't thought very hard
about these questions, so they may be trivial to answer, or they may
be hard.  The main point is, what can one say about the
``interaction'' between KS theorems and the tensor-product structure
of quantum mechanics?

\section{14 January 2001, \ ``{\Barnum} and KS''}

\bjb
I think the answer to your question is: yes. I say this on the basis
of the following reasoning (so if there is a flaw in the reasoning,
then I'm not sure):
\ejb

I think your reasoning is sound.  Thanks.  I'm just playing with
several ideas in my head about how to better pose a question about
what connection (if any) there is between KS theorems and the
existence of an information-disturbance tradeoff.  And the question
I asked you came up somewhere in the middle of that.  What I'm
grappling for is something like the following. (Before I go further,
though, please, please forgive my vagueness.)  Presumably I can
construct various {\it noncontextual\/} hidden variable theories with
an information-disturbance tradeoff property.  BUT, we know that
these noncontextual theories CANNOT match quantum mechanics in every
detail.  So the question is this, what general statements can be
made of their information-disturbance properties?  The guess of
course in my mind is that they will not be able to match quantum
mechanics in ALL I-D properties.  And that is because I hope
sincerely that quantum mechanics is nothing other than the
conjunction of all its I-D properties.

\section{18 January 2001, \ ``The Short Answer''}

OK, so I won't wait for the weekend.  Let me just try to give you a
short-ish answer and see if I can get away with it.

\bjb
What I'm really saying is this: You have embeddability in the 2-D
special case. But you still have the feature of an
information-disturbance trade-off. So perhaps this is not all there
is to quantum mechanics in the general case. Or perhaps in the
general case the information-disturbance trade-off is qualitatively
different from the situation in the 2-D case.
\ejb

Indeed, I have quite understood that the 2-D case might be a special
one.  I'm sorry I didn't mention that in my note to you.  And, in
fact, if what I were trying to get at didn't start until $d=3$, I
don't think I would be so bothered.  Remember the fascination I have
for {\Gleason}'s theorem?  So, it might appear that there is
something essential about quantum mechanics that does not kick in
until $d=3$.  In other words, I might also bank on that to fulfill my
vague information-disturbance thoughts.

But that's the cheap answer.  A deeper answer relies (1) on some
partially unpublished work, and (2) on some terrain in quantum
information that you're probably not all that familiar with (yet).
And the combination of these two things is going to make it
difficult to talk about this in email.  Let me give you the very
roughest of sketches, enough maybe to whet your interest, and then
I'm going to have to leave it at that for now.

The main point of departure is that whenever the words
``information'' and ``disturbance'' are used, they should not be used
in the fashion you did in your previous note to me.  Instead one
should harken back to the quantum cryptographic usage of the terms.
I tried to lay this out very carefully in my recent paper with
{\Jacobs}.  (It's all in the Introduction; so as far as this point is
concerned, all you need to read is the Introduction.)  The next step
in the process is to realize, as  {\Asher} and I did in our PRA '96,
that the optimal information-disturbance tradeoff comes about only
by a POVM-equipped Eve.  For a given amount of information,
projective measurements are generally too harsh.

So what?  Well, you also have to remember that I was never imbued
with the idea that projective measurements are somehow more
fundamental than general POVMs.  That's because it never seemed
convincing to me that measurements reveal ``properties'' and, thus,
that measurements should obey the algebraic properties one would
deem necessary for that.  So, I want to think of POVMs as every bit
as fundamental as anything else.  You say, ``To measure a
three-outcome POVM on a qubit, you MUST imagine interacting the
qubit with a larger ancilla.''  I say, ``Humbug!  If you admit any
mystery about standard von Neumann measurements, why lay such stock
in them for the POVM case?  How does adding the word `ancilla' take
any of the mystery out of POVMs?  Just take them as equally
fundamental and be done with it.''

So, let me ask, can I always simulate arbitrary POVM measurements
with a noncontextual hidden variable theory?  NO, of course not
(because I can't even simulate the von Neumann measurements).  But
what's interesting is that I cannot even do it when $d=2$.  (This
last statement is the partially unpublished part of what I'm telling
you.  I say ``partially'' because it is joint work with {\Caves} and
{\Renes} that has not been published.  When I showed it off to David
{\Mermin}, however, he faintly remembered having seen something like
that \ldots\ and then ultimately remembered Paul {\Busch}.  So, we
came across it independently, but we three will now never get any
credit for it. Too bad.)

Putting all this together, it is not obvious to me---and I'm willing
to bet it's wrong---that one can even simulate the curves  {\Asher}
and I found.  Or, at the very least, the curves for some similarly
defined, but more complicated problem.  So that's the sort of
direction I'm thinking in.  And, I think it gives hope to the
program.  But as I also said earlier, it wouldn't shock me either if
a fundamental change in the I-D question comes about when $d=3$.

Did that whet your interest?  Now the problem is, you're going to
ask me even more questions!  And I'm up to my knees in the day to
day fire-fighting that has come with my new job.  (Oops, I should
call it a ``more senior position.'')  So, if I end up being silent
for a while on things other than organizational, please don't take
offense.

\chapter{Letters to {\Carl}ton {\Caves}}

\section{08 April 1996, \ ``{\Hartle}/{\Coleman} Comments''}

I like the grass fire analogy: I agree we should stomp it out if we
can.  However, as with most debates, I imagine we'll just have to
wait until it falls the wayside \ldots\ or wait until someone gives
it a name like the Modern Interpretation of Quantum Mechanics with
Sugar on Top, so that everyone else will just discount it and move
on.

Do you know whether {\Coleman} has published any of this?  Wojciech
first told me about it at least a year ago (and told me I should
contact {\Coleman} concerning it to ``help your career along.'')  If
you have a copy of the paper, please send it my way.  I've been at a
little disadvantage in thinking about this because I cannot find my
copy of {\Hartle}'s paper and the library has been shut down since
Wednesday evening; I'll be able to get a new copy Tuesday morning. I
know I liked the paper very much after turning semi-Bayesian because
of what it said in words (even though I never agreed with the
mathematics); however it certainly did have a frequentist streak
running through it \ldots\ and that's what's bringing us back now.

I think what I'll do in this note is just write down various thoughts
as they come to me; I've got a few pieces of junk spread out on my
desk to prod me along.  I can see I'm going to repeat much that
you've already written me, but I want to have it written down for
myself.  Later we can see what comes of any of it.  (By the way, I
am sure that I will split many infinitives in this letter; I hope you
have the patience.)

{\bf Point 1)}\ \ I think there is much to be said for funneling the
problems of this point of view into their loose use and confusions of
``definitely'' and ``with probability 1.''  If I understand you
correctly, both {\Hartle} and {\Coleman} would like to take it as an
axiom of quantum theory that:  if a quantum system is prepared in a
state $|s\rangle$ which is a normalized eigenvector of the Hermitian
operator $\hat S$, i.e.,
\be
\hat S|s\rangle = s|s\rangle\;,
\ee
then the eigenvalue $s$ will {\it definitely\/} be the outcome of a
measurement corresponding to $\hat S$.  If, on the other hand,
$|s\rangle$ is not an eigenstate of $\hat S$, then what we can say
about the outcome (i.e., it's probability or whatever) is yet to be
determined.  In particular, they would like to determine just
exactly {\it what can be said\/} from the simple axiom above (the
``Eigenstate Axiom'').

They do this by talking (ostensibly) about what happens upon the
repeated measurement of $\hat S$ on identical preparations
$|\psi\rangle$ that are not eigenvectors of $\hat S$.  (Let's keep
things simple and suppose that $\hat S$ is $2\times2$ and its
eigenvalues and eigenvectors are labeled by $0,1$ and $|0\rangle$,
$|1\rangle$, respectively.)  Thus they consider an infinitely large
quantum system prepared in the pure state
\be
|\psi^\infty\rangle=|\psi\rangle|\psi\rangle|\psi\rangle\cdots\;.
\ee
Presumably what they {\it would like\/} to do---by defining
probability as frequency---is show that a measurement of the operator
\be
\hat S^\infty = \hat S\otimes\hat S\otimes\hat S\otimes\cdots
\ee
on $|\psi^\infty\rangle$ can {\it only\/} yield a ``completely
random'' string of outcomes $\vec{s}=s_1s_2s_3\cdots\;$ with the
``correct'' frequency of 0's and 1's, i.e.,
\bea
f_0&=&\lim_{n\rightarrow\infty}\frac{1}{n}\sum_{i=1}^n\delta_{0,s_i}
\;=\;|\langle0|\psi\rangle|^2\\
f_1&=&\lim_{n\rightarrow\infty}\frac{1}{n}\sum_{i=1}^n\delta_{1,s_i}
\;=\;|\langle1|\psi\rangle|^2\;.
\eea
However, instead of doing this, they move on to something else first.
{\Hartle} defines the ``frequency'' operator
\be
\hat F^{(n)}_0=\sum_{s_1\cdots s_n}\left(\frac{1}{n}\sum_{i=1}^n
\delta_{0,s_i}\right)|s_1\rangle\cdots|s_n\rangle\langle s_n|\cdots
\langle s_1|\;,
\ee
and similarly for $\hat F^{(n)}_1$. (The first summation in this is
taken over all possible strings of outcomes $\vec{s}_n=s_1s_2\cdots
s_n$ of length $n$.)  He then lets
\be
\hat F^\infty_0 = \lim_{n\rightarrow\infty}\hat F^{(n)}_0\;,
\ee
etc.  I presume---but certainly cannot know, since I have not seen
the paper---{\Coleman} defines a projector corresponding to the set
of random outcome strings in the following way.  Let ${\cal
S}^{(n)}_m$ be the set of all strings $\vec{s}_n=s_1s_2\cdots s_n$
of length $n$ that pass ``all'' randomness tests ({\it with respect
to the probability distribution induced by a measurement of $\hat
S$}, i.e., the numbers $|\langle0|\psi\rangle|^2$ and
$|\langle1|\psi\rangle|^2$) at confidence level $m$.  (There are
various ways of making this notion formal; I'll come back to this a
little later.)  Then take
\be
\hat\Pi^{(n)}_m = \sum_{\vec{s}_n\in{\cal S}^{(n)}_m}|\vec{s}_n
\rangle\langle\vec{s}_n|\;,
\ee
and finally let
\be
\hat\Pi_{\scriptscriptstyle{\rm rand}}=\lim_{m\rightarrow\infty}
\lim_{n\rightarrow\infty} \hat\Pi^{(n)}_m\;.
\ee
This projector should correspond to the subspace spanned by all truly
``random'' outcome sequences.

Let us assume, for the moment, that there are no real mathematical
difficulties in defining the operators above in their infinite limits
\ldots\ or, at the very least, that if there are difficulties in the
definitions I gave, they can be out-maneuvered by somebody smarter
than me.  {\Hartle} then goes on to show that
\be
\hat F^\infty_0 |\psi^\infty\rangle = |\langle0|\psi\rangle|^2\,
|\psi^\infty\rangle \hspace{5mm}\mbox{and}\hspace{5mm} \hat
F^\infty_1 |\psi^\infty\rangle = |\langle1|\psi\rangle|^2\,
|\psi^\infty\rangle\;.
\ee
{\Coleman} presumably shows
\be
\hat\Pi_{\scriptscriptstyle{\rm rand}}|\psi^\infty\rangle=
|\psi^\infty\rangle\;.
\ee

The conclusions {\Hartle} and {\Coleman} would like us to draw from
these nifty equations is that, indeed(!), a measurement of $\hat
S^\infty$ can {\it only\/} yield a ``completely random'' string of
outcomes $\vec{s}=s_1s_2s_3\cdots\;$ with the ``correct'' frequency
of 0's and 1's.  They base this on the Eigenstate Axiom even though,
as far as I can see, there is no completely clean cut connection
between measurements of the operator $\hat S^\infty$ and the
constructed operators $\hat F^\infty_0$, $\hat F^\infty_1$, and
$\hat\Pi_{\scriptscriptstyle{\rm rand}}$.

Putting that objection aside, however, this is still a {\it silly\/}
conclusion (as you've already pointed out).  For just let us
construct two other projectors:
\bea
\hat\Pi'&=&\hat\Pi_{\scriptscriptstyle{\rm rand}}+
|000\cdots\rangle\langle000\cdots| \\
\hat\Pi''&=&\hat\Pi_{\scriptscriptstyle{\rm rand}}-
|\mbox{rand}\rangle\langle\mbox{rand}|\;,
\eea
where the first projector contains one clearly nonrandom outcome
string, and the second has one component of the original
$\hat\Pi_{\scriptscriptstyle{\rm rand}}$ taken away.  What are we to
make of these?  Since, surely,
\be
\hat\Pi'|\psi^\infty\rangle=\hat\Pi''|\psi^\infty\rangle=
|\psi^\infty\rangle\;,
\ee
should we take it to mean that $000\cdots\;$ is---contrary to our
earlier conclusion---a possible outcome?  Or, instead, should we
take it that the excluded random string in $\hat\Pi''$ is {\it
not\/} possible after all?

The problem lies in asserting that something {\it definitely does
happen\/} contingent upon its having {\it probability 1\/} in
someone's mind's eye.  On this, I do agree.  What we have seen above
is a dangerous mix of ``sloppy'' frequentism and confidence that
quantum mechanics is about unpredictable thingies.  A strict
frequentist (such as von {\Mises} or {\Reichenbach}) would {\it
not\/} base his belief in a random string occurring in an infinite
measurement repetition on a probability-1 statement like the Law of
Large Numbers. Rather, he would only apply probability to phenomena
that is already ``known'' to be repeatable and which have definite
limiting relative frequencies (but are lawless in pattern
otherwise).  That is to say, the frequentist only applies
probability statements to individual terms in a random infinite
sequence or {\it kollectiv.}
\bq
 \hspace*{\fill} {\it ``Erst das
Kollectiv, dann die Wahrscheinlichkeit''\/}---R. von {\Mises}.
\eq

I used to be attracted to the idea that individual quantum
measurement outcomes (as opposed to classical coin tosses) really
were elements of kollectivs.  Apparently so was Paul {\Benioff}
[1,2]. The main thing I learned from that silly business was that if
one really wants this to be the case, then one cannot hope to find
evidence for it in the present structure of quantum theory. It has
to be taken as a {\it new\/} axiom---it is no more derivable from
quantum mechanics than the kollectiv is derivable from a theory in
which ``probability'' makes its appearance first. ({\Benioff} wrote
several papers---I cited two above, but there are many more---in
which he tried to do just that, i.e., replace some of the standard
axioms with ones about kollectivs, but I don't think anything came
of it.)

{\bf Point 2)}\ \ A few things about ``mathematical foppery.''  In
two of {\Benioff}'s papers (which are about standard quantum theory
instead of the stuff mentioned above) [3,4], he worries about
``maverick worlds'' in the {\Everett} interpretation. To get at the
questions he's concerned with, he does something that has a lot of
the flavor of what {\Coleman} seems to be talking about. Namely, he
plays with the mathematics of constructing a state like
$|\psi^\infty\rangle$, but which only has random outcome sequences
in its expansion.  He makes it pretty clear that there are a lot of
difficulties in defining such a thing rigorously.  If you're really
worried about whether Mr.\ {\Coleman}'s mathematics are on the up and
up, these papers are probably worth a second look.

Also in these papers, {\Benioff} points out that the notion of a
``random'' string is not a God-given concept: there are several
distinct notions of random string definable within the limits of the
phrase ``strings that pass {\it all\/} statistical tests.''  There
is the one due to {\MartinLof}, but there are also others (some of
which are also due to {\MartinLof} but less well-known) that
essentially depend on what kinds of oracles you allow your {\Turing}
machines to have.  {\Benioff} found this intriguing because he hoped
that quantum theory would give a means of defining the ``true''
notion of randomness.  Does {\Coleman} give any real justification
for the notion of randomness he uses?

One last thing that may be of real interest for the interpretational
question.  You say,
\bcc
He stated his results in terms of a spin-half particle and
measurements of $z$-spin, but he claims to have the general result,
and it's not hard to believe that he does. \ldots\ where randomness
is defined algorithmically (or, equivalently, in terms of all the
statistical tests of randomness).
\ecc
Let us restrict ourselves to the standard {\MartinLof} definition of
randomness in terms of statistical tests.  Then if we are talking
about randomness with respect to a fair coin toss there is indeed a
simple relation between {\MartinLof}'s definition and one in terms
of algorithmic complexity, namely an infinite string
$\vec{s}=s_1s_2\cdots s_n\cdots$\ is random if and only if there
exists a constant $c$ such that for all $n$
\be
K(s_1s_2\cdots s_n)\ge n - c\;.
\ee
(Here $K(\cdot)$ is the self-delimiting version of complexity of
{\Chaitin} and {\Levin}.)  However, if you try to define randomness
for a biased coin in the same way, things become much more sticky!
For instance, you might have thought that for a coin biased with
probability $p$ for heads, a string is random if and only if there
exists a constant $c$ such that for all $n$
\be
K(s_1s_2\cdots s_n)\ge n H(p) - c\;,
\label{bitter}
\ee
where $H(p)$ is the {\Shannon} information of the distribution
$\{p,1-p\}$.  But that ain't the case.  Instead it can be shown [5]
that $\vec{s}$ is random with respect to $\{p,1-p\}$ in the
{\MartinLof} sense if and only if there exists a constant $c$ such
that for all $n$
\be
K(s_1s_2\cdots s_n)\ge -\log\mu[s_1s_2\cdots s_n] - c\;,
\ee
where $\mu[s_1s_2\cdots s_n]$ is the {\it probability\/} for the
string $s_1s_2\cdots s_n$.

The point here is that even if one uses algorithmic complexity to
define what one means by randomness, one still has to start off with
the notion of probability and build on top of that.  If {\Coleman}
really has a general result {\it and\/} defines randomness in the
relatively standard way of {\MartinLof}, then he must make use of
the quantum probability expressions $|\langle0|\psi\rangle|^2$,
etc., at the outset \ldots\ and in a way that is not equivalent to
inserting a simple parameter as in Eq.~(\ref{bitter}).  That, it
seems to me, can hardly be viewed as deriving the quantum mechanical
probability law!

So, after all and independently of the general problem listed in
Point 1, I am skeptical of what {\Coleman} has to say.  Nevertheless,
this brings me back to {\Hartle}'s original work.

{\bf Point 3)}\ \  What can be salvaged of the Eigenstate Axiom
point of view?  There is something very interesting in {\Hartle}'s
paper if it can be made to work out in the proper way.  (This is an
idea inspired by a paper of {\Ballentine} [6], which itself is very
frequentist in its orientation.)

Normally the Law of Large Numbers can be made use of to sharpen up
imprecise statements about single events (i.e., that the event will
occur with such and such probability) into definite statements of
probability 1 (like the probability of such and such frequency is
one).  If we take the Eigenstate Axiom to be as you would like, i.e.,
\begin{quote}
{\bf Axiom}: If $|\psi\rangle=|s\rangle$, then a measurement of
$\hat S$ will reveal outcome $s$ with unit probability,
\end{quote}
then {\Hartle}'s demonstration does hint at something that would be
great.  Namely, that the structure of quantum mechanics is such that
it allows us to derive statements with less than unit probability
from something that itself is specified with probability one.  In
other words, the structure of quantum mechanics may be such as to
allow us to pursue the Law of Large Numbers in reverse order.  In
this way it would look like the MaxEnt principle in setting up priors
\ldots\ but from the Eigenstate Axiom alone and not the standard full
structure of the theory.  After all, it is intriguing that the
quantity $|\langle s|\psi\rangle|^2$ seems to pop out of his
mathematics from nowhere.

The problem, though, in carrying this out is in 1) doing it {\it
without\/} identifying probability and frequency, and 2) meeting my
old objection that $\hat S^\infty$ and $\hat F^\infty_s$ are not
related in a completely obvious way.

I'll leave you with that for now \ldots\ my wrist is really starting
to hurt.  I'm sure I'll have more to say after I get a chance to
read {\Hartle}'s paper (and maybe {\Coleman}'s).  Have any new
developments come about since your last note to me?  Keep me
informed.
\begin{enumerate}

\item
P.~{\Benioff}, ``Possible strengthening of the interpretative rules
of quantum mechanics,'' {\em Physical Review D}, vol.~7(12),
pp.~3603--3609, 1973.

\item
P.~{\Benioff}, ``Some consequences of the strengthened interpretative
rules of quantum mechanics,'' {\em Journal of Mathematical Physics},
vol.~15(5), pp.~552--559, 1974.

\item
P.~A. {\Benioff}, ``Finite and infinite measurement sequences in
quantum mechanics and randomness: {T}he {E}verett interpretation,''
{\em Journal of Mathematical Physics}, vol.~18(12), pp.~2289--2295,
1977.

\item
P.~{\Benioff}, ``A note on the {E}verett interpretation of quantum
mechanics,'' {\em Foundations of Physics}, vol.~8(9/10),
pp.~709--720, 1978.

\item
M.~van {\Lambalgen}, ``von {\Mises} definition of random sequences
reconsidered,'' {\em Journal of Symbolic Logic}, vol.~52(3),
pp.~725--755, 1987.

\item
L.~E. {\Ballentine}, ``Can the statistical postulate of quantum
theory be derived?---{A} critique of the many-universes
interpretation,'' {\em Foundations of Physics}, vol.~3(2),
pp.~229--240, 1973.
\end{enumerate}

\section{12 April 1996, \ ``Fopping Math''}

Last night, I dug into the black hole of quantum information that is
my file cabinet and found a couple more tidbits that might be of
interest to you.  The first is a paper by {\Farhi}, {\Goldstone} and
{\Gutmann} [1] that claims to do the argument of {\Hartle} ``right.''
The second is a paper by {\Mittelstaedt} [2] that I, unfortunately,
cannot trace the source of.  (Perhaps it was one of the conference
proceedings he edited and so it can be found in the library there.
Actually it's a pretty good article all around about the
``measurement problem.'')  I still haven't been able to get a copy
of {\Hartle}'s old article---the AJP's here don't go back that
far---I may have to go to McGill to get one.

To tell you the things I want to tell you, let me build a little
notation again.  First we start off with the state corresponding to
$n$ preparations
\be
|\psi^n\rangle=|\psi\rangle|\psi\rangle\cdots|\psi\rangle\;,
\ee
where
\be
|\psi\rangle=c_0|0\rangle+c_1|1\rangle
\ee
and $|0\rangle$,$|1\rangle$ are eigenvectors of the observable $\hat
S$ that we are talking about.  Also let
\be
|\psi^\infty\rangle=\lim_{n\rightarrow\infty}|\psi^n\rangle\;.
\ee
Define $\hat S^{(n)}$ and $\hat S^\infty$ similarly.  Recall the
definitions of the frequency operators $\hat F^{(n)}_0$, $\hat
F^{(n)}_1$, $\hat F^\infty_0$, and $\hat F^\infty_1$ from my last
note.

Now {\Farhi} and company point out, and I think rightly so (though
it's hard to tell without getting steeped in mathematics), that the
actual statement proven by {\Hartle}, i.e.,
\be
\lim_{n\rightarrow\infty}\left\|\hat F^{(n)}_s |\psi^n\rangle
-|\langle s|\psi\rangle|^2\,|\psi^n\rangle\right\|=0\;,
\ee
$s=0,1$, cannot be construed as an eigenvalue equation for the
operator $\hat F^\infty_s$.  They claim that this is really a
statement about finite $n$, saying:  for any $\epsilon>0$, there
exists a finite $n$ such that
\be
\left\|\hat F^{(n)}_s |\psi^n\rangle -|\langle
s|\psi\rangle|^2\,|\psi^n\rangle\right\|<\epsilon\;.
\ee
Therefore no conclusions based on the ``Eigenstate Axiom'' can be
drawn from it.

To ``remedy'' this situation, they then work to find frequency
operators ${\cal F}_s^\infty$ on the (nonseparable) Hilbert space
${\cal H}=H\otimes H\otimes H\otimes\cdots\;$ such that
\be
{\cal F}_s^\infty|\psi^\infty\rangle=|\langle s|\psi\rangle|^2\,
|\psi^\infty\rangle\;.
\ee
The idea being, of course, that the Eigenstate Axiom {\it really\/}
can be applied to this beast.  But, boy, you want to talk about
``mathematical foppery,'' you should see the lengths they go to here
to get at this.  (I have not attempted to disentagle it.)

In any case, that might have represented some clarification for me
if I had not seen quite a jumble of other inane statements throughout
the rest of the paper.  For instance, consider the following from
their ``Conclusions'' section.  (Emphases were added by me.)
\begin{quote}
To what extent is the system of an infinite number of copies
physically realizable and can the measurement of the frequency have
physical meaning?  This is like asking whether an infinite sequence
of coin flips can be physically realized and what does the Strong
Law of Large Numbers say in this case?  Clearly a coin cannot
actually be flipped an infinite number of times any more than
someone can count to infinity; yet we can make sensible statements
that refer to these operations.  The Strong Law of Large Numbers
says that if you did flip a coin an infinite number of times you
would, {\it with absolute certainty}, find the proportion of heads
to be one-half. Our quantum statement is that if you construct an
infinite number of copies of [the system] and measure the frequency
of the outcomes you would, {\it with absolute certainty}, find the
frequency to be [the quantum mechanical prediction].
\end{quote}

This, of course, is just the sort of problem with the interpretation
you've been pointing out all along.  However, I find this paragraph
even a little more telling: since these fellas don't have a proper
understanding of the implications of the classical SLLN, their
quantum conclusions seem doomed from the outset.

If they want to say that an outcome string with the ``correct''
frequencies will occur {\it with absolute certainty} just because it
is a member of a set of probability-1 \ldots\ and that an outcome
string with ``incorrect'' frequencies will NOT occur {\it with
absolute certainty} just because it is a member of a set of
probability-0 \dots, then they must give a criterion for why some
probability-0 sets are ``more probability-0'' than others.  (Recall
Mr.\ {\Orwell}.)  For, every individual outcome string in this event
space is, literally, of probability zero.  This is the sort of thing
that I called ``sloppy frequentism'' the other day.  Only some extra
criteria like von {\Mises} idea of the kollectiv can make this
silliness go.  Things can certainly be no better in the quantum
world.

How do you combat this sort of thing?  I haven't a clue: no more
than I would know how to turn a strict nonbeliever into a Bayesian.
And the more I think about it, the less I'm inclined to bother.

However, as I said last time, I am more than willing to re\"examine
what might be salvaged of {\Hartle}'s derivation.  In the words of
{\Mittelstaedt}, ``There is, however, no obvious interpretation of
this result.''  (Which, by the way, has been generalized by {\Ochs}
[3] to situations where the preparations are not absolutely
identical and may even be entangled.)

Among other things, though, I'm feeling pretty stupid about this
thing of the connection between $\hat F^{(n)}_s$ and $\hat S^{(n)}$.
What am I missing?  Let's lay this to rest.  If I think of a
measurement of $\hat S^{(n)}$ as {\it only\/} revealing it's
eigenvalues, then indeed there is a clear-cut connection between the
two operators: reading off the eigenvalues of $\hat S^{(n)}$ amounts
to reading off the eigenvalues of $\hat F^{(n)}_s$.  However, if I
think of the repeated measurement of $\hat S$ in more physical terms
of actually carrying out the measurements one by one \ldots\ then
there are no real degeneracies in its spectrum:  each outcome
corresponds to a {\it distinct\/} string of length $n$.  In this
sense, the projectors onto the spaces spanned by different
frequencies are a coarse-graining of the strings generated by a
measurement of $\hat S^{(n)}$.  To some extent, I see two different
beasts; how should I adjust my glasses?

I suppose even if you cannot satisfy me that the questions are the
same, it doesn't really matter.  I continue to be intrigued that the
standard probability expression $|\langle s|\psi\rangle|^2$ seems to
pop out of nowhere in {\Hartle}'s derivation.

Suppose we take the Eigenstate Axiom to be along the lines of:
\begin{quote}
{\bf Axiom}:\ \ For the measurement of a given Hermitian operator
$\hat S$, the closer (in the Hilbert space sense) the preparation
$|\psi\rangle$ is to being an eigenstate of $\hat S$, the more {\it
likely\/} the measurement will reveal the appropriate eigenvalue of
$\hat S$.
\end{quote}
Then, without anything else being said about the probability
interpretation, {\Hartle}'s result would reveal the quantity
$|\langle s|\psi\rangle|^2$ popping up in a very elementary way. I
sort of like that.  However, it still seems to me that you need
something more to get down to a probability assignment for the single
measurement case.  If one has a strong inclination or belief for one
set of frequencies over another in a large number of repetitions and
no other prior information, does that give him warrant to set a
prior that is independent, identically distributed, and matching the
expected frequency in the single measurement case?  It seems to me
that that is likely, and might actually be justified on Bayesian
grounds.  Or maybe even MaxEnt grounds---something like, independent
distributions for the individual measurements will lead to the
highest overall {\Shannon} info.

In this sort of program though, I'm also a little bothered by another
thing.  I am accustomed to thinking of the eigenvalues of measurement
operators like $\hat S$ as being more or less arbitrary labels for
the outcomes.  Why should anything be different in this case?  Could
there be some Cox-like freedom for rescaling probabilities (via
arbitrary functions of the ``plausibilities'') that we are missing
here?

Perhaps what is more important is that, for any basis
$\{|s\rangle\}$, the strings ``random'' with respect to the measure
$|\langle s|\psi\rangle|^2$ are close to $|\psi^\infty\rangle$,
while all other strings are not.  And this might be why a
{\Coleman}-type derivation might be better than one of a {\Hartle}
type. However, making the derivation rigorous is something else.
(Just look at {\Benioff}'s JMP paper for the difficulties in
expanding $|\psi^\infty\rangle$ in the right way so that the
nonrandom components can be excluded.)

But that leads me to ask to what extent {\Coleman} has really carried
this sort of thing out \ldots\ or, alternatively, to what extent
{\Benioff} {\it already\/} has carried it out (given this new
perspective).  It would really be great if you'd take a look at the
two old {\Benioff} papers on the {\Everett} interpretation.

Oh, finally as a side note, intrigued by these ideas, you can imagine
that I was elated when I ran across---completely by accident---Ref.\
[4].  The game that this guy tries to play is to drop the standard
axioms of probability and reconstruct them (much as Cox did in
another way).  In particular, he holds on to the Weak Law of Large
Numbers and a couple other simpler qualitative things and tries to
get all the rest (including Bayes' rule) out of it.  I haven't
studied the paper in detail yet, but its similarity (in the
classical domain) with {\Hartle}/{\Coleman} might be worth fleshing
out.

That's enough for now.

\begin{enumerate}

\item
E.~{\Farhi}, J.~{\Goldstone}, and S.~{\Gutmann}, ``How probability
arises in quantum mechanics,'' {\em Annals of Physics}, vol.~192,
pp.~368--382, 1989.

\item
P.~{\Mittelstaedt}, ``The objectification in the measuring process
and the many worlds interpretation.''

\item
W.~{\Ochs}, ``On the strong law of large numbers in quantum
probability theory,'' {\em Journal of Philosophical Logic}, vol.~6,
pp.~473--480, 1977.

\item
R.~E. {\Neapolitan}, ``A limiting frequency approach to probability
based on the weak law of large numbers,'' {\em Philosophy of
Science}, vol.~59, pp.389--407, 1992.

\end{enumerate}

\section{15 April 1996, \ ``No More Math''}

Ahh, like Baron Frankenstein, you create a monster and then want
nothing to do with it!  OK, I can see you're tiring of the
``{\Hartle}/{\Coleman}'' issue, and that's all right.  I just want to
put down some last thoughts; then I'll wait until you come back to
discuss it more.  Mostly, I'm afraid that you got the wrong
impression from my last note: I certainly do NOT want to see this
problem mathematized further.  I think that would be the wrong way to
tackle it.

The constructive idea that you should take from my last note (until
we meet again) is the following.  I think there may be something in
the mathematical piece of {\Hartle}'s original derivation that is
worth using for our own purposes.  Namely, that the quantitative
quantum mechanical probability expression may have its roots in a
much more qualitative statement \ldots\ just as Cox showed that the
standard probability axioms can be derived from more qualitative
statements. The qualitative statement in this case is not {\Hartle}'s
``Eigenstate Axiom'' but rather something much more Bayesian:
regardless of the size of the Hilbert space, the closer the system's
state vector is to an eigenstate of the Hermitian operator being
measured, the more likely that eigenstate will correspond to the
measurement outcome. Period.  Then {\Hartle}'s derivation
demonstrates that we can make ever more likely predictions about the
outcome of a measurement of the frequency operator.  Presumably, by
Bayesian methods, we can then use that knowledge as prior
information for setting up a probability assignment for the outcomes
in an individual trial.  That assignment will be the one the old
axioms of quantum theory told us it would be anyway.

That's it; that's the idea.  Have a good CEPI and I look forward to
hearing back from you when you have some time.

\section{21 December 1997, \ ``Falling Off the Earth''}

I'll send you the final version of the proposal I sent off to [a
gazillion colleges]. The last section has been significantly
restructured. You'll probably like it less, but---at least for the
present---I like it much more. I've presented it all much more
confidently, added some details about MaxEnt, and got even more
outlandish at the end.  (I guess now, not only have I come out of
the closet, but I've shut the door behind me.) Believe it or not,
this was done on {\Hideo}'s advice this time.  I've also gotten rid
of any hint of the long-term/short-term thing:  now the last section
looks to be purely motivational.

\subsection{{\Carl}'s Replies}

\bcc
\indent I am very impressed by your proposal.  My main warning is that
if I'm swept away, more sensible people will probably not be.
\ecc
\bcc
I just finished reading your revised proposal a bit more carefully.
It's just terrific, and as I said, deserves to be out there where
people can read it and appreciate it.
\ecc
\bcc
I understand fully your reasons for not wanting the proposal
displayed. If you change your mind at some point and display it on
your own home page, let me know so that we can link to it.  It's
such a terrific way of saying this stuff about using quantum
mechanics to do things instead of just putting up with its
limitations.  Of course, all of us theoretical physicists have long
been thankful that the world isn't classical, but that was just
because quantum mechanics gave us something to do.  Now there is a
better reason for preferring a quantum world.
\ecc

\noindent [NOTE:  With these flattering compliments, {\Carl}
suggested that I post this oddity of a research proposal to a
webpage---either to my own or to his Information Physics page at
UNM.  For various reasons, I declined at the time.  It won't be much
more out in the open in the middle of this book, but it seems
worthwhile to record it here.]

\subsection{The Structure of Quantum Information}

\noindent {\it Dedicated to a good teacher.}

\subsubsection{\protect\hspace*{.2in} I. Orientation}

The world we live in is well-described by quantum mechanics. What
should we make of that?  In a way, the answer to this question was
once less positive than it is today.  For although quantum theory is
a tool of unprecedented accuracy in predicting and controlling the
phenomena about us---and by way of that is the basis of our
technological society---the intellectual lesson we have come to
derive from it has been one largely of limitations.  The best place
to see this attitude at work is in a standard presentation of the
{\Heisenberg} uncertainty relations.  It is almost as if the world
were holding something back that we really had every right to
possess: the task of physics, or so it was believed, is simply to
sober up to this fact and make the best of it.

In contrast to this textbook lesson, the last five years have seen
the start of a significantly more positive, almost intoxicating,
attitude about the basic role of quantum mechanics.  This is
evidenced no more clearly than within the small, but growing
[\ref{Taubes96}], community of workers in {\it Quantum Information
Theory\/} [\ref{Bennett95}] and {\it Quantum Computing}
[\ref{Steane97}].\footnote{For quick reference, two recent articles
on the subject can be found in {\sl Physics Today}: Oct.\ 1997,
pp.~19--21, and Oct.\ 1995, pp.~24--30.  Some WWW links can be found
in John {\Preskill}'s ``Physics 229'' homepage at {\tt
http://www.theory.caltech.edu/people/preskill/ph229}. Also see
Oxford University's Centre for Quantum Computation homepage at {\tt
http://www.qubit.org/}.} The point of departure in both these
disciplines is not to ask what limits quantum mechanics places upon
us, but instead what novel, productive things we can do in the
quantum world that we could not have done otherwise.  In what ways
can we say that the quantum world is fantastically better than the
classical world?

The two most striking examples of this so far have been quantum
cryptographic key distribution [\ref{BBE92},\ref{Muller95}] and
{\Shor}'s quantum factoring algorithm
[\ref{DiVincenzo95},\ref{EJ96}]. In the case of the first example,
one sees that quantum mechanics allows two communicators to transmit
to each other a random cryptographic key in such a way that
eavesdropping on the transmission can be excluded out of hand.  This
is impossible in the classical world because there is no direct
connection between the information that can be gathered about a
physical system's state and the disturbances induced upon that state
in the process [\ref{Fuchs96a},\ref{Fuchs97a}].  Without the
indeterminism of quantum mechanics, two-party data transmissions
would remain forever vulnerable to clever or powerful eavesdropping
techniques.

In the case of the second example, one sees that algorithms designed
for computers built of unabashedly quantum components---that is,
components that can remain coherent with each other throughout the
computation---can factor large integers exponentially faster than
anything written for standard classical computers.  To give a quick
example of what this means in real terms, consider a 600-digit
number that is known to be the product of two (secret) primes.  The
number of computational steps required of a classical computer to
crack it into its two components is of the order of $10^{34}$. In
contrast, the corresponding number of steps for a quantum computer
is only $10^{11}$.  Quantum computing can give 23 orders of
magnitude greater efficiency in this problem!

These two examples are the most outstanding of the class, and there
is well-founded hope that they are the tip of a technological
iceberg.  However, I believe there is a similarly founded hope that
they are also the small tip of a {\em physical\/} iceberg. Looking
at quantum mechanics through the eyes of these two fields cannot
help but lead to greater and deeper---and perhaps the
deepest---insights into its structure and ultimate use.  These are
the insights that could poise physics for the great breakthroughs
that will surely come about next century, even in disciplines as
far-flung as quantum gravity and quantum cosmology.

But this is my grand vision.  Little will come of it if it is not
preceded by years of more realistic, more concrete exploration of the
{\it structure of quantum information}. This, as part of the
accumulating results of the communities just described, is the
subject of this research proposal.

To give some indication of the wider context that flows into the
specialization of Quantum Information and Quantum Computing, one
need only note that, by its very makeup, it must call upon the
expertise of standard communication theory, cryptography,
computation theory, number theory, signal processing, and various
branches of statistical mechanics.  Dreams of possible experimental
implementations have called upon the quantum optics community
[\ref{Turchette95},\ref{Furusawa98}], the ion trap community
[\ref{Monroe95},\ref{Hughes97}], the NMR spectroscopy community
[\ref{Cory96},\ref{Gershenfeld97}], and to a smaller extent
solid-state physics [\ref{Loss97},\ref{Kane98}].

The particular aspect of Quantum Information Theory that has been my
focus the last four years is closely allied to the well-established
tradition in mathematical physics pioneered by the likes of {\Holevo}
[\ref{Holevo73}], {\Lieb} [\ref{Lieb73}], {\Lindblad}
[\ref{Lindblad76}], and {\Uhlmann} [\ref{Uhlmann76}].  It is my
intent to strengthen and build upon the connections between that
tradition and this upstart field, which already in many ways is its
continuation. I hope this becomes apparent in the details that
follow.

\subsubsection{\protect\hspace*{.2in} II. Research Proposal}

My research interests might be described as {\em tria juncta in
uno}.\footnote{Not to worry, I won't pretend that I knew this phrase
before looking in my thesaurus! ... But when you learn something
like this, you've just got to use it!  Apparently this phrase is the
motto of ``The Most Honourable Order of the Bath,'' a particular
British order of chivalry.}  The conjunction of these three topics,
for the most part, exhausts what is presently meant by ``Quantum
Information Theory.''

$\bullet$ {\bf Sending Classical Information on Quantum Me\-chanical
Channels.} People encode ``classical'' infor\-mation---like the
stories in today's newspaper---into the states of quantum systems
for a simple reason:  to get it from one place to another.  Since
the world is quantum mechanical, this, in the last analysis, is
exactly what one always does in transmitting information.  Strangely
enough however, literally almost all of modern information theory
(as exhibited in the 44 existent volumes of {\sl IEEE Transactions on
Information Theory}) has ignored this fact in any but the most
trivial ways.

Once one takes it seriously that physical information carriers are
quantum mechanical, one can ask a whole host of questions that could
not have been asked before.  For instance, can it help the receiver
to collect many separate transmissions before performing the quantum
measurement required to decode them
[\ref{Holevo96},\ref{Schumacher97a}]? That is to say, can collective
quantum measurements on separate signals be more powerful than
individual measurements [\ref{Holevo79},\ref{Peres91}]?  Can it ever
help to entangle separate transmissions---as with
{\Einstein}-{\Podolsky}-{\Rosen} pairs---before sending them through
the channel [\ref{BFS97}]? Can one help evade a channel's noise by
encoding the signals in nonorthogonal quantum states
[\ref{Fuchs97b},\ref{Fuchs97c}], in spite of the fact that the
classical analog of this corresponds to sending noisy signals?

Of course, the answer to all these questions is ``yes.''  And this
is enough to demonstrate that these lines of thought are not trivial.
However, the work remaining before a theory as coherent as classical
information theory can emerge is legion.  In particular, the
counterpart of the most basic question of all classical information
theory---What is the capacity of a discrete memoryless
channel?---has yet to be solved.

The most exciting prospect of this set of questions for physics is
the potential it holds for giving new and unique and very strongly
motivated measures of ``correlation'' between two subsystems of a
larger whole.  {\Shannon}'s solution of the channel capacity question
brought with it a measure of correlation (the ``mutual
information'') of a generality greatly exceeding the scope of its
motivation [\ref{Cover91}].  Its physical applications have ranged
from information theoretic versions of the {\Heisenberg} relations
[\ref{Hall95}] to a final solution of the old Maxwellian demon
problem [\ref{Bennett82},\ref{Zurek89}].  One can expect no less for
a quantum measure of classical correlation.  In particular, the
importance of uniquely quantum measures of correlation for quantum
statistical mechanics has been emphasized recently by {\Lindblad}
[\ref{Lindblad91}] and {\Schack} and {\Caves} [\ref{Schack96}].

$\bullet$ {\bf Information Gain vs.\ Quantum State Disturbance in
Quantum Theory.}] The engine that powers quantum cryptography is the
principle that it is impossible to gather information about a
quantum system's unknown state without disturbing that system in the
process. (This is so even when the state is assumed to be one of only
two nonorthogonal possibilities.)  This situation is often mistakenly
described as a consequence of the ``{\Heisenberg} uncertainty
principle'' but, in fact, is something quite distinct
[\ref{Fuchs96a},\ref{Fuchs97a}] and only now starting to be studied
in the physical literature.  A more accurate account of the
principle is that it is a feature of quantum mechanics that rests
ultimately on the unitarity of the theory, and may be seen as a
quantitative extension of the so-called ``no-cloning theorem''
[\ref{Wootters82},\ref{BBM},\ref{Fuchs96b}].  In contradistinction,
the {\Heisenberg} principle concerns our inability to ``get hold'' of
two classical observables---such as a position and
momentum---simultaneously.  It thus concerns our inability to
ascribe {\it classical\/} states of motion to {\it quantum\/}
systems---that has very little to do with the issue of encoding
information in and retrieving information from the quantum states
themselves.

Because this way of looking at ``information gain'' and
``disturbance'' for quantum systems is itself purely quantum
mechanical and does not rely on antiquated classical notions, it
holds the possibility of giving the best understanding yet of those
things the founding fathers (like {\Heisenberg}, {\Pauli}, and
{\Bohr}) labored so hard to formulate.  But what is the unifying
theme?  What are the directions to take?  One can elaborate upon the
direction defined by practical quantum cryptography
[\ref{FGGNP},\ref{Slutsky97},\ref{Cirac97}] or one can take a more
direct route inspired by the original no-cloning theorem
[\ref{Buzek96},\ref{Gisin97},\ref{Bruss97}].  Each method is begging
for a more systematic account than has been afforded by these simple
preparatory explorations.

A novel approach, and one which I have turned my attention to
recently, is to seek out the connection between quantum entanglement
measures and the information--disturbance principle
[\ref{Bruss97},\ref{privamp},\ref{Schumacher97b}].  The main point
about this line of thought is that in the scenario of quantum
cryptography, any would-be eavesdroppers must become entangled with
the information carriers traveling between the legitimate users. Can
one read the tradeoff between information and disturbance directly
from something to do with entanglement itself?  Perhaps by a sort of
``entanglement conservation'' principle?  These are the sorts of
questions that first require progress in the next research topic.

$\bullet$ {\bf Quantifying Quantum Entanglement: Separating It from
Classical Correlation.} The preoccupation of classical information
theory is to make the correlation between sender and receiver as
high as possible.  This is what communication is about.  But it is
only part of the story in Quantum Information Theory. The quantum
world brings with it a new resource that senders and receivers can
share: quantum entanglement, the stuff
{\Einstein}-{\Podolsky}-{\Rosen} pairs and {\Bell}-inequality
violations are made of.  This new resource, of all the things
mentioned so far, is the most truly ``quantum'' of quantum
information.  It has no classical analog, nor might it have been
imagined in a classical world.

What is quantum entanglement?  It is {\it not\/} probabilistic
correlation between two parts of a whole.  Rather it is the {\it
potential\/} for such a correlation. \vspace{5pt} In a quick
portrayal:\\
{\it classical correlation}---
\bq
\noindent Alice and Bob entered a lottery for which they were the only
players. They have not opened their ``winnings'' envelopes yet, but
the messages in them say that one is the winner and one is the loser.
Which is which, they do not know---they only know the
correlation---but the answer is there, objectively existent, without
their looking.
\eq
{\it quantum entanglement}---
\bq
\noindent Alice and Bob will eventually perform measurements on the EPR
pair their envelopes contain and the outcomes {\em will\/} be
correlated.  However, before the measurements are performed, there
are no objectively existent variables already there. Different
measurements can and will lead to different correlations.
\eq

In a certain sense, entanglement is a kind of {\it all-purpose
correlation\/} just waiting to be baked into something real---a
quantum ``Martha White's Flour'' [\ref{Flatt}].  The uses for this
all-purpose correlation are manifold within Quantum Information
Theory. Beside the applications above [\ref{BBE92},\ref{BFS97}],
there is also quantum-state teleportation [\ref{BBCJPW}], quantum
superdense coding [\ref{BW},\ref{Mattle96}], error-correction for
quantum computers [\ref{Gottesman97}], entanglement-assisted
multi-party communication games [\ref{Cleve97}], better control of
atomic frequency standards [\ref{Bollinger96},\ref{Huelga97}], and
the list goes on.

The deepest set of questions here, and the largest focus of my
present research [\ref{Fuchs98a},\ref{Fuchs98b}], concern
quantifying this newly recognized physical essence in an
application-independent way
[\ref{dist1},\ref{dist2},\ref{Wootters97}]. As an example, take an
EPR pair, half of which has been transmitted through a noisy
(decohering) quantum channel.  Because of the noise, the final state
of this bipartite system is no longer a pure state: it is described
by a mixed state density operator.  Some of the correlation there is
still potential or all-purpose, but some---because the decoherence
has helped promote it to a more tangible status---is simply
classical correlation.  How do we quantify the amount of each? What,
if anything, is the exchange rate between the two?  With some time,
creativity, and hard work, we will one day have these issues under
control.

$\bullet\,\bullet$ {\bf Synthesis.} In some ways the project of
Quantum Information Theory can be likened to the beginning of
thermodynamics.  It is not our place to develop the question ``What
is heat, work, energy?'' but instead ``What is correlation,
indeterminism, entanglement?''  No informed judgment of the historic
question could have been made before the development of a
quantitative theory of thermodynamics, and it will be likewise with
our field.  Whereas the fruits of the old question were the
mechanical theory of heat and its corollary of atomism, we do not
yet have a firm grasp of where our field is leading.  It is clear,
however, that it is going somewhere and somewhere fast; at the very
least, its applied, technological innovations can neither be denied
nor safely ignored.

What is correlation, indeterminism, entanglement?  This is what the
three research areas above are trying to make quantitative.  Each
contains within itself a little seed of the others; each sheds light
on the structure of quantum information.

\subsubsection{\protect\hspace*{.2in} III. Wider Seas}

Eight years after the inception of classical information theory,
Claude {\Shannon}, its founder, warned [\ref{Shannon56}],
\bq
\noindent
Although this wave of popularity is \ldots\ pleasant and exciting for
those of us working in the field, it carries at the same time an
element of danger.  While we feel that information theory is a
valuable tool in providing fundamental insights into the nature of
communication ...\ it is certainly no panacea \ldots. Seldom do more
than a few of nature's secrets give way at one time.
\eq
History has borne {\Shannon} out:  information theory is not a
panacea.  But, cure-all or not, the field has had a great impact on
applications that can hardly be said to resemble the original one,
that of describing communication over noisy channels
[\ref{Shannon48}]. One need only look at information theory's
influence on fields as far ranging as biology, economics, and
psychology [\ref{Campbell82}], to see this point.

What is it that we can expect of {\it Quantum\/} Information Theory
once it is complete and coherent?  What more might it say about a
{\it few\/} of natures secrets?  With the reader's indulgence, I
will attempt to express some of my present views on the question.
These have to do with the ``grand vision'' and ``physical iceberg''
of the Introduction---the real sources of my day-to-day motivation.

The year 1957 is significant in physical thought because it marks the
penetration of information theory into physics in a systematic
way---into statistical mechanics in particular [\ref{Jaynes57}]. This
refers to the {\it Maximum Entropy\/} or ``MaxEnt'' program for
statistical mechanics set into motion by E.~T. {\Jaynes}
[\ref{maxent}]. With the tools of information theory, one was able
for the first time to make a clean separation between the purely
{\it statistical\/} and the purely {\it physical\/} aspects of the
subject matter.

Perhaps it would be good to present a mild example of this.  Because
of MaxEnt, a standard statistical mechanical ensemble---like the
canonical en\-sem\-ble---can finally be seen for what it really is:
an expression of the physicist's {\it state of knowledge\/}
(specified, of course, by the experimental parameters under his
control).  Though this reveals a subjective element in statistical
mechanics, the ensemble is not arbitrary.  Two physicists working on
a single experiment and possessing identical data---if true to their
states of knowledge---will derive the same distributions for the
system's variables. The {\it structure\/} of the canonical
distribution, with its exponential form, is due purely to the kind
of knowledge the experimenter possesses---in this particular case,
the expectation value of some observable and nothing else.  That is
to say, the canonical distribution's form is a theorem of the laws of
inference, {\it not physics}.  The physics of the system rests solely
in its Hamiltonian and boundary conditions.  This conceptual
separation between the physical and the statistical can be fruitful.
With it, one can, for instance, derive the second law of
thermodynamics in an almost trivial way [\ref{Jaynes65}].

In contrast to this, quantum theory is at its heart a statistical
theory of {\it irreducibly\/} statistical phenomena---this is the
great lesson of the {\Kochen}-{\Specker} theorems and the {\Bell}
inequalities [\ref{Mermin93}].  What can this possibly mean for the
issue just explored?  With due attention to the success of the
MaxEnt program in the last 40 years [\ref{Jaynes79}], one can hardly
feel it unreasonable to ask:  What part of the formal structure of
quantum mechanics is forced upon us by physics alone---i.e., that
the theory be about irreducibly statistical phenomena---and what
part is forced upon us as a consequence of the form {\it any\/}
theory must take in light of that subject
matter[\ref{Caves96},\ref{Fuchs97d}]?  George {\Boole} called
probability theory a ``law of thought'' because it specifies the
rules with which we should think when we come upon situations where
our information is incomplete [\ref{Boole},\ref{Jaynes97}].  What
part of quantum mechanics is simply ``law of thought,'' and what
part is {\it irreducibly\/} physics?

A mature Quantum Information Theory is likely to be uniquely
stationed to contribute to this question, or at least to test
whether anything might come of it.  The quantitative statements it
{\it will\/} possess for the in\-for\-ma\-tion--disturbance tradeoff
and the correlation--en\-tan\-glement dichotomy should be of just the
right flavor for such a thing.  Both threads explore the difference
between probabilities that can be improved upon because they
correspond to lack of knowledge and probabilities that are more the
nature of ``potentialities'' for which no improvement can be had.

Once this issue is settled, one may finally hope for a simple, crisp
statement of what our quantum theory is all about [\ref{Rovelli96}].
And once that is in hand, who knows what the limits might be?  To
place the issue within an historical context, one can speculate how
long it would have taken to stumble across general relativity if it
had not been for the compelling vision that {\Einstein} found lying
behind the Lorentz transformations.  The equations were there with
Lorentz, but the essence of it all---and the simple picture with
which progress could be made---came with {\Einstein}'s special
relativity.

Comparable to this opportunity for fundamental phys\-ics, one might
imagine a similar blossoming of opportunity for other endeavors. A
classification of quantum theory's content in the way suggested above
could distill mathematical structures that other,
extra-quantum-mechanical, efforts
[\ref{Pitowski89},\ref{Pitowski94},\ref{Segal98}] might use to their
advantage. After all, this is exactly the sort of thing that
happened with the MaxEnt program: its applications range from
observational astronomy to pharmaceutical studies to artificial
intelligence\footnote{A particularly cogent example of the use of
these methods in artificial intelligence can be found at Microsoft
Research Division's {\sl Decision Theory \& Adaptive Systems Group\/}
home\-page: \\
{\tt http://www.research.microsoft.com/dtas/}.} [\ref{maxent}]. Are
there fields beside quantum physics that encounter situations where
the maximal knowledge of something can never be made
deterministically complete
[\ref{Watanabe61},\ref{Peres82},\ref{Heilbron86}]?  If so, then they
will plausibly find novel use for the {\it
mathematics}\footnote{Please note that I {\it did\/} say ``the
mathematics of'' here.} of quantum physics and Quantum Information
Theory.

The main point that I would like to impress with this final
speculation---even allowing that the details above be taken with a
grain of salt---is that the structure of quantum mechanics is an
amazingly beautiful intellectual edifice.  It would be a shame if its
only application were for quantum mechanics itself.

\subsubsection{\protect\hspace*{.2in} IV. Summary}

There is a grand adventure in front of the physics community called
Quantum Information and Quantum Computing.  Its hallmark is to view
quantum theory in a way little before explored, in a way that
accentuates the positive.  Delimiting the structure of quantum
information may well be a key to great progress in fundamental
physics---but that may be some time in the coming.  In the mean time
there is much solid work to be done exploring entanglement,
information vs.\ disturbance, and the information-carrying
capacities of quantum mechanical systems.  This is my preoccupation;
this is the field of research I call my home.  I thank you for your
consideration.

\subsubsection{\protect\hspace*{.2in} V. References}

\begin{enumerate}
\footnotesize

\item
{\em Note 1}: This bibliography makes no attempt to be complete or
to consistently cite original references.  Its main purpose is to
give the reader a more detailed introduction to the issues discussed
in this text.

\item
{\em Note 2}: All ``LANL e-print'' listings refer to the Los Alamos
National Laboratory E-print Archive at \\ {\tt http://xxx.lanl.gov}.

\item\label{Taubes96}
G.~Taubes, ``All Together for Quantum Computing,'' Science {\bf
273}, 1164 (1996).

\item\label{Bennett95}
C.~H. {\Bennett}, ``Quantum Information and Quantum Computation,''
Phys.\ Tod.\ {\bf 48}, No.~10, 24 (1995).

\item\label{Steane97}
A.~M. {\Steane}, ``Quantum Computing,'' Rep.\ Prog.\ Phys.\ {\bf 61},
117 (1998).

\item\label{BBE92}
C.~H. {\Bennett}, G.~{\Brassard}, and A.~K. {\Ekert}, ``Quantum
Cryptography,'' Sci.\ Am.\ {\bf 267}, No.~10, 50 (1992).

\item\label{Muller95}
A.~Muller, H.~{\Zbinden}, and N.~{\Gisin}, ``Underwater Quantum
Coding,'' Nature {\bf 378}, 449 (1995).

\item\label{DiVincenzo95}
D.~P. {\DiVincenzo}, ``Quantum Computation,'' Science {\bf 270}, 255
(1995).

\item\label{EJ96}
A.~{\Ekert} and R.~{\Jozsa}, ``Quantum Computation and {\Shor}'s
Factoring Algorithm,'' Rev.\ Mod.\ Phys.\ {\bf 68}, 733 (1996).

\item\label{Fuchs96a}
C.~A. Fuchs and A.~{\Peres}, ``Quantum State Disturbance vs.\
Information Gain: Uncertainty Relations for Quantum Information,''
Phys.\ Rev.\ A {\bf 53}, 2038 (1996).

\item\label{Fuchs97a}
C.~A. Fuchs, ``Information Gain vs.\ State Disturbance in Quantum
Theory,'' Fortschr.\ Phys.\ {\bf 46}, 535 (1998).

\item\label{Turchette95}
Q.~A. {\Turchette}, C.~J. {\Hood}, W.~{\Lange}, H.~{\Mabuchi}, and
H.~J. {\Kimble}, ``Measurement of Conditional Phase Shifts for
Quantum Logic,'' Phys.\ Rev.\ Lett.\ {\bf 75}, 4710 (1995).

\item\label{Furusawa98}
A.~{\Furusawa}, J.~L. {\Sorensen}, S.~L. {\Braunstein}, C.~A. Fuchs,
H.~J. {\Kimble}, and E.~S. {\Polzik}, ``Unconditional Quantum
Teleportation,'' Science {\bf 282}, 706 (1998).

\item\label{Monroe95}
C.~Monroe, D.~M. Meekhof, B.~E. King, W.~M. Itano, and D.~J.
Wineland, ``Demonstration of a Universal Quantum Logic Gate,'' Phys.\
Rev.\ Lett.\ {\bf 75}, 4714 (1995).

\item\label{Hughes97}
R.~J. {\Hughes}, et al., ``The Los Alamos Trapped Ion Quantum
Computer Experiment,'' Fortschr.\ Phys.\ {\bf 46}, 329 (1998).

\item\label{Cory96}
D.~G. Cory, A.~F. Fahmy, and T.~F. Havel, ``Ensemble Quantum
Computing by NMR Spectroscopy,'' Proc.\ Natl.\ Acad.\ Sci.\ USA {\bf
94}, 1634 (1997).

\item\label{Gershenfeld97}
N.~A. {\Gershenfeld} and I.~L. {\Chuang}, ``Bulk Spin-Reso\-nance
Quantum Computation,'' Science {\bf 275}, 350 (1997).

\item\label{Loss97}
D.~Loss and D.~P. {\DiVincenzo}, ``Quantum Computation with Quantum
Dots,'' Phys.\ Rev.\ A {\bf 57}, 120, (1998).

\item\label{Kane98}
B.~E. Kane, ``A Silicon-Based Nuclear Spin Quantum Computer,''
Nature {\bf 393}, 133 (1998).

\item\label{Holevo73}
A.~S. {\Holevo}, ``Statistical Decision Theory for Quantum Systems,''
J.\ Multivar.\ Anal.\ {\bf 3}, 337 (1973).

\item\label{Lieb73}
E.~H. {\Lieb}, ``Convex Trace Functions and the
{\Wigner}-Yanase-{\Dyson} Conjecture,'' Adv.\ Math.\ {\bf 11}, 267
(1973).

\item\label{Lindblad76}
G.~{\Lindblad}, ``On the Generators of Quantum Dynamical
Semigroups,'' Comm.\ Math.\ Phys.\ {\bf 48}, 119 (1976).

\item\label{Uhlmann76}
A.~{\Uhlmann}, ``The `Transition Probability' in the State Space of a
$\,^*$-algebra,'' Rep.\ Math.\ Phys.\ {\bf 9}, 273 (1976).

\item\label{Holevo96}
A.~S. {\Holevo}, ``The Capacity of Quantum Communication Channel with
General Signal States,'' IEEE Trans.\ Inf.\ Theor.\ {\bf 44}, 269
(1998).

\item\label{Schumacher97a}
B.~{\Schumacher} and M.~D. {\Westmoreland},  ``Sending Classical
Information via Noisy Quantum Channels,'' Phys.\ Rev.\ A {\bf 56},
131 (1997).

\item\label{Holevo79}
A.~S. {\Holevo}, ``On the Capacity of Quantum Communication
Channel,'' Prob.\ Inf.\ Transm.\ {\bf 15}, 247 (1979).

\item\label{Peres91}
A.~{\Peres} and W.~K. {\Wootters}, ``Optimal Detection of Quantum
Information,'' Phys.\ Rev.\ Lett.\ {\bf 66}, 1119 (1991).

\item\label{BFS97}
C.~H. {\Bennett}, C.~A. Fuchs, and J.~A. {\Smolin},
``En\-tangle\-ment-Enhanced Classical Communication on a Noisy
Quantum Channel,'' in {\sl Quantum Communication, Computing and
Measurement}, edited by O.~{\Hirota}, A.~S. {\Holevo}, and C.~M.
{\Caves} (Plenum, New York, 1997).

\item\label{Fuchs97b}
C.~A. Fuchs, ``Nonorthogonal Quantum States Maximize Classical
Information Capacity,'' Phys.\ Rev.\ Lett.\ {\bf 79}, 1163 (1997).

\item\label{Fuchs97c}
C.~A. Fuchs, P.~W. {\Shor}, J.~A. {\Smolin}, and B.~M. {\Terhal},
``Quantum-Enhanced Classical Communication,'' to be submitted to
Phys.\ Rev.\ A; preliminary draft available upon request.

\item\label{Cover91}
T.~M. Cover and J.~A. Thomas, {\sl Elements of Information Theory},
(John Wiley \& Sons, New York, 1991).

\item\label{Hall95}
M.~J.~W. {\Hall}, ``Information Exclusion Principle for Complementary
Observables,'' Phys.\ Rev.\ Lett.\ {\bf 74}, 3307 (1995).

\item\label{Bennett82}
C.~H. {\Bennett}, ``The Thermodynamics of Computation---a Review,''
Int.\ J.\ Theo.\ Phys.\ {\bf 21}, 905 (1982).

\item\label{Zurek89}
W.~H. {\Zurek}, ``Algorithmic Randomness and Physical Entropy,''
Phys.\ Rev.\ A {\bf 40}, 4731 (1989).

\item\label{Lindblad91}
G.~{\Lindblad}, ``Quantum Entropy and Quantum Measurements,'' in {\sl
Quantum Aspects of Optical Communications}, edited by
C.~Bendjaballah, O.~{\Hirota}, and S.~Reynaud (Springer-Verlag,
Berlin, 1991).

\item\label{Schack96}
R.~{\Schack} and C.~M. {\Caves}, ``Information-Theoretic
Characterization of Quantum Chaos,'' Phys.\ Rev. E {\bf 53}, 3257
(1996).

\item\label{Wootters82}
W.~K. {\Wootters} and W.~H. {\Zurek}, ``A Single Quantum Cannot Be
Cloned,'' Nature {\bf 299}, 802 (1982).

\item\label{BBM}
C.~H. {\Bennett}, G.~{\Brassard}, and N.~D. {\Mermin}, ``Quantum
Cryptography Without {\Bell}'s Theorem,'' Phys.\ Rev.\ Lett.\ {\bf
68}, 557 (1992).

\item\label{Fuchs96b}
H. {\Barnum}, C. M. {\Caves}, C. A. Fuchs, R. {\Jozsa}, and B.
{\Schumacher}, ``Noncommuting Mixed States Cannot Be Broadcast,''
Phys.\ Rev.\ Lett.\ {\bf 76}, 2818 (1996).

\item\label{FGGNP}
C.~A. Fuchs, N.~{\Gisin}, R.~B. {\Griffiths}, C.~S. {\Niu}, and
A.~{\Peres}, ``Optimal Eavesdropping in Quantum Cryptography. I.
Information Bound and Optimal Strategy,'' Phys.\ Rev.\ A {\bf 56},
1163 (1997).

\item\label{Slutsky97}
B.~A. Slutsky, R.~Rao, P.-C.~Sun, and Y.~Fainman, ``Security of
Quantum Cryptography Against Individual Attacks,'' Phys.\ Rev.\ A
{\bf 57}, 2383 (1998).

\item\label{Cirac97}
J.~I. {\Cirac} and N.~{\Gisin}, ``Coherent Eavesdropping Strategies
for the 4-State Quantum Cryptography Protocol,'' Phys.\ Lett.\ A {\bf
229}, 1 (1997).

\item\label{Buzek96}
V.~{\Buzek} and M.~{\Hillery}, ``Quantum Copying: Beyond the
No-Cloning Theorem,'' Phys.\ Rev.\ A {\bf 54}, 1844 (1996).

\item\label{Gisin97}
N.~{\Gisin} and S.~Massar, ``Optimal Quantum Cloning Machines,''
Phys.\ Rev.\ Lett.\ {\bf 79}, 2153 (1997).

\item\label{Bruss97}
D.~Bru\ss, D.~P. {\DiVincenzo}, A.~{\Ekert}, C.~A. Fuchs,
C.~{\Macchiavello}, and J.~A. {\Smolin}, ``Optimal Universal and
State-dependent Quantum Cloning,'' Phys.\ Rev.\ A {\bf 57}, 2368
(1998).

\item\label{privamp}
D.~{\Deutsch}, A.~{\Ekert}, R.~{\Jozsa}, C.~{\Macchiavello}, S.
{\Popescu}, and A.~Sanpera, ``Quantum Privacy Amplification and the
Security of Quantum Cryptography Over Noisy Channels,'' Phys.\ Rev.\
Lett.\ {\bf 77}, 2818 (1996).

\item\label{Schumacher97b}
B.~{\Schumacher} and M.~D. {\Westmoreland}, ``Quantum Privacy and
Quantum Coherence,'' Phys.\ Rev.\ Lett.\ {\bf 80}, 5695 (1998).

\item\label{Flatt}
L.~{\Flatt} and E.~{\Scruggs}, ``The Martha White Theme,'' on {\sl
{\Flatt} \& {\Scruggs}: 1948--1959\/} (Bear Family Records, 1994).

\item\label{BBCJPW}
C.~H. {\Bennett}, G.~{\Brassard}, C.~{\Crepeau}, R.~{\Jozsa},
A.~{\Peres}, and W.~K. {\Wootters}, ``Teleporting an Unknown Quantum
State via Dual Classical and {E}instein-{P}odolsky-{R}osen
Channels,'' Phys.\ Rev.\ Lett.\ {\bf 70}, 1895 (1993).

\item\label{BW}
C.~H. {\Bennett} and S.~J. {\Wiesner}, ``Communication via One- and
Two-Particle Operators on {\Einstein}-{\Podolsky}-{\Rosen} States,''
Phys.\ Rev.\ Lett.\ {\bf 69}, 2881 (1992).

\item\label{Mattle96}
K.~Mattle, H.~{\Weinfurter}, P.~G. Kwiat, and A.~{\Zeilinger},
``Dense Coding in Experimental Quantum Communication,'' Phys.\ Rev.\
Lett.\ {\bf 76}, 4656 (1996).

\item\label{Gottesman97}
D.~{\Gottesman}, {\sl Stabilizer Codes and Quantum Error Correction},
Ph.~D. Thesis, California Institute of Technology, 1997; LANL e-print
{\tt quant-ph/9705052}.

\item\label{Cleve97}
R.~{\Cleve} and H.~{\Buhrman}, ``Substituting Quantum Entanglement
for Communication,'' Phys.\ Rev.\ A {\bf 56}, 1201 (1997).

\item\label{Bollinger96}
J.~J. Bollinger, W.~M. Itano, D.~J. Wineland, and D.~J. Heinzen,
``Optimal Frequency Measurements with Maximally Correlated States,''
Phys.\ Rev.\ A {\bf 54}, R4649 (1996).

\item\label{Huelga97}
S.~F. Huelga, C.~{\Macchiavello}, T.~Pellizzari, A.~K. {\Ekert},
M.~B. {\Plenio}, and J.~I. {\Cirac}, ``On the Improvement of
Frequency Stardards with Quantum Entanglement,'' Phys.\ Rev.\ Lett.\
{\bf 79}, 3865 (1997).

\item\label{Fuchs98a}
C.~H. {\Bennett}, D.~P. {\DiVincenzo}, C.~A. Fuchs, T.~{\Mor},
E.~Rains, P.~W. {\Shor}, J.~A. {\Smolin}, and W.~K. {\Wootters},
``Quantum Nonlocality without Entanglement'' Phys.\ Rev.\ A {\bf
59}, ?? (1999); LANL e-print {\tt quant-ph/9804053}.

\item\label{Fuchs98b}
D.~P. {\DiVincenzo}, C.~A. Fuchs, H.~{\Mabuchi}, J.~A. {\Smolin},
A.~{\Thapliyal}, and A.~{\Uhlmann}, ``Entanglement of Assistance,''
to appear in {\sl The 1st NASA International Conference on Quantum
Computing \& Quantum Communications (NASA QCQC'98)}, edited by C.\
Williams (Springer-Verlag, Berlin, 1998); LANL {\tt
quant-ph/9803033}.

\item\label{dist1}
C.~H. {\Bennett}, G.~{\Brassard}, S.~{\Popescu}, B.~{\Schumacher},
J.~A. {\Smolin}, and W.~K. {\Wootters}, ``Purification of Noisy
Entanglement and Faithful Teleportation via Noisy Channels,'' Phys.\
Rev.\ Lett.\ {\bf 76}, 722 (1996).

\item\label{dist2}
C.~H. {\Bennett}, D.~P. {\DiVincenzo}, J.~A. {\Smolin}, and W.~K.
{\Wootters}, ``Mixed State Entanglement and Quantum Error
Correction,'' Phys.\ Rev.\ A {\bf 54}, 3825 (1996).

\item\label{Wootters97}
W.~K. {\Wootters}, ``Entanglement of Formation of an Arbitrary State
of Two Qubits,'' Phys.\ Rev.\ Lett.\ {\bf 80}, 2245 (1998).

\item\label{Shannon56}
C.~E. {\Shannon}, ``The Bandwagon,'' IEEE Trans.\ Inf.\ Theor.\ {\bf
IT-2}, No.~3, 3 (1956).

\item\label{Shannon48}
C.~E. {\Shannon}, ``A Mathematical Theory of Communication,'' Bell
Sys.\ Tech.\ J.\ {\bf 27}, 379, 623 (1948).

\item\label{Campbell82}
J.~Campbell, {\sl Grammatical Man:~Information, Entropy, Language,
and Life}, (Simon \& Schuster, New York, 1982).

\item\label{Jaynes57}
E.~T. {\Jaynes}, ``Information Theory and Statistical Mechanics,''
Phys.\ Rev.\ {\bf 106}, 620 (1957).

\item\label{maxent}
A convenient source of information about this program can be found
in the dozen or so conference proceedings all published under the
same title, {\it Maximum Entropy and Bayesian Methods\/} (Kluwer,
Dordrecht).

\item\label{Jaynes65}
E.~T. {\Jaynes}, ``Gibbs vs.\ Boltzmann Entropies,'' Am.\ J.\ Phys.\
{\bf 33}, 391 (1965).

\item\label{Jaynes79}
E.~T. {\Jaynes}, ``Where Do We Stand on Maximum Entropy?,'' in {\sl
The Maximum Entropy Formalism}, edited by R.~D. Levine and M.~Tribus
(MIT Press, Cambridge, MA, 1979).

\item\label{Mermin93}
N.~D. {\Mermin}, ``Hidden Variables and the Two Theorems of John
{\Bell},'' Rev.\ Mod.\ Phys.\ {\bf 65}, 803 (1993).

\item\label{Caves96}
C.~M. {\Caves} and C.~A. Fuchs, ``Quantum Information:~How Much
Information in a State Vector?,'' in {\sl The Dilemma of {\Einstein},
{\Podolsky} and {\Rosen} -- 60 Years Later}, edited by A.~{\Mann} and
M.~{\Revzen}, Ann.\ Israel Phys.\ Soc.\ {\bf 12}, 226 (1996).

\item\label{Fuchs97d}
C.~M. {\Caves}, C.~A. Fuchs and R.~{\Schack}, ``Bayesian Probability
in Quantum Mechanics,'' to be submitted to Am.\ J.\ Phys. Preliminary
draft available upon request.

\item\label{Boole}
G.~{\Boole}, {\sl An Investigation of the Laws of Thought}, (Dover,
New York, 1958).

\item\label{Jaynes97}
E.~T. {\Jaynes}, {\sl Probability Theory:~The Logic of Science}. This
massive book was unfortunately not completed before Prof.\ {\Jaynes}'
death.  Preprints are available at {\tt http://bayes.wustl.edu/}.

\item\label{Rovelli96}
C.~{\Rovelli}, ``Relational Quantum Mechanics,'' Int.\ J.\ Theor.\
Phys.\ {\bf 35}, 1637 (1996).

\item\label{Pitowski89}
I.~{\Pitowsky}, ``From George {\Boole} to John {\Bell}---The Origins
of {\Bell}'s Inequality,'' in {\sl {\Bell}'s Theorem, Quantum Theory
and Conceptions of the Universe}, edited by M.~Kafatos (Kluwer,
Dordrecht, 1989).

\item\label{Pitowski94}
I.~{\Pitowsky}, ``George {\Boole}'s `Conditions of Possible
Experience' and the Quantum Puzzle,'' Brit.\ J.\ Phil.\ Sci.\ {\bf
45}, 95 (1994).

\item\label{Segal98}
W.~Segal and I.~E. Segal, ``The Black-Schole Pricing Formula in the
Quantum Context,'' Proc.\ Natl.\ Acad.\ Sci.\ USA {\bf 95}, 4072
(1998).

\item\label{Watanabe61}
S.~Watanabe, ``A Model of Mind-Body Relation in Terms of Modular
Logic,'' Synthese {\bf 13}, 261 (1961).

\item\label{Peres82}
A.~{\Peres} and W.~H. {\Zurek}, ``Is Quantum Theory Universally
Valid?,'' Am.\ J.\ Phys.\ {\bf 50}, 807 (1982).

\item\label{Heilbron86}
J.~L. {\Heilbron}, {\sl The Dilemmas of an Upright Man:~Max
{\Planck} as Spokesman for German Science}, (U.~California Press,
Berkeley, 1986), pp.~127--128.

\end{enumerate}

\section{22 December 1997, \ ``{\Bohr} was a Bayesian''}

It dawned on me recently that {\Bohr} was actually a Bayesian.
Anyway, if you view him that way, some of the crap he said
miraculously starts to make sense.

It's also dawned on me recently that I need to get to bed.
SuperShuttle arrives at 5:30 AM and it's already past 1:15!!  {\Kiki}
and I fly to Texas tomorrow, i.e., today.  However, I'd like to talk
to you Tuesday about the ``{\Bohr} from Bayes'' program (i.e., ``A
Bayesian Derivation of the Quantum Probability Law'').  E-mail me the
times you plan to be in the office Tuesday and I'll give you a call
when it is most convenient on me.  (Remember I'm on vacation!)

In the mean time, I'll forward on (for your holiday enjoyment) some
of the ramblings I sent to David {\Mermin}.  The beginnings of the
bald assertion [sic] above start up somewhere about Merminition
\ref{OldAsm10}.

\section{20 September 1998, \ ``Dropping In''}

I read the {\Beller} article in Physics Today.  In fact, I've read
several of her articles before:  she writes very well.  She of
course has a point about {\Bohr}'s intractable language; I've spent
many hours myself trying to make some sense of it all.  To the
people with less patience than me, I'm sure it's not obvious that
they should struggle to find some meaning there.  That's exactly why
someone has to get in and say something reasonable about (a
modern-day version of) the ``Copenhagen interpretation'' before
things get out of hand.  It's starting, for instance, to look like
the editor of {\sl Physics Today\/} has some Bohmian leanings: why
else would he/she allow the two {\Goldstein} articles and the
{\Beller} article within such proximity of each other?

\subsection{{\Carl}'s Reply}

\bq
I agree with you that it is very disturbing that {\sl Physics
Today\/} has run all those articles endorsing Bohmian mechanics.
It's caused me to get some books and articles about Bohmianism and
to think about it a little.

The main thing I got out of the {\Beller} article was that wonderful
``cock-a-doodle-doo'' quote from {\Einstein}, which shows that
{\Einstein} was not a naive realist.  He believed that the ONLY
reality we have is the one we construct from our perceptions and,
more sophisticatedly, from our scientific theories and models.  Our
everyday experience leads ineluctably to the notion that there is a
real world out there, in which we can often make reliable
predictions of the future based on past observation and which
proceeds on its way without much noticing our existence.  {\Einstein}
was quite right in maintaining that any fundamental scientific
theory ought to allow construction of that real world.  But, of
course, our perception of an everyday ``effective reality'' does not
mean that our theory of the microsopic world must have an objective
reality, since microsopic phenomena do not contribute directly to
our perception of an everyday effective reality.

I am confident that the Bayesian interpretation of quantum mechanics
is up to the challenge.  All probabilities are ignorance
probabilities.  There are certain situations where we are free to
ask whatever question we want to a microscopic system, and then our
questions and the answers we get help to create the Universe in
which we live.  For macroscopic phenomena, however, the laws of
physics, with their position-dependent potentials, combine with
decoherence to restrict the set of questions we can ask, thus giving
rise to an emergent reality.  The statistical predictions of quantum
mechanics within this emergent reality are just what we need to
construct an effective classical world, in which we make inferences
among alternatives based on the probabilities we assign; moreover,
we can often make reliable predictions based on past observations,
as in a deterministic, classical world.

This emergent reality is just what is needed for our everyday life
and also for the other sciences, such as biology, geology,
astronomy, etc. Bohmians, postmodernists, and {\Bohr} and other
physicists who tried to apply complementarity to social and
political questions---all make the basic mistake of
reductionism---i.e., thinking that our quantum-mechanical picture of
the microscopic level must naively correspond to our picture of the
macroscopic level.  As far as the higher, macroscopic level of
everyday life is concerned, the only function of the lower,
microscopic level of quantum mechanics is to give rise to the
emergent realistic structure of our classical world. It's simply
wrong to think that lack of realism at the microscopic level must
give rise to lack of realism at the macroscopic level.  This is the
reason that postmodernists can't use the lack of realism of quantum
mechanics to discredit objectivity and truth in our everyday lives
and in the study of history, politics, and culture.

Bohmians, it seems to me, make just the opposite mistake.  Convinced
that macroscopic reality can only come from microscopic reality,
they make up a microscopic reality for quantum mechanics.  This
microscopic reality is truly bizarre---an acausal web in which
everything in the Universe influences everything else.  Who needs
this Bohmian microscopic reality?  It is utterly unlike any kind of
reality that is demanded by our perceptions.  Moreover, it is
completely irrelevant to and unnecessary for the emergent reality
provided by quantum mechanics.  Why bother with it?

As you can see, I have been bothered not only by Bohmians, but also
by postmodernists.  If we're going to say that microscopically there
is no objective reality, then how do we avoid falling into the trap
of saying that everything is subjective?  The answer, it seems to
me, is that the absence of objective reality in quantum mechanics
does not invalidate the emergent reality of our everyday lives.

{From} the Bayesian view the two great questions---What basis? and
Which alternative within that basis?---are answered.  Decoherence
gives us the basis (I know there are problems here, but bear with
me), and once you have a basis, the notion that there is one actual
alternative within that basis is simply what the whole idea of an
emergent reality means.  In contrast, at the microscopic level,
where one can ask different questions, the idea that there is one
actual alternative for each question is inconsistent with the
statistical predictions of quantum mechanics.
\eq

\section{24 September 1998, \ ``Emerging Reality''}

I'm just about on my way to a long weekend in Texas.

I've just printed out your motivational note so that I can chew upon
it over the weekend.  In the mean time, would you try to say a
little more precisely what you mean by the phrase ``emergent
reality''?  Please, please, please.  (I honestly don't understand
the phrase.)  Also, what papers/books have you been reading about
Bohmianism?  I keep coming back to the idea that maybe there is a
way to punch a hole in it, like you suggested:  that would be
wonderful.  The only weak point that I know of is in getting the
initial distributions right, i.e., via some damping mechanism or
such (the first attempt was {\Bohm}, ``Proof that Probability Density
Approaches $|\psi|^2$ in Causal Interpretation of Quantum Theory,''
PRA {\bf 89}, pp.\ 458--466, 1953).  I wonder how good those
arguments are?

\section{27 January 1999, \ ``Bohmian Computers''}

It dawned on me today after talking to {\Carl} that I had forgotten
an old conversation I once had with none other than the famous Howard
{\Barnum}.  The place was a path near the Villa Gualino in Torino;
the time was about four years ago.  The subject was precisely Bohmian
computers.  It, like so many subjects in the field still to be
explored by the mediocre, had already been visited by the mind of
Howard {\Barnum}.  I remember him asking whether the existence of
Bohmian mechanics already proved that there existed a classically
efficient algorithm for factoring.

\section{19 February 1999, \ ``A Bright Young Man''}

I was taken aback, shocked even!  Can you believe what I heard one
of Jeff's graduate students mutter here:  ``Physics is the ability
to win a bet.''  The fellow's name is Joseph Robert {\Buck}, Jr.
Watch out for him; a proto-Bayesian is a good thing.

\section{02 March 1999, \ ``{\Caves} History''}

By the way, I noticed in reading your old paper ``In Defense of the
Standard Quantum Limit,'' that you had at sometime in the deep past
read {\Schroedinger}'s 1935 paper on entanglement (the one translated
into English by Trimmer).  If I can make a recommendation, I would
suggest you have a look at it again while rooting through those
{\Galavotti}/{\Jeffrey} papers.  The guy is really fixated on the
point that the thing that sets the quantum world apart from
classical one is not that maximal information is incomplete, but
that maximal information of the whole does not correspond to maximal
information of the parts.  (You may recall I made this point to you
before in a note titled, ``Oh Hibernia.'')  [See note to {\Ruediger}
{\Schack}, dated 25 January 1999.]

\section{03 May 1999, \ ``Complex H Space''}

One of your references should be Chapter 1 of that book {\sl
Quaternionic Quantum Mechanics\/} (I think it's by Steven {\Adler},
but you'll have to check the library or something for the exact
references.)

The other references I can think of are {\Wootters}' two paper below.
And there's at least one by John {\Wheeler} with some discussion of
the match between uniform distributions in probability space and
Hilbert space (it works for complex but not for reals).  {\Wheeler}'s
paper was in the special issue of the IJTP in 1982.  I think it was
titled ``The Computer and the Universe.''
\begin{enumerate}
\item
W.~K. {\Wootters}, ``Quantum mechanics without probability
amplitudes,'' {\em Foundations of Physics}, vol.~16(4),
pp.~391--405, 1986.

\item
W.~K. {\Wootters}, ``Local accessibility of quantum states,'' in {\em
Complexity, Entropy and the Physics of Information} (W.~H. {\Zurek},
ed.), (Redwood City, CA), pp.~39--46, Addison-Wesley, 1990. Santa Fe
Institute Studies in the Sciences of Complexity, vol. VIII.
\end{enumerate}

\section{10 June 1999, \ ``Gleasonization''}

\bcc
If we're radical probabilists instead of rational Bayesians, how can
[we] expect someone to adopt the quantum rule?
\ecc

I guess I agree with this.  That's why I wrote the sentence:
\bq
\noindent
Quantum mechanics, despite all its objectivist trappings, is a theory
of how we should behave in light of the facts the world presents
us---that is, in light of the answers the world gives to the
questions we ask.
\eq
in my last attempt at an introduction to our paper.  (Have a look at
it again when you get a chance.)

But I'm starting to get a little more dubious about the statement:
\bcc
That's why {\Gleason}'s theorem is the greatest triumph of Bayesian
reasoning: probs that are consistent with the Hilbert space structure
must be derived from a density operator.
\ecc

My worry comes from a couple of fronts.  The first is the way
{\Gleason}'s theorem is intimately tied to the field over which the
quantum mechanical Hilbert spaces are defined.  It's a consequence of
David {\Meyer}'s recent result (and Adrian's extension) that
{\Gleason}'s theorem does not hold for Hilbert spaces over a rational
field (real or complex).  (This is a question I had posed to Itamar
{\Pitowsky} a couple of years ago, and then to David \ldots\ but he
already knew the answer.) This new fact troubles me a little bit.

{\it Aside}:  Though, really, it is probably just telling us that
that kind of discrete structure doesn't capture the essence of
quantum mechanics.  It makes me curious what discrete structures
would still give rise to {\Gleason}. One idea that pops up---and one
directly tied to the importance of distinguishability---is whether
the theorem can be proved if one ``rationalizes'' not the field for
the vector space, but instead the {\em angles\/} between the vectors.
``The valid questions that quantum systems can be asked correspond to
a given standard orthonormal basis and all bases that can be gotten
from it by rational rotations.  Or perhaps by rotations that are
rational fractions of pi; this assumption might be more natural.  The
theory must then give probabilities for the outcomes.''---that's the
sort of thing I'm thinking.

The other front---and this one is more important---comes from the
suspicion that the assumption behind {\Gleason}'s theorem are not
purely Bayesian in motivation.  There's something extra there.  I
probably can't make this more clear than by tacking on a note I
wrote to Howard the other day.  [See note to Howard {\Barnum} dated
2 June 1999.] Probabilities to do with the Hilbert space structure,
yes. But why probabilities that are context independent (in the sense
below). That part strikes me as having nothing to do with
Bayesianism.

\section{10 September 1999, \ ``The Dangers of Probabilismo''}

\bcc
This seems somehow relevant to Chris's idea that Hamiltonians are
subjective, whereas I prefer to think that they are the objective
part of our description of the world.
\ecc

I believe I understand your worry (fear).  I too feel that there must
be some solid bedrock in quantum mechanics somewhere \ldots\ but I
haven't been convinced that the Hamiltonian is it.  The point is, as
best I can tell, each and every argument we have used for the
subjectivity of the wave function can also be used for the unknown
CPM.
\begin{enumerate}
\item
An unknown Hamiltonian can't be measured on a single shot.  If you
    want to know the Hamiltonian you can't ask the system it lives on,
    you have to ask the maker.
\item
The decomposition of the CPM is not unique.  (Somebody once told
    me that a similar argument was the most damning for the quantum
    state.)
\item
Let half of an EPR pair be the control bit entering in a controlled
    unitary gate.  The action on the target bit can be made to go one
    way or the other, according to our distant measurement.
\item
Probabilities are subjective, and quantum states are just compendia
    of probabilities.  The Hamiltonian's main (sole?) purpose within
    the theory is just to evolve those compendia.  That makes it look
    more like a law of thought than a law of nature (see note 5 below).
\end{enumerate}
And---now this one is a lot more speculative---5) Time evolutions are
linear.  Just, as you pointed out to me, this is something of an
analogy to Bayes' rule.  I wonder whether this suggests that
Hamiltonian evolution might be enforced solely by probabilistic
coherence a la Dutch book.

As {\Herb} likes to say, ``We need a little reality.''  But I think
it's pretty important that we hit the right one when we do.

\section{12 September 1999, \ ``CPM Extreme Points''}

Here's the reference (finally):
\bq
\noindent Seung-Hyeok Kye, ``On the convex set of completely positive
linear maps in matrix algebras,'' Math.\ Proc.\ Camb.\ Phil.\ Soc.\
{\bf 122}, 45--54 (1997).
\eq

\section{23 September 1999, \ ``The Ontological Tide''}

\bcc
When you deny the dynamics an ontological status, you will leave
everyone far behind, because you will be left with no ontology at
all.
\ecc

No, I wouldn't want to go that far.  And I don't think we have to.  I
can tell now, you've never quite gotten this information-disturbance
bit I've been preaching.  (It might not be right, but it does seem to
be a way out.)  So, let me try again:  read the little manifesto
below that I wrote up earlier this year.  [See note to {\Greg}
{\Comer} dated 22 April 1999.]  It drips of ontology \ldots\ but an
ontology that gets shunted away every time we try to get our grubby
hands on it. THAT statement is an ontological one in and of itself.

\section{23 September 1999, \ ``The Ontological Rip Tide''}

Continuing,
\bcc
Here's a thought.  I think there is no more power[ful] principle for
dynamics than the statement that we never lose information unless we
voluntarily throw it away.  Either classically or quantum
mechanically this means that we can lose information about a system
if it interacts with another system about which we don't have maximal
information.  The new ingredient quantum mechanically is
entanglement---{\Schroedinger}'s ``maximal information about the
whole is not maximal information about the parts.''  This means that
we can lose information about a quantum system if it interacts with
another system even though we have maximal information about the
whole.
\ecc

If you think about it, you might notice a similarity between this and
the manifesto I just sent you.  Here's how I put it to {\Todd}
{\Brun} a few months ago: \ldots\  [See note to {\Todd} {\Brun}
dated 8 June 1999.]

Be careful of the rip tide you're swimming.

\section{27 September 1999, \ ``Cleaning the Plate'' and ``Last Scrap''}

\bcc
On the ontology question, I wasn't convinced by your
information-disturbance document.  Look as hard as I could, I didn't
find the slightest hint of what the underlying ontology is supposed
to be.
\ecc

\begin{flushright}
Whereof one can not speak, thereof one should not speak.\\
\hspace*{\fill} --- L. Wittgenstein
\end{flushright}

\bcc
Scientists have a way of not liking being told that something is
outside their speculation.
\ecc

Thus my reason for testing the idea that information gathering and
information disturbance go hand in hand as the ``mechanism'' (a real
property of the world) to ensure that maximal information is not
complete.  That is the proposed mechanism for blocking the
speculation.  I had precisely this conversation with Lucien {\Hardy}
a couple months ago. The I vs.\ D ``mechanism'' seemed to placate
him a bit:  I drew strength from that.

\bcc
Once we get to this point, we proceed to the points in your second
note, where I am glad to see that we converge instead of diverging
and where I'm happy to concede that you have long appreciated the
point, and I am only now starting to catch up.
\ecc

I hope you didn't get the impression that I was worried about some
piddling issue of priority.  I only wanted to take the opportunity to
let your own words push you in the right direction.  The dates on the
passages I sent you were there because I cut and pasted them from my
master file ``Notes on a Paulian Idea'' (soon to be made available to
some of our sympathetic colleagues).

It is my plan to attain nirvana by becoming egoless.

\bcc
Information-disturbance is the fundamental principle---beyond just
chanting the slogan that ``maximal information is not
complete''---that says that you can go ahead and gather all the
information you want---i.e., draw from the well---but you must
produce a disturbance to be consistent with the slogan---i.e., the
well is more than just a source of information, it's also a destroyer
of information.
\ecc

I'm not sure I quite like the way you present things here.  Remember,
in my world it takes at least two scientists to tango \ldots\ and a
lot more than that to come to some consensus about a theory.  It is
information gathering and INFORMATION disturbance that accompany each
other.  But we might be in agreement about the broad sweep of things,
especially the part about consistency.

\section{24 March 2000, \ ``Title and Abstract?''}

How about the one below?  This was the talk I gave at MIT. I'll try
to remember roughly what I said then.

\bq
\noindent Title: Quantum Information, Quantum Channels \medskip

\noindent Abstract: Most physics students, with their first lesson on the
{\Heisenberg} uncertainty principle, are given a subliminal message:
quantum mechanics is a limitation.  The attitude is, ``Quantum
mechanics is something we deal with because we have to, but wouldn't
the world have been so much better if we could just measure a
particle's position and momentum simultaneously?''  This talk is
about the counterpoint to that attitude.  Recent advances in the
fields of quantum computation, quantum cryptography, and quantum
information theory show that the physical resources supplied by the
quantum world are anything but a limitation.  With these new
resources we can do things almost undreamt of before, from the secure
distribution of one-time pads for use in cryptography to the
factoring of large numbers with a polynomial number of steps.  The
magic ingredient in all this is something called quantum
information.  I will illustrate the subtle strangeness of this new
kind of information and the nice effects it buys with several
concrete examples drawn from my own work in quantum cryptography and
quantum channel-capacity theory.
\eq

\section{30 June 2000, \ ``Greece''}

In Greece right now.  Hitting naked breast overload, but otherwise
I'm OK.  Strangely, I've spent a lot of time with Roland {\Omnes} the
last couple of days, and he's starting to strike me as really OK. In
the right light, his view of QM (even including consistent
histories) may not be so different from ours.  It's been a most
useful time.  Too bad though that he still seems to be stuck in
thinking that taking measurement as a primitive in the theory is the
same thing as taking measurement as a primitive in the world.
(Though a few things seem to hint that he doesn't believe that
either.)

Thanks for the {\Jaynes} quotes too.  I'm hoping to get a chance to
read them and think about them on my four hour boat ride to Athens
tomorrow.

\subsection{{\Carl}'s Preply}

\bq
Motivated by Chris's reading of {\Jaynes}'s article, ``Predictive
Statistical Mechanics,'' I read the three articles that seem to have
his most direct statements about quantum mechanics.  The quotes I
mined follow.\medskip

\noindent From E.~T. {\Jaynes}, ``Probability in Quantum Theory,'' in {\sl
Complexity, Entropy and the Physics of Information}, edited by W.~H.
{\Zurek} (Addison-Wesley, Redwood City, CA, 1990), pp.~381--403:

\bq
{\bf Abstract:} ``For some sixty years it has appeared to many
physicists that probability plays a fundamentally different role in
quantum theory than it does in statistical mechanics and analysis of
measurement errors. It is a commonly heard statement that
probabilities calculated within a pure state have a different
character than the probabilities with which different pure states
appear in a mixture, or density matrix.  As {\Pauli} put it, the
former represents `\,\dots\,eine prinzipielle {\it Unbestimmtheit},
nicht nur {\it Unbekanntheit}'.  But this viewpoint leads to so many
paradoxes and mysteries that we explore the consequences of the
unified view, that all probability signifies only human
information.  We examine in detail only one of the issues this
raises: the reality of zero-point energy.''

{\bf p.~382:} ``Today we are beginning to realize how much of all
physical science is really only {\it information}, organized in a
particular way.  But we are far from unravelling the knotty
question: `To what extent does this information reside in us, and to
what extent is it a property of Nature?'\,''

{\bf p.~385:} ``Let me stress our motivation: if quantum theory were
not successful pragmatically, we would have no interest in its
interpretation.  It is precisely {\it because\/} of the enormous
success of the QM mathematical formalism that it becomes crucially
important to learn what that mathematics means.  To find a rational
physical interpretation of the QM formalism ought to be considered
the top priority research problem of theoretical physics; until this
is accomplished, all other theoretical results can only be
provisional and temporary.

``This conviction has affected the whole course of my career.  I had
intended originally to specialize in Quantum Electrodynamics, but
this proved to be impossible.  Whenever I look at any
quantum-mechanical calculation, the basic craziness of what we are
doing rises in my gorge and I have to try to find some different way
of looking at the problem, that makes physical sense.  Gradually, I
came to see that the foundations of probability theory and the role
of human information have to be brought in, and so I have spent many
years trying to understand them in the greatest generality.''

{\bf pp.~386--387:} ``{\Einstein}'s thinking is always on the
ontological level traditional in physics, trying to describe the
realities of Nature.  {\Bohr}'s thinking is always on the
epistemological level, describing not reality but only our
information about reality.  The peculiar flavor of his language
arises from the absence of all words with any ontological import;
the notion of a `real physical situation' was just not present and
he gave evasive answers to questions of  form: `What is really
happening?' \dots

``Although {\Bohr}'s whole way of thinking was very different from
{\Einstein}'s, it does not follow that either was wrong.  In the
writer's view, all of {\Einstein}'s thinking---in particular the EPR
argument---remains valid today, when we take into account its
ontological character.  But today, when we are beginning to consider
the role of information for science in general, it may be useful to
note that we are finally taking a step in the epistemological
direction that {\Bohr} was trying to point out sixty years ago.''

{\bf p.~387:} ``\,\dots Our present QM formalism is a peculiar
mixture describing in part laws of Nature, in part incomplete human
information about Nature---all scrambled up together by {\Bohr} into
an omelette that nobody has seen how to unscramble.  Yet we think the
unscrambling is a prerequisite for any further advance in basic
physical theory, and we want to speculate on the proper tools to do
this.

``We suggest that the proper tool for incorporating human
information into science is simply probability theory---not the
currently taught `random variable' kind, but the original `logical
inference' kind of James {\Bernoulli} and {\Laplace}.  For historical
reasons explained elsewhere, this is often called `Bayesian
probability theory.'\,''

{\bf p.~390:} ``We would like to see quantum theory in a similar way;
since a pure state $\psi$ does not contain enough information to
predict all experimental results, we would like to see QM as the
process of making the best predictions possible from the partial
information that we have when we know $\psi$.  If we could either
succeed in this, or prove that it is impossible, we would know far
more about the basis of our present theory and about future
possibilities for acquiring more information than we do today.''
\eq

\noindent From E.~T. {\Jaynes}, ``Predictive Statistical Mechanics,'' in
{\sl Frontiers of Nonequilibrium Statistical Physics}, edited by
G.~T. Moore and M.~O. Scully (Plenum Press, NY, 1986), pp.~33--55.

\bq
``We think it unlikely that the role of probability in quantum
theory will be understood until it is generally understood in
classical theory and in applications outside of physics.  Indeed,
our fifty-year-old bemusement over the notion of state reduction in
the quantum-mechanical theory of measurement need not surprise us
when we note that today, in all applications of probability theory,
basically the same controversy rages over whether our probabilities
represent real situations, or only incomplete human knowledge.

``If the wave function of an electron is an `objective' thing,
representing a real physical situation, then it would be
mystical---indeed, it would require a belief in psychokinesis---to
suppose that the wave function can change, in violation of the
equations of motion, merely because information has been perceived
by a human mind.

``If the wave function is only `subjective,' representing a state of
knowledge about the electron, then this difficulty disappears; of
course, by definition, it will change with every change in our state
of knowledge, whether derived from the equations of motion or from
any other source of information.  But then a new difficulty appears;
the relative phases of the wave function at different points have
not been determined by our information; yet they determine how the
electron moves.

``There is no way quantum theory could have escaped this dilemma
short of avoiding the use of probability altogether.  Not only in
Physics, but also in Statistics, Engineering, Chemistry, Biology,
Psychology, and Economics, the nature of the calculations you make,
the information you allow yourself to use, and the results you get,
depend on what stand you choose to take on this surprisingly
divisive issue: are probabilities `real'?

``But in quantum theory the dilemma is more acute because it does not
seem to be merely a choice between two alternatives.  The
`subjective' and `objective' aspects are scrambled together in the
wave function of an electron, in such a way that we are faced with a
paradox like the classical paradoxes of logic; whatever stand you
take about the meaning of the wave function, it will lead to
unacceptable consequences.

``To achieve a rational interpretation we need to disentangle these
aspects of quantum theory so the `subjective' things can change with
our state of knowledge while the `objective' ones remain determined
by the equations of motion.  But to date nobody has seen how to do
this; it is more subtle than merely separating into amplitudes and
phases.''

``As many have pointed out, starting with {\Einstein} and
{\Schroedinger} fifty years ago and continuing into several talks at
this Workshop, the Copenhagen interpretation of quantum theory not
only denies the existence of causal mechanisms for physical
phenomena; it denies the existence of an `objectively real' world.

``But surely, the existence of that world is the primary experimental
fact of all, without which there would be no point to physics or any
other science; and for which we all receive new evidence every waking
minute of our lives.  This direct evidence of our senses is vastly
more cogent than are any of the deviously indirect experiments that
are cited as evidence for the Copenhagen interpretation.''

``Now let's look at the mind-boggling problem from a different side.
A single mathematical quantity $\psi$ cannot, in our view, represent
incomplete human knowledge and be at the same time a complete
description of reality.  But it might be possible to accomplish
{\Bohr}'s objective in a different way.  What he really wanted to do,
we conjecture, is only to develop a theory which takes into account
the fact that the necessary disturbance of a system by the act of
measurement limits the information we can acquire, and therefore the
predictions we can make. This was the point always stressed in his
semipopular expositions.  Also, in his reply to EPR he noted that,
while there is no physical influence on $S$, there is still an
influence on the kinds of predictions we can make about $S$.''

``On deep thought, it will be seen that whenever we allow
probabilities to become `physically real' things, logical
consistency will force us, eventually, to regard the objects as
`unreal'.  If we are to reach {\Bohr}'s goal while at the same time
keeping our objects real we must recognize, with {\Laplace}, Maxwell,
and Jeffreys, that whenever we use probability it must be as a
description of incomplete human knowledge, as it was in classical
statistical mechanics.''
\eq

\noindent From E.~T. {\Jaynes}, ``A Backward Look to the Future,'' in
{\sl Physics and Probability:\ Essays in Honor of Edwin
T.~{\Jaynes}}, edited by W.~T. Grandy,~Jr. and P.~W. Milonni
(Cambridge University Press, Cambridge, England, 1993), pp.~261--275.

\bq
{\bf p.~269:} ``\,\dots throughout the history of quantum theory,
whenever we advanced to a new application it was necessary to repeat
this trial-and-error experimentation to find out which method of
calculation gives the right answers.  Then, of course, our textbooks
present only the successful procedure as if it followed from general
principles; and do not mention the actual process by which it was
found.  In relativity theory one deduces the computational algorithm
from the general principles.  In quantum theory, the logic is just
the opposite; one chooses the principle to fit the empirically
successful algorithm.''

{\bf pp.~270--271:} ``What has held up progress in this field for so
long?  Always our students are indoctrinated about the great
pragmatic success of the quantum formalism---with the conclusion
that the Copenhagen interpretation of that formalism must be
correct.  This is the logic of the Quantum Syllogism:

\begin{quote}
The present {\it mathematical formalism\/} can be made to reproduce
many experimental facts very accurately.
\begin{center}
{\bf Therefore}
\end{center}
The {\it physical interpretation\/} which Niels {\Bohr} tried to
associate with it must be true; and it is na{\"\i}ve to try to
circumvent it.
\end{quote}

\noindent
Compare this with the Pre-Copernican Syllogism:

\begin{quote}
The mathematical system of epicycles can be made to reproduce the
motions of the planets very accurately.
\begin{center}
{\bf Therefore}
\end{center}
The theological arguments for the necessity of epicycles as the only
perfect motions must be true; and it is heresy to try to circumvent
them.
\end{quote}

In what way are they different?  The difference is only that today
everybody knows what is wrong with the Pre-Copernican Syllogism; but
(from the frequency with which it is still repeated) only a
relatively few have yet perceived the error in the Quantum
Syllogism.''

{\bf p.~272:} ``One of the principles of scientific inference---which
has always been well understood by the greatest scientists---is that
it is idle to raise questions prematurely, when they cannot be
answered with the resources available.  For Isaac {\Newton} it would
have been foolish to raise questions that were not foolish for Erwin
{\Schroedinger} 250 years later; for Gregor Mendel it would have been
foolish to raise questions that were not foolish for Francis Crick
100 years later.  By `foolish' we mean `without hope of success'. Of
course, we all enjoy indulging in a little free speculation about
the future of science; but for scientists to expend their serious
professional time and effort on idle speculation can only delay any
real progress.''
\eq
\eq

\section{07 September 2000, \ ``Critical Letters and Reply''}

\bcc
Liked the ``effective reality,'' too.
\ecc

\noindent [NOTE:  This sentence of {\Carl}'s refers to my article with
 {\Asher} {\Peres}:  C.~A. Fuchs and A. {\Peres}, ``Quantum Theory --
Interpretation, Formulation, Inspiration:\ Fuchs and {\Peres}
Reply,'' Phys.\ Tod.\ {\bf 53}(9), 14, 90 (2000).]\medskip

But I had even managed to get that into the original:
\bq
\noindent We do not deny the possible existence of an objective reality
independent of what observers perceive. In particular, there is an
``effective'' reality in the limiting case of macroscopic phenomena
like detector clicks or planetary motion: Any observer who happens
to be present would acknowledge the objective occurrence of these
events. However, such a macroscopic description ignores most degrees
of freedom of the system and is necessarily incomplete.
\eq

Keep in mind, though, that when I speak of an effective reality I
think I mean something quite different from you.  [See ``{\Carl}'s
Reply'' to my note ``Dropping In,'' dated 20 September 1998.]  In
particular, what I have in mind seems to have nothing to do with the
particular properties of the Hamiltonian or dynamics. That's where
we part mostly.  My ``effective reality'' is, instead, a function of
our ignorance and the crudeness of our experimental intervention
capabilities.  It is about the idea that the
information--disturbance tradeoff disappears under such conditions.
The {\sl Physics Today\/} reply makes that a little more explicit:
\bq
\noindent
They could do that because this aspect of carbon is part of the
``effective reality'' quantum theory produces in some regimes of our
experience. Indeed, this ``effective reality'' forms the ground for
all our other quantum predictions simply because it is the part of
nature that is effectively detached from the effect of our
experimental interventions.  But, if one tries to push this special
circumstance further and identify an overarching ``reality''
completely independent of our interventions, then this is where the
trouble begins and one finds the {\it raison d'\^etre\/} of the
various ``interpretations.''
\eq

\section{13 November 2000, \ ``{\Bush} vs.\ {\Bohr}''}

From the {\sl New York Times\/} this morning:
\bq
\noindent Throughout the primaries and general election, the {\Bush}
campaign had been unusually confident, partly because Mr.\ {\Bush}
and his aides genuinely seemed to think they would win. But advisers
also said they were trying to convey a sense of inevitability to Mr.\
{\Bush}'s candidacy. Since the election, the {\Bush} team has done
much the same thing, striking the posture of victory, which helps
explain why Mr.\ {\Bush} spoke last week about his planning for the
transition and his aides leaked the names of potential top cabinet
members.
\eq
I do think the parallels between {\Bush}'s behavior and
consistent-history quantum mechanics are very real.

\section{01 February 2001, \ ``An Ed/Ellen Jaynesian''}

You are an Ed Jaynesian!  I just came to realize this in rereading
the quotes of his that you jotted down.
\bq
\noindent
To achieve a rational interpretation we need to disentangle these
aspects of quantum theory so the `subjective' things can change with
our state of knowledge while the `objective' ones remain determined
by the equations of motion.  But to date nobody has seen how to do
this; it is more subtle than merely separating into amplitudes and
phases.
\eq
(Saying Ed Jaynesian, I keep thinking of the Ellen Jaynesians in
{\sl The World According to Garp\/} and wonder if you too will
eventually cut out your tongue when you realize that Hamiltonians
are just the pure states of time evolution.)

\section{01 February 2001, \ ``My {\Jaynes}''}

On the other hand, I should classify myself a Jaynesian too.
Perhaps a mild Jaynesian, for my passion only extends so far as:
\bq
\noindent
We would like to see quantum theory in a similar way; since a pure
state $\psi$ does not contain enough information to predict all
experimental results, we would like to see QM as the process of
making the best predictions possible from the partial information
that we have when we know $\psi$.  If we could either succeed in
this, or prove that it is impossible, we would know far more about
the basis of our present theory and about future possibilities for
acquiring more information than we do today.
\eq

\section{26 February 2001, \ ``Death of a Salesman''}

I know the difference between science and philosophy.  You said that
you ``would never, ever believe that Hamiltonians are not the real
properties of systems.''  To the extent that I know you too, I know
you didn't mean that.  You only meant that {\it without\/} firm
scientific evidence placed upon your plate you would not believe
it.  That's my burden, I understand, not only with respect to you
but, more importantly, for judgment day.  Maybe I'll believe I'm
wrong ultimately, but if so, I think we still both have to ask
ourselves the source of the ``miracle'' that something unquestionably
a state of knowledge can masquerade so easily as a state of nature at
times. The evidence below is nothing new---the essential was in an
old email to you---but maybe it looks more respectable now.  [See
{\tt quant-ph/0012067}, ``Storage of quantum dynamics on quantum
states: a quasi-perfect programmable quantum gate,'' by G.~{\Vidal}
and J.~I. {\Cirac}.]

\chapter{Letters to {\Greg} {\Comer}}

\begin{flushright}
\baselineskip=12pt
\parbox{4.25in}{\baselineskip=12pt\small
\bq
\noindent The ghost of my father sat in front of me\\
\indent sprinkling salt into its beer.\\
The floor, wooden and sole soaked slick\\ I could just push my feet
and rub my fingers.\\ ``Give me more life,'' I whispered.
\medskip\\
\noindent I wanted to touch the flakes of rust\\
\indent on the cooler, mingled with sweat.\\
Bottle caps everywhere; it didn't understand\\
\indent the boredom.\\
My only sound was that of a screen door.\\ ``My son will be a
professor of physics in three years.''
\eq
}
\end{flushright}
\bigskip
\medskip

\section{14 February 1995, \ ``Turtle Wax''}

Let me wax just a little.  A few quotes by Wolfgang {\Pauli} on
quantum mechanical randomness and objective reality.  (Taken from the
English translation of a paper titled ``Matter.'')
\bq
\noindent
``Like an ultimate fact without any cause, the {\it individual\/}
outcome of a measurement is, however, in general not comprehended by
laws. This must necessarily be the case ...''
\eq
\bq
\noindent
``In the new pattern of thought we do not assume any longer the {\it
detached observer}, occurring in the idealizations of this classical
type of theory, but an observer who by his indeterminable effects
creates a new situation, theoretically described as a new state of
the observed system.  In this way every observation is a singling
out of a particular factual result, here and now, from the
theoretical possibilities, thereby making obvious the discontinuous
aspect of the physical phenomena.''
\eq
\bq
\noindent
``Nevertheless, there remains still in the new kind of theory an
{\it objective reality}, inasmuch as these theories deny any
possibility for the observer to influence the results of a
measurement, once the experimental arrangement is chosen.''
\eq
Objectivity grounded upon randomness!!  Hmm, I like that.

\section{04 March 1995, \ ``Saturday Slush''}

Let me give you another {\Pauli} quote \ldots\ as is my habit of
late. This one is from an article titled ``The Theory of Relativity
and Science'' and is really good.

\bq
In spite of this, {\Einstein} held firmly to the narrower concept of
reality of classical physics; from this point of view a description
of nature which permits single events not determined by laws was
bound to appear to him ``incomplete.''  He combined with this a
regressive longing, not indeed for the old mechanistic idea of the
point-mass, but for his geometrical conception of the field in the
general theory of relativity.  Motivating his attitude, he frankly
explained that to depart from the narrower reality concept of physics
before quantum mechanics seemed to him to be getting perilously
close to a point of view in which it is impossible to discriminate
sufficiently clearly between dream or hallucination and ``reality.''
As against this, the objective character of the description of
nature given by quantum mechanics has appeared to the rest of us to
be adequately guaranteed by the circumstance that its statistical
laws describe reproducible processes, and that the results of
observation, which can be checked by anyone, cannot be influenced by
the observer, once he has chosen his experimental arrangement.
\eq

Man o' man I like that.  In all my years of reading silly things
about quantum mechanics, I have never found anybody better than
{\Pauli} at putting forth the crucial point in such a clear fashion.
Even {\Wheeler} in his Game of Twenty Questions, didn't make it quite
this clear.  The world in some very real sense is a construct and
creation of thinking beings simply because its properties are so
severely tied to the particular questions we ask of it.  {\it But\/}
on the other hand, the world is not completely unreal as a result of
this; we generally cannot control the outcomes of our measurements.
It is precisely because the outcome of the individual event cannot be
determined by law that the world still has a trace of reality.  If
we actually could control the outcomes of measurements, {\it then\/}
the world would as well be a ``dream or hallucination.''  \ldots\ But
we can't!

Do you remember the following words from a poem I wrote last year:
\bq
\noindent
{\Einstein} must have foreseen the evil in our dear quantum.  I don't
envy the pain; the blood so very frightening.  No evil in $A$-bomb,
but rather the evil of no image; the evil of no substrate; the
burning people in Waco; my father tied to a bed.  The poor roaming
wisdom of E.
\eq
Well Mr.\ {\Pauli} confirms my suspicion of what {\Einstein} feared
most in quantum mechanics.

\section{07 March 1995, \ ``Please, Please No More!''}

One LAST time, for the sake of a little more clarity!  I promise I
won't send any more quotes along these lines unless I find something
significantly new. But, in the mean time, Mr.\ {\Pauli} again; this
time from ``Albert {\Einstein} and the Development of Physics.''

\bq
We often discussed these questions together, and I invariably
profited very greatly even when I could not agree with {\Einstein}'s
views.  ``Physics is after all the description of reality,'' he said
to me, continuing, with a sarcastic glance in my direction, ``or
should I perhaps say physics is the description of what one merely
imagines?''  This question clearly shows {\Einstein}'s concern that
the objective character of physics might be lost through a theory of
the type of quantum mechanics, in that as a consequence of its wider
conception of objectivity of an explanation of nature the difference
between physical reality and dream or hallucination become blurred.

The objectivity of physics is however fully ensured in quantum
mechanics in the following sense.  Although in principle, according
to the theory, it is in general only the statistics of series of
experiments that is determined by laws, the observer is unable, even
in the unpredictable single case, to influence the result of his
observation---as for example the response of a counter at a
particular instant of time.  Further, personal qualities of the
observer do not come into the theory in any way---the observation
can be made by objective registering apparatus, the results of which
are objectively available for anyone's inspection. Just as in the
theory of relativity a group of mathematical transformations
connects all possible coordinate systems, so in quantum mechanics a
group of mathematical transformations connects the possible
experimental arrangements.

{\Einstein} however advocated a narrower form of the reality concept,
which assumes a complete separation of an objectively existing
physical state from any mode of its observation.  Agreement was
unfortunately never reached.
\eq

\section{27 January 1996, \ ``Philo''}

The philosophy of it all my friend.  I just heard a song on the
radio that I had not heard since the Chapel Hill days---it drew my
thought to you.  I remember writing you almost exactly three years
ago about my arrival in the land of the quantum.  I'm writing you
from the very same machine that I did then.  So much is the same and
so much is different.

Philosophy in every turn I used to say.  Can you believe I've
secured a full time job playing with quantum mechanics.  I get payed
to explore it and pick it apart in any way I wish.  I get payed to
keep dreaming about this lovely little structure.

The present project is finding bounds on the minimal resources
required to transpose a quantum state from here to there.  It's just
another way to explore how quantum a set of little quantum thingies
are.  Paul {\Simon} is the moral guide.

When I get back to Montreal two projects are on the burner.  Another
calculation to gauge the tradeoff between information gain and
disturbance in quantum measurement.  And another attempt to derive
much of quantum theory from information theoretic principles---an
extension of {\Bill} {\Wootters} work as John {\Wheeler}'s graduate
student.

So many ways to profiteer from the madness of trying to build a
quantum computer.  That's OK though---there really is hope that all
the efforts won't be wasted along the way.

I'm having a cup of coffee; it's a little after 8:00 in the
evening.  {\Kiki} is in Montreal just going to bed about now.  The
world is turning beneath me and I feel it; it all seems very
mystical tonight.

Where am I going with this?  I don't know.  I just wanted to say
that the quantum is alive and well.  And that coming out here panned
out.  And that in some ways it's just the way it was three years ago.

I turn within myself.

\section{20 February 1996, \ ``The Mont-Royal Sweats''}

\bv
She said: \ How much does information weigh? \\
I said: \ How heavy is a kiss?
\ev

Thanks for the tidbits and epiphanies.

I do like it up here very much (except when I have to walk the dogs
every morning and every evening!)  We have a really big apartment;
it's about twice as big as our last house.  And for some reason it
has more the feel of an ``academic'' home---a place where thoughts
have a chance of being made.  It's a shame I'll have to move before
I know it; I was really looking forward to a two year stay in
Montreal.

I'm glad you're getting to teach a ``Modern Physics'' course.  (I
presume that means ideas about the {\Bohr} atom and a little special
relativity?)  Is this your first time in such an endeavor?  I would
like to think that when I have the same opportunity I'd break the
bounds of protocol \ldots\ and not mention even once any crap about
wave-particle dualism or any other such vague ideas. Rather I'd
focus on quantum mechanical randomness at the outset and not even
attempt to get past two and three-level systems.  But of course such
blasphemy will never be; I know that I'll be bound by what the other
professors in the department need.  Ahh, the dreams in my life.

{\Gilles} just wrote a semi-popular account of what new things are
happening in quantum computing; if it's any good (I haven't read it
yet), I'll send it your way.

This quantum information business, do you know what I like about
it?  I very much like it because it puts the focus just where it
should be:  that quantum theory is a theory of information and
predictability \ldots\ and probably nothing more.  It gives us new
handles for studying and manipulating this idea.  That's what I
like, and that's why I've decided to stick with it.  When I'm old
and grey, I'll come back to the deeper question of what's really
going on with the quantum phenomena themselves. ($\longleftarrow$
That's probably a lie; since when have I ever been able to stop
thinking about the deeper structure of the quantum?)

The silly theorem I'm trying my damnedest to prove right now is
another simple example of this sort of stuff.  I have a set of pure
quantum states $\psi_i$ on some $n$-dimensional Hilbert space that
occur with probabilities $p_i$.  Then I consider ``coding'' them in
any way I like onto the states of a $d$-dimensional Hilbert space
with $d<n$; i.e., I imagine making any one-to-one correspondence
between the $\psi_i$ and a similar set of vectors on the smaller
Hilbert space. These ``code'' vectors are later to be ``decoded'' by
an automatic device that obeys the laws of physics, i.e. has a
unitary interaction with the $d$-dimensional space.  The final output
is a set of new states $\rho_i$ (which are generally mixed because
we may ignore the final state of the device itself).  The question
is: what is the largest average inner product that can be made
between the $\psi_i$ and the $\rho_i$ by such a procedure?  I would
like to say that that number is bounded above by the sum of the
largest $d$ eigenvalues of the density matrix formed from the
ensemble of $\psi_i$'s \ldots\ but I just can't show it!!!  And it's
driving me crazy.

My pet idea of late is that the foundation of unitary dynamics for
quantum theory can be replaced with the idea that nonorthogonal
quantum states cannot be cloned or broadcast.  There is a certain
theorem originally due to {\Wigner} that makes this idea quite
plausible.  His theorem is that any evolution on Hilbert space that
preserves inner products must either be unitary or anti-unitary.
Cloning nonorthogonal quantum states requires that certain inner
products must decrease.  So if you outlaw cloning at the outset, then
you outlaw inner-product-decreasing evolutions.  Well it turns out
that if you outlaw such evolutions you also outlaw
inner-product-increasing evolutions (this I've proven, it's not
hard).  So all you're left with is inner-product-preserving
evolutions.  Given that and {\Wigner}'s theorem, quantum mechanics is
almost uniquely pinned down.  It's a very pleasant idea I think ...
because it is the no-clonability of nonorthogonal states that gives
quantum information theory all its umph.\footnote{NOTE:  I later
found a flaw in this argument.  The problem is, cloning
transformations do not describe enough kinds of
``inner-product-decreasing'' maps.  See C.~A. Fuchs, ``Information
Gain vs.\ State Disturbance in Quantum Theory,'' Fort.\ der Phys.\
{\bf 46}(4,5), 535--565 (1998) for a correct solution to the
problem.}

\section{27 February 1996, \ ``{\Kant} Cola''}

Just thought I'd take a little coffee break before I get started up
for the morning.  I've got to start going through (with a
fine-toothed comb) a first draft of a paper that Howard {\Barnum},
Ben {\Schumacher}, Richard {\Jozsa} and I are coauthoring.  (Ben
wrote the first draft of this one.)  The subject is on the full
fledged converse to the quantum coding theorem; namely, if less than
a von Neumann's entropy worth of qubits are sent per transmission,
then a quantum signal can be reconstructed with vanishingly small
error. $\longleftarrow$ I know that doesn't mean much to you, but
that's the subject.

The weather up here is very nice today.  It's a little below
freezing, but the sun is out and there's not a bit of wind.

{\Kiki} and I went into the Outr\`emont area Friday evening in
search of an interesting restaurant \ldots\ and what a find we
made!  Let me tell you a strange little story.  During the summer of
1985, I was reading a book by C.~F. von Weizsacker titled {\sl The
Unity of Nature}.  Most of the book was about quantum mechanics and
{\Kant}ian philosophy. Apparently it spurred me to have the following
dream. I was in a little hole-in-the-wall joint somewhere in Austin;
my old friend David was there, also John Simpson and Marshall Burns.
The place really stood out in my mind because of the Bohemian feel to
it:  dark, smoky, mystical almost. The night wasn't filled with much
of interest:  David only wanted to talk about getting drunk, John
only wanted to talk about finding a girl, and Marshall only wanted
to talk about philosophy.  In those days John didn't drink alcohol,
so, at some point, when he asked for a drink, I thought we'd be out
of luck.  But upon looking around, I saw a refrigerator in the
middle of the bar near the pool table.  We walked over to it and
took a look.  It was filled with all different sorts of vegetable
drinks.  John grabbed one, and I looked through it for something
more interesting.  At the very back, I found one lone can of
``{\Kant} Cola.''  That was written on the label, along with a small
portrait of Immanuel {\Kant}.  I opened the can, took a drink,
\ldots\ and, for a miraculous moment, I understood all the
intricacies of the world---I understood the necessity of quantum
mechanics.  When I came out of my trance, the can was empty and I
knew that I would never see the light again.  Then I awoke.  I was
so taken with this dream that the next day I sketched out the layout
of the joint and made a record of the dream.  That was over ten
years ago.

So back to the restaurant of Friday. The place was called ``City
Pub''; it was such a strange little place:  dark, smoky, mystical
almost.  The food was excellent---far better than it should have
been for the price.  Each option in the place was only \$4.99.  I had
steak, fries, and a vegetable.  {\Kiki} had potato soup, quiche,
fries, and veggie. They had a special on beer, three for the price
of one (so we had six). The music was some sort of strange mesh of
things that I suspect you'd only hear in some little bar in Germany
where everyone wears black.  Anyway, we had quite a time there.
However, just a little while before leaving I started to note how
similar this place was to the place in my dream 10 years ago.  I
told {\Kiki} the whole story. Then I looked around and---strangely
enough---there was a refrigerator in the middle of the room near the
pool table!  I was so taken by this that upon my way to the
restroom, I took a look into it.  What a disappointment: it only
contained beer.  However, the restroom did have a surprise for me.
In the middle of all the graffiti (about Qu\'ebec's hoped for
independence) was something written in bold black letters:
\begin{center}
\parbox{1.4in}{De nobis ipsis silemus\\
\hspace*{\fill} --- {\it E. {\Kant}}}
\end{center}
That made my evening. I wrote down the words so that I wouldn't
forget and went home to look up my old notes on the dream.  Sure
enough, there were similarities in the layout of the two places, and
moreover, I saw that the name of the original place in my dream was
``Hole in the Wall Pub.''  Very strange.  I asked {\Ruediger}
{\Schack} to translate the words for me, and he came up with ``About
ourselves we remain silent.''

That's the story.  I suppose I should be off to work.

\section{06 March 1996, \ ``Rub Me''}

How would I respond?  ``If upon learning of quantum mechanics, you
are not left with your head spinning \ldots\ then you haven't
understood a thing.''  (Paraphrase of {\Bohr})

A question quite similar to this is what got me interested in
quantum theory in the first place.  In my sophomore year of high
school, in my chemistry class, I saw some of these wonderful
pictures of electron clouds around a hydrogen nucleus.  And there
were lumps of the cloud that were disconnected.  So I asked the
teacher what is the meaning of this cloud.  She said. ``It's where an
electron can be.''  I said, ``Then, for any given atom, a more
accurate picture would only have one lump.  Because there's no way
then that the electron could get from one lump to the other.''  She
said, ``No, you're wrong.''  I said, ``No, I'm right.  Convince me
otherwise.''  She couldn't of course, and I thought, ``Ahh, what an
idiot.''  But then I went off to college and thought long and hard
about these silly quantum thingies \ldots\ and came to the conclusion
that I was after all the one who was wrong!  I've never quite gotten
out of the befuddlement.

Maybe the best thing I can do is quote Charlie {\Rasco}:  ``You just
have to realize that {\it electrons are not little fucking billiard
balls\/}!''

A better answer, perhaps, is what I actually believe.  In the
problem of the square well, like in all other quantum mechanical
problems, the state vector we ascribe to the system encapsulates
just exactly what we can predict about the outcomes of our probings
on the system and it quantifies how much we are surprised about the
outcome we actually find.  If we are honestly able to build an
honest-to-god square well potential and have a particle in it in its
first excited state (and we are able to do so repeatedly), then we
will never find the particle at the center.  Period.  If we ever
find it there, the quantum state we ascribe the system should be
updated to take into account that we have not been preparing the
system as we had said we would.  Any attempt to build a more
``classical'' picture beyond this---one that explains why some
particles are found to the left of the center and why some are found
to the right---is dangerous business.  Chances are well beyond the
99.999\% mark that there will be an inconsistency in any such fix-up
you try.  Just try it and see: you will find that (in one way or
other) you have sneaked in superluminal signalling into physics, or
broken the Second Law of Thermodynamics, or even worse, given up on
a world in which there is anything beyond what Allah wills.

\section{26 March 1996, \ ``Savagery''}

Silly, silly science; why did we ever get involved!

When I first started finding books on quantum mechanics in my
freshman year in high school, I thought, ``What dull stuff!''  I
preferred to devote my time to things about relativity, wormholes,
and black holes.  It was only after getting to college that my
interest in quantum mechanics took off.  It happened the first week
I was there, after reading Heinz {\Pagels}' {\sl The Cosmic Code}.
It was then that I started to take seriously that my old questions
in my chemistry class (that I told you of) didn't have such
satisfactory answers (and it wasn't just the ignorance of my old
chemistry teacher).

So the upshot is:  be patient, I guess.  Maybe they'll decide it's
interesting after they leave your nest.

\section{27 March 1996, \ ``Strange Comment''}

Here's an interesting comment I thought you might like to hear. I
was talking to one of the computer scientists interested in quantum
cryptography here the other day [Jeroen van de Graaf].  I made a
comment that the number of baryons in the universe is estimated to
be roughly $10^{80}$. He said (quite seriously), ``Is that all?''
``That's just an 80 digit number \ldots\ it can be factored on a
classical computer.''

\section{01 April 1996, \ ``Puppy Scoops''}

I read a wonderful little piece on {\Einstein}'s ontology yesterday
while sitting in a coffee shop---Chez Dick, I love the name!
(Actually I read a much more detailed account in {\Fine}'s book a
year or so ago, but had apparently forgotten the force of it.  The
source this time was Aage {\Petersen}'s book.)  The point is this.
Mr.\ E did {\it not\/} take the existence of an objective (real)
world independent of the knowers of it as an a priori given.  So,
though E was not a dyed-in-the-wool Baptist, he was not a
dyed-in-the-wool {\it realist\/} either.  His argument was better
than a simple straight-up {\it belief\/} in a real world.  He saw
clearly that:  it is {\it not\/} the case that there are no
ontological systems for coordinating and organizing our experiences
other than a ``real world.'' Rather, he only contended that the idea
of a real world {\it is\/} a simple explanation.  Moreover, at least
until the advent of quantum theory, it was an idea that served its
purpose very well---allowing all science hitherto to be based on it.
His only point was that he would be very hard pressed to give up the
idea of a ``real world'' and---as of yet---had not accumulated enough
evidence to force himself to do so.

\section{04 May 1996, \ ``You May Enjoy''}

For some reason or other I recieved the note attached below.  You
may enjoy.  I found particularly interesting the comment:
\bq
\noindent
Diffeomorphism invariance is, in my view, precisely the expression
of the statement that the theory may make no implicit reference to
reference frames outside of the system.
\eq
Hmm?  But what if quantum phenomena really require ``measurement''
to come into being?  I.e.\ something outside the system to register
their reality?

Ahh, it's like talking to a brick wall with some of these guys.

Anyway, my present attitude is summarized on p.\ 27 of  {\Asher}'s
book. ``Other authors introduce a wave function for the whole
Universe.  In this book, I shall refrain from using concepts that I
do not understand.''

\section{19 July 1996, \ ``Old Moods''}

I'm in one of those old ``want to derive quantum mechanics'' moods.
Gloomy rainy days always do that sort of thing to me---they're
always so conducive to deeper thoughts.  I have it stuck in my head
that ``quantum mechanics is a law of thought'' and the silly idea
won't go away.

Today, on a more mundane level, I've been working on whether
entanglement can be used to enhance the classical information
capacity of quantum mechanical channels.  The quantum is so fun.
I've done some decent preliminary work; now all I need to do is get
to Yorktown Heights (IBM Research) and get these guys to do a little
simulation and see what happens.

Always I come back to the silly questions about what is at the heart
of the matter.  Is it ``entanglement'', i.e. that a maximal state of
knowledge about a composite system is not a maximal state of
knowledge about its parts?  Is it that nonorthogonal quantum states
cannot be identified with certainty?  Is it that information cannot
be gained about nonorthogonal alternative states of a quantum system
without disturbing it? And, what do all these questions have to do
with the idea that the structure of quantum mechanics is a ``law of
thought'' and nothing more? (The physics is in assigning
Hamiltonians and boundary conditions.)

Rainy days and love songs.  I wrote Sam a longish note on all this
``law of thought'' business today.  It's like I really feel it
today, for one reason or the other.  [See note to Sam {\Braunstein},
dated 19 July 1996.]

I noticed this morning that yet another algorithm has come out for
something that can be done in an interesting way on a quantum
computer.  I like to see that sort of thing!  Yes indeedy.

What would the world be without our friend the quantum.  I wouldn't
want to imagine really; I suspect it would be a mess \ldots\ or
perhaps even nonexistent. But who really knows.  My friend Richard
{\Jozsa} thinks that all of quantum mechanics is hogwash.  He likes
working on quantum computing because it gives a good vehicle for
fleshing out all the interesting consequences of the theory.
However, if he had to take a bet, it would be that when we actually
attempt to build an interesting quantum computing device, we'll find
that it simply won't work.  QM will play out before then \ldots\ or
so he's willing to bet.

Me, I take the opposite religion \ldots\ that quantum mechanics is
here to stay, just as probability theory is here to stay.  Who would
you call conservative, me or Richard?

\section{16 August 1996, \ ``Hot Pavement and Slushy QM''}

Well, you know, I really knew that you probably weren't going to
become too enthralled by the variational problem.  Like so many of
my old girlfriends, you were just swept away by the lust of the
moment \ldots\ forgetting how hard it is to make a relationship go.
Tsk, tsk.  At the very least, I forgive you \ldots\ just as with my
old girl friends.

More seriously, do what you can, when you can, and just make sure
you're having fun.  I've been tied up with so many things myself.
{\Bennett}, {\Smolin}, and I have made really quite a discovery.  And
we've been trying to consolidate that into something even more
powerful, maybe even a bombshell.

The question is this: suppose you have a noisy fiber optic cable and
you need to transmit one bit of information down it.  Your resources
are two and only two photons.  Can it help in this simple
transmission problem to entangle the photons into an EPR pair before
sending them along their way?  That is, can the bit (0 or 1) be
identified better at the receiver if the photons are entangled first
at the transmitter. The answer is, {\it wonderfully}, yes it does
help. Entanglement can really help protect information as it's being
transported.  We now have explicit examples.  The next question is
whether we can up the overall information carrying capacity of the
channel via entanglement.  If this is true, then we'll have a
bombshell.

By the way, someone here has lent me a copy of a new book by Chris
{\Isham}, {\sl Lectures on Quantum Theory:\ Mathematical and
Structural Foundations}.  It looks to be really very good and very
concise, saying just what needs to be said and not much more.  It's
only 220 pages and I hear pretty cheap.  I think I would recommend
it to someone, as yourself, interested in getting a tight
introduction back into the game.

With that, let me leave you for the weekend, by quoting the first
sentence of the last paragraph of {\Isham}'s book.  ``The central
issue in all this is really the phenomenon of quantum entanglement,
and its striking contrast with the reductionist concepts of Western
philosophy.''

\section{18 August 1996, \ ``Morning Thoughts''}

A gloomy Sunday morning in Montr\'eal, coffee in hand, {\TonyBennett}
on CD. I'm reading a little book called {\sl Night Thoughts of a
Classical Physicist\/} and I came across something that made me think
of you.
\bq
\noindent
Voigt had decided on a career in physics rather than in music, since
a musician had to be absolutely first rate and a physicist could get
along on less.
\eq

\section{29 August 1996, \ ``Gerry {\Niewood} on Saxophone''}

I'm courageously entering this letter without any topic of
conversation planned whatsoever.  I'm feeling very strange tonight,
watching my family fall apart because of a very poorly written
will.  I don't think I'll talk about that though.

As you've probably guessed, I do have {\Simon} and {\Garfunkel} on in
the background:  {\sl Concert in Central Park}.  And of course,
quantum theory is on my mind too.  It's always on my mind when I'm
troubled.  It's so mystical, so mysterious.  I think I take
religious comfort in it.  I worked hard at it today, exploring how
to shuffle things about in Hilbert space---you know the problem.
``Just weary to my bones.''  I wish I could get a better handle on
the problem; it eludes me and I just don't know what to expect.  I
know what I want to be true, but I just don't know what to expect.

Almost a religion.  That concept has really been taking me lately:
I am a priest, a student of the holy scriptures.

\section{01 September 1996, \ and to David {\BakerD},
\ ``The Iconostasis''}

Presently, I am reading a biography of C.~G. {\Jung}; yesterday, I
came across a little tidbit that I found quite intriguing.  Let me
repeat it:
\bq
\noindent
The core of the Pueblo religion was that the deity sun needed the
assistance of his sons, the Pueblos, who lived on the roof of the
world. Their religious practices helped the Sunfather through his
daily course. They believed that if their religion were to die out
the sun would die, too, and the world be left in eternal night.
\ldots\ And he [{\Jung}] found some solace in the thought that the
central idea of the Pueblo religion---that God requires the
cooperation of his creatures to carry out and perfect His
existence---had long been familiar to him through the writings of
medieval Christian mystics like Meister Eckhart and through his own
meditations on the mystery of the God-Father-Son relationship.
\eq

As you both know, I've long been interested in the idea of a
``participatory universe'' ({\Wheeler}'s big $U$ with an eye at the
top looking at the other side) because of quantum mechanics.
However, I had not realized that a similar idea can be found in more
established religions \ldots\ much less in some older versions of
Christianity!

Do either of you have any thoughts or pointers on this?  I presume
{\Jung} wrote something up on this; I'll look that up.  How about
Joseph Campbell? Does he have much to say on the Pueblo Indians? Who
the heck was Meister Eckhart?

\subsection{{\Greg}'s Reply}

\bq
Joseph Campbell got interested in world religions in the first place
by studying the American Indians.  So yes, he has very much to say
about American Indian religions, and I assume he talks about the
Pueblos--but it has been so long since I read Campbell I can't
remember many particulars.  [\ldots]

Campbell has this wonderful picture in one of his books: a
photograph of a relief on a wall somewhere in India, I think.  The
relief has 4 brothers at the bottom, asleep, and Vishnu, the cosmic-
dreamer whose dream is the universe, at the top and also asleep.
Campbell's interpretation is this: by construction, the brothers
exist because of Vishnu's dream.  The amazing thing is Vishnu exists
because of the dream of the 4 brothers.  The first time I saw this I
immediately thought of {\Wheeler} and his capital U.  If you like, I
can give you a reference for Campbell's book.  I'm willing to bet
that this example from the Indian culture predates anything in
Christianity.  It would be interesting to know what the Dead Sea
Scrolls say, though, since they contain the earliest written records
of Christianity.

I seem to recall other cultures where the Sun God needs help to get
up and around the Earth each day, for instance, maybe the Egyptians
or Aztecs?  But, I don't know if they considered it a central idea
that the God would die if the religion died.
\eq

\section{03 September 1996, \ ``A Procrastinating Plunk''}

It's amazing how I can always find time to procrastinate when I have
to.  I told you I was reading {\Jung}'s {\sl Synchronicity\/}
yesterday, right?  Well, I did read it finally, and I must say, I
was quite disappointed.  In all, I think I like {\Pauli}'s meager
writings on the concept much more than {\Jung}'s.  In fact, I now
think what I had been running around calling ``{\Jung}'s''
synchronicity concept is really {\Pauli}'s: whatever it is that
{\Jung} himself is talking about now appears to be much closer to
ghosts and goblins than I had imagined.  It seems only to have been
a fancy way of talking about astrology, ESP, and
psychokinesis---things he was evidently taken by.

There were a few tidbits in the book that I really liked the sound
of, but they were pretty scarce on the whole.

\section{31 May 1997, \ ``Long Day''}

This week was a pretty good week.  As predicted, I did get to meet
{\Mermin} and had nice ample time to get to talk to him about quantum
mechanics.  He's struggling hard to formulate his own interpretation
of quantum mechanics; he figures that since he's over 60 now, he has
the right to do so.  For a while he was calling the set of thoughts
the ``Ithaca Interpretation of Quantum Mechanics,'' but that ran into
a little trouble from the locals; now he's jokingly calling it the
``75 Hickory Road Interpretation.''  I'm not sure what to make of it;
it's still a little too vague for me to formulate any strong
opinions.  The main point seems to be the idea that ``if we can sweep
all the problems of interpretation under a single rug called {\it
objective probability}, then that would be progress.''  Then the
issue would be to just figure out what the hell ``objective
probability'' could possibly be.  I'm very slightly sympathetic to
this, but I now pretty strongly believe that there can never be
``objective probability,'' and would rather he focused on searching
for a notion of ``objective indeterminism.''  For me, probability
encapsulates a state of knowledge; it's a good notion of
indeterminism that seems so hard to define.  Especially one that
seems so crucially dependent upon the information processing
capabilities of those beings that work to describe the world.

Something rather strange about {\Mermin} cropped up this week.  The
guy actually thinks that philosophers of the French
literary-criticism tradition (such as {\Derrida}) are worth wading
though.  Also that they likely have something to say of use in
understanding the quantum!  Of all places, I think I was least
likely to look there!!!

\section{09 June 1997, \ ``Dictionaries and Their Problems''}

How are you my friend?  It's been so long since I've written you
anything of substance, I almost wonder if I can still remember how!
Lately, I've once again taken to reading about {\Bohr}'s (and the
other founding father's) thoughts on the epistemological and
ontological lessons of quantum mechanics.  I suppose part of my
reason for getting back to these things is just a general tiredness
of looking at equations; maybe it's a form of
procrastination---papers need writing, papers need revising, papers
need refereeing, talks need preparing \ldots\ and I'm getting a
little tired of it all.

In any case, the exercise is having its own payoff.  Maybe I'll share
a little with you.  Remember I told you that {\Mermin} suggested that
{\Derrida}'s mumblings shouldn't be written off?  I guess I'm
starting to think he was right (though I have to admit that I
haven't yet read any of {\Derrida}'s own writings, only
commentaries). It seems that the focal point of {\Derrida}'s thought
centers around none other than your ``problem of the dictionary''!
Let me try to give you something of a flavor of how these things
might be connected to the quantum.  My starting point has been an
excellent essay by John {\Honner} titled, ``Description and
Deconstruction: Niels {\Bohr} and Modern Philosophy'' (found in {\em
Niels {\Bohr} and Contemporary Philosophy}, edited by Jan {\Faye}
and Henry J. {\Folse} (Kluwer, Dordrecht, 1994), pp.\ 141--151). I
hope you enjoy the quotes:

\bq
{\Derrida} undermines the notion that words and signs can capture
present experience:  our tracing of experience always discloses a
supplement, a `difference'.  This attack is equivalent to a
subversion of the notion of strong objectivity and correspondence
theories of truth.  For the deconstructionist, the foundations for
knowledge are never securely laid: words do not correspond exactly to
the world.  ``Presence'' can never present itself  to a present
consciousness, and hence experience is always and already constituted
as a text.  [I use `text' loosely here, of course, meaning any
collection of signs---discourse, mathematical equations, pictures,
poems, prose, drama, hand-waving---used to trace and express insight
and experience.]  A text is a collection of signs and any sign
presumes a presence which it represents, but the sign is not the same
as that which it represents. In signifying our awareness of a
presence a move is made from the presence to sign.  By the word
`presence' {\Derrida} is indicating something like substance,
essence, or object, but he rejects such `totalising' categories as
these, for such terms assume more about the presence than perhaps we
are entitled to assume.  The term may `trace' the presence, but a
remainder is always left over.
\eq

\bq
Speaking and writing are, according to {\Derrida}, `linear'
activities which lock us into space and time.  ``The great
rationalisms of the seventeenth century'', as {\Derrida} describes
them, fall into the trap of objectivity and neglect the timelessness
of self-presence.  The linearity of the words limits the conditions
for the use of language: ``If words and concepts receive meaning
only in sequences of differences, one can justify one's language,
and one's choice of terms, only within a topic [an orientation in
space] and an historical strategy.''  Here we have a curious
serendipity.  Our usage of words is tied, arguably, to the
reidentifiability of particular objects, which itself implies those
bastions of classical physics, the conservation of position and
momentum and an absolute space-time framework.  And it was precisely
these bastions that {\Bohr} attacked.  As I have argued elsewhere,
{\Bohr}'s fundamental arguments entail a provocative hint at a link
between the given character of ordinary language and a
deterministic-mechanistic view of the workings of nature.  For
{\Bohr}, classical physics is the inexorable result of the use of
language based on the identification of experienced material
particulars; or, vice versa, the use of language based on
identification of experienced particulars will ultimately lead to a
sense of the persisting presence and movement of material object in
space and time, and hence to principles of conservation, causal
change, and continuous space-time frameworks.
\eq

\section{23 September 1997, \ ``Airy Nothing''}

I'm nearing the end of the flight and feeling a little philosophical.
I hope you'll let me entertain you for a while.  Lately I've been
thinking about the airy nothings of quantum mechanics again.  It's
been a long time since I've done that to any extent---it's sort of
refreshing.

Indeterminism and entanglement.  The first is an old friend, that you
know.  The second, though, every day takes my heart a little more. In
a certain way, indeterminism couldn't live without entanglement: the
EPR argument would have triumphed over indeterminism if entanglement
hadn't {\it also\/} led to a necessary violation of {\Bell}
inequalities.  I believe in the ultimate indeterminism of quantum
mechanical measurement outcomes just because of the experimental
confirmation of {\Bell} inequality violations and the experimental
confirmation of Special Relativity.  I've said this to you before
(probably three years ago), but now it's starting to weigh on my mind
more heavily.  If I want to understand quantum indeterminism, then I
must also understand entanglement: the argument goes in just that
order.

Luckily for me, I think, the field of Quantum Information is
especially suited to that purpose.  Viewing entanglement as a new
resource is the main thing on everybody's mind.  In fact, I'm
starting to feel that the situation we're in can be likened to the
beginning of thermodynamics.  What is heat, energy, work?  No one
knew at the outset; some thought them fluids, some thought them vital
forces much like the soul, and so on.  However, one thing did become
clear eventually: no informed judgment on that fundamental question
stood a chance until there existed a quantitative theory of
thermodynamics.  Without that, one could have never come across the
mechanical theory of heat and the corollary of atomism that it led
to.

So what is this thing called entanglement?  What is its use?  That
we're just starting to understand.  If I had to put it in a phrase
right now, I would say it is ``all-purpose correlation.''  Alice and
Bob come to me and say, ``Give us a little correlation, something
that we can both have and no one else can possess.  We think we're
going to need it pretty badly tomorrow.''  I say, ``Sure, no problem,
just tell me which variables you'll be needing correlated and I'll do
the trick for you.''  They say, ``Sorry, we don't know which ones
we'll need correlated yet.  A lot of that will depend upon what we
actually encounter tomorrow.''

In the classical world, Alice and Bob would have been out of luck.
But because the world is quantum, I actually can do something for
them.  I can give them a little ``all-purpose correlation.''  And it
turns out that that stuff can be really useful for several tasks. (In
fact, we're finding ever more uses all the time.)

Thus, in a certain way, I'm starting to be impressed that
``entanglement'' shares a strong analogy to ``energy.''  Both fulfill
similar roles in our engineering endeavors:  they are ``all-purpose''
essences that can be used for various beneficial tasks.  Once we
understand that in real depth, I think we'll finally put a dent in
this question of ``How come the quantum.''

\section{12 November 1997, \ ``Mad Girl's Lovesong''}

I found this little poem on the web a minute ago.  I thought you
might enjoy.

\bv
``Mad Girl's Lovesong'' by Sylvia {\Plath}:\medskip
\\
I shut my eyes and all the world drops dead \\
I lift my eyes and all is born again. \\
(I think I made you up inside my head)\medskip
\\
The stars go waltzing out in blue and red \\
And arbitrary darkness gallops in \\
I shut my eyes and all the world drops dead\medskip
\\
I dream you bewitched me into bed\\
And sung me, moon-struck, kissed me quite insane \\
(I think I made you up inside my head)\medskip
\\
God topples from the sky, hell's fires fade \\
Exit seraphim and Satan's men \\
I shut my eyes and all the world drops dead\medskip
\\
I fancied you'd return the ay you said, \\
But I grow old and I forget your name \\
(I think I made you up inside my head)\medskip
\\
I should have loved a thunderhead instead \\
At least when spring comes they roar back again \\
I shut my eyes and all the world drops dead \\
(I think I made you up inside my head)
\ev

\section{19 November 1997, \ ``Poesy of the Quantum''}
\medskip

\begin{flushleft}
\parbox{2.9in}{
``Many are poets but without the name,\\ For what is poesy but to
create''\\
\hspace*{\fill} --- {\it Lord Byron}}
\end{flushleft}
If so, then we all be poets in this quantum world!

\section{25 November 1997, \ ``{\Flatt} and {\Scruggs}''}

By the way, thanks for the words to the Martha White Theme!  I don't
think I knew of the song's existence before that \ldots\ or at least
I don't remember it.  I was referring to an old Tennessee Ernie Ford
television commercial for the flour.

Because of this little gem that you gave me, I now make the following
citation in my research proposal:
\bq
\noindent
L.~{\Flatt} and E.~{\Scruggs}, ``The Martha White Theme,'' on {\sl
{\Flatt} \& {\Scruggs}: 1948--1959\/} (Bear Family Records, 1994).
\eq
What do you think those Princeton boys will make of this when they
see it?

\section{17 December 1997, \ ``It's a Wonderful Life''}

Good holidays to you.  This morning, as I was driving to work, it
dawned on me that roughly this day 10 years ago, I was conferred my
degrees at the University of Texas.  Time does fly.

It made me think of a little anecdote about John {\Wheeler} that I
heard from John {\Preskill} a few days ago.  In 1972 he had
{\Wheeler} for his freshman classical mechanics course at
Princeton.  One day {\Wheeler} had each student write all the
equations of physics s/he knew on a single sheet of paper.  He
gathered the papers up and placed them all side-by-side on the stage
at the front of the classroom.  Finally, he looked out at the
students and said, ``These pages likely contain all the fundamental
equations we know of physics.  They encapsulate all that's known of
the world.''  Then he looked at the papers and said, ``Now fly!''
Nothing happened.  He looked out at the audience, then at the
papers, raised his hands high, and commanded, ``Fly!'' Everyone was
silent, thinking this guy had gone off his rocker. {\Wheeler} said,
``You see, these equations can't fly.  But our universe flies.
We're still missing the single, simple ingredient that makes it all
fly.''

Merry Christmas.

\section{04 January 1998, \ ``Lazy Weekend''}

There's so much that needs doing, but still I feel compelled to a
lazy weekend.  I'm sitting here listening to the rain drizzle down
outside, listening to the Sunday blues radio show, and thinking of
you.  My two dogs are at my feet, contemplating the meaning of life
in their own small way.  {\Kiki} is preparing for her first day back
at school tomorrow: 28 kindergarteners freshly back from a half-month
of play \ldots\ ouch.

Lately I've been thinking about the program of Law Without Law again.
Perhaps it's just a fancy form of procrastination.  I keep dreaming
of the day when all this will become immaculately clear, and we will
have the start of a new physics.  I guess I've been saying this for
eight years, but it seems that it really must be just around the
corner:  I think we're almost at a point where the possibilities in
our world will open up like a blooming flower.

Ever more I am compelled to believe that the ontology of {\Wheeler}'s
``game of twenty questions, surprise version'' is not only a central
lesson of quantum theory, but actually the singular principle upon
which the detailed structure of the theory is built.  The ``fact''
that {\it my\/} information-gathering yields a disturbance to {\it
your\/} predictions is the only ``physical'' (or ontological)
statement that the theory makes; all the rest of the structure is
``law of thought'' subject to that consideration.  To put it another
way, quantum theory is a theory of ``what we have the right to say''
in a world where the observer cannot be detached from what he
observes.  It is that and nothing more.

The central issue then becomes: what are the further implications of
this ``lack of detachedness'' for observers?  Now that we know that
it is actually the essence of quantum phenomena, what can we do with
it?  Quantum cryptography is a nice applied example of that line of
thought.  But there's got to be so much more.  It seems to me that
we're almost poised in the same way that {\Einstein} was when he
finally formulated the physical/ontological observation that ``maybe
it's not coincidental or accidental that gravitational and inertial
mass are numerically the same.''  It was then just a question of
counting the time until something wonderful came out of its asking.

Anyway, thanks for lending me your ear this afternoon.  Did I send
you the notes I wrote on David {\Mermin}'s Ithaca Interpretation of
QM? I don't think I did \ldots\ or at least I can't find a record of
it. In case I didn't---and, in case you're interested(!)---I'll
forward them on to you following this note.  I say the same things
there that I did above, but perhaps in slightly more detail.  (See
especially the stuff from pages [$X$] through [$X+3$]---all the
stuff following Merminition \ref{OldAsm10}.)

\subsection{{\Greg}'s Reply}

\bq
I am amazed to receive again remarks about {\Wheeler}'s game of
twenty questions, for just last night I was exposed to something of
particular importance.  I was watching an interview of Paul
{\McCartney} by David {\Frost}.  Of course it was fascinating.  But
\ldots\ the most fascinating was this: {\Frost} was asking Paul about
the source of some of the Beatles most famous work.  Paul mentioned
that his most played song, Yesterday, came to him in a dream; almost
no effort on his part.  The killer was this: he likened the creation
of the Sgt.\ Pepper album to a Mike Lee---do you know this
guy?---play, where one does not start with a script!  You start off
by asking the actors (I suppose) ``\ldots\ are you a dentist, or a
Bob? And then you work it up from there.''  (The quote is from Paul
last night.) Sounds like twenty questions to me.  The actors randomly
decide who they are, and through self-consistency an undeniable plot
emerges. I think I was most excited by this because it was a Beatle,
and that now a Beatle was professing what may be a fundamental fact
about physical law, something the physics profession itself is
unwilling to admit.  Out-laws forever! Out-side-the-laws, through
their own designs, leading to laws.
\eq

\section{19 January 1998, \ ``Happy Quotes''}

I forgot, I was going to send on some quotations that I found on the
net the other day (on some homepage concerned with the existence of
God or something).  Anyway their pretty nice; I append them below.  I
especially like the last one by {\Heisenberg}.  It gets at exactly
what I've been trying to get at with my motto, ``{\Bohr} was a
Bayesian.''

Also, let me add to that one I heard around here the other day:
\bq
\noindent
``The measurement problem refers to a set of people.''
--- {\it {\Hideo} {\Mabuchi}}
\eq

\section{26 January 1998, \ ``Lunch Break''}

By the way, it dawned on me about mid-morning that my remark about
time was prompted by something I heard in the X-Files last night. A
time traveler said something like, ``Can you imagine the horror of a
world with no past and no future?''

General relativity and its four-dimensional manifold?  I've never
been able to shake the feeling that part of the story is still
missing.  Like {\Wheeler} said, ``The equations of physics that we
have can't fly \ldots\ but still our world flies!''  David {\Mermin}
dug up something interesting recently in Rudolf Carnap's
autobiography:
\bq
\noindent
Once {\Einstein} said that the problem of the Now worried him
seriously. He explained that the experience of the Now means
something special for man, something essentially different from the
past and the future, but that this important difference does not and
cannot occur within physics. That this experience cannot be grasped
by science seemed to him a matter of painful but inevitable
resignation.
\eq

I'll be willing to bet that {\Einstein} was wrong on this count,
just as he was with quantum mechanics.  (And that perhaps the two
wrongs weren't/aren't unconnected.)  It is not that the difference
{\it cannot\/} occur within physics, but rather that our {\it
present\/} physical description is not capable of doing it.

\section{07 February 1998, \ ``Getting Nerves''}

Thanks for your ``test-taking'' message of the other day.  It may not
be of much comfort to you now, but your pain is a source of thought
about {\it what is\/} and {\it what is not}, and, to that end, serves
us both.  You've caused me to think a little more than I would have
otherwise this week.

In case it's of use to you, I'll attach below a note I wrote my
friend {\Herb} ({\Bernstein}) alerting him to an article by Sylvan
{\Schweber}.  Take a look at it if you get a chance (and can find
it). You're not the only one trying to reconcile the stuff on the
inside with the stuff on the outside.

I wish you could have been at my colloquium though; I think it was
one of my best performances yet.  (And I think this is my first
physics colloquium \ldots\ unless the talk I gave at your school was
a colloquium.) They say I had an attendance of over 70 people; I just
know that the lecture hall was pretty full.  I tried very hard to
convey the feeling that physics is changing \ldots\ not because the
equations are changing, but because we are finally starting to feel
at home in the quantum world.  We are starting to understand that
quantum mechanics is a gift from God, not a devil in the machine.

\section{03 March 1998, \ ``Japanese Kitchen''}

I'm in Tokyo right now.  I just thought I'd drop in on you for a
moment to say hello.  Last night at dinner I had the nice opportunity
of spreading the word of our old friend John {\Wheeler} to our
Japanese colleagues.  I told them the story of the Game of 20 Qs
(surprise version).  This came up because Prof.\ {\Hirota} told me
that when asked why quantum communication devices can perform better
than classical ones, he answers that, ``It is because the quantum
world is `noncausal'.''  When I pressed him further to find out what
exactly he meant by the word ``noncausal'', it turned out that he was
essentially talking about 20 Questions (surprise version).  So that
was very pleasing.

{\Hirota}, by the way, is a great fan of {\Giacometti}.  He has his
work plastered all around the laboratory.

\section{31 March 1998, \ ``What a Lunchtime {\Bohr}''}

Remember I once wrote you the following lines:
\bq
\noindent
    There you have it in any case.  Concerning the ``free choice''
    in quantum measurement:  {\Dirac} voted ``nature,'' {\Heisenberg}
    voted ``man,'' {\Bohr} voted ``neither,'' and {\Wheeler} voted ``both.''
\eq
with the point being that I don't believe that history bears too much
hope that {\Bohr} really would have fallen into agreement with
{\Wheeler}'s game-of-twenty-questions ontology.

Anyway, I just ran across another quote that has me thinking along
the same lines again.  Let me record it:
\bq
\noindent
The circumstance that, in general, one and the same experimental
arrangement may yield different recordings is sometimes picturesquely
described as a `choice of nature' between such possibilities.
Needless to say, such a phrase implies no allusion to a
personification of nature, but simply points to the impossibility of
ascertaining on accustomed lines directives for the course of a
closed indivisible phenomenon.  Here, logical approach cannot go
beyond the deduction of relative probabilities for the appearance of
the individual phenomena under given conditions.
\eq
It is so difficult to figure out what the fellow is really saying.

\section{16 April 1998, \ ``Just the Way It Had To Be''}

Well I'm very pleased that you started off your last note with an
allusion to our wonderful quantum theory, but honestly there was no
potentiality here.  In this case, it has been clear for years that
the world was going to turn out just the way it did.  Your getting
tenure was an example of the classical world at its very best: if
the initial conditions are fixed and pure, the trajectory has no
choice but to be exactly what it is.  Your initial conditions made
themselves known that very first year we were together in North
Carolina.  It's quite hard for a seeker of the truth to hide in
anonymity.  Congratulations to you and your family!  Life now will
certainly be a little more secure for all of you.

But now since you started it, let me do pick up on your allusion.
What does the life of the academic hold?  I'm not so sure in all
detail---especially as I am not one myself---but a certain general
outline is clear.  It comes from something that the scientific world
could not properly take serious until the quantum.  L\'eon
{\Rosenfeld}, I think, said it best, ``We are not merely
contemplating the world, but acting upon it and thereby modifying
its course.''

\section{05 May 1998, \ ``Expected''}

Thank you for thinking of me when reading about Ed {\Jaynes}'s
demise. Sadly though, I had already expected it and come to grips
with the sadness.  I found out last week from Larry {\Bretthorst} at
Washington U. that Ed had just been removed from dialysis and wasn't
expected to live for more than a few days.  He was a truly great
man.  The most unfortunate thing for mankind is that he never got
the chance to finish his already massive book on ``probability
theory as extended logic.''  We, the children of this beautiful
idea, will have to forge ahead without our beacon.

\section{27 May 1998, \ ``My Head, My Head''}

Yikes my head is hurting right now.  But that's not why I write you.

I enjoyed your musings about strong gravitational fields and matter.
I wish I had something interesting to say back to you on that
account, but so far I'm empty.  I have had some musings lately on
matter, but I guess there's not much of a connection.  I keep toying
with the notion of matter---more specifically, energy---as a
``resource.''  That is to say, if I think of the resources of
classical (non-general-relativistic) physics required to get some
task done, then they all seem to boil down, in one form or other, to
one single, simple ``substance.''  Namely energy.  Be it elbow
grease, petroleum jelly, or electromagnetic jelly, they all boil
down to some manifestation of energy-expenditure.  Now, here's where
things get exciting.  With our new understanding of quantum
information theory, we have come to realize that ``entanglement'' is
a physical resource in its own right. One can use it for
teleportation, to reduce the communication complexity of some games,
for quantum cryptography, to protect signals in noise, etc., etc.,
etc.  This I think is a resource {\it sui generis}. I.e., it is not
reducible to energy.  And in this way, I think we have something
here truly distinct from classical physics. But on the other hand,
it does appear to be intimately tied to energy through {\Landauer}'s
erasure cost.  Namely, if I want to turn some non-concentrated
entanglement into a concentrated form (like pure EPR pairs), then I
generally have to perform some quantum measurements, and thus gather
some random noise from the quantum nothingness.  If I try to think
of this entanglement concentrator as a finite, but closed system, I
am stuck.  This is because its memory will eventually get full with
all the bits generated from the measurements.  Thus it ultimately
has to erase, and that links our erasure process back to
thermodynamics.  Thus entanglement is distinct from energy (unless
you tell me it couples to the gravitational field!!!), but still it
appears that one cannot get away from the use of energy to make it
useful in the first place.

But, as I say, my head hurts and this is only making it worse.

\section{18 July 1998, \ ``Short Note''}

It's Saturday morning and I'm sitting in the little village of
Benasque.  My second week here has just passed away and I'm starting
my third.  I've spent a lot of time collaborating with {\Carl}
{\Caves} trying to push our point of view about quantum mechanics to
its logical conclusion.  We are trying to derive that Hilbert space
has to be a *complex* vector space (rather than a real one) simply
from the idea that quantum states are states of knowledge and nothing
more.  So far, there is a piece that's working and a piece that's
not.  Keep your fingers crossed.

Time translation symmetry.  It's always struck me that beyond all
else, there is one fundamental fact of all our experience and that is
that time flows.  No law of physics captures this.  (I say this fully
aware of all the tracts you can find on the ``arrow of time'' in any
library.)  Physics is missing something, and something very
fundamental.  So I'm glad you're thinking about these things.  Keep
me filled in on all your thoughts.  In loose connection to this, I'll
forward you a note I wrote Rolf {\Landauer} a couple of weeks ago.  I
doubt that ``information is conserved'' in any useful sense:  every
time one reaches into the quantum well, a little information is
created.

\section{28 July 1998, \ ``Macaroni''}

The last time I was in Europe (last month), I met up with {\Charlie}
{\Bennett} as usual.  I don't know how it happened, but we got into a
conversation about coin tossing (real coin tossing, not the
philosophical issue).  {\Charlie} said, ``You know, there's a really
easy method for predicting someone else's tosses; I'll show you.''  I
said, ``Oh bull.''  Then he pulled a quarter from his pocket and gave
it to me.  I flipped; he called heads.  It was heads.  So I did it
again.  He called heads.  It was heads.  I said, ``Oh you just got
lucky again.'' ``Well, do it again,'' he says.  So I flip.  He calls
heads.  It was heads!!  In astonishment, I say, ``OK what's the trick
you have up your sleeve?  Did you give me a fixed coin?!?''  He
bursts out laughing, ``No, not at all.  One time in eight you get
lucky, and when you do, it's a really good joke!''

But still, synchronicity happens.  The evidence is our two stories.
I'll never forget mine; maybe you'll never forget yours.

\section{07 September 1998, \ ``The Quotable {\Pauli}''}

Well I'm a little angry because I can't get connected to the web this
morning and I have no intention of going in to the office on Labor
Day (I'm fooling myself into believing that I deserve a day of rest).
So who knows when you'll finally get this note.  But here it is
\ldots\ exactly what you've been waiting for:  The Quotable
{\Pauli}.

I include the first two quotes for the sake of definition, i.e., for
his meaning of ``detached observer.''  But the last quote, now the
last quote, that's the real reason I'm sending this note.  Three
cheers for {\Pauli}!  (By the way have I ever told you about my
secret desire to get a parrot?  I'd name it {\Pauli} and teach it to
say that it wants a cracker.)

\bq
\ldots\ [I]t seems to me quite appropriate to call the conceptual
description of nature in classical physics, which {\Einstein} so
emphatically wishes to retain, ``the ideal of the detached
observer.'' To put it drastically the observer has according to this
ideal to disappear entirely in a discreet manner as hidden spectator,
never as actor, nature being left alone in a predetermined course of
events, independent of the way in which the phenomena are observed.\\
\hspace*{\fill} --- Wolfgang {\Pauli}, in a letter to {\Bohr} (15
February 1955)
\eq
\bq
In the new pattern of thought [quantum mechanics] we do not assume
any longer the {\it detached observer} \ldots\ but an observer who by
his indeterminable effects creates a new situation, theoretically
described as a new state of the observed system.  In this way every
observation is a singling out of a particular factual result, here
and now, from the theoretical possibilities, thereby making obvious
the discontinuous aspect of the physical phenomena.\\
\hspace*{\fill} --- Wolfgang {\Pauli}, in ``Matter'' (1954)
\eq
\bq
Indeed I myself even conjecture that the observer in present-day
physics is still too completely detached, and that physics will
depart still further from the classical example.\\
\hspace*{\fill} --- Wolfgang {\Pauli}, in ``Phenomenon and Physical
Reality'' (1957)
\eq

\section{09 October 1998, \ ``Billiards and the Eight''}

Speaking of things common to us, I've started sitting in on Kip
Thorne's GR class.  I've only attended two lectures so I haven't
started to learn much of substance yet.  The thing that really
strikes me is the way these lectures have taught me something about
myself.  Before these lectures, I hadn't appreciated how far my
world-view has come to diverge from the one of classical physics.
Everything Kip says has such a foreign feel to me.  He speaks of
particles, fields, and world lines; he talks about what ``things''
are ``doing.''  Strange, really strange, how classical physics has
come to be so far from my heart.  Physics has become such a
``{\Wheeler}ian game of twenty questions'' for me that I guess I've
forgotten the old point of view.

That old point of view \ldots\ that the world was made of little
billiard balls.  At best, it's made of Magic Eight Balls.

\section{17 November 1998, \ ``A Break in Pace''}

And to comment on another thing in that same letter:  boy, how I wish
I could have been at that symposium on {\Jaynes}!  I am now Bayesian
through and through.  But, concerning quantum mechanics, I am a
Bayesian of a quite different flavor than usual.  Most thorough
Bayesians just don't believe quantum mechanics---{\Jaynes} didn't
believe quantum mechanics.  Witness this quote by A.~J.~M. {\Garrett}
(that I'm stealing from our paper):
\bq
The nondeterministic character of quantum measurement can, and
should, be taken to imply a deeper ``hidden variable'' description of
a system, which reproduces quantum theory when the unknown values of
the variables are marginalised over.  Differences in measurements on
identically prepared systems then represent differences in the hidden
variables of the systems.  Not to seek the hidden variables, as the
Copenhagen interpretation of quantum mechanics arbitrarily instructs,
is to give up all hope of improvement in advance, and is contrary to
the purposes of science.
\eq
This wave of thought is what {\Carl}, {\Ruediger} and I are trying to
set right in our paper ``Bayesian Probability and Quantum
Mechanics.''

\section{17 November 1998, \ ``Heresy!''}

{\Kimble} is really pounding on me to get that {\sl Physics Today\/}
proposal out, but I just can't let this confusion go!  I will not
have my name shamed!

\bgc
I am a bit confused about the passage on quantum mechanics.
\egc
Probably because I don't write nearly as clearly as I would like to
think that I do!

\bgc
Do I understand correctly that you now believe in some type of hidden
variables?
\egc
Yeaks!!!  Never, never give that thought another ounce of attention.
I couldn't, I wouldn't, I would never think that.  All the evidence
is in; it's only been hard for some to cope with the verdict!

\bgc
Or, do you believe that we still need to dig deeper into quantum
mechanics, contrary to what the Copenhagen teachings would have us
do?
\egc
Oh we certainly need to dig deeper, always.  But my understanding of
Copenhagen (in any of the 30 versions) has never been that they
commanded us to stop thinking.  {\Bohr} very much attempted to
express a line of thought that demonstrated---in a certain
sense---that quantum mechanics is the end of the line.  He just
wasn't so good at it.  He didn't think that it was an arbitrary
edict that the end of the line was here, as {\Garrett} would have us
think.

\bgc
Which wave of thought is it that should be expanded on?
\egc
I said ``set right'' meaning that it should be corrected, not
``expanded upon''!  This is where your confusion is coming from.

By saying that I am Bayesian through and through, I mean that I
understand that probability quantifies a state of knowledge:  It is
{\it never\/} a physical property outside of someone's head.  This
statement even applies to the probabilities generated by quantum
mechanics.  Now, most Bayesians---myself {\it not\/} included---also
take it that whenever one's state of knowledge is not maximal (i.e.,
one assigns non-0-1 probabilities to something), then that knowledge
can be improved upon. Moreover, one ought always look to improve it.
That is what {\Garrett} is trying to state clearly.  But, from my
point of view, this last statement is not compelled at all by the
Bayesian doctrine.  It is simply a throwback from classical thought,
where one always has it that nonmaximal information can be made
complete (with sufficient effort). In the quantum world this is just
not the case: maximal information is not complete and cannot be
completed!

\section{05 April 1999, \ ``Pasadena Visit''}

{\Charlie}, {\Herb} {\Bernstein}, and I had the most wonderful
conversation last night about the foundations of quantum mechanics.
{\Herb} and I are much alike, thinking that reality (in part) is
created by quantum questions and quantum answers.  {\Charlie},
though, is a strict many-worlder.  There were some good sparks a
flyin'.  Most interesting though was the deep similarity in some of
the things we were all saying.  For the first time, it sort of made
me believe that the only question is one of language.  In the near
future I hope to construct a little essay about this.

A little sad news came out of the discussion.  I mentioned that I
had wanted to write Rolf {\Landauer} again about his slogan
``Information is Physical'' to tell him about the parts I like and
the parts I don't, and to find out what his response is. {\Charlie}
said it wouldn't do any good:  Rolf is on his deathbed.  They've
found cancer in his brain, removed part of it, and given him only a
month or so to live.  I missed my chance with {\Jaynes}; now I've
missed my chance with Rolf.  The finiteness of life.

\section{11 April 1999, \ ``One I Like''}

\bq
\noindent
Is there no way out of the mind?

--- Sylvia {\Plath} (1932-63), U.S. poet, {\sl Apprehensions}.
\eq

\section{22 April 1999, \ ``Fuchsian Genesis''}

\bq
\noindent
In the beginning God created the heaven and the earth.  And the earth
was without form, and void; and darkness was upon the face of the
deep. And the Spirit of God moved upon the face of the waters. And
God said, Let there be light: and there was light.  And God saw the
light, that it was good; and God divided the light from the darkness.
And God called the light Day and the darkness he called Night.  And
the evening and the morning were the first day. \ldots\ [Day 2],
[Day 3], [Day 4], [Day 5] \ldots\ And God saw everything that he had
made, and behold, it was very good. And there was evening and there
was morning, a sixth day.  Thus the heavens and the earth were
finished, and all the host of them.
\eq

But in all the host of them, there was no science.  The scientific
world could not help but STILL be without form, and void.  For
science is a creation of man, a project not yet finished (and perhaps
never finishable)---it is the expression of man's attempt to be less
surprised by this God-given world with each succeeding day.

So, upon creation, the society of man set out to discover and form
physical laws.  Eventually an undeniable fact came to light:
information gathering about the world is not without a cost.  Our
experimentation on the world is not without consequence.  When {\it
I\/} learn something about an object, {\it you\/} are forced to
revise (toward the direction of more ignorance) what you could have
said of it.  It is a world so ``sensitive to the touch'' that---with
that knowledge---one might have been tempted to turn the tables, to
suspect a priori that there could be no science at all.  Yet
undeniably, distilled from the process of our comparing our notes
with those of the larger community---each expressing a give and take
of someone's information gain and someone else's consequent loss---we
have been able to construct a scientific theory of much that we see.
The world is volatile to our information gathering, but not so
volatile that we have not been able to construct a successful theory
of it.  How else could we, ``Be fruitful, and multiply, and replenish
the earth, and subdue it?''  The most basic, low-level piece of that
understanding is quantum theory.

The {\sl speculation\/} is that quantum theory is the unique
expression of this happy circumstance:  it is the best we can say in
a world where {\it my\/} information gathering and {\it your\/}
information loss go hand in hand.  It is an expression of the ``laws
of thought'' best molded to our lot in life.  What we cannot do
anymore is suppose a physical theory that is a direct reflection of
the mechanism underneath it all:  that mechanism is hidden to the
point of our not even being able to speculate about it (in a
scientific way).  We must instead find comfort in a physical theory
that gives us the means for describing what we can {\it know\/} and
how that {\it knowledge\/} can change (quantum states and unitary
evolution).  The task of physics has changed from aspiring to be a
static portrait of ``what is'' to being ``the ability to win a
bet.''\footnote{The nice phrase ``physics is the ability to win a
bet'' is due to J.~R. {\Buck} (a grad student at Caltech) circa 19
February 1999.}

This speculation defines the large part of my present research
program.

Why do I say all this?  Because I wanted to say it to myself and use
you as the other side of my brain.  Today I'm going to lunch with
Roger {\Penrose} and there is no doubt that the topic of the
interpretation of quantum mechanics will come up.  How best to
express what needs expression?  That's the question on my mind this
morning:  this is iteration \#1.

\section{27 June 1999, \ ``Raining Down in Cambridge''}

\bigskip

\begin{flushright}
\baselineskip=12pt
\parbox{2.5in}{\baselineskip=12pt
``Anyone who considers arithmetical methods of producing random
digits is, of course, in a state of sin.''}\medskip\\
 --- {\it John von Neumann}
\end{flushright}
\bigskip

As you can see, I'm in Cambridge now.  In fact, I'm looking out my
window to all the rain that's coming down---what else would you
expect?  Man, I love it here!  I haven't felt so productive in
ages.  {\Carl} {\Caves} and I are on the tail of a really good
theorem (useful to interpreting quantum mechanics in a Bayesian way)
and the old English pubs just can't be beat.  The Isaac {\Newton}
Institute too is just wonderful:  I have never been to a place more
conducive to thinking.  Any place that actually puts chalk boards in
the john so that you can continue your discussions in there if need
be, well \ldots\ is just alright in my books!!!  Ben {\Schumacher}
and I have been working on some quantum coding theorem stuff.  And
Vladimir {\Buzek} and I have been able to some extent to put into
equations some of the ideas in the ``genesis'' note I sent you.  (He
calls that note my ``manifesto.'')  All these things are very
exciting for me.

You asked a question in your last note.  Well, I really like that
quote of von Neumann.  It's true.

\bgc
How is it that the mathematical formalism of quantum
mechanics---which I take here to be simply the solving of a partial
differential equation---can model for me random results?
\egc
The continuous time-evolution part of the theory concerns how our
state of knowledge changes when we're not shoving a system about to
see how it will react (i.e., we're not asking it a question, we're
not making a measurement on it).

\bgc
Does randomness enter when we take the output of the partial
differential equation and interpret that output as a probability?
\egc

Yep, that's the whole story.  The outcomes of our measurements are
governed by no laws (much less differentiable ones).  The probability
is there because there is ignorance or impredictability \ldots\
sometimes more so, sometimes less so.  But there's always some.  The
randomness in quantum mechanics is that no matter how much
information we have about how a system will react to our prods, it is
never enough. Maximal information is not complete \ldots\ that's the
slogan of the Albuquerque interpretation.

You, by the way, once asked about whether the information-disturbance
point of view about quantum mechanics would nullify the many-worlds
point of view.  I never answered you.  Sorry about that.
Unfortunately I don't think it will.  But part of that is because the
many-worlds point of view is much less well-defined than it's
proponents would have us believe.  The real question is which point
of view about quantum mechanics is the most ripe for leading to new
things.  Which one will carry us through to a greater appreciation of
the possibilities in our world?  The many-worlds point of view is a
dead end in that regard. Understanding that the world is, at its
core, malleable to our hands?  I think that that will ultimately be
the greatest gift this century has to give to the next.

\section{28 August 1999, \ ``No Podunks Here''}

\bgc
It's amazing at times the synchonization that occurs.  Just last
week, Bill {\Thacker} was looking around for an explanation for why
one needs complex vector spaces.  He found one book that said it was
because one wants wavelike behavior, but linearity in the
Hamiltonian, which therefore requires an i in the {\Schroedinger}
equation (so that if one ``squares'' the {\Schroedinger} equation,
then the appropriate sign will come out in front of the time
derivatives). It will be very interesting to see what you and
{\Caves} have to say.
\egc

Interesting synchronicity with Mr.\ {\Thacker}.  Below I'll attach
some references I sent to {\Carl} for the use of his funding agent;
Bill will also be able to make good use of them.  The discussion in
Steven {\Adler}'s book is useful and good, but I'm not too impressed
by the arguments there.  The one he likes the most is this.  If you
suppose that time evolutions must be unitary, then the generators of
the unitary transformations (the ``Hamiltonians'') cannot be
observables in real-Hilbert-space QM.  But I say, ``So what?  Why
need they be observable?''  I don't think {\Wheeler}'s favorite
argument (actually it's one of {\Wootters}'s, but never published by
he himself) is so good either; though I fluctuate on that.  The
argument that {\Caves}, {\Schack} and I have is much more akin to
the {\Wootters}90a argument listed below.  At least the two
arguments have the same mathematical source. The basic effect comes
about from this:  for a Hilbert space of dimension $d$, the
dimensionality of the vector space of Hermitian operators over the
base space is $d^2$ for complex Hilbert spaces, but only $d(d+1)/2$
for real ones.  This means that the two theories have quite
different behaviors as concerns the conglomeration or concatenation
of separate systems.  In the first case, the operator space of the
tensor product is isomorphic to the tensor product of the original
operator spaces.  In the second case, the operator space of the
tensor product is ``larger'' than the tensor product of the operator
spaces.  Now we are able to give an information theoretic reason why
one should want the former over the latter:  it's actually required
if one takes it seriously that quantum states are states of
knowledge and NOT states of nature.

The way I see it is this.  It seems to me that the only reasonable
understanding of the wave function is that it is information and
nothing else.  It is a symbol that stands for what one knows, the
most one can know \ldots\ and makes no commitment to ontology
outside of the information-disturbance relations that one can derive
from the mathematical structure that is built around it.  But one
cannot stop there, as our Copenhagen forefathers did.  (That's why
there's so damned much confusion in the field.)  Instead, we in the
quantum information field are obliged to give an information
theoretic reason for each and every axiom of quantum mechanics. When
we find an axiom or theorem for which we can't do that---my personal
favorite for a candidate is information-disturbance---then we will
have identified the crucial ontology in the theory.  That's my
take.  And I think this new theorem (we call it the quantum de
{\Finetti} theorem) is a small step in that direction.

\section{25 November 1999, \ ``Thanks and Giving''}

I hope you and your family have a wonderful Thanksgiving today.

My latest problem is that I've got it in my head that if I'm going
to make any real contribution to physics, it will be vicariously,
through guidance to some good graduate students.  I have a big
research program planned out now, one that I know I can't pull off
by myself \ldots\ and one that I know has enough open ends that lots
of outside inspiration will be needed (the inspiration of some hard
thinking students).  This firmed up in me during my teaching at the
summer school in Italy this summer.  I started to collect this set
of groupies there, and I have to tell you it did wonders for the
ego.  I started to think (for the first time), ``You know, I could
do this.  I have a role in this community.''

But let me go back to this dream before taking off.  You know
{\Asher} and I are writing this thingy for {\sl Physics Today}.
Well, we were having a bit of a fight about the meaning of the
wavefunction.  He likes to say that wavefunction is just a symbol
that stands for a system's preparation.  I on the other hand like to
say that it stands for the information we have about how a system
will react to measurement.  So I got a little smart with him.  In
one passage he had written that if we know a quantum state then we
can prepare as many copies of that state as we wish.  I retaliated
by saying the following:
\bq
\noindent
Just for one last emphasis: suppose a pure state were a pretty good
approximation to what we could say about all the observables
associated with the radius of the universe. How would we produce
arbitrarily large numbers of that state?
\eq

Ultimately I won, but one thing he said in retaliation to my
retaliation caught my eye.  I love it!  I leave you with a quote of
it for wishes of a happy Thanksgiving.
\bq
\noindent
{\bf ANSWER:}\\
It's just a matter of money.  As {\Archimedes} answered to the king:
give me firm support, and I shall move the Earth.
\eq
The lesson of the quantum is that the world can be moved.

\section{17 December 1999, \ ``Trinity''}

\bgc
Forgive my ignorance, but isn't there some famous nuclear explosion
site with ``Trinity'' in the name?
\egc

Yes, the Trinity test was the first test of an atomic, July 1945.
July 15, I think.  The bomb was a plutonium bomb.  They tested a
plutonium bomb, because by that time there was overwhelming
confidence that the uranium bomb would work.  Trinity is out in the
White Sands missile base---I've been there once (they only open it
to the public two days per year).  At my house, because of the old
fellow that I lived with one summer Jack Aeby, I have a shard of
glass that was formed from the sand below that explosion.  It is a
mineral that was unknown to the world before the explosion; it has
an official name:  trinitite.  When you touch it, you really do find
yourself thinking, ``I am become death, the destroyer of worlds.''

\section{23 January 2000, \ ``Darling Quotes''}

Let me send you some quotes from David Darling's book.  You'll see
the connection to what you wrote me in your holiday letter.

From D.~Darling, {\sl Equations of Eternity:  Speculations on
Consciousness, Meaning, and the Mathematical Rules that Orchestrate
the Cosmos}, (Hyperion, New York, 1993):
\bq
\indent
The interface between mathematics and everyday reality appears sharp
and immediate at this point: one sheep, one finger, one token;
another sheep, another finger, another token, and you can take away
tokens or add them, as you can with your fingers. The tokens---the
numbers---are just abstracted fingers; the operations for dealing
with the tokens are just the abstracted raising or lowering of the
fingers.  You make a one-to-one correspondence between the tokens and
whatever it is you want to reckon, and then forget about the fingers.

At first, it seems clear from this that mathematics must be somehow
already ``out there,'' waiting to be discovered, like the grain of
the stone.  One sheep add one sheep makes two sheep.  Two sheep add
two sheep makes four sheep.  That is certainly the practical end of
the matter as far as the shepherd and the merchant are concerned. But
already, even in this most simple mathematical maneuver, something
strange has happened.  In saying ``one sheep add one sheep'' we seem
to be implying that any two sheep will always be identical.  But that
is never the case.  Physically, the first sheep is never exactly
equal to the second: it may be a different size, have different
markings.  It takes only one molecule to be out of place between the
two, and they are not identical.  Indeed, because they are in
different places they are inevitably not the same on that basis
alone.  We have extracted a perceived quality to do with the
sheep---namely, their ``oneness,'' their apartness---and then merged
this quality by means of another abstraction---the process of
addition. What does it mean, physically, to ``add'' things?  To put
them together?  But then what is ``putting together'' two sheep?
Placing side by side, in the same field---what?

All this may seem like nit-picking. But on the contrary, it brings us
back to the central mystery---the relationship between the inner and
the outer, the world of the rational mind and the world ``out
there.'' In the physical world, no two sheep are alike. But, more
fundamentally, {\it there are no ``sheep.''}  There are only some
signals reaching the senses, which the left brain combines and then
projects as the illusion of a solid, relatively permanent thing we
call a sheep.

Like all objects, sheep are fictions: chimeras of the mind. It is our
left hemispheres, having through natural selection evolved this skill
for extracting survival-related pieces of the pattern, that trick us
into seeing sheep, trees, human beings, and all the rest of our
neatly compartmentalized world.  We seek out stability with our
reasoning consciousness, and ignore flux.  We shut our eyes to the
continuous succession of events if those events seem not to
substantially affect the integrity of what we see.  So, through this
classifying and simplifying approach we make sections through the
stream of change, and we call these sections ``things.''  And yet a
sheep is not a sheep.  It is a temporary aggregation of subatomic
particles in constant motion---particles which were once scattered
across an interstellar cloud, and each of which remains within the
process that is the sheep for only a brief period of time. That is
the actual, irrefutable case
\eq
and
\bq
\indent
When you think, the cosmos thinks---not in some nebulous, poetical
sense, but literally. Your brain is a product and a process of the
cosmos. When you look out and form pictures in your mind and try to
make sense of those pictures, you are the universe trying to make
sense of itself. Forget that we sometimes imagine ourselves to be
tiny bipeds on a ball of rock somewhere in the awful depths of
space-time doing cosmically trivial things such as changing a diaper,
or mowing a patch of grass, or brewing the next cup of coffee. The
incredible and undeniable fact is that we are the thinking, reasoning
components of the universe---a realization that makes it slightly
less astonishing that our brains might have some role to play at the
cosmic level.
\eq

\section{25 January 2000, \ ``Fear''}

Have we made progress in our 11 years of discussion?  I've certainly
made huge gobs of progress from the process.  I started out 11 years
ago thinking (or believing in the back of my mind at least) that we
were just inches away from a final understanding of the universe.
Today I think we're not even close, haven't even scratched the
surface of ``reality.''  That's why I wrote John {\Preskill} (the 10
September 99 note), ``\ldots I don't see physics as an expression of
our great knowledge of the world \ldots\ but more accurately as an
expression of our great ignorance.''  Seeing what needs to be done is
progress!

\section{27 January 2000, \ ``Chili Cheese Fries''}

\bgc
I think what impressed me about {\sl Pi} [the movie] was how much I
disagreed with the main premise---that ``mathematics is the language
of nature.''
\egc

You know, maybe I should see it again, now that these issues are so
much clearer in my mind.  (I think I saw it in Pasadena about three
years ago.)  Reading that article by {\Banville} and almost
simultaneously having a conversation with {\Bill} {\Wootters} about
the limitations of a dog's understanding of the world made a big
turning point in my life.  [See note to {\Herb} {\Bernstein}, dated
2 January 1999.]

\section{07 February 2000, \ ``The Fine Line''}

\bgc
He has suggested that I put my work on the lanl archive.  I am
reluctant to do this because I failed to get the work published in
good refereed journals; I think it sets the wrong precedent.
\egc

My buddy Jeff {\Nicholson} and I have had long fights about this
issue.  He stands more where you do now; I tend toward the opposite
extreme.  Your career has been punished simply because you had some
bad luck of the draw in having truly awful referees.  Why should
that propagate?

We had a similar situation in our field.  Stephen {\Wiesner} wrote a
wonderful little paper circa 1971 that essentially contained the
first quantum cryptography protocol.  Moreover it contained a whole
new (and important!) classical cryptographic protocol called ``one
out of two oblivious transfer.''  He submitted it to the {\it IEEE
Transactions on Information Theory}.  (His father was an important
information theorist and actually president of MIT.)  The paper was
rejected in a scathing way.  {\Wiesner} stuck it in his desk, tuned
in, turned on, and dropped out.  The paper had to wait until 1984 to
be revived by {\Charlie} {\Bennett} and {\Gilles} {\Brassard}.  These
days, {\Wiesner} lives in a hut in the desert somewhere in Israel.
Not completely cause and effect here, I'm sure, but things just
didn't need to go that way.  Now this year, there's some hope that
{\Wiesner} will win the Quantum Communication award that {\Shor} and
{\Kimble} won last year.

If I were you, I would stick the paper on the archive.  Or better
yet, have X submit it for you, with a note explaining that it was
submitted and rejected by Journal Y, along with the date of that
event.

\section{20 February 2000, \ ``Rabbits on the Moon''}

I've just come into my office out the crisp night air of Haifa and
an early sighting of a beautiful full moon.  And like it has been
the case for the last two years, but never before that, I couldn't
help but marvel at the rabbit on the moon:  his head and ears in the
top left corner, and his bushy tail in the bottom right.  A Japanese
friend mentioned its existence two years ago; the legend is part of
their culture.  Now it's always there whenever I look.  Where was it
for the previous 32 years of my life?  When I stop to think about
this, I just can't shake the idea that so much of the order we see
in the world around us is simply placed there by ourselves.  For so
much of my life the man in the moon was a perfectly adequate
metaphor---now I can't see it at all.

The world is an organism, the ancient civilizations said.  No, the
world is a mechanism, {\Laplace} said:  I can see no organism.  Now
that my eyes are opened, what could be clearer?  The world is a
\underline{\phantom{substrate}}, quantum information said.  Now that
I see it, I can never turn back.

\section{25 June 2000, \ ``Waxing Surfboards''}

I'm sitting on my balcony waiting for the Grecian sunrise and
contemplating my finite existence.  I thought about titling this
note "Waxing Philosophical," but somehow the surfboards intruded
into my mind more forcefully.  The scene is quite beautiful:  to my
left are some typically Greek-styled beach houses, straight in front
of me is a little island peeking over the waters, and above that is
the shining sliver of the morning's moon.

Yesterday I met quite an interesting guy, Arkady {\Plotnitsky}.  He's
in the English Dept. faculty at Purdue University, but his sideline
is on interpreting {\Bohr}.  And from what I could tell yesterday,
he's quite accomplished in that respect:  there are a load of
similarities between our points of view.  I had tried to read his
book "Complementarity" about three years ago, but gave up in
frustration.  In fact, I had even contemplated making fun of the
book during one of my journal club talks at Caltech as one of the
prime examples of the new science war gibberish.  When I told David
{\Mermin} that story, he quickly defended {\Plotnitsky} and told me
that I really should take him seriously, that he is impressive (much
more so in person than in his book).  And David's really right on
this count.

I don't believe I'm going to see the real sunrise; it looks like
it'll be coming up over the hill to my left.

Waxing surfboards philosophical.

\section{28 July 2000, \ ``Great Quote''}

\bq
\noindent
If triangles made a god, they would give him three sides. \medskip

--- Charles de Montesquieu (1689-1755), French philosopher, lawyer,\\
\indent \phantom{---} {\sl Lettres persanes}, letter 59 (1721).
\eq

\section{10 August 2000, \ ``Heavy Information''}

\bgc
In the meantime, Bekenstein once wrote a paper with Schiffer where
they pose the question if information had ``real'' physical
qualities. For instance, does it weigh? I can get the reference if
you are interested.
\egc

Yeah, I would like those references if you can dig them up.  I think
there were a few papers.  I remember the consensus of {\Caves} and
the Great W (that's what {\Carl} calls Bill Unruh) was that there was
something wrong in those papers---they had technical reasons.

I myself don't have a strong feeling that ``information'' should have
any physical (perhaps a better word would be ontologic) properties.
For, whenever information is a meaningful concept, we have
explicitly put ourselves into the picture.  The information,
whenever there is any, resides in our heads, not in the objects of
our attention.  However, I do think that information-theoretic
questions can go a long way toward elucidating properties of the
critters we use as information carriers.  For instance, that
ultimate cap on how much I can predict versus how much you can
predict if we imagine the system in a measurement context:  is that
telling us that there is a property we should be mindful of when we
contemplate using that object for getting something done?  (Well,
yes, both in quantum cryptography and quantum computing: that
``sensitivity to the touch'' is a {\sl resource\/} that doesn't exist
in the classical world, and it can do things for us if we use it
properly.)  Is this extra something that quantum systems have,
something we've been missing (however subtle) in our accounting of
how objects affect the spacetime around us?

Yeah, I do want to see the Bekenstein-Schiffer papers again.

\section{11 August 2000, \ ``The Mathematics of Experiment''}

\bgc
Is it reasonable to say that the experimental process is also a form
of mathematics?  (See, now that you opened the window of
possibilities, I'm trying to ease my conscience.)  Actually, I now
think of mathematics as just a search for, or creation of, logical
structures.
\egc

I like your last sentence.  The only way I would change it is to
substitute for ``logical structures'' something like ``structures
congenial for our reasoning in light of the information we dredge up
from the world.''  (It's longer winded but it better captures my
present attitude.)

\section{18 September 2000, \ ``A Non-Randy Non-Bugger''}

\bgc
So, this thought is keeping me from my grant: Our laws of physics
are not about the real world, never will be?  They're only rules
that describe our interface with the real world?  And what would it
matter if even the real world of {\Einstein} existed?  Because it's
not about the ``real world'' but rather how we are allowed to
interface with it, and perhaps it with us?
\egc

I was especially intrigued by this point of yours.  My friend {\Herb}
{\Bernstein} also likes to make it (over and over).  By complete
chance I happened to read a very nice expression of the same thought
the other day in an article on a ``feminist approach to teaching
quantum mechanics.''  The article was quite good actually.  But I've
been thinking about writing the author to say that what she has
written has nothing to do with a ``feminist reading of {\Bohr}.''
It's just a {\it careful\/} reading of {\Bohr} \ldots and I'm
qualified to say that as I have {\it both\/} the same reading of
{\Bohr} {\it and\/} I'm likely one of the most chauvinistic men
she's ever met.  How do you think she'd react?

Anyway, here's the passage below.  The part that's especially like
your comment is in the last paragraph.\medskip

\noindent K.~{\Barad}, ``A Feminist Approach to Teaching Quantum
Physics,'' in {\sl Teaching the Majority:\ Breaking the Gender
Barrier in Science, Mathematics, and Engineering}, edited by S.~V.
Rosser (Teachers College Press, New York, 1995), pp.~43--75.

\bq
\indent
The {\Newton}ian worldview is compatible with an objectivist
epistemology, in which the well-prepared mind is able to produce a
privileged mental mirroring of the world as it exists independently
of us human beings.  That is, what is ``discovered'' is presumed to
be unmarked by its ``discoverer.''  The claim is that the scientist
can read the universal equations of nature that are inscribed in
[God's] blackboard:  Nature has spoken.  Paradoxically, the objects
being studied are given all the agency, even and most especially
when they are seen as passive, inert objects moving aimlessly in the
void.  That is, these cultureless agents, existing outside of human
space-time, are thought to reveal their secrets to patient observers
watching and listening through benignly obtrusive instruments.
Notice that agency is not attributed to human beings; once all
subjective contaminants have been removed by the scientific method,
scientists simply collect the pure distillate of truth.

The {\Newton}ian worldview is still so much a part of contemporary
physics culture that it infects the teaching of post-{\Newton}ian
physics as well.  That is, the stakes are so high in maintaining the
mirroring view of scientific knowledge that quantum physics is
presented as mysticism.
\eq
and
\bq
\indent
Notice that particular experimental arrangements can be used to give
more or less definite meaning to each of the complementary
variables, but due to the lack of object-instrument distinction
\ldots\ it is not possible to assign the value obtained to the
object itself.  The ``property'' being measured in a particular
experimental context is therefore not ``objective'' (that is, a
property of the object as it exists independently of all human
interventions), but neither is it created by the act of measurement
(which would belie any sensible meaning of the word {\it
measurement\/}).  {\Bohr} speaks of this ``interaction'' between
``object'' and ``instrument'' as a ``phenomenon.''  The properties
then are properties of phenomena.  That is, within a given context,
classical descriptive concepts can be used to describe phenomena,
our intra-actions within nature.  (I use the term {\it
intra-action\/} to emphasize the lack of a natural object-instrument
distinction, in contrast to {\it interaction}, which implies that
there are two separate entities; that is, the latter reinscribes the
contested dichotomy.  \ldots\ That is, the ambiguity between object
and instrument is only temporarily contextually decided; therefore,
our characterizations do not signify properties of objects but
rather describe the intra-action as it is marked by a particular
constructed cut chosen by the experimenter (see {\Barad}, 1995 for
more details).

The notion of ``observation'' then takes on a whole new meaning
according to {\Bohr}:  ``[B]y an experiment we simply understand an
event about which we are able in an unambiguous way to state the
conditions necessary for the reproduction of the phenomena'' (quoted
in {\Folse}, 1985, p.~124)\@.  According to the analysis of the
previous section, this is possible because, in performing each
measurement, the experimenter intervenes by introducing a
constructed distinction between the ``object'' and the ``measuring
device'' (e.g., deciding whether the photon is part of the object or
the instrument).  The claim is that unambiguous, reproducible
measurements are possible through the introduction of constructed
cuts.  Notice that ``[n]o explicit reference is made to any
individual observer'':  Different observers will get the same data
set in observing any given phenomenon.  Therefore, reproducibility,
not some {\Newton}ian notion of objectivity denoting observer
independence, is the cornerstone of this new framework for
understanding science.

For {\Bohr}, the uncertainty principle is a matter of the inadequacy
of classical description.  Unlike the ``mirroring''
representationalism inherent in the
{\Newton}ian-Cartesian-Enlightenment framework of science, scientific
concepts are not to be understood as describing some independent
reality.  A post-{\Newton}ian framework sees these constructs as
useful (i.e., potentially reproducible) descriptions of the entire
intra-action process (the phenomenon, which is context dependent by
definition), not of an isolated object.  The implications of this
finding are profound.  In {\Bohr}'s own words:
\bq
The extension of physical experience in our own days has \ldots\
necessitated a radical revision of the foundation for the
unambiguous use of elementary concepts, and has changed our attitude
to the aim of physical science.  Indeed, from our present
standpoint, physics is to be regarded not so much as the study of
something a priori given, but rather as the development of methods
for ordering and surveying human experience. ({\Bohr}, 1963, p.~10)
\eq
In other words:
\bq
These facts not only set a limit to the extent of the information
obtainable by measurements, but they also set a limit on the meaning
which we may attribute to such information.  We meet here in a new
light the old truth that in our description of nature the purpose is
not to disclose the real essence of [physical objects] but only to
track down, so far as it is possible, relations between the manifold
aspects of our experience. ({\Bohr}, 1963, p.~18)
\eq
\eq
and
\bq
\indent
{\Bohr}'s philosophy of physics involves a kind of realism in the
sense that scientific knowledge is clearly constrained, although not
determined, by ``what is out there,'' since it is not separate from
us; and given a particular set of constructed cuts, certain
descriptive concepts of science are well-defined and can be used to
achieve reproducible results.  However, these results cannot be
decontextualized.  Scientific theories do not tell us about objects
as they exist independently of us human beings; they are partial and
located knowledges.  Scientific concepts are not simple namings of
discoveries of objective attributes of an independent Nature with
inherent demarcations.  Scientific concepts are not innocent and
unique.  They are constructs that can be used to describe ``the
between'' rather than some independent reality.  (Why would we be
interested in such a thing as an independent reality anyway?  We
don't live in such a world.)  Consideration of mutually exclusive
sets of concepts produces crucial tensions and ironies, underlining
a critical point about scientific knowledge:  It is the fact that
scientific knowledge is socially constructed that leads to reliable
knowledge about ``the between''---which is just what we are
interested in.  This shifting of boundaries deconstructs the whole
notion of identity:  Science can no longer be seen as the end result
of a thorough distillation of culture.  There is an author who marks
off the boundaries and who is similarly marked by the cultural
specificities of race, history, gender, language, class, politics,
and other important social variables.  Reproducibility is not a
filter for shared biases.  In stark contrast to the objectivist
representationalism that is usually transmitted to students, the new
framework inspired by {\Bohr}'s philosophy of physics is robust and
intricate.  In particular, there is an explicit sense of agency and
therefore accountability.  And so I refer to this {\Bohr}-inspired
framework, which shares much in common with central concerns in
contemporary feminist theories, as ``agential realism.''
\eq

\section{06 October 2000, \ ``The Evolution of Thought''}

\bgc
Remember my epiphany from a month ago?  Well, it has now prompted
the following response: If wave functions don't collapse, because
they don't exist, then why is there the Shr\"odinger equation, i.e.\
why do these things that don't exist evolve?
\egc

My state of mind exists, and it evolves (even when I'm gathering no
new information).  Schr\"o\-dinger evolution, I think, ultimately
just reflects the differing rates at which internal clocks run (the
system's and mine).  The deeper questions are why linearity?  Why
unitarity?  I'm not very satisfied with all the answers I've heard
to date.

\section{14 October 2000, \ ``On the Mark''}

Just a very quick note.

\bgc
Or is linearity perhaps a necessary consequence of a description of
nature that is not about nature, but rather about how the interface
with nature?
\egc

That's exactly the direction I've been heading, and the sort of
thing I had been wanting to write to you.  So you've anticipated me
and saved me a little trouble.  Let me try to give you a smidgen of
the flavor.  Ben {\Schumacher}, {\Ruediger} {\Schack} and I have a
little argument for the linearity of time evolutions that goes like
this.

Suppose you and I walk into a room and agree that we have a
gazillion (i.e., an infinity of) copies of some physical system. The
only thing is we disagree about the overall quantum state of the
lot.  I say it is $\rho$; you say it is $\sigma$.  (This means
nothing more than that we differ in our probability assignments for
how the systems will react to our potential measurements.)  The
question is under what conditions can we be sure to converge in
opinion if you start making measurements system by system and I look
over your shoulder to see the results too.  {\Caves}, {\Schack} and I
answered that question last year with the ``quantum de {\Finetti}
theorem'' (still need to publish the damned thing):  remember it
only worked for complex, not real, vector spaces.  It states that
the necessary and sufficient condition that we will ultimately
converge in opinion is that our initial density operator assignments
have the property of exchange symmetry.  (Also we need the good
Bayesian assumption that nothing ever be assigned probability
absolutely zero; there should always be some $\epsilon$ no matter
how tiny.  But that's a technical point that doesn't concern the
present discussion.)  What this means is that if we interchange any
two of the gazillion systems, my state should evolve from $\rho$ to
$\rho$ and yours should evolve from $\sigma$ to $\sigma$.  This
gives some operational meaning to the statement that all the systems
are the same.  But beside that, you should also see that in order to
come to a tighter agreement we have to have at least some initial
agreement.  Otherwise no amount of empirical evidence will sway us
otherwise.  Anyway, these are the necessary and sufficient
conditions.

Now let us ask what might be a reasonable form for the notion of
``identical physical evolutions'' in this context?  I say it is
this:  if we have conditions so that we can converge in opinion
before the evolution takes place, then we should also be able to
converge in opinion after the time evolution.  Technically, time
evolution should preserve exchange symmetry.  Also if we time evolve
our converged opinion it should be identical to what we would
converge to after the time evolution.  Well, using the quantum de
{\Finetti} theorem, what B, S, and I proved is that these conditions
imply linearity.  Linearity can be viewed as coming from the idea
that science is about our converging in opinion.

We still don't have a good argument for unitarity (or more carefully
complete positivity at the level of density operators), but that's
the sort of direction I'm heading.  There'll eventually be a way to
crack it:  I'm confident.

OK, now I need to go whack some weeds and write some recommendation
letters for an Oxford boy (a Rhodes scholar even, but I try not to
hold that against him).

\section{14 October 2000, \ ``More Evolution''}

\bgc
Pretty simple questions, right?!?  I guess you get to ask me what
happens to my "worldview" if the equivalence principle or causality
or spacetime are discovered not to exist, right?
\egc

I'd be especially curious to know what would happen if we had no
equivalence principle.  How strange would our world be then?  What
would break?

OK, now I really go whack weeds.

\section{16 October 2000, \ ``More Linearity''}

\bgc
Thanks for your replies!  I was pretty much shocked that you've
actually got some proofs going on the question of linearity.
\egc

Oh, I've got a load of things like that that are still unpublished.
That's why my collaborators are starting to hate me.  X, for
instance, gets especially miffed that I have an extra outlet to
bring fame and fortune to myself that he doesn't:  I go to loads of
meetings and talk about all these things (privately and when I'm on
the stage) and they sort of become folklore knowledge.  I really
need to work hard to remedy that. But getting at the bottom of
things is so much more interesting than writing things up!!

So let me spend a little bit of time this morning trying to corrupt
you.  Lately (or maybe for years) I've got it in my head that the
time evolution part of quantum mechanics is the least interesting
part of the theory.  In fact---I conjecture---it is so
disinteresting as to be identical with classical Hamiltonian time
evolution \ldots\ {\it when\/} couched in terms of the Liouville
equation.  The italicized {\it when\/} is important.  What I'm toying
with is that it would be interesting to give a direct derivation of
the Liouville equation from fundamental principles (and forget the
usual track of starting with Hamilton's or Lagrange's equations and
{\it then\/} moving to Liouville's).  That is to say, what general
principles can we give that would uniquely pin down the form of the
evolution of probability distributions over a phase space?  Then we
would see that once {\Newton} agreed that physics should be
scientific (in the sense that we would all ultimately come into
agreement) and that the ``observables'' in the theory should be $x$
and $p$, then he had absolutely no choice but to write down the
precise equation that he did, i.e., $F=ma$.

One thing that is already clear to me is that the argument I
presented in the last email also works for classical physics:  one
just relies on the classical de {\Finetti} theorem where we relied on
the quantum one.  So the classical Liouville evolution must be
linear if it is going to preserve convergence of opinion in
exchangeable situations.  But that's still a far cry away from
getting the full Liouville equation.

So, I find myself wondering what extra principles are needed even
there.  Here's one idea that crossed my mind last year or so.  I'll
just cut and paste from my samizdat:  [See notes to  {\Asher}
{\Peres}, dated 23 September 1999, 26 September 1999, and 27
September 1999.]

My description of the {\Wigner} theorem in that is not quite right.
Here's what I should have said:
\bq
If we have a bijection from unit vectors to unit vectors on a
Hilbert space that preserves all [absolute values of] inner
products, then [one can redefine the overall phase assignments of]
that mapping [so that it becomes] a unitary linear map or an
antiunitary antilinear map.  If we further suppose that mappings are
continuously connected to the identity, then we are left with the
unitaries.
\eq

Oh, here's another entry, this time written to {\Ruediger} {\Schack}:
[See note to {\Ruediger} {\Schack}, dated 22 September 1999.]

Anyway, would you be interested in thinking about this problem with
me?  You certainly have a greater knowledge of classical mechanics
than I do.  I'm not quite sure that ``overlap preservation'' will do
the trick.  (Is that completely equivalent ``volume preservation?''
Doesn't one need more than that?)  Maybe one needs something more
than one needed in the quantum case.  (For instance, I suspect
overlap preservation for all functions, not just normalized
probability distributions will do the trick---because then one has a
full vector space not just a cone---but that wouldn't be interesting
from the physical side in this case.)  So the question is, if extra
assumptions are needed, can one give them a knowledge-theoretic
justification?

Finally a vaguer question, but one that I know will pique you is: if
this does work out, how does the result mesh with general
relativity.  Can one generally give a Liouville form for the
{\Einstein} equations?  (Probably in the foliable cases for sure.
But what about the rest?)  Can we learn something about the
equivalence principle from considerations such as this?  (That's the
sort of thing I was trying to push you toward yesterday.)

\section{16 October 2000, \ ``Three References''}

Glad to hear your positive response.  Have a look at the
introduction and conclusions sections of any of these papers and
tell me what you think.  ({\Uhlhorn}'s paper might be the most
notable in that he relaxes {\Wigner}'s assumptions significantly but
still recovers the result---that's the sort of thing I'm banking on
in the classical case.  But you might feel the most comfortable with
Bargmann's for a first shot.)
\begin{enumerate}
\item
V.~Bargmann, ``Note on {\Wigner}'s Theorem on Symmetry Operations,''
J. Math.\ Phys.\ {\bf 5}, 862--868 (1964).

\item
C.~S. Sharma and D.~F. Almeida, ``A Direct Proof of {\Wigner}'s
Theorem on Maps Which Preserve Transition Probabilities between Pure
States of Quantum Systems,'' Ann.\ Phys.\ {\bf 197}, 300-309 (1990).

\item
U.~{\Uhlhorn}, ``Representation of Symmetry Transformations in
Quantum Mechanics,'' Arkiv F\"or Fysik {\bf 23}, 307--340 (1963).
\end{enumerate}

\section{29 October 2000, \ ``{\Poincare} Singularities''}

Have a think about this little passage from an essay {\Poincare}
wrote sometime before 1912 (when he died).  What do you think he
would have made of the big bang or black holes? \medskip

\noindent
From H.~{\Poincare}, ``The Evolution of Laws,'' in his book {\sl
Mathematics and Science:\ Last Essays (Derni\`eres Pens\'ees)},
translated by J.~W. Bolduc, (Dover, New York, 1963), pp.~1--14:

\bq
\indent
Mr.~Boutroux, in his writings on the contingency of the laws of
Nature, queried, whether natural laws are not susceptible to change
and if the world evolves continuously, whether the laws themselves
which govern this evolution are alone exempt from all variation.
\ldots\ I should like to consider a few of the aspects which the
problem can assume.
\eq
and
\bq
\indent
In summary, we can know nothing of the past unless we admit that the
laws have not changed; if we do admit this, the question of the
evolution of the laws is meaningless; if we do not admit this
condition, the question is impossible of solution, just as with all
questions which relate to the past. \ldots

But, it may be asked, is it not possible that the application of the
process just described may lead to a contradiction, or, if we wish,
that our differential equations admit of no solution?  Since the
hypothesis of the immutability of the laws, posited at the beginning
of our argument would lead to an absurd consequence, we would have
demonstrated {\it per absurdum\/} that laws have changed, while at
the same time we would be forever unable to know in what sense.

Since this process is reversible, what we have just said applies to
the future as well, and there would seem to be cases in which we
would be able to state that before a particular date the world would
have to come to an end or change its laws; if, for example, our
calculations indicate that on that date one of the quantities which
we have to consider is due to become infinite or to assume a value
which is physically impossible.  To perish or to change its laws is
just about the same thing; a world which would no longer have the
same laws as ours would no longer be our world but another one.
\eq

\section{27 November 2000, \ ``A Lesson in Physics?''}

From today's {\sl New York Times}:
\bq
This view of what judges do when they interpret statutes may prove a
hard sell even for Justice Scalia, who wrote in a concurring opinion
in a 1991 case, James Beam Distilling Company v.\ Georgia: ``I am not
so naive (nor do I think our forebears were) as to be unaware that
judges in a real sense `make' law. But they make it as judges make
it, which is to say as though they were `finding' it -- discerning
what the law is, rather than decreeing what it is today changed to,
or what it will tomorrow be.''
\eq
A lesson in physics?

\section{15 January 2001, \ ``Good Turns of Phrase''}

I just ran across this and I wanted to get in my hard-disk archive.
So, I pick on you. \medskip

\noindent From:  Richard {\Rorty}, ``Phony Science Wars,'' [Review of {\sl
The Social Construction of What?} by Ian {\Hacking}], {\sl Atlantic
Monthly}, November 1999.

\bq
The stalemate that {\Hacking} brilliantly describes but does not try
to break is between many scientists' intuition of the inevitability
of quarks and many philosophers' suspicion that the claim of
inevitability makes sense only if the idea of the intrinsic
structure of reality makes sense. This teeter-totter between
conflicting intuitions is, {\Hacking} rightly says, a genuine
intellectual problem. Which answer one gives to his third question
-- about the source of the stability of the most reliable bits of
science -- is likely to be a matter of which side of the seesaw has
most recently descended.

These alternating intuitions have been in play ever since Protagoras
said ``Man is the measure of all things'' and {\Plato} rejoined that
the measure must instead be something nonhuman, unchanging, and
capitalized -- something like The Good, or The Will of God, or The
Intrinsic Nature of Physical Reality. Scientists who, like Steven
Weinberg, have no doubt that reality has an eternal, unchanging,
intrinsic structure which natural science will eventually discover
are the heirs of {\Plato}. Philosophers like {\Kuhn}, Latour, and
{\Hacking} think that Protagoras had a point, and that the argument
is not yet over.

The most vocal and inflamed participants in the so-called science
wars are treating the latest version of this fine old philosophical
controversy as a big deal. In the very long run, perhaps, it will
prove to be one. Maybe someday the idea of human beings answering to
an independent authority called How Things Are in Themselves will be
obsolete. In a thoroughly de-Platonized, fully Protagorean culture
the only answerability human beings would recognize would be to one
another. It would never occur to them that ``the objective'' could
mean more than ``the agreed-upon upshot of argument.'' In such a
culture we would have as little use for the idea of the intrinsic
structure of physical reality as for that of the will of God. We
would view both as unfortunate and obsolete social constructions.
\eq

\chapter{Letters to Charles {\Enz}}

\section{12 June 1999, \ ``Interest in {\Pauli}''}

I have come across a reference to a paper of yours that I have not
been able to obtain.  May I ask of you to send me a reprint of it?
The paper is:
\bq
\noindent
C. P. {\Enz}, ``Wolfgang {\Pauli} and the Role of the Observer in
Modern Physics,'' in {\sl Philosophy of the Natural Sciences:
Proceedings of the 13th International Wittgenstein Symposium},
edited by P.~Weingartner and G.~Schurz (Verlag
Holder-Pichler-Temsky, Vienna, 1989), pp.~110--119.
\eq

I have read several of your other papers on {\Pauli}'s thought and
have been uniformly pleased with each.  I am especially interested
in his ideas concerning the demise of the ``detached-observer''
notion within quantum theory.  (I would also be interested in any of
your other works on the subject that may have been written since your
1992 article, ``Wolfgang {\Pauli} between Quantum Reality and the
Royal Path of Dreams.'' This is the latest one that I am aware of.)

About myself, I am a theoretical physicist at Caltech who
specializes in quantum information theory and quantum computing.  I
am well acquainted with your colleague Nicolas {\Gisin} at the
University of Geneva.

\section{21 August 1999, \ ``One More Article?''}

Upon my return from Cambridge a couple of weeks ago, I found in my
mailbox your package full of wonderful papers!  Thank you so much.
I have gobbled them all down except the one written in German
(unfortunately I don't know a word of German).

I wonder if I can impose upon you one more time.  I noticed a
citation to still another of your papers of which I was not aware:
``The wave function of correlated quantum systems as objects of
reality'' in a book with a Finnish title (Helsinki University Press,
1996), pp.\ 61-76.  May I obtain a copy of that also?  (If you
happened to write it in \TeX\ or \LaTeX, you could just email the
file and I could view it that way; I am much more interested in the
thoughts than in having an official reprint.)

My deep interest in {\Pauli} comes from his interest and thought
about the ``engaged observer.''  In particular part of my research
program is to see how far the idea that the engaged observer is not
only consistent with the structure of quantum theory but actually
compels that structure (i.e., the standard axioms of the theory).
For the technical side of this research, I take as my starting point
many of the concepts and techniques we have learned from quantum
cryptography and quantum information theory.  But on the
philosophical side I have {\Pauli}, {\Fierz}, yourself and perhaps
some other disciples that I am not yet aware of.

\section{09 June 2000, \ ``Another Request''}

I wonder if I might trouble you again with another request for your
papers on Wolfgang {\Pauli}.  (Do you remember my earlier requests
that you so nicely fulfilled?)  My house was one of the unfortunate
ones that burned in Los Alamos during the forest fire last month.
Consequently my wife and I lost almost everything we owned,
including all our books and all our papers.

If you could replenish my supply of all your papers, I would be ever
so grateful.

I know that I had at least the following papers of yours on
{\Pauli}, but I believe there were several more that I had not yet
recorded into my computer.

\begin{enumerate}

\item
C.~P. {\Enz}, ``W.~{\Pauli}'s Scientific Work,'' in {\sl The
Physicist's Conception of Nature}, edited by J.~Mehra (D.~Reidel,
Dordrecht, 1973), pp.~766--799.

\item
C.~P. {\Enz}, ``The Space, Time and Field Concepts in Wolfgang
{\Pauli}'s Work,'' in {\sl Symposium on the Foudations of Physics,
Joensuu, 1987}, edited by P.~J. {\Lahti} and P.~{\Mittelstaedt}
(World Scientific, Singapore, 1987), pp.~127--145.

\item
C.~P. {\Enz}, ``Wolfgang {\Pauli} and the Role of the Observer in
Modern Physics,'' in {\sl Philosophy of the Natural Sciences:
Proceedings of the 13th International Wittgenstein Symposium},
edited by P.~Weingartner and G.~Schurz (Verlag
Holder-Pichler-Temsky, Vienna, 1989), pp.~110--119.

\item
C.~P. {\Enz}, ``Book Review: {\sl Beyond the Atom:  The Philosophical
Thought of Wolfgang {\Pauli}}.~By K.~V. {\Laurikainen}.,'' Found.\
Phys.\ {\bf 20}, 1025--1028 (1990).

\item
C.~P. {\Enz}, ``Wolfgang {\Pauli} between Quantum Reality and the
Royal Path of Dreams,'' in {\sl Symposia on the Foundations of Modern
Physics 1992: The Copenhagen Interpretation and Wolfgang {\Pauli}},
edited by K.~V. {\Laurikainen} and C.~Montonen (World Scientific,
Singapore, 1992), pp.~195--205.

\item
C.~P. {\Enz}, ``The Wavefunction of Correlated Quantum Systems as
Objects of Reality,'' in {\sl Vastakohtien todellisuus: Juhlakirja
professori K. V. {\Laurikainen} 80-vuotisp\"aiv\"an\"a}, edited by U.
Ketvel, et al. (Helsinki U. Press, 1996), pp.~61--76.

\end{enumerate}

\chapter{Letters to Henry {\Folse}}

\section{17 January 2001, \ ``First Contact''}

I am writing this letter to find out if you are interested in
attending a conference in V\"axj\"o, Sweden this summer titled
``Quantum Theory: Reconsideration of Foundations.''  I do this
because I have read a few of your papers and parts of your book on
{\Bohr}, and I have been extremely impressed.  By trade, I am a
practitioner of the new field of quantum information theory and
computing---and indeed the main session I am organizing at the
V\"axj\"o meeting will be about quantum foundations reconsidered in
that light---but your description of {\Bohr}'s thought meshes so well
with what I'm seeking in physics that I'd like to get you there too.
PLEASE NOTE that this is not exactly a real invitation yet:  I am
still lobbying for a more philosophical contingency at the meeting.
But, I would very much like to hear your reaction just as soon as
possible.  If things work out, and you would like to come, I will
recontact you very soon with further details.

The main reason I want you at our meeting is that I have this ``madly
optimistic'' ({\Mermin} called it) feeling that Bohrian-Paulian ideas
will lead us to the next stage of physics.  That is, that thinking
about quantum foundations from their point of view will be the {\it
beginning\/} of a new path, not the end of an old one.  I think you
find this kind of thought attractive too, so I think the conference
will be quite a natural vacation for you.

In this regard, allow me to place several pieces of (used!)
information below.  (Please don't be offended by this:  It just
seemed like the simplest way to get it to you quickly.)  What you
will find is the following. [\ldots]

\section{07 February 2001, \ ``Heartfelt Thanks''}

Yesterday I received the package you sent me.  I can't express how
grateful I am.  I never imagined that you would be sending such a
wealth of material!  Indeed I didn't even know that you had more
than three or four articles (outside your book) on the subject!

As I promised, I should have them digested before I meet you.  I am
so happy that you're coming to the meeting.

\section{16 February 2001, \ ``Is It Complete?''}

On top of the last question I sent you, I wonder if I can ask
another favor of you?  Could you scan down the list below and note
whether it is complete with respect to your quantum writings?  If
any glaring mistakes jump out at you, please let me know.  If you
could also fill me in with the details of the last two entries, that
would be great.

\begin{enumerate}

\item
H.~J. {\Folse}, ``The Copenhagen Interpretation of Quantum Theory and
Whitehead's Philosophy of Organism,'' Tulane Stud.\ Phil.\ {\bf 23},
32--47 (1974).

\item
H.~J. {\Folse}, ``The Formal Objectivity of Quantum Mechanical
Systems,'' Dialectica {\bf 29}, 127--?? (1975).

\item
H.~J. {\Folse}, ``A Reinterpretation of Democritean Atomism,'' Man
and World {\bf 9}, 393--417 (1976).

\item
H.~J. {\Folse}, ``Complementarity and the Description of
Experience,'' Int.\ Phil.\ Quart.\ {\bf 17}, 377--399 (1977).

\item
H.~J. {\Folse}, ``Quantum Theory and Atomism:\ A Possible Ontological
Resolution of the Quantum Paradox,'' Southern J. Phil.\ {\bf 16},
629--640 (1978).

\item
H.~J. {\Folse}, ``{\Kant}ian Aspects of Complementarity,''
{\Kant}-Studien {\bf 69}, 58--66 (1978).

\item
H.~J. {\Folse}, ``{\Plato}nic `Atomism' and Contemporary Physics,''
Tulane Stud.\ Phil.\ {\bf 27}, 69--88 (1978).

\item
H.~J. {\Folse}, Jr., ``Complementarity, {\Bell}'s Theorem, and the
Framework of Process Metaphysics,'' Process Studies {\bf 11},
259--273 (1981).

\item
H.~J. {\Folse}, {\sl The Philosophy of Niels {\Bohr}:~The Framework
of Complementarity}, (North-Holland, Amsterdam, 1985).

\item
H.~J. {\Folse}, ``Complementarity and Scientific Realism,'' in {\sl
Foundations of Physics:\ A Selection of Papers Contributed to the
Physics Section of the 7th International Congress of Logic,
Methodology and Philosophy of Science}, edited by P.~Weingartner and
G.~Dorn (Verlag H\"older-Pichler-Tempsky, Vienna, 1986), pp.~93--101.

\item
H.~J. {\Folse}, ``Niels {\Bohr}, Complementarity, and Realism,'' in
{\sl PSA 1986:\ Proceedings of the Biennial Meeting of the
Philosophy of Science Association, Vol.\ I}, edited by A.~{\Fine} and
P.~Machamer (Philosophy of Science Association, East Lansing, MI,
1986), pp.~96--104.

\item
H.~J. {\Folse}, ``Complementarity and Truth,'' trascript of talk
given at Valamo Monastery in Finland (1987).

\item
H.~J. {\Folse}, ``Niels {\Bohr}'s Concept of Reality,'' in {\sl
Sym\-posium on the Foundations of Modern Physics 1987: The Copenhagen
Interpretation 60 Years after the Como Lecture}, edited by
P.~{\Lahti} and P.~{\Mittelstaedt} (World Scientific, Singapore,
1987), pp.~161--179.

\item
H.~J. {\Folse}, ``Realism and the Quantum Revolution,'' in {\sl
Abstracts of the 8th International Congress of Logic, Methodology,
and Philosophy of Science, Vol.~4, Part I}, (Inst.\ of Philosophy of
the Academy os Sciences of the USSR, Moscow, 1987), pp.~199--200.

\item
H.~J. {\Folse}, ``Complementarity and Space-Time Description,'' in
{\sl {\Bell}'s Theorem, Quantum Theory and Conceptions of the
Universe}, edited by M.~Kafatos (Kluwer, Dordrecht, 1989),
pp.~251--259.

\item
H.~J. {\Folse}, ``What Does Quantum Theory Tell Us About the
World?,'' Soundings {\bf 72}, 179--205 (1989).

\item
H.~J. {\Folse}, ``{\Bohr} on {\Bell},'' in {\sl Philosophical
Consequences of Quantum Theory:\ Reflections on {\Bell}'s Theorem},
edited by J.~T. Cushing and E.~McMullin (U. Notre Dame Press, Notre
Dame, IN, 1989), pp.\ 254--271.

\item
H.~J. {\Folse}, ``{\Bohr}'s Framework of Complementarity and
{\Pauli}'s Philosophy,'' in {\sl Kohti uutta
todellisuusk\"asityst\"a. Juhlakirja professori Laurikaisen
75-vuotisp\"aiv\"an\"a} (Towards a New Conception of Reality.\
Anniversary Publication to Professor {\Laurikainen}'s 75th Birthday),
edited by ?? (Yliopistopaino, Helsinki, 1990), pp.~91--99.

\item
H.~J. {\Folse}, Jr., ``Complementarity and the Description of Nature
in Biological Science,'' Bio.\ Phil.\ {\bf 5}, 211--224 (1990).

\item
H.~J. {\Folse}, ``Laudan's Model of Axiological Change and the
{\Bohr}-{\Einstein} Debate,'' in {\sl PSA 1990:\ Proceedings of the
Biennial Meeting of the Philosophy of Science Association, Vol.\ I},
edited by A.~{\Fine}, M.~Forbes, and L.~Wessels (Philosophy of
Science Association, East Lansing, MI, 1990), pp.~77--88.

\item
H.~J. {\Folse}, ``Metaphysical Awakening in Philosophy of Quantum
Physics:\ A Review Article,'' Int.\ Stud.\ Phil.\ {\bf 23}, 89--98
(1991).

\item
H.~J. {\Folse}, ``Complementarity and Our Knowledge of Nature,'' in
{\sl Nature, Cognition and System II: Current Systems-Scientific
Research on Natural and Cognitive Systems}, Volume2:\ On
Complementarity and Beyond, edited by M.~E. Carvallo (Kluwer,
Dordrecht, 1992), pp.~51--66.

\item
H.~J. {\Folse}, ``The Environment and the Epistemological Lesson of
Complementarity,'' Environmental Ethics {\bf 15}(4), 345--353 (1993).
The abstract to this article reads:

\item
H.~J. {\Folse}, ``{\Bohr}'s Framework of Complementarity and the
Realism Debate,'' in {\sl Niels {\Bohr} and Contemporary
Philosophy}, edited by J.~{\Faye} and H.~J. {\Folse} (Kluwer,
Dordrecht, 1994), pp.\ 119--139.

\item
A.~Dotson and H.~Folse, ``Bearers of Properties in the Quantum
Mechanical Description of Nature,'' Int.\ Stud.\ Phil.\ Sci.\ {\bf
8}, 179--194 (1994).

\item
H.~J. {\Folse}, ``Essay Review:\  Niels {\Bohr} and the Construction
of a New Philosophy,'' Stud.\ Hist.\ Phil.\ Mod.\ Phys.\ {\bf 26},
107--116 (1995).

\item
H.~J. {\Folse}, ``The {\Bohr}-{\Einstein} Debate and the
Philosopher's Debate over Realism versus Anti-Realism,'' in {\sl
Realism and Anti-Realism in the Philosophy of Science}, Boston
Studies in the Philosophy of Science, Vol.~169, edited by R.~S.
Cohen, R.~Hilpinen, and Q.~Renzong (Kluwer, Dordrecht, 1996),
pp.~289--298.

\item
H.~J. {\Folse}, ``Ontological Constraints and Understanding Quantum
Phenomena,'' Dialectica {\bf 50}, 121--136 (1996).

\item
H.~J. {\Folse}, ``Realism, Idealism, and Representation in the
Description of Nature,'' in {\sl Avartuva ajatus:\
Julkaisutoimikunta}, edited by U.~Ketvel and ?? (Luonnonfilosofian
seuran, Espoo, 1999), pp.~73--77.

\end{enumerate}

\section{09 March 2001, \ ``The Right Choice''}

Sorry for my long delay in getting back to you.  I'm off to Japan in
about five hours for two weeks, and *trying* to get things tied up
here before leaving has made the week pretty hectic.

Anyway, the most important message I wanted to tell you is that I'm
gettin' damned happy I invited you to V\"axj\"o.  Reading your papers
has been a really pleasant experience.  I think I hit my first dozen
last night at about this time.  (But maybe the bigger question is,
why am I up at this time?!?)  Below, I'll place my {\Folse}
compendium as it stands---every word typed in lovingly with my own
little fingers!  Of course, I have a few quibbles with some things
I've read, but I think I'll save my comments until I've read the
complete body of work.  27 years is a long time, and you could well
have changed your mind about some things:  I'll give you the benefit
of the doubt for now.

\bhf
I should certainly warn you that very few in this community agree
much with my reading of {\Bohr}.
\ehf

I could care less about that:  I like it (or most of it), and that's
all that counts for me.  Besides, I've read {\Bohr} myself---fairly
carefully I've always thought---and your view significantly
coincides with my memory of that.

\bhf
I'm more concerned about my lack of knowledge of anything about
information theory.
\ehf

You need not be too concerned, but of course it wouldn't hurt you to
do a little reading on the side if you've got some time.  Somewhere
below, you wrote:
\bq
\noindent
In describing the phenomena of observational interactions, quantum
theory describes them as being caused by the interactions between
the observing systems and microsystems.  The fact that we can form
no representation, no mechanical picture, of the atoms on which the
mechanistic description of the phenomenal world is based, hardly
reveals that we are ignorant of what these entities {\it are}.
Rather it testifies to what we have learned about them -- that they
cannot be so represented -- through explorations of the atomic
phenomenon in which their strange behavior is revealed to human
experience.
\eq

It is my strongest opinion that the great fruits of quantum
information and computing get at precisely this point \ldots\ and in
spades!  The point is that this ``nonrepresentability'' in actual
fact boils down to a positive statement rather than a negative one.
So it would do you well to learn a little about our field.  (And,
luckily, most of what we do is not abstruse stuff:  it's just basic
quantum mechanics, viewed mostly from a new point of view with a new
set of goals in mind.)  Where to start?  Maybe a good place would be
three ``recent'' {\sl Physics Today\/} articles:
\begin{enumerate}
\item  {\Gottesman} and Lo, ``From Quantum Cheating to Quantum Security,'' PT
November 2000 (don't have the page numbers)

\item  {\Preskill}, ``Battling Decoherence: The Fault-Tolerant Quantum
Computer,'' PT June 1999, p.~24

\item  {\Bennett} ``Quantum Information and Computation,'' PT October 1995,
p.~24 (interesting coincidence)
\end{enumerate}
If you get that far, let me know, and I'll suggest a couple of
really {\it mild\/} technical articles that'll be worth their weight
in gold in insight.

But why do I think it would do you well?  Because I think you have
an honest heart.  And, because while I believe {\Bohr} and his gang
certainly started to point us in the right direction, I think we
have a long, long technical way to go before we can claim a
particularly deep understanding of the quantum structure.  Here's
how I put it in exasperation to David {\Mermin} once:
\bq
What's your take on this passage?  Can you make much sense of it?
What does he mean by ``providing room for new physical laws?''  What
``basic principles of science is he talking about?''  What five pages
of derivation are lying behind all this business?

It nags me that {\Bohr} often speaks as if it is clear that the
structure of quantum theory is derivable from something deeper, when
in fact all the while he is taking that structure as given.  When did
he ever approach an explanation of ``Why complex Hilbert spaces?''
Where did he ever lecture on why we are forced to tensor products for
composite systems?  It's a damned shame really:  I very much like a
lot of elements of what he said, but as far as I can tell all the
hard work is still waiting to be done.
\eq

The issue in my mind is {\it not\/} to {\it start\/} with complex
Hilbert space, unitary evolution, the tensor product rule for
combining systems, the identification of measurements with Hermitian
operators, etc., etc., and {\it showing\/} that {\Bohr}'s point of
view is {\it consistent\/} with that.  Instead it is to start with
{\Bohr}'s point of view (or some variant thereof) and see that the
{\it precise\/} mathematical structure of quantum theory {\it must\/}
follow from it.  Why complex instead of real? Why unitary, rather
than simply linear? Indeed, why linear?  Why tensor product instead
of tensor sum?  And, so on. When we can answer {\it these\/}
questions, then we will really understand complementarity.

I'm banking my career on the idea that the tools and issues of
quantum information theory are the proper way to get at this program.

OK, I've got to get some sleep.  I have this dream that I'm going to
get work done all the way to Japan.  But if I don't get some sleep,
I'll certainly be kidding myself.  (By the way, I will most
certainly be in email contact while I am away, so write if you
wish.  But don't send any viruses!)

\section{14 March 2001, \ ``The Archive Again''}

\bhf
If you're planning to visit the archive at the {\Bohr} Institute, you
should get in touch with the head archivist, Dr.\ Finn Aaserud. When
I last visited, they weren't really set up for the ``drop in''
visitor.
\ehf

Thanks again for the Archive information.  Ben {\Schumacher} and I,
as we were hiking down the old Hakone trail, were getting more and
more excited about the possibility of spending two days in the
archive after Sweden.  Can you tell me how to get in touch with the
person you speak of?

Ben told me a wonderful story yesterday.  Somehow the new {\sl Star
Wars\/} movie came up in the conversation ({\Charlie} {\Bennett} and
John {\Smolin} were also in the conversation).  I said I had never
seen the movie. John said it was awful.  Ben said it was wonderful.
I suspect I'd have to side with John, just seeing what I've seen of
the previews and reviews, but the surprising thing was that Ben said,
``When the movie was over, I was really depressed.''  ``For about 10
days after the movie, I was really depressed because I wanted to be
a Jedi.  To learn the ways of the Force.  I'd never have a chance to
live that wondrous life, and that depressed me.''  But then he went
on to say, ``But about two months later I was in London and I saw the
play {\sl Copenhagen}.  And that got me going all over again.  And it
almost depressed me, {\it but\/} then I realized, `Wait!  I am a
Jedi!' Learning the ways of the quantum world {\it makes\/} me a
Jedi!'\,''

\chapter{Letters to Bob {\Griffiths}}

\section{10 February 1997, \ ``Sunday Afternoons and Consistent
Histories''}

I've finally completed to my satisfaction your challenge of taking a
plunge into ``consistent histories.'' And I've had great fun!  In
particular, I read all or most of each of the following:\medskip\\ 1.
``Consistent Histories and the Interpretation of Quantum Mechanics,''
\\
2. ``Quantum Interpretation Using Consistent Histories,''
\\
3. ``Correlations in separated quantum systems: A consistent history
analysis of the EPR problem,''
\\
4. ``The Consistency of Consistent Histories: A Reply to
d'Espagnat,''
\\
5. ``A Consistent History Approach to the Logic of Quantum
Mechanics,''
\\
6. ``Consistent Histories and Quantum Reasoning,''
\medskip\\
I don't think I've read everything, but I'll trust that I haven't
missed anything too significant---at least for a novice education in
the subject.

I say I had great fun because, before this, my only introduction to
consistent histories had been through various talks by {\GellMann},
{\Hartle}, and {\Zurek}.  And, I can't say I ever found any of those
satisfactory---they all made it a very confusing muddle to me.  I
never could figure out the point, i.e., what actually was being
proposed as a ``solution'' to the quantum interpretation problem.
Consequently, I often just fell asleep or thought about other things
or both.  In contrast, I found all your papers logically crisp.

I now understand what you mean when you say that consistent histories
is a very natural extension of the standard quantum mechanics.  Apart
from the ``ontological aspects'' of the interpretation, as far as I
can tell, there is nothing questionable about it whatsoever.  It
appears to me to be a valuable tool, and I regret not having
practiced with it before.

What I mean by this particularly is that I have known for quite some
time how to play a certain game when I restrict my attention to a
single orthogonal set of projectors on a Hilbert space.  For
instance, I know that I can build a device to ``clone'' the quantum
states associated with that set; more than that, I know that I can
build a device to ``broadcast'' any density operator diagonal in that
representation.  Basically, I can play the games that I might have
imagined playing in the classical world.  What I have learned from
your papers is that there are a whole host of new games of
``classical inference'' within (consistent) ``frameworks'' that I
should contemplate.  And that may well lead to more economical ways
of viewing problems than I could have anticipated with textbook
quantum mechanics. Better yet, it may lead to new problems!  (I think
you and Chi-Sheng{\index{Niu, Chi-Sheng}} are carrying that aspect
of it forward quite nicely.)

All that said, my main purpose in writing this note is to learn even
more---especially about those aspects of the ``interpretation'' that
I am not so disposed to agree with.  I hope you'll bear with me,
especially in those cases when I will---of necessity---be fairly
vague. (It can be very difficult to debate a point when you feel that
your opponent is wrong, but you also know that you yourself do not
know what is right!)

As I say, I believe your consistency conditions and the examples you
give for what can be done with them are technically unassailable.
Concerning the ``foundational value'' of consistent histories,
however, I am not completely enthusiastic.  First let me state
outright that I am fairly agnostic about most things: interpretations
of quantum mechanics are no exception.  I am perfectly willing to
work both sides of the railroad track, so to speak.  I've seen how
believing in ``many worlds'' helped David {\Deutsch} search for
interesting applications of quantum parallelism.  I've seen how
believing that quantum theory is likely to be secondary and derived
from some deeper theory of communication helped lead {\Bill}
{\Wootters} to the quantum teleportation idea.  I've now seen how
consistent histories led you and Chi-Sheng{\index{Niu, Chi-Sheng}} to
some nice things.  The point is I don't have too much reason to be
too hard-headed about any one point of view, nor too much passion,
but I would like to express my troubles.

Several times over in your papers you say things similar to one line
I'll quote from your last paper:
\begin{quote}
The principal thesis of the present paper is that the major
conceptual difficulties of non-relativistic quantum theory \ldots\
can be eliminated, or at least tamed, by taking a point of view in
which quantum theory is a fundamentally {\it stochastic\/} theory,
in terms of its description of the time development of a physical
system.
\end{quote}
Presumably what you mean by this is that, within a {\it single\/}
consistent family of histories---or framework as you call it---one
and only one history is actually the case. Quantum theory ascribes a
``weight'' or probability to each history to signify that even
complete knowledge of the initial and final conditions, $D$ and $F$,
does not carry enough information for us to predict precisely which
history within the family is actually the case.  A consequence of
this is that quantum theory concerns an indeterministic state of
affairs.  What I mean by the phrase ``is actually the case'' is the
common everyday usage of the notion, i.e. something that is an
objective reality and observer-independent. ``Regardless of the two
opposing rulings, it is a fact that O. J. Simpson either did or did
not commit the murders; we simply may never know which is actually
the case.''

In equal measure you then state a fundamental requirement of the
consistent histories interpretation:
\begin{quote}
A meaningful description of a (closed) quantum mechanical system,
including its time development, must employ a single framework.
\end{quote}
And finally you qualify the consistent histories formalism and
interpretation with a discussion of the following sort.
\begin{quote}
\ldots\ once we have agreed that quantum mechanics is a stochastic
theory in which the concept of ``true'' corresponds to ``probability
one'', then precisely because probabilities (classical or quantum)
only makes sense within some algebra of events, the truth of a
quantum proposition is necessarily labeled, at least implicitly, by
that algebra, which in the quantum case we call a framework.  The
existence of incompatible quantum frameworks is no more or less
surprising than the existence of non-commuting operators representing
dynamical variables \ldots
\end{quote}

Let me use these quotes and my extra---but presumably
mild---interpretive work above as a point of departure. I believe you
are dead on the mark with the very last statement, i.e., that the
existence of incompatible frameworks is no more or less surprising
than the existence of incompatible (noncommuting) observables in
standard quantum mechanics. Moreover, I completely endorse the
systematic scheme consistent histories gives us for making
Wittgenstein's famous statement precise in the quantum context,
``Whereof one cannot speak, thereof one must be silent.''  Whenever a
statement of the form $A\wedge B$ cannot be formed, it is simply
meaningless.

What I cannot see is how consistent histories makes the
``objectivity'' of ``quantum properties'' more palatable or
believable than the single event formalism that we learn in standard
quantum mechanics.  That is to say, if I take the quantum
interpretation problem to be that of answering the questions
\begin{enumerate}
\item
Where do the unpredictable events in quantum phenomena come from?
\item
Are they always ``there'' independent of the framework I employ in my
description of the system?
\end{enumerate}
then I do not see how consistent histories has moved us any closer to
a solution. I don't even see how it has ``tamed'' the problem.

I do not say these things because I am {\it not\/} willing to revise
the set of ``logics'' I might entertain for reasoning about our world
(as was apparently the case with d'Espagnat).  The problem I have is
that I do not see these changed rules of logic as a {\it call\/} for
revising my intuition for what can be said of the ``real'' in the
world (or the ``True'' to use your terminology), as you suggest your
reader do.  Rather, I view that as exactly the nub of the problem for
the ``stochastic'' interpretation of quantum theory you endorse.
(Here I am using stochastic in the sense I defined above, which I
think is an adequate version of what you have in mind.) If
proposition $A$ is True in one framework, and proposition $B$ is True
in an incompatible framework, then I do not see the meaninglessness
of the proposition $A\wedge B$ as a call for revising my notion of
``being the case.''  Rather, it does push me very hard to understand
the (incompatible) True propositions as something other than
completely objective states of affairs.

Essentially all I'm saying here is that I don't see how consistent
histories has tamed any of the strangeness of the logic you refer to;
your papers provide no convincing arguments in this regard (or I am a
poor reader!).  I believe that it is precisely this strangeness about
quantum events (i.e., decompositions of the identity operator) that
has brought so many to interpret them as referring to something {\it
other\/} than completely independent and objective states of affairs.
This is the sort of thing that led to what I understand as the
``Copenhagen interpretation'': the meaning of the propositions $A$
and $B$ are defined solely in terms of the experimental procedures
required for eliciting their truth values; the proposition $A\wedge
B$ is meaningless precisely when there exists no experimental
procedure for eliciting the truth values of $A$ and $B$
simultaneously.

I believe I understand your concern that so much of quantum mechanics
has, in the past, been tied conceptually to the notion of
measurement. It would be a very strange world indeed that floated
around in superposition until Henri Becquerel and his laboratory
appeared in one branch of the wave function!  More seriously, the
notion of a real world, independent of what we think and see, is
beyond a doubt {\it almost\/} indispensable for our coordination in
this world. As I once wrote my friend {\Greg} {\Comer}:
\begin{quote}
{}From where and of what utility comes this notion of an ``objective
world'' independent of man, woman, animal, and plant? My opinion is
that (in the end) ``objective reality'' is posited for nothing more
than to have a device for coordinating the various experiences common
to all those who communicate.  Mark Twain once wrote, ``If you tell
the truth, you don't have to remember anything.'' It seems to me that
that pretty much sums it up.  The objective truth saves us from
having to make sure that the stories we tell are consistent; it saves
us from having to remember all aspects of the past.  It gives us a
means for determining whether someone's behavior is insane.  It gives
us a means by which to determine the guilt or innocence of an accused
murderer.  Is there any other real motivation for positing an
objective reality?  I can think of none \ldots\ but why should I, the
reasons listed above should be powerful enough for anyone involved in
the sciences.
\end{quote}
The problem of course with a measurement-based interpretation of
quantum mechanics is that much of this ``simplest view of the world''
seems to fade away.  The fighting of that, I believe, may have been
your motivation for seeking out consistent histories.

If this is the case, then I share your troubles.  However, it seems
to me that I haven't yet seen a convincing way of getting past these
things by revising our notion of {\it conjunction}, i.e., $\wedge$,
for ``objective'' but incompatible propositions.  I, in my own way,
have tried to grapple with these difficulties by trying to make
sharper what quantum probabilities are about.  For instance, {\Carl}
{\Caves} and I in our paper ``Quantum Information: How Much
Information in a State Vector'' (I gave you a copy) tried to promote
the following idea.
\begin{quote}
``Quantum probabilities'' are ignorance probabilities; they express
ignorance about the outcomes of potential measurements.  {\it What is
different in quantum physics is not the status of probabilities, but
rather the nature of the alternatives.}  In classical physics,
probabilities are concerned with {\it actualities}: ``One of the
alternatives actually occurs, but since I don't know which, I assign
probabilities based on what I do know.''  The probabilities that
describe intrinsic quantum unpredictability---the ``quantum
probabilities''---express ignorance about {\it potentialities} that
are actualized by measurement: ``I know one of these alternatives
{\it would\/} occur if I enquired about that set of alternatives, but
since I don't know which, I assign probabilities based on what I do
know.''
\end{quote}
(For a more extended discussion of this, see the bottom of page 22
through the top of page 26 of that article.) I would change the
language of some of this now if I could---{\Carl} likes the word
``potentiality'' more than I do---but I think the intention is clear.
One would like to divorce the language of quantum mechanics from
measurement {\it per se}.  Our attempt in this passage was through
emphasizing that quantum probabilities are about ``relational
contingencies'' more than actualities: they are about what would
occur if such and such situation were at hand.

This, at least formally, is not so different from what I saw many
times over in your papers on consistent histories: the physicist
chooses {\it any\/} framework consistent with his initial and final
data, $D$ and $F$; thereafter he can derive probabilities for the
elements within that framework.  The real difference appears only to
come in the ``ontological status'' we would give those elements.  I
think some of your very nice examples to do with incompatible
frameworks---like the one to do with Eqs.~(3.4)--(3.8) of Ref.~(2)
above---will allow me to sharpen up my phrase ``such and such
situation at hand'' substantially.  However, I don't see how my new
found knowledge about consistent histories will change my point of
view drastically.  (I welcome you to try to convince me otherwise!)

The problem with the focus on contingencies---which is the point of
view I am happiest with right now---is that you will immediately ask,
``What of the world before physicists were here making their
measurements?  What element in quantum theory reflects the clockwork
of the world that ticks away independently of man?''  The only answer
I am capable of giving right now is, ``I am not sure {\it anything\/}
in quantum mechanics reflects that.''  Given my point of view, I
would think that the hypothetical questions I just gave you are much
like, ``What in Bayesian probability theory reflects the operation of
the weather before weathermen appeared on the scene?''  Perhaps it's
much like {\Heisenberg} once said:
\begin{quote}
If the system is closed, \ldots\ [it] is then represented by a vector
in Hilbert space.  The representation \ldots\ so to speak, contains
no physics at all.
\end{quote}
When I'm feeling skeptical I sometimes think that perhaps there's not
a heck of a lot more to it than your ``quantum bicycle riding.''

On a side note, let me talk about one more thing of an even-more
philosophical flavor than the things we've discussed so far; after
that, I'll return to some further technical comments about consistent
histories.  In your last PRA paper, you ask somewhat rhetorically at
the beginning, ``Do quantum measurements reveal pre-existing
properties of a measured system, or do they in some sense create the
properties they reveal?''  Near the end of it you ask, ``\ldots\ does
quantum theory itself specify a unique framework? \ldots\ if the
answer is `no', \ldots\ [does this somehow imply] that physical
reality is influenced by the choices made by the physicist?'' The
truth of the matter is, to a nonnegligible extent, I like to think
that there is some truth to the wilder of these alternatives. Quantum
indeterminism is certainly one of the most wonderful things found in
this century.  If this indeterminism had not been of such a variety
as to allow us more free reign over our destiny or destruction, I
think I would be quite disappointed in it.  When my wife and I have a
philosophical discussion on just this point, I like to say that the
invention of the electric coffee maker is just as real a creation as
anything else in the world; if she can accept that, then why can she
not accept that a photon's polarization (or what we have a right to
say about it) is also created by our actions?  I am not joking when I
think these may be somewhat on the same footing.  The question then
turns to whether quantum theory always requires an intervention on
man's part for the properties of the world to come into being.  In
that regard, I am much, much more skeptical!!  (I refer you back to
my talk of clockworks above.)  However, just to emphasize again, I
think there is more than a grain of truth in John {\Wheeler}'s
``Game of 20 Questions, Surprise Version'' and it is the quantum
muddle that first led us to take it seriously.  (If this doesn't
completely frighten you off, I can send you a couple of short little
pieces of my correspondence that makes these ideas clearer.)

Whew, this note is growing!  Let me try to wrap it up soon.  Back to
more technical points.  Several times in your papers you have made
mention of ``quantum logic'' and contrasted your system to that.
Always, however, you point to systems like Birkhoff and von Neumann's
to do with orthomodular lattices.  It doesn't appear that you are
aware that there is still another approach to ``quantum logic'' that
is very much of the flavor of the system you construct in the
consistent history interpretation.  These systems are sometimes known
as ``partial Boolean algebras''; a simple introduction to what is
known about them can be found in R.I.G. Hughes book {\it The
Structure and Interpretation of Quantum Mechanics\/} (pp.~192--194).
In a nutshell,
\begin{quote}
A partial Boolean algebra is thus a set of Boolean algebras pasted
together in a consistent way, so that, where two or more Boolean
algebras overlap, their operations agree with each other.
\end{quote}
These objects are much like logical versions of a manifold in general
relativity: instead of being locally Lorentzian, they are locally
Boolean.  For instance, when two propositions are not contained
within compatible Boolean algebras their conjunction is simply deemed
meaningless, just as in your system. Apparently this logical
structure predates the more famous Birkhoff and von Neumann one by
several years, in the work of Strauss.  Hughes also wrote a review
article which is pretty good, J.~Symbolic Logic {\bf 50}:558--566
(1985). For the same reason that I find your system interesting, I
think these systems and the questions logicians are asking of them
are quite nice. For instance, one can ask whether the set of partial
Boolean algebras is isomorphic to the set of finite dimensional
Hilbert spaces. That is, once one has decided that physics has to do
with partial Boolean structures, is one led ineluctably to a Hilbert
space description of physical phenomena?

On a different subject, I think it would be very nice to see if the
consistent history idea can be made to be more general than it
already is.  By this, I do {\it not\/} mean something like {\Isham}
suggests, i.e., to use non-tensor-product histories; that sort of
generalization strikes me as taking away the prettiness of the
history idea. Instead, as you can probably guess, I am thinking of
using more general decompositions of the identity for the elementary
events in your histories.  In particular, it would be very nice if
something like the consistency conditions could be worked out for
general POVMs.  My inclination is that POVMs can be viewed as
measurements that are every bit as elementary as von Neumann
measurements.  If that is the case, then I don't see why they can't
also be fit somehow into the consistent history framework for closed
systems.  Do you have any thoughts on this?

Finally in Section~6.2 of your original article, you question the
uniqueness of your ``weight'' function.  I think that is an
intriguing question.  As you probably know, Jim {\Hartle} in his 1968
AJP article, has a very nice pseudo-derivation of the {\Born}
probability rule.  (In my opinion, he somewhat confuses
``probability'' with ``frequency''; however, his train of thought
does appear to lead nicely from qualitative statements to the
quantitative rule---as you say of the {\Everett} efforts to the same
effect, the starting point is not completely devoid of probability
concepts.  {\Carl}, Howard {\Barnum}, and I plan to write an article
putting in our two cents worth on this subject soon.)  In any case,
have you attempted to pin down your weight function by similar
techniques?

I guess that's more than enough for now.  My left hand is hurting
pretty badly!  I hope you find some of this commentary useful. Your
reactions, good or bad, would be appreciated.  In particular, if you
can give me some sounder reasoning for the ``objectivity'' of your
histories, I hope you will attempt it.  I hope you don't write me off
as a real quack!

\chapter{Letters to Adrian {\Kent}}

\section{30 August 1998, \ ``Desiderata''}

I just read over your four desiderata for an acceptable
``\"uber-quantum theory'' for the third time.  (When I finish this
note, I'll read them over for the fourth time \ldots\ at the very
least for simile's sake.)  They're not so uninspired! (\ldots\ to use
your words, that is.)  But I do hope you will keep your promise of
writing a longer exposition.

In the mean time, let me ask you for one point of clarification about
the connection between your Desideratum \#1 and my latest happy
thoughts.

\bak
1.  A theory in which some elementary, and precisely defined,
components constitute the basic statements from which propositions
about an external and objective reality can be built.  Maybe
trajectories a la {\Bohm}, maybe collapse events a la GRW, maybe
events of more general form a la consistent historians, maybe the
evolution of the classical gravitational field in some (unknown)
theory that links classical metric and quantum matter.  Maybe
something quite different from any of the above.  But, at any rate,
something which allows us to talk about what happened in the past,
in the absence of observers, and which allows us to resolve the
measurement problem by replacing the vague notion of measurement by
something precise.
\eak

As I tried to convey to you in Spain, I believe that I am basically a
realist at heart.  The evidence seems conclusive (to me, at least)
that there were things in this world long before there were
physicists grappling with the quantum measurement problem.  There
were things in this world long before there was intelligent life at
all.  However---and this may be where we disagree the most---I don't
see why ``reality'' should be construed as something absolutely
static.  It is the essence of a nondeterministic world that there is
real change within it.  But I think it goes much further than that.
Because I see the coffee maker as having no less of an ontological
status than a rock (or an electron for that matter), I am struck with
the happy thought that man's actions do contribute---in part!---to
the actual construction of reality.

Part of my quest for the last three years has been to find some
precise counterpart to all that within quantum theory itself.  And I
surely think it is there.  The most clear-cut example of it is in the
existence of quantum key distribution schemes.  As I see it, these
examples are trying to tell us something very deep about the story of
quantum theory; I think they capture the ontological content of the
theory.  It is this idea that I would like to sharpen up to the point
of being useful.

Here is my question.  Suppose in a few years I could come up with a
clear, precise statement of what it is that I'm trying to get at.
Something in essence that takes away the vagaries of the statement:
``The world in which we live happens to have a funny property.  It is
that {\it my\/} information gathering about something you know,
causes {\it you\/} to loose some of that knowledge \ldots\ and this
happens even in the case that you know all my actions precisely.
Physical theory, and quantum mechanics in particular, is about what
we can say to each other and what we can predict of each other in
spite of that funny property.''  Would that constitute something that
fulfills your Desideratum \#1?  It would be my claim that Hilbert
space vectors, hermitian operators, unitary operators, and even
consistent histories for that matter, all lie on the side of the
subjective in the theory:  they are all pieces of the theory that
concern what we can ask, what we can say, and what we can predict.
(They have no ontological content \ldots\ and it is a sad thing that
people have been searching for so much ontology in them for so long.)
On the other hand, the ``funny property'' does strike me as having
some ontological content.  And it does have a (rather ethereal)
counterpart in the theory.  Would pinpointing that precisely meet
your Desideratum \#1?  Especially, if in excising the vagary in
quotes above, we will be forced to acknowledge measurement as a
primitive for the theory? (Please note the distinction between a
primitive for the theory and a primitive for the world in itself.)

\section{30 August 1998, \ ``Well Look At That''}

Just a short note this time.  Right after writing you about coffee
makers and electrons earlier today, I ran across the following little
passage in {\Heisenberg}'s book {\sl Physics and Beyond}.
\bq
A few hundred yards away, a large liner was gliding past, and its
bright lights looked quite fabulous and unreal in the bright blue
dusk. For a few moments, I speculated about the human destinies being
played out behind the lit-up cabin windows, and suddenly Wolfgang's
questions got mixed up with it all.  What precisely was this steamer?
Was it a mass of iron with a central power station and electric
lights?  Was it the expression of human intentions, a form resulting
from interhuman relations?  Or was it a consequence of biological
laws, exerting their formative powers not merely on protein molecules
but also on steel and electric currents?
\eq
The coincidence was too much for me to pass up, and so I found myself
copying the words into the computer.  (You know how I am.)

\section{02 September 1998, \ ``Vague and Vagary''}

Thanks for the appraisal.  Let me chew upon it for a while.
Certainly clarity is what I strive for \ldots\ and you're helping me
in that regard. You're right, I never do refer to ``consciousness.''
That word frightens me.  I don't even really have an interest in what
it means. So I would hope that I'm {\it not\/} trying to build a
picture of the world based on it---I don't think I am.  Besides that,
though, I do worry a little bit that you and I sometimes use the same
words when we mean different things:  I'll try to sort that out.

I apologize for the continued use of ``slogans'' in summarizing what
it is that I'm trying to get at:  unfortunately, they are all I've
got right now!!  But let me leave you with one more before I go to my
chewing.  It's part of the conversation I'm having with you in my
head right now, and I'm having a hard time staying quiet; maybe, just
maybe, it's the pithiest version yet for expressing what I mean by
saying that I'm a ``realist.''  As I see it:  Quantum theory is
nothing beyond WHAT WE HAVE THE RIGHT TO SAY in a world where
information gathering necessarily causes disturbance.  (Disturbance
to what?  To each other's descriptions, nothing more.)  Where's the
reality in this? Well it has to be there, or it would be awfully hard
to imagine how it is that we come to intersubjective agreement in an
only-partially predictable world.  But that reality, we have every
right to believe, must be even more numinous than {\Kant}'s noumenon!

\bak
I wouldn't query this if I didn't know your fondness for linguistic
discussions.  But since I do: do you really mean ``vagaries'' here?
Rather than, say, ``vagueness''.
\eak

Vagary: a whimsical, wild, or unusual idea or notion.  (Webster's
Encyclopedic Unabridged)  Try again and tell me if it still doesn't
make sense.  The last thing I want to do is be vague (though I'm
often forced to be!).

\section{17 January 1999, \ ``Cigar and the Greeks''}

On another subject, over the weekend I read {\Schroedinger}'s
``Nature and the Greeks,''  ``Science and Humanism,'' and ``Mind and
Matter.'' That helped me recall that you once placed this little
quote at the end of one of your letters to me:

\bq
\noindent In fragment D 125 \ldots\ [{\Democritus}] introduces the intellect
in a contest with the senses.  The former says ``Ostensibly there is
colour, ostensibly sweetness, ostensibly bitterness, actually only
atoms and the void''; to which the senses retort: ``Poor intellect,
do you hope to defeat us while from us you borrow your evidence?
Your victory is your defeat.''  You simply can not put it more
briefly and clearly.
\begin{flushright}
 --- Erwin {\Schroedinger}, ``Nature and the Greeks''
\end{flushright}
\eq

I didn't understand then why you put it there, and I guess I don't
even understand it now.  What struck you about this?  I've gotten the
impression over and over again that your inclination is that of a
``naive realist'' (no disrespect intended).  Have I missed something?

\section{20 January 1999, \ ``Clarity''}

\bak
Nature and the Greeks, yes, I'm glad it is propagating. Wonderful
book, even though he does go off on extended rambles before pulling
himself and his argument together at the end, just when you'd almost
abandoned hope.

Am I a naive realist?  I really should get my philosophical
self-definitions straight --- except that I know that e.g.\ where
consciousness is concerned I fit none of the standard philosophical
positions, most of which seem to me elaborately silly.

But well, to try to answer: it seems to me that consciousness is a
genuine phenomenon, that qualia are at least as real as tables, and
that anyone genuinely interested in understanding the world ought to
be at least as interested in the inner world as the outer.  Moreover,
that while consciousness obviously seems to inhere in particular
material objects (brains), its properties are not, according to our
present understanding, reducible to known facts about the material
world.  It's pretty much a complete mystery. So I'm on the side of
the senses, and of {\Schroedinger}, on this one
--- unless they press further and start to deny the existence of
the atoms and the void (which I hadn't understood them as doing ---
what was your impression?).
\eak

Thanks for clarifying where you lean for me.  My initial impression
of the {\Democritus} fragment was that it was pressing further (at
least for rhetorical purposes).  But now that I've read your point
of view, I see that it need not have been.  So I guess it's hard to
say where it was really going without looking at the context
surrounding it. Have you read an extended version of the fragment in
another source?

Actually, the more I sit her thinking about it, the more I think your
reading is likely to be on the right track.  {\Democritus} ultimately
invented a very fine set of atoms with which to compose the
consciousness.  Atoms understanding atoms:  silly or not, it was a
physical mechanism for explaining cognition.

\section{20 January 1999, \ ``Atomic Souls''}

\bak
Does your account of {\Democritus}'s very fine cognitive atoms (which
is new to me but seems --- making obvious allowances for era --- to
be in characteristic good taste) come from original, translation, or
commentary --- and if t or c, can you recommend?
\eak

You can find a discussion of them in {\Schroedinger}'s book itself.
It's in Chapter 6, ``The Atomists'', more particularly ``main
feature'' number five (pp.\ 78--80 of the Canto edition).

\section{03 February 1999, \ ``Noumenal Mental Atoms''}

\bak
By the way, I still think, contra {\Schroedinger}, that
{\Democritus}' picture of the soul as  built of finer atoms is a
rather good one
--- for its time, of course, of course.   D. saw the mind-body
problem as a real problem, saw (apparently) that it has no very good
resolution in his atomic model of the material world, but saw also
that it could have a physical solution if you accept that mind and
matter are both reducible, but to qualitatively different things.  A
characteristically sharp insight, maybe, not (as S. suggests) a
lapse of taste brought about by the metaphor of the soul as breath.

And --- though it's desperately unfashionable to say so, and I trust
won't be taken as a wildly naive comment: I do appreciate how strong
the counterarguments are --- we still are not absolutely sure
{\Democritus} is wrong on this.   (Are we?  Are you, I mean?)
\eak

Did {\Democritus} have the right intuition?  Actually I sort of hope
he did too.  I wouldn't really want my world made of two
fundamentally different things.  But my realism is much closer to
{\Kant}'s than yours (appears to me to be):  if mind and matter are
made of the same stuff, then I suspect they'll both be just as
noumenal.

\section{05 February 2001, \ ``The Eerie Parallel''}

\noindent [NOTE:  The following Kentism refers to a letter consisting
solely of the exclamation, ``Thanks for the pointers. But I'm not
quite sure what functionalism is!!'']

\bak
If I understand right, functionalists believe that statements about
mind states (conscious thoughts and perceptions, if you prefer) are
equivalent to statements about physical actions.  For example, on
this view, to be angry is precisely the same thing as to display
(perhaps subtle and well hidden) signs of anger.

It's a view which has little to recommend it except a pleasing sense
of answering a deep question with no work, and so naturally has
become very widely held and respected among philosophers of mind. In
this, it eerily parallels the {\Everett} interpretation. It is only
fitting that the two should be combined into a grander exercise in
question-begging.
\eak

At times in the past I've found myself wanting to be able to write
like Mark Twain, and like William {\James}.  Both had this way of
putting things---very different ways---that made me shiver from
seeing the truth in their thoughts.  Today I found myself wanting to
be able to write like you!

\chapter{Letters to Rolf {\Landauer}}

\section{15 February 1998, \ ``Law Without Law \ldots\ Or
Something Like That''}

This morning I read one of the papers you gave me last week
(``Information is Inevitably Physical'') and was reminded of your
interest in the idea that the laws of physics may not be written in
stone once and for all \ldots\ but instead may be contingent upon the
particular details of the universe itself \ldots\ and indeed may be
evolutionary \`{a} la {\Peirce}, {\Wheeler}, Lee {\SmolinL}, etc.  At
least one version or other of this kind of thing---not Smolin's
though(!)---has intrigued me for some time (essentially since
meeting John {\Wheeler} in 1985).  It's been sort of a hidden hobby,
somewhat connected with separating the wheat from the chaff in
quantum mechanics.  My personal writings on the subject are now
pretty voluminous, though not yet tight enough in substance that I
show them to too many:  most are in the form of emails to colleagues
such as you.

Anyway, if you're interested, perhaps I'll compile a small compendium
and send it your way when I get back from Tokyo March 15---I'm
traveling essentially every moment until then.  Maybe this will help
us open a dialog to pinpoint where some real progress might be made.

In the mean time let me tip you off to another article on the subject
that I ran across recently:  S. S. {\Schweber}: ``The Metaphysics of
Science at the End of a Heroic Age,''  in {\em Experimental
Metaphysics}, edited by R. S. Cohen, M. Horne, and J. Stachel
(Kluwer, Dordrecht, 1997), pp.\ 171--198.  Among other things, it is
nice in that it contains a good bibliography of several previous
efforts in the direction (that at least I wasn't aware of).

\subsection{Rolf's Reply}

\bq
By all means send me your thoughts; especially if I do not have to
guarantee, ahead of time, how much energy I devote to them.

I am not sure, either, how much energy I can devote to tracking down
all the related literature. I started saying my stuff in 1967, in an
article which was published in in a journal which is far from
obscure, and has been reprinted (admittedly in an obscure journal).
I have the delusion that it is the job of later arrivals to make
contact with my work, rather than the other way around. But I am not
inflexible about that.
\eq

\section{06 July 1998, \ ``Evolution and Physics''}

I remembered a few days ago that I had promised to start something of
a dialog with you about evolutionary notions of physical law.

The thing that brought this back to mind was a nice endorsement I
got from {\Carl} {\Caves} on some of the stuff I've written.  (I'll
place a piece of his note below in the case that it might give you
the fortitude to wade through some of my silliness.)

As I see it, there is a deep affinity between a key thought in your
writings and the one concept for which I have the greatest hope to
finally give us a clear view of what quantum theory is all about.  I
agree with you that physical law should be limited in its form by the
information-processing capabilities of physical systems.  The
ultimate reason for this is simply that ``physical law,''---i.e., the
equations and rules our students read in their textbooks and apply to
great merit---is itself a construct of information processing
systems.  (In this case, the information processing system is ``us,''
the community of physical scientists, but it need not have been so:
it could have been Deep Blue for all I care.)  There is the world
independent of us, and then there is our attempt at describing that
world in the most accurate fashion.  The two things are different,
and I don't see any need to make apologies for that fact.  However, I
think there is more to it than an ultimate ``discretization'' of the
present physical laws that you so often allude to.  This is where my
views about the essence of quantum theory come into the picture.  To
a very great extent, I believe that your edict about the connections
between information-processing limitations and physical law are
already well taken into account by present-day quantum mechanics.
(Don't get me wrong, though:  this is certainly not to say that we
don't have still a long way to go in forming a completely consistent
physics \ldots\ in particular, in the manner you suggest.)

The already-discovered information-processing limitation has to do
with a funny property of the world we happen to live in.  It is this:
{\it my\/} information gathering about a given physical system will
generally disturb {\it your\/} description and predictions for that
selfsame system. Nonetheless we, as communicating beings, must come
to a consistent description of what we see and know about that
physical system.  This, I ever more firmly believe, is the essence of
the quantum mechanical formalism.  Quantum theory is nothing more
than Bayesian-like reasoning in a world with such a funny property.
(This I see as a large chunk of my personal research program:
clarifying, delineating, and searching for holes in this point of
view.)  Perhaps \ldots\ to put it in a more amusing way \ldots\ what
I am saying is that I wouldn't be surprised if the whole edifice of
quantum theory couldn't be constructed from the singular fact that
``quantum cryptography exists.''  That is to say, Hilbert spaces, the
inner product rule, entanglement, and unitary evolution all from that
clean, simple idea {\Bennett} and {\Brassard} were the first to make
some currency of.

Fundamentally, I think this point of view is quite liberating.  For I
think it indicates that physics is far from the closed book that our
quantum cosmologist friends (in their great lack of imagination and
ultimate chutzpah) would have us believe.  Rather than knowing all
of physics, it seems more likely to me that we know almost nothing
at all about it.  In fact, the only thing that we've really got a
firm hold of is the most basic idea of all:  in learning about our
world, we change its course.  Now let's get to the real work of
understanding and making use of the greater import of this small
piece of knowledge.

Well I think that in a nutshell is where I stand in relation to
you.  If you've got any comments on fruitful approaches I'd like to
hear them.  I had planned on putting together a complete compilation
of my murmurings on this subject for you. But I discovered that
that's more of a task than I'm willing to undertake presently.  (As
you can see by the length of this note, I write a lot, probably much
more than I should.)  What I think I'll do instead is send you the
compilations that I had already sent {\Carl}. The first, most clear,
and---I think---most interesting set of thoughts is connected to
several letters I wrote to David {\Mermin} and a few other fellows.
The best part of that actually starts up on page [$X$] and goes on
to the end at page [$X+13$].  [NOTE:  This probably refers to the
things between Merminitions \ref{OldAsm10} and \ref{NewAsm} in the
present volume.]  The second thing in this connection is one of my
long letters to  {\Asher} {\Peres}; the most interesting stuff there
comes after Asherisms \ref{OldIsm7} and \ref{OldIsm10}.  Finally
I'll send a letter I wrote Bob {\Griffiths} explaining what I see as
the emptiness of the ``consistent histories'' business as anything
over and above textbook quantum theory.  All of these passages
tackle in one way or the other the idea that physics and the world
it describes can evolve in a nontrivial and interesting sense, and
this all has to do with the information-disturbance tradeoff
principle and information-processing limitations that you like to
point out. All the documents are written in standard \LaTeX.  If for
whatever reason you have trouble getting them printed out, please
let me know and I'll try to get you fixed up in another way.

\subsection{Rolf's Reply}

\bq
Maxwell and Boltzmann, {\Bohr} and {\Einstein}, sharpened up their
perception via exchanges. In the day and age of phone-mail and
e-mail that has largely disappeared. Most of our colleagues will
only write things down for publication; and I thought I was the only
one who still went beyond that (I just sent out some dozen or so
copies of an e-mail note re shot noise reduction in mesoscopic
samples). But you outdo me by many orders of magnitude. I am getting
ready to leave the lab. for a few weeks, and giving you only a hasty
first reply. I will also send you the latest version of what you
call ``discretization'' (not an accurate label). Your views have much
more in common with John {\Wheeler}'s ``Participatory Universe'' than
with mine.  No, there is no way that present day q.m., utilizing
ordinary continuum math., can possibly be taking my notions into
account. You believe that there is ``the world independent of us'',
and if put into such a strong anthropocentric form, I might agree.
But I do not think that in science it makes sense to discuss objects
whose behavior cannot be observed or predicted. Your view that the
world is doing its own thing, to the full precision allowed by our
laws of physics, even though we cannot ever check that, is not
exactly ``wrong'', but it is a matter of religion, not science.

You seem to understand ``consistent histories'' and I may need to get
back to you on that, some day, in my non-converging attempt to catch
on to what is right and wrong there. But first I need to do more
homework.
\eq

\section{28 July 1998, \ ``Precision''}

Let me comment on one of your points.  I'll try to have something
more intelligent to say about your thoughts after I see your new
article.  [You say,]

\bq
\noindent Your view that the world is doing its own thing, to the full
precision allowed by our laws of physics, even though we cannot ever
check that, is not exactly ``wrong'', but it is a matter of religion,
not science.
\eq

Indeed I do think it is my view that the world is ``doing its own
thing'' \ldots\ but that should be qualified.  For it doesn't seem to
me that the ``doing its own thing'' before information processing
units evolved and the ``doing its own thing'' after they evolved are
precisely the same.  The point of the idea of an evolutionary
universe is that it evolves:  if the course of that evolution were
written in stone at the beginning of time, then there would hardly be
any reason to invoke a notion of evolution.  In any case, I don't
believe that there's anything in the writings I sent you that would
imply your phrase ``to the full precision allowed by our laws of
physics.'' You're fighting a battle with someone else in your mind
there; it's not me.  I don't see any reason to imagine a one-to-one
correspondence between {\it our\/} laws and the stuff out there.
You're right, my view does have some affinity with John
{\Wheeler}'s:  I wouldn't doubt that the stuff out there ultimately
has no laws at all.

\subsection{Rolf's Reply}

\bq
Thanks for the clarification. I am not yet on a first name basis
with evolutionary universe concepts; that barrier may take some time
to disappear.
\eq

\section{08 October 1998, \ ``Info is Physical''}

I have room in a very tight conference proceedings paper that I'm
writing to make one citation to your phrase, ``Information is
Physical.''  Do you have a favorite paper that I should cite?
Should I make the citation to your earliest mention of the phrase?
Or should I make the citation to what you think is the clearest
statement of it?  Please make the decision for me and send me the
correct reference.  My paper is going off to the editors tomorrow,
so if you could send me a quick note that'd be great!

\subsection{Rolf's Reply}

\bq
The concept, but not that phrase, first appeared in R.~L. {\sl IEEE
Spectrum\/} vol 4, issue \#9, pgs.\ 105--109 (1967). (Like for {\sl
Physics Today}, the page numbering restarts with every issue.) The
exact wording PROBABLY first showed up in my 1991 {\sl Physics
Today\/} paper. But it seems best to cite an early or a very recent
paper. Two are on their way into print. One that is likely to appear
within a few weeks: R.~L. in {\sl Feynman on Computation 2}, ed.\ by
A.~J.~G. Hey, Addison Wesley, Reading (1998?). The title of that one
``Information is Inevitably Physical''. About 11 days ago at a
session in Helsinki I complained to my audience: I have gotten a
fair amount of acceptance for that phrase, but not for the message
attached to it. But thanks for checking.

P.S. I'll mail both papers, even though they will arrive too late.
But if you take an instant dislike to the one you used, you can
scratch it at galley proof time.
\eq

\chapter{Letters to {\Hideo} {\Mabuchi}}

\section{22 November 1997, \ ``Martha White and Her Flour''}

\noindent [Note:  Words supplied by {\Greg} {\Comer}.  See notes to  {\Asher}
{\Peres}, dated 17 October 1997, {\Greg} {\Comer}, dated 25 November
1997, and {\Carl}ton {\Caves}, dated 21 December 1997.]\medskip

\bv
Martha White Theme Song \medskip
\\
(Open with a Banjo break by Earl {\Scruggs})\\
(Next, come in with Lester {\Flatt} singing \ldots)\medskip
\\
Now you bake right,\\
(other band members respond:) ah ha,\\
With Martha White, \\
(band response:) yes Mamm, \\
Goodness gracious, good and light, \\
With Martha White.\medskip
\\
Now you bake better biscuits, cakes, and pies,\\
With Martha White Self-Risin' Flour,\\
(band response:) that one all-purpose flour,\\
With Martha White Self-Risin' Flour \\
You've done all right! \medskip
\\
(Finish with an Earl {\Scruggs} banjo solo.)
\ev

\section{05 October 1998, \ ``Oh Magic Eight Ball''}

{\Democritus} said, ``All the world is but Atom and Void.''  I've
decided (with {\Kiki}'s help) that I much more like the metaphor,
``All the world is but Magic Eight Balls and Questions.''  See the
resemblance?  You couldn't have atom without void; but you could have
void without atom. You couldn't have Questions without Magic Eight
Balls; but you could have Magic Eight Balls without Questions.  The
sole remainder of archaic atomism is that ``atoms'' be
``repositories.''  Not repositories of ``properties.''  But
repositories of answers to the questions we might ask.  To keep these
sentences from being contradictory, those answers can't exist before
we ask them (else they be properties).  I'm sure all this has left
you cold:  but I'm in the middle of one of the fuzziest feelings I've
had yet.

\section{01 July 2000, \ ``Destinkifiers''}

Thanks for those nice compliments.  I've been meaning to write you
for quite some time, but there was always something:  my email
contact has gone down drastically in general.

Anyway congratulations on that big award.  I was much impressed by
your speculation on the Caltech webpage on how you might spend it. I
say ``here, here'' for a little creativity in our field.

I'm in Greece right now, just finished with the NATO meeting.
Tomorrow morning I leave for Capri (the QCMC conference), and then
finally join {\Kiki} in Munich at the end of the week.  My talk was
pretty successful in Mykonos; I was pretty happy with it.  For Capri
I'm going to make a completely new one, this time based on the stuff
I did with Kurt {\Jacobs}.  I've decided the best way to say what
I've been hoping to get at.  Question:  ``If the wavefunction isn't
real, then what is it that IS real about a quantum system?''  Answer:
``The locus of all information-information tradeoff curves that one
can draw for such a system.''  (I've decide to stop calling it
information-disturbance because it conveys bad imagery and
preconceptions.  The disturbance is to information, so why not just
make it explicit.)

OK, gotta get up tomorrow at 4:30.  Take care \ldots\ and spend that
money wisely!

\chapter{Letters to David {\Mermin}}

\section{18 September 1996, \ ``Theorem II''}

I encountered your ``Ithaca Interpretation'' paper this morning on
the {\tt quant-ph} archive \ldots\ and, I must say, I've been walking
around with a nice feeling since.  There are some things in it that
I like very much!

Your Theorem II (or a very slight variation of it) has indeed been
proven before, by {\Bill} {\Wootters}.  I'll attach the citations
below along with part of a note {\Bill} once sent me.
(Interestingly, one can also make a cut between real, complex, and
quaternionic Hilbert spaces based on such a theorem.)

About your Theorem I, if I am not mistaken, it also has a history
that predates the references you gave.  I believe Richard {\Jozsa}
once told me that a fellow named Hadjisavvas found it well before
HJW and {\Gisin}.  If you are interested you can contact Richard
directly; his e-mail address is \ldots.

As I say, I enjoyed your paper very much!  (Presently) I think my
only point of departure in what you say is about the issue of
``objective probability.''  {\Carl} {\Caves} and I (in still another
paper cited below which can be found on {\tt quant-ph}, pp.\ 22--26
in particular) have advocated the view that even the probabilities of
quantum theory should be interpreted in a Bayesian or subjective
way.  We prefer to say that it is the ``indeterminism'' of quantum
mechanics that is ``objective,'' and not the probabilities
themselves.  One always assigns probabilities based on incomplete
information; it is just that in quantum physics ``maximal
information is not complete.''

If you have any comments on our (somewhat garbled) thoughts there, I
would very much appreciate hearing them.  Also, you may find the
various references to ``objective probabilities'' in it useful.  (If
you would like a much more extensive list; I can compile that for
you \ldots\ the exercise would also be useful for me: I have an
amorphous pile of things in my file cabinet at home.)

I meet {\Carl} in Dallas Monday for a flight to Japan \ldots\ so I'll
have several hours at my disposal to make him think a bit more
philosophically again.  I'll show him your paper then \ldots\ and
that'll also give me a chance to read it still more carefully.

\begin{enumerate}
\item
W.~K. {\Wootters}, ``Quantum mechanics without probability
amplitudes,'' {\em Foundations of Physics}, vol.~16(4),
pp.~391--405, 1986.

\item
W.~K. {\Wootters}, ``Local accessibility of quantum states,'' in
{\em Complexity, Entropy and the Physics of Information} (W.~H.
{\Zurek}, ed.), (Redwood City, CA), pp.~39--46, Addison-Wesley, 1990.
Santa Fe Institute Studies in the Sciences of Complexity, vol. VIII.

\item
C.~M. {\Caves} and C.~A. Fuchs, ``Quantum information: How much
information in a state vector?,'' to appear in {\em Sixty Years of
EPR} (A.~{\Mann} and M.~{\Revzen}, eds.), (Ann.\ Phys.\ Soc., Israel)
1996. Also {\tt quant-ph/9601025.}
\end{enumerate}

{From} {\Bill} {\Wootters}, June 7, 1996:
\bq
By the way, my favorite distinction between real, complex, and
quaternionic states is hinted at in my paper ``Local Accessibility of
Quantum States,'' in {\sl Complexity, Entropy, and the Physics of
Information}.  The issue there is this: given a measurement scheme
that is set up to ascertain the states of the {\it parts\/} of a
composite system, how well does it determine the state of the
whole?  For a complex space, the measurement scheme gives you just
what you need to determine the state of the whole system.  For a
real space, it gives you less than what you need, and for a
quaternionic space, it gives you more than what you need.  (Ben
{\Schumacher} called this the Goldilocks principle.)  Just in case
you feel like verifying that statement and you actually have time to
do so, the function $g(N)$ in my paper--which is the number of real
parameters necessary to specify an arbitrary mixed state in $N$
dimensions--is $(N^2 + N - 2)/2$ for a real space, $N^2 - 1$ for a
complex space, and $2N^2 - N - 1$ for a quaternionic space.  (You
probably knew that anyway.)
\eq

\section{13 October 1997, \ ``Caltech and Propensities''}

I've heard rumors that you'll be coming out to Caltech for a visit
soon.  If so, when?  What are the dates?  I'd like to make sure that
I'm here then.

By the way, I'm much more versed in ``propensity'' / ``objective
probability'' stuff than I was this summer.  If you come, I'd like to
talk about such things again (in connection to quantum mechanics).
Plus, I still owe you that bibliography.

\section{09 December 1997, \ ``The Philosophy of `Nothing-More'-ism''}

\ldots\ perhaps it's appropriate for a paper about Correlation
Without Correlata.  (If you rotate and reflect those initials, by
the way, they're my old graduate advisor's.)

Anyway, did you realize you use the phrase ``nothing more'' 11 times
in this paper.  By the time you're finished, there's not a lot left
to quantum theory \ldots\ but I guess that's the intention.

\section{10 December 1997, \ ``Deconstructing Ithaca''}

I feel as if I'm caught in a tornado of deconstructionism, with your
friend {\Derrida} outside my window---pointed nose, broomstick, and
all---laughing insanely.  I look at your phrase ``correlation without
correlata'' and see a story of quantum mechanics as purely
knowledge-theoretic in character.  You look at the same phrase and
see a call to arms for objective probability (propensity).  We are
reading different stories, both presumably consistent, but still
almost antithetical.  The shocking thing is that it's the same text!

Anyway, advisable or not, I decided to read your paper for a third
time! And then I decided to read the original IIQM paper again.  And
after that I went back to {\Everett}'s original relative state paper
and {\Wheeler}'s assessment of it.

Now it is time to reach into myself: I think I really am ready to
write now.  So standby, prepare for goulash.  Things should slowly
start to trickle in through the night and into tomorrow.

\section{10 December 1997, \ ``Game and Format''}

I don't think I told you, but the name of my laptop on our local
network here is ``tychism.''  If you look up that word in a Webster's
dictionary, you'll find, ``(in the philosophy of {\Peirce}) the
theory that chance has objective existence in the universe.''  If,
on the other hand, you look in {\Peirce}'s essay, ``Evolutionary
Love'' ({\sl The Monist}, 1893), you'll find that he draws a fine
distinction between the usage of four words in all:  tychism,
tychastic, tychasticism, and tychasm.  The last of these refers to
the actual operative principle or mechanism in the world that
tychism is concerned with. If you will, it is the correlatum of the
sentence, ``Our world is governed by tychism.''

I've been thinking about how I'm going write all these things that
I'd like to write to you.  When I played the game with  {\Asher}---as
you saw last week---I'd highlight a passage of his paper, dub it an
``Asherism \#$x$'', and then make some comment.  It seems to me that
I should do something no less grandiose for you.  But I wouldn't want
you to feel cheapened or secondary by my doing exactly the same.
(You can see where this is leading.)

Here's what I'll do.  I'll comment on your paper in the same way I
did before, but now the quotes will each be designated ``Merminition
\#$x$.''

What I plan to do is send you separate installments with each
Merminition as I write it.  When the process is complete I'll lump
them all into one long file and send it again.  But these are the
plans. Right now, I'm off to dinner.  Then from 8:30 to 9:30 I have
my favorite TV shows of the week.  We'll see what happens after that.

\section{14 December 1997, \ ``An Idle Thought''}

\bdm
Meditating on your insistence that I simply have to say more about
the relation of IIQM to many worlds, I'm struck by the fact that I
am also told by {\Griffiths} that I simply have to say more about its
relation to consistent histories, and by {\Gottfried} that I simply
have to say more about its relation to traditional descriptions of
measurement and the problematic character of ``observables'' as
described, for example, in his book.

If everybody (I exaggerate) from all these mutually incompatible
schools thinks that I am so close to simply restating what they have
known all along, then maybe I really am on to something!
\edm

It seems to me that that's probably a bad sign!  I suspect that when
we finally understand ``what quantum mechanics is trying to tell us''
it will be rather clean and decisive.  If it helps, I'll be your
beacon of opposition:  I'm more inclined to the view that the quantum
formalism has almost nothing to say about {\it Reality} \ldots\ only
that {\it my\/} information-gathering questions/measurements disturb
{\it your\/} predictability.  At least that's the only firm
ontological lesson that I've been able to draw from the theory.  But
like you \ldots\ and unlike {\Griffiths} \ldots\ I am not 100\%
convinced that I'm on the right track---that's why I've been willing
to study your paper.  In any case, I don't see how we can reconcile
our two antithetical views.  So there.

\bdm
But bear in mind that I am not out to convince the world that
this is the answer. (I'm not convinced myself.)  I only want to
persuade a few smart people that this might be an useful way to think
about things.
\edm

I have born that in mind \ldots\ or at least, looking at myself
through my own eyes, it looks like I have.  My only point was that
maybe you should go after a few of the ``dumb'' ones too.

\section{17 December 1997, \ ``Exercise in Anti-Lucidity''}

The document is getting a little unwieldy, and I think I've reached a
convenient breaking point.  So I've decided I'll send on now what
I've written so far.  [See note to David {\Mermin}, titled ``Big Bowl
of Anti-Lucidite,'' dated 17 December 1997.]  I'd still like to say a
few things about what you've written on the modal interpretations
and the consistent histories stuff, your distaste of
anthropocentrisms, and, most importantly, on objective
probabilities.  But I need a bit of a break, and it's getting hectic
here:  Sam {\Braunstein} arrived Sunday. I do want to complete this
project and have these things sitting in the archive of my mind.
Perhaps, with a good finger crossing, I'll have some more in your
mailbox with the weekend.  But certainly move on and put your paper
on the archive in the mean time if you wish.

The document that is about to arrive includes some of the things that
I've already sent you.  However, I edited those old things slightly
so that they more accurately portray what I really think \ldots\ and
where I have some doubts.

In all, looking back over the multitude of words, I see that I don't
say nearly as much as I had wanted to.  (Make sure you understand me
properly on this:  I say a lot of words, but those words don't say a
lot!)  In any case, I did put an honest effort into it; I tried very
much to be unbiased and to learn from you.  And I did learn a lot
from this exercise \ldots\ though maybe the fruits will be
slow-growing in a recalcitrant Copenhagenist.  We shall see.  Any
reactions you have on what I've written will be most welcome.

\section{10--17 December 1997, \ ``Big Bowl of Anti-Lucidite''}

\subsection{A. My Outlook}

My strongest and most basic agreement with you is this.

\bdm
The IIQM does not emerge from a general view of the world out of
which quantum mechanics is extracted; the strategy is rather to take
the formalism of quantum mechanics as given, and to try to infer
from the theory itself what quantum mechanics is trying to tell us
about physical reality.
\edm
This attitude, along with the paper's title, is what I like above
all else with your efforts.  The firmest thing of all in quantum
mechanics is the formalism itself.  It's hard to see how there can
be any better starting point for a deeper understanding than that.
In addition to this, however, I really, really do like the business
you emphasize of focusing on correlations.  I have always liked
points of view that subordinate the state vector to probabilistic or
correlative statements.  And I think you are doing grand justice to
clarifying the similarities between ``measurement interactions'' and
general ``correlating interactions.''  Viewing the {\Wootters} and
HJW theorems as statements of some foundational importance is a key
insight, and it strikes me as a particularly fruitful strategy for
getting at something real.

Because you write so clearly on all these things---and you are
careful to state what you know and what you don't know---this, in my
eyes, is going to be an important paper.  I am glad that you're going
to push this to the point of publication.

But, what is my outlook about the ultimate content here?  I've said
it before, I'll say it again:
\bq
\noindent
I feel as if I'm caught in a tornado of deconstructionism, with your
friend {\Derrida}\footnote{I had promised you a reference to a decent
article on the connections between {\Bohr} and {\Derrida}'s thoughts.
It is: J.~{\Honner}, ``Description and Deconstruction: Niels {\Bohr}
and Modern Philosophy,'' in {\sl Niels {\Bohr} and Contemporary
Philosophy}, edited by J.~{\Faye} and H.~J. {\Folse} (Kluwer,
Dordrecht, 1994).} outside my window---pointed nose, broomstick, and
all---laughing insanely.  I look at your phrase ``correlation
without correlata'' and see a story of quantum mechanics that hints
of being purely knowledge-theoretic in character.  You look at the
same phrase and see a call to arms for objective probability
(propensity). We are reading different stories, both presumably
consistent, but still almost antithetical.  The shocking thing is
that it's the same text!
\eq

I hope some of the things I say below will help elucidate this.
Failing that, I hope it helps me identify how I am trying to {\it
force\/} quantum mechanics to ``emerge from a general view of the
world out of which [it] is [to be] extracted.''  You do realize that
I go to such lengths in writing things down, most selfishly, so that
I can look at what I'm thinking.  Rarely do I really and completely
do it for the designated recipient of the letter!  I hope you will
indulge me in my self-education.  You are my point of departure.

\subsection{B. Teleported Goulash}

\subsubsection{\protect\hspace*{.2in} I. Immediate Goulash}

I'll start with the two points I tried to emphasize to you the other
day.  If I can make any strong suggestions for the present paper, it
is these two.  The remainder of my comments, concern more the
substance of the IIQM than the content of anything you'll be putting
on the web next week.

\bdm
The central (and possibly fatal) conceptual difficulty for the
IIQM is the puzzle of what it means to insist that correlations {\it
and only\/} correlations have physical reality.
\edm

This point is buried on page six of a 42 page paper.  To be sure, you
do close the paper with a weak reiteration of the point:  ``Whether
this is a fatal defect of the IIQM, or whether it is a manifestation
of the primitive state of our thinking about objective probability,
remains to be explored.''  But, I guess I want to take everything
you've written pretty seriously (regardless of whether it is or is
not on the mark) \ldots\ and I want other people to do the same.

Therefore, I think it would be immensely useful to have a summary at
the end of the paper delimiting---in the most precise way that you
presently know---the problems, prospects, and challenges of the
Ithaca Attitude.  You do after all say over and over again that the
project is not complete.  In what ways can a young researcher hope to
make a contribution to this (possibly) final turn in our
understanding quantum mechanics?  And what will the completion of
this project give us beside a warm fuzzy feeling that there are no
longer any mysteries? Will its completion mean that we, the physics
community, can finally go full-head tilt in exploring quantum
cosmology?  Will it possibly give us any insight into dealing with
something so mundane as Kitaev's nearest-four-neighbor-interaction
lattice model for a quantum computer?  Of course, I don't mean for
you to address these particular questions specifically, but some
inspirational words would be most useful.  I believe it would make
this paper a more lasting contribution.

\bdm
It is at this point that the IIQM comes closest to the many-worlds
extravaganza.  The original paper of {\Everett} said nothing
whatever about many worlds.  {\Everett} took the view that the
quantum state ought to be understood only in a {\it relational\/}
sense, but the relations he emphasized were not as straightforward
as the subsystem joint distributions emphasized here.
\edm

I think that I have a strong feeling now for how {\bf
IIQM}$\;\ne\;${\bf {\Everett}}, but that comes after some relatively
long thought and (most importantly) after some personal contact with
you.  The general reader may not have the opportunity for either of
these methods of clarification.  About the confusion in general,
though, clearly there is something amiss here:  otherwise, you would
not have had people both at Caltech and Montr\'eal making the
comments that they did.  It seems to me that your discussion of
{\Everett} should be expanded substantially.  And it needs to come
somewhere in the paper {\it long\/} before page 22.

The things you have to fight hard against dispelling are the
suspicious similarities between the two trains of thought.  Both you
and {\Everett} purport to have {\it only\/} unitary evolution.  Both
you and {\Everett} have a universe in which all possible
``somethings''---whatever they be---simultaneously exist, even though
their (naive) derivatives are mutually inconsistent.  In your case it
is the joint probabilities that are simultaneously existent, though
the conditional probabilities they seemingly give rise to cannot be.
In {\Everett}'s case it is all possible relative-state
decompositions of the state vector that simultaneously exist; it is
the facticity of the ``worlds'' so-derived that cannot be imagined
to fit within a single reality.  (That is, unless you go so far as
the extravaganza that you say.)

The main thing that you have to be on your guard against---and I'm
trying to get it across that you've yet to do it sufficiently---is
the loose associations that people have already drawn in their minds.
Such a thing is so easy to do when there are no equations involved in
a paper!  Almost anyone who is going to read this paper is likely to
have a long history of having read (at least cursorily) things about
the many-worlds interpretation.  And their dissection of your ideas
is not going to be helped in any way by the many things that have
already been said in the literature.  Let me give you two strong
examples of where the trouble is going to come from.  I'll just pull
two quotes out of {\Everett}'s original paper and {\Wheeler}'s
assessment\footnote{By the way, I learned the most amazing thing in
digging up these papers yesterday.  The particular volume of Reviews
of Modern Physics containing them displays back-to-back {\it
eight}(!) papers from John {\Wheeler}'s group.} of it (that appears
immediately afterward):
\bq
{\bf {\Everett}:} Thus with each succeeding observation (or
interaction), the observer state ``branches'' into a number of
different states.  Each branch represents a different outcome of the
measurement and the {\it corresponding\/} eigenstate for the
object-system state.  All branches exist simultaneously in the
superposition after any given sequence of observations.
\eq
and
\bq
{\bf {\Wheeler}:} Another way of phrasing this unique association of
relative state in one subsystem to states in the remainder is to say
that the states are correlated.  The totality of these correlations
which can arise from all possible decompositions into states and
relative states is all that can be read out of the mathematical
model.
\eq
The first quote comes darned close to contradicting what you say in
this Merminition.  The second quote (especially the last sentence
thereof) comes darned close to saying, ``correlation without
correlata.''  (Or, at least to the uninitiated it does.)  The main
thing is to be long-winded on these points:  it can't hurt.

\subsubsection{\protect\hspace*{.2in} II. Big Bowls of Goulash}

\bdm
According to the IIQM the only proper subjects for the physics of
a system are its correlations.  The physical reality of a system is
entirely contained in (a) the correlations among its subsystems and
(b) its correlations with other systems, viewed together with itself
as subsystems of a larger system.
\edm

With correlata excluded from the realm of physical discourse, what is
it that determines the identity of these subsystems?  What mechanism
is left for drawing some kind of line in the sand?  Can you see my
worry about this?  Take the four dimensional Hilbert space ${\cal
H}_4$.  There is a continuous infinity of ways of decomposing it into
something of the form ${\cal H}_2\otimes{\cal H}_2$\@. Presumably you
want to single out one particular cut in this way as being physically
real.  (You give some hint of this on the next page with your phrase,
``A possible further requirement \ldots\ .'') So, is that cut just
built in at the outset?  Physical reality consists of a Hilbert
space, a fixed particular decomposition into elementary tensor
product spaces, and a vector on that Hilbert space \ldots\ oops,
can't say vector \ldots\ must say the complete set of all
correlations.  I don't suppose this is any real big deal.  But it
does impose one fixed absolute structure above and beyond the Hilbert
space itself.  If there were correlata, then the existence of the cut
would be of no conceptual difficulty:  that's what the world is, just
a lot of separate thingies.  But if you banish the correlata, why do
you not banish the fixed cut while you're at it?  What compels you to
keep it?

\bdm
\label{OldAsm5}
The crucial formal property of a resolution into subsystems is
that all {\bf observables\/} associated with one subsystem must
commute with all {\bf observables\/} associated with any other
distinct subsystem.
\edm

Note the emphases that I've added to ``observables'' in this
Merminition.  Doesn't it bother you that so much of what you are
trying to formulate relies on the old ``anthropocentric''\footnote{I
put the scare quotes here because I don't really think that speaking
in terms of measurement and observables is overtly anthropocentric.}
concepts that you're trying to do away with in the first place?  If
we have no correlata, and we don't truly have any need of
measurement for understanding what quantum theory is about, then
what is it that singles out ``observables'' as the relevant concept
with which to construct an interpretation?

Perhaps I haven't amassed enough evidence for this complaint yet, but
every time I see something in physics or philosophy that has to
strain so to get off the ground, I think of a piece of a Paul
{\Simon} tune.  And it's going through my head right now, so I might
as well say it:
\bv
I don't know why I spend my time\\
Writing songs I can't believe\\
With words that tear and strain to rhyme.
\ev
But let me try to go into more detail on this point, instead of
fixating on a diatribe.

\bdm
\label{OldAsm6}
By correlations among subsystems, then, I mean nothing more
than the \ldots\ [probability distributions] \ldots, at any given
time, \ldots\ [derivable from] \ldots\ system observables (hermitian
operators) that consist of products over subsystems of individual
subsystem observables.
\edm

Allow me to try to say more clearly what I was trying to tell you at
the chalk board the other day.  Consider a universe described by
${\cal H}_2\otimes{\cal H}_2$.  ``Physical reality'' in this
universe consists, according to your prescription, of all joint
probability distributions $p(\hat a_i,\hat b_j)$ that can be derived
from some fixed {\it pure\/} quantum state $\hat\rho$ via
\be
p(\hat a_i,\hat b_j)={\rm tr}\!\left(\hat\rho(\hat a_i\otimes\hat
b_j) \right)\;,
\ee
where the $\hat a_i$ and the $\hat b_j$ both form complete sets of
orthonormal projectors.  Thus, physical reality is captured by the
mathematical model of a set with a certain structure---in this case,
the set of joint probability distributions derivable from the
antiquated, ``anthropocentric'' notion of (standard, von Neumann)
measurements.  But why that particular structure and not another?

You shore up your argument for this by pointing out that if one knows
all these probabilities, then one can automatically infer the quantum
state $\hat\rho$.  But you don't need nearly that much structure to
recover the quantum state.  Indeed you only need consider the joint
probability distributions formed from three complementary observables
on the A side and three complementary observables on the B side.  You
don't need the gross overkill of a continuous infinity of observables
on both sides.  So why not pick your six favorite local observables
and call the set of nine probabilities distributions derivable
thereof physical reality?

Well the reason you don't is that, though it doesn't take so many
observables to determine a state, there are nevertheless loads of
other observables that are measurable \ldots\ when we insert an
external observer into the game.  So you go to textbook quantum
mechanics and pull out the Hermitian operators as the relevant set
with which to construct your interpretation.

But why stop there?  External observers can play lots of games with
this toy 4-D Hilbert-space universe.  In particular, they can imagine
measuring completely general POVMs on the A and B sides,
respectively, and build up joint distributions in that more
ecumenical way.  So why not take this as the starting point for an
interpretation?  This physical reality would be immensely larger in
scope than one based on Hermitian observables.

I know you may think it silly that I bother over this issue.  But
mostly I want to emphasize the somewhat arbitrary feel and
construction that your present ``correlation without correlata''
(CWC) has.  How can you better justify this picture of reality that
you're trying to draw?  Why, if you are searching for deep reality,
are you playing the role of a complemento-anti-{\Heisenberg}?  You
remember the story of how {\Heisenberg} came across the uncertainty
relations: he decided to let the theory tell him what could and could
not be measured. You wish---it looks like---to use the theory to tell
you what is measurable and then use that in turn to tell you what is
real.  If that's the case, and you want to de-anthropocentrize the
theory, shouldn't there be a better way to go from theory to reality?

\bdm
\label{OldAsm7}
\ [{\Wootters}'] theorem follows immediately from three facts:
\ldots\ (3) The algorithm that supplies observables with their mean
values is linear on the algebra of observables.
\edm
Have you thought about how the assumptions behind {\Gleason}'s
theorem might be weakened to help out your cause?\footnote{I know
you said you spoke to {\Gleason} recently, but it didn't sound to me
like your question to him was of this flavor.  What was it?  Can you
explain the idea a little more thoroughly?} For instance, one
could/should, based upon your sought-after interpretation, consider
the following.

The standard scenario behind {\Gleason}'s theorem is this:
\bq
The questions that I may ask of a physical system correspond to the
set of all orthonormal bases for a given Hilbert space.  The task of
physical theory is to provide a probability distribution over the
answers to those questions, i.e., the basis elements in each
orthonormal basis.  What is the most general form these probability
distributions can have?
\eq
For two-dimensional Hilbert spaces, this question doesn't pin down
anything interesting.  For three-dimensional and above, however, as
you know, one gets back the familiar {\Born} rule {\it for some\/}
particular density operator.  This warrants us to say that the set of
valid ``physical states'' for a system corresponds to the set of
density operators over some Hilbert space.

Just to repeat the logic behind this:  one starts with the set of
valid questions (measurements), and says that it is the task of
physical theory to give probabilities to the answers (outcomes).
{}From that, one derives the structure of the state space.

What can we do in the same ilk for the set of universes
crudely\footnote{I say ``crudely'' here because I don't want to
specify the set of physical states yet.} described by ${\cal
H}_3\otimes{\cal H}_3$ \ldots\ when physical reality consists of
``correlation without correlata?''  Since this is a nine-dimensional
Hilbert space, and $9\ge3$, we could snap {\Gleason}'s theorem on it
like a hand tool.  But that wouldn't be very interesting.  For, after
all, a basis consisting of vectors entangled across these two spaces
corresponds---using language you don't condone---to a set of
correlata with respect to a third system, one we're not allowing in
our toy universe.

However, could it possibly \ldots\ wonderfully \ldots\ be the case
that if we said, ``it is the task of physical theory to provide a
probability distribution for each orthonormal {\it tensor-product\/}
basis,'' then that would be enough to recover the full theorem? You
see it's not immediately clear that one could hope to get such a
thing.  In the way {\Gleason}'s theorem is proved presently, one of
the crucial first steps is to show that the ``frame function'' must
be continuous on the set of bases.  But here there is no possibility
of that---a probability distribution need not even be defined for a
basis of entangled vectors.

This is probably just a silly thought---it seems pretty unlikely that
it even stands a chance of being true---but it might be worth mulling
over.  In particular, if it isn't true, it seems you should ask: why
is it that the old question/measurement-based foundation can go so
far (through {\Gleason}'s theorem), but the CWC foundation isn't
powerful enough to get back the {\Born} rule without assuming it at
the outset?  That is to say, why does the CWC foundation start at the
{\Wootters} theorem and not earlier?

\bdm
This is another familiar tale.  The IIQM shifts the way it is
told, by emphasizing that the state of a non-trivially correlated
subsystem is never pure: the state of the specimen evolves
continuously from a pure state through a sequence of mixed states
into the ``post-measurement'' mixed state \ldots\ at the moment the
measurement interaction is terminated.  If at that stage one wishes
to regard the state of the specimen as undergoing an abrupt change,
it is at worst a collapse from a mixed state viewed in this
fundamental way, to the same mixed state viewed under the ``ignorance
interpretation''.  Since the internal correlations of the specimen
are exactly the same regardless of which view you take, the collapse,
if one chooses so to regard it, is rather ethereal.
\edm

I rather like this passage, especially the last two sentences.  From
my favorite point of view, the collapse is, by its very nature,
pretty darned ethereal.  It alone is the piece of quantum phenomena
that cannot be governed by physical law.  What else could you expect
of it {\it but\/} to be ethereal?

But more substantially, what is it in the IIQM that warrants the {\it
particular\/} transition from ``fundamental'' mixed state to
``ignorance'' mixed state that you are imagining here?  That is to
say, what is it in the IIQM that tells us the appropriate way to view
collapse is in terms of an orthogonal mixture of vectors (the
eigenvectors of the post-interaction density operator)?  What about
all those other decompositions of the density operator assured by the
{\Hughston}-{\Jozsa}-{\Wootters} theorem?  Yes, the measurement
interaction was set up just to elicit this possible picture of
things, an ignorance-based mixture of reliably distinguishable
states.  But remember, you wish to have neither ``real'' correlata
nor ``real'' observers out there to define a fixed set of questions
of which the measurement is about.

I guess the point I'm making here isn't so different from the point
(I had forgotten) that you make on the page following this
Merminition. But it seems to me even worse than you had
thought---this is because of all the extra density-operator
decomposition freedom.  Thus it seems to be that you are left with
one more onus for the conscious observer that you exclude from the
problem.  His role is to first define a decomposition and, second,
to play the measurement-problem equivalent of the separation between
past and future.

\bdm
\label{OldAsm9}
There is thus no quantum measurement problem for the {\it
internal\/} correlations of the specimen or the apparatus. \ldots\
The measurement problem survives only in the {\em specimen\/}-{\em
apparatus correlations} \ldots\ .
\edm

I still have a funny feeling about this that I'm trying to put my
finger on.  I suppose the problem goes all the way back to the main
declaration of your paper, ``Correlations have physical reality; that
which they correlate, does\footnote{Two points of grammar that I've
been meaning to mention.  (1) Shouldn't the pronoun `that' be taken
to be plural here?  So you would use `do' in the sentence?  (2) If
you look in a standard Webster's dictionary, you'll find 99.9\% of
all words starting with ``non'' nonhyphenated. Therefore, this has
been my standard.  In particular, I speak of nonorthogonal states,
nondemolition measurements, etc.  I'm not sure where physicists have
gotten the habit of hyphenating so much.  Is this an older protocol
that has fallen out of favor with the new dictionaries?} not.''  The
problem seems to stem from my willingness to think of {\it
entanglement\/} itself more as a physical ``reality'' than you, as
something different than the sum total of correlations.  Or, at
least, I like to think of it as a valuable currency in its own right.
You seem to only allow correlation to sit in that lofty spot.

The thing that's going on here is that ``correlation'' is linear in
the specimen's density-operator decomposition, whereas all reasonable
definitions of mixed-state entanglement that I know of, are not. For
instance, consider a specimen density operator
$\rho^{\rm\scriptscriptstyle AB}$ that arises from a tripartite pure
state $|\Psi^{\rm\scriptscriptstyle ABC}\rangle$ just as you wish:
$\rho^{\rm\scriptscriptstyle AB}= {\rm tr}_{\rm\scriptscriptstyle C}
|\Psi^{\rm\scriptscriptstyle ABC}\rangle\langle
\Psi^{\rm\scriptscriptstyle ABC}|$. Now focus on a particular
pure-state decomposition ${\cal E}=\{p_i,\Pi_i^{\rm\scriptscriptstyle
AB}\}$ for that density operator:
\be
\rho^{\rm\scriptscriptstyle AB} =\sum_i
p_i\Pi_i^{\rm\scriptscriptstyle AB}\;.
\label{McGeehee}
\ee
If we denote the partial traces of the states in this decomposition
by $\rho_i^{\rm\scriptscriptstyle A}= {\rm tr}_{\rm\scriptscriptstyle
B}\Pi_i^{\rm\scriptscriptstyle AB}$, then the (average) entanglement
$E({\cal E})$ of the decomposition $\cal E$ is
\be
E({\cal E})=\sum_i p_i S(\rho_i^{\rm\scriptscriptstyle A})=-\sum_i
p_i{\rm tr}(\rho_i^{\rm\scriptscriptstyle A}\log_2
\rho_i^{\rm\scriptscriptstyle A})\;.
\label{Sutton-Taylor}
\ee
Pretty clearly, this quantity depends on the particular decomposition
that is under consideration.

\bq
{\bf Aside:} Quantities like this find quite some use within quantum
information theory.  For instance, suppose Alice, Bob, and Charlie
share this tripartite pure state.  Moreover suppose that, though
Alice and Bob don't possess a pure state on their own, they need to
use their part for quantum-state teleportation.  Charlie, being their
friend, decides to help as much as he can, but suppose he's quite
some distance away and can't transport his system to them.  With that
constraint, the best Charlie can do is perform a measurement so that
A and B have a pure state conditioned on his outcome (as far as he is
concerned).  Then, if he transmits that classical information to A
and B, they can make use of it to teleport more effectively than they
could otherwise.

Now, if Charlie is smart, he'll try to force the pure state between
them to have as high of an entanglement as possible.  By the HJW
theorem, for any decomposition ${\cal E}$, there is a measurement he
can perform to generate it. Since Charlie can't generally predict the
outcome of his measurements, the best thing he can do is to try to
maximize his {\it expected\/} entanglement for Alice and Bob.  His
best expected entanglement we call the {\it entanglement of
assistance}\footnote{ This quantity (and what is known about it)
hasn't yet made a public debut.  This is something that I will
eventually write up in one of six papers I told you about!}
\be
A(\rho^{\rm\scriptscriptstyle AB})=\max_{\cal E} E({\cal E})\;.
\label{eq:ehelpdef}
\ee
The operational reason that this particular average is important is
because if Alice, Bob, and Charlie actually have many, many copies of
the state $|\Psi^{\rm\scriptscriptstyle ABC}\rangle$, this controls
the ultimate number of (perfectly entangled) singlets that Alice and
Bob can get with Charlie's assistance.
\eq

Where am I going with this?  The main point is that because
entanglement is nonlinear, {\it as far as Charlie is concerned\/} the
average entanglement between Alice and Bob will change dependent upon
which measurement he performs.  In this way, it is quite unlike the
``correlation'' between Alice and Bob that you speak of.  Now because
of this and your Desideratum \#5, it cannot be---in your
terminology---an objectively real quantity.

I think that may be fine with me:  nothing to do with the quantum
state is too objectively real if you ask me.  But still, it feels a
little funny that a quantity so close to correlation \ldots\ or
better yet, one that provides a measure of all-purpose correlation
\ldots\ isn't allowed quite the same objectivity status as its more
primitive, classically-oriented cousin.  Any thoughts on this?

\bdm
\label{OldAsm10}
If we leave conscious beings out of the story and insist that
physics is only about correlation, then there is no measurement
problem in quantum mechanics.  This is not to say that there is not a
problem. But that problem is not a problem for the science of quantum
mechanics but an everyday question of life:  it is the problem of the
nature of conscious awareness.\footnote{Grammar check:  too many
``buts.''}
\edm

The scary thing is that we don't differ so much on this point!  But
my take on this is that the thing really being emphasized here is:
quantum mechanics itself has nothing to say about the {\it real\/}
measurement problem.  The {\it real\/} measurement problem is
precisely that quantum mechanics does not address the ``mechanism''
that brings about the individual outcome that we and all the fellows
we talk to perceive.\footnote{Please don't confuse this statement in
the way that Bob {\Griffiths} did.  I am not exhibiting in it any
desire for a hidden-variable theory to underlie quantum mechanics.
I am perfectly happy with the indeterminism of the theory.}

The other day, you and John {\Preskill} scoffed at me when I said
that I am inclined to take ``measurement'' as a primitive of the
theory. Well, there is a sense in which that is exactly what you are
doing here!  Let he who has not sinned cast the first stone.  For the
purpose of clarifying quantum theory, you leave {\it this\/} notion
of measurement unanalyzed; it is taken as a given.  Albeit for the
world you are trying to construct, it is an incidental primitive
\ldots\ one that is not to be confused with anything inside physics
proper.  I, on the other hand, see it not at all as incidental:  it
seems to me that it is the big red flag that's telling us just
exactly what quantum theory is about.  The terms in the theory
correspond to what we can ask about the world and what we can predict
of the answers that it gives us.  When there were no people (or
artificially intelligent Dell laptop computers), there were no wave
functions.  Period.

But everyone wants an ontological lesson from the theory.  I guess I
am no exception.

\bdm
\label{OldAsm11}
It is a another remarkable feature of quantum mechanics (not
shared with classical physics, where external correlations are always
possible) that the totality of all possible internal correlations is
enough to determine whether or not any non-trivial external
correlations are possible.
\edm

This statement may be the only firm ontological lesson that can be
gathered from quantum mechanics.  It's the one that powers the
{\Pauli} quotes at the beginning of these notes, though said in such
different language.  It's the principle that lies behind quantum
cryptography.\footnote{C.~A. Fuchs, ``Information Gain vs.\ State
Disturbance in Quantum Theory,'' to appear in Fortschritte der
Physik.  Also in LANL archive, {\tt quant-ph/9611010}.}  The lesson
is this: {\it My\/} information-gathering measurements (about systems
that you have some knowledge of) disturb {\it your\/} predictability
(over them).

This ``lesson'' is constructed or couched in epistemological terms,
but it is ontological in nature.  It is a statement about a
significant {\it property\/} of the world.  Confronted with the task
of describing, manipulating, and studying a world with this property,
one is led to ask with what consistent means such a thing can be
carried out?  How can one hope to describe a world in which the
{\Kant}ian noumena are even further removed from observable phenomena
than {\Kant}'s lofty philosophical arguments could have dreamt of?

That's the sort of question that really piques my interest.  One has
to wonder whether such a thing---much more carefully
stated(!)---could pin down quantum theory as the only reasonable
answer.  Quantum theory, that is, with all its attendant silence
about measurement itself and its troublesome concept of entanglement.
Could it be that {\Bohr} was not so far off the mark when he
said,\footnote{I haven't yet confirmed the exact wording of this
quote:  it comes from {\Bohr}'s last interview (with {\Kuhn}), but I
was only able to get it second-hand.} \bq
\noindent
They have not that instinct that it is important to learn something
and that we must be prepared to learn something of very great
importance. \ldots\  They did not see that it [the complementarity
description of quantum theory] was an objective description, and that
it was the only possible objective description.
\eq
That's one thing that I'd really like to know.  More than that I
suppose, I'd like to know what one can do with such a wonderful piece
of newfound knowledge.

But still, you ask:

\bdm
Why should the scope of physics be restricted to the artificial
contrivances we are forced to resort to in our efforts to probe the
world?  Why should a fundamental theory have to take its meaning from
a notion of ``measurement'' external to the theory itself?  Should
not the meaning of ``measurement'' emerge from the theory, rather
than [the] other way around?  Should not physics be able to make
statements about the unmeasured, unprepared world?
\edm

I still see no completely satisfactory answer to these questions, but
I do strongly think that so much of this is forced upon us by the
``ontological lesson'' above.  Physics, and quantum mechanics in
particular, does say something about the unmeasured, unprepared
world:  the level of abstraction just seems to be one above where
you'd like to see it.  If you ever catch yourself saying the phrase,
``information is physical,'' like so many in the field of quantum
information do, then you will have gone a long way toward buying into
this.  I say the mantra of ``information gain vs.\ state
disturbance'' every morning with breakfast and know that I feel a lot
better because of it.

So, with this, let me try to summarize my present feelings about CWC:

\bdm
But to insist that physics is exclusively about measurement, is
unnecessarily to relegate to an inferior ontological status the more
general correlations among arbitrary subsystems.
\edm

I must say, I completely agree with this and, still, completely
disagree with it.  I have no problem with your feeling that the
conventional idea of measurement and more general entanglement---I am
reluctant to use the phrase ``general correlations''---are of the
same ontological status.  But that ontological status is zip, zero,
triviality.  Entanglement, either from a measurement interaction or a
general, nonteleological interaction, {\it to me\/} signifies the
information that is {\it potential\/} in one system about another.
That information can be activated by the supplement of a set of
correlata.  How this supplementation comes about is part of the {\it
real\/} measurement problem spoken of above \ldots\ the part you are
willing to sweep under the rug of the question of consciousness.

Correlation without correlata, as an ontological statement?  Can I
buy it yet?  Not yet.  As part of a larger story that we're still
getting hints of?  Yes, I think so.

\subsection{C. Pause}

So, where does this leave me?  Strangely, perhaps too predictably,
not so far from where I was on 5 December 1995, just over two years
ago.  On that day, I wrote my friend {\Greg} {\Comer} a summary of my
position in the form of a little table that I'll reproduce on the
next page.  Expressed in a fairly poetic way, you'll find something
that's not so far from your ``correlation without correlata.''  I
think it expresses fairly clearly why I've been drawn a little to
your paper \ldots\ and also why I've been repelled a little by your
paper.  Quantum theory still strikes me as a theory of ``what we have
the right to say'' in a world where the observer is no longer
detached.  If anything, what you've written here has helped
strengthen that impression.

But that's enough for now.

\section{07 January 1998, \ ``Cloud-Cuckoo-Land''}

After having my first cup of coffee this morning, I took your new
Section IX to that place where I do my best thinking.  And I found
that I liked what I was reading.  A little later I looked at the new
additions again and found that I still liked them.  So I guess that
means they're OK.

You do now say more clearly those things that needed saying the first
time around.  The only change that I might make if I were in your
shoes would be to add a little blurb that goes something like:
\bq
\noindent
Whereas the essence of IIQM can be described as ``correlation without
correlata,'' the essence of {\Everett} can be described as
``correlata without correlation.''
\eq
Indeed one of the standard research projects for the Everettistas
over the years has been to somehow {\it derive\/} the standard
{\Born} probability rule from the relative-state formalism itself.
{\Everett} tried his hand at it in the original paper \ldots\ then
{\Graham} in his PhD thesis tried to fix that up \ldots\ then
{\Benioff} in several papers tried to refine that \ldots\ and then,
finally, {\Deutsch} in his papers tried to argue that the project
needed to be given up and that the {\Born} rule should be taken as a
postulate for describing the {\it fraction\/} of worlds with one
property or other.  The main point is that in simple {\Everett},
there are no ``correlations'' in the classical, statistical sense
because there are no joint {\it probability\/} functions lying
around with which to work.  You might cash that in for something.

Also, let me comment that, in my heart, I still think you are being
too soft on Consistent Histories.  As far as I can tell, all the
mathematical, objective-sounding coating that they put on their cake
boils down to something every bit as {\it ad hoc\/} as {\Bohr}'s
metaphysical ramblings on complementarity.  {\it Why\/} are the
single-time ``events'' in a history one of a set of orthogonal
projectors? {\it Why\/} are the ``weight functions'' with which we
derive the consistency requirements just exactly the ones given by
the {\Born} rule in the first place?  These questions are never
asked by {\Griffiths}. If he were deadly serious about wanting to
ascribe a man-independent reality to a system at every moment of
time, why does he restrict himself to those structures already
existent in the old ``meas-

\pagebreak

\begin{center}
\begin{tabular}{|ll|} \hline
& \\
What is a quantum system?        & A line drawn in the sand.\\
& \\
What is a quantum state?         & What little we can say about
what's on the\\
& other side.\\
& \\
What is unitary evolution?       & Our way of saying that once we
know everything\\
& about our side, there can be no contribution\\
& to our state of knowledge about the other side\\
& without crossing the line to take a look.\\
& \\
What is wave function collapse?  & Something not so very different in
spirit than\\
& Bayes' rule for updating probability\\
& assignments in the light of new evidence.\\
& \\
What is a measurement model?     & A description of the particular
way Bayes'\\
& rule must be modified when we attempt to\\
& get some new information from the other side.\\
& \\
Modification to Bayes' rule?     & This is another way of expressing
what we\\
& have known so long about quantum phenomena.\\
& Information gathering measurements disturb\\
& quantum states.\\
& \\
Is the disturbance mechanical?   & No more so than Bayes' rule.  The
space of\\
& quantum states is just as ethereal as the\\
& space of probability assignments.\\
& \\
What is the sand the lines are   & It is not spacetime.  Spacetime
apparently\\ drawn in?                        & has nothing to do
with it.  The perfect case\\
& in point is the EPR thought experiment.  This\\
& is why E thought it to be the final blow.\\
& \\
Do we really need a sand in      & Probably not.  Did we need an
ether for the\\ which to draw the lines?         & field?  Did we
need a caloric for heat?  Did\\
& we need a phlogiston for the flame?\\
& \\
Ok then.                         & $+++++++++$ \\
& \\
What is a quantum system?        & A line drawn \ldots\\
& \\
What is a measurement outcome? $\quad$   & What we find when we ask a
question, fully\\
& prepared to make use of our modified Bayes'\\
& rule (measurement model) to describe the\\
& system afterward.\\
& \\
Where does it come from?         & The other side of the line.\\
& \\
\hline
\end{tabular}
\end{center}

\pagebreak

\noindent urement-interpreted'' formalism?  Let me give an
example. For instance, why not suppose that the elementary events
can always be drawn from a set of projectors that {\it do not\/}
give rise to a {\Kochen}-{\Specker} paradox?  That is to say, why not
base the notion of a ``framework'' on this larger set of events? For
a three-dimensional Hilbert space, one might be able to get away
with eight or nine {\it possible\/} events at each shot in time,
only one of which is {\it real\/} from the consistent histories
point of view.  (I say eight or nine because I do not know what the
ultimate record in the {\Mermin}-{\Peres}-Kernaghan
small-vector-number war will be.)

One last thing that just dawned on me.  \ldots\

Wait, wait, one more last thing:  I defer discussion of objective
probability once again.  Of the lot, that's the hardest thing to talk
about without writing a whole paper in the mean time.

\section{15 January 1998, \ ``Buttered Toast''}

\bdm
Could I impose on your sound literary and physical instincts once
more?  I've taken your advice to heart yet again, and put [\ldots]
\edm

Boy, you really know how to butter a guy up!

Anyway, I have looked at your new section, but I'm not exactly sure
what I want to say about it.  I am glad that you've added it, but I
wish it were longer!  I think that your saying ``the paper is already
too long'' carries no force at this point.  It seems to me that if a
potential reader is willing to get so far into the paper as to read
Section XII, then he's already demonstrated a substantial commitment
\ldots\ so substantial, in fact, that he's likely to want a little
more (like I did).  Three or four more pages won't kill him.  On the
other hand, I understand that you're losing steam \ldots\ and
something of the present order is likely to be the best that I can
hope from you.  So I am glad that you added what you did:  it does
help make the ending a little less lame.

A couple of points.  I did find the first paragraph a little
difficult to read.  The sentence starting with ``This shifts the
terms \ldots'' is almost Bohrian in its topology.  Either that's a
good sign or a bad sign, but I tend to suspect the latter.  Also in
the final sentence, you say, ``This question \ldots\ has not \ldots\
been asked.''  What you mean to say is something different: as you
point out above, it has been asked but the answer has always been
sought in a different way.  Beside that, overall, the section is a
little choppy \ldots\ but I guess there's no avoiding that if you
want to keep the length down (while still expressing the same
thoughts).

\bdm
I suspect our unfathomable conscious perceptions are going to
have to enter the picture, as a way of updating the correlations.  To
acknowledge this is not to acknowledge that ``consciousness collapses
the wave-packet'' but it is, I suspect, to admit that quantum
mechanics does not offer a picture of a clockwork world of eternally
developing correlation (described by ``the wave-function of the
universe''), but a phenomenology for investigating what kinds of
correlations can coexist with each other and for updating current
correlations and extrapolating them into the future.
\edm

This paragraph intrigues me, of course.  It is at that point that
your opinion and mine seem to converge most closely.  But along with
that, I have to wonder what it is that's keeping us so far apart
otherwise.  This remains a mystery to me that I want to understand
much better.

Your paragraphs \#4 and \#5 caused me a little narcissistic twinge.
You actually address one of the technical points in my long note!
Wow!  But alas, you address it only by edict:  that was a little
disappointing.  ``Isn't it plain to everyone that `measurement' is an
anthropocentric concept? \ldots\ But of course `system' \ldots\ now
`system' \ldots\ no there's nothing wrong with having a objective
preordained {\it cut\/} even when there are no correlata.''  Oh
well, we do what we can.  (This, by the way, induced me to carry the
narcissism a little further this morning: I actually reread my long
note to you. Looking back at it---this morning, at least---I decided
that I sort of like it after all.  Even I got something new after
reading it a second time.  Did you read it a second time?)

That's about it; that's all I can think to say.  You've emptied me
again.  I really have enjoyed working through and thinking about the
Ithaca point of view.  Maybe in the next couple months, while writing
the paper with {\Caves} and {\Schack} on Bayesianism (and thinking
about IIQM again), I'll be able to come to some decent perspective
on it.

Oh yes, and connected to that, one final thing \ldots\ one that got
my dander up:
\bdm
P.S.  Did you know that Sidney {\Coleman} claims to have made
respectable the introduction of probabilities into {\Everett} by his
adaption of an old argument of {\Hartle}?
\edm
``Claims'' is the key word here.   I would say that I am more than
aware of the argument:  all versions of it, including {\Coleman}'s
(as eventually published by {\Gutmann}) and {\Benioff}'s (which
preceded {\Coleman} by 20 years), make their ostensive progress by
slipping the notion of probability in from the back door.  After
slipping that by, the only thing they end up doing is proving
various versions of ``the law of large numbers'' by a long and
circuitous route. Dispelling some of the ``respectability'' of those
claims is what this paper I'm writing with {\Caves} and {\Schack}
will be about and work very hard to make clear.

\section{15 January 1998, \ ``Thanks''}

One last comment prompted by your last comment:
\bdm
Glad to hear you think {\Coleman} is just the law of large numbers.
That's what it sounded like to me to.  When he talked about it he
freely admitted that probability had to be put in by hand.  What
seemed to be the issue was where it got put in and how natural the
putting in was.  There was also stuff about randomness and
noncomputability that sounded different, which didn't make much sense
to me.
\edm

This is what I meant by ``various versions'' of the LLN.  This
randomness business is really just a refinement of that \ldots\ and
nothing more.  This remains true even when one evokes words and
concepts from algorithmic information theory.  Whenever you're ready
for some points of departure for reading about these things, let me
know:  I can provide you with some citations ({\Benioff}, van
{\Lambalgen}'s review article, etc.)

\section{15 February 1998, \ ``{\Landau}, {\Lifshitz}, Correlata, and {\Pauli}''}

The other day I (accidentally!) got embroiled again in a discussion
of the foundations of quantum mechanics.  In particular I was making
fun of the naivet\'e of taking state vectors as having a one-to-one
correspondence with reality.  My sparring partner responded with,
``Well then, what would {\it you\/} take in the theory as
corresponding to a reality?''  In a sort of desperation I pointed to
the middle part of a picture that we had already drawn on the board:
two square blocks connected by a few squiggly lines.  With my finger
on the lines, I exclaimed ``entanglement!''  \ldots\ But don't you
even think {\it ``Yes!!\ Correlation without correlata!''}\ \ldots\
despite your coming to my mind in just that moment.  For then I
quickly followed with, ``But only as a sort of shorthand for a more
fundamental situation:  when one part of the world tries to obtain or
leave an imprint on another, both are indeterministically disturbed
in the process.  The apparatus of quantum mechanics appears to be the
only available method for reasoning in light of that situation.''

One day you'll have to respond to me why you dislike that line of
thought so much (which, in any case, is not so much original in me as
it is in {\Pauli}, in his version of Copenhagen).
\bq
\noindent
``What we learn about is not nature itself, but nature exposed to our
methods of questioning.''  \hspace*{\fill} {---Werner~{\Heisenberg}}
\eq
\bq
\noindent
``All that is solid melts into air, all that is holy is profaned, and
man is at last compelled to face with sober senses his real
conditions of life and his relations with his kind.'' \\
\hspace*{\fill} ---Karl Marx
\eq

\section{14 March 1998, \ ``{\Pauli} in Ithaca''}

Have you ever seen this:

\bq
Quite independently of {\Einstein}, it appears to me that, in
providing a systematic foundation for quantum mechanics, one should
{\it start\/} more from the composition and separation of systems
than has until now (with {\Dirac}, e.g.)\ been the case.  --- This is
indeed --- as {\Einstein} has {\it correctly\/} felt --- a very
fundamental point in quantum mechanics, which has, moreover, a
direct connection with your reflections about the {\it cut\/} and
the possibility of its being shifted to an arbitrary place.
\eq

It comes from a letter written by {\Pauli} to {\Heisenberg} in the
wake of the EPR paper.  I thought this might intrigue you with your
Ithaca attitude.  In case you know German better than I---I swiped
this piece of the translation from another source---and you are
interested in reading further, the place to look is in letters
numbered 412, 413, 414, and 415 in {\sl Wolfgang {\Pauli}: Scientific
Correspondence with {\Bohr}, {\Einstein}, {\Heisenberg} a.o.}, Vol.
2, edited by Karl von Meyenn (Springer, 1985).

\section{29 March 1998, \ ``Ithacan Monadology''}

Lest you think I've long since forgotten the Ithaca II paper and am
not continually mulling it over and debating it about in my head, let
me bring it up once again.  I just ran across the following
interesting line in an otherwise wacky article by Julian Barbour:

\bq
What distinguishes [{\Leibniz}'s philosophy] from Berkelian idealism
is the most radical (and least well explained \ldots)\ element in the
Monadology---the assertion that the perceptions of any one monad are
nothing more and nothing less than the relations it bears to all
other monads.
\eq

What think you?  Have you ever given any thought to the connection or
disconnection between ``C w/o C'' and Mr.\ {\Leibniz}?

By the way, whatever happened to the {\Landau}/{\Lifshitz} stuff you
were going to send me?  Or did you come to an opinion on it
yourself?  For your other question (i.e., about the renaming of
{\Wootters}' Theorem), I still don't have a strong opinion:  maybe
``state-vector local reconstruction theorem'' or just ``local
reconstruction theorem'' or ``local representation theorem.''  I
guess if you wanted to be really bold you could call it ``The
Equivalence Principle.'' :-)

I think to some extent I have a new objection or twist on the
credence I would give this IIEP.  It is this:  who says---and why do
they say it---that my ability to reconstruct a wave function from
solely local measurements is so all-fired important?  In order to
give the IIEP any operational significance, as you know, we must be
able to make many many local measurements on many many identically
prepared systems.  But why is the issue of the {\it efficiency\/} of
our measurements ignored in your interpretation?  That is to say, if
we worry about such matters, it seems to me to become more suspect
that the ``correlations'' (as you like to call them) are the whole
story. Nonlocal measurements will in general be more efficient than
solely local ones.  And this is not simply due to the fact that there
are nonclassical correlations (i.e., {\Bell} inequality violations)
for entangled states.  I'm thinking in particular about a certain
malicious example that {\Bennett} and company and I are writing up
right now.  (Hopefully it'll be on the net in a couple or three
weeks---the manuscript is already sitting at 21 pages.)  In this
case, Alice and Bob each have possession of a three-dimensional
Hilbert space, and are assured that the quantum state in front of
them is one of a set of nine {\it orthogonal\/} product states
$|\psi_i\rangle|\phi_i\rangle$.  Because the nine states are
orthogonal, Alice and Bob can in principle identify the unknown
quantum state with a {\it single\/} measurement.  But the twist is
this, the set of $|\psi_i\rangle$ and the set of $|\phi_i\rangle$ are
each maliciously chosen so as not to be orthogonal sets.  And in
particular, they are chosen so that one can prove that a nonlocal
measurement is required if one wants to completely recover the
identity of the unknown quantum state.  No ingeniously clever set of
local measurements (including POVMs) and classical communication can
take the place of a nonlocal measurement in this example.  What in
the Ithaca Interpretation gives warrant for ignoring the issue of
efficiency of state identification in asserting the equivalence
between ``correlations'' and ``state vectors''?  With the issue of
efficiency taken into account, doesn't it seem like there is more to
a state vector than ``C w/o C''?

Anyway aside from this negative comment, I do believe that I have a
much firmer understanding of what I see as important in your efforts.
(I.e., I'm closer to being on the verge of identifying what it is
that resonates most with my happy neo-Copenhagen predispositions.)
With some luck I'll write you up a little ditty on it while I'm in
Munich next week.  (My wife and I are going to visit her parents, and
I'm hoping to spend the time being more philosophical than usual.)

\section{17 April 1998, \ ``References''}

\bdm
Fire away.  Just be polite.
\edm

Of course, I'll be polite:  I am a Texan foremost, and therefore a
gentleman!  \ldots\  Actually I should have used those ``irony
quotes'' you suggested.  The reference I'll make to your papers will
be mostly likely be in passing.  The main concern of the paper I'm
writing is with the Bayesian way of looking at quantum probabilities,
deconstructing the {\Hartle}-{\Coleman} stuff, and [trampling on] the
philosophical notions of objective probability (von {\Mises}'
frequency, {\Popper}'s propensity, and Lewis's chance).  I'd
probably like to attack your notion of {\it objective\/} probability
too, but I can't figure out what your notion is!

Thanks for sending on the {\Landau}--{\Lifshitz} stuff.  I saw the
following citation in the April 1 issue of J. Mod.\ Optics the other
day:  ``[ for such and such \ldots\ ] see the textbook series of
{\Landau} and {\Lifshitz}.  Admittedly, we do not know which volume
should be consulted, but our Russian colleagues are positive that
everything can be found in {\Landau} and {\Lifshitz}.''

By the way, you know it has dawned on me that for all the 57
kilobytes(!)\ of notes that I wrote on your ``correlation without
correlata'' you've made hardly any effort to reply to anything in
detail.  Either that means that you think that what I had to say was
pretty foolish or \ldots\ more than likely \ldots\ you think it even
more foolish than that!  If I had to boil it down to one thing I
would like the king to give me audience on, I would say that that
would be the stuff addressed in Merminitions \ref{OldAsm5} and
\ref{OldAsm6}. You want to turn to the structure of the theory to give us
one ontological statement, but from my perspective it doesn't look
as if it gives anything unique in that regard.  We can make
{\Wootters} theorem go through with a set of observables much
smaller than the ones you use in your derivation in the appendix; we
can make it go through with a set much larger.  What in the theory
gives you the right to say ``Ouh, this is just right!''\ when you
get to the complete set of (local, product) Hermitian-operator
observables?

\section{17 April 1998, \ ``How Do I Sleep?''}

I'm back again.  I was thinking earlier that, among other things, I
know you must ask yourself, ``How can a devout Copenhagenist sleep at
night?''  Surely they must have awful nightmares with all the silly
anti-realistic metaphysics they carry around in their heads!  \ldots\
Well, anyway, I've given my best shot at exposing the ontological
lesson of quantum theory---the reason there's not too much worry for
nightmares---and yet it seems to make no impression on my friends. So
let me try once again, this time with a relatively eloquent passage
that I found just today.  It's from a 1957 paper of L\'eon
{\Rosenfeld}; I'll scan it in for you.

Sleep well.

\bq
\begin{center}
The Curse of Positivism
\end{center}
\medskip

According to our critics, the epistemological point of view of
quantum theory undermines the sound belief in the reality of the
external world, in which all physical thinking is rooted, and opens
the door to the barren doctrine of positivism:  we are no longer
concerned with things, but only with the way to speak about things;
science is degraded from a quest for truth to a verbal exercise.
Dogmatic assertions that certain questions are meaningless bar the
way to further inquiry; inability to understand the riddles of the
quantum is hidden in the clouds of mystical renouncement.

This picture would be alarming if it were true.  However, it is just
another dream, a nightmare perhaps, of our critics.  Obviously, it
cannot result from a serious assessment of the immense broadening and
enrichment of scientific thought which has resulted from the endeavor
to formulate the laws of atomic phenomena.  In fact, it is based on
the most futile casuistics:  the critics diligently excerpt from the
writings in which the principles of quantum theory are discussed
isolated sentences on which they put arbitrary interpretations.  No
wonder that they should find (as they freely confess) some difficulty
in 'understanding {\Bohr}':  which, incidentally, does not prevent
them from branding him as a positivist.  There is no difficulty, at
any rate, in understanding the critics' philosophy and exposing its
unscientific character.

The realization of the mutual limitations imposed upon the use of
classical concepts by the conditions of observation has forcefully
reminded us of our own position in the world, and of the function of
science in relation to this position.  We are not merely
contemplating the world, but acting upon it and thereby modifying its
course.  Accordingly, the scientific description of the phenomena is
fundamentally concerned with the interaction of external agencies
with the human observer; or, at least, in the narrower domain of
physics, with material systems under the latter's control.  The mode
of description of classical physics appears, from this point of view,
as a special case of wide validity, in which the quantitative effects
of the interaction between observational devices and observed systems
may be neglected:  which does not mean, however, that the presence of
these observational devices is not just as essential as in quantum
theory for the very definition of the physical concepts.

It must again be stressed that there is in this view of the nature of
science no arbitrary element:  it is just an explicit statement of a
situation which has always existed, even though it was not always so
clearly recognized.  Certainly, it puts an emphasis unknown to the
outdated materialistic metaphysics of the nineteenth century on the
active role of the observer in defining the phenomena:  but in so
doing, it brings the whole structure of science nearer to reality, in
closer conformity with our real relationship to the external world.

There is an undeniable similarity between the epistemological
conclusion drawn in such a straightforward, unambiguous way from the
peculiar character of the quantum laws, and the insistence of the
early positivists on the essential part played by our sensations in
determining our knowledge of the external world.  This only means
that, to that extent, the early positivist movement was a healthy
reaction against the shallow metaphysics of mechanistic materialism.
But why should scientists be made responsible for the later
positivists' blundering into a metaphysics of their own?  No
scientist would accept the extreme positivist contention that there
is nothing more in statements about phenomena than the conceptual
expression of relations between sensations:  he would maintain that
such statements refer primarily to real processes of the external
world; our mental representation of these processes being itself, of
course, subject to definite laws depending upon our sensorium.

To point out that certain relations between classical concepts cease
to be meaningful in quantum theory has none of the sinister
implications fancied by our critics:  it is a plain statement of
fact, founded in a law of nature.  The words `renouncement' or
`resignation' often used in this context are ambiguous in their
emotional connotation:  renouncement may be felt as privation or as
liberation. Some critics seem to take the invitation to
`renouncement' as an attempt on their personal freedom:  the right to
indulge in metaphysical dreams is not disputed; only, this activity
is not science.
\eq

\section{20 April 1998, \ ``{\Socrates} and {\Rosenfeld}''}

\bdm
What do you mean, I didn't respond?  I added major chunks to the
paper in response to your criticisms.  That kid had every right to be
proud to know you.   Whereof I had no answers, thereof I remained
silent, if you know what I mean.

But as for {\Wootters}' theorem (now called the SSC theorem, by the
way
--- I'm not going to keep changing its name every time I get a new
email) I don't care if it takes less than subsystem correlations to
determine the state.  The crucial thing is that it doesn't take more
(as it would if one were restricted to a Hilbert space over the
reals).
\edm

Thank you for your extra sentence of explanation.  (Oops, forgot
those irony quotes again.)  More to the point---and more
sincerely---thank you for the three nice emails.  I'm starting to
feel like I really am a grain of sand in your shell.  Whereof one
cannot speak, thereof one should think harder.

Funny that you should find in {\Socrates} a new poster boy for
IIQM\@. Just the other day I wrote  {\Asher} {\Peres}:  [See note
dated 5 April 1998.]

The {\Rosenfeld} quote came from:  L.~{\Rosenfeld},
``Misunderstandings about the Foundations of Quantum Theory,'' in
{\sl Observation and Interpretation: A Symposium of Philosophers and
Physicists}, edited by S.~K\"orner (Academic Press, New York, 1957),
pp.~41--45.

\section{21 April 1998, \ ``${\bf C^*}$ Hunkies''}

\bdm
\noindent
So if you take the conventional view (which I remember from listening
to $C^*$ algebra types in ancient days) that the state is just a
catalog of all possible mean values, then there's nothing to favor
complex over real scalars in the underlying Hilbert space. But if you
take the SSC view that the state is a catalog of the non-trivial
subsystem correlations, then there is.  That has to (be trying to)
tell us something. \edm

I agree that there must be something deep going on with this.  But it
seems to me that the $C^*$ people could always feel better about
themselves because they had {\Gleason}'s theorem to back them up.
That is to say, if they hadn't had {\Gleason} lying around, they
would have had to have given some extra justification for why they
chose this particular type of catalog (i.e., the state) over
another.  But what about Ithaca? Take a look again at my silly idea
following Merminition \ref{OldAsm7}.  Wouldn't you need something
like that if you wanted to feel as comfortable as the $C^*$ people?

\section{07 May 1998, \ ``A Rug and A Sweep''}

Didn't you use some analogy like ``sweeping the problems of quantum
theory under the rug of objective probability'' in Ithaca I or II\@?
If you did, I can't find the words ``rug'' or ``sweep'' anywhere in
there at all now---can you fulfill my need to quote accurately by
pointing me to the right place \ldots\ or telling me that I just
imagined their use?

For history's sake, let me tell you about another little discovery I
made a couple of days ago.  I found the main piece of the
{\Hughston}-{\Jozsa}-{\Wootters} thingy that you rederive in Ithaca
I---namely the piece about unitary reshuffling (both if and only if)
\ldots\ but not the part about creating ensembles from a
distance---in an old paper by Ed {\Jaynes}.  Let me give you the
citation, E. T. {\Jaynes}, ``Information Theory and Statistical
Mechanics.\ II'' Phys.\ Rev.\ {\bf 108}, 171--190 (1957).

Point 3.  In Germany last month I read your debate with {\Stapp}.
Hate to admit, but I had a lot of trouble reading it:  that could
either reflect on the writing, or the state of my head but I think
it was the writing this time.  Anyway, I was amazed, among other
things, about the trouble you went to to illustrate the meaning of
{\Bohr}'s phrase, ``Of course there is in a case like that just
considered no question of a mechanical disturbance \ldots\ \@. But
\ldots\ there is essentially the question of an influence on the
very conditions which define the possible types of predictions
regarding the \ldots\ behavior of the system.''  Amazed, maybe,
because I was already pretty happy with it.  Well, while amazement
is running high, I just found today that Ed {\Jaynes} also didn't
think the phrase was clear at the outset. If you'd like to see his
rendition in terms of Bayesian language, take a look at E.~T.
{\Jaynes}, ``Clearing Up Mysteries
--- The Original Goal,'' in {\sl Maximum Entropy and Bayesian Methods
(Cambridge, England, 1988)}, edited by J.~Skilling (Kluwer,
Dordrecht, 1989), pp.~1--27.  [To wit:]
\bq
The spooky superluminal stuff would follow from Hidden Assumption
(1); but that assumption disappears as soon as we recognize, with
Jeffreys and {\Bohr}, that what is traveling faster than light is
not a physical causal influence, but only a logical inference. Here
is {\Bohr}'s quoted statement (italics his):
\bq
\noindent
Of course there is in a case like that just considered no question of
a mechanical disturbance of the system under investigation during the
last critical phase of the measuring procedure. But even at this
stage there is essentially the question of {\it an influence on the
very conditions which define the possible types of predictions
regarding the future behavior of the system.\/}
\eq
After quoting these words, {\Bell} added: ``Indeed I have very little
idea what this means.'' And we must admit that this is a prime
example of the cryptic obscurity of {\Bohr}'s writings. So~--~with
the benefit of some forty years of contemplating that statement in
the context of his other writings~--~here is our attempt to translate
{\Bohr}'s statement into plain English:
\bq
\noindent
The measurement at $A$ at time $t$ does not change the real physical
situation at $B$; but it changes our state of knowledge about that
situation, and therefore it changes the predictions we are able to
make about $B$ at some time $t^\prime$. Since this is a matter of
logic rather than physical causation, there is no action at a
distance and no difficulty with relativity [also, it does not matter
whether $t^\prime$ is before, equal to, or after $t$].
\eq
Again we see how {\Bohr}'s epistemological viewpoint corresponds to
Bayesian inference, and could have been expressed precisely in
Bayesian terms. However, {\Bohr} could not bring himself to say it
as we did, because for him the phrase ``real physical situation'' was
taboo.
\eq

There's much in the article that I don't like.  {\Jaynes}, like most
good Bayesians, didn't quite believe in quantum mechanics---where
there is probability, there is ignorance about something objectively
existent \ldots\ yuk---but he does hit the {\Bohr} part of the
discussion on the mark (as far as I am concerned).

Point 4.  I had a dream about you a few nights ago.  I was sitting on
a park bench in Ithaca.  You were jogging by and stopped for a moment
to talk.  The significant thing that stood out about you was that you
were wearing white tennis shoes, white shorts, and now had blond
hair. Ithaca---I've never seen it in real life---looked like
something out of ``It's a Wonderful Life.''  We spoke for a moment on
reality and Bayesian probability, and then I awoke.

\section{08 May 1998, \ ``A Bug in the Soup''}

Well, I'm always happy to get credit for something(!)\ \ldots\ but I
have no idea what you are talking about.  Feel free to refer away
\ldots\ as long as no one gets the idea that I'm an Ithacan.
(Ithacan: that species of people marked by a regressive desire to
believe the mysteries of quantum mechanics can be relieved with the
formulation of an ``objective'' notion of probability.  Compare
{\Einstein}'s regressive desire to relieve the mysteries with some
version of classical field theory.  And compare {\Pauli}'s constant
use of the word ``regressive'' when speaking of {\Einstein}.)

Yeah, I hope you do take a look at that {\Jaynes} article.  Also,
there's another article where he bashes on quantum mechanics.
 I'm sure he boobalates somewhere in there (given my
religious love of the quantum and its mysteries), but I haven't yet
made the effort to pinpoint these troubles.  I'll do that next week
with {\Ruediger}. Please, please, please, send me any thoughts or
comments you might have.  I was especially pleased to find that
{\Jaynes}'s had some very vague desire to build up (something like,
{\it but not identical to}---that's where he screws up) quantum
mechanics as a system for inductive and deductive inference under
the situation that information-gathering measurements cause
disturbances.  [For instance, he says:]
\bq
Put most briefly, {\Einstein} held that the QM formalism is
incomplete and that it is the job of theoretical physics to supply
the missing parts; {\Bohr} claimed that there are no missing parts.
\ldots\ if we can understand better what {\Bohr} was trying to say,
it is possible to reconcile their positions and believe them both.
Each had an important truth to tell us. \ldots\

\ldots\ When {\Einstein} says QM is incomplete, he means it in the
ontological sense; when {\Bohr} says QM is complete he means it in
the epistemological sense.  \ldots\

Needless to say, we consider all of {\Einstein}'s reasoning and
conclusions correct on his level; but on the other hand we think that
{\Bohr} was equally correct on his level, in saying that the act of
measurement perturbs the system being measured, and this places a
limitation on the information we can acquire and therefore on the
predictions we are able to make.  The issue is merely whether this
limitation is as great, and has the same quantitative form, as
{\Bohr} supposed.  \ldots\

{\Bohr} had no really cogent reason for his postulate that the
limitations on the ability of the QM formalism to predict are
also---in complete, quantitative detail---limitations on what the
experimenter can measure; this seems to us an outstanding example of
the Mind Projection Fallacy.
\eq

I would have liked to believe that I was the only person to have that
epiphany.  But then I finally realized that that is what {\Bohr} was
talking about.  I guess I don't mind making a smaller
contribution---especially if it's in a form that people can
understand when it's finished \ldots\ and not debate it about on the
pages of AJP, for instance.  Irony quotes, irony quotes.


\section{04 September 1998, \ ``The Bohrazine Shuffle''}

After all these years, for some reason today I finally had the heart
to wade through {\Bohr}'s reply to {\Einstein}, {\Podolsky}, and
{\Rosen} again. Wow, what tough going!  I wonder if the man had any
clue just how horribly he wrote?  Not surprisingly, once again I
hardly got anything from the article.  {\it However}, there was one
little thing that made my toes tingle.  It's a small passage that
comes just after the famous one you see quoted everywhere---I think
even you quoted it recently.  (I'll enter it in below for you.) What
I found most intriguing about it is that it {\it seemed\/} to be
talking to what I see as the most elegant---and promising for
progress(!)---solution to all our quantum troubles.

Now, the word ``seemed'' above is where you come into this picture.
Being the {\Bohr} scholar that I know you are, I wonder if you can
shed some light on what he might be saying.  And whether he
({\Bohr}) is saying anything that---vague though it is---might be of
use in steering me.

Lest you have forgotten my point of view, before moving on to
{\Bohr}, let me remind you of it with a couple of summaries I wrote
Adrian {\Kent} a few days ago: \medskip

 [See notes to Adrian {\Kent} dated 30 August and 2 September
1998.]
\medskip

With that in mind, let me stress once again that, for me, it is not
just enough to say these things and then feel comfortable with
myself.  I want to see quantum mechanics come out as the end product
of these vague proclamations, not as the starting point.  (Then, when
I'm finished with that, it'll be time to move on to some really
revolutionary stuff \ldots\ but that's a different story.)  Anyway,
if I want to go the route of relying on much of the hard work of
{\Gleason} and {\Kraus}, this leaves me with having to justify the
following five things from the information-disturbance foundation
described above:
\begin{enumerate}
\item  Why are information-gathering measurements restricted to being
orthogonal bases in a Hilbert space?
\item  Why is there a noncontextuality assumption about the
probabilities of overlapping basis vectors?
    [{\Gleason} uses 1 and 2 to derive that all measurement
probabilities can be summarized by a density operator.  This gives us
quantum mechanical states.]
\item  Why is that Hilbert space above complex?
    [I know you think your ``correlation without correlata'' gives a
good explanation of that, but I think a strict interpretation of
quantum states as states of knowledge (and nothing more) gives an
even better explanation---there's some chance I'll be writing this up
this fall (or in a personal letter to you before then).]
\item  Why must the evolution of our states of knowledge be a linear
mapping (from density operators to density operators)?
    [{\Kraus} needs only assumption 4 to show that unitarity can always be
invoked at some sufficiently high level.]
\item  Why is the Hilbert space of a composite system the tensor
product of the more primitive Hilbert spaces (and not some other
magical combination of the primitive spaces)?
\end{enumerate}

My focus for this note is Question 1 above.  In essence, the question
is:  How can an ``information-disturbance foundation'' force
``complementarity'' upon us?  I.e., how can such a foundation force
upon us the restricted (mutually exclusive) measurements described in
Question 1?

OK\@.  Now I hope you're ready for {\Bohr}'s cryptic passage.

\bq
Of course there is in a case like that just considered no question of
a mechanical disturbance of the system under investigation during the
last critical stage of the measuring procedure.  But even at this
stage there is essentially the question of {\it an influence on the
very conditions which define the possible types of predictions
regarding the future behavior of the system}.  Since these conditions
constitute an inherent element of the description of any phenomenon
to which the term ``physical reality'' can be properly attached, we
see that the argumentation of the mentioned authors does not justify
their conclusion that quantum-mechanical description is essentially
incomplete.  On the contrary this description, as appears from the
preceding discussion, may be characterized as a rational utilization
of all possibilities of unambiguous interpretation of measurements,
compatible with the finite and uncontrollable interaction between the
objects and the measuring instruments in the field of quantum theory.
In fact, it is only the mutual exclusion of any two experimental
procedures, permitting the unambiguous definition of complementary
physical quantities, which provides room for new physical laws, the
coexistence of which might at first sight appear irreconcilable with
the basic principles of science. It is just this entirely new
situation as regards the description of physical phenomena, that the
notion of {\it complementarity\/} aims at characterizing.
\eq

I'm especially interested in the part above starting with ``On the
contrary \ldots''\@.  It looks to me like he is wanting to say that:
measurements causing disturbance {\it causes\/} complementarity
(i.e., the mutual exclusivity of certain measurements).  (Be aware
however:  I fully realize that he is NOT talking about something so
mundane/mechanical as trying to get the {\Heisenberg} relation from a
{\Heisenberg} microscope-like picture.)  What's your take on this
passage? Can you make much sense of it?  What does he mean by
``providing room for new physical laws?''  What ``basic principles of
science is he talking about?''  What five pages of derivation are
lying behind all this business?

It nags me that {\Bohr} often speaks as if it is clear that the
structure of quantum theory is derivable from something deeper, when
in fact all the while he is taking that structure as given.  When did
he ever approach an explanation of ``Why complex Hilbert spaces?''
Where did he ever lecture on why we are forced to tensor products for
composite systems?  It's a damned shame really:  I very much like a
lot of elements of what he said, but as far as I can tell all the
hard work is still waiting to be done.  Is he hinting above that he
has a magical way of seeing where ``orthogonal bases for allowed
measurements'' come from? Is he hinting that it comes from
``information vs.\ disturbance''?

Any thoughts you have will be most welcome.

\section{08 September 1998, \ ``I'll {\Bohr} You to Tears''}

Thanks for the long reply to my query.  That's the kind of email I
like to get.  I found myself mulling over it the whole long holiday
weekend. I reread Chapters 13 and 14 in your book:  now I see I can
blame my frustration on an inheritance from you!  I first read these
chapters about 5 years ago and I guess I had forgotten them a little.
I also read again the introduction to your long article in Chapter 12
and washed it all down with a rereading of {\Pais}'s Chapter 19 and
Section 25c in his {\Bohr} and {\Einstein} biographies.  [I should
have chosen a better drink though:  I found {\Pais} pretty annoying.]

Did you have any luck with {\Plotnitsky}?  I'll try to dig up his
book again and have a look at it sometime this week.

But, let me presently add a few comments to your comments.

\bdm
\label{NewAsm}
You could use some interpreting yourself, you know.
\edm
Yeah I do know that!!!  That's why I like to associate with clear
thinkers like you.  At least as far as this aspect of my career
(i.e., dabbling in the foundations) goes, I want to hone things to
immaculate clarity before saying them to someone who doesn't already
know me.

\bdm
\bq
\noindent\rm
\ldots\ {\it my\/} information gathering about something you know,
causes {\it you\/} to loose some of that knowledge \ldots\
\eq
This is an admirable way to start thinking about the problem, but
there has to be a better way to put it even at this crude level.  I
mean my knowledge is my knowledge.  I can't lose it. \edm You are
indeed right: I apologize.  I was priding myself on being pithy in
that formulation, but I see I went too far.  The trouble comes from
my usage of the words ``know'' and ``knowledge'' \ldots\ which should
be taken to be, more accurately, ``can predict'' and
``predictability.'' PLEASE indulge me {\it two\/} last times, and
then I promise I'll leave you alone \ldots\ that is, if you give some
small evidence that you did indulge me!

1)  A much better technical statement of what I'm trying to express
above can be found in Section III of my paper ``Information Gain vs.
State Disturbance in Quantum Theory'' ({\tt quant-ph/ 9611010}),
Fort.\ der Phys., {\bf 46}, pp.\ 535--565 (1998).  It is crucial to
my point of view that there be at least {\it two\/} players and {\it
two\/} quantum states in the game.  Appreciating that will, I think,
to some extent negate your questions:
\bdm
Does gathering some information disturb other information? Or is it
just in the nature of the information that was gathered that it no
longer makes sense to contemplate the content of the other
information?
\edm
Please read the cited section in my paper, and then come back for the
following comment.

I think one of the troubles in our founding fathers' discussions is
that they continually focused their attention on {\it one\/} observer
making measurements on a quantum system described by {\it one\/}
(known) quantum state.  This led them to say things---in language
similar to some of the specimens in your note---like, ``The gain of
knowledge by means of an observation has as a necessary and natural
consequence the loss of some other knowledge.''  ({\Pauli}) \
Without at least a second player in the game, those gains and losses
hardly seem to be sensible concepts to me: they can only refer to
the observer's attempt to ascribe one or another classical picture
to the quantum system in front of him.  Since we know---from
{\Bell}'s argument and the religion of locality---that it is not
reasonable to assume that those classical variables (correlata) are
there and existent without our prodding, it is hard to call the
revelation of a measurement outcome a ``gain of knowledge.''  What
did you learn about the world that was there before your looking?
Nothing. However, throw a second player into the game and that
situation changes. Those random quantum outcomes now have something
{\it existent}, some unknown truth, that they can be correlated with.
The revelation of an outcome really can correspond to a ``gain of
knowledge,'' but you need at least two information processing units
in the world for that to be the case.

Disturbance?  The founding fathers like to say things like, ``For
every act of observation is an interference, of undeterminable extent
\ldots'' But again, this is a muddled concept when the discussion is
restricted to {\it one\/} observer and {\it one\/} (known) quantum
state.  This is true for exactly the same reason as above:  there
just ain't no correlata already there to disturb.  (Nice of me to use
the word correlata just for you!)  So why did they even bother with
the word disturb?  It's because they had semi-classical ideas in
their head. But, still, trying to fix the picture by saying that
wave-function collapse will do the job is no good either:  if one is
talking about a {\it single\/} (known) quantum state, there is no
necessary collapse upon measurement.  I can build a device that
``measures'' any observable you wish while still regenerating the
known quantum state in the process.  However, just as before, throw a
second player and two nonorthogonal states into the game and
everything changes.  Now one observer (the eavesdropper) can't help
but impart a disturbance to at least one of the two quantum
states---that is, if she puts herself in the position to gather some
information about the other observer's state of knowledge (i.e., the
quantum state he prepared). Moreover, in this context, the phrase
``of undeterminable extent'' remains meaningless \ldots\ or
superfluous at best.  If Alice knows everything that Eve is doing,
then the disturbance will be perfectly determinable---that's what my
paper is about---but that doesn't mean the disturbance disappears.
The disturbance arises because the eavesdropper's equipment
necessarily becomes entangled with the system it is measuring.
(Sometimes I think there is nothing to the concept of entanglement
beyond this.)

But all this, everything I've just said, is derivable from quantum
mechanics.  It's part of the answer to your question, ``What is
quantum mechanics trying to tell us?''  It's an example of quantum
mechanics teaching us when and how to use the words ``information''
and ``disturbance'' properly.  As I've told you, though, I'd like to
see the tables turned.  But for that I have to take on an attitude
that I think is a little foreign to you.  In your original Ithaca
paper you speak of the minimal requirements for a quantum mechanical
universe:  it is, you say, two qubits---two things to have
correlation without correlata.  I, however, am more afraid to go that
far, i.e., to some final/overarching ontological statement.  Instead,
the most I think I'm willing to ask is, ``What are the minimal
requirements for a physical {\it theory}?''  And there, I think the
answer is two ``theory makers'' and a physical system.  (It's hard to
get rid of that damned Catholic Church!)  From my perspective, a {\it
theory\/} never stands a chance of giving a complete description of
reality---for that would require some nasty self-reference by at
least one of the theory makers.  However from a theory, one can still
hope to intuit {\it some\/} (safely assumed) ontology.

2)  This leads to my second request for an indulgence.  In a second
mailing, I'll send you something I wrote Rolf {\Landauer} along these
lines.  It doesn't say much more than I've already said above, but
maybe it says it from a slightly different slant.  Reading it may
help the ideas click on \ldots\ or click off \ldots\ for you.  In
either case, it may help you provide me with some useful criticism.
(That would be good enough small evidence that you did indulge me.)

There, that concludes my painful requests for indulgence.  As far as
making some real progress---not words(!)---toward actually turning
those tables, that's a more difficult story.  But, I don't think it
is a story without hope for progress.  I give an (unfortunately,
fairly wimpy) example of this in Section VI of the above-said paper:
it might interest you a little. I think a much more important tack,
though, is to first get straight how quantum probabilities are purely
subjective probabilities---for measurement outcomes, not for existent
but unknown properties, not for correlata(!)---and see how that
already pins some things down about the theory.  This is where you
and I differ most on the Ithaca call.  Once that is carried out, one
can come back to ask why it is that we're stuck with subjective
probabilities \ldots\ and that's where information vs.\ disturbance
comes in.

So let me finally move on to some of your other comments.

\bdm
The ``funny property'' might be intersubjective correlations, you
know \ldots\
\edm
It's funny to watch the two of us try to squeeze everything into our
own pet language!  In any case, I do hope you realize that despite
my spitting back the words ``information'' and ``disturbance'' in
each and every email to you, I do work very hard to listen to what
you're saying. ``How do I correlate/anti-correlate what he's saying
with what I'm saying?''---this I try to do every time.

In that vein, I wanted to read your big Ithaca paper one more time
over the weekend, but I was thwarted by my broken printer.  It's too
darned hard to read something that long on a computer screen.  If I
ever get back to the office, I'll try to give the final version a
read and then come back to your {\Bohr}-interpretation work again.

\bdm
I haven't read {\Kant} since I was a freshman in college. Do I have
to?
\edm
All I was trying to say here is that {\Kant} argued that, though it
must be there, we can never know the thing in itself (the noumenon).
``Objective reality is forever beyond our reach.''  And he was even
able to argue this in a classical world!  Do I believe the
argument?  I don't know; I don't even remember it---I haven't read
him since I was a freshman in college either.  But, it does seem to
me, if there is anything worth salvaging in his stuff, we'll find
that we have it heaped even higher in a quantum world.  (It's like
having second quantization.) I tell you what:  I'll pledge to read
his Prolegomena again---it's only 163 pages---if you will.  It'd
probably do us both some good for party jokes.  Lately I've been
trying to read the neo-{\Kant}ian {\Schopenhauer} because {\Pauli}
had great respect for him \ldots\ and supposedly saw something of
quantum mechanics in his philosophy \ldots\ but it'd probably do me
some good to have a better background before tackling him.

\bdm
Who is {\Kraus}, by the way?  What should I read by or about him?
\edm
I took the liberty to fix your spelling mistake.  Karl {\Kraus} was
a mathematical physicist.  He was the first to prove that a
completely positive linear map on density operators can always be
viewed as a unitary operation on a larger space followed by a partial
trace. (Recall that a positive linear map is any map that takes
positive operators to positive operators.  It is completely positive
if, when extended by the identity operation on any tensor product
space alongside the original, it continues to preserve positivity
even for entangled density operators.)  This theorem is quite nice
from the point of view of foundations:  for it helps to show that
unitarity can be viewed as a convenient, rather than essential,
description of time evolutions.  From the point of view of ``quantum
states as states of knowledge and nothing more'' the criterion of
complete positivity is more natural than unitarity.  Why assume such
a detailed thing as unitarity just to say that states of knowledge
must go to states of knowledge?  If one could get rid of the
linearity assumption, then it would be even more natural.

\pagebreak

\bdm
\bq
\noindent\rm
2) Why is there a noncontextuality assumption about the probabilities
of overlapping basis vectors?
\eq
What's that? \edm I like to split the premise for {\Gleason}'s
theorem into two assumptions. 1)  It is the task of physical theory
to assign probabilities to the outcomes of all {\it valid\/}
measurements on a system.  The set of all valid measurements on a
system corresponds to the collection of all complete sets of
orthogonal one-dimensional projectors over some Hilbert space. 2)
All such probabilities can be derived from a single function (a
``frame function'') on the set of one-dimensional projectors.

It is Assumption 2 that I am calling the ``noncontextuality
assumption.''  The probability for a given projector-as-outcome does
not depend upon the particular set of projectors it is considered
embedded in.  Given that we know that our measurements don't simply
reveal pre-existing (but unknown) properties, my question is, ``Who
ordered that?''

By the way, it is of some interest to note that  {\Asher} says in his
paper, ``An Experimental Test for {\Gleason}'s Theorem'' (Phys.\
Lett.\ A, 163, 243--245, 1992) that {\it complex\/} Hilbert spaces
are required for the proof. That is not true.  Also, I think his
Assumption (3) is overkill; it is rather a consequence of the
theorem:  one need not start out by assuming that two-dimensional
projectors even correspond to valid questions.  His equation (4) is a
mild consequence of the definition of a frame function.  I take as my
bible of all things {\Gleason}-like:  Itamar {\Pitowsky}'s paper,
``Infinite and Finite {\Gleason}'s Theorems and the Logic of
Indeterminacy'' (J.\ Math.\ Phys., {\bf 39}, 218--228, 1998).

\bdm
I don't see any connection between the passage and your Question
1, so I guess that's my (disappointing) answer.  On the other hand
(as noted above) I find the passage irritatingly obscure, so perhaps
it provides room for your interpretation. [I.e.,]
\bq
\noindent\rm
It looks to me like he is wanting to say that: measurements causing
disturbance {\it causes\/} complementarity (i.e., the mutual
exclusivity of certain measurements).
\eq
\edm
I guess I was taking as my cue the following
rearrangement/interpretation of {\Bohr}'s words:
\bq
\noindent
On the contrary this description [i.e., the quantum mechanical one]
\ldots\ may be characterized as a rational utilization of all \ldots\
measurements [i.e., information gathering], compatible with the
finite and uncontrollable interaction between the objects and the
measuring instruments [i.e., necessary disturbance] \ldots\ \@.  In
fact, it is only the mutual exclusion of any two experimental
procedures, permitting the unambiguous definition of complementary
physical quantities, which provides room for new physical laws
[i.e., quantum theory itself, assuming it being built upon the
``quantum postulate'' that he so often speaks of, and assuming the
validity of equating with ``finite and uncontrollable interaction'']
\eq

That's about it.  I just mostly wondered if you had some insight.  My
attempt at understanding the passage was certainly more amateurish
and self-serving than your effort.  I suspect my shifting of his
words came about mostly because they were so confusing: it's so much
easier to impose your own reading on the scripture.  Also I may have
been subliminally influenced by some old readings of {\Pauli} (which
I also reread this weekend).  Check out this piece of a letter from
{\Pauli} to {\Bohr}, dated 15 February 1955:
\bq
\noindent
\ldots\ [I]t seems to me quite appropriate to call the conceptual
description of nature in classical physics, which {\Einstein} so
emphatically wishes to retain, ``the ideal of the detached
observer.'' To put it drastically the observer has according to this
ideal to disappear entirely in a discrete manner as hidden spectator,
never as actor, nature being left alone in a predetermined course of
events, independent of the way in which the phenomena are observed.
\ldots\

In quantum mechanics, on the contrary, an observation {\it hic et
nunc\/} changes in general the ``state'' of the observed system in a
way not contained in the mathematically formulated {\it laws}, which
only apply to the automatical time dependence of the state of a {\it
closed\/} system.  I think here on the passage to a new phenomenon by
observation which is technically taken into account by the so-called
``reduction of the wave packets.''  As it is allowed to consider the
instruments of observation as a kind of prolongation of the sense
organs of the observer, I consider the impredictable change of the
state by a single observation---in spite of the objective character
of the result of every observation and notwithstanding the
statistical laws for the frequencies of repeated observation under
equal conditions---to be {\it an abandonment of the idea of the
isolation (detachment) of the observer from the course of physical
events outside himself.}

To put it in non-technical common language one can compare the role
of the observer in quantum theory with that of a person, who by its
freely chosen experimental arrangements and recordings brings forth a
considerable ``trouble'' in nature, without being able to influence
its unpredictable outcome and results which afterwards can be
objectively checked by everyone.

Probably you mean by ``our position as detached observers'' something
entirely different than I do, as for me this new relation of the
observer to the course of physical events is entirely {\it
identical\/} with the fact that our situation as regards objective
description in ``this field of experience'' gave rise to the demand
of a renewed vision of the foundation for ``the unambiguous use of
our elementary concepts,'' logically expressed by the notion of
complementarity.
\eq
If the last passage of that is not trying to equate some notion of
disturbance with complementarity, I'll eat my hat.  But, of course,
these are {\Pauli}'s words and not {\Bohr}'s.

\bdm
{\em ``In fact, it is only the mutual exclusion of any two
experimental procedures, permitting the unambiguous definition of
complementary physical quantities,''}

Don't you think ``definition'' is crucial here? ``Physical
quantities'' have no ontological status (there are no correlata);
instead they are simply ``defined''. What defines them is an
``experimental procedure''
\edm
Yes I do think ``definition'' is crucial here.  And just for the
reason you cite.  That much I think we can give him credit for.

\bdm
Complementarity comes first.  He became very down on saying that
measurements cause disturbance (after EPR).  I read him as saying
that complementarity opens things up (provides room) by allowing you
to contemplate apparently incompatible things (irreconcilable with
the basic principles of science) secure in the knowledge that only
one of them can be DEFINED.
\edm
Yeah, I agree with this too.  However, remember {\it his\/} sentence,
``But even at this stage there is essentially the question of {\it
an influence on the very conditions which define the possible types
of predictions regarding the future behavior of the system}.''  It
is exactly this notion of ``disturbance'' that I was invoking in my
longish discussion above: a disturbance, or at the very least a {\it
change}, in predictability \ldots\ something not so far away in
flavor from Bayes rule for updating probabilities in classical
statistics.  It is this notion of disturbance that I think will
survive, one that he could not be so averse to.

---------------

There, I think that's all the comments on comments I can muster up
right now.  I'm getting pretty tired.  But I do want to quickly tell
you about one rushed {\Bohr}-related interpretive idea I've just
had:  I don't have the stamina to verify it presently, but I want to
record it.  Last month, when  {\Asher} was visiting, he said
something that really struck me. It was something like, ``I learned
enough in my education as an engineer to know that one can build
precision instruments from very crude tools. You don't have to have
a precision instrument to build a precision instrument.''  His
immediate application for this principle was in making fun of the
word reality:
 {\Asher}'s view is that ``reality'' (certainly within quotes) is the
crude tool with which we build the precision instruments of
scientific concepts.  \ldots\ But never mind the particular
application of the aphorism.  Instead, before going to bed, I want to
bring the aphorism into conjunction with {\Bohr}'s article
``Discussions with {\Einstein} \ldots'' in the {\Schilpp}
volume---the article {\Pais} called ``[{\Bohr}'s] finest expos\'e on
complementarity.''  As I recall, I was pretty dissatisfied with that
article.  I may be wrong, but I thought all of {\Bohr}'s arguments
were pretty darned circular.  The game was always the same:  {\Bohr}
would take a measuring device, {\it assume\/} that some uncertainty
relation held for its parts, and then go on to argue that that
instrument couldn't be used to perform a measurement on the quantum
system more accurately than allowed by the uncertainty relations!
His arguments seemed hardly the triumph that so many commentators
portray them to be.  But maybe the surprise, the triumph, was just
in the disarming of  {\Asher}'s aphorism in this context. I wonder
if that's a plausible excuse for the ``{\Bohr}-{\Heisenberg}
tranquilizing philosophy''?

\section{09 September 1998, \ ``Unbohrlievable!''}

You just won't believe this.  Apparently in 1949, while reviewing his
old papers, {\Bohr} must have had trouble understanding the selfsame
passage from his EPR reply that you and I have been talking about!
For he writes in the {\Schilpp} paper, just after quoting the
passage,
\bq
Rereading these passages, I am deeply aware of the inefficiency of
expression which must have made it very difficult to appreciate the
trend of the argumentation aiming to bring out the essential
ambiguity involved in a reference to physical attributes of objects
when dealing with phenomena where no sharp distinction can be made
between the behavior of the objects themselves and their interaction
with the measuring instruments.
\eq

\section{07 October 1998, \ ``Carnap''}

I see my last long note left you speechless \ldots

Anyway, I ran across something this morning that reminded me of a
conversation with you.  It's from the notes of a 1992 conference in
Helsinki.  Abner {\Shimony} says, ``I was a pupil of Carnap.  He said
that \ldots''  Remember I once told you that I had heard {\Shimony}
say that he was Carnap's student:  this seems to confirm it.  As I
recall, you disagreed, thinking it was another famous philosopher.
Now we both know.

\section{08 October 1998, \ ``Oh Mighty {\Mermin}''}

Boy you are a popular guy!  Take a look at this conclusion I just
wrote for my contribution to the QCM conference proceedings.

\bq
\noindent {\bf CONCLUSION.} \ What is the essence of quantum theory?
What crucial features of the phenomena about us lead ineluctably to
just this formalism? These are questions that have been asked since
the earliest days of the theory.  Each generation has its answer;
ours no doubt will find part of it written in the language of quantum
information. What is striking about the newest turn---the quantum
information revolution---is that it provides a set of tools for this
analysis from {\it within\/} quantum theory.  The example of the
tradeoff between information and disturbance in quantum
eavesdropping is typical.  Words about ``measurements causing
disturbance'' have been with us since 1927, but those always in
reference to outdated, illegitimate classical concepts.  The time is
ripe to consider turning the tables, to ask ``What is quantum
mechanics trying to tell us?'' \verb+\refnote{\cite{Mermin98}}+ \ Why
is the world so constituted as to allow single-bit information
transfers to be disturbed by outside information-gatherers, but
never {\it necessarily\/} more so than by an amount
$D_{\rm\scriptscriptstyle @MI}\approx0.067$?  Why is the world so
constituted that binary preparations can be put together in a way
that the whole is more than a sum of the parts, but never more so
than by $Q\approx0.202$ bits?  The answers surely cannot be that far
away.
\eq

On another note about your popularity, I received a paper from Adan
{\Cabello} today titled ``Quantum correlations are not local elements
of reality'' (to appear in PRA).  Since you are thanked for
``feedback'' in the back, I presume you've seen at least a
preliminary version of it.  I had seen one too.  In fact, I wrote
him a letter about it \ldots\ somewhat discouraging his attempt at
publication.  I don't think his point is oh so good as to warrant a
paper---we should reserve those trees for something a little more
substantial.  Anyway, I'm pretty sure his argument didn't faze you;
that it wouldn't is what I tried to express in my letter to him. Now
that the cat is out, let me forward on the comments I wrote him.  I
was wondering whether you really do agree with the points I made to
him.  Not to worry, a short reply will suffice!

{\Cabello} stuff in next message.  [See note to Adan {\Cabello},
titled ``{\Mermin} and His Correlata,'' dated 16 August 1998.]

\section{13 December 1998, and to several others, \ ``Little
Historical Point''}

\noindent [WARNING: The present note is factually inaccurate, as
David points out immediately below.]

\noindent To my friends with a little interest in the history of
quantum mechanics,\bigskip

Let me share a delightful little piece of information that I ran
across this morning.  Though, perhaps I should say that I ``became
cognizant'' of it this morning; for I know that I have read over
this passage on at least two other occasions.

I think we are all accustomed to crediting David {\Bohm} with being
the first to speak of a finite-Hilbert space version of the EPR
gedankenexperiment.  (At least I know that I do, and both  {\Asher}
and David do in their books.)  But look at this little passage from
the {\Born}-{\Einstein} letters.  It comes from a letter from
{\Born} to {\Einstein}, dated 9 May 1948:

\bq
Let me begin with an example.  A beam of light falls on to a plate
of doubly refracting crystal, and is split into two beams.  The
direction of polarization of one of the beams is determined by a
measurement: it is then possible to deduce that that of the second
beam is perpendicular to the first.  In this way one has been able
to make a statement about a system in a certain part of space as a
result of a measurement carried out on a system in another part of
space.  That this is possible depends on the knowledge that both
beams have originated from one beam which has passed through a
crystal; in the language of optics, that they are coherent.  It
seems to me that this case is closely related to your abstract
example, which is apparently connected with collision theory.  But
it is simpler and shows that such things happen within the framework
of ordinary optics.  All quantum mechanics has done is to generalise
it.
\eq

Isn't that neat?

\subsection{David's Reply}

\bq
I remember that passage as yet another example of {\Born}'s having
completely missed the point of EPR in all his arguments with
{\Einstein}.

I don't think it has anything to do with two-particle entangled
states. Nor does it have anything to do with coherence.  It seems to
me all {\Born} is saying is that if you have a birefringent crystal
(in a fixed orientation) then if you determine the horizontal
polarization of one of the output beams then you can be quite sure
that the polarization of the other will be in the perpendicular
direction.

I don't think that's significantly different from saying if you have
two streams of water flowing into a bathtub and you determine that
one of them is the cold water then you can be quite sure the other
one is hot.

Quite aside from entanglement playing no role, it also has nothing to
do with what's bothering {\Einstein}, since the two beams already
have their character (ordinary and extraordinary --- or horizontally
and vertically polarized) before any measurement is made on either of
them.

Am I missing something?
\eq

\section{13 December 1998, \ ``Missing Something''}

\bdm
Am I missing something?
\edm

I don't know: Maybe I was missing something myself.  I took the
license to think that {\Born} was not actually talking about a beam
of light, but a beam of light so attenuated as to be a single photon,
say.  (I don't think this is too much license.)  But then a
birefringent crystal wouldn't do the trick, would it?  We'd only
have the photon coming out one way or the other \ldots\ not two
entangled photons.  (At best we could say that the photon is
entangled with the crystal.)  You don't suppose I can take so much
license as to assume that {\Born} was talking about some nice
downconverting crystal, could I?  Guess not. Hmm.

Well, in any case, when he returns to the abstract discussion, it
seems to me that he is indeed talking about entangled states, as
{\Einstein} had attempted to steer him.  He just seemed to disagree
that there was anything mysterious in them as the two systems had
interacted in the past.  Do you agree with this?  Do you have the
whole passage there with you?  But then why did he---being a hell of
a lot better physicist than me---say the silly things that I
misinterpreted above?

\bdm
Quite aside from entanglement playing no role, it also has nothing
to do with what's bothering {\Einstein}, since the two beams already
have their character (ordinary and extraordinary --- or horizontally
and vertically polarized) before any measurement is made on either
of them.
\edm

This is sort of beside the historical issue, but \ldots\ \  If you
give me the first amount of license above, then I'm not so sure I
agree with this.  Each photon makes a ``random'' decision (though
obeying the probability law) about which way to go; it's hard to say
that it's polarization (as determined by the axis of the crystal)
existed beforehand.

Are we making progress?  Do you agree with these things?

Sorry for the sloppy scholarship!  (I got carried away with the
excitement of the morning.)

\section{13 December 1998, and to several others, \ ``I Whig-ed Out''}

You may remember the Whig historians: they ``analysed the events and
ideas of the past from the point of view of the present rather than
trying to understand the people of the past on their own terms.'' In
that sense, I committed an utter Whigism this morning, and I
apologize.  In my over-enthusiasm for finding a new historical
nugget, I read more into {\Born}'s passage than I should have.  In
particular, when my eyes saw ``doubly refracting crystal'', my brain
said ``down converting crystal.''  As David {\Mermin} pointed out to
me right away, that passage is just yet more evidence that {\Born}
never quite got what {\Einstein} was talking about with his
gedankenexperiment:  {\Born}'s example doesn't make any sense at all
as a reply to {\Einstein} (whose previous letter was indeed about
entanglement and his reasons for not believing that quantum
mechanics is complete).  A poor birefringent crystal wouldn't give
us any entanglement between two beams of light (some fraction of the
photons go here, some fraction go there):  {\Born} didn't seem to
notice the difference.

Sorry for the false alarm.

\section{27 December 1998, \ ``Executive Summary''}

I just finished skimming Kurt {\Gottfried}'s paper, ``Is the
Statistical Interpretation of Quantum Mechanics Implied by the
Correspondence Principle?'' ({\tt quant-ph/9812042}).  I'm not sure I
know what to make of it; I'm not sure it added anything to the
compulsion I already feel for the ``statistical interpretation'' of
the theory (where by that I mean ``quantum states as states of
knowledge'' much in analogy to the role I see for classical Liouville
distributions).  But I did notice in the end that you were thanked
for ``asking several pointed questions.''  If you have a moment,
could you give me an executive summary of what you think of the
paper and maybe ask some of those pointed questions again.

I hope you're having a nice holiday, and not spending too much time
working on your fences again (as I recall you were last year at this
time).  My in-laws are visiting us in California; we're having a
wonderful time doing basically nothing but eating and drinking.

\subsection{David's Reply}

\bq
I can't offer you much to go on beyond what Kurt says in the paper
itself.  He's posing the question whether, if you just gave a
classical physicist the {\Schroedinger} equation, he could figure out
the probability interpretation without any further help.

My primary ``pointed question'' of his first draft was simply to say,
over and over again, whenever he claimed to have extracted a
probability, ``probability of WHAT?''.  The ultimate result of this
prodding was that he retreated from a version in which a probability
density in configuration space emerged naturally, to the one you see,
in which only probabilities for internal degrees of freedom are, in
some sense, natural.

He's driven by some long conversations he had with John {\Bell} a
couple of months before {\Bell} dropped dead, where {\Bell} claimed
that in some deep way QM lacked a naturalness that all classical
theories possessed --- that there was a qualitative difference in the
``interpretation problem'' --- that indeed, there either was no
``interpretation problem'' for classical physics or that it was
trivial, while for QM something profound was missing.  Kurt never
felt this in his bones (the way you and I do) and this paper is (a)
an attempt to articulate why he feels this way but also (b) an
acknowledgment that winning his argument with {\Bell} is not going to
be as straightforward as he once thought.

I'm not sure he would agree with this ``executive summary'' but
that's more or less how I see it.

We celebrated the first anniversary of the half-mile fence a few
days ago.  No deer have got onto the premises and the rhododendra
have been liberated from their cages and remain uneaten.

Hope your digestion survives the holidays.
\eq

\section{11 January 1999, \ ``Chewin' the Fat''}

\bdm
My primary ``pointed question'' of his [{\Gottfried}'s] first draft
was simply to say, over and over again, whenever he claimed to have
extracted a probability, ``probability of WHAT?''.
\edm
I never did reply to you on this.  That's a good pointed question. I
presume you've noticed that it's the same one you ask---in one form
or another---to me (and other Copenhagenists like me) over and over.
Probabilities of the ANSWERS to the QUESTIONS we can ask, I say. Who
could want more from a physical theory?  Yeah, yeah, I know you want
more \ldots\ but I suspect you can't get it.

In this connection, by the way, I read a very nice article by
{\Anton} {\Zeilinger} the other day, ``On the Interpretation and
Philosophical Foundation of Quantum Mechanics,'' that lays out what
he thinks is the very most important feature of quantum mechanics.
You can find the article at:  {\tt
http://info.uibk.ac.at/c/c7/c704/qo/philosop.html}. It looks like his
hopes and dreams aren't so different from mine.

\section{11 January 1999, \ ``Chewin' the Cud''}

\bdm
Thanks for the {\Zeilinger} reference.  I hadn't seen it.  At a
glance it looks very reasonable, but I will have to look further
before pronouncing {\Anton} a closet Ithacan.
\edm

I'm having a hard time believing you said this!!!!  Is my
understanding of the Ithaca Attitude so far off the mark?  Or, is
this just a sign that {\Anton} should be in politics rather than
physics?  Can he make us both believe we have an ally?

Seriously, I read that article on three separate occasions; the one I
mentioned to you this afternoon was only the last of the lot.  I was
so taken by his speculation that quantum mechanics might be a hint of
something even bigger---via the recognition that quantum observers
are not ``detached'' (or at least their instruments cannot be)---that
I had little daydreams of trying to secure a postdoc with him in case
all else fails in my job search.  (I hear he's going to Vienna this
year; sounds like an awfully nice place to take a postdoc.)

If you find upon another reading that you still think {\Zeilinger}
``reasonable'', please write me again.  I want to know more details
about what you like about that article.

\section{17 January 1999, \ ``My Homework''}

I took your assignment over the weekend and read {\Schroedinger}'s
``Nature and the Greeks and Science and Humanism'' (all in one Canto
volume) and also his little book ``Mind and Matter.''  You were
right:  it was quite enjoyable reading.  I guess I'll say that even
though in my diligent search for Ithacan ideas I found so little.  I
presume you were most happy about pages 54 and 92--98.  This issue,
by the way, he calls the ``principle of objectivation'' and devotes a
chapter to it in the latter book above.  Have you read it?  I think
it may be more directly relevant to your quest than Nature and the
Greeks.

If I could say I liked one idea in N \&\ G a lot, it would be the one
expressed on pages 72 and 73:  the real world is by definition that
which can be distilled from our common experiences (that and nothing
more).  But, of course, my liking of that is self serving.

How's that reading of {\Kant}'s Prolegomena coming along?  (Cf.\ my
note dated 8 September 98, ``I'll {\Bohr} You to Tears.'')

Laugh at you?  Never!  I submit this note as evidence.

\section{21 January 1999, \ ``Schr\"odingerization''}

As you could tell from my last note, you helped spur my interest in
{\Schroedinger}.  \ldots\ Well, that's not completely true I guess.
My interest in him was mostly spurred by reading his paper ``The
Present Situation in Quantum Mechanics'' (1935) and the discussion of
entanglement therein which I think is particularly deep (aside from
his silly remarks about relativistic stuff at the end).  But you did
give me another point of contact for his thought.  Let me now give
you another point of contact.

The book is:  William T. Scott, {\sl Erwin {\Schroedinger}: An
Introduction to His Writings} (University of Massachusetts Press,
Amherst, 1967).  It's not beautiful and elegant in its composition
like {\Schroedinger}'s writings, but it does give an introduction to
lots of things Mr.\ S wrote and lots of good references.  In
particular, you might be interested to know that Michael Polanyi (in
apparently several books) has also discussed the thing S calls the
``principal of objectivation.''

\section{21 January 1999, \ ``More Stupendous Literature Man''}

Let me send you a little more in connection with your (probably
minor) interest in {\Schroedinger}.  This is a letter I sent off to
{\Caves} and {\Schack} for a completely different purpose.  [See
note to {\Ruediger} {\Schack}, dated 21 January 1999, titled
``Stupendous Literature Man.''] But I started to wonder what you
might think of {\Schroedinger}'s ``dangerous'' {\Wheeler}-like
tendencies: recall {\Wheeler}'s big U with an eye atop it to
represent the universe. Have a look at the {\Schroedinger} quote
below.

No need to reply if you're not moved.

\section{28 January 1999, \ ``The Propaganda Cocktail''}

On another subject: your prodding me into reading {\sl Nature and the
Greeks\/} sent me into a wild {\Schroedinger} spiral.  I am
surprised I shunned him all these years; I now think he actually
came up with quite a bit of thought provoking material.  Have you
studied all of his 1935 papers on entanglement carefully?  That in
conjunction with his apparent belief in indeterminism before quantum
mechanics (1916--1922) makes for a really interesting mix.  (In that
connection have you read Paul {\Hanle}'s article, ``Indeterminacy
Before {\Heisenberg}: The Case of Franz {\Exner} and Erwin
{\Schroedinger}''?)

\section{31 January 1999, \ ``{\Schroedinger} Note''}

Thanks for the encouragement and the advice.

\bdm
I don't know if I've seen all the 1935 {\Schroedinger} papers on
entanglement, but the ones I did look at were very impressive.
\edm

The one that I think is particularly elegant and insightful is this:
J.~D. Trimmer, Proc.\ Am.\ Philos.\ Soc.\ {\bf 124}, 323--338
(1980).  For some reason it is listed under the translator's name
rather than the original author.  (It can also be found in the
{\Wheeler}-{\Zurek} volume on quantum measurement.)  The paper is
advertised as the ``{\Schroedinger} cat'' paper, but it's really
about entanglement through and through.

\section{21 May 1999, \ ``A {\Bohr}, a Door, and a \ldots''}

Let me give you the coordinates to that {\Honner} article again:
it's listed below along with a sampling from it in a letter I wrote
{\Greg} {\Comer} almost two years ago.  [See letter to {\Greg}
{\Comer} dated 09 June 1997, ``Dictionaries and Their Problems.'']
(How long have I known you now?) I read the article again yesterday,
and didn't like it nearly as much as I had thought I had.  But I am
very happy I picked up that book again:  it's a marvelous collection
of papers trying to make sense of the old man.  This time around, I
was especially impressed with Henry {\Folse}'s article (of course
because I think it reminds me of myself).  And I'm slowly going to
try to make my way through some of the other articles that defend
{\Bohr} as a peculiar form of realist (instead of as an idealist,
positivist, or instrumentalist, etc.). I think that depiction of him
is what captures best the yearning of my own heart.

\section{08 June 1999, \ ``I'm Not Alone''}

I thought you might enjoy the little passage I scanned in today.  Pay
particular attention to the last paragraph where {\Pauli} expresses
what he means by complementarity.  [Letter from {\Pauli} to {\Bohr},
dated 15 February 1955]
\bq
\indent
In quantum mechanics, on the contrary, an observation {\it hic et
nunc\/} changes in general the ``state'' of the observed system in a
way not contained in the mathematically formulated {\it laws}, which
only apply to the automatic time dependence of the state of a {\it
closed\/} system.  I think here on the passage to a new phenomenon by
observation which is technically taken into account by the so-called
``reduction of the wave packets.''  As it is allowed to consider the
instruments of observation as a kind of prolongation of the sense
organs of the observer, I consider the unpredictable change of the
state by a single observation---in spite of the objective character
of the result of every observation and notwithstanding the
statistical laws for the frequencies of repeated observation under
equal conditions---to be {\it an abandonment of the idea of the
isolation (detachment) of the observer from the course of physical
events outside himself}.

To put it in non-technical common language one can compare the role
of the observer in quantum theory with that of a person, who by his
freely chosen experimental arrangements and recordings brings forth a
considerable ``trouble'' in nature, without being able to influence
its unpredictable outcome and results which afterwards can be
objectively checked by everyone.

Probably you mean by ``our position as detached observers'' something
entirely different than I do, as for me this new relation of the
observer to the course of physical events is entirely {\it
identical\/} with the fact that our situation as regards objective
description in ``this field of experienced'' gives rise to the demand
of a renewed revision of the foundation for ``the unambiguous use of
our elementary concepts,'' logically expressed by the notion of
complementarity.
\eq

\section{13 June 1999, \ ``Crackpots in Quantum Land''}

You're speaking to one!  I've just discovered something this morning
that I have to make a confession to.  Recall the little {\Pauli}
passage I sent you the other day?  Well, I had already sent it to
you!  (Which means of course that I had already read it and scanned
it into the computer at least once before!)  What a crackpot!  (I'm
sure you'd already caught me \ldots\ right?~\ldots\ you just didn't
want to embarrass me \ldots\ right?)  Anyway, if you've got nothing
better to do, have a look at the old note [dated 8 September 1998, a
long one titled ``I'll {\Bohr} You To Tears''].  It builds up the
context of the passage much better than the last rendition.

\section{15 October 1999, \ ``In Transit''}

How did your quantum information course go?  I was amused by your
remarks on the spelling ``qubit.''  {\Carl} {\Caves} and I once put
the following footnote in one of our papers:
\bq
{\it Qubit\/} is a shorthand for the minimal quantum system, a
two-state quantum system, that can carry a bit of information.
Logically, if one wishes to give a special name to the minimal
physical system that can carry a bit, one should do so for both
classical and quantum two-state systems, calling them perhaps c-bits
and q-bits.  We are reluctant to use the neologism ``qubit,'' because
it has no standard English pronunciation, the ``qu'' being pronounced
as in ``cue ball,'' instead of as in ``queasy.'' We prefer ``q-bit,''
but acquiesce in the use of ``qubit,'' which has attained a degree of
general acceptance.
\eq

\section{15 October 1999, \ ``More In Transit''}

\bdm
Your samizdat package just [arrived] and I look forward to perusing
it (and to anything else you have to tell me, especially if it is
surprising).
\edm
Thank you for teaching me the word samizdat!  If you have any
thoughts on my wacky train of thought, I would like to hear.  The
goal is, of course, some new equations and a lot less words.  So the
word samizdat fits perfectly.  I'd especially look forward to
hearing any of your thoughts on the later things I wrote Howard
{\Barnum} and John {\Preskill}.  I felt that I was especially lucid
in explaining the consistency of the state-vector-is-knowledge point
of view there.

\bdm
Hey, if you guys also think it should be q-bit or qbit or Qbit (I
thought I was alone) it's not too late to fight back.  (The
precedent, by the way, was set by Qtips.)
\edm
No, it's too late.  We wrote that in 1996 and the quantum info world
has certainly gotten a lot larger since then.  Everyone uses qubit,
including the press.

\bdm
Will you be at {\Gilles}'s meeting in Montr\'eal in December?  (I'm
trying to decide whether to go myself.)
\edm
Please do come.  I keep giving {\Gilles} subliminal messages that we
should organize a small ``Quantum Foundations in the Light of Quantum
Information'' meeting.  (He's interested.)  You can help reinforce
that.  I will be there from (late) Dec 7 through the end (Dec 11 or
12, I can't remember).  I'm showing up late because the week before
that, I'll be at Naples at a ``Chance in Physics'' conference.  In
case you're interested in that, it may not be too late.  And there
should be some interesting characters there.  Check out
http://www.mathematik.uni-muenchen.de/~bohmmech/chance.htm . You
won't by chance be at the APS meeting in Waterville, MA on Nov 5 and
6, will you?  I'll be there too.

I guess I'll give a talk on our quantum de {\Finetti} theorem at
{\Gilles}' meeting.  It's about how the notion of an ``unknown
quantum state'' (a phrase you see everywhere in quantum information)
is probably only meaningful in complex Hilbert spaces, not reals.
That might intrigue you.  In part, it's sort of a more malicious
version of the phenomenon {\Bill} uncovered.

\section{16 December 1999, \ ``{\Peierls} and Commutivity''}

 {\Asher} and finally came to some kind of agreement on our Physics
Today piece \ldots\ what a relief!!  Did I tell you the other day
that I had scanned in that {\Peierls} article that we were talking
about?  I scanned it in because  {\Asher} had wanted a copy for
something to do with the PT article I just mentioned.

Anyway, I thought you might like a copy too.  It's placed below.
[See note to  {\Asher} {\Peres}, titled ``Miracles of Office
Technology,'' dated 21 November 1999.] (In it, for  {\Asher}, I had
placed some commentary about how {\Peierls} wasn't being consistent
with his language, etc.) I just reread the part about commuting
density operators.  I'm still not quite sure what he is saying, but
there may be some connection with one of my better technical
results. Have a look at: H.~{\Barnum}, C.~M. {\Caves}, C.~A. Fuchs,
R.~{\Jozsa}, and B.~{\Schumacher}, ``Noncommuting Mixed States
Cannot Be Broadcast,'' Phys.\ Rev.\ Lett.\ {\bf 76}(15), 2818--2821
(1996). (There's a slightly extended/smoother discussion in my PhD
thesis, which you can find on {\tt quant-ph}. Also G\"oran
{\Lindblad} rederived the result from some more standard $C^*$
algebra results in Lett.\ Math.\ Phys.\ last year---but honestly, I
couldn't understand his proof.)

Anyway the result is a generalization of the no-cloning theorem that
explores when an automatic device can create a (correlated) copy of
a density operator:  it can if and only if the set of possible
density operators are restricted to a commuting set.  Perhaps
{\Peierls} had that kind of thing in mind.  I.e., thinking about
people's knowledge as being represented by density operators, and
our ability to share information one with the other having to do
with attempting to ``broadcast'' those density operators into other
people's heads.

Tell me what you think about this interpretation.

\section{17 December 1999, \ ``Reality of the Symbol''}

\bdm
The Eco quote, for what it is worth, is from Umberto Eco, {\em
{\Kant} and the Platypus}, Chapter 1 (``On Being''), section I.I
(Semiotics and the Something) which begins: [no scanner]
\bq
\noindent \rm Why should semiotics deal with this something? Because
one of the problems of semiotics is to say whether and how we use
signs to refer to something, and a great deal has been written on
this. But I do not think that semiotics can avoid another problem:
What is that something that induces us to produce signs?
\eq
It's the part after the colon that I like, though I've got to admit,
seeing it above in black and white, I guess it could be grinding
almost anybody's axe.  But for me the sign is the quantum state.
Haven't yet read Eco's answer to his question.  He can get pretty
difficult, though.
\edm

Thanks for sending the Eco quote.  Now it's safely in my quantum
archive.  I did a search on the word ``semiotics'' in my samizdat,
but couldn't find it anywhere:  so you must have gotten a
misimpression from something else I wrote.  I do have a couple of
interesting tidbits on the ``reality of the symbol'' that you might
be interested in.  One is the note titled ``A {\Fleck} of {\Fleck}''
on page 49; the other is a note titled ``Echoes'' on page 183.

I think in the next note, I'll send you a copy of the final version
of the thingy I wrote with  {\Asher}.  It's changed quite a bit since
what you last saw.  I would like to get your comment as to what you
see is deficient in the point of view.  I'm toying with the idea of
writing a much extended version of it for the proceedings of this
Bohmian meeting that I just returned from.  The title would be
``Knowledge About What?'' and I think I would be much less timid
about making statements about the ``what'' than we were in the {\sl
Physics Today\/} article:  the great lesson of the quantum is that
the world is so ``sensitive to the touch'' that we cannot
(conceptually) distill a free-standing ``reality'' from our
experiences.  What a wonderful property for the world to have!
Anyway, in that regard, I wonder if you wouldn't mind looking at one
other thing I put together:  it's a note titled ``{\Penrose} Tiles''
starting on page 165.  In it I express the kernel of the new
language I want to push for:  forget ``measurements'' and
``outcomes,'' quantum mechanics is about ``acts'' and
``consequences.''  Those ideas too would go into the Naples
proceedings paper.

I want to continue thinking about the {\Peierls} thing you brought to
my attention.  The more I keep thinking about it, the more it all
resonates with some other stuff I've been thinking about for the
last month.  I'm searching for a notion of ``macroscopic'' within my
``information gain vs.\ information disturbance'' (don't forget the
adjectives ``information'') foundation for quantum mechanics. Here's
the direction.  Suppose I describe some system by either $\rho_0$ or
$\rho_1$ but you just don't know which.  So you perform a
measurement on the system to try to gain some of the same
information I have, i.e., you try to guess which quantum state I'm
using.  If the two states are pure and noncommuting, then we know
that there will be some inevitable disturbance to my description
because of your intrusion---the minimal disturbance can even be
quantified purely as a function of your information gain.  But what
happens if the two states are very, very mixed (almost the identity
matrix)?  I suspect your information gathering will cause little
disturbance to my description:  it's just a question of quantifying
this now.  If it works out as I suspect, then I think I know what
``macroscopic'' means.  Take the tree outside my window.  I know
some things about it, but in the grand scale of things I know very,
very little.  If I were to assign a quantum state to it, it would be
a very mixed quantum state.  Suppose you've narrowed down my
assignment (just by a pure guess, no measurement) to two
possibilities, and now you intrude on the system to gather some
further information about my precise choice.  Your information
gathering won't cause much disturbance to my description.  And that
is what means ``macroscopic'':  when I know very little about a
system, so little that your information gathering {\it need not\/}
disturb my description, then the system is macroscopic {\it with
respect to me}. That's the idea.  If the system is also macroscopic
with respect to you, then we can cross-check our information, and the
system (with respect to both of us) will be no worse for the wear. I
should work to try to quantify this.

\section{26 January 2000, \ ``Note \#2 -- Foundations in Montr\'eal''}

\bdm
P.P.S.  Have you seen a little book by {\Schroedinger} called ``My
View of the World''?  It contains two extended essays on the nature
of objective reality, one in 1925 (before his Equation!) and one in
1960. They're surprisingly similar.  They turn upside down the
notion that objective reality is a valid inference from the fact of
intersubjective agreement.  He argues instead (I think) that
intersubjective agreement is a manifestation of the unity of all
consciousness, which is also what creates [the illusion of] objective
reality.  A kind of global mass-solipsism.  But very beautifully put.
Makes me think I should read {\Spinoza}, which I never have done.  So
the IIQM is being tugged in one direction by QIP and quite another by
goddamed mysticism.  It's really delicious that Schr"odinger is the
hero of those who believe {\Bohr} abandoned hard-headed rationality.
\edm

Regarding the big S, I have indeed read that book.  After you
suggested reading ``Nature and the Greeks,'' I read everything by him
that I could get my hands on.  Now that you mention it though, with
such an extreme form of idealism working in the background of his
thoughts, I can't quite remember why he still had so many troubles
with Copenhagenish kinds of views about quantum mechanics.  Do you
have any insight?

\section{20 March 2000, \ ``Invitation Coming Finally''}

{\Gilles} has asked me to ask you about one further possibility:
Jeffrey {\Bub}.  Do you know him?  Do you have an opinion?  What's he
like personally?  I have never met him.  I am a little against him
in that his book gives him the appearance of having a more strongly
set opinion on the foundations than I perceive from the rest of us.
The point of this meeting is that I don't think the set opinions
existing presently are very adequate.  {\Gilles}, on the other hand,
is for him because he is Canadian (and that sort of thing may be
important for the source of our funds), but also because he has had
dinner with him once and enjoyed his company. And of course there is
the issue that he does represent a school of thought in the
foundations that is not directly reflected in any of the other
potential participants.  It is because of this that I could easily be
swayed to plop him into the invitation list.\footnote{I simply could
not resist including this very personal paragraph from the note:  It
more than anything else in the book demonstrates the dangers of the
close-mindedness I am predisposed to (and that I battle every day).
See Chapter \ref{BubChapter} ``Letters to Jeffrey {\Bub}'' to
understand this remark!}

Your candid opinion will be useful to the both of us.

By the way I enjoyed your {\sl Physics Today\/} column this month.
If the issue with {\Bub} hadn't come up, I would have written you a
note titled ``Ginger or Mary Ann?''  If you've ever watched
Gilligan's Island, then you'll know what I'm talking about.  I was
always a fan of Mary Ann; my best friend was always a fan of
Ginger.  I just can't believe that this issue of beauty and elegance
in science goes any deeper than that.  And I get really annoyed by
the opinion-Nazis who tell me otherwise.  Strangely Richard {\Jozsa}
and I had a heated debate on just this issue last week in Tokyo.

\section{23 March 2000, \ ``Can You Complete This?''}

\bdm
Rereading your article with  {\Asher} I was struck by your (not
entirely convincing) assertion that it's OK for Cat's density matrix
to be either fruit or cake while Erwin's still is a superposition.
\edm

That's because you're not entirely convinced that the wavefunction
is an epistemological beast.  You don't yet really believe that it
has no ontological content.  Too bad.

If it helps, I believe I explained this ``striking assertion'' a
little more clearly in some notes to Howard {\Barnum}.  Read ``It's
All about Schmoz'' and ``New Schmoz Cola'' on pages 10 through 15 of
my compilation.  If not more clearly, at least I can claim to have
said things more vividly.  The {\Fierz} article that is referred to
can be found starting on page 168.  [See note to {\Ruediger}
{\Schack}, titled ``{\Penrose} Tiles,'' dated 29 August 1999.]

I'm sorry I myself don't have time to think about your question (or
even read it in detail actually) until late next week.  I'm in a bit
of a panic getting ready for my lectures in Scotland.  But I will
get back to you soon thereafter.

\section{03 April 2000, \ ``{\Gleason}, For the Record''}

I'm flying back to ABQ from Boston in some of the roughest wind I've
ever been in!  Yeaks.

Anyway, I just wanted to write this memory down for my record, and I
thought you might enjoy a little bit of it.  It turns out that
Andrew {\Gleason} was at the AMS meeting that I just attended.  I
told Mary Beth {\Ruskai} how I would like to catch a glimpse of him,
but that I was too shy to meet him \ldots\ not having anything to
say. Well, she introduced me nevertheless.  At first all I could
think to say was, ``I don't really have anything to say, other than
that I really love your theorem.  It was a real pleasure to work
through its proof.'' He said, ``Which theorem?''  So I said the 1957
one on quantum mechanical probability measures.  Then somehow my
automatic mouth kicked in and I started talking.  I told him about
how David {\Meyer} had shown that it doesn't hold for vector spaces
over rational number fields.  And then I asked him his opinion about
whether it might not still hold when the field is only weakened from
the full set of reals down to the algebraic numbers.  Somehow this
led to a discussion of {\Kochen}-{\Specker} and I told him about the
minimal known examples in $R^3$.  I explained the idea of the whole
thing, and he said(!), ``Is that what that is about?  That you just
can't make that kind of coloring?  I never realized that that's all
they were talking about.  Can you give me a reference where I can
find some of these minimal examples?''  So I gave him a reference to
{\Peres}' book, and we parted.

He struck me as such a gracious sort of person.  It really made my
day.  Rarely am I in awe, like some people in the presence of a
movie star, but this time I really felt it.

I hope you're working on your problem set for FILQI.

\section{05 April 2000, \ ``Commutivity''}

I finally had a chance to sit down and reread the {\Peierls} passage
you recommended and also your accompanying note.  When I had last
looked at his piece in Physics World, I too thought he was up to
something interesting.  But now, after looking at his {\sl More
Surprises\/} passage and thinking a little more carefully , I
understand why you're having such difficulty sussing it all out:
{\Peierls} is just trivially wrong here.  People ascribe noncommuting
density operators to the same system all the time---it's called
quantum cryptography.  Just consider the B92 protocol.  Alice
ascribes one of two possible nonorthogonal (pure) states to the
system she just prepared.  Eve, if she is aware of Alice and Bob's
protocol, can do no better than ascribe a density operator that is
the simple mixture of the two possibilities.  This mixed-state
density operator will commute with neither of Alice's possible
preparations.  Moreover, this will generally remain true even after
Eve has done some eavesdropping \ldots\ and even after Alice has
taken into account that Eve has done some eavesdropping.

I general I think {\Peierls} just has the whole idea backward (from a
Paulian point of view, that is).  In his Physics World article, he
writes:  ``\ldots\ there are limitations to the extent to which their
knowledge may differ.  This is imposed by the uncertainty
principle.''  It seems to me, people can believe whatever they
want---I like the word ``belief'' better than ``knowledge.''  There
are no limitations to that.  The unique thing about the quantum
world is that it never lets people's opinions get so aligned that
they can jump to the conclusion of a free-standing reality.  This is
captured by the fact that the best that can be said of almost all
measurements can only be couched in statistical language:  there are
no certainties for almost all observables even when one has maximal
knowledge (i.e., a pure state).

I'll put some further thoughts on this below, in the form of a draft
that I'm presently writing on information and disturbance.  If
you've got the stamina to get all the way to the paragraph starting
with ``It is at this point \ldots'' then you might see what I'm
talking about when I say that {\Peierls} has it backward.  (By the
way, be merciful on my English:  this is only a draft, and I'm not
even very far into it yet.)  [See C.~A. Fuchs and K.~{\Jacobs},
``Information Tradeoff Relations for Finite-Strength Quantum
Measurements,'' {\tt ArXiv:quant-ph/0009101}.]

\subsection{David's Reply}

\bq
P.S.  {\Peierls} is sometimes wrong, but never trivially wrong.  (He
is my {\Bill} {\Wootters}, you know.)   I'll think more about your
remarks when peace returns.
\eq

\section{05 April 2000, \ ``{\Pauli} Understood Entanglement''}

Yes, by FILQI I did mean the Montr\'eal meeting.

Also I forgot to reply a little more to this:
\bdm
Rereading your article with  {\Asher} I was struck by your (not
entirely convincing) assertion that it's OK for Cat's density matrix
to be either fruit or cake while Erwin's still is a superposition.
\edm

Authority certainly won't help with you, but I was struck the other
night in reading {\sl Niels {\Bohr} Vol.\ 6}---I've never been able
to get hold of that volume before---by another letter from {\Pauli}
to {\Bohr}. This one was dated 17 October 1927!  Among other things,
it says the following:
\bq
Further down on page 5 where the `proper reduction of the spatial
extension of the fields' is mentioned, one could perhaps still add
something for the sake of clarity.  This is of course just a point
which was not quite satisfactory in the {\Heisenberg} paper; there
the `reduction of the wavepacket' seemed a bit mysterious.  Now it
should of course be emphasized that such reductions first of all are
not necessary when all the measuring instruments are {\it
included\/} in the system.  In order to be able to describe the
results of observation theoretically at all, one must ask what can
be said about one {\it part\/} of the total system on its own.  And
then one sees as a matter of course the complete solution---that the
omission of the instruments of observation in many cases (not
always, of course) may formally be replaced by such discontinuous
reductions.
\eq

Isn't that cute?

\section{06 April 2000, \ ``Mist of the Future''}

\bdm
I would like nothing better than a convincing demonstration that it's
all about knowledge and only knowledge, but the ways in which the
knowledge of different people can and cannot be interrelated \ldots\
It seems to me that you, being it's-only-about-knowledge-ish, have to
say a little more about this. There remains work to be done.
\edm

The first sentence in this quote shocks me actually.  How can I
forget desideratum \#1 in your original Ithaca paper?  But of the
second two sentences, you should know that you have my acute
sympathy.  What do you think page 2 of my ``Notes on a Paulian Idea''
is all about?  [See page ?? of the present volume.]  I think my main
role in life lately---second only to being a father---has been to
drum up support and turn our community to just this issue.  There
remains not only work to be done, but A LOT of work to be done.  And
it's a gut feeling admittedly, but the fruits will be great.

\section{20 July 2000, \ ``Zing!''}

I promised you a report on {\Plotnitsky}, and now maybe I finally
have some time to do it.  As you predicted, I did find him extremely
enjoyable to talk to.  He does know his {\Bohr}, and is certainly my
philosophical cousin.  I spent a lot of time in Mykonos with him, on
the beach and at dinners, and here and there otherwise.  I
challenged him to write an essay comparing and contrasting {\Pauli}
and {\Bohr}'s thoughts on QM:  he took the challenge!  So I'm looking
forward to seeing that, or at least talking about that with him.

Thanks for sending me your Firing Line article.  It didn't irritate
me at all; I enjoyed it.  However, it does reveal that you are far
more Platonic than I.  (Or is it, me?   I must admit that this is
one rule of grammar I never remember.) {\Bill} {\Wootters}, Richard
{\Jozsa}, Peter {\Shor} and I went to dinner one evening in Capri and
the conversation ended up degenerating into one about the
foundations of quantum mechanics and the foundations of
mathematics.  It seems that Peter wouldn't admit the real existence
of the number continuum. Richard protested (in his usual way) that
that was just silly, and started to try to build up a hierarchy of
ideas.  ``Without saying, the natural numbers exist.''  Peter and
{\Bill} both agreed quickly; I didn't.  Richard was incensed!
``Well, then are you going to tell me that the number three doesn't
exist!?!? What about the three glasses in front of you?''  I just
replied, ``There's no sense in which those objects are the same
without our talking about them.  If I weren't here to make the
judgement that they're {\it roughly\/} the same, what principle of
nature would do it for me?''  He found me boring (in his usual way)
and left the subject.

Let me give you yet another suggestion for reading (that'll you'll
probably ignore 2/3 of \ldots\ you told me you only read about 1/3 of
what I suggest).  I really enjoyed this one on my flight home the
other day.
\bq\noindent
H.~J. {\Folse}, ''Niels {\Bohr}'s Concept of Reality,'' in {\sl
Symposium on the Foundations of Modern Physics 1987: The Copenhagen
Interpretation 60 Years after the Como Lecture}, edited by
P.~{\Lahti} and P.~{\Mittelstaedt} (World Scientific, Singapore,
1987), pp.~161--179.
\eq
As always, I almost surely liked it because it was saying something
I wanted to hear.

Somehow I feel that I had an epiphany in Mykonos.  Do you remember
the parable of ``Genesis and the Quantum'' from my Montr\'eal problem
set?  [See note to {\Gilles} {\Brassard}, titled ``Problem Set,''
dated 15 May 2000.] And do you remember my slide of an empty black
box with two overlays. The first overlay was of a big $|\psi\rangle$
(hand drawn in blue ink of course).  I put the slide of the box up
first, and said ``This is a quantum system; it's what's there in the
world independent of us.'' Then I put the first overlay on it and
say, ``And this symbol stands for nothing more than we know of it.
Take us away and the symbol goes away too.''  I then remove the
$|\psi\rangle$.  ``But that doesn't mean that the system, this black
box, goes away.''  Finally I put back up the $|\psi\rangle$ over the
box, and the final overlay. This one says: ``Information/knowledge
about what?  The consequences of our experimental interventions into
the course of Nature.''

Well, now I've made another overlay for my black box slide.  At the
top it asks, ``So what is real about a quantum system?''  In the
center, so that it ends up actually in the box, is a very stylistic
version of the word ``Zing!''  And at the bottom it answers, ``The
locus of all information-disturbance tradeoff curves for the
system.''  In words, I (plan to) say, ``It is that zing of the
system, that sensitivity to the touch, that keeps us from ever
saying more than $|\psi\rangle$ of it.  This is the thing that is
real about the system.  It is our task to give better expression to
that idea, and start to appreciate the doors it opens for us to
shape and manipulate the world.''  What is it that makes quantum
cryptography go?  Very explicitly, the zing in the system.  What is
it that makes quantum computing go?  The zing in its components!

Anyway, I'm quite taken by this idea that's getting so close to
being a technical one---i.e., well formed enough that one might
check whether there is something to it.  What is real of the system
is the locus of information-disturbance (perhaps it would be better
to say ``information-information'') tradeoff curves.  The thing to do
now is to show that Hilbert space comes about as a compact
description of that collection, and that it's not the other way
around.  As I've preached to you for over two years now, this idea
(though it was in less refined form before now) strikes me as a
purely ontological one \ldots\ even though it takes inserting an
Alice, Bob, and Eve into the picture to give it adequate
expression.  That is, it takes a little epistemology before we can
get to an ontological statement.

I looked back at your original Ithaca Interpretation paper, and I'll
be bold enough to say that this idea satisfies all your most
important desiderata:  (1), (2 suitably modified), (3), and (5).

Part of this, by the way, is why I liked so much {\Folse}'s paper.
Also, believe it or not, for a moment while reading it I thought I
could finally ``SEE'' correlation without correlata.  (Not lying.)
But then I thought I liked the phrase ``Interaction without
Interactoids'' even more.  My wife just thought I was being silly.
Maybe you will too.

PS.  I did take the job at Bell Labs.  I wonder if they really know
what they've gotten themselves into?

\section{20 July 2000, \ ``{\Schroedinger} from a Different Angle''}

Me again.  Let me see if I can troll your good memory for a
reference I'm looking for.  It's something {\Schroedinger} said, but
where?  (I would be able to find it if my library hadn't burned up.)

This is a passage from a paper by Ed {\Jaynes}.  He writes:
\bq
\noindent [The Copenhagen interpretation] denies the existence of an
``objectively real'' world.

But surely, the existence of that world is the primary experimental
fact of all, without which there would be no point to physics or any
other science; and for which we all receive new evidence every
waking minute of our lives.  This direct evidence of our senses is
vastly more cogent than are any of the deviously indirect
experiments that are cited as evidence for the Copenhagen
interpretation.

Perhaps our concern should be not with hidden variables, but hidden
assumptions; not only about the theory, but about what we are
measuring in those experiments.  Consider a cascade decay
experiment.  As soon as we say something like ``In this experiment
we observe two photons emitted from the same atom,'' we have already
assumed the correctness of a great deal of the theory that the
experiment was supposed to test.  This initial stacking of the cards
then affects how we analyze the data. \ldots

Do we need more hidden variables?  Perhaps eventually, but maybe our
immediate problem is the opposite; first we need to get rid of
some.  To define the state of a classical particle we must specify
three coordinates and three velocities.  Quantum theory, while
denying that even these degrees of freedom are meaningful and
claiming to ``remove unobservables,'' replaces them with an infinite
number of degrees of freedom defining a continuous wave field.  To
specify a classical wave field, we need one complex amplitude for
each mode; for a quantized field we need an infinity of complex
amplitudes for each mode.

So perhaps quantum theory, far from removing unobservables, has
introduced infinitely more mathematical degrees of freedom than are
actually needed to represent physical phenomena.  If so, it would
not be surprising if a few infinities leak out into our calculations.
\eq

The part in particular that I'm interested in is the last two
paragraphs.  I remember {\Schroedinger} saying something so similar
somewhere.  Where was it?

Thanks for any help you can give!

\section{23 July 2000, \ ``Great {\Garrett} Quote''}

\bq
\noindent AUTHOR'S NOTE:  In my {\sl Physics Today\/} article ``Quantum
Theory Needs No `Interpretation'\,'' with  {\Asher} {\Peres}, there
is a passage that says the following:
\bq
\noindent Contrary to those desires, quantum theory does {\it not\/}
describe physical reality.  What it does is provide an algorithm for
computing {\it probabilities\/} for the macroscopic events
(``detector clicks'') that are the consequences of our experimental
interventions. \ldots
\eq
This passage (usually taken out of context) has been found
distasteful by several readers.  At least one of those readers was
Alvaro {\Carvalho}, and at least one reader of {\Carvalho} was the
iconic many-worlds activist David {\Deutsch}.\footnote{It probably
goes without saying, but perhaps I should have said ``many-worlds
activist and Star Trek fan.''  See {\tt
http://www.qubit.org/people/david/David.html} for details.} I know
this because Michael {\Nielsen} forwarded to me {\Carvalho}'s posting
of 12 July 2000 and {\Deutsch}'s reply of 16 July 2000 from the
electronic bulletin board {\tt Fabric-of-Reality@egroups.com}.
{\Carvalho} writes:
\bq
For those who have not read the Letter: ``Quantum theory needs no
'interpretation'\,'' by C. Fuchs and  {\Asher} {\Peres} (Physics
Today -March 2000), here are some short excerpts (with
comments\ldots): [\ldots]

``Contrary to those desires, quantum theory does not describe
reality \ldots'' I wonder what it can possibly describe. Is there
anything else beyond reality?
\eq
While {\Deutsch} responds:
\bq
No, but that's not what they think. They think it describes our
observations, but that we are not entitled to regard this as telling
us anything about a reality beyond our observations. Why? Just for
the Bohring old reason that they don't like the look of the reality
that it would describe, if it did describe reality. Why? -- I have
many speculations, but basically I don't know. I don't understand
why.

It's sad enough when cranks churn out this tawdry old excuse for
refusing to contemplate the implications of science, but when highly
competent physicists -- quantum physicists -- dust it off and proudly
repeat it, it's a crying shame.
\eq
In bemusement, I forwarded these quotes to David {\Mermin}, who
responded with:
\bdm
Funny, there was this English Bayesian, Tony {\Garrett}, who said
more or less the same thing about anybody who had given up the
search for hidden variables.  And I suppose there's a sense in which
many worlds, insofar as it can be made coherent, is the ultimate
hidden variables theory.
\edm
This comment spurred the following note on my part.\medskip
\eq

Indeed I know about {\Garrett}.  I have a great quote from him in the
passage below.  In fact, I wrote this passage to combat people just
like him.  I thought you might enjoy it as whole (along with the
quote, that is).

When am I ever going to publish all this crap I've written? \bigskip

\noindent ---------------------
\bq
The first opposition we must tackle is not the physicist but the
average Bayesian himself.  Bayesians generally tend to be ``naive''
philosophical realists. Their frame of mind goes something like this:
``There is a real world out there.  We use probability to quantify
our ignorance of it.  When we have maximal information about it, we
have no ignorance.  Consequently, all probability assignments we
might make in that situation should be of the trivial sort, simple
Kronecker deltas---0 if false, 1 if true. In contradistinction, when
there is ignorance, we should seek to alleviate it.  By definition,
when we have ignorance our information is nonmaximal, and therefore
stands to be completed.''\footnote{In this connection, it is probably
worth noting a widespread misconception of non-Bayes\-ians. It is
sometimes said that Bayesians have dangerous tendencies toward
philosophical idealism. Nothing could be farther from the truth: this
passage tries to show that.  The confusion seems to come about
precisely because Bayesians support a {\it subjective\/} notion of
probability rather than a objective one.  Somehow, this is seen
instead as an endorsement of a ``subjective reality.''  The Bayesian
is simply careful to point out that there is the world and then there
is what we know about it.  The two categories should not be confused.
In particular, the Bayesian insists that probability should be
recognized as inhabiting the latter category only; this says nothing
negative about the existence of the former.}

Because of this, most Bayesians have trouble coming to grips with the
fundamental indeterminism posed by quantum mechanics.\footnote{Hints
of this are peppered throughout many of the papers in Ref.~[1]. See
also, in particular, Refs.~[2,3].}  It makes no sense within {\it
their\/} framework.  Perhaps this point of view has been expressed
most clearly by A.~J.~M. {\Garrett} [4]:
\begin{quote}
\baselineskip=11pt
The nondeterministic character of quantum measurement can, and
should, be taken to imply a deeper `hidden variable' description of
a system, which reproduces quantum theory when the unknown values of
the variables are marginalised over.  Differences in measurements on
identically prepared systems then represent differences in the
hidden variables of the systems.  Not to seek the hidden variables,
as the Copenhagen interpretation of quantum mechanics arbitrarily
instructs, is to give up all hope of improvement in advance, and is
contrary to the purposes of science.
\end{quote}
But then again, most Bayesians are relatively far removed from the
countless attempts that have been made to find a {\it reasonable\/}
hidden-variable model underneath quantum mechanics [5,6]. The
indifference of most physicists to yet further attempts is far from
arbitrary.  The strong constraints placed by the {\Bell} and
{\Kochen}-{\Specker} theorems are quite compelling in this respect
[7, Chaps.\ 5--7].  What practicing physicist would really want to
bother with a hidden variable theory of the Bohmian type, which is
nonlocal to its core [8]? Its defining feature appears to be no more
than this:  more difficult mathematics than the standard theory,
while at the same time making no further predictions that can
actually be tested.  Hardly a bargain.  All this is done for the
sole sake of an easily pointed to ``reality'' underneath the
statistical predictions---a ``reality'' so far removed from common
experience that it is practically unrecognizable [9,10].

This turn is not for us.  It need not be.  The open-eyed Bayesian has
no use for hidden variables in the quantum context.  Our strength
actually lies in adopting the selfsame language of {\Einstein}
already introduced. The quantum mechanical description---the quantum
state---is in fact not a {\it complete\/} description of the quantum
system. {\it However}, we temper this by adding that, at the same
time, we staunchly reject {\Einstein}'s assessment of the situation
[11]:
\begin{quote}
There exists [\ldots] a simple psychological reason for the fact
that this most nearly obvious interpretation is being shunned.  For
if the statistical quantum theory does not pretend to describe the
individual system (and its development in time) completely, it
appears unavoidable to look elsewhere for a complete description of
the individual system; in doing so it would be clear from the very
beginning that the elements of such a description are not contained
within the conceptual scheme of the statistical quantum theory.  With
this one would admit that, in principle, this scheme could not serve
as the basis of theoretical physics.
\end{quote}
The theory prescribes that no matter how much we know about a quantum
system---even when we have {\it maximal\/} information about
it---there will always be a statistical residue.  There will always
be questions that we can ask of a system for which we cannot predict
the outcomes.  {\it In quantum theory, maximal information is not
complete and cannot be completed.}

To think that this is an arbitrary rejection of the principles of
science is to work within a prejudice imposed by classical physics.
This statement is actually a physical postulate, perhaps not so
different in character than that the speed of light is constant in
all reference frames.  It breaks an old tradition, but that need not
be a blemish.  The clarity the ``$c=\mbox{const}$'' assumption
brought to the already-existent Lorentz transformations cannot be
questioned [12].

\begin{enumerate}
\small

\item
R.~D. Levine and M.~Tribus, eds., {\sl The Maximum
Entropy Formalism}, (MIT Press, Cambridge, MA, 1979); C.~R. Smith
and W.~T. Grandy, Jr., eds., {\sl Maximum-Entropy and Bayesian
Methods in Inverse Problems}, (D.~Reidel, Dordrecht, 1985); J.~H.
Justice, ed., {\sl Maximum Entropy and Bayesian Methods in Applied
Statistics}, (Cambridge U. Press, Cambridge, 1986); J.~Skilling,
ed., {\sl Maximum Entropy and Bayesian Methods:\ Cambridge, England,
1988}, (Kluwer, Dordrecht, 1989); P.~F. Foug\`ere, ed., {\sl Maximum
Entropy and Bayesian Methods:\ Dartmouth, U.S.A., 1989}, (Kluwer,
Dordrecht, 1990); B.~Buck and V.~A. Macaulay, eds., {\sl Maximum
Entropy in Action:\ A Collection of Expository Essays}, (Clarendon
Press, Oxford, 1991); W.~T. Grandy, Jr.~and L.~H. Schick, eds., {\sl
Maximum Entropy and Bayesian Methods:\ Laramie, Wyoming, 1990},
(Kluwer, Dordrecht, 1991); C.~R. Smith, G.~J. Erickson, and P.~O.
Neudorfer, {\sl Maximum Entropy and Bayesian Methods:\ Seattle,
1991}, (Kluwer, Dordrecht, 1992); A.~Mohammad-Djafari and
G.~Demoment, eds., {\sl Maximum Entropy and Bayesian Methods:\
Paris, France, 1992}, (Kluwer, Dordrecht, 1993); G.~R. Heidbreder,
{\sl Maximum Entropy and Bayesian Methods:\ Santa Barbara,
California, U.S.A., 1993}, (Kluwer, Dordrecht, 1996); J.~Skilling
and S.~Sibisi, {\sl Maximum Entropy and Bayesian Methods:\
Cambridge, England, 1994}, (Kluwer, Dordrecht, 1996); K.~M. Hanson
and R.~N. Silver, {\sl Maximum Entropy and Bayesian Methods:\ Santa
Fe, New Mexico, U.S.A., 1995}, (Kluwer, Dordrecht, 1996).

\item
E.~T. {\Jaynes}, ``Clearing Up Mysteries -- The Original Goal,'' in
{\sl Maximum Entropy and Bayesian Methods (Cambridge, England,
1988)}, edited by J.~Skilling (Kluwer, Dordrecht, 1989), pp.~1--27.

\item
E.~T. {\Jaynes}, ``Probability in Quantum Theory,'' in {\sl
Complexity, Entropy and the Physics of Information}, edited by W.~H.
{\Zurek} (Addison-Wesley, Redwood City, CA, 1990), pp.~381--403.

\item
A.~J.~M. {\Garrett}, ``Making Sense of Quantum Mechanics: Why You
Should Believe in Hidden Variables,'' in {\sl Maximum Entropy and
Bayesian Methods (Paris, France, 1992)}, edited by
A.~Mohammed-Djafari and G.~Demoment (Kluwer, Dordrecht, 1993),
pp.~79--83.

\item
F.~J. Belinfante, {\sl A Survey of
Hidden-Variables Theories}, (Pergamon Press, Oxford, 1973).

\item
M.~{\Jammer}, {\sl The Philosophy of Quantum Mechanics: The
Interpretations of Quantum Mechanics in Historical Perspective},
(John Wiley, New York, 1974).

\item
A.~{\Peres}, {\sl Quantum Theory: Concepts and Methods}, (Kluwer,
Dordrecht, 1993).

\item
D.~{\Bohm} and B.~J. Hiley, {\sl The Undivided Universe: An
Ontological Interpretation of Quantum Theory}, (Routledge, London,
1993); J.~T. Cushing, {\sl Quantum Mechanics: Historical Contingency
and the Copenhagen Hegemony}, (U. of Chicago Press, Chicago, 1994).

\item
J.~T. Cushing, A.~{\Fine}, and S.~{\Goldstein}, eds., {\sl Bohm\-ian
Mechanics and Quantum Theory:\ An Appraisal}, (Klu\-wer, Dordrecht,
1996).

\item
M.~O. Scully, ``Do {\Bohm} Trajectories Always Provide a Trustworthy
Physical Picture of Particle Motion?,'' Phys.\ Scripta {\bf T76},
41--46 (1998).

\item
A.~{\Einstein}, ``Remarks Concerning the Essays Brought Together in
this Co-Operative Volume,'' in {\sl Albert {\Einstein}:\
Philosopher-Scientist}, third edition, edited by P.~A. Schilpp (Open
Court, La Salle, IL, 1970), pp.~665--688.

\item
C.~{\Rovelli}, ``Relational Quantum Mechanics,'' Int.\ J.\ Theor.\
Phys.\ {\bf 35}, 1637--1678 (1996).

\end{enumerate}
\eq

\section{05 October 2000, \ ``Bad {\Gleason} Mornings''}

\bdm
\bq
\em \noindent As far as Q foundations are concerned, some pretty
interesting things are starting to happen again.  (Though you know
I've got this pathetic backlog of things to write up---one of these
days somebody needs to write that stuff up, de {\Finetti}, Dutch
book, etc.) Thinking about a POVM version of {\Gleason}'s theorem
has been immensely useful.  (Remember it was one of the problems in
my Montr\'eal problem set?)  In particular, with POVMs taken as the
base for a notion of frame function, one can even make {\Gleason}
work for two dimensions! We're writing this one up immediately,
making PRL the target.\footnote{This describes work of Joseph
{\Renes}, Kiran Manne, {\Carl}ton {\Caves}, and myself.}
\eq
I'm eager to see whatever you've come up with.  I seem to remember
something about {\Gleason}'s theorem and POVM's by somebody named
{\Busch} some time within the last year.  Am I making that up?
\edm

Boy, you know how to spoil a morning.  If anybody, it's Paul
{\Busch}. I'm gonna have to do a search on it when I get into work.
The result is simply this:  If we think of the basic elements in
quantum measurement theory to be the POVMs (i.e.\ resolutions of the
identity into positive operators), then there is essentially only
one way to assign probabilities to the measurement outcomes in a
noncontextual way.  That is to say, if I suppose a function $f$ from
positive operators to the real interval $[0,1]$ such that $\sum_i
f(E_i) = 1$ whenever the $E_i$ form a POVM, then the theorem says
there must exist a density operator $\rho$ such that $f(E_i)= tr
\rho E_i$. What is cute about this is that it works even in 2-D,
whereas {\Gleason}'s standard theorem doesn't.

Oh, I hope Paul didn't do it already!

\section{12 October 2000, \ ``Turning Heads \ldots\ At Least One''}

Thanks for bringing my attention to Paul {\Busch}'s paper.  Yep, it
pretty much looks like he did it.  We have some extra stuff, but
he's taken most of the wind from our sails.    It seems like I'm
always a little late.

But look at this below.  It's a note from Jeffrey {\Bub}, and it
thrills me to no end.  [See ``Jeff's Reply'' to my note ``Commit the
Bit!,'' dated 30 September 2000.]  I think I take particular
indulgence in this because the guy has pretty much spent his life on
the opposite side of the fence from me---that is, trying to find a
free-standing reality within quantum mechanics.  He even went so far
as to write a book on the subject. So to turn his head some strikes
me as an immense triumph for the power and reasonableness of the
Paulian idea.  I used to tell {\Carl} {\Caves} that if we could just
convert David {\Mermin} and John {\Preskill} (two very rational
skeptics) to our way of thinking, that would be a great victory. But
maybe this is even better.

\section{03 November 2000, \ ``Do You Know?''}

Do you know who first used the terms ``separable,'' ``nonseparable,''
and ``inseparable'' to describe quantum systems?  I have a faint
memory that it might have been D'Espagnat, but I'm not sure why I
have that faint memory.\footnote{Anyone who knows the answer, please
do write me.}

\section{23 April 2001, \ ``New Posting''}

I just made the next posting.

Since my last writing to you, the thing has mainly changed by my
beefing up the {\Bennett} chapter.  In particular, there are two
notes added there mostly for the purpose of saying two sentences
(embedded within them).  [See notes to {\Charlie} {\Bennett} titled
``{\Emma} Jane Fuchs'' and ``{\Emma} Jane,'' dated 13 January 1999
and 5 June 1999 respectively.]  Recontemplating them while taking a
shower, I was lead to recontemplate the main (and enduring)
confusion about my point of view on quantum mechanics. There seems
to be a great tendency to think that when one says ``the wave
function is about knowledge, only knowledge'' one is placing the
observer outside of the physical world.  But it is exactly the
opposite of that!  And I just can't see why no one can see it.  We
are forced to the quantum {\it description\/} precisely because
there is nothing special about us (beyond the specialness of the
world itself).  In stark contrast, it is the ideal of classical
physics that makes the observer irrevocably special.

Now, I must work on ``The Quantum State of a Laser Field'' for a
while to convey the illusion that I am a physicist.  I'll get back
on the samizdat this afternoon.

\subsection{David's First Reply}

\bq
Congratulations, you're almost to {\Schroedinger}!  All you have to
do is change the last sentence to
\bq
\noindent We are forced to the quantum {\it description\/} precisely because
there is nothing special about the world (beyond the specialness of
us).
\eq
\eq

\subsection{David's Second Reply}

\bq
Funny you adding the note you sent me last week about the exact
opposite of placing the observer outside of the world.

You never replied to my comment on it: that you were getting close to
Schr\"odingerian mysticism.  I suspect the only difference (if it is
a difference) is that you say the observer is in the world while
{\Schroedinger} --- this is a cartoon version of what he says but is
hard to resist in the current context --- says that the world is in
the observer.  Both of you say (with {\Pauli}) that they cannot be
separated.

But setting my growing taste for mysticism aside, do you really want
to add a letter that comments on the preparation of the text itself?
\eq

\section{26 April 2001, \ ``Fire of the Soul''}

\bdm
You never replied to my comment on it: that you were getting close to
Schr\"odingerian mysticism.  I suspect the only difference (if it is
a difference) is that you say the observer is in the world while
{\Schroedinger} --- this is a cartoon version of what he says but is
hard to resist in the current context --- says that the world is in
the observer.  Both of you say (with {\Pauli}) that they cannot be
separated.
\edm
God, I loved that!  Can I put it in the samizdat??!?

\chapter{Letters to David {\Meyer}}

\section{11 May 1999, \ ``SD and KS'' and ``Even More Relevant''}

How about we plan on my dropping by sometime during the week of May
24--28.  You guys choose the exact dates (either one or two nights,
as you said).  I'll probably bring my wife and little girl down with
me.  They'll drop me off to play with you, and go off to play a
little on their own.

I have a question about your rational-point anti-KS result.  I'm
starting to wonder how robust the idea is.  The way you pose the
problem is to consider the points on the sphere with rational
cartesian coordinates.  But are you sure that nothing changes under
rather natural coordinate changes?  For instance, imagine working in
spherical coordinates for vectors on the unit sphere in $R^3$.  If we
restrict ourselves to rational angles $\theta$ and $\phi$, the
cartesian coordinates so generated will almost always be
irrational.  And this discrete set of vectors is not just a rigid
rotation of the ones you consider.  Can we find subsets of vectors
within this set that cannot be consistently colored?  You write,
``Surely the meaning of quantum mechanics should not rest upon such
non-experimental entities.  But, at least in the three dimensional
arena for the {\Kochen}-{\Specker} theorem it does \ldots''  But the
notion of a rational point for the experimentalist has to be
specified with respect to something natural within the experimental
procedure:  who is to say it is not the rational angles that are
accessible to the experimentalist instead of the rational cartesian
coordinates on the sphere?

Am I overlooking something simple?  (Likely.)  Anyway, regardless of
the philosophizing, it would be interesting to know if there is a
fundamental distinction between these two distinct sets of ``rational
points'' with respect to colorability.

Maybe an even more relevant question might have been not to consider
the rational angles, but instead those that are given by rational
fractions of $\pi$.  What happens then?  This indeed seems much more
to me like a reasonable constraint to put on the
gedanken-experimentalist.

\section{13 November 1999, \ ``What Did vN Not Know?''}

I quote from page 836 of the original Birkhoff and von Neumann paper:

\bq
One conclusion which can be drawn from the preceding algebraic
considerations, is that one can construct many different models for
a propositional calculus in quantum mechanics, which cannot be
differentiated by known criteria.  More precisely, one can take any
field $F$ having an involutory anti-isomorphism satisfying $Q4$ (such
fields include the real, complex, and quaternion number
systems\footnote{\ldots\ Conversely, A. Kolmogoroff \ldots\ has shown
that any projective geometry whose $k$-dimensional elements have a
{\it locally compact topology} relative to which the lattice
operations are continuous, must be over the real, the complex, or the
quaternionic field.}), introduce suitable notions of linear
dependence and complementarity, and then construct for every
dimension-number $n$ a model $P_n(F)$, having all the properties of
the propositional calculus suggested by quantum-mechanics.
\eq

Stephen {\Adler} in the pages of his book that I copied long ago,
writes ``\ldots the subject was further explored in an important
article by Finkelstein, Jauch, and Speiser (1959).''  Unfortunately
I don't have a more complete reference than that.  I'll talk to
{\Adler} about all this when I meet him in Naples a couple of weeks
from now.

\section{13 November 1999, \ ``A Couple More Things''}

Jauch [Synthese {\bf 29} (1974) 131--154] writes, ``It is known that
every proposition system $\cal L$ admits a representation in a linear
vector space with coefficients from the real, complex, or quaternion
fields.  \ldots\ Under some mild additional restrictions one can show
that the coefficients of the Hilbert space are the complex number
field $C$.''  He cites C. Piron, ``Axiomatique Quantique,'' Helv.\
Phys.\ Acta {\bf 37} (1964) 439--467, for that.  But he also writes
in a footnote, ``The question of the number field remained for a long
time beyond empirical test.  The recent work by Gudder and Piron
[`Observables and the Field in Quantum Mechanics,' J. Math.\ Phys.\
{\bf 12} (1971) 1583--1588] is the best one can do.''

\chapter{Letters to Jeff {\Nicholson}}

\section{06 June 1996, \ ``Cyclic Never-Endings''}

Yes, yes, I like Montreal!  Does that finally answer your question?
The only thing I really miss about the states is having a swarm of
good physicists around me.

\bjn
You know, I really miss listen to you babble about quantum
mechanics.  Nobody around any more to tell me that the moon isn't
there when I'm not looking; you even had me believing for a while.
\ejn
Damn, if I had only stayed there a bit longer \ldots perhaps you
would have remained a true believer!

\section{22 August 1996, \ ``Laws of Thought''}

[Referring to the phrase ``Quantum Mechanics is a Law of Thought.'':]
\bjn
I was just reading about the mathematician George {\Boole}, the guy
who invented Boolean Algebra.  In 1854 he published a book called
{\em An Investigation of the Laws of Thought, on which are founded
the mathematical theories of logic and probabilities}.  Is that where
your lovely little phrase comes from?
\ejn

Yes indeedy, that's exactly where it comes from.  Quantum Mechanics
is a Law of Thought.  God I like the ring of that.

\section{22 August 1996, \ ``Plagiarists''}

\bjn
I am disappointed in you.  I never realized that you plagiarized that
from poor ol' {\Boole}. Can't think up your own catch phrase, eh?

I think the phrase bothers me far less when applied to something like
Boolean algebra.  It still gives me the heebies when you refer to
real live physical systems with such terminology.
\ejn

Hey, what right do you have calling me a plagiarist?!?!  I can hardly
believe that {\Boole} knew anything of quantum mechanics in 1854!
Please reevaluate!

And, sir, by the way, I am not referring to ``real live physical
systems with such terminology.''  I am referring to the mathematical
structure of quantum mechanics.  Some of the best examples of real
live physical systems I've ever seen have been in your lab \ldots\ and---%
I promise you---try as I might each time I would visit you there, I
never could see a Hilbert space sitting on your optics table.  I
looked and looked; never could find the damned thing.  Never saw a
state vector either.  Just silly little arrangements of mirrors and
cavities.

\section{09 October 1997, \ ``Pickin', Scratchin', and Grinnin'\,''}

Thanks for the long note.  I promised you a long one back, didn't I?
Oops, I lied.  But maybe I'll say a couple of things while I have my
first quarter cup of coffee.  Things are fine out here in Lava Lamp
Land.  {\Kiki} is busy, frustrated, and happy with her job: she now
has a class of 24 little kindergartners.  Me, I, myself have been a
little apathetic lately---don't ignore the ``a''---but I still dream
of finding the ultimate philosophy.  It has something to do with
``information'' you know.  (Lucky for you, my coffee's too hot to
drink.)  I've been thinking about the similarities between
``introspection'' and the quantum mechanical measurement situation
lately---they're really a lot alike.  The maximum information I can
attain about what I am thinking can never be complete information
\ldots\ that is, in the sense of allowing me complete predictability
over my next action.  The reason is that complete information about
one's own present thought requires self-reference and an infinite
regress. ``OK, I am thinking about whether to stay in bed or get up.
Wait, I'm thinking about my thinking of it.  Doh!  I can't stop this.
Will I get up or won't I?''  At some point the action just
happens---the causal mechanism that brought it about is usually
filled in slightly later, after the action has actually occurred.
Anyway, one has to wonder whether the mathematics of quantum
mechanics---I didn't say the ``physics of''!---might be useful in
getting at a precise description of the process.

\section{23 November 1997, \ ``Long Note on a Sunday Morning''}

This morning I caught myself reading Jean {\Baudrillard} (in the five
hours I couldn't think of anything to say in).  He's one of those
weird Frenchies who---even though he knows no quantum
mechanics---thinks nothing is real.

\section{10 March 1999, \ ``Schnockerization''}

I tell you what, I'll do a little schnockerizing with you tonight.
Tell me how {\it my\/} grammar fares!

The biggest fear I had to deal with when I was alone was exactly the
feeling of permanence that came with it.  If somehow, someway I
could have heard a whisper ``You will meet the girl of your dreams,
your wife-to-be in two years.'' If I had better understood the
finiteness of my time alone, I could have coped so much better---in
fact, I don't think it would have been a problem at all.  It was the
fear of the permanence of the emptiness that really set me off.  I
guess I've always thought it is a load of crap that one has to ``be
for themselves'' before they can ``be for someone else.''  My view of
man has always been quite overtly trinitarian: man, woman,
(potential) child.  We make each other, define each other, and give
each other meaning.

In that vein, let me tell you one of my favorite acknowledgements
ever in a scientific paper.  It was {\Schroedinger} in his 1935 paper
``The Present Situation in Quantum Mechanics'':  ``My warmest thanks
to Imperial Chemical Industries, London, for the leisure to write
this article.''  {\Schroedinger} could not have been {\Schroedinger}
without the support of the community.  And so the same for us all.

\chapter{Letters to Michael {\Nielsen}}

\section{22 February 1996, \ ``Goo-Losh''}

Mmm, ghoulash \ldots\ is that how you spell it?  Thanks for the tip.
I haven't heard the reaction of {\Bennett} and company on this
business yet, but when I do, I'll send a report your way.

I've been thinking about {\Kochen}-{\Specker} theorems lately;
they're very nice you know \ldots\ especially the ones with small
numbers of vectors.  Implications for ``quantum information''?
There must be! In particular, why can't one go much further in this
effort to prove the theorem with smaller and smaller numbers of
vectors?  Why is there no proof of this property (i.e.\ no
noncontextual hidden variables for the outcomes of standard von
Neumann measurements), the very second you admit a measurement that
is nonorthogonal to the others?  Just food for late night thoughts.

\section{18 March 1996, \ ``The Commitments''}

It's a good movie; did you ever see it?

Suppose you need some cash and so you go to your Automated Teller
Machine (ATM).  You insert your bank card and the machine asks you
for your Personal Identification Number (PIN).  It does this for the
obvious reason that it wants to know that you are indeed who you say
you are before it hands over the money.  That's a good safeguard.
However suppose, as really did happen a year or two ago, someone had
set up a fake ATM in place of the real one.  By inserting your bank
card into it, you have revealed your account number.  By typing in
your PIN, you have revealed how to make transactions on that
account.  The (dishonest) maker of this machine is now capable of
withdrawing money from a real ATM machine anytime he wishes.

Is there a way of circumventing this sort of thing?  Yes, in fact
there are several ways---but they are all based on a problem that is
computationally hard (classically).  For instance one scheme is
based on the user choosing two large prime numbers instead of a PIN.
The user only reveals the product of the primes to the bank.  Then
it can be shown that there is a way the user can interact with the
ATM so that he reveals no information about his primes and yet the
ATM can verify that the user is indeed who he says he is.  This is a
very handy little protocol \ldots\ so long as large numbers cannot be
factored effectively.  (By the way, the rubric for these sorts of
protocols is ``zero knowledge proof.'')

Well it turns out that as long as you have a secure ``bit
commitment'' scheme you can implement a zero knowledge proof.  The
point is that bit commitment has loads of (practical) applications.
Now it would be very nice to build an analogous scheme for which the
security of it relied on the validity of quantum physics \ldots\
instead of some mathematically hard problem.  That is the idea
behind trying to build a quantum bit commitment scheme.

Whether such a thing is really possible, I do not know.  Dominic
{\Mayers} is now running around here saying that they're not \ldots\
but I have not been able to understand his reasoning yet.

I think philosophically the existence of such things would be very
nice.  I used to think that the existence of quantum key distribution
was important to Prof.\ {\Wheeler}'s mumblings on quantum mechanics.
He likes to say that it is the ``{\it community\/} of communicators''
that is at the base of what is real in the world.  This is different
from ``idealism'' and ``solipsism'' because it relies crucially on
the existence of many individuals, all communicating with each other.
The world is not just a big thought.  Well what good would it be to
say that the world consists of individuals if there were nothing in
physics to allow them completely private communication?  If everyone
could, in principle, listen in to everyone else's conversation, then
the world could hardly be said to be composed of individuals at a
basic level.  This sort of thought excited me about quantum key
distribution.  A similar set of thoughts excites me about quantum
bit commitment.  What good would it be to say that the world
consists of individuals if there were no means to protect one's
identity from emulation by others?  (Of course this idea could also
be implemented with QKD, but dissimilar examples are good.)

So I think settling the question of quantum bit commitment is not
only interesting from the practical side but also from the wispier
side.

Entropy measurements on individual systems?  I need a Latin phrase
similar to ``non sequitur'' to answer that.

What am I doing researchwise?  Nothing that I'm supposed to be
doing!  I still haven't been able to really bring myself back to the
dissertation stuff \ldots\ but every day I look at it a little.  I've
written a couple of very nice referee reports.  I worked an
inordinate amount of time trying to turn 6 into 1 \ldots\ to no
avail. The last couple weeks have been filled with giving lectures on
inference-disturbance tradeoff.  (Lectures go extremely slowly here;
the first two were essentially a mini-course on quantum mechanics.)
Also I've spent a lot of time trying to work things out further for
two-state systems, i.e., I'm trying to fill in the gaps that were
only filled with computational work in the paper with  {\Asher}.

Philosophically, I've been thinking about {\Kant} again.  (Ask Jeff
{\Nicholson} the story of {\Kant} Cola.)  I'm about half way through
a book by C.~F. von Weizs\"acker on {\Kant} and Quantum.  Quantum
Theory as a law of thought---nothing excites me more!  Before that I
read a book on G.~E. Moore and the Higher Sodomy; didn't glean much
from it though.  Let's see, after this book, the plan is to go back
to Russell's writings on ``logical atomism.''

{\Kiki} continues on her rampage to culture me.  You should see, we
have a very ``cultured'' apartment this time round.  She very much
likes it up here.  I do too actually.  The only thing I miss is a
good used bookstore.

\section{01 April 1996, \ ``Mo' Better Sophistry''}

I thought it was time to unload some thoughts.  I hope you don't mind
being chosen ``designated driver.''  Please just make sure I make it
home safely.

Oh, by the way, before I get started, could you do me a favor and
enquire about the meaning of the term ``three-lock box.''  It comes
from an old Sammy {\Hagar} tune.  A good source is probably either
Chris{\index{Hains, Chris}} or Amy{\index{Sponzo, Amy}}. (Actually,
taking the mean of the answers would be even better.)

I'm just back from a two week foray into the netherworld of
cryptology; God save my soul!  Maybe I did learn something and didn't
waste my time, but that still remains to be seen:  I'm not sure. In
the mean time, however, this may give some fodder to the old
question of quantum measurement.

The idea of a ``bit commitment'' is, in the end, a very simple one.
Alice would like to give Bob some evidence that she is committed to
a bit, 0 or 1, and that she will not change her mind, but she would
like to do so in such a way that Bob can gain no information about
the bit she actually chose.  A simple classical way to enact such an
idea is for Alice to write the bit on a piece of paper, place it in
a safe, and give the safe to Bob without the key.  When Alice would
like to reveal that she was actually committed to the bit, she gives
Bob the key.  He opens the safe and sees the bit.  As long as there
is no intricate mechanism for transferring a last-minute change of
mind from the key to the paper, this can be safely deemed
``evidence'' that Alice did indeed commit.  Moreover, as long as Bob
cannot break into the safe, he can gain no information about Alice's
bit.

The way such a thing is done in practice is that Alice hides her bit
in a mathematically hard problem, working under the assumption that
Bob does not have the computational power to break it.  The problem
with this, along with the simple example of above, is that one has
to work under the assumption that Bob does not have the power to
break into the container that Alice puts her bit in.

Does quantum theory give a way to get around this difficulty?  At
first sight it might appear that quantum mechanical thingies are a
great place to hide information.  If Alice prepares a quantum system
in one of two nonorthogonal states and gives it to Bob, then there
is no way he can recover the full bit---that's something of an
example.  As another (pseudo-)example, Alice can dump a bit into EPR
correlations by way of teleportation.  This is very much something
that looks like an unbreakable safe: without the device that Alice
used to prepare the state that was teleported, or the two classical
bits arising from measurement of the Bell basis, there is nothing
that Bob can do to recover the information from his element of the
EPR pair.

However, the idea of Quantum Bit Commitment calls for something even
stronger.  Alice {\it must not\/} be able to reverse her commitment,
and yet Bob {\it must note\/} be able to recover any information
whatsoever about the bit she committed to.  These two requirements
appear to be mutually exclusive in the quantum world---{\it IF\/} I
believe Dominic {\Mayers} general attack (and I think I do, though
I'm still not 100\% convinced, only 99.8\%).  I suppose my purpose
for writing this note is to say, {\it viewed in this way}, the
impossibility of quantum bit commitment is something we---as
physicists---should have known long ago \ldots\ and if the problem
hadn't been mired in cryptological notation, we would have
discounted it at the start.

\bq
\noindent ``No quantum phenomenon is a phenomenon until it is an
observed phenomenon, brought to close by an irreversible act of
amplification.'' --- J.~A. {\Wheeler}.
\eq

Well, it's not exactly that, but it's very much of the same flavor.
Alice and Bob play with two systems $A$ and $B$ and the game is the
following.  Alice ``commits'' to a bit by preparing the total system
$AB$ in one of two distinct states.  She may not have had initial
control over $B$, but that seems to be irrelevant.  At the end of the
day, after the ``commit phase'' has been carried out, Alice can only
diddle with system $A$, Bob can only diddle with system $B$.  The
requirements for a bit commitment are this.  After the commit phase
there must be no unitary evolution that Alice can perform on $A$
alone that will change the two states into each other.  And there
can be no measurement that Bob can perform on $B$ alone that reveals
any information about the commitment.  These two requirements are
mutually exclusive.  If there is no tangible evidence on Bob's side
after the ``commitment'' (i.e.\ evidence capable of revealing at
least to some extent which bit was actually committed to), then there
is no real coupling between Alice and Bob and she can reverse without
in any way coming into contact with $B$ again.  That's what it boils
down to as I see it.\footnote{Looking back over this paragraph for
the first time in five years, I have to wonder what on earth I was
thinking?  Still it intrigues me that I {\it might\/} have been
thinking something that---with a sufficient amount of
modification---{\it might\/} even be sensible.  I have no qualms
with intellectual alchemy!}

\section{04 September 1996, \ ``One Mo' Book''}

Here's another book that I just heard about that may also be of
interest to you.  Apparently it anticipated the many-worlds
interpretation in 1931 (according to Chris {\Isham}).  It's {\sl Many
Dimensions\/} by Charles Williams (Faber, London, 1968).

\section{16 May 1999, \ ``Purifications''}

How's this?  If you look at H.~Araki and E.~H. {\Lieb}, ``Entropy
Inequalities,'' Commun.\ Math.\ Phys.\ {\bf 18}, 160--170 (1970),
you'll find in Lemma 4, the following statement:
\bq
\noindent
Let $\rho^1$ be a density matrix on $H^1$.  Then there exists a
Hilbert space $H^2$ and a pure state density matrix $\rho^{12}$ on
$H^1\otimes H^2$ such that $Tr^2\rho^{12}=\rho^1$.
\eq
After that follows the proof without any further attribution.
Figuring that these guys were probably pretty well-established by
then, this might well be the first appearance of a purification.

Wait \ldots\ no I'm probably wrong about that.  I bet the first place
was: D.~Bures, ``An Extension of Kakutani's Theorem on Infinite
Product Measures to the Tensor Product of Semifinite
{$w^*$}-Algebras,'' Trans.\ Am.\ Math.\ Soc.\ {\bf 135}, 199--212
(1969).

I'd momentarily forgotten about that one.  Well, anyway, maybe
somewhere between the two of these papers lies the truth.  (Is that
a bad pun?)

\section{23 August 1999, \ ``A Convergence of Sorts''}

Friday morning as I was walking in to the office, I caught myself
singing an old Paul {\Simon} tune that you may not know.  It's called
Kathy's Song.  Let me pull up the words for you; I hope you'll take a
slow moment to read them.

\bv
I hear the drizzle of the rain\\ Like a memory it falls\\ Soft and
warm continuing\\ Tapping on my roof and walls.\bigskip\\

And from the shelter of my mind\\ Through the window of my eyes\\ I
gaze beyond the rain-drenched streets\\ To England where my heart
lies.\bigskip\\

My mind's distracted and diffused\\ My thoughts are many miles away\\
They lie with you when you're asleep\\ And kiss you when you start
your day.\bigskip\\

And a song I was writing is left undone\\ I don't know why I spend my
time\\ Writing songs I can't believe\\ With words that tear and
strain to rhyme.\bigskip\\

And so you see I have come to doubt\\ All that I once held as true\\
I stand alone without beliefs\\ The only truth I know is
you.\bigskip\\

And as I watch the drops of rain\\ Weave their weary paths and die\\
I know that I am like the rain\\ There but for the grace of you go I.
\ev

My singing actually had something to do with you in that it set my
mind to remembering a conversation you and I had had in one of the
Cambridge pubs.  I was thinking of the one concerning what is most
important in life, and, in particular, the contrast of your thoughts
with those of David {\Deutsch}'s.  This led me to think about the
things that {\Deutsch} is likely to be willing to define as
``truth,'' in contrast to the things that Paul {\Simon} would.  What
does {\Simon} take to be the bedrock of our world, and what does
{\Deutsch}?  And to which of these camps am I closer at this time in
my life?

Then I remembered something I had written {\Carl} and {\Ruediger}
just a couple of days earlier.  [See note to {\Ruediger} {\Schack}
dated 18 August 1999.]

I held that thought, mulling it over and over trying to refine it,
and then---by happy circum\-stance---happened to read two wonderful
articles by Silvan {\Schweber} over the weekend.  The first was:
S.~S. {\Schweber}, ``Physics, Community and the Crisis in Physical
Theory,'' in {\sl Physics, Philosophy, and the Scientific
Community}, edited by K.~Gavroglu, J.~Stachel, and M.~W. Wartofsky
(Kluwer, Dordrecht, 1995), pages 125--152.  The second was:  S.~S.
{\Schweber}, ``The Metaphysics of Science at the End of a Heroic
Age,'' in {\sl Experimental Metaphysics: Quantum Mechanical Studies
for Abner {\Shimony}, Vol.~1}, edited by R.~S. Cohen, M.~Horne, and
J.~Stachel (Kluwer, Dordrecht, 1997), pages 171--198.  (If you want
to copy them, I'll have them in my office for the next couple of
days \ldots\ that is, once I come in again.)  Let me pluck a piece
out of the first one that I thought was significant [pages 142--144].

\bq
For {\Einstein} physics became a retreat from the world, and he
found a monastery at the Institute for Advanced Study in Princeton.
Meaning was given by his faith in the comprehensibility and the
comprehension of nature.  More specifically, {\Einstein} believed
``in the simplicity, i.e., the intelligibility of nature.''

Although {\Bohr} did appreciate {\Einstein}'s position, for him
science could not be an escape from the world.  In fact for him
there is no escape from the world.  There is no way to step outside
of it and obtain a God's eye view of it.  We are always on the
inside and whatever we know of the world we have mediated and
constructed.  He believed that all scientific understanding---all
understanding---required that one test and risk one's convictions and
prejudgments in and through encounters with others and in particular
in encounters with what is radically new and alien.  For {\Bohr} the
practice of science exhibited a commitment to an underlying moral
order. Scientists form an ideal society that relies on the virtues of
honesty, tolerance, trust, truthfulness, and cooperation.

Without their moral commitment to be truthful and trustworthy members
of this society that guarantees tolerance for the views of others and
the verification of their claims, no important statements of fact
could see the light of day and be stabilized.  It was always
self-evident to {\Bohr} that science is a human activity, and
constitutes a collaborative and cooperative enterprise.  In fact,
{\Bohr} had a vision that the scientific community could serve as a
paradigm for a possible future international order.  {\Einstein}
would have mostly agreed---but he also believed that the communal
aspect of science emphasized by {\Bohr} could be transcended by
individual acts of transcendence.

Ultimately, the difference between the two is the following:  Deep
down---although a child of the Enlightenment, {\Einstein} was really
a gnostic priest.  He believed that he could read the mind of his
{\Spinoza}n God and discover the laws that govern the objective
physical world out there.  He believed in Truth with a capital T.
For {\Bohr} science does not determine the truth, truth with a
capital T---rather there is convergence toward truth and the
understanding of physical reality.  On the basis of their
experiments---their queries to nature---scientists constantly
reorder and reinterpret external reality, and these imaginative
reorderings must be presented to the members of the community for
their critical assessment by means of language.  And because
language is a uniquely human activity---our distinctive (though
imprecise) means to communicate with one another and convince one
another---and a unique means to stabilize communities without
recourse to physical violence---language is a central and constant
concern in {\Bohr}'s philosophical gropings.

{\Bohr}'s conception of community, as well as those of truth and
reality have close affinity to the views of {\Peirce}.  {\Peirce}'s
notion of reality combines a social conception, that ``the real is
the idea in which the community ultimately settles down'' with the
notion that the real is what it is regardless of what anyone may
think of it. For {\Peirce}, the scientific enterprise is the model
for how a society of free inquiring individuals can generate
discourse and cooperation without chaos.  What gives the actions of
individuals their coherence is their decision as agents to aim
toward a definite goal.  We are all at work building and maintaining
a scaffolding that will give stability and meaning to our lives and
to the communities in which we share our lives.  Science is part of
that scaffolding; just as are literature, music, the visual arts,
and the other activities we engage in.  Science is a linkage between
individualism and coordinated public choices and actions.

So too for {\Bohr}.  {\Bohr}'s conception of scientific activities in
fact nurtures a wider vision of human interactions.  Understanding
the origin of the universe, its possible demise, unravelling the
constitution of matter in its simplest manifestations and
comprehending the rich and complex diversity matter can assume,
understanding the limitations of mathematical theorizing and gaining
insights into why it has been so effective in describing nature---are
all part of our collective scaffolding---irrespective of the kind of
scientific practitioners we are.  Equally importantly, science is a
piece of the scaffolding that makes for bridges to the members of
other civilizations and cultures.  Science, in fact, is that piece of
what may be called the collective mind of the human race that is
independent of place or individuality.  And only in building together
and in maintaining together the scaffolding can we guarantee
stability.
\eq

So what?  Nothing really, just that I'd probably buy you a beer at
the Athenaeum if you're seeing the same sort of convergence that I
am.

\section{25 August 1999, \ ``Late Night Threads''}

Your spelling was impeccable.  The running thread (the convergence)
was meant to be:  1) There is no truth with a capital T within
science.  2) Truth (with a small t) is actually grounded in the fuzzy
rather than the precise.  ``Having been trained as an engineer, I see
nothing mysterious about the fact that I can use very clumsy
imprecise tools to manufacture delicate, high precision
instruments.''

About the cat, a side point.  One thing that I was wanting to get at,
but couldn't think of some good words for at the time is this.  The
``property'' of being alive versus being dead is not a property
intrinsic to the cat alone.  It never was.  Life is now ``restored''
routinely in ERs across the nation.  In the ultimate limit of
``maximal knowledge'' of the patient (i.e., a pure quantum state for
its description), the doctor would always be able to restore life.

I personally feel the luckiest when I have a little fuzzy truth.  I
wonder if Paul {\Simon} still thinks that of his time with Kathy?

\section{14 July 2000, \ ``Quantum Theory Does Need an
Interpretation''}

Thanks for sending the evidence of Mr.\ {\Carvalho}'s short attention
span.  He writes [regarding my {\sl Physics Today\/} article with
 {\Asher} {\Peres}],
\bq
\noindent
\bq
\noindent ``\ldots We set up this or that experiment to see how natures
reacts. We have learned something new when we can distill from the
accumulated data a compact description of all that was seen and an
indication of which further experiments will corroborate that
description \ldots''
\eq
This reminds me of Bertrand Russell's chicken. They also learned a
lot from experiment. According to this description of the purpose of
experiment, where does explanation fit in? I guess inductivism is not
dead after all \ldots
\eq
But if he would have just read the remainder of the paragraph
starting one sentence further down, he might have had some pause to
think.  Why would a follower of pure inductivism write such a thing
as this:
\bq
\noindent
If from such a description we can {\it further\/} distill a model of
a free-standing ``reality'' independent of our interventions, then
so much the better. Classical physics is the ultimate example in
that regard. However, there is no logical necessity that this
worldview always be obtainable. If the world is such that we can
never identify a reality independent of our experimental activity,
then we must be prepared for that too.
\eq

Like I wrote  {\Asher} just yesterday, as we were working on these
replies for {\sl Physics Today\/} (Mr.\ {\Carvalho} got that wrong
too):
\bq
\noindent
I am not pushed to the rejection of a free-standing reality in the
quantum world out of a predilection for positivism.  I am pushed
there because that is the overwhelming message quantum THEORY is
trying to tell me.
\eq

But then, you probably already knew that of me \ldots

\section{09 August 2000, \ ``Big Red X's''}

Let me tell you a good one.  When I was in Greece last month, I
started out my talk by saying:  ``The last couple of days I've
noticed that most talks start out at a pretty heuristic level, with
lots of pictures and motivation, etc., and slowly become more and
more technical.  But, by the end of the talk all of the speakers are
really pressed for time and the pace becomes phenomenal.  It's kind
of an uphill battle as equation after equation flashes across the
screen.  So, today, I've decided to give you a downhill slide:  I'll
start out with some technical results and derivations, slowly using
more and more pictures as I go, and by the end---if I have
time---I'm hoping the talk will degenerate into a discussion on the
foundations of quantum mechanics.''

I got a good laugh.  And true to form, at the beginning while I was
talking about things like Wojciech's ``predictability sieve,'' Ben's
``distinguishability sieve,'' and my ``mutual information sieve''
all the young guys were taking notes pretty thoroughly and all the
old guys were nodding off.  But by the end of the talk, when I
changed to honest-to-god foundation stuff, all the young guys put
down their pencils, and all the old guys became livid.  It was quite
a scene.  I guess I riled them with my first slide (for that portion
of the talk).  It had at the top ``Decoherence Studies'' and at the
bottom ``Quantum Foundations'' and it had an arrow running down from
the top phrase to the bottom phrase.  And I said, ``Everyone tells
me that decoherence has something to do with quantum foundations.
But I have to admit, try as I might, and as much as people tell me
that they're saying something convincing, I just don't see it.  And
I have tried to understand.  So either I'm really thick, or my
colleagues just aren't being as convincing as they think they are. I
think it's the latter.''  Then I put up an overlay that had a big
red X.  ``I think decoherence studies like the ones I've just showed
you, important as they are, have nothing to do with the foundations
of quantum mechanics.''  You should have seen the explosion!  (But
I'm used to that sort of thing: my house burnt down.)

\chapter{Letters to  {\Asher} {\Peres}}

\section{05 June 1997, \ ``Painting My Stripes''}

I am just returning from the library, where---completely by
accident---I ran into and read an entertaining little article:
A.~{\Peres}, ``{\Einstein}, {\Goedel}, {\Bohr},'' Foundations of
Physics {\bf 15}(2), 1985, 201--205.  I was especially entertained
by your closing line, ``Any attempt to inject realism in physical
theory is bound to lead to inconsistencies.''  Do you have any other
writings (beside the AJP paper with {\Zurek}) where you expand upon
the ideas presented in this?

In connection to this, I have another question.  If I recall
correctly, you once told me that you finally came to grips
(``interpretationally,'' that is) with quantum mechanics in 1986.
(Perhaps I have the year wrong, but that's part of the question.)
Would you be willing to expand upon that point?  In particular, was
your arrived-at comfort founded upon the line I quote above?

I suppose I ask these questions for several reasons, perhaps the
strongest of which is a renewed dilettante interest in the
foundations of quantum theory.  Among other things, I am trying to
plan (or dream up!) a talk for the Hull conference in September to be
titled something like:  ``Quantum Information Theory's Implications
for the Foundations of the Quantum as a Whole.''

My own pet ``idea for an idea'' (as {\Wheeler} would say) is that the
quantum theoretical de\-scrip\-tion---along with its Hilbert-space
structure---is forced upon us when the information we have about a
physical situation/system is of a certain variety.  In this way,
quantum theory is not so different from Ed {\Jaynes}' molding of
classical statistical mechanics and the maximum entropy principle.
It just so happens in the latter case that the prior information is
in the form of an expectation value for some physical quantity that
can be assumed objective and independent of measurement: this prior
information along with a small few desiderata lead uniquely to the
canonical probability assignment.  In the case of uniquely quantum
problems, we start with a different kind of prior information.  The
{\it hope\/} is that, upon pinpointing the nature of that information
(and again a few small desiderata), one would see that a
``wavefunction assignment'' is the natural (and unique) way of
summarizing what we know and what we can predict.

{}From these remarks, perhaps you can see my attraction to your
little article.  Please fill me in!

\section{16 October 1997, \ ``Royal Order of Recalcitrant
Positivists''}

I'm perfectly happy with the comment you'd like to make.  So use it,
please.  (In fact it's a little flattering almost.)

I was intrigued by one of the sentences in the {\Bub} quote that you
sent me.  Namely his,
\bq
\noindent The disturbance terminology itself suggests the existence of
determinate values for observables, prior to measurement, that are
disturbed \ldots\
\eq
I thought I had seen something similar before, so this morning I dug
and dug until I found it.  It was {\Braunstein} and {\Caves} in the
paper ``Wringing out Better {\Bell} Inequalities''; I like it much
better than {\Bub}'s:
\bq
\noindent In quantum mechanics the ``disturbance'' (quotes because one
should think twice before speaking of a disturbance if there are no
objective properties to disturb) comes from using amplitude logic
instead of probability logic (wave-function collapse instead of
Bayes's theorem).
\eq

Is {\Bub}'s book any good?  As I recall, {\Bub} did work in
{\Bohm}-like hidden variable theories, but also with things to do
with ``quantum logics'' (which have always struck me as
overly-elaborate empiricist or Copenhagen-type frameworks).  Both
ends of the spectrum, so to speak. Reading him was a hard go---so I
gave up a few times.  What's the title of this new book?  Maybe I
should wait for your review to be finished, and you can send me that.

\section{17 October 1997, \ ``Energy \ldots\ Really''}

\bap
Not energy but ENTROPY! (when observed separately by Alice and Bob).
\eap

No, energy \ldots\ really.  Really.  Energy is the ``tangible'' in
classical thermodynamics; entropy is the ``intangible.''  I realize
the analogy may not be so strong that I might take it outside the
poetic license of a personal letter---that's one of the reasons I
qualified the letter as philosophical-sounding and left it
indented---but there may be just a little to it.  Energy is a sort of
``Martha White's All-Purpose Flour'' for physics.  Just as flour may
be thought of as a potential bread, biscuit or cake, energy is a
common ``ingredient'' that, if we wish, can be transformed into
inertial mass, electromagnetic field intensity, or motion itself.  It
was in this way that I was thinking of entanglement. ``Entanglement''
is not ``non-classical correlation'' itself, but rather the {\it
potential for correlation}.  By choosing their measurements
appropriately and adjusting their actions, Alice and Bob can build
up---within limits of course---just the right type of correlation
that they need.  Quantum cryptography is one example of this:  Alice
and Bob would like to have a perfectly correlated binary string.
However, teleportation is another example:  there the ``unknown
state'' preparer---Charlie say---would like future measurement
outcome statistics on Bob's particle to be as he pleases. I think the
things {\Bennett}, {\Smolin}, and I call ``entanglement-enhanced
classical communication'' may show this even more strikingly.

For whatever it's worth, that's the sort of thing I was really
thinking.

\section{20 October 1997, \ ``A Little More Reality''}

Thank you for giving me the opportunity to compile this set of notes:
even if they turn out to be of little use to you, going through this
exercise has been of great use to me!  It has helped me re-gel my
opinions on the subject, so to speak.  And that, if nothing else,
will be of use to me for a project---tentatively titled ``A Bayesian
Reconstruction of Quantum Probabilities''---that {\Ruediger}
{\Schack} and I are presently involved in.

First off, let me express just how much I really have enjoyed reading
and contemplating your manuscript.  You and I are clearly on the same
wavelength when it comes to ``quantum philosophy.''  Perhaps there's
no greater pleasure than to read something that you yourself think
you are thinking!

\subsection{Harsh Criticisms}

But, now on to the scathing criticisms (\ldots\ though I promise not
to belabor them too much).  My biggest worry about your draft is
that, {\em maybe to too large of an extent}, it is more a systematic
account of your views on the interpretation of quantum mechanics than
a review of {\Bub}'s book.  Ultimately you're the one to decide on
that---you're the writer---but that is the impression I get.  This,
for instance, is indicated in your curt dismissal of {\Bub}'s
dominant program (i.e., attempting to translate various quantum
interpretations into statements about lattices and sublattices):
\bap
\ldots\ {\Bub} \ldots\ endeavors to assign numerical values to
observables, or truth values to lattices of propositions. These are
notions that have been borrowed from the classical world, and I
don't see why quantum reality, whatever it is, has to be described
in terms of observables or lattices of propositions.
\eap
Indeed what you say is true.  However, given that you, in the end,
say,
\bap
In conclusion, there is much to learn from {\Bub}'s book, and I
recommend it without hesitation to all those who are interested in
the foundations of quantum theory.
\eap
you really don't give the reader much to base the validity of that
opinion on.  Why is this book recommendable if so very much of it is
devoted to something that can be dismissed in a single sentence? (I
do realize you spent some time saying good things about it, but I
think the things you praised were things {\em he\/} likely saw as
tangential for building his main argument.)

My main disappointment in this respect is the very little attention
you pay to what he clearly thinks of as the backbone of the book: the
uniqueness theorem for the ``no collapse'' interpretations.  What is
this theorem?  Couldn't you at least give the reader of the review a
little of the flavor of it?  Even just a small extra paragraph might
make up for this.  Also, I think it would be very useful for you to
give an opinion---one way or the other---as to what you think of his
attempt to view {\Bohr}'s interpretation as one of these ``no
collapse interpretations'' that his theorem covers.  It appears to
me at least that, though he may not go into great detail about
{\Bohr}'s rebuttal of EPR, he does spend quite some effort in
Chapter 7 trying to make sense of {\Bohr} in his language.  In
particular, I think he does make the appearance of actually agreeing
with your remark:
\bap
{\Bohr} also was sometimes quite elusive. He never explicitly
treated a measurement as an interaction between two quantum systems,
and he could thereby completely elude the measurement problem, as
formulated by {\Bub}.
\eap
You can see this in the last full paragraph on p.~195 in addition to
the thought being basically spread here and there throughout the
chapter.

I don't want to give you the impression that I understand this book
better than I actually do:  I've only really given it the very
slightest of skimmings.  For the most part, I was spurred to notice
these things by testing your manuscript.  Nevertheless, with
hindsight, I can say that I---as a reader---would have liked to see
you touch on this stuff at least a little.  (Plus, who knows, going
into that might have helped out the things you would like to say
about {\Bohr}.)  Ultimately, I think---assuming that all the
theorems in the book are on the up and up---the main use of this
work for the conceptualist may be in giving an effective way of
lumping the main features of these ``mutations'' of quantum
mechanics (as you call them) into a single description.  This may be
of use in the relativistic domain where one might hope to find
something testable (in the spirit of a {\Bell} inequality) for
discrediting these modifications.  {\Bohm}'s, for instance, might be
a good test case.

\subsection{Typo's and Englo's}

But that's enough negative criticism.  Let me now get on to some more
positive comments.  First a point of English.  In the last sentence
of the last full paragraph on the first page, you write ``{\Bub}'s
book gave me an opportunity of understanding why.''  That's
nonstandard. I would say ``for'' instead of ``of'', or I would say
``to understand'' instead of ``of understanding.''  Second, a
typographical error: you forgot the word ``problem'' after
``observational'' in the long {\Bohr} quote on page 8.

Also---upon some reflection---I think your ``recalcitrant
positivists'' quote might convey something of a dangerous flavor. See
discussion in the next section.

\subsection{Foods for Thought [{\em sic\/}]}

Finally let me comment on some of the many sentences in your paper
that caused me to think a little.  These remarks may not be of direct
relevance for improving your exposition, but like I've already told
you, that wasn't my only reason for reading the manuscript.

\bap
To interpret quantum theory, to explain its meaning, to make it
understandable, the way has been shown to us: we have to {\it
translate\/} the abstract mathematical formalism in such a way that
`we can tell others what we have done and what we have learned and
[this] must be expressed in unambiguous language with the terminology
of classical physics.'
\eap

I've always liked this very much, but in certain ways it has always
haunted me a little.  To show you the ghost, let me quote a little of
a letter from {\Schroedinger} to {\Bohr}, dated 13 October 1935.
\bq
As a matter of fact I did not really want to talk about this point,
at least not with the purpose that you should reply, but rather about
something else.  You have repeatedly expressed your definite
conviction that measurements must be described in terms of classical
concepts. For example, on p.\ 61 of the volume published by Springer
in 1931: ``It lies in the nature of physical observation, that all
experience must ultimately be expressed in terms of classical
concepts, neglecting the quantum of action.'' And ibid.\ p.~74 ``the
invocation of classical ideas, necessitated by the very nature of
measurement.'' And once again you talk about ``the indispensable use
of classical concepts in the interpretation of all measurements.''
True enough, shortly thereafter you say: ``The removal of any
incompleteness in the present methods of atomic physics \ldots\ might
indeed only be effected by a still more radical departure from the
method of description of classical physics, involving the
consideration of the atomic constitution of all measuring
instruments, which it has hitherto been possible to disregard in
quantum mechanics.''

This might sound as if what was earlier characterized as inherent in
the very nature of any physical observation as an ``indispensable
necessity,'' would on the other hand after all just be a, fortunately
still permissible, convenient way of conveying information, a way we
presumably sometime will be forced to give up. If this were your
opinion, then I would gladly agree. However, the subsequent stringent
and clear comparison with the theory of relativity makes me doubt
whether, in what I just said, I have understood your views correctly.
\ldots\

However that may be, there must be clear and definite reasons which
cause you repeatedly to declare that we {\em must\/} interpret
observations in classical terms, according to their very nature.
Whenever you say that, you state it so definitely and clearly, in the
indicative, without any reservations like ``probably,'' or ``it might
be,'' or ``we must be prepared for,'' as if this were the uttermost
certainty in the world. It must be among your firmest
convictions---and I cannot understand what it is based upon.

It could not be just this point (about which you talked so
insistently to me already in 1926): that our traditional language and
inherited concepts were completely unsuited to describe the phenomena
with which we are confronted now.  Because, in the course of the
development of our science (and mathematics), from its earliest
beginnings to the situation at the end of the nineteenth century,
this was certainly the case over and over again. If the break with
the old traditions seems greater now than ever before, then we should
take into account that a particular time perspective is responsible
for forming the impression that {\em that\/} development in which we
ourselves take part, stands out as being more important and more
essential than earlier ones, which we cite only from history, and
whose stages we get to know mostly in reverse order. In fact, it is
often difficult for us to imagine {\em earlier\/} ways of thinking.
And although the difficulty of such a historical step {\em back\/}
actually speaks most eloquently of {\em how\/} significant [the step]
must have seemed to the pioneers at their first advances, still now
and then we cannot avert the feeling: ``Incredible that, up to then,
people were so narrow-minded!'' Here, the underestimation of the time
perspective shows itself most clearly.

Thus I think that the fact that we have not adapted our thinking and
our means of expression to the new theory cannot possibly be the
reason for the conviction that experiments must always be described
in the classical manner, thus neglecting the essential
characteristics of the new theory. It may be a childish example [but
I use it] only to say briefly what I mean: after the elastic light
theory was replaced by the electromagnetic one, one did not say that
the experimental findings should be expressed---just as before---in
terms of the elasticity and density of the ether, of displacements,
states of deformation, velocities and angular velocities of the ether
particles.

Forgive my long-windedness. What I mean to say is: whether you
couldn't make this point completely clear in the more detailed paper
you announce in your Phys. Rev. note: Why do I [{\Bohr}] emphasize
again and again that according to the very nature of a measurement,
it can only be interpreted classically?  And above all: Is this a
temporary resignation, or can we somehow recognize that we will
never get beyond it?
\eq
{\Bohr}'s reply of 26 October 1935 gave very little help in that
respect: it simply repeated over again the same {\em seemingly\/}
``purely logical'' argument.
\bq
On this point I must confess that I cannot share your doubts. My
emphasis of the point that the classical description of experiments
is unavoidable amounts merely to the {\em seemingly obvious fact that
the description of any measuring arrangement must, in an essential
manner, involve the arrangement of the instruments in space and their
functioning in time, if we shall be able to state anything at all
about the phenomena}.  The argument here is of course first and
foremost that in order to serve as measuring instruments, they cannot
be included in the realm of application proper to quantum mechanics.
\eq

(By the way, these quotes are coming from {\em Niels
{\Bohr}:~Collected Works}, vol.\ 7.)  I think Schr\"o\-dinger has a
point that just can't be ignored.  The gist I would like to give
this, I think, is that the term ``classical description'' is an
awfully loaded term: we have grown accustomed to thinking of it in a
very technical sense that precludes Aristotelian mechanics, for
instance.  What {\Bohr} was more particularly trying to get at is
that all measurements must be describable with respect to the
presumption of some stable, predictable, repeatable phenomena.
That's the important aspect of his edict: it is exactly that that
allows us to communicate the experimental set-up to our fellow
observers.  It does not depend in particular on the exact
equational/conceptual form of our ``classical physics.''  It's that
the language has to be based upon a certain primitive form of
stability.  It so happens that the present form of classical
mechanics is our best description of such stable phenomena, but it
need not have been so for the formulation of the edict.  For
instance, General Relativity which succeeds Newtonian Mechanics
could have been discovered in 1964 and it would not have made a bit
of difference to {\Bohr}'s declaration of 1935.

\bap
{\Bohr} was very careful and never claimed that there were in nature
two different types of physical systems.  All he said was that we
had to use two different (classical or quantum) {\em languages\/} in
order to describe different parts of the world.
\eap

Indeed you are right about this as far as I can tell from my various
readings of {\Bohr} (and some of the other Copenhagenists).  So, I'm
very glad you say this so clearly!  So many are apt to screw it up in
just the way you hint \ldots\ and it generally annoys me!  For your,
entertainment I'll attach some quotes from the ``new orthodoxy'' in
that regard.

{\Zurek} wrote in {\em Physics Today\/} (October 1991, p.~36):
\bq
The first widely accepted explanation of how a single outcome emerges
from the many possibilities was the Copenhagen interpretation,
proposed by Niels {\Bohr}, who insisted that a classical apparatus is
necessary to carry out measurements.  Thus quantum theory was not to
be universal.  The key feature of the Copenhagen interpretation was
the dividing line between quantum and classical.  {\Bohr} emphasized
that the border must be mobile, so that even the ``ultimate
apparatus''---the human nervous system---can be measured and analyzed
as a quantum object, provided that a suitable classical device is
available to carry out the task.

In the absence of a crisp criterion to distinguish between quantum
and classical, an identification of the ``classical'' with the
``macroscopic'' has often been tentatively accepted.  The inadequacy
of this approach has become apparent as a result of relatively recent
developments:  A cryogenic Weber bar must be treated as a quantum
harmonic oscillator even though it can weigh a ton.  Nonclassical
squeezed states \ldots\
\eq

{\GellMann} and {\Hartle} in their article ``Quantum Mechanics in the
Light of Quantum Cosmology'' (in {\em Complexity, Entropy and the
Physics of Information}, edited W.~H. {\Zurek}, 1990) write:
\bq
The familiar rule of the ``Copenhagen'' interpretations described
above is external to the framework of wavefunction and
{\Schroedinger} equation.  Characteristically these interpretations,
in one way or another, assumed as fundamental the existence of the
classical domain we see all about us.
\eq

And {\Hartle} in his article ``Excess Baggage'' (in {\em Elementary
Particles and the Universe}, edited by John H. Schwarz, 1991) wrote:
\bq
All singled out the measurement process for a special role in the
theory.  In various ways, these authors were taking as fundamental
the manifest existence of the classical world that we see all about
us.  That is, they were taking as fundamental the existence of
objects which do have histories obeying classical probability rules
and, except for the occasional intervention of the quantum world as
in a measurement, obey deterministic classical equations of motion.
This existence of a classical world, however it was phrased, was an
important part of the Copenhagen interpretations for it was the
contact with the classical world which mandated the ``reduction of
the wave packet.''

The Copenhagen pictures do not permit the application of quantum
mechanics to the universe as a whole.  In the Copenhagen
interpretations the universe is always divided into two parts: To one
part quantum mechanical rules apply.  To the other part classical
rules apply.
\eq

These quotes may not completely pin the tail on the donkey, but
definitely they're written in such a way as to be apt to cause a
belief such as you described.

\bap
\label{OldIsm6}
I wonder whether it is my idea that the apparatus must accept both
quantum and classical descriptions, or someone else said that before
me. Do you know?
\eap

Consider the following passage in David Cassidy's biography of
{\Heisenberg}, {\em Uncertainty:~The Life and Science of Werner
{\Heisenberg}\/} (page 261).

\bq
{\Heisenberg} made the same point differently in his unpublished
manuscript and published lecture. [Cassidy is here talking about
{\Bohr}'s reply to EPR.]  Any experiment entails two types of laws:
those applying to the laboratory measuring apparatus, which, as
{\Bohr} tirelessly argued, are thoroughly classical in nature, and
those pertaining to the atomic phenomena to be studied, which are
thoroughly quantum mechanical in nature. Each set of laws holds in
its own domain and is precise and nonstatistical. Statistics enters
only with the researcher's measuring experiment. Since every
experiment attempts to bridge the gap between the classical
laboratory domain and the quantum world of the atom, the experimental
apparatus introduces a ``cut'' or interface between the two domains.
Because physical laws are precise on either side of the cut, the
statistical characteristics of quantum mechanics enter with and along
the cut; that is, on the performance of the measurement.  A
deterministic completion of quantum mechanics must occur along the
cut. But if new determining variables are introduced along the cut
and the cut is subsequently moved, ``then \ldots\ a contradiction
between the lawlike consequences of the new properties and the
[precise] relationships of quantum theory will be unavoidable.'' A
deterministic completion of quantum mechanics is not possible, after
all.
\eq

The argument, whatever the original really was, is probably stated
somewhat loosely here---for instance the author completely botches
the EPR argument a few pages before this---but the passage may still
convey the essentials.  Simply substitute ``description'' or
``language'' for ``law'' and it starts to sound pretty nice.  The
main point in my quoting this is that certainly by 1935, the issue of
making the domain of classical description fluid or mobile was well
understood by the Copenhagen circle.

The idea of the mismatch in, but necessary use of, two languages and
the lack of a translating dictionary may be all yours.  However it
sounds awfully a lot like the passage above.  {\Heisenberg} uses
``cut''; you use ``apparatus.''  In any case, it's likely that you've
given it a more precise or modern spin than the founding fathers
could have.

\bap
\label{OldIsm7}
{\Bub} wants to get rid of these fictitious observers, in spite of
the fact that they appear to be quite harmless.  (They are like the
ubiquitous observers who send and receive signals in textbooks on
special relativity.)
\eap

I'm not so sure I agree with this.  It's not so obvious to me that
they are completely as ``harmless'' as their special relativistic
cousins.  In fact, they form part of the excitement of the quantum
world for me.  The difference as I see it is this.  The observers in
special relativity serve the role of defining various perspectives on
a single underlying reality, the spacetime manifold.  Their existence
serves to delimit all the possible points of view that a real
observer might take for splitting space and time into separate
concepts.

On the other hand, the quantum observers---with their free choice
over various incompatible measurements---serve to define just what it
is that can be taken to be real.  That is to say, they define the
referent of the measurement record (as you say pretty clearly on page
3).  Without these observers, or measurement records, there is no
underlying reality.  There is no analogue of the spacetime manifold
in this case.

I'm not ashamed to admit that I'm perfectly willing to subscribe to
(at least mild forms of) {\Wheeler}'s observer-participancy.  It's
the puckishness of these new kind of ``observers'' that {\Wheeler}
was talking about when he said, as quoted in {\Bub} p.~191 (my
emphasis),
\bq
\noindent
In broader terms, we find that nature at the quantum level is not a
machine that goes its inexorable way.  Instead, what answer we get
depends on the question we put, the experiment we arrange, the
registering device we choose.  We are inescapably involved in
bringing out {\em that which appears to be happening}.
\eq

This is not idealism or solipsism---it is simply a recognition of
what one must deal with because of the ``no-go theorems'' (modulo
nonlocal theories).  {\Pauli} said it like this in his essay
``Matter,''
\bq
\noindent
\ldots\ This takes into account the observer, including the apparatus
used by him, differently from the way it was done in classical
physics, both in Newtonian mechanics and in Maxwell-Einsteinian field
theories.  In the new pattern of thought we do not assume any longer
the {\em detached observer}, occurring in the idealizations of this
classical type of theory, but an observer who by his indeterminable
effects creates a new situation, theoretically described as a new
state of the observed system.  In this way every observation is a
singling out of a particular factual result, here and now, from the
theoretical possibilities, thereby making obvious the discontinuous
aspect of the physical phenomena.

Nevertheless, there remains still in the new kind of theory an {\em
objective reality}, inasmuch as these theories deny any possibility
for the observer to influence the results of a measurement, once the
experimental arrangement is chosen.
\eq
and like this in his essay ``Albert {\Einstein} and the Development
of Physics,''
\bq
We often discussed these questions together, and I invariably
profited very greatly even when I could not agree with {\Einstein}'s
views.  ``Physics is after all the description of reality,'' he said
to me, continuing, with a sarcastic glance in my direction, ``or
should I perhaps say physics is the description of what one merely
imagines?''  This question clearly shows {\Einstein}'s concern that
the objective character of physics might be lost through a theory of
the type of quantum mechanics, in that as a consequence of its wider
conception of objectivity of an explanation of nature the difference
between physical reality and dream or hallucination become blurred.

The objectivity of physics is however fully ensured in quantum
mechanics in the following sense.  Although in principle, according
to the theory, it is in general only the statistics of a series of
experiments that is determined by laws, the observer is unable, even
in the unpredictable single case, to influence the result of his
observation---as for example the response of a counter at a
particular instant of time.  Further, personal qualities of the
observer do not come into the theory in any way---the observation can
be made by objective registering apparatus, the results of which are
objectively available for anyone's inspection. Just as in the theory
of relativity a group of mathematical transformations connects all
possible coordinate systems, so in quantum mechanics a group of
mathematical transformations connects the possible experimental
arrangements.

{\Einstein} however advocated a narrower form of the reality concept,
which assumes a complete separation of an objectively existing
physical state from any mode of its observation.  Agreement was
unfortunately never reached.
\eq

Why did I say this forms part of the excitement of the quantum world
for me?  For whatever it's worth, let me include a little essay I
wrote to my old friend {\Greg} {\Comer} on 15 April 1996 (titled
``Old Twenty Q's'').
\bq
\noindent
\begin{verse}
C:  \ (While buttering the toast \ldots)  I really can't see why you
    contend that the world, in its fundamentals, is independent of
    us, what we ask of it, and what we do with it.

K:  \ It's just ridiculous to think that we're that God-like.  The
    world is what it is.  It was here long before we were, and it'll
    be just what it is long after we're gone.  This is just nonsense.

C:  \ But what about this toaster?  It is most certainly a free
    creation of man.  It is quite real.  (Ouch!)  You can't possibly
    tell me that it is something innate in the world independent of
    man.  If you can't believe this for a toaster, why on earth do
    you have so much trouble with quantum mechanics?

K:  \ Bring your coffee; let's eat.
\end{verse}
\bigskip

\noindent My friend,
\medskip

I gained a little insight the other day, something new.  Now it's
time to tell it to my alter ego.  {\Wheeler}, like so many others,
has always said that his understanding of the quantum came straight
from the mouth of {\Bohr}.  But now I don't think so.

Recall the Game of Twenty Questions (surprise version).  {\Wheeler}
walked into the room and started probing the other participants for a
word.  First asking one yes/no question, then another, and another,
until there were 20 in all.  Each person in the audience---playing a
dirty trick on our friend---could answer completely at random so long
as their answer was consistent with the answers to the previous
questions.  At the end of the questions, {\Wheeler} himself answered
with a relatively random choice:  a concept, a word, the only one he
could think of consistent with the 20 previous questions and answers
\ldots\ ``cloud.''  The audience thought and thought, each individual
checking his mind for consistency.  Finally it emerged that ``cloud''
was OK!  The audience rolled in laughter, and so the ``cloud'' came
into existence.

{\Wheeler} quite literally took this game as a basis for an ontology.
Nature plays the part of the audience; the experimenter
(communicator) plays the part of the poor sucker that was the butt of
the joke.  Each quantum measurement is a little act of creation---a
little bit of it is on the part of nature with its seemingly random
answer; a little bit of it is on the part of man in freely choosing
the particular question to ask.  If all of the (random) creation were
on the part of nature, one would hardly expect the nice world we
have. If all the creation were on the part of man, everything we hold
dear would vanish into the ethereal mist of a dream.

That's the idea.  I still like it very much and, in fact, I like it
more and more every day.  However, contrast this with {\Bohr}.  This
is something I know that I've read before, but it never quite made
the same impression on me.

\bq
[At the Solvay meeting in 1927] an interesting discussion arose about
how to speak of the appearance of phenomena for which only
predictions of statistical character can be made.  The question was
whether, as to the occurrence of individual effects, we should adopt
a terminology proposed by {\Dirac}, that we were concerned with a
choice on the part of `nature,' or, as suggested by {\Heisenberg},
we should say that we have to do with a choice on the part of the
`observer' constructing the measuring instruments and reading their
recording. Any such terminology would, however, appear dubious
since, on the one hand, it is hardly reasonable to endow nature with
volition in the ordinary sense, while, on the other hand, it is
certainly not possible for the observer to influence the events
which may appear under the conditions he has arranged.  To my mind,
there is no other alternative than to admit that, in this field of
experience, we are dealing with individual phenomena and that our
possibilities of handling the measuring instruments allow us only to
make a choice between the different complementary types of phenomena
we want to study.
\eq

I snatched this quote from Aage {\Petersen}'s book ``Quantum Physics
and the Philosophic Tradition'' (which is worth reading), but it also
appears in the {\Schilpp} volume on {\Einstein}.  There you have it
in any case.  Concerning the ``free choice'' in quantum measurement:
{\Dirac} voted ``nature,'' {\Heisenberg} voted ``man,'' {\Bohr} voted
``neither,'' and {\Wheeler} voted ``both.''  The last point of view
is, as I've written you before, quite similar to things in
{\Pauli}'s writings.
\eq

\bap
The fundamental dilemma in the quantum formalism is not there [i.e.,
in assigning truth values to lattices of propositions]. It is that
quantum mechanics permits the occurrence of all possible events (with
definite probabilities), but in our consciousness there is only one
world.
\eap

Do you really believe there is a dilemma, i.e., a problem that seems
to defy a satisfactory solution?  This seems to go against the grain
of the way you've presented so much else in this essay.  Or are you
using this word in order to make easy connection to {\Bub}'s words
and points of view?

\bap
\ldots\ a completely different problem is whether we can consider
just a few collective degrees of freedom of the universe, such as its
radius, mean density, total baryon number, etc., and apply quantum
theory to these degrees of freedom, which do not include the observer
and other insignificant details. This seems legitimate: this is not
essentially different from quantizing the magnetic flux and the
electric current in a SQUID, and ignoring the atomic details. You may
object that there is only one universe, but likewise there is only
one SQUID in my laboratory.
\eap

I really like this passage a lot!  Where on earth the community got
the idea that it must ``include the observer in the wave function''
(as {\Wheeler} would say) in order to do any sensible quantum
mechanical calculations for cosmology, I don't know.  It's very nice
the way you pinpoint the problem here.

Also I like very much the essentially Bayesian viewpoint you take
here!  ``There is only one SQUID in my laboratory.''  One does not
need large ensembles of repetitive preparations to make sense of
quantum probabilities.

\bap
\label{OldIsm10}
For the opposite opinion of some recalcitrant positivists, see Fuchs
and {\Peres} (1996).
\eap

I liked your earlier, less committal (but still funny) quote better.
The reason for my backing out on this is that I'm not sure how to
read the ``opposite'' in this sentence---I think it is a little
ambiguous and I'm not sure I want to be too associated with a
possibly outlandish interpretation.

Would you consider either restoring the original (i.e., simply
without the word ``opposite'') or trying something like:  ``To see
how two recalcitrant positivists actually define `disturbance' in
quantum mechanics see Fuchs and {\Peres} (1996).''?

Of course the important thing is that the ``disturbance'' is not to
any ``dynamical variable'' as {\Bub} would like to say, but rather to
Alice's predictability of Bob's measurement outcomes.  That's the
whole content of it.

You know looking at a few more passages of {\Bohr}, like
\bq
The essential lesson of the analysis of measurements in quantum
theory is thus the emphasis on the necessity, in the account of the
phenomena, of taking the whole experimental arrangement into
consideration, in complete conformity with the fact that all
unambiguous interpretation of the quantum mechanical formalism
involves the fixation of the external conditions, defining the
initial state of the atomic system concerned and the character of the
possible predictions as regards subsequent observable properties of
that system. Any measurement in quantum theory can in fact only refer
either to a fixation of the initial state or to the test of such
predictions, and {\em it is just the combination of measurements of
both kinds which constitutes a well-defined phenomenon.}
\eq
and
\bq
As a more appropriate way of expression I advocated the application
of the word {\em phenomenon\/} exclusively to refer to the
observations obtained under specified circumstances, including an
account of the whole experimental arrangement. In such terminology,
the observational problem is free of any special intricacy since, in
actual experiments, all observations are expressed by unambiguous
statements referring, for instance, to the registration of the point
at which an electron arrives at a photographic plate. Moreover,
speaking in such a way is just suited to emphasize that the
appropriate physical interpretation of the symbolic
quantum-mechanical formalism amounts only to predictions, of
determinate or statistical character, pertaining to individual
phenomena appearing under conditions defined by classical physical
concepts.
\eq
it dawns on me that quantum cryptography (the BB84 setup, say) might
just provide a wonderful vehicle for explaining and making clear
{\Bohr}'s notion of ``quantum phenomenon.''  And it, of course, has
the added advantage of having questions of ``information'' and
``disturbance'' built into the very reason for defining the game in
the first place.

I don't know if you recall, but I once told you that I would really
like to write something for the {\sl American Journal of Physics\/}
that explains well what is going on in the question of information
gain versus disturbance in quantum theory.  It now strikes me that
one could incorporate so much of this discussion (from your paper and
these notes) into a relatively nice exposition of both {\Bohr} and
information/disturbance (an article that has ``something old,
something new'').  Would you be interested in seeing if we could
tackle this in an effective way?

\bap
{\Bub}'s goal is to liberate the quantum world from its dependence on
observers. \ldots\ The tacit assumption made by {\Bub} \ldots\ is
that the wave function is a genuine physical entity, not just an
intellectual tool invented for the purpose of computing
probabilities.
\eap

Finally let me leave you on a happy note.  Recall how I scolded you
above for your ``curt dismissal'' of {\Bub}'s lattices?  Well the
truth of the matter is that he wasn't any more sympathetic,
explanatory, or observant of our views!  Just look at pages 38--39:
\bq
The only argument for introducing a special mode of state transition
characterizing measurement processes is that ``experience only makes
statements of this type: an observer has made a certain (subjective)
description; and never any like this: a physical quantity has a
certain value.''  This echoes {\Bohr}'s remark \ldots\ that the task
of physics is not ``to find out how nature {\em is},'' but rather
``what we can {\em say\/} about nature.''  But where is the argument
that quantum mechanics, as opposed to classical mechanics, is a
theory about subjective descriptions or what we can say about
nature?  After all, quantum mechanics is, in effect, a non-Boolean
generalization of classical mechanics, and classical mechanics
manages quite nicely to be a theory of how nature {\em is}, in
principle, by ascribing values to physical quantities.
\eq
Perhaps after-all your article is a deserved ``touch\'e''!

\section{22 October 1997, \ ``Honesty''}

\bap
The idea of the ``cut'' is not due to {\Heisenberg}, but already
appears in von Neumann's book (1932). However, this is not quite the
same as my way of formulating this issue. von Neumann, and others
after him, considered a sequence of apparatuses, described by
quantum mechanics up to some point, and the following apparatuses
were treated classically. My contention is that the SAME apparatus
must get both descriptions, at different stages. Was it really my
personal chutzpah to have this idea?
\eap

I understand your contention (or, at least, think I do).  However,
please read the passage I sent you again:  there is more being said
there than what von Neumann had to say.  It may not be---and probably
{\em is not}---exactly what you have to say, but it is distinct from
my understanding of von Neumann (which mimics what you said above).
Cassidy says, pretty explicitly, ``the experimental apparatus
introduces a `cut' or interface between the two domains [i.e., the
quantum and the classical].''  I am, in particular, thinking of
``interface'' in the computer science sense as something like
Windows95 or Sun OpenWin that is half computer program, half scratch
pad for my thoughts.  If I am not mistaken by the larger context of
Cassidy's remarks, I think that is the sense in which Mr.\
{\Heisenberg} was using it too.  He speaks of the ``cut'' as if it
were a recepticle in which one might place hidden variables, but
because of its mobility this would cause a contradiction.

In any case, other than these loosely related things, I don't think I
have heard your idea before.  (That is, other than reading it in your
book two years ago!)  I've certainly never seen it posed in a
quantitative way by anyone else.

\section{23 October 1997, \ ``Tyche or Moira?''}

It must have been an interesting combination of the headache I had
today, curiosity, and just plain luck that lead me to the right place
in the library.  I ran across the following in an article by
{\Shimony} (AJP, vol.\ 31, p.\ 755, 1963):

\bq
{\Bohr} is saying that from one point of view the apparatus is
described classically and from another, mutually exclusive point of
view, it is described quantum mechanically. In other words, he is
applying the principle of complementarity, which was originally
formulated for microphysical phenomena, to a macroscopic piece of
apparatus.  This application of complementarity on a new level
provides an answer to the difficulty regarding macrophysical objects
which confronted ``positivism of higher order'': the macro-physical
object has objective existence and intrinsic properties in one set of
circumstances (e.g., when used for the purpose of measuring) and has
properties relative to the observer in another set of circumstances,
thereby evading the dilemma of choosing between realism and idealism.
Two important conclusions follow from this discussion. \ldots\  The
second conclusion is that {\Bohr}'s extension of the principle of
complementarity beyond its original function of reconciling
apparently contradictory microphysical phenomena is not gratuitous,
as critics have often claimed. The internal logic of his position
requires the application of complementarity to macroscopic measuring
instruments.
\eq

It still doesn't capture what you said, namely,
\bq
\noindent
The peculiar property of the quantum measuring process is that we
have to use {\it both\/} descriptions for the {\it same\/} object:
namely, the measuring apparatus obeys quantum dynamics while it
interacts with the quantum system under study, and at a later stage
the same apparatus is considered as a classical object, when it
becomes the permanent depositary of information. This dichotomy is
the root of the quantum measurement dilemma: there can be no
unambiguous classical-quantum dictionary.
\eq
but it comes closer to the flavor.  In particular, the passage
captures there being no classical-quantum dictionary.  However it
does not completely address the dual use of the measuring apparatus
as such---it only says ``in another set of circumstances.''  Maybe
that is not explicit enough.  In any case, though, I am coming to
understand you a little better.

With that, I sign off on this issue.

\section{15 December 1997, \ ``Hay--{\Peres} Comments''}

I think it may be harder to comment on your paper than I had thought
it would be.  I take it that your largest concern is how a referee
might react to the section on ``von Neumann's cut'' and also the
concluding section.  Even after reading it over twice now, I find it
difficult to estimate.  Very likely it will be the toss of a coin as
to whether a referee will like it or not.  In any case, if the
calculation is sound, I don't see how a referee can deny that it was
worthwhile doing.  If you get unlucky enough to get a
holier-than-thou follower of the many-worlds idea, then it seems
likely that you'll get a hand-spanking for your ``antiquated'' ideas
and some suggested modifications, but it'll still be accepted.  (Do
you actually ever get papers rejected?)

I am concerned by the growing abandonment of standard (Copenhagen)
quantum mechanics that I see all around me.  Actually, I mostly see
that in the theorists; the experimentalists around here seem OK with
the idea that the state vector describes what one knows about their
lab table and nothing more.  Everywhere I turn, I see theorists
climbing onto the many-worlds bandwagon for no well-defined
scientific reason.  It just seems to give them a happier, stable
feeling---at least that's all I can see.  I worry about how to combat
this in an effective way, but I haven't found one yet.  Talking to
you, {\Mermin}, and {\Preskill} has been helpful in that respect,
but I have a long way to go I think.  (Notice that I purposely
placed you three in just that order:  you being on the far-left, and
John being on the far-right.)

\section{29 January 1998, \ ``Strange Surprise''}

 Have you looked at {\Mermin}'s latest tractatus on {\tt
quant-ph}? His aspirations for an (objective) ``interpretation'' of
quantum mechanics are a little annoying, but if you can put a
distaste for that aside, there are a couple of nice points in it
(relying on {\Wootters}' theorem). Aside from that---in a way that he
hadn't intended, of course---I think the paper makes a strong case
for the standard interpretation that you, for instance, wrote about
in your last book review.  Somehow, we've got to get out of all these
interpretational quandaries and look for that aspect of the standard
interpretation that is ripe for building upon. So that we can move on
to much bigger, much better physics.  I wish I could find the key for
convincing good guys like {\Mermin} of that point of view.

\section{05 April 1998, \ ``Other Things''}

\bap
After two months, they sent the paper to a decoherentist, who
demanded to mention decoherence. It's surely better than a
many-worldist, or a Bohm-trajectorist, etc.
\eap

First off, congratulations to you and Ori Hay on the new paper!  You
have, of course, told Ori about the coveted {\Einstein} Number that
he now possesses?!  Are you so sure, though, that the decoherentists
are more desirable than the Everettistas and the Bohmians?  I find
them all equally unpalatable \ldots\ though each for a different
specific reason.  My blanket complaint is that they all hope to find
ontology where there is none.  I, by the way, have been working on a
new way to express what I find most positive about my
neo-Copenhagenist views.  Namely, we find a lesson in quantum theory
not so different from the one taught by the great {\Socrates}:  we
are only in a position to learn when we realize that we are fully and
completely ignorant. The version of the theory with the least
ontological commitment is certainly the one poised for the greatest,
most revolutionary progress next century.  (That is, once the next
{\Einstein} \ldots\ (perhaps a future son or daughter of mine?)\
\ldots\ is born!)

Speaking of numbers and lineage, etc., there's a little story that I
haven't told you.  When I won the Michelson prize I started to look
for a good way to open my colloquium and it dawned on me that I had
once heard John {\Wheeler} (my academic great-grandfather) describe
his Ph.D. advisor as ``the great optical physicist \ldots''\ though I
couldn't remember the precise name.  So I hoped very hard that
whoever he was, he might actually be Michelson's student.  Alas, it
wasn't to be!  But I ended up learning something in the process of my
search that was really interesting to me.  {\Wheeler}'s advisor was
Karl {\Herzfeld}, who first taught in Munich (where he and
{\Sommerfeld} gave birth to {\Heitler}), then at Johns Hopkins
(where he gave birth to {\Wheeler}), and finally at the Catholic
University of America.  The thing that I love to know is that, as
best one can trace, {\Herzfeld} may have been the first person to
introduce the term ``quantum mechanics'' into the literature!!
{\Born} thought that he had done it himself, but his first usage of
it was in a paper of 1924. Lorentz preceded that in 1923 with ``the
mechanics of quanta.'' However, {\Herzfeld} had a paper of 1921
titled with the German equivalent of ``Quantum Mechanics of Atomic
Models.''  (These last details come from a footnote in Mehra and
Rechenberg, vol 4.) {\Herzfeld} by the way obtained his Ph.D. in
Vienna in 1914 but I haven't been able to trace who his advisor was.

\section{01 December 1998, \ ``Here Comes the Judge''}

\bap
The measuring process is an external intervention in the dynamics of
a quantum system.
\eap

I know that I've already expressed what I'm about to say several
times, but let me just set things off to a good start again: I really
like this turn of phrase!  For years now, I have believed exactly
what you say, namely,
\bap
[I]t is preferable not to use the word ``measurement'' which suggests
that there exists in the real world some unknown property that we are
measuring.
\eap
but I had never invented such a clever phrase to express it.  For a
long time instead I made use of the word ``creation'' in my
tract-like emails on the subject.  In some ways, I still
think---taking my cue from John {\Wheeler}---that that word captures
a certain central truth of quantum theory, but your word is more
encompassing.  I think it captures better the idea that quantum
measurement is a double-edged sword, learning {\it and\/} creating.

But I wouldn't be doing my scholarly duty, if I didn't remind you of
some passages of Niels {\Bohr}, whose words you seem to have so much
respect for.  In {\Bohr}'s 1949 article ``Discussion with
{\Einstein} on Epistemological Problems in Atomic Physics,'' he
wrote:
\bq
[At the Solvay meeting in 1927] an interesting discussion arose about
how to speak of the appearance of phenomena for which only
predictions of statistical character can be made.  The question was
whether, as to the occurrence of individual effects, we should adopt
a terminology proposed by {\Dirac}, that we were concerned with a
choice on the part of `nature,' or, as suggested by {\Heisenberg},
we should say that we have to do with a choice on the part of the
`observer' constructing the measuring instruments and reading their
recording. Any such terminology would, however, appear dubious
since, on the one hand, it is hardly reasonable to endow nature with
volition in the ordinary sense, while, on the other hand, it is
certainly not possible for the observer to influence the events
which may appear under the conditions he has arranged.  To my mind,
there is no other alternative than to admit that, in this field of
experience, we are dealing with individual phenomena and that our
possibilities of handling the measuring instruments allow us only to
make a choice between the different complementary types of phenomena
we want to study.
\eq
And, more to the point, in his 1958 article ``Quantum Physics and
Philosophy: Causality and Complementarity,'' he wrote:
\bq
In the treatment of atomic problems, actual calculations are most
conveniently carried out with the help of a {\Schroedinger} state
function, from which the statistical laws governing observations
obtainable under specified conditions can be deduced by definite
mathematical operations.  It must be recognized, however, that we are
here dealing with a purely symbolic procedure, the unambiguous
physical interpretation of which in the last resort requires a
reference to a complete experimental arrangement.  Disregard of this
point has sometimes led to confusion, and in particular the use of
phrases like ``disturbance of phenomena by observation'' or
``creation of physical attributes of objects by measurements'' is
hardly compatible with common language and practical definition.

In this connection, the question has even been raised whether
recourse to multivalued logics is needed for a more appropriate
representation of the situation.  From the preceding argumentation it
will appear, however, that all departures from common language and
ordinary logic are entirely avoided by reserving the word
`phenomenon' solely for reference to unambiguously communicable
information, in the account of which the word `measurement' is used
in its plain meaning of standardized comparison.
\eq

Do you see these passages as meshing well with your new terminology?
I don't really, but then again I don't feel that {\Bohr} was all that
right when he came to these points.  I think it might be useful for
your ultimate reader to add some words about your opinion on this.

\bap
As a concrete example, consider the quantum teleportation scenario.
The first intervention is performed by Alice: she has two
spin-$1\over2$ particles and she tests in which {\Bell} state these
particles are.
\label{Naugahyde}
\eap
In which state they {\it are\/}?!?!  There's got to be a better way
of putting this \ldots\ especially one more consistent with your
whole view of the measurement process.

\bap
I do not want to use the word ``histories,'' which has acquired a
different meaning in the writings of some quantum theorists.
\eap
What happened to your nice word ``chronicle''\@?  I liked it a lot.
Can't you find some way to reinstate it?  I guess I liked it because
it always reminded me of a passage that I like to quote from
{\Pauli}. In a letter to Markus {\Fierz} in 1947, he wrote:
\bq
\noindent
Something only really happens when an observation is being made
\ldots\,.  Between the observations nothing at all happens, only time
has, `in the interval,' irreversibly progressed on the mathematical
papers!
\eq

\bap
Quantum mechanics is fundamentally statistical, and any experiment
has to be repeated many times (in a theoretical discussion, we shall
imagine many replicas of our gedanken\-experiment).
\eap
\bap
Each one of these records has a definite probability, which is
experimentally observed as its relative frequency among all the
records that were obtained.
\eap
\bap
Note that the ``detector clicks'' are the only real thing we have to
consider. Their probabilities are objective numbers and are Lorentz
invariant.
\eap

As a disciple of the Reverend Bayes, you should know that I strongly
dislike all these expressions.  A good Bayesian would say that
probability quantifies a state of knowledge.  It is that and that
alone, meeting an operational verification {\it only\/} through a
subject's betting behavior.  In particular, probability has no {\it a
priori\/} connection to the frequency of outcomes in a repeated
experiment.  It matters not whether that repetition is real or,
instead, imaginary and virtual.

Let me just try to drive this home with two of the simplest possible
examples.

(1)  Consider a coin for which you have no reason to believe that a
head will occur over a tail in a toss.  Now imagine that you flip
that coin an infinite number of times, tabulating the number of heads
and tails.  Do you really believe that a frequency of precisely 1/2
{\it must\/} occur in the infinite limit?  Answer this question to
yourself very honestly.  What is to bar you from getting a head as
the outcome in literally every single toss?  What is there in the
coin to favor a random-looking sequence to a nonrandom one? Nothing.
To say something like, ``Well the nonrandom-looking sequences have
probability zero'' is just to beg the question.  For one thing, you
have to invoke the concept of probability again to even say it.  For
another, even if you allow yourself that, it still carries no force:
just take the set of nonrandom-looking sequences and add to it any
sequence that you would consider a valid random one (i.e., one that
you believe could be generated by a repeated coin toss).  That set
still has probability-measure zero, but now you would have to say
that that means the random-looking sequence could not be generated by
your imaginary coin.  The point:  probability has no direct
connection to frequency.

(2)  Consider now the case where I toss a coin repeatedly for you. I
assure you strongly that the coin is weighted 80--20 heads vs.\ tails
{\it or\/} 20--80 , but I steadfastly refuse to tell you which way it
goes.  What probability would you ascribe to the outcome of a head
upon my first toss?  Fifty-percent of course. The point is, my
probability ascription is not your probability ascription.  As I toss
the coin ever more, if you are rational, your probability ascription
will {\it very likely\/} approach mine, but there are no absolute
assurances.  The point:  again probability has no direct connection
to frequency.  But also, two perfectly rational people can make
different probability assignments to the same experiment without
being inconsistent with each other.  There is nothing objective about
a probability assignment per se.

If you'd like to understand better this point of view, then I'd
strongly suggest the wonderful collection of essays, ``Studies in
Subjective Probability,'' edited by Henry~E. {\Kyburg}, Jr., and
Howard~E. Smokler (Wiley, NY, 1964).  Also there is a later edition
of the book with a few other essays.  For a much more in-depth study
I couldn't recommend anything more than Ed {\Jaynes}' book,
``Probability Theory: The Logic of Science.''  It unfortunately may
never be published properly, but preprints of the parts that were
written can be found at the ``Probability Theory as Extended Logic''
web page, {\tt http://bayes.wustl.edu/}.

\bap
[P]robabilities are objective numbers \ldots\,. \ On the other hand,
wave functions and operators are nothing more than abstract symbols.
They are convenient mathematical concepts, useful for computing
quantum probabilities, but they have no real existence in Nature.
\eap

I think it is really cute the way you cite {\Stapp} here!

But again I want to come back to the last point above.  A good
Bayesian already knows that probabilities cannot be taken to be
objective numbers, existent ``out there'' in nature.  And a good
quantum physicist similarly knows that the wave function is {\it
not\/} something ``out there'' existent in nature.  The similarity of
these two points of view is not an accident: it had to be the case!
For, from one point of view, the wave function is nothing more than a
compendium of probabilities. It may be the case that ``unperformed
measurements have no outcomes,'' but it is not the case that
unperformed measurements have no probabilities!  If we specify the
probabilities of the outcomes for every {\it potential\/}
measurement, then we have specified the wave function.  There cannot
be a difference in the objectivity levels of these two mathematical
objects:  they are both abstract symbols that summarize our states of
knowledge.  Their empirical meaning comes about precisely in how a
rational being would behave in the light of that knowledge.

There is a famous quote in Bayesian probability theory due to Bruno
de {\Finetti}.  It starts out the two volumes of his book.
\bq
\small
\noindent
My thesis, paradoxically, and a little provocatively, but nonetheless
genuinely, is simply this:
\begin{center}
PROBABILITY DOES NOT EXIST.
\end{center}
The abandonment of superstitious beliefs about the existence of
Phlogiston, the Cosmic Ether, Absolute Space and Time, \ldots, or
Fairies and Witches, was an essential step along the road to
scientific thinking. Probability, too, if regarded as something
endowed with some kind of objective existence, is no less a
misleading conception, an illusory attempt to exteriorize or
materialize our true probabilistic beliefs. \normalsize
\eq
In contrast, my paper ``Bayesian Probability in Quantum Mechanics''
with {\Caves} and {\Schack}, opens up with the following lines:
\bq
\small
\noindent
My thesis, paradoxically, and a little provocatively, but nonetheless
genuinely, is simply this:
\begin{center}
QUANTUM STATES DO NOT EXIST.
\end{center}
The abandonment of superstitious beliefs about the existence of
Phlogiston, the Cosmic Ether, Absolute Space and Time, \ldots, or
Fairies and Witches, was an essential step along the road to
scientific thinking. The quantum state, too, if regarded as something
endowed with some kind of objective existence, is no less a
misleading conception, an illusory attempt to exteriorize or
materialize the information we possess.\\
\hspace*{\fill} --- {\it the ghost of Bruno de {\Finetti}}
\normalsize
\eq
I don't know of any better way to express it than this.

\bap
The physical evolution that leads to Eq.~(1) is the following. The
intervener receives a quantum system in state $\rho$ and adjoins to
it an auxiliary system (an {\it ancilla\/}) in a known state
$\rho_a$. The composite system, in state $\rho\otimes\rho_a$, is
subjected to a unitary transformation \ldots\
\eap

In my present way of speaking, I like to make this sound much less
absolute.  I usually say that this is only one {\it representation\/}
of a superoperator.  It is the superoperator that is the only thing
that need be given; anything else is just one way or another of
thinking about it.  I guess the point is I see no reason to give
unitary evolutions and von Neumann measurements a special status in
the axiomatics of the theory.  POVMs and superoperators are perfectly
good (and much less specific) starting points for the theory. You can
certainly do as you please, but it seems to me that there is a lot to
be gained from this point of view.  For instance, it seems to me to
de-emphasize the point of view that Charlie {\Bennett} labels ``the
Church of the Larger Hilbert Space'' (which is, as he admits, a
euphemism for the many-worlds interpretation).

\bap
If we wish to consider only the states just before and after the
intervention, without entering into the detailed dynamics of the
intervention, the result appears as a discontinuous jump of the wave
function (often called a ``collapse'').  Clearly, this jump is not a
dynamical process. It results from the introduction of an ancilla,
followed by its deletion, or that of another subsystem. The jump is
solely due to abrupt changes in our way of delimiting what we
consider as the quantum system under study.
\eap

I agree with you to some extent here, but not the whole way.  The
abrupt change that comes about when we conceptually delete a
subsystem from a larger composite system corresponds to a {\it
partial trace}.  Where does the {\it random\/} jump come from that
corresponds to the measurement outcome label $k$?  I've heard you say
things like this many times, i.e., that the statistics in quantum
mechanics comes about because of a mismatch in two languages, the
quantum and the classical---your talk at QCM was just one
example---but how does one see that you're not just talking about a
partial trace here?  I am perfectly willing to take
measurement/intervention as a primitive of the theory---it's the very
thing that gives the theory meaning---but you seem to want to go
further, to have it fall out of something more primitive, namely the
act of drawing a conceptual line in an overall unitary dynamics.  I
guess I still don't see it.

\bap
Summing over them is like saying that peas and peanuts contain on the
average 42\% of water, instead of saying that peas have 78\% and
peanuts 6\% [25].
\eap

I loved the citation here!  You don't think there might be a
connection between {\Stapp} and the USDA, do you?

\bap
Between these two events, there is a ``free'' (that is,
deterministic) evolution of the state of the quantum system. What
distinguishes such a free evolution from an intervention is that the
latter has unpredictable output parameters, for example which one of
the detectors ``clicks,'' thereby starting a new chapter in the
history of the quantum system. As long as there is no such a
branching, the evolution will be called {\it free\/}, even though it
may depend on external classical fields, that are specified by the
classical parameters of the preceding interventions.
\eap

I like this distinction, but I cannot completely agree with it. Take
any trace-preserving completely positive map that is not unitary,
i.e., most any quantum channel like the amplitude damping channel or
the depolarizing channel will do.  It gives rise to no ``branching''
changes in the quantum state, but I think one would be hard pressed
to call it a ``free'' evolution.

I overheard {\Herb} {\Bernstein} once say something like, ``Of course
there are two kinds of evolution in quantum mechanics; there are the
times when we learn something and then there are the times when we
learn nothing.''  The line struck me.  I know that this is something
like what you're trying to get at, but the cut isn't between unitary
and nonunitary then.

\bap
Quantum mechanics asserts that during a free evolution the quantum
state undergoes a unitary transformation.
\eap

This needs cleaning up exactly because of the point above.  By the
way, note that there is a typo in the sentence following Eq.~(7).

\bap
Note that ${\rm Tr}(\rho_f)={\rm Tr}(\rho'_f)$ is the joint
probability of occurrence of the records $k$ and $\mu$. This is the
only observable quantity in this experiment. \ldots\ \ {\Einstein}'s
principle of relativity asserts that both descriptions given above
are equally valid. Formally, the states $\rho_f$ (at time $t_f$) and
the state $\rho'_f$ (at time $t'_f$) have to be Lorentz transforms of
each other.
\eap

I am afraid that this might be construed to mean that the only way
the equality of the traces can be satisfied is if $\rho_f$ and
$\rho'_f$ are ``Lorentz transforms'' of each other.  Surely that's
not true and that's not what you mean.  Maybe you could clarify this
a bit.

\bap
As a further simplification, let all the $U$ and $V$ operators be
unit matrices, so that the two particles are really free, except at
the intervention events.
\eap

What do you mean by the phrase ``really free'' here?  Is it a joke?

\bap
We thus have to accept that unit matrices of any order may also be
legitimate Lorentz transforms of each other.
\eap

So I guess what you're meaning is that you want this problem to
partially define what is meant by the very term ``Lorentz
transform?''  I'm looking forward to understanding the business about
Green's functions much better.  After that I'll come back for some
more substantial comments on everything in this section following
Eq.~(10).

\bap
Quantum nonlocality has led some authors to suggest the possibility
of superluminal communication.
\eap

Notice the correction to {\it your\/} spelling mistake \ldots\ ahem.
Grin.  Anyway, in this connection let me point you to a very nice
article by a philosopher:  J.~B. {\Kennedy}, ``On the Empirical
Foundations of the Quantum No-Signalling Proofs,'' Phil.\ Sci.\ {\bf
62}, 543--560 (1995).  The point he makes---and I think validly
so---is that any so-called ``proof'' that quantum mechanics cannot
support superluminal signalling by using entanglement is essentially
circular.  It's not that the ``proofs'' are invalid, but essentially
that there was no use in doing them in the first place: no-signalling
is much more of an {\it axiom\/} than a result of the standard
structure of quantum mechanics.  {\Kennedy} argued by way of digging
into the historical references, finding for instance von Neumann's
original motivation for introducing the tensor product rule for
combining Hilbert spaces---it was essentially to block the
possibility of superluminal signalling!  You might enjoy reading the
paper.

\bap
The classical-quantum analogy becomes complete if we use statistical
mechanics for treating the classical case. The distribution of the
bomb fragments is given by a Liouville function in phase space. When
Alice measures ${\bf J}_1$, the Liouville function for ${\bf J}_2$ is
instantly altered, however far Bob is from Alice. No one would find
this surprising, since it is universally agreed that a Liouville
function is only a mathematical tool representing our statistical
knowledge.  Likewise, the wave function $\psi$, or the {\Wigner}
function, which is the quantum analogue of a Liouville function, are
only mathematical tools for computing probabilities. It is only when
they are considered as physical objects that superluminal paradoxes
arise.
\eap

Indeed you have captured the point here.  Part of this, i.e., that
the proper comparison is between state vectors and Liouville
distributions, is what {\Carl} and I were striving to convey in our
paper for Nathan {\Rosen}'s festschrift---so I'm already very
sympathetic to this.  (Of course I know that you already made this
point long ago in your paper ``What is a State Vector?'')  More
particularly, though, this is one thing that Steven van {\Enk} and I
had planned to write in a silly little paper---titled ``Entanglement
is Super \ldots\ but not Superluminal!''---for our contribution to a
book on superluminal signaling!  I hope you won't mind the overlap.

The thing that holds so many up from immediately grasping this point
of view is the {\it difference\/} between the quantum and classical
states of knowledge (i.e., those things that are changing
instantaneously as you say).  Classically, the probability is
attached to an existent property.  ``The particle has a position; I
just don't know it.  I capture what little bit I do know with a
probability assignment.''  Quantumly, the probability is attached to
{\it potential\/} measurement outcomes and nothing more.  How is it
that my probability assignment for some potential measurement outcome
can change from 50-50 to complete certainty when there is no sense in
which that measurement outcome can already be said to be ``out
there'' without the performance of a measurement?  That's the thing
that has people perplexed and, I think, is the source of their
confusion about using entanglement for communication.

\bap
In our approach, there is no `delayed choice' paradox. It is indeed
quite surprising that JAW, who is a hard core positivist, introduced
this `delayed choice' idea.
\eap

Sometimes it's pretty hard to figure out what precisely it is that
John is trying to say.  Perhaps you shouldn't be overly harsh on him.
In some of his discussions of ``delayed choice'' he makes it very
clear that he is not talking about affecting the ``past'' itself
(whatever that might mean) but instead ``what we have the right to
say about it.''  Combine that with one of the phrases he is fond of
saying, i.e., ``The past exists only insofar as it is recorded in the
present,'' and I think you will have to position him closer than not
to the positivist you used to think of him as.

\section{16 December 1998, \ ``Nonlinear CPMs''}

I think the question you are starting to ask is a good one.  In fact,
one of the big things on my (thinking) wish list for the coming year
is to get a better handle on the {\it physical\/} assumptions
underlying complete positivity.  Weinberg's evolution operator is
trivially not a completely positive map because---in the standard
usage of the term---completely positive maps can only be {\it
linear}.  There is a good open question about the proper way to
define nonlinear generalizations of complete positivity.  I had a
conversation about this with {\Holevo} at QCM and he said it would
be a mess, the difficulties arising somehow because the tensor
producting operation is explicitly a linear operation.  I didn't
understand him very well, but I have wanted to come back to think
about it.

In a certain sense, complete positivity is relatively natural from
the Bayesian viewpoint of quantum mechanics.  (The Bayesian
viewpoint, by the way, is what I'm starting to call my way of
thinking about quantum mechanics:  the term will make more sense to
you once you read the opus that {\Caves} and {\Schack} and I are
writing on it.)  Whatever physical evolution is, it had at the very
least take states of knowledge to states of knowledge.  That {\it
plus\/} linearity gives complete positivity.  And, once one has that
one is also warranted in thinking about unitarity in the Church of
the Larger Hilbert space.  (Unitarity is a theorem, not a starting
point!)  But what compels linearity within the Bayesian view?  That I
don't understand yet.  The standard story of the Karl {\Kraus} types
is that linearity comes from the reasonable requirement that the
evolution of a mixture be a mixture of the evolutions of the
independent states.  But that is not so motivated within the Bayesian
view where the quantum state is not solely a recipe (as you like to
call it), but rather a state of knowledge full stop. (A recipe is a
special case of a state of knowledge.)  Things to ponder over \ldots\
though I suspect the issues won't make too much sense to you until
you read our paper.

There is perhaps one thing interesting to read presently in the
literature on this.  A paper by Czachor and Kuna, ``Complete
positivity of nonlinear evolution: A case study,'' PRA {\bf 58}, July
1998, 128--134.  I think their point is that the one nonlinear
generalization to CPMs that the mathematicians have proposed is not a
physically reasonable generalization.  But I have not read the paper
yet.  Part of lectures in Torino will be on this subject.  (I am
lecturing on quantum channels, i.e., the details of CPMs.)

\section{26 December 1998, \ ``The Deep Intervention''}

Well I have finally had some time to read through your new draft of
RQM completely.  I am not sure, however, what I should compose as
comment yet---I think I will chew upon it for a while, before saying
anything more.  In the mean time, let me share a passage I just
entered into my computer by way of my ``voice recognition system.''
  These came from a recent paper on {\tt quant-ph} by Doug
{\Bilodeau}, ``Why Quantum Mechanics is Hard to Understand.''  In his
introduction he says, ``In this paper I present the conceptual system
which I think gives the clearest understanding of what it is we are
doing when we use quantum mechanics.''  I think he fails at that, as
I cannot figure out what he is talking about for much of the paper.
{\it But}, there were some parts of the paper that struck me as
particularly deep.  The passage below was one of them.  It concerns
another use of the term ``intervention,'' one that I am particularly
fond of (but is maybe a little more risky than the meaning you
introduce in your paper).  I hope you enjoy it as much as I did.

\bq
A thing is historical insofar as it is objective (can be observed and
treated as an object).  It then enters into the realm of recordable
objective occurrences which can be ordered in historical space and
time. It is dynamical insofar as it is defined as an abstract element
of the dynamical theory which explains causal relationships between
objects. \ldots\

Imagine that we could see the universe as omniscient external
observers, all space and time at once, and that what we ``see'' is a
tangle of intersecting particle world-lines (cf.\ Ch.\ 1 of Misner,
Thorne, and {\Wheeler}).  We might detect some patterns which would
constitute physical rules or laws in some sense, but it would be
quite difficult or impossible to know whether we had found all the
important patterns, or to distinguish significant relationships from
accidental ones.  Even more difficult would be to translate this
omniscient description into the kinds of relationships and laws which
would be observed by the huge clumsy bunches of world-lines which
constitute ourselves.

When we set out to investigate Nature, we are not like that external
omniscient observer at all.  We look for relationships and patterns
in the behavior of objects we know.  We want to find out---does this
kind of object always behave this way under these circumstances?  The
phrases ``this kind,'' ``this way,'' and ``these circumstances''
imply the ability to abstract relevant or significant features from
what are really unique events.  They also imply that we can find or
(even better) set up many instances of these typical situations.  The
result is that the concepts we develop to describe physical phenomena
depend not only on what we can observe, but also on what we can do.

To say that A affects or causes or influences or interacts with B
implies a counterfactual:  If A had been different, B would have been
different, too.  The most convincing way to establish a connection is
to ``wiggle'' some parameter in A more or less randomly and then
observe the same odd pattern showing up in some property of B.  If I
want to know whether a wall switch controls a certain light, I can
flip the switch on and off and observe whether the light follows my
actions. There is always the possibility that the light is being
controlled by someone else or goes on and off spontaneously; but if I
put the switch through a very irregular and spontaneous sequence of
changes and the light still follows along, then the probability of a
causal connection is very high (barring a conspiracy to deceive the
experimenter).

Physical theory is possible because we {\it are\/} immersed and
included in the whole process---because we can act on objects around
the us. Our ability to intervene in nature clarifies even the motion
of the planets around the sun---masses so great and distances so vast
that our roles as participants seems insignificant.  {\Newton} was
able to transform {\Kepler}'s kinematical description of the solar
system into a far more powerful dynamical theory because he added
concepts from Galileo's experimental methods---force, mass,
momentum, and gravitation.  The truly external observer will only
get as far as {\Kepler}.  Dynamical concepts are formulated on the
basis of what we can set up, control, and measure.
\eq

\section{08 April 1999, \ ``Delayed Entanglement''}

When I first entered the University of Texas---coming out for the
first time from my bucolic upbringing in the small town of Cuero,
TX---one of the things that impressed me most was the level of
intellectualism I found in the graffiti on the bathroom walls.  A
popular graffito at the time was:  ``God is love.  Love is blind. But
Ray Charles is blind.  So Ray Charles is God?''  (Ray Charles, in
case you don't know, is a famous blues singer and piano player.)
Despite the miscarriage of the syllogism, this was still light years
ahead of what I normally read on the bathroom walls in my old home
town!

Today while I was having a look at your new paper, I couldn't help
but think about a new bathroom syllogism (i.e., an equally miscarried
one) that I've been playing with for the last few weeks.  ``Knowledge
is power.  Power is a resource.  But entanglement is a resource.  So
entanglement is knowledge?''  OK, as advertised, there's no syllogism
here.  But the lesson is sound:  ``Entanglement is a kind of
knowledge and that knowledge can be a resource.''

The overall lesson of course is that knowledge is a resource.  Actual
entanglement is only surrogate to that.  And that is what your
example brings out nice and clearly.  Nonclassical correlations can
arise when one has a {\it nonclassical\/} state of knowledge, be that
knowledge present entanglement or instead delayed entanglement.

Let me just make a few almost trivial comments.

2)  I don't really like the sentence you end Section 2 with.  You say
``Alice and Bob find experimentally that each one of the four subsets
consisted of maximally entangled pairs.''  They don't really find
that; they find post-selected measurement statistics consistent with
measurements statistics that could have arisen from entangled pairs.
I think there is an important point that shouldn't be misrepresented
here. There is never entanglement in the more common sense, and yet
there is a {\Bell} inequality violation.  ``Nonlocality without
Entanglement'' --- where've we heard that before?  It's a completely
differently flavored example, but the slogan still applies here (a
little at least).

That's about it.  As a general comment, I didn't find the example
very surprising.  But I've been your student for a long time.  You've
already taught me to ``clearly understand quantum mechanics and
firmly believe in its correctness.''  Nevertheless I think the
example will still turn some heads---Richard {\Jozsa} (who doesn't in
his heart really believe QM) might be one of them.  It's crisp and
clean, and therefore not easily forgotten.  It's good pedagogy.

\section{10 April 1999, \ ``My Turn''}

A few weeks ago you pointed out a typo in my CV.  Now it is my turn
to return the favor.  In the publication list you sent me, you list
the following article:  A. {\Peres}, ``{\Schroedinger}'s Immortal
Cat,'' Found.\ Phys.\ {\bf 17} (1988) 57.  That should be Vol.\ 18
instead.

The title intrigues me, but I have no clue what the article is about.
The way I came across it was that I read your article ``On
Quantum-Mechanical Automata'' (commenting on a paper by David
{\AlbertD}), and was intrigued by your phrase, ``There are, however,
some distinguished theorists who claim that irreversibility \ldots\
is not fundamental \ldots, in brief, that human ignorance or
weakness have no role in physics.  Although I may disagree on this
point [15] I shall tentatively adopt this approach just for the sake
of argument \ldots'' That sent me to your CV to see what the
mysterious Ref.~[15] was.  It was an article titled ``The
Physicist's Role in Physical Laws.''  And I thought, while I'm in
the library, I might as well see what other (previously unseen)
{\Peres} tidbits I can find. Hence my finding that I could not find
the article mentioned above.

In any case, I liked your anti-{\AlbertD} article very much.  It
seems to me that it (along with your 1982 paper with {\Zurek})
contains a point that is in need of much more exploration.
(Particularly the point you make in the last three paragraphs of the
paper.)  In my heart I feel that you are on the right track with
that, but you have only scratched the surface.  In that connection,
by the way, I have given your new appendix to RQM a first reading.
But I want to withhold my comments until I can read it again more
carefully, and have some time to construct a proper commentary.
There is something about it that troubles me, something that is
incongruous with the anti-{\AlbertD} article I just mentioned.  I
will need some time to articulate that.

About the mysterious Ref.~[15], the title itself almost sent me into
Pavlovian salivation.  So I dropped everything to find it.  I made a
copy of it for my files, but still I came back from the library with
my head hung low.  It was that---certainly through no fault of your
own---the paper just didn't say what I had wanted it to say!!  I
can't have all my wishes, can I\@?  Of course I very much like part
of the sentiment you express at the beginning, ``The physicist's role
as an observer \ldots\ appear[s] \ldots\ in the formulation of the
physical laws themselves.''  A large part of the foundation of my
research program (as you know by now) is to give some substance to
just this thought. But my tack on it is that the structure of
quantum mechanics itself is just this sort of thing:

``In the beginning God created the heaven and the earth.  And the
earth was without form, and void \ldots\ [Day 1], [Day 2], [Day 3],
[Day 4], [Day 5] \ldots\ And God saw everything that he had made, and
behold, it was very good.  And there was evening and there was
morning, a sixth day.  Thus the heavens and the earth were finished,
and all the host of them.'' But in all that, there was still no
science.  For science must of its nature be a creation of man---it
is the expression of man's attempt to be less surprised by this
God-given world with each succeeding day.  So the society of man
slowly but surely set out to discover and form physical laws.
Eventually a wonderful fact came to light:  information gathering
about the world is not without a cost.  When I learn something about
an object, you are forced to revise (toward the direction of more
ignorance) what you could have said of it.  And yet from our
comparing our notes (and our comparing our notes with those of the
larger community) we are still in a position of constructing a
scientific theory of the things we observe.  The world is volatile
to our information gathering, but not so volatile that we have not
been able to construct a successful theory of it.  The {\it
speculation\/} is that quantum theory is the unique expression of
that happy circumstance:  it is the best we can say in a world where
information gathering and disturbance go hand in hand.

But I must get back to more serious things:  my own JMO article needs
finishing.

\section{10 May 1999, \ ``Eat My Words''}

Also I found one very interesting piece of history that I guess I had
forgotten.  Karl {\Kraus} in his paper, ``General State Changes in
Quantum Theory'' [Ann.\ Phys.\ {\bf 64}, 311--335 (1971)], starts the
introduction with:
\bq
The state of quantum systems may be changed by external
interventions.\footnote{Familiar examples are the state change due to
an ideal measurement (``reduction of the wave packet''), or the state
change produced by the action of an external field.  In the
{\Heisenberg} picture used here, the state is constant in the
absence of such external interventions.}  Haag and
Kastler\footnote{R.~Haag and D.~Kastler, J.~Math.\ Phys.\ {\bf 5}
(1964), 848.} have discussed a particular case of such state changes
in algebraic quantum theory. They called them operations.  The same
notion has been used by Pool\footnote{J.~C.~T. Pool, Comm.\ Math.\
Phys.\ {\bf 9} (1968), 118.} in a lattice-theoretic framework.
\eq

You'll note in particular that he used the word ``intervention'' too.
I really like that word much more than ``measurement.''  (I hope also
you've reinstated the word ``chronicle'' into your latest version.)

\section{19 May 1999, \ ``I Finally Decohere''}

\bap
In the next section of this article, the notion of measurement will
be extended to a more general one: an {\it intervention\/}. This is
the interface of the classical (Boolean) world and the quantum world.
\eap

Do you really mean to say that the notion of intervention is ``more
general'' than measurement?  Or, instead that it is what we have
always had, but now stated in a crisper language? The word
``measurement'' is one that fits more within the classical world
view: it seems to me, part of your great contribution is to point out
that limitation in a nice manner.

On a different point, what should the cryptic use of the word
``Boolean'' be taken to mean in the reader's mind?  I like its
insertion there, but I'm not sure what to make of it.

\bap
{\Bell} pointed out that the notion of measurement, or observation,
was logically inconsistent in a genuine quantum world.
\eap

I don't like this phraseology.  It is my working assumption that the
world really is {\it genuinely quantum}.  But by that I mean that the
best description I can make of the world's workings must make use of
the quantum formalism.  What you mean however (I think) is that
{\Bell} was considering a kind of world where the wave function is
reified so to speak:  a world in which there is an overall wave
function controlled by unitary dynamics that is a state of nature,
not a state of knowledge.

I think a simple substitution of a different adjective than
``genuine'' would do the trick for me.

\bap
The probabilities of the various outcomes of an intervention can be
predicted by using a suitable theory, such as quantum theory.
\eap

These ``outcomes'' are the ``creations'' that you speak of later?
That is probably a good way to put it.  David {\Mermin} once asked
something like:  ``Where there is subjective probability, there is
ignorance.  What are quantum probabilities to signify our ignorance
of if the ignorance cannot be of objective properties?''  The
outcomes of our interventions, I say!

\bap
Note that the ``detector clicks'' are the only real thing we have to
consider. Their probabilities are objective numbers and are Lorentz
invariant. On the other hand, wave functions and operators are
nothing more than abstract symbols. They are convenient mathematical
concepts, useful for computing quantum probabilities, but they have
no real existence in Nature.
\eap

I very much agree with the first sentence of this.  But I don't like
(or agree with) the distinction you draw thereafter. If wave
functions or density operators are not objective entities, then
neither are the probabilities we calculate from them through
{\Born}'s rule. This is the Bayesian point of view: probabilities are
subjective numbers to aid in our decision making and gambling (in the
case of physics, gambling with the world).  Period.  It is
inconsistent to take one as objective but not the other.  And there
are perfectly adequate arguments to show that neither is objective,
i.e., agent independent.

\bap
Our partial ignorance is not a sign of weakness. It is fundamental.
If everything were known, acquisition of information would be a
meaningless concept.
\eap

I like this point very much.  In fact I had wanted to have a
discussion along these lines with Rolf {\Landauer} before he fell
ill. ``What can you mean by your phrase `Information is Physical'?
Was information physical before there were physicists doing
physics?  I get the feeling that that is not what you mean.  But
then why call what you are talking about `information'?'' I never
got the chance to have the discussion.

\bap
Clearly, quantum jumps are not dynamical processes. They result from
the introduction of an apparatus, followed by its deletion or that of
another subsystem. The jump in the quantum state occurs because there
are abrupt changes in our way of delimiting the object we consider as
the quantum system under study.
\eap

Regardless of the formalism you've introduced and the discussion of
decoherence you've given up to this point, I still don't believe
you've given a convincing argument for this point.  From my point of
view, the random jump that we find upon measurement is simply not
contained within the quantum formalism.  This is not a blemish to the
theory:  it is just a fact.

\bap
Consider the evolution of the quantum state in the Lorentz frame
where intervention $\sf A$ is the first one to occur and has outcome
$\mu$, and $\sf B$ is the second one, with outcome $\nu$. Between
these two events, nothing actually happens in the real world. It is
only in our mathematical calculations that there is a deterministic
evolution of the state of the quantum system. This evolution is {\it
not\/} a physical process.
\eap

I like this very much.  It reminds me of a quote of {\Pauli} that I
often repeat.  In a 1947 letter to {\Fierz}, he wrote:
\bq
\noindent
Something only really happens when an observation is being made
[\ldots].  Between the observations nothing at all happens, only time
has, `in the interval,' irreversibly progressed on the mathematical
papers!
\eq
Also it reminds me of quote of {\Heisenberg} that I just read
recently. Henry {\Stapp} in a recent paper quotes him as saying:
\bq
Since through the observation our knowledge of the system has changed
discontinuously, its mathematical representation also has undergone
the discontinuous change, and we speak of a `quantum jump'.

A real difficulty in understanding the interpretation occurs when one
asks the famous question: But what happens `really' in an atomic
event?

If we want to describe what happens in an atomic event, we have to
realize that the word `happens' can apply only to the observation,
not to the state of affairs between the two observations.
\eq

\bap
What distinguishes the intermediate evolution between interventions
from that due to the intervention itself is that the latter has
unpredictable output parameters, for example which one of the
detectors ``clicks,'' thereby starting a new chapter in the history
of the quantum system.
\eap

Did you want to say ``history'' or ``record'' here?  I like history
better---it has a nicer flow---but you had earlier said that you
wouldn't use the word history.

\section{02 June 1999, \ ``9906006''}

In all, I'm not sure what to make of {\Meyer}'s result. Strictly
speaking---without placing an interpretation on the mathematics---it
is that all the orthonormal triads in $\mathbf{R}^3$ with rational
Cartesian coordinates can be consistently colored.  One could ask a
similar question about rational-angled spherical coordinates, and
maybe the result changes drastically, but {\Meyer} didn't do that.
Or one might consider sets for which the $\theta$ and $\phi$ for the
spherical coordinates are exclusively rational fractions of $\pi$,
but {\Meyer} didn't do that either.  Instead, he wants to derive a
{\it physical\/} conclusion from the result based on the fact that
the rationals are dense in the reals:  thus the issue of experimental
verifiability.

But I'm not too sure what it even means to speak of an experimental
test of the {\Kochen}-{\Specker} theorem.  We can never measure all
these overlapping triads simultaneously.  (Didn't {\Cabello} and
{\GarciaAlcaine} once address the issue of the
``experimentalization'' of KS? --- that memory just came back to me.)

In my mind, if there's anything interesting to be learned here it is
that the character of quantum mechanics might change drastically if
we were to imagine the Hilbert spaces of the theory to be vector
spaces over the field of (complex) rational numbers, rather than the
full set of complex numbers.  Several in the past ({\Adler},
{\Wootters}, Stuekelberg, etc.) have pointed out oddities in the
theory if the base field were taken to be the reals instead of the
complex numbers. But I think this---{\Meyer}'s paper---may be the
first to consider the oddities of the theory if posed over the
rationals.

\section{26 August 1999, \ ``A Contextual Thought''}

The following thought just popped to mind; let me pass it by you.
Would you call our ``nonlocality without entanglement'' result a
``contextuality'' result in the tradition of the BKS idea?  We didn't
say that in the paper, but it seems that it has a bit of the flavor.
{\Charlie} instead wrote the following in the paper:

\bq
In what sense is a locally immeasurable set of states ``nonlocal?''
Surely not in the usual sense of exhibiting phenomena inexplicable by
any local hidden variable (LHV) model. Because the $\psi_i$ are all
product states, it suffices to take the local states $\alpha_i$ and
$\beta_i$, on Alice's and Bob's side respectively, as the local
hidden variables.  The standard laws of quantum mechanics (e.g.
Malus' law), applied separately to Alice's and Bob's subsystems, can
then explain any local measurement statistics that may be observed.
However, an essential feature of classical mechanics, not usually
mentioned in LHV discussions, is the fact that variables
corresponding to real physical properties are {\em not\/} hidden, but
in principle measurable.  In other words, classical mechanical
systems admit a description in terms of local {\em unhidden\/}
variables.  The locally immeasurable sets of quantum states we
describe here are nonlocal in the sense that, if we believe quantum
mechanics, there is no local unhidden variable model of their
behavior.  Thus a measurement of the whole can reveal more
information about the system's state than any sequence of classically
coordinated measurements of the parts.
\eq

It seems that one can explain nonlocality without entanglement with a
local hidden variable model \ldots\ but only via hidden variables
that are necessarily disturbed by the measurement process.  That's
something like a contextual hidden variable theory, isn't it?  I
wonder how the {\Clifton}-{\Kent} model fares with respect to this?

Do you think there's anything to this?

\section{23 September 1999, \ ``End of the Day''}

\bap
Yesterday in my mail I got two related items. One was an e-mail from
the PRA editor, asking my opinion on proposed changes in PACS:
03.65.Bz, which includes now too many papers, would be split into 10
subsections. One of them is ``Foundations and interpretations of
QM''. I objected to an explicit mention of ``interpretations'' and
Bernd Crasemann agreed with me. The other item was {\sl Physics
Today\/} of August, with a paper by our consistent historian friends
Bob and Roland (I have not yet read it).

This follows a couple of papers on Bohmian trajectories, and led me
to wonder whether the PT editors have lost their mind and I (or we)
should try to educate our colleagues and explain that quantum
mechanics needs no interpretation.
\eap

You've heard the story that Rudolf {\Peierls} refused to use the
words ``Copenhagen interpretation.''  For according to him, ``the
Copenhagen interpretation IS quantum mechanics.''

I haven't read the {\Griffiths}/{\Omnes} article yet.  But you do
know that this subject has been getting under my skin lately.  Let me
read it over the weekend and see what I think \ldots\ though I'm
almost sure what my conclusions will be.  What about we perhaps
write a letter to the editor of PT concerning it \ldots\ if we turn
out to have similar opinions.  Or were you thinking of a larger
project?

Now to something serious.  I know you know {\Wigner}'s marvelous
theorem on symmetries; I think you described it in your book.  If we
have a bijection from unit vectors to unit vectors on a Hilbert
space that preserves all inner products, then that mapping must
either be a unitary linear map or an antiunitary antilinear map.  If
we further suppose that mappings are continuously connected to the
identity, then we are left with the unitaries.

Here is my question.  Can something of the same flavor be proven
within classical physics?  What I am thinking of is this.  Start with
a phase space, and consider the space of all normalized probability
densities over that space.  (These functions all live within the
positive cone of the vector space of ALL square-integrable functions
over the phase space.)  Let us suppose that whatever dynamics is, it
must be a bijection on the probability densities that preserves
overlaps.  Will that be enough to give a {\Wigner}-like theorem for
classical physics:  i.e., that all allowed evolutions must satisfy a
Liouville equation.

Have you seen this proven anywhere rigorously?  If so, can you give
me a reference?

\section{24 September 1999, \ ``More {\Peierls}''}

\bap
Have you a reference for that phrase? I asked Ady {\Mann} who was
{\Peierls}'s post-doc, and he promised to try to find.
\eap

That's a good question:  what is the origin of that phrase?  I know
that I have seen it in print, but presently I don't recall from
where.  The first time I used it was in a letter to my friend {\Greg}
{\Comer} on 26 July 1993.  I know that because I did a quick look-up
yesterday:  I was thinking for a moment that {\Rosenfeld} had said
it, and I wanted to make sure I was correct.  To my surprise I found
{\Peierls}.  (And by the way, I just discovered that I misspelled
{\Peierls} yesterday; I'm probably the fault of your doing the same.)

\section{26 September 1999, \ ``{\Wigner}''}

\bap
I have no immediate answer about {\Wigner}'s theorem for Liouville
functions. The proof given by {\Wigner} is for discrete Hilbert
spaces, and Liouville's equation has a continuous spectrum. I assume
that you are familiar with Koopman's theorem (pp~317-319 in my
book). I'll think about it, but I don't see the physical motivation.
\eap

Essentially what I am asking for is a converse to Koopman's theorem.
The physical motivation is simple:  it would mean that {\Newton}'s
laws are almost content free \ldots\ that they almost couldn't have
been anything else once one has established that what classical
physics is about is the evolution of probability distributions over
phase space $(x,p)$.  To say that one knows a system is isolated and
evolving according to some intrinsic dynamical law is to say that
one is neither losing information about the system nor gaining
information about the system.  A way of mathematizing that would
seem to be simply the statement that the overlaps of any two unknown
probability distributions should not change with time.

Maybe this sort of question is tackled in Arnold's book on classical
mechanics?  Or perhaps in some of Prigogine's work?

Anyway, that's the idea.

\section{27 September 1999, \ ``contra-Koopman''}

Many thanks for sending the further references about the Liouvillian.

\bap
Anyway, your idea of an anti-Koopman theorem is probably correct from
the point of view of mathematics.  I am not yet convinced by the
physical motivation, so that you should make it clearer.
\eap

I don't presently know how to say it much clearer than I already
have.  I will work on that.  In the mean time, though, let me get
 {\Asher} {\Peres} to say a few words about the subject:

\bq
\noindent
\ldots\ a quantum state is {\it not\/} the analog of a point in the
classical phase space.  The classical analog of a quantum state is a
{\it Liouville probability density}.  [p.~347, {\sl Quantum Theory:\
Concepts and Methods}]
\eq
and
\bq
\noindent
Since there are analogies between classical and quantum mechanics,
why not try to use quantum methods for solving classical problems.''
[p.~317, {\sl Quantum Theory:\ Concepts and Methods}]
\eq

It seems to me that we just have potentially (if all works out) a
knew way of thinking about Hamiltonian dynamics.  It comes not so
much from the ``mysterious'' (to me at least) Principle of Least
Action, but rather from the requirement that statistical
distinguishability never increase.

While I'm quoting from great men, let me also call Max {\Born} to my
aid until I can say things better myself.

In his lecture ``In Memory of {\Einstein}'' he writes:
\bq
{\Einstein}, however, stuck to his opinion.  \ldots\ \ It was in
fact a matter of a fundamental difference in our views of nature.
\ldots\ \ For years afterwards I kept turning over in my mind the
philosophy which was behind my theory, and then gave a very brief
summary of it in the Festschrift in honor of {\Heisenberg}'s 60th
birthday.  What it boils down to is that scientific forecasts do not
refer directly to `reality,' but to our knowledge of reality.  This
means that the so-called `laws of nature' allow us to draw
conclusions from our limited approximate knowledge at the moment on
a future situation which, of course, can also be only approximately
described.
\eq
And more pertinently, in his essay ``Is Classical Mechanics In Fact
Deterministic?'' he writes:
\bq
To summarize, we may say that it is not the introduction of the
indeterministic statistical description which places quantum
mechanics apart from classical mechanics, but other features, above
all the concept of the probability density as the square of a
probability amplitude $P=|\Psi|^2$; the phenomenon of probability
interference results from this, and therefore it is impossible to
apply without modification the idea of an `object' to the mass
particles of physics: the concept of physical reality must be
revised.  This however is beyond the scope of these elementary
considerations.
\eq

What I'm thinking is that a {\Wigner}-like theorem is part of the
``indeterministic statistical description'' that is common to both
classical and quantum mechanics.  It is just another way of helping
to clarify where the two theories diverge.

\section{18 December 1999, \ ``KS Crypto''}

\bap
We are working on qutrit cryptography -- it's not completely silly
and quite amusing, because related to KS theorem.
\eap

That sounds quite interesting actually.  I look forward to seeing
your results.  I have often wondered whether there is an interesting
``phase transition'' in the information vs.\ disturbance curves when
one goes from doing q-crypto with a colorable set of bases to doing
q-crypto with a noncolorable set.  Just as a first guess, one might
suspect that a noncolorable set of vectors would be intrinsically
more secure:  though with so many bases in the protocol the data
rates would be quite atrocious.

\section{19 December 1999, \ ``State of Confusion''}

Many thanks for the PostScript file and the note.  However, now I'm
really confused.  I don't understand the motivation for considering
tri-colorings, etc., etc.  Yesterday when I wrote you, I thought you
were talking about the standard KS colorings.  No need to bother
with me any more:  I'll just wait until your paper appears on {\tt
quant-ph} or if you send a review copy to me beforehand.

\section{27 December 1999, \ ``Fuchs, but not {\Laflamme}''}

About Steven van {\Enk}'s comment on David {\Meyer}:  it is about his
paper on quantum games.  Steven's comment, as you will read, is
pretty much on the mark.  As I recall, he made three points.  1)
{\Meyer} restricts himself to two-level systems, so in principle
there is nothing quantum mechanical about his example. Simple
two-level systems can always be simulated by an underlying
hidden-variable theory.  2) {\Meyer}'s example could have just as
well been constructed using classical wave mechanics.  In quantum
computing, of course, there is the extra issue of
scaling---classical wave computers have an exponential scaling of
resources (energy, space) that quantum computers don't---but for
{\Meyer}'s quantum game that is not an issue.  {\Meyer} claims to
have shown something interesting at the simple qubit level.  3)  This
one, I think, is the most devastating and at the same time the most
obvious.  In {\Meyer}'s game, one player must abide by the rules of
classical physics while the other one has the full freedom of
quantum mechanics.  It is no wonder that one player can win all the
time!  There is nothing here that is not in ANY two-player game
(regardless of physics):  if one player obeys the rules, while the
other CHEATS then of course the latter will always win.  van {\Enk}
shows an explicit example of this in the classical context:  it is
precisely {\Meyer}'s game, but with one player having the extra power
to stand a coin on its edge, rather than just flipping it.  That
player, of course, can always win \ldots\ and there is nothing
quantum mechanical about this example.

One interesting tidbit about van {\Enk}'s comment and {\Meyer}'s
rebuttal.  {\Meyer} thanks both Raymond {\Laflamme} and me for
discussions, while van {\Enk} simply thanks me.  Steven, for a while,
toyed with the idea of writing, ``I thank Chris Fuchs, but not
Raymond {\Laflamme}, for discussions.''  (That made me chuckle,
though he certainly didn't mean it as an insult.  Steven doesn't
even know {\Laflamme}.)  Anyway, while they both thank me, I am on
Steven's side with this one.

\section{9 January 2000, \ ``KS Cryptoland''}

I had a chance to give your paper with Helle a brief ``once over.''
Let me just make a couple of comments.

1)  I very much liked the mutually unbiased basis example.  It has
so much pretty symmetry, it makes one wonder if it might not be
(easily) amenable to a full analysis for the information-disturbance
tradeoff as we did in FGGNP (and I believe Dagmar did for the
three-basis protocol).

2)  However I was a little dissatisfied with your so-called KS
analogue.  I think it would be much more interesting to look at how
well a real KS example fairs.  By that, I mean something like the 33
vector, 16 basis example that you found, or the 31 vector, 17 basis
example that Conway and {\Kochen} found.  In those cases, there can
be no underlying noncontextual hidden variable carrying the key
information between Alice and Bob.  What is the connection between
that and the information-disturbance tradeoff that quantum mechanics
supplies?  Is there any connection?  I think these questions are
quite interesting from a fundamental point of view.

Would you be interested in collaborating on questions of this sort
during my visit to Haifa?  Perhaps also with Helle or whoever else
is interested?

Tomorrow I go to Albuquerque for my weekly collaboration with
{\Caves}.  Tuesday, however, I plan to start searching for tickets
for my visit to you.

\section{09 March 2000, \ ``{\Petra}''}

\bap
Curiously, there is another student who is going to work with me by
remote control. Her name is {\Petra} Scudo, and she just got a MSc
from Pavia. She has family in Israel and came to visit me just before
your visit (she came with her mother and little sister). Chiara
likes her and promised to help guiding her until she arrives here in
the fall. The project I have in mind for her involves data analysis,
probably by MaxEnt methods. Where can she read about Bayesian
statistics? Once you mentioned to me a good book but I don't
remember which one.
\eap

The very best reference is:
\bq\noindent
E.~T. {\Jaynes}, {\sl Probability Theory:~The Logic of Science}.
This already-huge book was, very unfortunately, never finished before
Prof.~{\Jaynes}' death.  An incomplete draft dated 13 May 1996 is
available on the World Wide Web at {\tt http://bayes.wustl.edu/}.
\eq
Beside that though, she can find a lot of information in the
following series of books:
\bq\noindent
R.~D. Levine and M.~Tribus, eds., {\sl The Maximum Entropy
Formalism}, (MIT Press, Cambridge, MA, 1979); C.~R. Smith and W.~T.
Grandy, Jr., eds., {\sl Maximum-Entropy and Bayesian Methods in
Inverse Problems}, (D.~Reidel, Dordrecht, 1985); J.~H. Justice, ed.,
{\sl Maximum Entropy and Bayesian Methods in Applied Statistics},
(Cambridge U. Press, Cambridge, 1986); J.~Skilling, ed., {\sl
Maximum Entropy and Bayesian Methods:\ Cambridge, England, 1988},
(Kluwer, Dordrecht, 1989); P.~F. Foug\`ere, ed., {\sl Maximum Entropy
and Bayesian Methods:\ Dartmouth, U.S.A., 1989}, (Kluwer, Dordrecht,
1990); B.~Buck and V.~A. Macaulay, eds., {\sl Maximum Entropy in
Action:\ A Collection of Expository Essays}, (Clarendon Press,
Oxford, 1991); W.~T. Grandy, Jr.~and L.~H. Schick, eds., {\sl Maximum
Entropy and Bayesian Methods:\ Laramie, Wyoming, 1990}, (Kluwer,
Dordrecht, 1991); C.~R. Smith, G.~J. Erickson, and P.~O. Neudorfer,
{\sl Maximum Entropy and Bayesian Methods:\ Seattle, 1991}, (Kluwer,
Dordrecht, 1992); A.~Mohammad-Djafari and G.~Demoment, eds., {\sl
Maximum Entropy and Bayesian Methods:\ Paris, France, 1992}, (Kluwer,
Dordrecht, 1993); G.~R. Heidbreder, {\sl Maximum Entropy and Bayesian
Methods:\ Santa Barbara, California, U.S.A., 1993}, (Kluwer,
Dordrecht, 1996); J.~Skilling and S.~Sibisi, {\sl Maximum Entropy and
Bayesian Methods:\ Cambridge, England, 1994}, (Kluwer, Dordrecht,
1996); K.~M. Hanson and R.~N. Silver, {\sl Maximum Entropy and
Bayesian Methods:\ Santa Fe, New Mexico, U.S.A., 1995}, (Kluwer,
Dordrecht, 1996).
\eq

\section{07 April 2000, \ ``Why Fidelity?''}

I finally(!)\ had a chance to read your note ``why fidelity?''  Has
Dagmar responded to you?  I would like to hear her thoughts.

\bap
It is customary to evaluate the quality of a quantum transmission, or
cloning, or other set of operations by the ``fidelity'' of the
result. What is the justification for that, other than the
simplicity of the formula? It seems to me that the use of fidelity
as a criterion is completely arbitrary.
\eap

I think you are on the mark here:  it's pretty much arbitrary as far
as I can tell too.  It seems to me that distinguishability measures
or closeness measures should always be motivated by the context of
the problem when one can do that.  If a problem (like optimal
cloning) doesn't lead uniquely to one measure over another, then
perhaps one should reexamine one's motivation for looking at the
problem in the first place.

In our FP96 paper, we considered mutual information as a measure for
quantifying what Eve wanted and fidelity for quantifying how well
she was hidden.  I think the latter of these wasn't too bad.  The
fidelity in this case corresponds exactly to the probability of
Alice and Bob detecting Eve.  On the other hand, mutual information
may have been a little ad hoc on our part.  What if Eve had wanted
to optimize her ability to guess the identity of the signal and not
just sharpen the posterior probabilities as much as possible?  Then
she would have used Helstrom's techniques instead of {\Holevo}'s.
What should she really do for the somewhat related cryptographic
problem.  Probably still something different.  (I'm sure Tal {\Mor}
has two or three different answers for that!)

\bap
In general, I'll want to have a classical criterion for any process,
even if quantum notions are used at intermediate steps. As Niels
{\Bohr} told us long ago, the account of any experimental procedure
has to be done in a classical language, because we should tell
``what we have done and what we have learned''.
\eap
I like this philosophy.  But even classically there is no unique
solution independent of the context. And thus I don't see the
following as cut and dried as you do:
\bap
Thus, if the input is a discrete set of signals, the natural measure
for quality of transmission (or of identification) is the resulting
mutual information.
\eap

There has been some work in the classical information theory
literature trying to (rigorously) justify the mutual information as
an interesting quantity outside of the channel capacity scenario.
But none has been completely successful in my eyes.  Here are a
couple of references.
\begin{enumerate}
\item
D.~V. Lindley, ``On a measure of the information provided by an
experiment,'' {\em The Annals of Mathematical Statistics}, vol.~27,
pp.~986--1005, 1956.
\item
``A new interpretation of Information Rate'' by J.~L. Kelly, Jr.\
{\em The Bell System Technical Journal}, page 917, July 1956.
\end{enumerate}

\section{22 April 2000, \ ``Cats and Kitties''}

Recently you wrote me:
\bap
Another subject about which I had been thinking is our old friend the
Cat. Of course there is no pure  $|\mbox{living}\rangle$  state. The
initial state is represented by a density matrix, or rather a class
of density matrices (if we remove one atom from the cat, it still is
a living cat). The ``complementary'' set of matrices -- whatever this
means -- represents dead cats. How can we clone a cat? We don't need
to reproduce all the details. Only a restricted set of
characteristics are important. I don't know how to quantify that.
\eap

Let me just ramble a bit about this.  I think the answer to your
question, at least partially, lies in our old ``no-broadcasting''
paper [PRL {\bf 76}, 2818 (1996), also see {\Lindblad}, Lett.\ Math.\
Phys.\ {\bf 47}, 189 (1999)].  There the question was, when can an
``unknown'' density operator be cloned in a suitably general
way---i.e., the kind of way one usually thinks of when one is
thinking of a classical record being made.  The answer was---perhaps
not surprisingly (though it was really hard to work out)---if and
only if the ``unknown'' density operator is drawn from a known
commuting set.

Once that is in hand, it seems to me, the main question becomes
this.  Why is it that most states that we are willing to ascribe to
a class of living beings are commuting, or at least, awfully close
to commuting?  Wojciech---I'm sure---would like to think that it has
something to do with some kind of environmentally induced
superselection rules (and thus depends crucially on the
characteristics of the Hamiltonia around us).  But I suspect it has
almost nothing to do with that.  Instead I would be willing to bet
it's this.  Whenever we have a system so large and so complex that
we would be willing to call it ``living,'' we will most surely know
so little about it that our density operator assignment will be very
close to being proportional to the identity operator.  All such
states are very nearly commuting, and therefore, very nearly
broadcastable.

That is the direction I would like to flesh out more fully at some
point in the future.

By the way, let me give you another tip for a paper that has
something to say which I perceive as (very) loosely connected to
this issue.  P.~{\Busch} and J.~Singh, ``L\"uders Theorem for Unsharp
Quantum Measurements,'' Phys.\ Lett.\ A 249, 10 (1998).  It's kind of
trivial result, but you might enjoy it in a way.  (I did.)

Finally, let me send you the little review I wrote on {\Wigner}'s
thoughts on reproduction and cloning.  It is published at Dan
{\Gottesman}'s Quick Reviews in Quantum Computation and Information
webpage: {\tt http://quickreviews.org/qc/}.  [See note to Daniel
{\Gottesman}, dated 3 November 1998.]

\section{24 April 2000, \ ``Fuzzy Cats''}

Thanks for the long note.  I will have a thorough look at it
tomorrow.  Most of today I will be preparing and then giving a
lecture at LANL.  Let me just make a quick comment to an early
paragraph.

\bap
You wrote that living things would be represented by commuting
density operators, and mentioned environmentally induced
superselection rules (what's that?).
\eap
I mentioned that Wojciech {\Zurek} would say that it had something to
do with environmentally induced superselection rules.  I certainly
didn't endorse that, as I made clear.  So the two ideas shouldn't be
lumped back together.  In any case, you ask what is EISR about? Only
Wojciech's main research program in quantum foundations for the last
20 years!  (See PRD {\bf 26}, 1862 (1982) for the phrase, and Phil.\
Trans.\ R. Soc.\ Lond.\ A {\bf 356}, 1793 (1998) for its latest
mutation.)

\bap
Here I must disagree: take a living cat, and translate it by 0.1
nanometer. It's still a living cat and is not orthogonal to the
preceding one.
\eap
I said ``commuting or almost commuting'' (the latter phrase being
something awaiting a suitable quantification).  I never said
anything about the states being orthogonal.

But I'll be back for a more thorough read of your note later.

\section{12 July 2000, \ ``Two Answers''}

You've asked me at least two questions in my backlog of email.
\bap
Were the conferences you attended interesting? Any new discovery?
\eap

Actually both were fairly interesting.  At Mykonos, I met one person
of real note, Arkady {\Plotnitsky}.  David {\Mermin} had told me
that I would enjoy his company and it was quite true.  He was a real
oddity at this meeting.  He is a Professor of English at Purdue
University but is absolutely enamored with {\Bohr}'s writings.  I
felt that his understanding of {\Bohr} was quite deep.  In case you
have seen him in print (he wrote a book titled {\sl
Complementarity\/}), I know that he doesn't look so good there.  I
couldn't understand a word of his book and gave up very quickly;
but, I know him enough now to know that that is a misimpression of
his real intellect.  I challenged him to write an essay comparing
and contrasting {\Bohr}'s and {\Pauli}'s ideas on the foundations of
quantum mechanics, and he has accepted the challenge.  So I am
enthusiastic to see the result of that.

Another person I took the time to get to know better in Mykonos was
Roland {\Omnes}.  I think that was quite fruitful too, though he
remains a little of an enigma to me.  His ideas seem strangely
closer to Copenhagen than his {\sl Physics Today\/} article with
{\Griffiths} reveals.

At Capri, I think the most exciting thing was a talk by Nicolas
{\Gisin}.  He seems to have found an interesting classical
``analogue'' to entanglement, and even to bound entanglement.  I
think his paper with Stefan {\Wolf} (which I believe is on {\tt
quant-ph}) is well worth studying.

\section{24 August 2000, \ ``Expenses, Reimbursements, and
Memories''}

Thank you so for sending me the {\Enz} article on {\Pauli}.  The fire
has not only caused me to retabulate my old properties.  It has also
made me painfully aware of how fragile all my thoughts are, and how
like anything associated with a computing machine they should be
backed up.  To that end, I have also been working very hard to
recall everything in my various file cabinets.  One of those
cabinets was devoted to what I call the Paulian idea.  I've done a
good bit of a reconstruction job on it, recollecting papers and the
like, but this time I'm making sure to make it more electronically
accessible.  In the next mail, I'll send you the fruits of that.  I
deem that the project is about $1/4$ of the way done.  When it is
done completely, I will look for a more permanent repository.

\section{01 November 2000, \ ``Catching Up''}

Thank you for your long letter.  I'm glad to hear that you have
family in Morristown; it never ceases to amaze me how small the
world is.

I have a funny anecdote to tell in that regard.  One morning during
my last week at Los Alamos, I was getting out of my car when I saw
that Stirling {\Colgate} was getting out of his just three further
down the row.  I knew who Stirling was, but I had never met him.  We
both started to walk toward the Theory Building.  His nose was
buried in a scientific paper, and he was walking quite slowly.  I
sped up to overtake him, but strangely he sped up too; he never
looked up from the paper.   Now walking beside him, I sped up once
again.  And so did he!  Finally, after an uncomfortable moment, he
said, ``If I walk beside you, I can read a little more easily.  You
steady me.'' I said, ``Oh, like a seeing-eye dog.''  We then walked
further, but I was uncomfortable with the situation.  Finally, I
couldn't help but break the silence.  I said, ``Funny, I was just in
a town last week where seeing-eye dogs are trained.''  Stirling said,
``Morristown, New Jersey.''  I said, ``Yeah, how did you know?!?!''
He said, ``My dad started that service.''  In shock, I said
``Really?''  He said, ``Well, my dad and Mrs.\ Eustice.''  So it is a
small world.

Thank you again for the offer of sending some books.  You will find
a willing recipient in me!  I'll place both my addresses below; you
can send them to either one you wish.

One of these days I will have to reply to {\Todd}. There is nowhere
{\it anywhere\/} in the Cathy-Erwin example where a {\it single\/}
system is given two distinct pure state assignments.  He is confused
on this point for the simple reason that he hasn't absorbed our main
point: Quantum states are states of knowledge, not states of nature.
One of these days I will break down and formulate a reply.

But now I've got to get to more serious things.

\chapter{Diary of a Carefully Worded Paper:
More Letters to  {\Asher} {\Peres}}

\begin{flushright}
\parbox{1.73in}{
``Partners in crime again?''\smallskip\\
\hspace*{\fill} --- \it CAF to AP \\
\hspace*{\fill} 29 September 1999 }
\end{flushright}

\section{Introduction}

Irving John {\Good} once titled a paper, ``46,656 Varieties of
Bayesians.'' With little stretch of the imagination, one can believe
that a similar paper on the interpretations of quantum mechanics
exists somewhere out there too.  And that may be true even though
46,656 is greater than the whole membership of the American Physical
Society! It is no wonder that taking a stand on quantum foundations
is treacherous business. Nevertheless, in the March 2000 issue of
{\sl Physics Today},  {\Asher} {\Peres} and I flouted the usual
rewards of a calm life and did just that. Our article was titled,
``Quantum Theory Needs No `Interpretation'.''

This chapter is about the evolution of that article and a later
``reply to critics.'' When I was an undergraduate at the University
of Texas, Bryce {\DeWitt} made a great impression on me by saying
something I have not forgotten: ``We learn mathematics so that we
don't have to {\it think\/} when we do physics.'' My paper with
{\Asher} contained no mathematics.  We had to think, and think very
hard. The fruits of that labor, for my part, were in a greatly
enhanced keenness for seeing through various foundational issues.
How I wanted to to share that newfound keenness with the reader! But
it was difficult, if not impossible, to convey such an aspect of the
paper-writing process in the very paper itself. I could not give the
reader as much as I would have liked.

I ask, ``What substitute can I give for the lucky gamble I put myself
through: that is, having had enough (unfounded) faith to go down one
path and then overcome enough trials to build an overpowering
confidence that the path taken was the {\it right\/} path?'' There
is no full-scale substitute, but here I wish to further the point of
view  {\Asher} and I put forward in those articles by opening the
curtains on the process of writing itself. After a reprint of the
final article and the ``reply to critics,'' I give a play-by-play
account of (predominantly) my side of things as the two articles were
hammered out.  In this chapter, I have opted to be unusually open,
revealing many letters that are only tangential to the actual words
in the articles:  they set the scene, and thus form part of the
realit{\it t\/}y behind the science that {\Herb} {\Bernstein} so
often emphasizes. The interest in this, however, is not only for the
historian of science, but for the physicist with the nagging feeling
that perhaps all is well with quantum mechanics after all: Look at
the trials we faced, and see the way we hammered. The deeper physics
will be found by taking quantum mechanics at face value, rather than
resorting to the tools of science fiction to slay an imagined
beast.\footnote{See note to David {\Mermin}, dated 23 July 2000}

\subsection{A. Quantum Theory Needs No `Interpretation'}

\begin{center}
Quantum Theory Needs No `Interpretation' \\
Christopher A. Fuchs and  {\Asher} {\Peres} \\
\small {\sl Physics Today\/} {\bf 53}(3), 70--71 (2000)
\end{center}

\bq
\small Recently there has been a spate of articles, reviews, and
letters in {\sc Physics Today} promoting various ``interpretations''
of quantum theory (see March 1998, page 42; April 1998, page 38;
February 1999, page 11; July 1999, page 51; and August 1999, page
26). Their running theme is that from the time of quantum theory's
emergence until the discovery of a particular interpretation, the
theory was in a crisis because its foundations were unsatisfactory
or even inconsistent. We are seriously concerned that the airing of
these opinions may lead some readers to a distorted view of the
validity of standard quantum mechanics. If quantum theory had been
in a crisis, experimenters would have informed us long ago!

Our purpose here is to explain the internal consistency of an
``interpretation without interpretation'' for quantum mechanics.
Nothing more is needed for using the theory and understanding its
nature. To begin, let us examine the role of experiment in science.
An experiment is an active intervention into the course of Nature:
We set up this or that experiment to see how Nature reacts. We have
learned something new when we can distill from the accumulated data
a compact description of all that was seen and an indication of
which further experiments will corroborate that description. This is
what science is about. If, from such a description, we can {\it
further\/} distill a model of a free-standing ``reality'' independent
of our interventions, then so much the better. Classical physics is
the ultimate example of such a model. However, there is no logical
necessity for a realistic worldview to always be obtainable. If the
world is such that we can never identify a reality independent of
our experimental activity, then we must be prepared for that, too.

The thread common to all the nonstandard ``interpretations'' is the
desire to create a new theory with features that correspond to some
reality independent of our potential experiments. But, trying to
fulfill a classical worldview by encumbering quantum mechanics with
hidden variables, multiple worlds, consistency rules, or spontaneous
collapse, without any improvement in its predictive power, only
gives the illusion of a better understanding. Contrary to those
desires, quantum theory does {\it not\/} describe physical reality.
What it does is provide an algorithm for computing {\it
probabilities\/} for the macroscopic events (``detector clicks'')
that are the consequences of our experimental interventions. This
strict definition of the scope of quantum theory is the only
interpretation ever needed, whether by experimenters or theorists.

Quantum probabilities, like all probabilities, are computed by using
any available information. This can include, but is not limited to
information about a system's preparation. The mathematical
instrument for turning the information into statistical predictions
is the probability rule postulated by Max {\Born}.$^1$ The
conclusiveness of {\Born}'s rule is known today to follow from a
theorem due to Andrew {\Gleason}.$^2$ It is enough to assume that
yes-no tests on a physical system are represented by projection
operators $P$, and that probabilities are additive over orthogonal
projectors. Then there exists a density matrix $\rho$ describing the
system such that the probability of a ``yes'' answer is ${\rm
tr}(\rho P)$. The compendium of probabilities represented by the
``quantum state'' $\rho$ captures everything that can meaningfully
be said about a physical system.

Here, it is essential to understand that the validity of the
statistical nature of quantum theory is not restricted to situations
where there are a large number of similar systems. Statistical
predictions do apply to single events. When we are told that the
probability of precipitation tomorrow is 35\%, there is only one
tomorrow. This tells us that it is advisable to carry an umbrella.
Probability theory is simply the quantitative formulation of how to
make rational decisions in the face of uncertainty.

We do not deny the possible existence of an objective reality
independent of what observers perceive. In particular, there is an
``effective'' reality in the limiting case of macroscopic phenomena
like detector clicks or planetary motion: Any observer who happens
to be present would acknowledge the objective occurrence of these
events. However, such a macroscopic description ignores most degrees
of freedom of the system and is necessarily incomplete. Can there
also be a ``microscopic reality'' where every detail is completely
described? No description of that kind can be given by quantum
theory, nor by any other reasonable theory. John {\Bell} formally
showed$^3$ that any objective theory giving experimental predictions
identical to those of quantum theory would necessarily be nonlocal.
It would eventually have to encompass everything in the universe,
including ourselves, and lead to bizarre self-referential logical
paradoxes. The latter are not in the realm of physics; experimental
physicists never need bother with them.

We have experimental evidence that quantum theory is successful in
the range from $10^{-10}$ to $10^{15}$ atomic radii; we have no
evidence that it is universally valid. Yet, it is legitimate to
attempt to extrapolate the theory beyond its present range, for
instance, when we probe particle interactions at superhigh energies,
or in astrophysical systems, including the entire universe. Indeed, a
common question is whether the universe has a wavefunction. There
are two ways to understand this. If this ``wavefunction of the
universe'' has to give a complete description of everything,
including ourselves, we again get the same meaningless paradoxes. On
the other hand, if we consider just a few collective degrees of
freedom, such as the radius of the universe, its mean density, total
baryon number, and so on, we can apply quantum theory only to these
degrees of freedom, which do not include ourselves and other
insignificant details. This is not essentially different from
quantizing the magnetic flux and the electric current in a SQUID
while ignoring the atomic details. For sure, we can manipulate a
SQUID more easily than we can manipulate the radius of the universe,
but there is no difference in principle.

Does quantum mechanics apply to the observer? Why would it not? To
be quantum mechanical is simply to be amenable to a quantum
description. Nothing in principle prevents us from quantizing a
colleague, say. Let us examine a concrete example: The observer is
Cathy (an experimental physicist) who enters her laboratory and
sends a photon through a beam splitter. If one of her detectors is
activated, it opens a box containing a piece of cake; the other
detector opens a box with a piece of fruit. Cathy's friend Erwin (a
theorist) stays outside the laboratory and computes Cathy's
wavefunction. According to him, she is in a 50/50 superposition of
states with some cake or some fruit in her stomach. There is nothing
wrong with that; this only represents his knowledge of Cathy. She
knows better. As soon as one detector was activated, her
wavefunction collapsed. Of course, nothing dramatic happened to her.
She just acquired the knowledge of the kind of food she could eat.
Some time later, Erwin peeks into the laboratory: Thereby he
acquires new knowledge, and the wavefunction he uses to describe
Cathy changes. From this example, it is clear that a wavefunction is
only a mathematical expression for evaluating probabilities and
depends on the knowledge of whoever is doing the computing.

Cathy's story inevitably raises the issue of reversibility; after
all, quantum dynamics is time-symmetric. Can Erwin undo the process
if he has {\it not yet\/} observed Cathy? In principle he can,
because the only information Erwin possesses is about the
consequences of his potential experiments, not about what is
``really there.'' If Erwin has performed no observation, then there
is no reason he cannot reverse Cathy's digestion and memories. Of
course, for that he would need complete control of all the
microscopic degrees of freedom of Cathy and her laboratory, but that
is a practical problem, not a fundamental one.

The peculiar nature of a quantum state as representing information
is strikingly illustrated by the quantum teleportation process.$^4$
In order to teleport a quantum state from one photon to another, the
sender (Alice) and the receiver (Bob) need to divide between them a
pair of photons in a standard entangled state. The experiment begins
when Alice receives another photon whose polarization state is
unknown to her but known to a third-party preparer. She performs a
measurement on her two photons-one from the original, entangled pair
and the other in a state unknown to her-and then sends Bob a
classical message of only two bits, instructing him how to reproduce
that unknown state on his photon. This economy of transmission
appears remarkable, because to completely specify the state of a
photon, namely one point in the {\Poincare} sphere, we need an
infinity of bits. However, this complete specification is not what is
transferred. The two bits of classical information serve only to
convert the preparer's information, from a description of the
original photon to a description of the one in Bob's possession. The
communication resource used up for doing that is the correlated pair
that was shared by Alice and Bob.

It is curious that some well-intentioned theorists are willing to
abandon the objective nature of physical ``observables,'' and yet
wish to retain the abstract quantum state as a surrogate reality.
There is a temptation to believe that every quantum system has a
wavefunction, even if the wavefunction is not explicitly known.
Apparently, the root of this temptation is that in classical
mechanics {\it phase space\/} points correspond to objective data,
whereas in quantum mechanics {\it Hilbert space\/} points correspond
to quantum states. This analogy is misleading: Attributing reality to
quantum states leads to a host of ``quantum paradoxes.'' These are
due solely to an incorrect interpretation of quantum theory. When
correctly used, quantum theory never yields two contradictory
answers to a well-posed question. In particular, no wavefunction
exists either before or after we conduct an experiment. Just as
classical cosmologists got used to the idea that there is no
``time'' before the big bang or after the big crunch, so too must we
be careful about using ``before'' and ``after'' in the quantum
context.

Quantum theory has been accused of incompleteness because it cannot
answer some questions that appear reasonable from the classical
point of view. For example, there is no way to ascertain whether a
single system is in a pure state or is part of an entangled
composite system. Furthermore, there is no dynamical description for
the ``collapse'' of the wavefunction. In both cases the theory gives
no answer because the wavefunction is not an objective entity.
Collapse is something that happens in our description of the system,
not to the system itself. Likewise, the time dependence of the
wavefunction does not represent the evolution of a physical system.
It only gives the evolution of our probabilities for the outcomes of
potential experiments on that system. This is the only meaning of the
wavefunction.

All this said, we would be the last to claim that the foundations of
quantum theory are not worth further scrutiny. For instance, it is
interesting to search for minimal sets of {\it physical\/}
assumptions that give rise to the theory. Also, it is not yet
understood how to combine quantum mechanics with gravitation, and
there may well be important insight to be gleaned there. However, to
make quantum mechanics a useful guide to the phenomena around us, we
need nothing more than the fully consistent theory we already have.
Quantum theory needs no ``interpretation.''

\bigskip\noindent{\bf References}\frenchspacing
\begin{enumerate}\itemsep -2pt
\item M. {\Born}, Zeits. Phys. {\bf37}, 863 (1926); {\bf38}, 803 (1926).

\item A. M. {\Gleason}, J. Math. Mech. {\bf6}, 885 (1957).

\item J. S. {\Bell}, Physics {\bf1}, 195 (1964).

\item C. H. {\Bennett}, G. {\Brassard}, C. {\Crepeau}, R. {\Jozsa}, A. {\Peres}, and
W. K. {\Wootters}, Phys. Rev. Letters {\bf70}, 1895 (1993).
\end{enumerate}
\eq

\subsection{B. Quantum Theory -- Interpretation,
Formulation, Inspiration}

\begin{center}
Quantum Theory -- Interpretation,
Formulation, Inspiration:\ Fuchs and {\Peres} Reply \\
Christopher A. Fuchs and  {\Asher} {\Peres} \\
\small {\sl Physics Today\/} {\bf 53}(9), 14, 90 (2000)
\end{center}

\bq
\small Like Paul {\Harris}, we had an old friend in Alexandria. His
name was {\Euclid}. When we asked him whether his famous books {\it
Elements\/} needed an interpretation, he answered categorically:
``Absolutely not! Geometry is an abstract formalism and all you can
demand of it is internal consistency. However, you may seek material
objects whose behavior mimics the theorems of geometry, and that
involves interpretation. For example, light rays might be considered
analogous to straight lines.'' More than 2000 years later, we met a
much younger friend in G\"ottingen and asked him eagerly,
``Bernhard, have you really found a new, improved interpretation of
Euclidean geometry?'' His answer was no less categorical:
``Riemannian geometry is not at all a new way of presenting
{\Euclid}'s work. It is a broader formalism, having Euclidean
geometry as a limiting case. Just reinterpreting Euclidean geometry
without introducing radically new features would have been an
illusion of progress. If some day people find that light rays do not
behave as {\Euclid}'s straight lines, my geodesics may turn out to
be a good description of them.''

We do not claim that contemporary quantum theory is the final word
for a description of nature; this should be clear from the last
paragraph of our ``Opinion'' essay. There may some day be
indications that our description of nature by means of vectors in a
complex linear space is insufficient to represent {\it
experimental\/} evidence. A more general theory may then be needed
to extend the present formalism. However, as Daniel {\Styer}
correctly points out, the various ``interpretations'' of quantum
mechanics cannot be distinguished through {\it experiments}. If they
could, they would not be new interpretations, but proposals for new
theories.

{\Styer} compares these ``interpretations'' with the various
formulations of classical mechanics: Lagrangian, Hamiltonian,
Liouvillian, and others. The analogy is not correct. Quantum
mechanics also has different formulations, such as those due to
{\Heisenberg} (close in spirit to Hamilton), {\Schroedinger} (close
to Liouville), and Feynman (which relies on a classical Lagrangian).
All these formulations are mathematically equivalent and the choice
of one of them for solving a particular problem is a matter of
convenience. On the other hand, the so-called interpretations of
quantum theory introduce new concepts, such as an infinity of
parallel worlds, without any experimental support nor any benefit to
the theorist who performs calculations to be compared with
experiment. These gratuitous additions to quantum theory are the
true analogs to {\Ptolemy}'s epicycles in {\Harris}'s story.

Stanley {\Sobottka} apparently wishes to retain some objective status
for the quantum wavefunction. He writes that interference phenomena
always suggest that physical waves are interfering, and that ``our
understanding is greatly enhanced if we assume that the pattern is
caused by actual, physical waves.'' However, {\Sobottka} himself has
reservations about this assumption, which is untenable for
higher-order interference effects involving two or more particles.
This is a problem Wendall {\Holladay} runs into immediately when he
says, ``The wavefunction of the four outer electrons in the ground
state of the carbon atom produces a tetrahedral structure in
Euclidean three-dimensional space that undergirds the observed
tetrahedral structure of the diamond crystal. This is an objective
fact about the physical world \ldots\ .'' The truth is that the
wavefunction of the four outer electrons lives in a 12-dimensional
space, while our tangible physical world has only three dimensions.
This example (contrary to its intended purpose) is an excellent one
for showing that the wavefunction is a mathematical tool, not a
physical object.

{\Holladay} asks rhetorically, ``How could a theory that does not
describe physical reality give such accurate results for the magnetic
moment of the electron and correctly predict the existence of
antiparticles?'' Similarly we could ask: How can ordinary probability
theory give such reliable results in the gambling house, knowing as
it does nothing whatsoever about the dynamics of the roulette wheel?
The probabilistic predictions we make in gambling would work just as
well whether the ultimate underlying physics were deterministic or,
instead, indeterministic.$^1$ The point is that a theory need make no
direct reference to reality in order to be successful or to be
perfectly accurate in some of its predictions. Probability theory is
a prime example of that because it is a theory of how to reason best
in light of the information we have, regardless of the origin of
that information. Quantum theory shares more of this flavor than any
other physical theory. Significant pieces of its structure could
just as well be called ``laws of thought'' as ``laws of physics.''
However, this does not preclude quantum theory from making {\it
some\/} predictions with absolute certainty. Among these predictions
are the quantitative relationships between physical constants such
as energy levels, cross sections, and transition rates that
{\Holladay} mentions.

The fallacy in {\Holladay}'s presumptive question is common to people
not accustomed to quantum lines of thought.  He makes no distinction
between nature, which we try to understand, and the description of
our experimental interventions into it. Accepting a distinction
between these concepts requires only that we humble ourselves before
nature, something our present scientific community is often
reluctant to do. On the other hand, attempting to identify the two
concepts discloses nothing more than a prejudice for a method that,
in the past, seemed to work well in the classical world.

When experimenters have similar information, they should make---on
the rules of quantum mechanics---similar predictions and draw
similar conclusions. By this account, quantum mechanics is a
scientific theory without rival. We feel comfortable in saying that
diamonds have tetrahedral symmetry because, on any number of
occasions, various experimenters have checked this with great
accuracy. They could do that because this aspect of carbon is part
of the ``effective reality'' quantum theory produces in some regimes
of our experience. Indeed, this ``effective reality'' forms the
ground for all our other quantum predictions simply because it is
the part of nature that is effectively detached from the effect of
our experimental interventions.  But, if one tries to push this
special circumstance further and identify an overarching ``reality''
completely independent of our interventions, then this is where the
trouble begins and one finds the {\it raison d'\^etre\/} of the
various ``interpretations.''

{\Todd} {\Brun} and Robert {\Griffiths} point out that ``physical
theories have always had as much to do with providing a coherent
picture of reality as they have with predicting the results of
experiment.'' Indeed, {\it have always had}. This statement was true
in the past, but it is untenable in the present (and likely to be
untenable in the future). Some people may deplore this situation,
but we were not led to reject a free-standing reality in the quantum
world out of a predilection for positivism. We were led there
because this is the overwhelming message quantum theory is trying to
tell us.

The main point of disagreement we have with {\Brun} and {\Griffiths}
is about the existence of a wavefunction of the universe that would
include {\it all\/} its degrees of freedom, even those in our brains.
We assert that this would lead to absurd self-referential paradoxes.
Therefore, it is necessary to restrict the discussion to a
(reasonably small) subset of the dynamical variables. {\Brun} and
{\Griffiths} ask, ``Can we only describe the Big Bang, or an
exploding supernova, in terms of the light that reaches our
telescopes?'' We never demanded such a restriction. We did not claim
that only what is directly observed exists. There is much more to
say about astrophysical phenomena than just describing the light that
originates from them. Yet, their description cannot be so detailed
as to include every particle involved in their observation, such as
those in the retina of the observer, in the optic nerves, in the
brain cells, etc. A limit must be put somewhere between the object
of our description and the agent that performs that description.
Quantum theory can describe {\it anything}, but a quantum description
cannot include {\it everything}.$^2$

We agree with {\Brun} and {\Griffiths} that the violation of
{\Bell}'s inequality by quantum theory is not a proof of its
nonlocality. Quantum theory is essentially local. {\Bell}'s discovery
was that any {\it realistic\/} theory that could mimic quantum
mechanics would necessarily be nonlocal. Near the end of his life,
{\Bell} was indeed inclined to seek such a theory, bearing traces of
realism and nonlocality. We do not rule out that such an extension
of quantum theory may some day be produced, but no one so far has
achieved this goal in a useful fashion, nor is an extension required
for a clear understanding of the quantum phenomena about us.

We surely agree with {\Brun} and {\Griffiths} that ``in science, one
cannot rule out alternatives by fiat; one must evaluate them on their
merits.'' We do not find any merit in the various alternatives that
were proposed to the straightforward interpretation of quantum
theory: It is a set of rules for calculating probabilities for
macroscopic detection events, upon taking into account any previous
experimental information. {\Brun} and {\Griffiths} may think this a
``straitjacket,'' but it prevents the endless conundrums that arise
solely from shunning quantum theory's greatest lesson---that the
notion of experiment plays an irreducible role in the world we are
trying to describe.

\bigskip\noindent{\bf References}\frenchspacing
\begin{enumerate}\itemsep -2pt

\item P. A. {\Hanle}, {\it Indeterminacy before {\Heisenberg}:~The Case of
Franz {\Exner} and Erwin Schr\"o\-dinger}, Hist.\ Stud.\ Phys.\ Sci.\
{\bf 10}, 225 (1979).

\item A. {\Peres} and W. H. {\Zurek}, {\it Is quantum theory universally
valid?\/} Am.\ J.\ Phys.\ {\bf50}, 807 (1982).
\end{enumerate}
\eq

\section{29 September 1999, \ ``The Next Move''}

\bap
You just heard from Eugen {\Merzbacher}. What shall we do next? How
busy will you be in the near future? I shall be ``normally'' busy,
teaching my course on foundations of QT, and doing research as usual.
\eap

I would classify myself as ``normally'' busy too.  {\Carl} and I are
working on a National Science Foundation proposal, but I should be
able to rise above that before my move to Los Alamos.  Also there's
the quantum de {\Finetti} theorem paper to be finished, but I'm quite
sure I can multi-task between you and {\Carl}.  Perhaps the biggest
thing to get in our way, is that I am likely to be significantly out
of commission from October 7--18.  I must drive our dogs to New
Mexico ({\Kiki} and {\Emma} will fly), then I will be spending
several days helping {\Kiki} get all the furniture arranged, our bank
accounts and other affairs opened, and build a fence in the yard to
contain the dogs.

I think the {\sl Physics Today\/} article is a nice opportunity and
we should run with it if we get a chance.  Too many of our close
colleagues are going the wayside:  our efforts may not change their
opinions, but we may help suppress the snowballing effect with
younger students.  And that has to be worth it.  I'm very glad you
had the idea of writing to Eugen.  (BTW, I also received a very nice
personal letter from him this morning; it appears that he has been
watching my career a bit from afar \ldots\ and has asked me to give a
colloquium at his university.)

\bap
My idea, for starting this opus, was to collect excerpts of my book
and articles, with nice slogans (``Unperformed experiments have no
results''), and let you add your own bibliographical sources --
surely you have much more, if you include writing from third parties
-- and then let you try to sort all that raw material into definite
subjects to work on. Or have you a better idea?
\eap

No, I think this is an excellent idea and a good starting point.  I
myself would like to find a way of (gently) inserting my provocative
slogan, ``Quantum states do not exist,'' and also I would like to
cover a little bit of the in-principle reversibility of measurement
in much the same way in the letter I wrote to Howard {\Barnum} (that
I forwarded to you):  I was very pleased that that discussion finally
turned his head a bit.  [See notes to Howard {\Barnum}, dated 30
August 1999 and 5 September 1999.] I'm sure we can come to some
agreement about substitutes for the words ``know'' and ``knowledge''
that you found disagreeable in that exposition. Also if we could
figure out a way of saying a word or two about quantum information
(much like Eugen did in his letter to {\Benka}), I think that it
would help to give an air of legitimacy and forward-looking feel to
all that we have to say.

\bap
Also, though Eugen was very helpful, we should have ASAP direct
contact with PT, before we start writing anything more than an
outline (a kind of abstract). Then we must ask them whether they want
any particular format, about the deadline, what they are willing to
reveal about the other ``feature'' and so on.
\eap

That is a good idea too.  Since you are the senior among us---and
anything an editor would say to you would likely be more
binding---would you take the task of initiating that?  I think it
would be most useful to understand who the authors of the other
``feature'' are.  That could also help us gauge what to say, to
understand whether it will ever materialize, and, if we know them,
well perhaps we could contact them directly for further
information.  (An off the cuff guess would be that {\Anton}
{\Zeilinger} might be writing something.)

Partners in crime again?

\section{01 October 1999, \ ``One More Round''}

Thank you for getting the PT process started by writing such a nice
detailed letter:  I think the article's already starting to shape up
in my mind.  Below I'll place the modifications I made; you should
have a look at them before we send the letter off to Eugen.  I think
you'll find most of the modifications minor, only adding a little
more detail to what you already said.  In a couple of cases I
Americanized the English a little bit.  I also removed the
``alternative medicine'' remark that Eugen started in order to keep
in line with his comment to {\Benka} that we would not write a
polemic.

The largest change you'll notice is that I deleted the original item
\#3.  As you know, I am Bayesian through and through now, and
Bayesians are pretty allergic to ensembles.  ``The world is given to
us only once.''  I decided to delete the passage rather than
attempting to modify it now because I don't think the deletion will
change the flavor of the letter enough that {\Benka} will even notice
its absence.  In the mean time, it will buy us some time to come to
agreement among ourselves:  I think we can do that and will probably
both learn a little bit in the process.

Tomorrow afternoon I'll try to write you a more detailed explanation
of why I disagreed with item \#3.

If you are OK with the letter as it is presently, feel free to send
it off to Eugen yourself.  Otherwise, I will send it off first thing
tomorrow morning or review any further modifications you make over
the evening.

\section{01 October 1999, \ ``We Are Converging''}

I made only a couple more minor style changes (that I would have
done last night if I were completely awake), so I went ahead and
sent it to Eugen.

\bap
I think that we completely agree that QM applies to a single system,
such as the Universe or the SQUID mentioned in \#6.
\eap

Absolutely.  About the semantics, I'll try to come back to that this
afternoon or sometime over the weekend.

\section{05 October 1999, \ ``Sorry So Late''}

I'm sorry to have been out of contact the last couple of days.  For
some reason my dial-up networking decided to quit on me this
weekend, and things just got too busy for me to get in to the
office.  I made all the changes you suggested except:

\bap
\#6:  delete ``But,'' in line 2 and the comma at the end of line 6.
\eap

I left the ``But'' but added ``standard'' in the paragraph:
\bq
\noindent Many feel the need to go to conceptual extremes with
quantum mechanics because they want to do quantum cosmology.  But,
standard quantum theory places no obstacle to that.  Is there a
``wave function of the universe?''  If we consider just a few
collective degrees of freedom, such as the radius of the universe,
its mean density, total baryon number, etc., we can apply quantum
theory to these degrees of freedom, which do not include ourselves
and other insignificant details.  This is just as when we quantize
the magnetic flux in a SQUID and ignore the atomic details.  One may
object that there is only one universe, but likewise there is only
one SQUID in our laboratory.  For sure, we can manipulate that SQUID
more easily than we can manipulate the radius of the universe.
Still, that SQUID is unique.
\eq
The reason I did that was to try to set some contrast.  I have
noticed that most people who go to the extreme of the {\Everett}
interpretation do so because they say that one cannot do quantum
cosmology with out it. Taking the ``But'' changed the meaning of the
thought, I believe. In any case, this is only a proposal letter:  so
I hope you will not mind my not bouncing off you once more.  We can
reserve endless iterations about meaning for the real thing!

I think we can have some fun writing an article like this:  I'm
looking forward to it.

\bap
Even a joint publication is conceivable, depending of what he
intends to write.
\eap

I think we will have a stronger position for our view if we're able
to get it represented by two articles in a single issue.  So I'm a
bit reluctant to join forces outside of a mutual proofreading and
encouragement, etc.  Certainly, though, I think we should contact
{\Anton} before we're ready to set down to write something.

Ok, now I must get all the final things packed up before the moving
people come tomorrow.  Tomorrow is reserved for their packing
everything in boxes.  The truck comes to be loaded the following
day.  So I may be out of email commission for a few days:  at least
until Sunday or Monday.

\section{03 November 1999, \ ``Explanation of My Absence''}

Below is a partial explanation of why you haven't heard from me.  I
feel pretty weak right now \ldots\ so I'll leave it at that.  I say
we start up our efforts on the opinion piece next Tuesday.  Have you
looked at {\Bub}'s criticism---recently on {\tt quant-ph}---of your
(our) view.  Studying that (mildly) may help us fend off similar
criticisms to our opinion piece.

\section{07 November 1999, \ ``A Sunday the Way I Like to Spend It''}

Today I had the chance to do a little work around the house, take
{\Emma} and the dogs to the park, listen to the morning jazz show on
the radio, and, in between, to read two papers.  Now this is the way
I like to spend a Sunday!

The first paper was your ``Karl {\Popper} and the Copenhagen
Interpretation.''  Of course, I agree on most counts and liked it a
lot.  (I spotted one typo:  the first sentence of the last paragraph
of page 3 in the {\tt quant-ph} version.  You're missing an ``it''
near the beginning.)  I especially found comfort in discovering that
you and I (and {\Bohr}) seem to agree on the following point:
\bap
Note that {\Bohr} did not contest the validity of counterfactual
reasoning. He wrote:
\bq
\rm \noindent Our freedom of handling the measuring instruments
is characteristic of the very idea of experiment \ldots\ we have a
completely free choice whether we want to determine the one or the
other of these quantities \ldots\
\eq
Thus, {\Bohr} found it perfectly legitimate to consider
counterfactual alternatives.  He had no doubt that the observer had
free will and could arbitrarily choose his experiments.  However,
each experimental setup must be considered separately.
\eap

I say I found comfort in that because recently Howard {\Barnum},
{\Charlie} {\Bennett}, and John {\Preskill} all have told me (on
separate occasions) that I'm being mystical when I say things like
that.  They seem to see me as somehow trying to buck the scientific
worldview, which angers me.  I simply say that I can't understand
what experimental science (or even science at all) is without that
simple presupposition---that we actually do meaningful experiments,
one or another or another \ldots\ that we actually learn things by
actively pursuing to learn them.  I think I replied most forcefully
to John {\Preskill}.  Let me paste that in, in case there's a phrase
or two in it that we might want to use in our PT article.  [See
letter to John {\Preskill}, dated 8 September 1999, titled ``Smug
Note.'']

The other paper I read today was Jeffrey {\Bub}'s ``Quantum Mechanics
as a Principle Theory'' ({\tt quant-ph/9910096}).  That wasn't a
waste of time:  in learning his motives, I think I come to a better
understanding of our position.  He devotes the last section of the
paper (titled ``Instrumentalism'') to criticizing you and van
{\Kampen} (but mostly you).  The only thing he has going for him is a
question, ``Knowledge of what?''  My answer would be terse:  what we
are speaking of is our probabilistic knowledge of the (experimental)
consequences of our (experimental) {\it interventions\/} into the
course of Nature.  It was only sloppy thinking about the nature of
classical physics that ever led us to (foolishly) believe that we
could have more.  Perhaps I shall compose a letter to {\Bub} and send
it directly.

Tomorrow morning early I drive to Albuquerque for my standard Monday
collaboration with {\Carl} {\Caves}.  Tuesday, I should be ready to
do something useful with respect to our PT project.  Let me give one
word of opinion now though:  I think we should lay a little low on
talking about {\Bohr} and Copenhagen directly.  As you make it clear
in your paper that I just read, there's not even a consensus on what
{\Bohr} actually said:  two people can have two completely different
readings of his words.  It has been my experience that sometimes
when people hear the words {\Bohr} or Copenhagen they just get riled
and refuse to listen to anything further.  I think we should stick
with the factual situation that quantum mechanics presents us
with---much in the style of the letter we sent to {\Benka}---and let
the simple force of the proper way of thinking carry itself.  But I
probably don't need to express this:  you're probably already ahead
of me on this.

Let me tell you a cute anecdote in that connection:  I'll just cut
and paste from a letter I wrote {\Carl} following the Baltimore
conference.  (Notes within notes within notes \ldots\ kind of like
{\Ezekiel} with his wheels.)
\bq
\noindent In general, by my reckoning, the talk went very, very well.
This time, I focused more on the overall program ``quantum states as
states of knowledge'' than on the technicalities of the proof \ldots\
and it really seemed to pay off in this audience.  I was absolutely
surprised at the number of heads that seemed to turn to these
ideas.  Even when people disagreed (like Nicolas {\Gisin} and Philip
{\Pearle} did), they seemed to come away as if they had heard a fresh
idea.  As I wrote {\Herb} {\Bernstein} earlier today, [See note to
{\Herb} {\Bernstein} date 17 August 1999, titled ``A {\Bernstein} Off
the Earth?'']
\eq

OK, I go to bed now:  it is a long drive to ABQ and back tomorrow.
I hope you are feeling better (and that this letter didn't exhaust
you).  Please give Aviva my best wishes.  Tell her that the two
miniature rugs she gave us have found a temporary home above our
fireplace.  (They will have to moved to a more permanent and safe
place before we actually build a fire in it!)

\section{12 November 1999, \ ``Embryo.tex''}

What a dreadful couple of days I've had!  The troubles began
yesterday morning when we discovered [\ldots]

I have created a file titled ``Embryo.tex'' containing the embryo of
the article you sent me and will hopefully start some modifications
to it today.  Now, though, I must go to the Lab for the Friday
Quantum Lunch.  Wish me more luck than I've had in the last two days.

\section{13 November 1999, \ ``Putting the Finger''}

Well I procrastinated a bit from the proper part of our project
today, but maybe I came out the better for it.  I finally sat down
and read all the relevant things in {\sl Physics Today\/} that I
could find:  the {\Goldstein} article, the {\Griffiths} and {\Omnes}
articles, that article by Mara {\Beller} on the Sokal hoax [PT,
Sept.\ 98, pp.\ 29--34] in which she makes a strong stand for
Bohmianism, and the letters replying to {\Goldstein} [PT, Feb.\ 99,
pp.\ 11--13, 15, 89--90, 92].

The only sensible thing written in all this lot was {\Anton}
{\Zeilinger}'s letter contra {\Goldstein}.  {\Goldstein}'s own
writings were, by far, the worst representatives.  As far as I could
tell, his 10 pages worth of article said almost nothing \ldots\
other than perhaps weakly expressing what he {\it wants\/} of
quantum theory! And this guy is just way over confident with
himself.  Listen to what he writes on the first page of his first
article:
\bq
\noindent
Many physicists pay lip service to the Copenhagen interpretation,
and in particular to the notion that quantum mechanics is about
observation or results of measurement.  But hardly anybody truly
believes this anymore---and it is hard for me to believe that anyone
really ever did.
\eq

I dread meeting him two weeks from now in Naples.

The {\Griffiths} and {\Omnes} article also only helped confirm the
conclusions I've already drawn on consistent histories.  Despite the
overtures RBG{\index{Griffiths, Robert B.}} and RO{\index{Omn\`es,
Roland}} (and JBH{\index{Hartle, James B.}} and MGM{\index{Gell-Mann,
Murray}}) make, they haven't in any way taken the observer out of
quantum mechanics, nor extended the theory beyond its old form.  Let
me put my finger on it in the most rhetorical way I can.  To do
this, just step back for a moment to standard quantum mechanics and
fix a specific von Neumann measurement on some system.  Then, when
performed, the observer will find one and only one of the possible
outcomes (the outcomes, of course, are taken to correspond to a
fixed set of orthogonal vectors).  As long as that measurement is
fixed and we never speak (counterfactually) about another, we can
play AS IF one of those outcomes actually ``was the case'' all
along, independent of our having performed the measurement.  In such
a simplistic situation, one can act AS IF ``unperformed measurements
have outcomes.''  Or, another way to say this is that as long as a
single measurement is fixed, we can imagine that the quantum state
has nothing deeper to say than a classical probability distribution
would have had for a fixed set of outcomes in some probabilistic
trial.  Quantum mechanics only starts to make its mark when we {\it
contemplate\/} simultaneously various noncommuting von Neumann
measurements.

All the Consistent Historians do is extend the simple case above to
a family of ``histories''---strictly speaking, a history is nothing
more than a projector on a tensor product of many copies of the
original Hilbert space.  When the histories all commute, and we
refuse to speak (counterfactually) about any other noncommuting set
of histories, then we can always act AS IF one of the histories is
truly the case (i.e., existent, real, objective,
observer-independent).  But this is no surprise and nothing deep.
What can be said about two noncommuting sets of histories from their
point of view?  ``Nothing,'' RBG{\index{Griffiths, Robert B.}} and
RO{\index{Omn\`es, Roland}} say, ``such an inquiry is meaningless.''
But that is just as in standard quantum mechanics, where we never
speak of measuring simultaneously position and momentum.  AND THAT
IS ALL THERE IS TO CONSISTENT HISTORIES. There's nothing new here
that wasn't in the old theory.  In particular, if our friends would
just have the mental nerve to broach the questions you and I do on a
regular basis---about distinct noncommuting sets of
observables---then they would be stuck in the same lot as us \ldots\
i.e., being forced to the conclusion that unperformed measurements
have no outcomes.

But I rant and rave.  Tomorrow I have the feeling I will be able to
turn some of this energy to creative purposes.  Sundays are
generally good for me in that way.  The other day you wrote:

\bap
If the title of our essay is ``QM needs no interpretation'', our task
is to explain why indeed no interpretation is needed.  Not that the
various interpretations ({\Bohm}, {\Everett}, consistent histories)
are wrong -- they probably are just complicated reformulations of the
theory -- they simply are redundant.
\eap

I agree with you completely on this.  (Don't let my words above lead
you to think that I'll come out swinging a battle axe for
RBG{\index{Griffiths, Robert B.}} and RO.  My head will be much more
sensible tomorrow.)

\section{15 November 1999, \ ``Have a Good Trip''}

In the next note, I'll send a little bit of progress to the
manuscript.  Please note that it is titled ``Embryo.tex'' \ldots\ so
everything I have written is certainly still in its formative
stages.  It is not too late to scrap it all and start again if you
find that you do not agree with what I have written.  There are so
many similarities in our thoughts, I am sure we will in one way or
other be able to find a middle ground if need be.

The main changes/additions of mine can be found in the first five
paragraphs and in the final paragraph.  I spent most of my time
today with the beginning, trying to set up a general point of view.
What I plan to do next is tie in, in a more seamless fashion, what
you have already written (both in the manuscript and in your
supplementary notes).

One point that we may need some private side discussion on before we
set it in stone, is captured among other places in your sentence:
\bap
The notion ``state'' refers to a {\it method of preparation,\/} it is
not an intrinsic property of a physical system.
\eap

In general I have noticed in this manuscript that you lean more
heavily on the word ``preparation'' than we did in our letter to
{\Benka}.  (In fact, I can't find any mention at all of the word
``preparation'' in that letter.)  Unless I misunderstand your usage
of the word, it may actually be a little too anthropocentric even for
my tastes.  The problem is this:  consider what you wrote in the
paragraph about the wave function of the universe.  It seems hard to
me to imagine the wave function of those degrees of freedom which we
describe quantum mechanically as corresponding to a ``preparation.''
Who was the preparer?

It is for this reason that {\Carl} {\Caves} and I prefer to
associate a quantum state (either pure or mixed) solely with the
compendium of probabilities it generates, via the {\Born} rule, for
the outcomes of all potential measurements.  And then we leave it at
that.  Knowing the preparation of a system (or the equivalence class
to which it belongs) is one way of getting at a set of such
probabilities.  But there are others ways which surely have almost
nothing to do with a preparation.  An example comes about in quantum
statistical mechanics:  when the expected energy of a system is the
only thing known, the principle of maximum entropy is invoked in
order to assign a density operator to the system.  There may be
someone beside me in the background who knows the precise
preparation of the system, but that does not matter as far as I am
concerned---my compendium of probabilities for the outcomes of all
measurements are still calculated from the MaxEnt density operator.

To help ensure that I was not jumping to conclusions on your usage
of the term, I reread today your paper ``What is a state vector?''
[AJP {\bf 52} (1984) 644--650].  There was a time when I agreed with
everything you wrote there (in fact, I think it was the first paper
with which I got to know you).  But as of today at least, I think a
more neutral language as in our letter to {\Benka} is more
appropriate.

Please let me know your thoughts when you have time.  If you prefer
we can wait until you return from your conference.  If, on the other
hand, you have no strong disagreement with me, I will continue
working on the manuscript Tuesday morning and perhaps we will have a
first draft by Wednesday.

Please have a safe trip and give my regards to {\Anton}.

\section{17 November 1999, \ ``Holding Pattern''}

By the way, I decided to wait upon your return to write any more on
our project.  Once I hear your opinions, I'll reengage.  I hope you
are enjoying being 2000 light years from home.

\section{18 November 1999, \ ``Gamete Actually''}

Thanks for ``deprosing'' me a little!  I know that I sometimes have
a tendency to go overboard on my first drafts---luckily I usually
trim them down myself on second and third readings.  For the most
part, I am pleased with your results.  I'll place the draft below
with a few yet further changes and the voicing of a little
disagreement here and there.  I won't go too deeply tonight as I
haven't seen the other things you have in mind (and it is late for
me given that I got up this morning at 2:30 and really never fell
back asleep).  I'll use the same system you did, adding footnotes
when relevant.

\subsection{Parsing the Paper}

\noindent {\bf Quote from Draft:} Recently there has been a spate
of articles, letters, and reviews in {\sl Physics Today\/} promoting
various ``interpretations'' of quantum theory. The running theme in
these is that from the time of quantum theory's inception until the
emergence of a particular interpretation, the theory was in a crisis
because its foundation was ????.
\smallskip\\
{\bf Comment:} Looking back at what I wrote here before, i.e.,
``mystical muddle of words,'' I see that it was a bit flowery.
However, I don't think the phrase ``logically inconsistent''
captures the right notion. For instance I know that {\Omnes} would
say that he is only extending the Copenhagen interpretation as it is
``incomplete.'' The main criticism of most these guys seems to me to
be that Copenhagen is just a lot of words without substance and that
exalting measurement as the primary process of quantum theory verges
on the mystical. I think we should work a little harder for a better
phrase here, but one does not come to me right now.
\bigskip\\
\noindent {\bf Quote from Draft:} An experiment is in its essence an
active intervention into the course of Nature: we set up this or that
experimental situation so as to understand the consequences of these
interventions.
\smallskip\\
{\bf Comment:} I agree with you that, ``We should have a defensive
strategy: avoid loose statement that can easily be attacked, if they
are not essential to the argument.''  What I was shooting for
though---and I think it is essential---was a kind of ``positivism
without positivism.''  I am afraid that the simple word
``understand'' can be construed in too many ways, and it may be
attacked on that account. In fact, I remember well a conversation
with David {\Mermin} in which he took the point of view that ``to
understand a phenomena'' is to pinpoint the {\it reality\/} that
gives rise to it.  So I think it kind of important to further
emphasize the distinction between positivistic/operational
understanding and the classical worldview.  Can you agree to the
version of the sentence that I reinstated?
\bigskip\\
\noindent {\bf Quote from Draft:}
We do not deny the existence of an objective reality independent of
what observers may perceive. However, this reality, whatever it may
be, is not described by quantum theory, nor by any other theory
known to us. John {\Bell} formally showed that any objective theory
giving experimental predictions identical to those of quantum theory
would necessarily be nonlocal. It would have to encompass all the
degrees of freedom of the Universe. As {\Bell} eloquently puts it,
``separate parts of the world would be deeply and conspiratorially
entangled, and our apparent free will would be entangled with
them.''.
\smallskip\\
{\bf Comment:} I am a little worried that these three sentences about
{\Bell} will open up a can of worms. The Bohmians should agree, but I
am afraid that the Consistent Historians, the Everettistas, and
GRWians will all come out of the woodwork for the attack. They'll
cry fowl, and say that we've just never taken the time to understand
their theory.  So I'm a little afraid to make these statements.  Are
they essential?
\bigskip\\
\noindent {\bf Quote from Draft:}
Indeed, a natural question is whether the universe has a wave
function. There are two ways of understanding this. If this ``wave
function of the universe'' has to give a complete description of
everything, including our own brains, we encounter bizarre
self-referential paradoxes. These paradoxes are not in the realm of
physics. Experimental physicists never have to bother about them. On
the other hand, if we consider just a few collective degrees of
freedom, such as the radius of the universe, its mean density, total
baryon number, etc., we can apply quantum theory to these degrees of
freedom which do not include ourselves and other insignificant
details.  This is just as when we quantize the magnetic flux in a
SQUID and ignore the atomic details.  One may object that there is
only one universe, but likewise there is only one SQUID in our
laboratory.  For sure, we can manipulate that SQUID more easily than
we can manipulate the radius of the universe.  Still, that SQUID is
unique.
\smallskip\\
{\bf Comment:} Before that, we have to explain why it is legitimate
to apply statistical reasoning to a single system. You're starting to
sound awfully Bayesian!! Congratulations.

\section{19 November 1999, \ ``The Nut House''}

\bap
I am sending now all that I wrote. It's a hodge-podge of paragraphs
in random order. I have no time to reread that, because we now have
dinner with my granddaughter and her parents. Please forgive any
insanities in that text. It's just raw material, as you know.
\eap

If that's temporary insanity, it ain't bad!  I quickly skimmed the
first three pages, and am very much starting to like the flow.
Hopefully I will get a chance to do some minor things on the draft
tonight.  But tomorrow I should be able to do some more serious work
with it.  (Today I am working with {\Carl} on our project.)

Have fun with your granddaughter.

\section{19 November 1999, \ ``End O' Day''}

I give up.  I was hoping to have a real look at your draft before
the day was out, but it doesn't look like I'm going to be able to do
it.  {\Kiki} (and {\Emma}) keep saying ``Come downstairs, it's Friday
night!'' (What is your secret?!?!)

\section{20 November 1999, \ ``Almost Fetal''}

Well I put in a solid day's work on our project today.  I was hoping
to completely finish it by the evening, but it has just become too
difficult and now {\Kiki} is calling me to dinner.  Tomorrow morning
I will rise early and continue the pace.  I've made many
modifications (almost all minor), tweaking this and that and working
toward getting it under our size limitation (max 1800 words) without
compromising the integrity.  Please give me another day, and then I
will send you the next iteration so that it can be your turn again.

I think its really starting to take shape.

\section{21 November 1999, \ ``Miracles of Office
Technology''}

That article by {\Peierls} is indeed quite good.  {\Plaga} was right.
In some places, {\Peierls}' writing could have just been ours
(perhaps more accurately I should have said it the other way
around).  It dawned on me that it wouldn't take any time to scan it
in for you. So I'll paste it below.  There may be some nonsense in
there: Optical Character Recognition isn't perfect yet and I didn't
reread it completely.  Still, here and there I have placed some
footnotes expressing my reactions to what he said.

\noindent --------------------

\begin{center}
In Defence of ``Measurement'' \\
Rudolf {\Peierls} \\
\small {\sl Physics World}, January 1991, pp.~19-20
\end{center}

\bq
\small In a stimulating article ({\sl Physics World\/} August p33)
the late John {\Bell} professed dissatisfaction with the foundations
of quantum mechanics as usually presented, particularly in connection
with the so-called ``collapse of the wavefunction'' as a result of a
measurement. He agreed that {\it for all practical purposes\/} the
use of quantum mechanics by qualified practitioners leads to well
defined answers which, where they can be checked, agree with
experiment.

However, he regarded it as necessary to have a clearly formulated
presentation of the physical significance of the theory without
relying on ill-defined concepts. I agree with him that this is
desirable, and, like him, I do not know of any textbook which
explains these matters to my satisfaction. I agree in particular that
the books he quoted do not give satisfactory answers (I assume that
they are fairly quoted; I have not re-read them).

But I do not agree with John {\Bell} that these problems are very
difficult. I think it is easy to give an acceptable account, and in
this article I shall try to do so. I shall not aim at a rigorous
axiomatic, but only at the level of the logic of the working
physicist.

In my view the most fundamental statement of quantum mechanics is
that the wavefunction, or more generally the density matrix,
represents our {\it knowledge\/} of the system we are trying to
describe. I shall return later to the question ``whose knowledge?".
It is well known that we have to use a wavefunction if we have a
``pure state'' i.e.~if our knowledge of the system is complete, in
the sense that any further knowledge is barred by the uncertainty
principle.\footnote{I like this point of view about the pure state
except for the loose term ``uncertainty principle.''  I think instead
our information--disturbance relations hint at the right sort of
rigor that one would need for putting some flesh on this sentence.
But that defines the large part of my research program.  Also
{\Caves} and I would advocate the term ``maximal knowledge'' here in
place of ``complete knowledge'' (preferring to say that when one is
forced to probabilistic predictions it is always because of
incomplete knowledge).} Failing such complete knowledge we must use
a density matrix, which therefore contains both quantum and classical
ignorance. The wavefunction is a special case of a density matrix,
and I shall here talk about ``density matrix'' when I mean
``wavefunction or density matrix".

More precisely, while the time variation of the density matrix is
given by {\Schroedinger}'s equation, the initial values represent
knowledge usually obtained from observations. (There are not always
measurements; for example, if an atom has been for a reasonable time
in free space we know it must be in its ground state.)

Our knowledge is not fixed, but may increase or decrease. It
increases if further observations are made; it decreases if the
system is disturbed by external factors which we cannot control.
There is nothing new in this. In classical physics our knowledge may
increase and decrease in the same way. The only difference is that in
quantum mechanics we have to be specific about what we know, because
our possible knowledge is confined by the uncertainty principle. In
classical physics there is no reason in principle why we cannot know
everything about the system and we usually argue as if we did. But in
a practical situation our knowledge may increase or decrease as
indicated.

In quantum mechanics any increase in our knowledge is usually
accompanied by a decrease in some other respect, because of the
uncertainty principle.\footnote{Again I think the term ``uncertainty
principle'' causes trouble here because it makes one think that the
knowledge is about a {\it preexisting\/} $x$ or $p$ independent of
our interventions.  See footnote toward the end of the article.} This
applies particularly when we are concerned with a ``pure state". Then
we can gain no new information (other than confirming what we know
already) without losing some of the existing information.

Once this significance of the density matrix is understood, it is
clear that upon a change in our knowledge the density matrix must
change. This is not a physical process, and we certainly cannot
expect it to follow from the {\Schroedinger} equation. It is just the
fact that our knowledge has changed, and thus must be represented by
a new density matrix.

When I refer to ``observation", this term has its common-sense
meaning, The observation usually (but not necessarily) involves an
apparatus which interacts with the system in question, and which
produces a signal (visible, audible, or other) which we can
recognise, and which is correlated with the variables of the system.
{\Bell} quoted the view of {\Landau} and {\Lifshitz} (and therefore
of {\Bohr}) that the apparatus must necessarily obey classical
physics. In my view this is not correct. It is of course true that
our senses are macroscopic, and that the instruments we find
convenient are also macroscopic and in practice classical. But this
is a practical point, not one of principle. The sensitivity of the
human eye is almost sufficient to detect a single photon. If some
experimentalist has sufficient vision to see one photon, the
observation of that photon might perfectly well serve as a
measurement.

The apparatus usually consists of a chain of correlated events. I
have elsewhere ({\Peierls} 1979, 1985) discussed as an example the
observation of a spin component of a spin-~ atom by a Stern-Gerlach
magnet. The first step, the passage through the inhomogeneous
magnetic field, sets up a correlation between the spin component and
the position of the atom. It is not yet a measurement; we have not
yet gathered any information. This requires determining the position
of the particle, i.e. in which part of the split beam it travels. To
find this out, we may use a counter, but again this conveys no
information---and nothing collapses!---until we find out whether the
counter has been activated. We can obviously pursue this chain: the
counter will be part of an electrical circuit, the circuit will
operate a digital recorder, we may read this recorder by means of the
light it reflects into our eye, etc. Each step is correlated with the
preceding ones and therefore with the spin component of the particle.
Each step keeps both options open until we ``see'' the result, and
then we revise our density matrix.

Because of the uncertainty principle we cannot acquire knowledge of,
say, the $z$ component of the spin without losing what information we
had previously about, say, the $x$ component. Is this happening in
the first step, the passage through the magnet? At first sight this
looks likely, because information about $s_x$ is contained in the
phase relation between the components of the wavefunction belonging
to $s_z=+\frac{1}{2}$ and $s_z=-\frac{1}{2}$. Since the beams
corresponding to the two $s_z$, values are now split, they do not
overlap and do not interfere, so their phase relationship is not
observable. However, the information is not irretrievably lost. By
arranging a further magnet we could recombine the two beams and
observe their phase relation (thereby foregoing the possibility of
observing $s_z$). We do finally lose the ``forbidden'' information
when we ``see'' the atom in one of the beams. We then have to replace
our density matrix by one containing only the one $s_z$ value, so
there is no interference.

As long as we do not ``see'' the atom in the beam, the reconstruction
of the seemingly lost information is troublesome, but easy to
visualise. At the next stage, i.e.~after the counter, it becomes much
more involved. Since the density matrix now contains the variables of
the counter, interference requires not only that the two atomic beams
be made to overlap, but in addition that there be an overlap between
the density matrices for the activated and unactivated states of the
counter. The observation of the phase relation therefore requires an
operator capable of deactivating the counter coherently. This is
possible in principle, but in practice prohibitively difficult. As we
go further down the chain of connections involved in our
``measurement", this difficulty gets worse.

This is the origin of the belief that the apparatus makes the
off-diagonal matrix elements of the density matrix disappear. In most
cases that is true ``for all practical purposes", but not in
principle. The off-diagonal matrix elements disappear only when we
know the result of the measurement.

The ``system'' to which we apply our description can be as large as
we like, including the whole world if we want. However, if we make
the system too large, the amount of information we can obtain is
relatively small, so that the density matrix is made up mostly of
parts proportional to the unit matrix (which denotes complete
ignorance) and it becomes hard to do any useful physics. In any case
the ``system'' cannot include the mind of the observer and his
knowledge, because present physics is not able to describe mind and
knowledge (and it is not obvious that this is a proper subject for
physics).\footnote{Sound familiar?}

The objection is sometimes made: ``How can one apply quantum
mechanics to the early Universe, when there were no observers
around?'' The answer is that the observer does not have to be
contemporaneous with the event. We can, from present evidence, draw
conclusions about the early Universe, the classical example being the
cosmic microwave background. In this sense we are observers. If there
is a part of the Universe, or a period in its history, which is not
capable of influencing present-day events directly or indirectly,
then indeed there would be no sense in applying quantum mechanics to
it.

That leaves the question: whose knowledge should be represented in
the density matrix? In general there will be many who may have some
information about the state of a physical system.\footnote{Here is an
example where he is being a bit sloppy.  ``Information about the
state of the system'' is not a well defined concept as we know. His
problem comes from slipping back into objectivist language:  he is
here acting as if a ``state'' is an objective property of the system.
In my view we ought to always be strict:  Knowledge of what?
Information of what?  {\it Probabilistic knowledge of how a system
will react to our external interventions!}} Each of them has to use
his or her density matrix. These may differ, as the nature and amount
of knowledge may differ. People may have observed the system by
different methods, with more or less accuracy; they may have seen
part of the results of another physicist. However, there are
limitations to the extent to which their knowledge may differ. This
is imposed by the uncertainty principle. For example if one observer
has knowledge of $s_z$, of our Stern-Gerlach atom, another may not
know $s_x$, since the measurement of $s_x$ would have destroyed the
other person's knowledge of $s_z$ and vice versa. This limitation can
be compactly and conveniently expressed by the condition that the
density matrices used by the two observers must commute with each
other.

John {\Bell} referred to two alternative interpretations of quantum
mechanics, that of de {\Broglie}-{\Bohm} (BB), and that of
{\Ghirardi}-{\Rimini}-{\Weber} (GRW). As far as I know the BB scheme
reproduces all predictions of quantum mechanics. A decision can
therefore be made only on aesthetic grounds. I must confess that the
scheme, with both hidden variables and probability rules, seems to me
exceedingly ugly, but of course one cannot argue about this. I have
not studied the implications of the GRW scheme in detail, but I
believe that there must be cases where it makes predictions differing
from those of quantum mechanics, which would be observable in
principle.
\begin{enumerate}
\item
J S {\Bell} 1990 Against ``measurement'' {\sl Physics World\/} August
33-40
\item
R {\Peierls} 1979 {\sl Surprises in Theoretical Physics\/} Princeton
section 1.6 (Some of the points made in this article will also be
discussed in a forthcoming volume, {\sl More Surprises in
Theoretical Physics}, Princeton)
\item
R {\Peierls} 1985 Observations in Quantum Mechanics and the
``Collapse of the Wave Function'' in {\sl Symposium on the
Foundations of Modern Physics\/} World Scientific
\end{enumerate}
\eq

\section{21 November 1999, \ ``Kathie and Erwin''}

I understand Erwin, but why Kathie?  Is that one of your
granddaughters?

\section{22 November 1999, \ ``A Baby Kicks?''}

I bet you're already awake and have already looked at your email
(and wondered, ``Where is that Fuchs?'').  I'll send you two files
momentarily.  The first is the latest draft including detailed
footnotes explaining why I did what I did, etc.  The second is the
same draft with all the footnotes stripped out so that you may have
an easier time editing if you decide to pick up where I left off
(for the most part).

It's starting to sound pretty good to me, but that could simply be
from the weariness of reading it over and over.  By my editor's
count we have presently 1849 words.  But the editor gets a little
confused by \TeX\ commands, etc., so the real count should be
somewhat below that (maybe 1750).

I think the transition between the penultimate and ultimate
paragraphs is still a little abrupt, but I couldn't think of a
slicker way to pull it off.  Maybe you'll have better luck.

Also, I would still really like to say just a little something about
Erwin's in principle possibility of reversing Kathie's superposition
eating.  The reason is I've seen too many people think that our
point of view precludes such a thing.  But I realize space
limitations are tight and I don't see much else that I would like to
cut out.

\subsection{Parsing the Paper}

\noindent {\bf Quote from Draft:} To start, let us assess the goals
of experimental science. An experiment is an active intervention into
the course of Nature: we set up this or that experiment to see how
Nature reacts to our prods.
\smallskip\\
{\bf Comment:} I started thinking that I had overused the word
``intervention.'' And besides I think this captures what I had
wanted to say better.  It adds the right amount of emphasis and
twist away from ``realist'' thinking.
\bigskip\\
{\bf Quote from Draft:} Understanding is achieved when we can
distill from the accumulated data a compact description of all that
was seen and an indication of which further experiments will help
corroborate that description. This is what science is about. If from
our understanding we can {\it further\/} distill a model of a
free-standing ``reality'' independent of our interventions, then so
much the better. Classical physics is the ultimate example in that
regard. However, there is no logical necessity that such a further
step always be obtainable. If our world is so ``sensitive to the
touch'' that we can never identify a reality independent of our
experimental intrusions, then we must be prepared for that too.
\smallskip\\
{\bf Comment:} I decided I liked the word ``sensitive'' better than
``ticklish.'' Also I did a little more ``intervention'' trimming.
\bigskip\\
{\bf Quote from Draft:} Contrary to these aspirations, quantum
theory does not describe physical reality. It only provides an
algorithm for computing probabilities for the macroscopic events
(``detector clicks'') we choose to associate with a measurement
process.
\smallskip\\
{\bf Comment:} Predicting $\rightarrow$ computing.  It's more
Bayesian that way.  I chose to use the word ``choose'' because of a
very nice paper I once read by some guys named Hay and {\Peres}.
\bigskip\\
{\bf Quote from Draft:} These clicks may be viewed as the
consequences of our experimental interventions.
\smallskip\\
{\bf Comment:} I threw this sentence in so that it would be clear
how ``macroscopic events'' makes a connection to our previous
discussion.  I think such a thing is necessary.
\bigskip\\
{\bf Quote from Draft:} This strict definition of the scope of
quantum theory is the only interpretation ever needed by
experimenters and theorists alike.
\smallskip\\
{\bf Comment:}  I didn't want us theorists left out.
\bigskip\\
{\bf Quote from Draft:} The probabilities spoken of here are
computed by means of {\it any\/} information that may have been
gathered previously, including, but not limited to, information
about a system's preparation.
\smallskip\\
{\bf Comment:}  I think it is really necessary that we get away from
the notion of ``a state {\it defined\/} purely as a preparation or
recipe'' if we want to talk about a very general situation like the
wavefunction of the radius of the universe, etc., later in the text.
That is to say, we need to be a little more {\Peierls}-ian in our
statement.  Also if we want to be able to say ``this strict
definition is the only interpretation ever needed by experimenters''
then we'll most certainly have to do something like this.  Just
think of an engineer who must worry about the thermodynamics or
statistical mechanics of some very fine microdevice.  As a first
characterization of his design, he will most likely assign a quantum
state based solely on some knowledge of the running temperature or
mean energy of the system. In the latter case, he will most likely
use the principle of maximum entropy for his quantum-state
assignment. He generally won't have the luxury of associating a
recipe or preparation with his quantum state. If a technician later
comes along and presents the engineer with a detailed analysis of
the energy flow into the device from the reservoir in which it sits,
then so much the better.  The engineer will just update his state
assignment to be in accord with his new information and forget about
his crude earlier methodology.
\bigskip\\
{\bf Quote from Draft:} The precise latitude with which such
information can be incorporated into quantum statistical predictions
finds its formal expression in the probability rule postulated by
Max {\Born}. The conclusiveness of this rule is made sharp by a
remarkable theorem due to Andrew {\Gleason}.
\smallskip\\
{\bf Comment:} I wanted to stress that {\Gleason} really puts a cap
on things.  And by the way, the theorem is indeed remarkable. I have
walked through every step of its proof, and it was no easy task!
\bigskip\\
{\bf Quote from Draft:} So long as one assumes that yes-no tests on
a system are represented by projection operators $P$ on a Hilbert
space and that probabilities are additive over orthogonal
projectors, then there exists a density matrix $\rho$ so that the
probability of a ``yes'' answer is ${\rm tr}(\rho P)$.
\smallskip\\
{\bf Comment:} I tried to make this sentence a complete statement of
the theorem, skipping only that dimensionality greater than two is
required (which doesn't seem so essential for an ``opinion''
article).  And another ``by the way'': in your book, p.~190, you
write, ``The premises needed to prove that theorem are \ldots,
supplemented by reasonable continuity arguments.'' That is not true.
One of the things that makes that makes the {\it mathematics\/} that
{\Gleason} did so remarkable is that he {\it did not\/} assume
continuity!  He proved it!  Now any physicist (except our friend
{\Kent}) would think that continuity is perfectly reasonable. But it
still shocks me that Hilbert space is such a nice place that
continuity need not even be assumed.
\bigskip\\
{\bf Quote from Draft:} We do not deny the existence of an objective
reality independent of what observers perceive. However this
reality, whatever it may be, is not described by quantum theory nor
by any reasonable theory known to us.
\smallskip\\
{\bf Comment:} I changed ``other'' to ``reasonable'' because most of
our antagonists claim to have just that.
\bigskip\\
{\bf Quote from Draft:} John {\Bell} formally showed that any
objective theory giving experimental predictions identical to those
of quantum theory would necessarily be nonlocal. It would have to
encompass all the degrees of freedom of the Universe, and as {\Bell}
put it, ``separate parts of the world would be deeply and
conspiratorially entangled, and our apparent free will would be
entangled with them.''
\smallskip\\
{\bf Comment:} I struck the phrase ``it would be beyond human
understanding'' and replaced it with parts of what you had written
before.  {\Bohm}'s theory, for instance, may be ugly and add features
extraneous to quantum mechanics, but I don't see that it makes the
world beyond human comprehension.
\bigskip\\
{\bf Quote from Draft:}  We can, on the other hand, have a
description of an ``effective'' objective reality in the limiting
case of macroscopic phenomena such as planetary motion, but such a
description ignores most degrees of freedom for the system and is
necessarily incomplete.
\smallskip\\
{\bf Comment:} I rearranged this sentence somewhat so that it read
better for me. But I am not sure that I completely agree with what
you are thinking here, so let me withhold complete judgement until a
later draft. The main problem is this.  Take the
information--disturbance relations from our 1996 PRA. It seems to me
that they capture a gross aspect of reality (suspending for the
moment our phrase ``whatever reality is''). Really I just mean that
all observers will agree that the information--disturbance
phenomenon exists. But that phenomenon has nothing to do with an
incomplete description. Instead it seems more an integral part of
the reason that we cannot get at a world with ``a free-standing
reality independent of our interventions.'' From this point of view
a phenomenon is ``macroscopic'' precisely when the tradeoff is weak
between information gain and state disturbance for various observers
trying to come to agreement on the predictions they can make about
potential further interventions on their system. That in general
can/will happen when we are only looking at certain gross aspects of
a system, for instance by ignoring most degrees of freedom.  I
understand that this is too advanced a topic for this essay.  But I
want to make sure that I don't sin by writing something I don't
believe.  So I wonder if there might be some way of adding two or
three extra words that would make me feel comfortable?
\bigskip\\
{\bf Quote from Draft:} Here it is essential to understand that the
statistical nature of quantum theory does not restrict its validity
to situations where there is a large number of similar systems.
Statistical predictions do apply to single events. When we are told
that the probability of precipitation tomorrow is 35\%, there is
only one tomorrow. This tells us that it is advisable to carry an
umbrella.
\smallskip\\
{\bf Comment:} I removed: ``Statistical predictions such as the
above one are prepared by using available information, which is
incomplete, and considering various hypotheses for the missing data.
We thus create a conceptual ensemble of initial conditions. For each
one, we model the future dynamical evolution, and thereby obtain a
probabilistic forecast, such as a weather forecast.''
\bigskip\\
{\bf Quote from Draft:} Probability theory is the formal
quantification of how to make rational decisions in the face of
uncertainty; this carries over as much to quantum phenomena as it
does to anything else.  When one makes a probability statement
concerning a quantum measurement outcome, one is essentially making
a {\it bet\/} about what will be seen.  As long as the probability
calculus is used, an adversarial gambler can never force the bettor
to a sure loss.
\smallskip\\
{\bf Comment:} If, on the other hand, it is ignored, there exists a
bet leading to a sure loss {\it even\/} in the single trial. This I
believe is the best, clearest, and most famous ``operational''
definition for probability within the Bayesian approach, and is why
I always say to you that one need never deal even with ``conceptual
ensembles'' to define probability. (Besides, the conceptual ensemble
approach is inconsistent as I tried to explain to you once before.)
The formal argument I am alluding to in the text is called the
``Dutch book argument'' and can be found in {\sl Studies and
Subjective Probability}, Second Edition, edited by H.~E. {\Kyburg}
and H.~E. Smokler (Krieger Publishing, Huntington, NY, 1980). If you
wish, I can send you a summary of the argument that {\Ruediger}
{\Schack} wrote up in \LaTeX; I also have some notes in \LaTeX\ that
{\Carl} {\Caves} wrote up. Alternatively, if you wish to read Bruno
de {\Finetti}'s famous article in French (where the argument was
first promoted), have a look at {\sl Annales de l'Institut Henri
Poincar\'e} {\bf 7} (1937), 1--68.
\bigskip\\
{\bf Quote from Draft:} The peculiar nature of a quantum state is
best illustrated by the fact that although an unknown quantum state
cannot be duplicated because of the no-cloning theorem it can
nevertheless be faithfully teleported.
\smallskip\\
{\bf Comment:} I really like the idea of saying some things about
teleportation and extracting something deep or expository from it,
but I had a hard time trying to figure out what you were getting at
here.  Moreover, I have to admit of being a little suspect that we
yet understand what it is that is peculiarly quantum mechanical
here. On the one hand, take two nonorthogonal Liouville
distributions; by Koopman's theorem there is a good sense in which
they cannot be cloned (this issue is what spurred my no-broadcasting
paper.  On the other hand take the paper by {\Cerf}, {\Gisin}, and
Massar, {\tt quant-ph/9906105}.  If instead of entanglement, Alice
and Bob are correlated via a some local hidden variable theory, they
can still teleport an unknown Liouville distribution for those hidden
variables for only 2.19 bits on average.  It's not quantum
teleportation for sure, but what's the difference of principle that
would make quantum teleportation illustrative of ``the peculiar
nature of the quantum state?'' Though I'll be trimming this passage
out of the concise version of the draft I'm also sending you, I did
make some changes in this paragraph.  So I'll go ahead and leave it
all in this footnoted version.
\bigskip\\
{\bf Quote from Draft:} It may seem paradoxical that in order to
completely specify the polarization of a photon, we need three real
numbers (represented by one point in the {\Poincare} sphere), but to
teleport that state from one photon to another with the help of a
previously shared entangled pair, only two bits of classical
information have to be transferred.
\smallskip\\
{\bf Comment:} Sentences combined.  I didn't understand what you
were wanting to express by entanglement ``contains no information at
all.''
\bigskip\\
{\bf Quote from Draft:} The difference is that if the state of the
photon is explicitly known, we can produce an arbitrarily large
number of photons with that state.
\smallskip\\
{\bf Comment:} Just for one last emphasis: suppose a pure state were
a pretty good approximation to what we could say about all the
observables associated with the radius of the universe. How would we
produce arbitrarily large numbers of that state?
\bigskip\\
{\bf Quote from Draft:} On the other hand, the unknown state that is
teleported disappears from the particle to which it ``belonged,''
and still remains unknown, even though we are sure (in an ideal
experiment) that another particle acquires that state.
\smallskip\\
{\bf Comment:} I put quotes around ``belonged'' because that was a
pretty blatant slip into objectivist language. Do you see what I
mean?

I hope that you now agree that the following passage was superseded
by all the previous discussions.  ``It thus appears that a ``state''
is like a recipe. Once it is defined, we can prepare as many systems
as we wish according to that state. Indeed, the simplest way to
understand the notion of state is to relate it to a preparation
protocol. However, it is not always possible to do so. When we
observe photons that originated in a distant star, no one was there
to prepare them millions of years ago. How shall we define such a
naturally prepared state? A more general (though more abstract) way
of defining a state is to specify the probabilities of the outcomes
of all potential observation processes. These probabilities
determine the best strategy for our future actions.''
\bigskip\\
{\bf Quote from Draft:} It is curious that some people are willing
to abandon the objective nature of the world we see around us with
its meaningful experiments and detector clicks, and yet work so hard
to retain the abstract quantum state as something real.  There is a
temptation to believe that each quantum system has a wavefunction
even if no one knows it.
\smallskip\\
{\bf Comment:} Stripped out ``The latter may not be known to any
physicist; if its value is needed for further calculations, one
would have to make reasonable assumptions about it, just as in
classical statistical physics. However, conceptually, the state
vector of any physical system would have a well defined, objective
value.''
\bigskip\\
{\bf Quote from Draft:} Most likely, the root of that temptation is
that in classical physics one has a ``state space,'' the points of
which can be assumed to correspond to an objective reality, and in
quantum mechanics one again has a ``state space'' (only this time a
Hilbert space).
\smallskip\\
{\bf Comment:} The roots of realism. I suspect that if one were to
do a thorough study of the Aristotelians, on would find that they
had no such compulsion.

\section{23 November 1999, \ ``Airport Time''}

I'm waiting at the Albuquerque airport for my mom and stepfather to
arrive.  They'll be staying with us until Friday when I depart for
Naples and they depart back to Texas.  This is only my mom's fourth
flight ever so, and she dreads flying like the dentist.  So I keep
my fingers crossed that they'll emerge from the plane confident and
comfortable.  If all is on time, they'll be here in 15 minutes.

So chances are, I won't be able to touch our draft again until
tomorrow evening.  So I hope you can be patient with me!  But I
would like to have it all tidied up before my European trip.

\bap
I restored the teleportation paragraph (in a more appropriate
location) because I think it's important.
\eap

I very much endorse trying to find some way of making quantum
information relevant to understanding quantum mechanics in its
entirety.  So I'm not put off by this as I tried to express.  The
main hurdle is to get me to understand the significance of what you
wrote.  Perhaps after looking at your new version, I'll feel more
comfortable.  Do you have a rejoinder for why the ``classical
teleportation'' of {\Gisin} doesn't make the same point?  (I'm too
weary to open up your bigger files right now; it's easier to just
babble my own thoughts.)  Of course, maybe classical teleportation
does make the same point (only quantitatively different):
probability distributions are states of knowledge without objective
reality just as quantum states are.  I will think about this more.

\bap
I think that {\Gleason}'s proof necessitates some continuity
assumptions.
\eap

No, {\Gleason} really does NOT need to assume continuity, that is
what makes his theorem so remarkable!  As I said, I've worked through
every step of it (in {\Pitowsky}'s simplified version).  Have a look
at:
\begin{enumerate}
\item
R.~Cooke, M.~Keane, and W.~Moran, ``An Elementary Proof of
{\Gleason}'s Theorem,'' Math.\ Proc.\ Camb.\ Phil.\ Soc.\ {\bf 98},
117--128 (1981).

\item
I.~{\Pitowsky}, ``Infinite and Finite {\Gleason}'s Theorems and the
Logic of Indeterminacy,'' J. Math.\ Phys.\ {\bf 39}, 218--228 (1998).
\end{enumerate}

Even {\Gleason}'s original paper did not assume continuity of the
frame functions:  it proved it.  The most difficult part of the
theorem is proving continuity:  it turns out to just be a wonderful
property of $R^3$ (the reals in 3-D).  {\Meyer}'s example on {\it
rational\/} vector spaces does not contradict this.  On rational
spaces, one cannot make the same set of moves (and build up the same
graph structure) as {\Gleason} did.

My mom's plane is here.

\section{25 November 1999, \ ``Apologies''}

Thank you for the Thanksgiving day wishes.  I apologize for my
hiatus.  I had hoped to get work done with my mother here, but it
just hasn't happened.  I hope you won't mind waiting too much for
one more day.  I will send you the next draft from the Chicago
airport tomorrow afternoon.  I must run now.

\section{26 November 1999, \ ``Too Late to Abort?''}

I'm in the Chicago airport waiting for my flight to Naples.  In the
next note I will send you the latest draft.  I feel like we are
converging:  I didn't make so many changes this time.  The draft
contains a hybrid notation for the changes: as you, I used \P, \P\P,
and \P\P\P, but for me the number of \P's represented simply (my
subjective assessment of) the magnitude of the change.  Also I put a
few footnotes.

I have no idea whether I will be able to make email contact from the
conference.  If you make no major changes, perhaps it would be best
for you to submit to {\Benka} and get the process started with PT.
You and I can always agree upon the minor tweakings later.  If you do
make significant changes though, perhaps we should risk it that I
will have some email contact and I will be able to review what you
have done.

By the way, I have no problem if we ask our colleagues for their
opinions.

The Naples conference is a resort hotel on the island of Ischia.
Next Saturday through Tuesday morning I go to Rome to visit
Francesco {\DeMartini}.  Tuesday evening through Dec.\ 12 I will be
in Montr\'eal.  So the only days I could possibly be out of complete
email contact is my time in Naples.

\subsection{Parsing the Paper}

\noindent {\bf Quote from Draft:} To start, let us assess the goals
of experimental science. An experiment is an active intervention
into the course of Nature: we set up this or that experiment to see
how Nature reacts.  We have learned something new when we can
distill from the accumulated data a compact description of all that
was seen and an indication of which further experiments will
corroborate that description.  This is what science is about. If from
such a description we can {\it further\/} distill a model of a
free-standing ``reality'' independent of our interventions, then so
much the better. Classical physics is the ultimate example in that
regard. However, there is no logical necessity that this worldview
always be obtainable. If the world is such that we can never
identify a reality independent of our experimental interventions,
then we must be prepared for that too.
\smallskip\\
{\bf Comment:} I thought there were too many ``interventions'' so I
went to ``intrusions'' even though that didn't seem wholly
satisfactory (I couldn't find another good synonym).  But I don't
like ``tests'' in this position either: to me (and I'm guessing
others) there word ``test'' conveys a sense of revealing a truth
value---just precisely the sort of thing we want to get away from
here.  When we test our students, we ascertain whether they have
studied or not.  So I reinstated ``interventions''; presently at it
doesn't seem like the word has been overused and I like the sentence
better now.
\bigskip\\
{\bf Quote from Draft:} The thread common to all the nonstandard
``inter\-pretations''\,---\,which actually are alternative
theories\,---\,is the desire to create a theory with features
corresponding to some reality independent of our experiments.\P\ As
we see it, trying to fulfill a yearning for a classical worldview by
inserting into the theory extra hidden variables, extra worlds,
extra kinds of time evolution, or extra consistency rules, while not
at the same time improving its predictive power, is utterly
counterproductive.
\smallskip\\
{\bf Comment:} This gives a stronger feeling that these extra things
are foreign and should be viewed as such.
\bigskip\\
{\bf Quote from Draft:} These probabilities are computed by using
any available information. This includes, but is not limited to,
information about a system's preparation. The apparatus for turning
that information into statistical predictions is formally expressed
by the probability rule postulated by Max {\Born} at the time quantum
mechanics was invented.
\smallskip\\
{\bf Comment:} I didn't like ``how to convert'' in this place
because it seemed far too specific. Generally there is no clean cut
algorithm for turning all the gathered information into a quantum
state assignment; that is still something of an art.  This is what I
tried to capture the last time when I used the phrase ``the latitude
with which such information can be incorporated.'' The quantum rule
only sets the limits; the rest is the experimentalist's art.
\bigskip\\
{\bf Quote from Draft:} We do not deny the possible existence of an
objective reality independent of what observers perceive.
\smallskip\\
{\bf Comment:} I cannot agree with either of your two new phrases:
``in particular, the `detector clicks' are definitely objective'' and
``they remain real even if no observer is there.'' The problem is
that, as we say later, ``detector clicks'' come about when we leave
something out of the picture, when our information is incomplete. If
we leave the observer out of the picture, I don't know what a
detector click is. We are on safer ground if we leave ``objective
reality independent of our interventions'' alone.  Part of what you
are hoping to get at might be better described as an
``intersubjective reality'' \ldots\ but still I think we should leave
it all alone.  I rearranged the part of your new stuff that I could
agree with to near the bottom of the paragraph.
\bigskip\\
{\bf Quote from Draft:} The peculiar nature of a quantum state as a
``state of knowledge'' can be illustrated by the teleportation
process.
\smallskip\\
{\bf Comment:} If we say it like this, I think I can accept it. The
{\Gisin} and company ``classical teleportation'' was pretty powerful
for me. Also inserting this phrase helps make connection to the
previous paragraph.
\bigskip\\
{\bf Quote from Draft:} In order to completely specify the
polarization state of a photon, we need three real numbers (one
point in the {\Poincare} sphere). Yet to teleport that state from one
photon to another\,---\,with the help of a previously shared
entangled pair\,---\,only {\it two bits\/} of classical information
have to be transferred. The difference is that when a preparer knows
the state of a photon, he can produce arbitrarily many further
photons with the same state.
\smallskip\\
{\bf Comment:} You wrote, ``{\bf ANSWER} It's just a matter of
money. As {\Archimedes} answered to the king: give me a firm support,
and I shall move the Earth.''  {\it I didn't find that answer
adequate!!}
\bigskip\\
{\bf Quote from Draft:} On the other hand, the unknown state that is
teleported disappears from the original photon.  It remains unknown
to the teleporter even though, with respect to the original
preparer, another photon acquires precisely that state.
\smallskip\\
{\bf Comment:} I thought it important that we made more explicit
that there are two points of view here.  Hence I was explicit about
what came from what point of view.
\bigskip\\
{\bf Quote from Draft:} It is curious that some well intentioned
theorists are willing to abandon the objective nature of
``observables,'' and yet wish to retain the abstract quantum state
as a surrogate reality.
\smallskip\\
{\bf Comment:} Why don't we just try it without the phrase ``(that
is, potentially observable properties)?''  The thing I'm having
trouble with is the word ``properties''---I no longer think of
quantum systems as having properties independent of our
interventions.  You made the mistake of inventing a word that I
could finally feel comfortable with (i.e., interventions), and now
you have to pay the consequences!!
\bigskip\\
{\bf Quote from Draft:} Of course, all this is {\it not\/} to say
that there are no fundamental questions still to be explored within
quantum theory. One can ask for better clarification of the features
in our world that compel us to this theory instead of another or any
number of other questions.  However, to make quantum theory a useful
guide to the phenomena about us and a self-contained theoretical
edifice, we need nothing more than we already have. Quantum theory
needs no ``interpretation.''\P\P\P

\section{07 December 1999, \ ``Took Care of Bureaucracy''}

You'll notice in one of my notes either before or after this one
that I did the final paper work for your symposium next year.  You
had previously asked for titles and abstracts for my talks; you will
find them within that message.

I feel that I {\it should\/} have been working on our paper today
while flying, since I was able to retrieve the draft yesterday.
However, I found that it was easier to get my head around the many,
many bureaucratic matters that had piled up in my email absence; so I
spent my energy in that direction.

My interaction with {\DeMartini} was quite fruitful and I am glad
that I went.  I got to see many nice Alice and Bob's on the
experimental tables!  And I was able to propose various experiments
that he might perform related to work that I had done with
{\Charlie} {\Bennett} and John {\Smolin}.  Also there is an
experiment that one can imagine for quantum state tomography related
to this quantum de {\Finetti} theorem of ours.  I was really pleased
to find that both Enrico {\Fermi} and Bruno de {\Finetti} had
actually been on faculty at that university.

However the meeting in Naples might have been a waste of the
taxpayer's money!  (Tax money from the European Union, that is, not
the US\@!  My funding was completely covered by the organizers).  The
meeting was advertised as science, but as far as I could tell most
of the attendants might have just as well belonged to the priestly
set and wore white robes.  The two main contingencies were the
followers of GRW and the Bohmians.  (The Bohmians were the larger.)
In all their talks and discussions, it seemed to me that they said
nothing and listened to nothing.  I was the lone hold-out at the
conference for a Copenhagen-like sensibility.  I gave a talk trying
to show that quantum information theory only deepens our
understanding in that regard, and, in fact, may be regarded as a
much more effective way of exploring the foundations of quantum
mechanics.  I first gave a little of a philosophic spiel, then I
focused on three significant examples (teleportation, our
information-disturbance relations, and the BDFMRSSW nonlocality
without entanglement).  My message fell on deaf ears!  And the thing
that really surprised me is how so few of them were even interested
in the physical phenomena themselves despite foundational issues.  A
particular thorn in my side was David {\AlbertD}:  he is arrogant and
mostly worthless as far as I could tell.  (But he did have a
penchant for hearing himself speak!)  I found him most annoying when
he branded our point of view about quantum mechanics---precisely the
point of view in our {\sl Physics Today\/} article---as
unscientific!  (I wanted to say, ``Then why is it that I am employed
as a physicist, but you are employed as a philosopher?'')

But there were some small highlights to the meeting.  I had many
chances to talk to Adrian {\Kent}, Jeremy {\Butterfield}, Orly
{\Shenker}, Chris {\Dewdney} (who is a Bohmian, but with a more open
mind), and Jos {\Uffink} (I especially enjoyed his input).

I now continue flying for several hours.  (I am stuck in the Chicago
airport for over four hours I believe.)  I will try to make some
progress on our project if I can fight the exhaustion.  In any case,
you should expect a revised draft from me no later than Thursday.

\section{11 December 1999, \ ``Productivity''}

This conference has turned out to be quite a productive affair. It's
so good to be back with real scientists.  I've been making changes
here and there to our draft when I've had a chance.  I think we're
getting into the final leg.  However, I am sorry to keep you waiting
so long.  Tomorrow I will be in planes and airports most of the day
(Montr\'eal, Chicago, Dallas, Albuquerque); I should be able to put
all my final touches on the draft during that time.

Do you have any plans to go to the Benasque meeting this summer?  It
falls straight on top of QCM.  (I have also heard that Sandu is
organizing a meeting during that same time, but I don't know where.)

Today, I meet with the mathematician Mary Beth {\Ruskai}.  (Her
career was made by proving with {\Lieb} the property of quantum
entropy known as strong subadditivity.  It had been an open question
for 10 years before that.)

Last night it snowed here.  Everything is beautiful.  I'm starting
to get so homesick for {\Kiki} and {\Emma}.  (That always happens to
me when I see a beautiful scene and she is many miles away.)

\section{12 December 1999, \ ``Thinking It's a Toddler''}

Now I'm starting to think its a toddler.  Let us see what you
think.  I just did a precise count on the words and it came to 1790
words in the text alone (i.e., not counting title, biographies, or
references).  That's cutting the limit very close.  But I think the
draft is starting to be very good now, something we will both proud
of some time from now.

The biggest changes you will notice are in the teleportation
paragraph and the closing paragraph.  I have thoroughly documented
all my thoughts and reasons for making changes in footnotes.

Let me bring up one more change that I have not yet inserted.  You
wrote:

\bap
Incidentally, {\Vaidman} also wrote to me that Cathy's digestion was
not an irreversible process, and I privately agreed with him on this
point, but explained that we didn't want to open a second front, and
we had constraints on the length of the text.
\eap

I understand that you don't want to open up a new front, but I think
{\Vaidman}'s reaction will be the general one.  I have in the past
heard Howard {\Barnum}, John {\Preskill}, and David {\DiVincenzo} all
say precisely the same thing.  I think it is much more likely that by
not saying anything about the issue we will by default open up a new
front.  We could of course defend ourselves when we write our reply
to critics, but don't you think it would be so much better if we
expressed the complete consistency of our point of view in the
article proper?  I think the article would be better for it, and
demonstrate too that we have a well thought out position.

I'm sure we could fix the problem with just two sentences.  For
instance, something like:
\bq
\noindent
If Erwin had chosen not to take a look that would be fine too.  With
so much information about Cathy and her digestion, he could in
principle reverse the whole chain, wiping Cathy's stomach and mind
of all memory of her tasty treat.
\eq

The problem is that that's 41 more words.  But surely we could try
to push the limits a bit.  We could just wait and see what happens
after the editors start wielding their knife.  I am inclined to do
that.  In my mind, the manuscript has now reached a certain stage of
completeness that I wouldn't want to see it toyed with (at least
voluntarily on our sides).  I like the way it reads now.

Please do think about this.  (I've been pushing for this point since
we started the project:  Lev's criticism has only reactivated my
worry.)

OK, now I must act like a good husband and spend some time with
{\Kiki}.  I'll be in touch with a report of {\Mermin}'s further
thoughts sometime tomorrow.

\subsection{Parsing the Paper}

\noindent
{\bf Quote from Draft:}
The thread common to all the nonstandard ``inter\-pretations''\,%
---\,some would say alternative theories\,---\,is the desire
to create a theory with features corresponding to some reality
independent of any potential experiments.
\smallskip\\
{\bf Comment:} I thought ``look like'' was too wishy-washy. You
might remember that I had once deleted this whole phrase because I
thought it would cause trouble, but then you reinstated it.  After
you did that, I actually grew to like it a lot.  It is simply a
mistake to think that MWI is the same theory as the quantum
mechanics you and I use; it is a contentless structure, whereas
quantum theory is a meaningful guide to our experiments and
engineering. What could it mean to say I ``choose'' this or that
experiment in MWI? Why not give up on life and just let the
universal wavefunction carry us along?  I have no great trouble with
creating controversy with this article. We should be brave and say
what needs to be said. If you are not averse to reinstating your old
stronger statement, I am not; I certainly prefer your old
formulation over my present one. Damn the torpedoes, I say, full
speed ahead!  As you said before, we can come out kicking again in
our inevitable ``reply to critics.''
\bigskip\\
{\bf Quote from Draft:} As we see it, trying to fulfill a yearning
for a classical worldview by inserting into the theory extra hidden
variables, extra worlds, extra kinds of time evolution, or extra
consistency rules, while not at the same time improving its
predictive power, is \smallskip utterly counterproductive. \\ {\bf
Comment:} Now, here's where I would weaken the grip. It is our {\it
opinion\/} that it is {\it utterly counterproductive\/} to attempt
to insert a classical worldview.  (I know from last week in Naples
that {\Duerr} and {\Goldstein}, for instance, think it is extremely
productive: at the very least it gets them enough funding to hold a
conference every two years.) I know you don't like the phrase ``as
we see it,'' but I think we need something here that carries the
same flavor. Also notice that a sentence starts with ``however''
just four sentences above---it conveys a little too much a
see-sawing feeling to the read to me.
\bigskip\\
{\bf Quote from Draft:} Quantum probabilities, like all
probabilities, are computed by using any available information.  In
the quantum case this includes, but is not limited to, information
about a system's preparation. The mathematical instrument for
turning that information into statistical predictions is the
probability rule postulated by Max {\Born} \smallskip at the
invention of quantum mechanics. {\bf Comment:} Good point.  So I
went one step further and made it clear that we are speaking of a
``mathematical instrument.'' Also I shortened the sentence just
slightly, by rearranging the ending phrase to flow a little easier.
\bigskip\\
{\bf Quote from Draft:} The conclusiveness of {\Born}'s rule is known
today to follow from an important theorem due to Andrew {\Gleason}.
So long as one assumes that yes-no tests on a physical system are
represented by projection operators $P$, and that probabilities are
additive over orthogonal projectors, then there exists a matrix
$\rho$ such that the probability of a ``yes'' answer is ${\rm
tr}(\rho P)$. The compendium of probabilities represented by the
``quantum state'' $\rho$ captures everything that can meaningfully
be said about the physical properties of the system.
\smallskip\\
{\bf Comment:} I still don't know what the phrase ``depending solely
on that system (not on the apparatus that probes it)'' can possibly
mean.  If one of the author's can't understand the phrase, then the
innocent readers will surely have trouble.  The quantum state cannot
depend on the system at all, as I understand the meaning of the word
``depend.'' This is because, as we have already agreed, the quantum
state is not an {\it objective\/} property of the system itself.
It's assignment depends only on the information that has been
gathered (by whatever means). So then what does it mean to say that
the quantum state depends on the system? Well certainly it does, if
by that one means the dimensionality of the density matrix:
presumably the dimensionality of the system of interest is an
``objective'' property.  But I can't see anything beyond that.
Please explain what you are trying to get at in more detail, or
let's just drop the phrase.  The present wording seems perfectly
adequate to me and less likely to create confusion.
\bigskip\\
{\bf Quote from Draft:} The peculiar nature of a quantum state as
representing information is strikingly illustrated by the quantum
teleportation process. In order to teleport a quantum state from one
photon to another, the sender (Alice) and the receiver (Bob) prepare
a pair of photons in a standard entangled state. The experiment
starts when Alice receives another photon whose state is unknown to
her. She performs a measurement on her two photons, and then sends
Bob a {\it two bit\/} classical message instructing him how to
exactly reproduce the unknown state on his photon. This economy of
transmission is remarkable, because to completely specify the
polarization state of a photon, we need three real numbers (one
point in the {\Poincare} sphere), that is, an infinity of bits. The
difference is that as far as Alice and Bob are concerned, nothing
whatsoever changes about Bob's system:  they start off maximally
ignorant about it and remain maximally ignorant about it when the
process is complete. The two bits of classical information transfer
only signify the cost of transferring the preparer's information
about one of Alice's systems\,---\,i.e., the ``unknown''
state\,---\,to Bob's previously correlated system. Nothing physical
is transferred.
\smallskip\\
{\bf Comment:} I understand that Eugen endorsed this paragraph with
the words ``much clearer now,'' but I have to admit that I still do
not get the point you are trying to make here. (It should always be
the job of the author's to be more critical of the words than the
casual reader.) For instance, I really don't see what it has to do
with the remarkableness of teleportation that if we have a system in
a pure state, we might be able to make many {\it other\/} copies of
that state.  It just doesn't click for me---and that's a minimal
requirement.  So let me try to rewrite the passage so that I can
understand the import of it. However, I'll keep the old words here
so that we can reuse parts of them if need be:  `` The difference is
that when we know the state of a photon, we can produce arbitrarily
many photons with that state. On the other hand, the unknown quantum
state that is teleported disappears from Alice's laboratory. It is
also unknown to Bob who receives it. Yet, the original preparer who
knows that state can verify that it is correctly transferred to
Bob's \bigskip photon.''

\noindent {\bf Quote from Draft:} It only gives the evolution of our
probabilities for the outcomes of potential experiments on the
system.
\smallskip\\
{\bf Comment:} In agreement with Eugen, I think it is best to
minimize the usage of the loaded term ``measurement.'' Hence I put
``experiment'' into this sentence since we had defined it precisely
previously.  Note also that I deleted the last sentence (that I had
added in the previous draft).  I decided that it didn't have such a
transition value after all.
\bigskip\\
{\bf Quote from Draft:} All this said, we would be the last to say
that the foundations of quantum theory were not worth further
scrutiny.  For instance, it may be quite useful to search for
various minimal sets of physical assumptions that give rise to the
theory. Since it is not yet understood how to combine quantum
mechanics with relativity and gravitation, there may well be insight
to be gleaned there. However, to make quantum mechanics a useful
guide to the phenomena about us and a self-consistent theory, we
need nothing more than we already have. Quantum theory needs no
``interpretation.''
\smallskip\\
{\bf Comment:} OLD PARAGRAPH: All this is {\it not\/} to say that
there are no fundamental questions still to be explored within
quantum theory. How to combine it with relativity and gravitation is
not yet satisfactorily understood. Another issue is whether any
quantum state or operator that we know to write can be realized, in
principle, experimentally. However, to make quantum mechanics, as we
know it, a useful guide to the phenomena about us and a
self-contained theoretical edifice, we need nothing more than we
already have. Quantum theory needs no ``interpretation.''
\smallskip\\
{\bf Comment:} How about a compromise here?  The thing I didn't like
about your examples is that both seemed to be more about
difficulties in {\it application\/} of quantum theory, rather than
its interpretation.  I think it is important that the reader
understand that we are not doomsayers about the
interpretation---that is to say, we are not completely rigid in the
way that {\Bohr} and {\Rosenfeld} were. A substantial number of
{\Peres}'s papers in the last 15 years have been precisely about
clarifying the foundations.  It has been a fruitful pursuit
precisely because the foundations are not yet a closed book. Flesh
and bones still need to be added even to Copenhagen-like ideas. So I
made what I was trying to get at a little more concrete by changing
the wording and adding two references of serious works. Then I gave
an even more concrete reason for pursuing that: namely precisely
your point about relativity and gravitation. Now I like the ending
even better; I hope you will agree.

\section{14 December 1999, \ ``Fear of Numbers''}

Please don't be frightened by the Annotated Toddler.  Despite the
number (and size) of footnotes, I think we are reaching rapid
convergence.  I think essentially everything is now stable (at the
very least readily negotiable) except the paragraphs on
teleportation and reversibility.  I expect though that we'll also
have all that cleared up soon, probably tomorrow.  I've given them
my best shot again.  Let's see what you think about this iteration.

This afternoon, {\Kiki}, {\Emma}, and I trudged through the snow to
find a Christmas tree.  For \$10, Los Alamos County gives a permit
that lets you cut anything on the mountain that is not over 10 feet
tall.  It was quite fun, and I can see that it will turn into a
family tradition.  We bundled up {\Emma} quite well (in a miniature
ski suit) and she seemed to take in every scene.  The only problem
now is that tonight she has developed a deep cough (though it is
probably unrelated).  We are a little worried, however, because
whooping cough is going around; we just read in the paper this
morning that 42 cases have been reported in the last month.  {\Kiki}
will take {\Emma} to the doctor tomorrow.

Thanks, by the way, for the stories about your grandchildren.  I
sometimes do silly things like rub a rubber coaster over {\Emma}'s
hair and then watch the hair rise from the static.  {\Kiki} says,
``She's not a science experiment!''

\subsection{Parsing the Paper}

\noindent
{\bf Quote from Draft:} The thread common to all the nonstandard
``inter\-pretations'' is the desire to create a new theory with
features corresponding to some reality independent of our potential
experiments.
\smallskip\\
{\bf Comment:} ``Any'' $\rightarrow$ ``our.''
\bigskip\\
{\bf Quote from Draft:} These attempts to fulfill a classical
worldview by inserting into quantum mechanics extra hidden
variables, extra worlds, extra kinds of time evolution, or extra
consistency rules, while not at the same time improving its
predictive power, only yield the illusion of a better
understanding.
\smallskip\\
{\bf Comment:} I took out the phrase ``a yearning for.'' The
sentence seems simpler this way without changing its meaning. Also I
replaced ``into the theory'' with ``into quantum mechanics'' because
I didn't want there to be any ambiguities about which theory we are
referring to.  Also, by the way, I think this now boils down to a
much subtler way of saying that the other ``interpretations'' are
not quantum mechanics---so we get our cake and eat it too.
\bigskip\\
{\bf Quote from Draft:} The compendium of probabilities represented
by the ``quantum state'' $\rho$ captures everything that can
meaningfully be said about a physical system.
\smallskip\\
{\bf Comment:} I am ashamed to admit that the phrase ``physical
properties of the system'' has made it past my sieve in all these
iterations.  This is precisely the sort of objectivist language that
we are trying to move past with this article: unperformed
measurements have no outcomes, systems have no ``properties.'' There
are only the macroscopic consequences of our potential
interventions. We need to be consistent on this point.
\bigskip\\
{\bf Quote from Draft:} Cathy's story inevitably raises the issue of
irreversibility. As quantum dynamics is time-symmetric, can Erwin
undo the process? Again, why not?  For this, Erwin would need
complete control of all the degrees of freedom in Cathy's
laboratory, but that is a difficulty of practice, not of principle.
The main point is that the information Erwin possesses has to do
with the potential consequences of his experiments.  The information
is not about what is ``really there.'' If Erwin performs no
experiments, there is no reason he cannot reverse Cathy's
evolution.  If, on the other hand, he intervenes on the her unitary
evolution, then he forfeits the capability: he too will have become
entangled with the laboratory and all its contents.
\smallskip\\
{\bf Comment:} OLD SENTENCES: ``This is obviously far beyond Erwin's
ability. He has no choice but to consider Cathy as a macroscopic
object that is part of an ``effective'' reality, as defined above.''
\smallskip\\
{\bf Comment:} Boy, do we diverge in opinion on this one!! The
problem is not at all one about Erwin using a less than complete
description (the kind of thing that creates the macroscopic reality
of detector clicks). In fact that would be inconsistent with our
previous explanation about Cathy being in a 50/50 superposition.
There is nothing to stop us from imagining that Erwin has absolutely
maximal information about Cathy's laboratory, i.e., that he
describes it all via some large entangled PURE state. The problem
instead stems only from a holdover of the objectivist training of
most physicists. The knowledge Erwin possesses has to do with the
potential consequences of his experimental interventions on Cathy's
laboratory. It doesn't mean anything more than that.  In particular,
it has nothing to do with what is ``really there,'' whatever that
might mean (i.e., like Cathy really eating the cake, or really
eating the fruit). That is the only consistent position. If Erwin
doesn't intervene on Cathy's laboratory with a ``measurement'', then
there is no reason that he cannot reverse the evolution of
everything in her laboratory.  What was your Archimedean quote? (Now
I'm being facetious.)  If on the other hand, he does intervene on
the laboratory with a measurement then he has forfeited all rights
to reverse her evolution:  he has now become entangled with that
system (to use a concept from the Church of the Larger Hilbert
Space).  We can only completely control the evolutions of systems to
which we are completely external.

I propose that we condense one or two sentences from what I just
wrote above in this footnote to fill the rest of the paragraph.  Let
me try my hand at it. I will do this in a tentative way, that you
will likely want to make clearer.  The main point is that we need to
get the argument right.

Anyway, for extra reference, let me record here how I put it to
Howard {\Barnum} when he had the same problem as Lev: [See the note
to Howard {\Barnum}, titled ``It's All About Schmoz,'' dated 30
August 1999.]
\bigskip\\
{\bf Quote from Draft:} The peculiar nature of a quantum state as
representing information is strikingly illustrated by the quantum
teleportation process. \ldots\  The conundrum is solved by realizing
that as far as Alice and Bob are concerned nothing changes about
Bob's photon: they end up describing it by the same density matrix
they started with, the completely depolarized state. The two bits of
classical information serve only to transfer the preparer's
information to be from being {\it about\/} the original photon to
being {\it about\/} the one in Bob's possession. This happens
because of the previously established correlation between Alice and
Bob's photons.
\smallskip\\
{\bf Comment:} I have trouble agreeing with the sentences you wrote
here. You write ``this complete specification is not transferred.''
But it IS, for the preparer! It is just a question of with respect
to whom that the specification is transferred.  That is the thing
that needs to be made clear to give the example of teleportation any
relevance to our quest.  In any case, I don't think the unprepared
reader will be able to pick this up from the description you wrote.
So let me try rewriting this again---somewhat along the lines of my
proposal yesterday---but this time I'll be careful not to introduce
any technical jargon that the whole physics community does not know.
You were right that my last attempt was not so good; I hope to do
better this time. OLD SENTENCE: ``However, this complete
specification is not transferred: the unknown quantum state that is
teleported disappears from Alice's laboratory, and it is also
unknown to Bob who receives it. Yet, the original preparer who knows
that state can verify that it is correctly transferred to Bob's
photon.''

\section{14 December 1999, \ ``Lost Toddlers''}

I see from your pile of notes to me that somehow you didn't receive
my toddlers until long after I had sent them.  It is strange how
email can disappear for so many hours.  You just have to wonder
where it is sitting (or bouncing) all that time.

{\Emma} coughed a deep cough all night.  Looking back at the note I
wrote you last night, I realized that I didn't mention that the 42
cases where all in this small county in the mountains.  I think it
is unlikely that she has whooping cough (it sounds to be more a
problem of chest congestion), but still I worry.  The curse of new
parenting?

I forgot to answer your question about the abstract of the
abstract.  What you have written is just fine.  I gave a slightly
more technical version of that talk in Montr\'eal Saturday.  I was
the second-to-last talk of the conference, so I didn't really expect
people's full attention, but still it flowed quite smoothly.  I was
left believing a counterfactual:  if the audience had not been
exhausted, then the talk would have been received well!  Actually, I
got some nice questions from Dan {\Gottesman}, Hoi-Kwong {\Lo}, and
Michael {\Nielsen}.  The main thing that made me think the talk
wasn't a waste was that David {\DiVincenzo} came up to me afterward
and said that he now understood quite clearly what the problem was
all about.  That was a triumph!

Aha, I just saw you have sent me some more notes.  I will go to them
now.

\section{14 December 1999, \ ``Training Pants?''}

Another day, another draft.  I again follow my usual protocol:  one
draft with footnotes, one draft without.  The draft with footnotes
gives all the details of what I was thinking as I made changes.  I
keep my fingers crossed that we will reach complete agreement soon.
The main thing that makes me feel giddy is that with Physics Today,
our potential readership may be on the order of 50,000 people or
more!  You know that John {\Wheeler} had this idea that things become
more real as they become acknowledged by more and more members of
the ``community of communicators.''  I would hate one of the things
we said to become real if later we decided we didn't like it!

Once we finish this project, I hope to have a look at the
{\Brukner}/{\Zeilinger} paper that you didn't recommend.  I contacted
{\Anton} once by email saying that I thoroughly enjoyed his web
article ``On the Interpretation and Philosophical Foundation of
Quantum Mechanics.'' (You can find it on his homepage.)  He wrote me
back that he had something much more serious in the works, something
that would rigorize the point of view he expressed there.  It's an
article that appeared in Foundations of Physics last year (or maybe
this year); I suspect the PRL you didn't recommend is a follow-on to
that.  Well anyway it was not very good, and I couldn't believe that
the same man had written such an empty paper.  A curious fellow he
must be.

Tonight {\Kiki} and I decorate the Christmas tree we cut yesterday.
We've already placed it in the stand:  so far, it looks marvelous.
Neither of us are religious of course, but we still have faith in
Santa Claus.

\subsection{Parsing the Paper}

\noindent
{\bf Quote from Draft:} Our purpose here is to explain the internal
consistency of an ``inter\-pretation without inter\-pretation'' for
quantum mechanics. Nothing more is needed for using the theory and
understanding its nature.  To start, let us examine the role of
experiment in science.
\smallskip\\
{\bf Comment:} The word ``define'' seemed too haughty to me in this
context:  I thought it conveyed a bit of the flavor that we view
ourselves as the arbiters of what science should be. Anyway, the
more I thought about this sentence, the more talking about the
``goals'' didn't seem right either.  So I reconstructed the sentence
completely.  I think this one hits the mark a lot better. By the
way, completely as an aside, you might enjoy the historical article
(I did): M. {\Jammer}, ``The Experiment in Classical and in Quantum
Physics,'' in {\sl Proceedings of the International Symposium:
Foundations of Quantum Mechanics in the Light of New Technology},
edited by S. Kamefuchi (Physical Society of Japan, Tokyo, 1984),
pp.~265--276.
\bigskip\\
{\bf Quote from Draft:} The thread common to all the nonstandard
``inter\-pretations'' is the desire to create a new theory with
features corresponding to some reality independent of our potential
experiments. But, trying to fulfill a classical worldview by
inserting hidden variables, parallel worlds, nonlinear evolutions,
or consistency rules into quantum mechanics, while not at the same
time improving its predictive power, only yields the illusion of a
better understanding.
\smallskip\\
{\bf Comment:} At first I put a ``however'' here, but the word
seemed overused. Then I tried ``but'' and it seemed so much better:
it seems to give the sentence a much more accusatory feel. And I
like it this way.  My only worry was that you would not like a
sentence starting with ``but.'' So I looked in my {\sl American
Heritage}, and lo and behold it said the following: ``{\it But\/} may
be used to begin a sentence at all levels of style.'' I'm keeping my
fingers crossed that you'll like it too.
\bigskip\\
{\bf Quote from Draft:} Yet, it is legitimate to attempt to
extrapolate the theory beyond its present range, for instance when
we probe particle interactions at superhigh energies, or in
astrophysical systems, including the entire Universe. Indeed, a
common question is whether the universe has a wave\-function. There
are two ways to understand this. If this ``wave\-function of the
universe'' has to give a complete description of everything,
including ourselves, we get the same meaningless paradoxes as
above.
\smallskip\\
{\bf Comment:} Somehow the word ``same'' seems essential for
emphasis to me.  Take {\Vaidman} as an example:  he rejects
Bohmianism as a solution to the quantum measurement ``problem,'' but
embraces MWI. He sees one solution as better than the other, and
somehow doesn't see that they are both ridiculous.
\bigskip\\
{\bf Quote from Draft:} Does quantum mechanics apply to the
observer?  Why would it not?  To be quantum mechanical is simply to
be amenable to a quantum description. Nothing in principle prevents
us from quantizing a colleague, say.
\smallskip\\
{\bf Comment:} This small point may only have to do with our
difference in our culture or ages or the literature we read, etc.,
but adding a little extra phrase like ``say'' or ``for instance'' to
me somehow gives this sentence the flavor of being tongue-in-cheek.
Without it, the joke seems to fall a little flat.
\bigskip\\
{\bf Quote from Draft:} Cathy's story inevitably raises the issue of
reversibility since quantum dynamics is time-symmetric. Can Erwin
undo the process if he has not yet observed Cathy?  A superficial
reading of the story might lead some to think not.  That
misconception arises from forgetting that the information Erwin
possesses has only to do with the consequences of his potential
experiments.  The information has nothing to do with what is
``really there.''  If Erwin has performed no observation, then there
is no reason he cannot reverse Cathy's digestion and memories. Of
course, he would need complete control of all the degrees of freedom
in Cathy's laboratory, but that is a practical problem, not a
fundamental one.
\smallskip\\
{\bf Comment:}  YOUR LAST VERSION:  For this, he needs complete
control of all the degrees of freedom in Cathy's laboratory. His
difficulty is practical, not fundamental. In practice, Erwin has of
course only incomplete control and he must consider Cathy as a
macroscopic object that is part of an ``effective'' reality, as
defined above. (If Erwin observes Cathy, he forfeits the reversal
capability: he becomes entangled with her and only a
``superobserver'' can reverse the process.)
\smallskip\\
{\bf Comment:} OK, it looks as if we still have a disagreement on
some points:  some practical and some presentational. Let me try to
separate those points and be as clear as I can be. Your point about
what Nathan {\Rosen} said is well taken. So, I hope we can come to
some agreement on this (and thus label it physics):  I think our
readers will definitely benefit if we do it right.

1)  There is a sense in which this paragraph is completely redundant.
We have already said quite clearly that the wavefunction is
information about the consequences of interventions, not about what
is ``really there.''  Therefore no one should have the right to think
that our example about Cathy and Erwin implies that quantum mechanics
should become irreversible at some sufficiently high level. I believe
this may have been the reason you were reluctant to include this
subject in the first place.  However the plain fact is that people do
seem to misunderstand the point. Conversations and email with LV, HB,
JP, DPDiV, and possibly CHB, attest to that. The point is that people
don't listen so readily when they already have some preconceived
notion in their head, no matter how sincere they are. So as I deem
it, the main goal of this paragraph is to reiterate the point of
``knowledge about what'' in a slightly new context, that of
reversibility.

2)  You wrote in your accompanying note:  ``We don't disagree at all,
but were concerned by different issues. It is quite obvious that
after Erwin performs his measurement, there is no return to Cathy's
$|hungry\rangle$ state. The question is whether this is still
possible before Erwin measures.''  Actually, your second sentence was
not the foremost issue on my mind:  I added a discussion of that
because other people (like {\Barnum}) often speak sloppily of the
observer reversing his own measurement---that phrase is an oxymoron.
The foremost issue, as I see it, is to push people away from thinking
that our story implies an in-principle {\it irreversibility\/} before
Erwin measures, i.e., the same concern as you.  The only reason
people think that what we are talking about is irreversible is (in my
experience) because they have not yet fully absorbed the lesson that
quantum mechanics is not about some free-standing reality.

3)  I cannot agree with your sentence ``In practice, Erwin has of
course only incomplete control and he must consider Cathy as a
macroscopic object that is part of an `effective' reality, as defined
above.''  Previously we defined ``effective reality'' as having to do
with ignoring most degrees of freedom and considering a description
that is necessarily incomplete.  Of course, both of those things
could enter into our discussion of Cathy and Erwin's experiment, but
that is not the essential point as I see it.  We did not introduce an
incomplete description previously (i.e., in the first paragraph about
Cathy and Erwin) and I think it would strongly cloud the relevant
issues if we do it now. If we had done it previously we would have
had no right to talk of 50/50 superpositions---instead our whole
discussion would have had to be in terms of significantly mixed
density operators. The solution to {\Vaidman}'s question does not
come about because of Erwin's technical inability.  We should stick
to our guns: the theory is not about our knowledge of a free-standing
reality. Therein lies the solution to {\Vaidman}'s conundrum.

4) I did not like your introducing the phrase ``and only a
`superobserver' can reverse the process.''  {\Vaidman} would like to
think that we edge upon acceptance of MWI, and I think that would
only fuel his opinion.  I know that I mentioned in my last note to
you something about the Church of the Larger Hilbert Space, but I
would like to keep that short-hand private.  I don't like the Church;
the last I want to do is give it a public endorsement by acting as if
there were always a superobserver available.  If so, then why not
speak of the superobserver of the whole universe and his wavefunction
for it?

5)  Given all that, let me try to take another shot at closing the
paragraph.  If we still can't come to agreement (after perhaps your
next iteration), then maybe we should follow {\Rosen}'s dictum.
\bigskip\\
{\bf Quote from Draft:} The peculiar nature of a quantum state as
representing information is strikingly illustrated by the quantum
teleportation process. In order to teleport a quantum state from one
photon to another, the sender (Alice) and the receiver (Bob) prepare
a pair of photons in a standard entangled state. The experiment
starts when Alice receives another photon whose polarization state
is unknown to her.  She performs a measurement on her two photons,
and then sends Bob a classical message of only two bits, instructing
him how to reproduce the unknown state on his photon. This economy
of transmission appears remarkable because to completely specify the
state of a photon, namely one point in the {\Poincare} sphere, we
need an infinity of bits.  However, the conundrum is only apparent.
As far as Alice and Bob are concerned nothing changes about Bob's
photon; they describe it by the same density matrix throughout the
whole process. The two bits of classical information serve only to
transfer the preparer's information to be from being {\it about\/}
the original photon to being {\it about\/} the one in Bob's
possession. This is not magical: it comes about by the previously
established correlation between Alice and Bob.
\smallskip\\
{\bf Comment:} WORDS OF THE LAST DRAFT:  ``However, quantum
teleportation does not involve such an explicit specification.
Neither Alice nor Bob can know the state that is teleported. Yet, the
original preparer who knows that state can verify that it was
correctly transferred.''
\smallskip\\
{\bf Comment:} Believe it or not, I probably spent over an hour or
an hour and a half composing the sentences that I put in this place
last time.  That, of course, does not mean that they should not be
discarded if they don't convey the proper information!  But I did
think hard about what it would take to make the significance of the
example clear to myself, and then how to translate that to the
layman.  I just can't believe that the uninitiated reader will
understand what you are really getting at if the point is made in
such a terse manner:  I use myself as an example, I didn't
understand it at first.

You write, ``quantum teleportation does not involve such an explicit
specification.''  But, again, it does for the original preparer; it
only doesn't for Alice and Bob.  That point needs to be made clear.
You write, ``Neither Alice nor Bob can know the state that is
teleported.'' True, but what does that have to do with ``strikingly
illustrating the peculiar nature of a quantum state as representing
information'' in the context of teleportation?

Finally, if one does not emphasize that the effect comes about
because of the previously established correlation---the resource the
preparer uses to update his knowledge of Bob's system---then
teleportation does indeed look like some mysterious superluminal
transportation of the state.  And that annuls the point of the
example.

Please give my words from yesterday (modified a fair amount now) one
more shot and see if you cannot find a kernel of a good explanation
within them.

\section{15 December 1999, \ ``Insanity and Insomnia''}

I just quickly skimmed the notes you sent me.

\bap
When you wrote ``extra kinds of time evolution'', I thought of the
two-time formalism of {\Aharonov} and {\Vaidman}, who claim they
have a better way of understanding QM. Now you wrote instead
``nonlinear evolutions''.
\eap

This was insanity.  Actually, all along, I had been thinking of
stochastic collapses of the GRW type (since {\Goldstein} had
mentioned them in his article).  Why I wrote nonlinear evolutions
yesterday to describe that, I don't know.  I like your new
sentence.  Should we though keep connection with the PT articles we
are criticizing by substituting ``stochastic evolutions'' for
``multiple times''?  I don't believe the {\Aharonov}-{\Vaidman} ideas
were mentioned in any of the PT articles.  In any case, I get the
impression that the GRW wackiness is more ``mainstream'' than the AV
wackiness.

I'll look into the draft in more detail later (after I wake up).

{\Emma}'s cough has subsided a bit, but it's still here.  The doctor
yesterday gave her a thorough examination and said that there are no
signs of whooping cough, only deep chest congestion.  Strange though
that whooping cough is running so rampant in Los Alamos since all
children are required to be immunized against it.  It suggests to me
that perhaps there is a new strain that is resistant to the old
immunization.

Now I go back to bed (since after all I am asleep).

\section{15 December 1999, \ ``One Last Stumble''}

Well, it does look like we're narrowing down.  The only stumbling
block for me is still the teleportation paragraph.  As the last two
sentences are presently worded, I still don't see how the reader
will come to the intended realization, i.e., your changes seem like
much more than MSC [Minor Style Changes] to me.  Perhaps I'm just
being stubborn and somehow have blinded myself to a significantly
simpler formulation---that is a real possibility, I'm not
discounting it.  I wonder what you think about the following.  Might
we see if Eugen would be willing to make a comment on the two
opposing versions?  He has thought a little about the issue, but is
still much more likely than us to be representative of the average
readers of PT. I could send him a note today, and if he responds by
tomorrow we wouldn't be any worse off.  If he does not respond
quickly, we can just do something with whatever we have---I'll
volunteer now to let you have the last word on the subject---and
send it to {\Benka}.

I don't want to be completely brickheaded.  I realize that in this
writing I have been quite stubborn about getting my way \ldots\ and
you probably don't deserve so much heartburn.  I don't want you in
the end wishing that you had just written the comment yourself and
been done with it.  I just want to feel comfortable with the clarity
of our exposition to the extent to which it is in my power.

I'll go ahead and send this note off now, in case you are nearing
bedtime.  In the next note I'll write a detailed commentary of why I
don't think your teleportation changes are just MSC.

\section{15 December 1999, \ ``Detailed Commentary''}

Let me try to expose where I see differences of substance between
what you wrote and what I write.  After that I will propose one last
version of the paragraph, to see if we can both agree upon it.

I can tell that in the end we are in agreement about the physics and
{\it our\/} understandings of the phenomenon.  Our only difference
may be in what we wish the reader to take away from the paragraph.
In fact we only differ in the description of the closing part of the
paragraph.

Here are your words:
\bq
\noindent However, the two bits of classical information serve only to
transfer the quantum state from being that of the original photon to
being that of Bob's photon, while neither Alice nor Bob know that
state.  Only the original preparer who knows it can verify that it
was correctly transferred.
\eq
Here were mine:
\bq
\noindent However, the conundrum is only apparent.  As far as Alice and
Bob are concerned nothing changes about Bob's photon; they describe
it by the same density matrix throughout the whole process. The two
bits of classical information serve only to transfer the preparer's
information to be from being {\it about\/} the original photon to
being {\it about\/} the one in Bob's possession. This is not
magical: it comes about by the previously established correlation
between Alice and Bob.
\eq

\noindent ------------

{\bf 1)}  In the first sentence you say, ``the two bits \ldots\ serve
to transfer THE quantum state.''  The word ``the'' implies that there
is one and only one state for that system.  But that is not the
case.  This, in my eyes, is why ``the peculiar nature of a quantum
state as representing information is strikingly illustrated by
quantum teleportation.''  There is the state Charlie, the preparer,
assigns to his original system, $|\psi\rangle$.  Then there is the
state that Alice and Bob assign to that system.  If they have no
prior information about Charlie's preparation, they will assign the
completely depolarized state, $\rho=\frac{1}{2}I$.  One is mixed and
one is pure, to be sure, but from our point of view---as endorsed by
{\Gleason}'s theorem---they are two equally valid states.

Then there is the state that Charlie originally assigns Bob's
system.  Because he knows that Alice and Bob share a maximally
entangled state, his assignment is the completely depolarized
state.  Alice and Bob too assign the completely depolarized state to
Bob's system.  So for this aspect of the experiment, there is one
state.

After the teleportation process is complete, Charlie updates his
assignment to Bob's system to be $|\psi\rangle$.  However, Alice and
Bob continue to describe that system via $\rho=\frac{1}{2}I$.  Thus
at the beginning and ending stages of the game, there is not one
quantum state in the picture, but rather two.

To present the idea in the way you did, it seems to me, only
endorses the bad mode of thought that has caused us to write this
article in the first place.  What right do Alice and Bob have to say
that there is a quantum state there and they just don't know it?
For all they know, Charlie could have given them half of an
entangled pair to teleport.  Or maybe there is no Charlie at all and
the photon has come from the three degree microwave background
radiation.

As useful as the idea of the ``unknown quantum state'' is, the
phraseology is a holdover from the idea that systems have ``states''
independent of the scientists speaking about them.  That is the
point of view we are trying to fight in this paper.  I realize that
we can't rectify the whole situation in a simple opinion piece, but
there is no reason to fuel the fire either.

{\bf 2)}  In the second sentence you say, ``Only the original
preparer who knows it can verify that it was correctly
transferred.''  Again I don't feel that this sufficiently emphasizes
that it was ONLY the preparer's description that was transferred and
nothing physical beyond that.  The wording you used is much more of
the feel that there is a state THERE, and Charlie can verify it if
he wishes, but nevertheless the raw fact is that there is a state
there.  Again, why help fuel the fire that is consuming our
community?

It is just a question of getting away from bad language.  Remember
my footnote (\#4) about {\Peierls}, i.e., the last one inserted into
the paper I scanned for you?  The major drawback about his little
piece was that every now and then he would revert to objectivist
language:  he wasn't as consistent as he could have been.  Sometimes
he would say that the quantum state stands for what we know, and
then sometimes he would speak about having information about THE
state of a system.  This is what I'm trying to get at:  I want us to
do better than that.

If we are going to be consistent, and educate the community a bit,
we should not speak of ``it being correctly transferred'' in any way
that hints of it being done independently of the preparer who knows
it.

{\bf 3)}  You don't like ``conundrum'' and ``magical.''  That's fine:
I can change that, they weren't essential.  The main point to me is
that after all our effort to de-objectify the quantum state, your
language in the teleportation example only seems to re-objectify
it.  I want us to be careful about that.  For instance, I know for a
fact that Richard {\Jozsa} takes that language very seriously:  he
thinks that our way of looking at the quantum state is just hogwash
(he told me again last week in Montr\'eal).  He, like his old
graduate advisor Roger {\Penrose}, likes to think that systems have
objective states and that sometimes they evolve unitarily and
sometimes they collapse.  Most importantly for the point at hand
though, he thinks that quantum teleportation endorses this view---he
told me so. This, I believe, comes about partially because people
have been fixated by the inappropriate phrase ``unknown quantum
state.''

{\bf 4)}  You didn't like my phrase ``As far as Alice and Bob are
concerned nothing changes about Bob's photon.''  But that is the
only position one can take, unless one believes that there is more
that we can know about a system than the quantum state one ASSIGNS
to it (based on any previously gathered information).  If you really
want to take that position, then we would have to retract our
earlier statement, ``The compendium of probabilities represented by
the ``quantum state'' $\rho$ captures everything that can
meaningfully be said about a physical system.''  I for one, however,
don't want to retract that statement.  Now, maybe there is an issue
that perhaps I shouldn't have inserted the ``nothing changes''
phrase into the actual text.  But you must see the force of the
point:  with respect to Alice and Bob, nothing changes about Bob's
system \ldots\ even if they perform a unitary operation on it.  Their
compendium of probabilities for all measurements remains the same,
and that is all that can be meaningfully said about THAT system.  It
is true that their state assignment for the Charlie-Alice-Bob system
changes in the teleportation process, but that change refers to a
composite system, not the singular system in Bob's possession.  We
have to have the courage of our convictions.

{\bf 5)}  There is nowhere in your discussion an emphasis on why
Charlie's state can be transferred from Alice to Bob after only one
of four potential unitary transformations has been performed.  This
is what I was trying to get at with my sentence, ``It comes about by
the previously established correlation between Alice and Bob.''  If
Charlie hadn't ``known'' something about Bob's system through the
previously established entanglement, then he would have never been
able to transfer his ``state of knowledge'' (quantum state) onto that
system.

{\bf 6)}  You didn't like my usage of the word ``about'' \ldots\ but
that's not overly essential.  In our discussion of {\Gleason}'s
theorem, we used the word ``describe'' (after your convincing me to
put it there) so it doesn't seem out of place to me to use it in this
context.

{\bf 7)}  Putting this all together, let me now give the passage one
more shot.

\noindent ------------

I'm sorry, I wasn't able to complete this note this morning.  I got
tied up in meetings at the lab.  In the mean time I've noticed that
you have sent me a few more notes.  I will study them and then try
to incorporate your further comments into the redraft of the
paragraph I was going to write.

\noindent ------------

NEW PROPOSAL FOR TELEPORTATION PARAGRAPH:
\bq
The peculiar nature of a quantum state as representing information
is strikingly illustrated by the quantum teleportation process. In
order to teleport a quantum state from one photon to another, the
sender (Alice) and the receiver (Bob) need a pair of photons in a
standard entangled state. The experiment starts when Alice receives
another photon whose polarization state is unknown to her, though
known to some preparer in the background. She performs a measurement
on her two photons, and then sends Bob a classical message of only
two bits, instructing him how to reproduce the unknown state on his
photon. This economy of transmission appears remarkable because to
completely specify the state of a photon, namely one point in the
{\Poincare} sphere, we need an infinity of bits.  However, the
disparity is merely apparent.  The two bits of classical information
serve only to transfer the preparer's information, i.e., his {\it
state}, to be from describing the original photon to describing the
one in Bob's possession. This can happen precisely because of the
previously established correlation between Alice and Bob.
\eq

\noindent ------------

OK, so what do you think of that?  I took away conundrum and
magical.  Also I stopped mentioning that ``as far as Alice and Bob
are concerned, nothing changes about Bob's system'' in a way that I
feel will not compromise my fidelity.  True as I believe it is, I
will back down that it is essential for making a point here.  A few
other things to notice.  1) I introduced the preparer in the same
sentence that ``unknown state'' is introduced.  I still don't like
that phrase (unknown state), but now is probably not the time to
buck the vernacular too hard.  (I will do that in the colloquium at
Technion.)  By introducing the knower at precisely the same spot, I
could stomach it a bit better.  2) I reiterated that state and
information are precisely the same thing.  3) I stuck to my guns
that the emphasis should be that it is the preparer's information
that is transferred from one system to the other.  Saying the word
``state'' over and over in this context without reminding the reader
that it is information we are talking about would have been
defeatist.  4) I emphasized that teleportation can be done for the
economical rate of only two bits because of the previously
established correlation between Alice and Bob's photons.

All these points, I think, are essential to me.  And looking back at
all that you wrote today, I don't think that you will be in any
essential disagreement with all this.  Aside from the ``nothing
changes'' stuff, it was really just a question of emphasis that kept
us apart.

Chances are I'll wake up with insomnia and read your notes early
tomorrow.  Thanks for having so much patience with me.

\section{16 December 1999, \ ``Our Baby Leaves Home''}

Well, we did it!  I've just sent the manuscript to {\Benka} and
carbon copied that letter to you.  I didn't make any further changes
to the manuscript; so it is the one you sent me this morning.

Looking back at the manuscript one last time, I thought, ``Wow,
that's pretty good.''  And we didn't once mention Copenhagan!

For the Naples conference proceedings, I'm going to write a much
extended version of our little piece titled ``Knowledge About
What?'' The plan is to address some of the issues where we (rightly)
feared to tread in the {\sl Physics Today\/} article.  In particular,
I'm hoping to hammer out some of my ideas that what quantum
mechanics is about is the best intersubjective agreement we can come
to in a world that is ``sensitive to our touch.''  As it takes
shape, I will consult with you about your opinion.

I never did tell you what David {\Mermin} told me:  he said the copy
editors at PT will be ruthless about changing things in our paper in
nonsensical ways.  So, despite what {\Benka} told us, we should be on
our toes.  On a scientific note, David said that in the last year he
has become much more inclined to our point of view about quantum
mechanics.  Now that's progress!

\section{22 January 2000, \ ``Big Old Jet Airliner''}

Now I am on my way home.  The conference was quite fun.  It started
when, just as I arrived, {\Bill} {\Wootters} arrived too.  We went to
dinner together, and had a nice time trying to derive the Hamilton
equations in classical physics from the principle that statistical
distinguishability of phase space distributions remain constant.  It
wasn't so hard!  I think we succeeded in broad outline.  When I have
a chance, I will try to make it completely rigorous.  Then there
were several very good talks at this meeting:  Daniel {\Gottesman},
{\Gilles} {\Brassard}, {\Charlie} {\Bennett} (he gave a talk that I
had never seen before), {\Bill}'s talk (it was just excellent), Ben
{\Schumacher}, Nolan {\Wallach}.  One sad thing:  Paul {\Benioff}
became ill this morning (I don't know with what) and had to go to the
hospital.  His talk was cancelled.

You asked about our anti-paper.  It is funny:  in {\Schumacher}'s
talk he had a slide titled ``Fuchs's Interesting Non-Property'' in
which he discussed a certain aspect of the {\Holevo} bound.  Anyway,
I haven't heard anything from {\Benka}.  Do you think I should send a
cordial note to him?

\section{28 January 2000, \ ``{\Benka}''}

I've now written {\Benka}.  I do hope he sends the proofs soon so
that the article will actually appear in the March issue.

Many thanks for {\Emma}'s birthday wishes:  I've already passed them
on to her.  (She woke up in such a good mood this morning.)

\section{01 February 2000, \ ``Us vs.\ Them''}

By the time you read this, you will no doubt have seen the fax that
{\Benka} sent us.  Below I'll tabulate the changes to our manuscript
that I was able to catch.  (I suspect I missed some:  there were so
many despite the sugar-coating he put in the letter!)  Accompanied
with the tabulation, I'll also put down my opinion of the change.

Here we go:\bigskip

\noindent {\bf Us:} To start, let us examine the role of experiment in
science.
\\
{\bf Them:} To begin, let us examine the role of experiment in
science.
\\
{\bf Opinion:} None. \bigskip
\\
{\bf Us:} An experiment is an active intervention into the course of
Nature: we set up this or that experiment to see how Nature reacts.
\\
{\bf Them:} An experiment is an active intervention into the course
of Nature: we set up this or that experiment to see how Nature
reacts.
\\
{\bf Opinion:} You'll notice no difference between these two
passages. However, let me take this opportunity to point out that in
the remainder of the paper they always capitalized the first word
following a colon.  I don't care about the capitalization.  But at
least we were consistent:  They should be too!! \bigskip
\\
{\bf Us:} Classical physics is the ultimate example in that regard.
However, there is no logical necessity that this worldview always be
obtainable.
\\
{\bf Them:} Classical physics is the ultimate example of such a
model. However, there is no logical necessity for a worldview to
always be obtainable.
\\
{\bf Opinion:} I don't like what they've done!  Our wording was on
the mark: Classical physics is a ``compact description'' AND, in that
case, one CAN distill a model of a free-standing reality.  But even
there, there is no necessity that has to distill such a model:  this
was even pointed out by Immanuel {\Kant} so long ago.  The second
sentence, as they rewrote it, is just nonsense. \bigskip
\\
{\bf Us:} If the world is such that we can never identify a reality
independent of our experimental activity, then we must be prepared
for that too. \\
{\bf Them:} If the world is such that we can never identify a reality
independent of our experimental activity, then we must be prepared
for that, too. \\
{\bf Opinion:} Don't know why they inserted a comma.  Looks funny to
me \ldots\ but maybe we have to pick our battles. \bigskip
\\
{\bf Us:} It only provides an algorithm for computing {\it
probabilities\/} for the macroscopic events (``detector clicks'')
that are the consequences of our experimental interventions. \\
{\bf Them:} What it does is provide an algorithm for computing {\it
probabilities\/} for those macroscopic events (``detector clicks'')
that are the consequences of our experimental interventions. \\
{\bf Opinion:} I don't like their construction at the beginning of
the sentence:  It's like listening to fingernails scratching on a
chalkboard.  Why did they change ``the macroscopic events'' to
``those''?  More than likely that's their philosophy spilling over
into our paper.  They want to imagine that there are macroscopic
events independent of our experimental interventions:  That's why
they changed the qualifier. \bigskip
\\
{\bf Us:} This strict definition of the scope of quantum theory is
the
only interpretation ever needed both by experimenters and theorists.\\
{\bf Them:} This strict definition of the scope of quantum theory is
the only interpretation ever needed, whether by experimenters or
theorists.\\
{\bf Opinion:} None. \bigskip
\\
{\bf Us:} Quantum probabilities, like all probabilities, are
computed by using any available information. This includes, but is
not limited
to, information about a system's preparation. \\
{\bf Them:} Quantum probabilities, like all probabilities, are
computed by using any available information, not just information
about a system's preparation.
\\
{\bf Opinion:} Can't stand it.  It blurs two ideas that were
distinct. While reinstating our sentence, I would however like to
take the opportunity to modify it to the following:  ``This can
include, but is not limited to, information about a system's
preparation.'' Again the issue is the old one that we debated so
much:  Some systems have no preparations (in the usual anthropic
sense). \bigskip
\\
{\bf Us:} The mathematical instrument for turning that information
into statistical predictions is the probability rule postulated by
Max {\Born}. \\
{\bf Them:} The mathematical instrument for turning the information
into statistical predictions is the probability rule postulated by
Max {\Born}.
\\
{\bf Opinion:} None.\bigskip
\\
{\bf Us:} Here it is essential to understand that the statistical
nature of quantum theory does not restrict its validity to
situations where there is a large number of similar systems.\\
{\bf Them:} Here, it is essential to understand that the validity of
the statistical nature of quantum theory is not restricted to
situations
where there are a large number of similar systems.\\
{\bf Opinion:} None. \bigskip
\\
{\bf Us:} In particular, there is an ``effective'' reality in the
limiting case of macroscopic phenomena like detector clicks or
planetary motion: their objective occurrence would be acknowledged
by any observer who happens to be present.\\
{\bf Them:} In particular, there is an ``effective'' reality in the
limiting case of macroscopic phenomena like detector clicks or
planetary motion: Any observer who happens to be present would
acknowledge the objective occurrence of these events.\\
{\bf Opinion:} None. \bigskip
\\
{\bf Us:} John {\Bell} formally showed that any objective theory
giving experimental predictions identical to those of quantum theory
would
necessarily be nonlocal.\\
{\bf Them:} John {\Bell} formally showed that any objective theory
giving experimental predictions identical to those of quantum theory
would necessarily be non-local.\\
{\bf Opinion:} How many times do I have to complain about this to
people in editorial positions, i.e., people who are supposed to know
the English language better than I!!!  Look in any modern dictionary
and count the number of words that start with ``non'' AND are
hyphenated. The fraction is minuscule (unfortunately it is not
zero).  Why do people in the sciences want to hyphenate when no one
else does?\bigskip
\\
{\bf Us:} It would eventually have to encompass everything in the
Universe, including ourselves, and lead to bizarre self-referential
logical paradoxes.\\
{\bf Them:} It would eventually have to encompass everything in the
universe, including ourselves, and lead to bizarre self-referential
logical paradoxes.\\
{\bf Opinion:} So they don't have the same reverence for the Universe
that we do \ldots probably another symptom of being a journal editor!
In any case, we weren't completely consistent on our capitalization
of this (as exhibited in the very next paragraph).\bigskip
\\
{\bf Us:} Yet, it is legitimate to attempt to extrapolate the theory
beyond its present range, for instance when we probe particle
interactions at superhigh energies, or in astrophysical systems,
including the entire Universe.\\
{\bf Them:} Yet, it is legitimate to attempt to extrapolate the
theory beyond its present range, for instance, when we probe particle
interactions at superhigh energies, or in astrophysical systems,
including the entire universe.\\
{\bf Opinion:} None. \bigskip
\\
{\bf Us:} On the other hand, if we consider just a few collective
degrees of freedom, such as the radius of the universe, its mean
density, total baryon number, etc., we can apply quantum theory to
these degrees of freedom which do not include ourselves and other
insignificant details.\\
{\bf Them:} On the other hand, if we consider just a few collective
degrees of freedom, such as the radius of the universe, its mean
density, total baryon number, and so on, we can apply quantum theory
to these degrees of freedom that do not include ourselves and other
insignificant details.\\
{\bf Opinion:} I think I like their rendition better in this
case.\bigskip
\\
{\bf Us:} According to him, she is in a 50/50 superposition of states
with some cake or fruit in her stomach.\\
{\bf Them:} According to him, she is in a 50/50 superposition of
states
with some cake or some fruit in her stomach.\\
{\bf Opinion:} None.\bigskip
\\
{\bf Us:} From this example, it is clear that a wavefunction is only
a mathematical expression for evaluating probabilities and depends on
the knowledge of whoever is computing them.\\
{\bf Them:} From this example, it is clear that a wavefunction is
only a mathematical expression for evaluating probabilities and
depends on
the knowledge of whoever is doing the computing.\\
{\bf Opinion:} I prefer their formulation in this case too.\bigskip
\\
{\bf Us:} Cathy's story inevitably raises the issue of reversibility
since quantum dynamics is time-symmetric.\\
{\bf Them:} Cathy's story inevitably raises the issue of
reversibility;
after all, quantum dynamics is time-symmetric.\\
{\bf Opinion:} I suppose I'm fine with this one.\bigskip
\\
{\bf Us:} Of course, he would need for that complete control of all
the microscopic degrees of freedom of Cathy and her laboratory, but
that is a practical problem, not a fundamental one.\\
{\bf Them:} Of course, for that he would need complete control of
all the microscopic degrees of freedom of Cathy and her laboratory,
but that
is a practical problem, not a fundamental one.\\
{\bf Opinion:} None.\bigskip
\\
{\bf Us:} In order to teleport a quantum state from one photon to
another, the sender (Alice) and the receiver (Bob) need a pair of
photons in a standard entangled state.\\
{\bf Them:} In order to teleport a quantum state from one photon to
another, the sender (Alice) and the receiver (Bob) need to divide
between them a pair of photons in a standard entangled state.\\
{\bf Opinion:} Darn, wouldn't you know that after we spent so very
much time coming to agreement on this paragraph, they would come in
and demolish it!!!  I suppose I'm OK with this sentence presently
(unless you propose something better).\bigskip
\\
{\bf Us:} The experiment starts when Alice receives another photon
whose polarization state is unknown to her, but known to some
external preparer.\\
{\bf Them:} The experiment begins when Alice receives another photon
whose polarization state is unknown to her, but known to some
external {\bf [or third-party?  It's not Bob, is it?]} preparer. \\
{\bf Opinion:} To make them happy, why don't we change the sentence
to this? ``The experiment begins when Alice receives a photon whose
polarization state is unknown to her, but known to its external
preparer---a third party in the protocol.''\bigskip
\\
{\bf Us:} She performs a measurement on her two photons, and then
sends Bob a classical message of only two bits, instructing him how
to reproduce that state on his photon.\\
{\bf Them:} She performs a measurement on her two photons---one from
the original, entangled pair and the other newly received---and then
sends Bob a classical message of only two bits, instructing him how
to reproduce that state {\bf [Which state? \ldots]} on his photon.\\
{\bf Opinion:} Why don't we try this?  ``She performs a measurement
on her two photons---that is, on one from the entangled pair and the
photon that was just received---and then sends Bob a classical
message of only two bits, instructing him how to reproduce the
unknown state on his photon.''\bigskip
\\
{\bf Us:} The two bits of classical information serve only to
convert the preparer's information, from describing the original
photon to
describing the one in Bob's possession.\\
{\bf Them:} The two bits of classical information serve only to
convert the preparer's information, from a description of the
original
photon to a description of the one in Bob's possession.\\
{\bf Opinion:} I'm OK with this.\bigskip
\\
{\bf Us:} It is curious that some well intentioned theorists are
willing to abandon the objective nature of physical ``observables,''
and yet wish to retain the abstract quantum state as a surrogate
reality.\\
{\bf Them:} It is curious that some well-intentioned theorists are
willing to abandon the objective nature of physical ``observables,''
and yet wish to retain the abstract quantum state as a surrogate
reality. \\
{\bf Opinion:} They got us on this one.  Both my {\sl American
Heritage\/} and my {\sl Webster's\/} dictionaries have
``well-intentioned'' hyphenated.\bigskip
\\
{\bf Us:} There is a temptation to believe that each quantum system
has a wavefunction even if the latter is not known explicitly.\\
{\bf Them:} There is a temptation to believe that every quantum
system has a wavefunction, even if the wavefunction is not explicitly
known.\\
{\bf Opinion:} OK.\bigskip
\\
{\bf Us:} This analogy is misleading. On the contrary, attributing
reality to quantum states leads to a host of ``quantum paradoxes.''\\
{\bf Them:} This analogy is misleading:  Attributing reality to
quantum
states leads to a host of ``quantum paradoxes.''\\
{\bf Opinion:} ???? \bigskip
\\
{\bf Us:} The latter, when correctly used, never yields two
contradictory answers to a well-posed question.\\
{\bf Them:} When correctly used, quantum theory never yields two
contradictory answers to a well-posed question.\\
{\bf Opinion:} None.\bigskip
\\
{\bf Us:} In particular, no wavefunction exists before we start an
experiment, nor after its completion.\\
{\bf Them:} In particular, no wavefunction exists either before or
after
conduct an experiment.\\
{\bf Opinion:} ???? \bigskip
\\
{\bf Us:} Also, the theory has no dynamical description for the
``collapse'' of the wavefunction.\\
{\bf Them:} Again, the theory has no dynamical description for the
``collapse'' of the wavefunction.\\
{\bf Opinion:} This was a dumb change.  Makes no sense; why would we
say ``again''?  Why don't we propose:  ``Furthermore, the theory has
no dynamical description for the ``collapse'' of the
wavefunction.''\bigskip
\\
{\bf Us:} Its collapse is something that happens in our description
of the system, not to the system itself.\\
{\bf Them:} Collapse is something that happens in our description of
the
system, not to the system itself.\\
{\bf Opinion:} None.\bigskip

Also they changed our biographies a bit.  I think I will ask for the
capitalization in ``director-funded'' back as that is actually the
job title.  As I say, I probably missed some of their changes.
You'll probably miss some too.  Let's just hope that there not big
enough to make a difference.

Tomorrow morning I go to ABQ for my weekly collaboration with
{\Caves} (I usually go on Mondays, but this week I was recruited to
give their journal club talk and so I traded days).  Nevertheless I
may get a chance to work on your next iteration, or compose a letter
to {\Benka}, while I'm there.

Oh, almost forgot.  My preference for the cartoon is strongly toward
the first one.  I tried them out on a few people here (Leslie
{\Weaver} our secretary, Kurt {\Jacobs} a postdoc, and Tanmoy
{\Bhattacharya} a staff member) and they all voted likewise.

\section{02 February 2000, \ ``Cartoon URGENT''}

Yes, please do send a fax of the cartoon when you can.  I will
compose a letter to {\Benka} this evening (gently) asking if he would
be willing to consider a cartoon from your source.

Now we have our lunchtime meeting.

\section{03 February 2000, \ ``Sorry''}

Sorry I haven't gotten back in touch with you.  It turned out that I
got in quite late last night.  I will conglomerate our comments in
some acceptable way and send them to {\Benka} before the day is out.

\section{04 February 2000, \ ``Running in Circles''}

It has turned out that {\Kiki} needs me to help her in Albuquerque
today (she's going to pick up her parents but also getting some
small furniture).  So I will drop by the lab and pick up the fax
before we depart.  Then on the ride to ABQ I will compose the letter
to {\Benka} and send it off when I arrive.

I believe I was confused about the cartoon you were sending.  There
was one sitting on my desk yesterday:  ``Then a Miracle Occurs.''
With that one, one of the most interesting coincidences occurred.
The boy in the office next to mine, walked in wearing a shirt with
just that cartoon emblazoned on it!  I told him (mistakenly) that
you had wanted that cartoon for our article.  He said, ``Maybe
that's not a good choice:  everyone knows this cartoon.''  (His
shirt made that quite believable.)

So hopefully I can retrieve the fax when I go in to the office in a
few minutes.

\section{04 February 2000, \ ``Oops''}

I'm sorry I forgot to carbon copy you the letter I wrote to {\Benka}!
Here it is below.  I took some liberties to modify slightly some of
your suggestions---in all cases, I think your opinion won't be too
much different than mine (so there shouldn't be much to worry about).

Let's now keep our fingers crossed that they will abide by our
wishes, and that this project will be done with.\bigskip

\noindent --------------- \bigskip

\noindent Hi Steve,

Thanks for the great job in editing you guys did.  The paper is
indeed generally clearer now.  Of the 31 changes we detected, we
only have an issue with a few of them.  I will detail these below.
The protocol will be this.  I will first list the complete sentence
or sentences you have in the present draft that we disagree with,
then I will list what we would like it or them changed to, and then
finally I will give an explanatory remark if necessary.\bigskip

\noindent {\bf You:} CHRIS FUCHS, previously the Lee DuBridge Prize
Postdoctoral Fellow at Caltech, is now a director-funded fellow at
Los Alamos
National Laboratory.\\
{\bf Us:} CHRIS FUCHS, previously the Lee DuBridge Prize Postdoctoral
Fellow at Caltech, is now a Director-Funded Fellow at Los Alamos
National Laboratory.\\
{\bf Remark:} As  {\Asher} wrote me (and which is actually true),
``\,`Director-Funded Fellow' must have capitals, just as the other
titles in our biographies. This denotes a prestigious award, not a
budgetary item.''\bigskip
\\
{\bf You:} An experiment is an active intervention into the course of
Nature: we set up this or that experiment to see how Nature reacts.\\
{\bf Us:} An experiment is an active intervention into the course of
Nature: We set up this or that experiment to see how Nature reacts.\\
{\bf Remark:} The first letter after the colon should be capitalized
to be consistent with all the other instances of the same in the
text. (Thanks for teaching us this rule!)\bigskip
\\
{\bf You:} Classical physics is the ultimate example of such a
model. However, there is no logical necessity for a worldview to
always be obtainable.\\
{\bf Us:} Classical physics is the ultimate example in that regard.
However, there is no logical necessity that this worldview always be
obtainable.\\
{\bf Remark:} Please reinstate the original wording here.  It was
chosen quite carefully to be exactly what it is.  The way we see it
is this: Classical physics is, on first pass, nothing more than a
compact description.  We are lucky that we can further distill a
MODEL of a free-standing reality from it IF WE WISH TO GO TO SUCH A
METAPHYSICAL LEVEL.  So in changing the words, you have changed the
meaning of what we wanted to convey.  Also you'll note that in your
second sentence in this cluster, there is clearly a necessary
qualifier missing from ``worldview.''\bigskip
\\
{\bf You:} What it does is provide an algorithm for computing {\it
probabilities\/} for those macroscopic events (``detector clicks'')
that are the consequences of our experimental interventions.\\
{\bf Us:} What it does is provide an algorithm for computing {\it
probabilities\/} for the macroscopic events (``detector clicks'')
that are the consequences of our experimental interventions.\\
{\bf Remark:} We wrote ``It only provides''---actually quantum theory
does other things too, so we understand the motivation for your first
change. However, please change back the ``those'' to a ``the.''
 {\Asher} finds this construction more palatable, and I think it better
conveys my overall belief:  ``those'' conveys a flavor that there are
things that qualify as ``macroscopic'' beyond the consequences of our
experimental interventions.  That does not reflect what I want
conveyed.\bigskip
\\
{\bf You:} Quantum probabilities, like all probabilities, are
computed by using any available information, not just information
about a
system's preparation.\\
{\bf Us:} Quantum probabilities, like all probabilities, are
computed by using any available information. This can include, but
is not
limited to, information about a system's preparation.\\
{\bf Remark:} This is an issue that we had much debate about as we
constructed our draft.  Please instate the wording above (which is
very slightly different from our original wording).  It conveys the
idea that some systems have no preparations (in the usual anthropic
sense). Nevertheless, we make do with whatever information we have
in making our predictions.  That set of thoughts is absent in your
formulation.\bigskip
\\
{\bf You:} John {\Bell} formally showed that any objective theory
giving experimental predictions identical to those of quantum theory
would
necessarily be non-local.\\
{\bf Us:} John {\Bell} formally showed that any objective theory
giving experimental predictions identical to those of quantum theory
would
necessarily be nonlocal.\\
{\bf Remark:} We've never seen a dictionary (English or American)
that hyphenates nonlocal.  In fact the vast majority of all words
starting with ``non'' have no hyphenation; it only appears to be a
habit that physicists have gotten into.\bigskip
\\
{\bf You:} On the other hand, if we consider just a few collective
degrees of freedom, such as the radius of the universe, its mean
density, total baryon number, and so on, we can apply quantum theory
to those degrees of freedom that do not include ourselves and other
insignificant details.\\
{\bf Us:} On the other hand, if we consider just a few collective
degrees of freedom, such as the radius of the universe, its mean
density, total baryon number, and so on, we can apply quantum theory
to these degrees of freedom, which do not include ourselves and other
insignificant details. \\
{\bf Remark:} As  {\Asher} wrote, ``They replaced `these degrees of
freedom which' by `those degrees of freedom that'. The original
meaning, as I understood it, was `these degrees of freedom, which
\ldots' (that is, the degrees of freedom that were just mentioned
and which have the following property \ldots). The new text may be
incorrectly interpreted as including ALL the degrees of freedom that
do not include ourselves. Please restore `these degrees of freedom,
which \ldots' with a comma before `which' so that the text is
unambiguous.''\bigskip
\\
{\bf You:} Let us examine a concrete example: The observer is Cathy
(an experimental physicist) who enters her laboratory and sends a
photon through a beam splitter.\\
{\bf Us:} Let us examine a concrete example: The observer is Cathy
(an experimental physicist) who enters her laboratory and sends a
photon
through a beamsplitter.\\
{\bf Remark:} ``Beamsplitter'' is usually written as a single word
in the optical literature.\bigskip
\\
{\bf You:} The experiment begins when Alice receives another photon
whose polarization state is unknown to her, but known to some
external preparer.\\
{\bf Us:} The experiment begins when Alice receives another photon
whose polarization state is unknown to her, but known to a
third-party
preparer.\\
{\bf Remark:} You wanted clarification.\bigskip
\\
{\bf You:} She performs a measurement on her two photons---one from
the original, entangled pair and the other newly received---and then
sends Bob a classical message of only two bits, instructing him how
to reproduce that state on his photon.\\
{\bf Us:} She performs a measurement on her two photons---one from
the original, entangled pair and the other with a state unknown to
her---and then sends Bob a classical message of only two bits,
instructing him how to reproduce that unknown state on his photon.\\
{\bf Remark:} You wanted clarification.\bigskip
\\
{\bf You:} Again, the theory has no dynamical description for the
``collapse'' of the wavefunction.\\
{\bf Us:} Furthermore, the theory has no dynamical description for
the ``collapse'' of the wavefunction.\\
{\bf Remark:} ``Again'' makes no sense in this context.\bigskip

That's pretty much it.  Concerning the cartoon, neither of us like
the bottom one you sent us (i.e., the one with the caption, ``Well,
uh, when a daddy's wave function \ldots'').  I also took a small poll
at the lab (five people) and found that the top cartoon (i.e., the
one with the caption, ``What do you mean \ldots'') was the preferred.
And that reflects my preferences too.   {\Asher}, however, is
concerned that the cartoon may not be ``quantum mechanical'' enough
and has therefore suggested one from the book ``What's so funny about
science?'' (cartoons by Sidney {\Harris}, from {\sl American
Scientist}, reproduced by Kaufmann, Los Altos, CA, 1970 \ldots\
1977).  The scene is a mechanic working on a car with the car owner
overlooking him. The caption reads, ``Actually I started out in
quantum mechanics, but somewhere along the way I took a wrong turn.''

 {\Asher} writes:  ``If {\sl Physics Today\/} has a copyright agreement
with {\sl American Scientist\/} (or Kaufmann), this would be the best
solution. If this cannot be arranged quickly enough, then let them
put `cause and effect'.''  This would be agreeable to me, though as
I say my own preference is for the first cartoon you supplied.  (It
seems sufficiently appropriate to me actually, since the decision
between one outcome or the other in a quantum measurement has no
``cause'' describable within physics.)  Thus I leave the issue to be
between you and  {\Asher} as far as that goes.  I will abide by
either ruling.
\bigskip

\subsection{ {\Asher}'s Reply}

I am about to leave for a visit to my grandson (and the rest of the
family, near Tel-Aviv). He just had his third birthday. I'll print
out your letter to {\Benka} and I'll study it later. Thank you for
giving a quantum interpretation to the cartoon ``cause and effect''.
Perhaps they can keep the cartoon, but change the caption into
something like ``Do you mean this resulted from a quantum
fluctuation?''

\section{05 February 2000, \ ``Only Trouble Spot''}

Let me start with an apology:  I apologize for opening a can of
worms.  It concerns this part of the paper:
\bq
\noindent If from such a description we can {\it further\/} distill a
model of a free-standing ``reality'' independent of our
interventions, then so much the better.  Classical physics is the
ultimate example in that regard.  However, there is no logical
necessity that this worldview always be obtainable.  If the world is
such that we can never identify a reality independent of our
experimental activity, then we must be prepared for that, too.
\eq

{\Benka} wrote:
\bq
\noindent Your wording is ambiguous, which led to the trouble in the
first place.  Specifically, you know what ``that regard'' refers to,
but no one else does.  To be clear, the passage needs to be longer.
Here's an attempt:
\bq
\noindent \ldots\ Classical physics is the ultimate example of such a
description leading (if we so desire) to a model of reality.
However, there is no logical necessity for any worldview to be
obtainable from a scientific description. Physics is not
metaphysics. If the world is such that \ldots
\eq
Please get back to me by Monday.
\eq

I don't know about him, but when I read the phrase ``in that regard''
I immediately back up to the previous sentence and check how the
present sentence meshes with it.  Despite a serious effort for
empathy with him, I can't see his problem.  The passage seems
crystal clear (and pretty well written) to me.  (I like what we've
written; I just hate to see it changed.)

But it's too late now:  Pandora's box has been opened.  You were
sure right, when you wrote, ``Our formulation was better, but I fear
making too many changes in proof, because of the risk of further
errors.''  The question is, how can we please him with the minimal
number of changes to our text.  Let me propose the following change
to the passage (specifically in the second and third sentences).
\bq
\noindent If from such a description we can {\it further\/} distill a
model of a free-standing ``reality'' independent of our
interventions, then so much the better.  Classical physics, for
instance, is the ultimate example of a theory reducible in that
way.  However, there is no logical necessity that this kind of
worldview always be obtainable.  If the world is such that we can
never identify a reality independent of our experimental activity,
then we must be prepared for that, too.
\eq

In this formulation, I tried to sidestep two things.  1)  The
expectation of good literacy on the part of the reader.  The second
sentence is now more redundant than it needed to be, so he/she won't
have to back up to the previous sentence to understand its intent.
2)  {\Benka}'s apparent misunderstanding of the word ``worldview.''
His attempt of a rewrite seems to indicate that he thinks we don't
have one, not simply that ours is different from the kind one finds
in classical physics.

For your convenience, let me put the whole paragraph together:
\bq
Our purpose here is to explain the internal consistency of an
``interpretation without interpretation'' for quantum mechanics.
Nothing more is needed for using the theory and understanding its
nature.  To begin, let us examine the role of experiment in
science.  An experiment is an active intervention into the course of
Nature:  We set up this or that experiment to see how Nature
reacts.  We have learned something new when we can distill from the
accumulated data a compact description of all that was seen and an
indication of which further experiments will corroborate that
description.  This is what science is about.  If from such a
description we can {\it further\/} distill a model of a
free-standing ``reality'' independent of our interventions, then so
much the better.  Classical physics, for instance, is the ultimate
example of a theory reducible in that way.  However, there is no
logical necessity that this kind of worldview always be obtainable.
If the world is such that we can never identify a reality
independent of our experimental activity, then we must be prepared
for that, too.
\eq

It's a little less literary now, but I think it might fulfill
{\Benka}'s requirements without compromising my integrity.  What a
pain to have to play these lawyer games!!

I hope you are having a good day with your grandson.  We will likely
give {\Kiki}'s parents a tour of the city today, showing them the
volcano that created these mesas, {\Oppenheimer}'s old home, etc.

Presently I am reading a very nice book by Arthur {\Zajonc}---{\sl
Catching the Light:\ The Entwined History of Light and Mind}---in
preparation for my interview.  I'm quite impressed that an
experimenter has such a literary command and extensive knowledge of
history!  If his personality seems to hold up like his writings,
he's bound to be a joy to work with.

\section{05 February 2000, \ ``Also''}

I forgot to tell you, beside that complaint of {\Benka} he said the
following:
\bq
\noindent By the way, all the other changes are fine except number 8.
You're stuck with ``beam splitter.''
\eq

So if we just get past the hurdle that the last note concerned \ldots

\section{05 February 2000, \ to Steve {\Benka}, cc:~Asher, ``Beaten''}

\noindent Dear Steve,\medskip

After long discussion  {\Asher} and I have come to agreement on our
one remaining disagreement with you.  Let us go with the rewrite
you've already proposed, except with the addition of one extra word
(which I'll write with in all caps below):
\bq
\noindent \ldots If, from such a description, we can {\it further\/}
distill a model of a free-standing ``reality'' independent of our
interventions, then so much the better.  Classical physics is the
ultimate example of such a model.  However, there is no logical
necessity for a REALISTIC worldview to always be obtainable. If the
world is such that we can never identify a reality independent of
our experimental activity, then we must be prepared for that, too.
\eq

Let me also convey another message made by  {\Asher}:
\bq
\noindent
Thank you for giving a quantum interpretation to the cartoon ``cause
and effect''.  Perhaps they can keep the cartoon, but change the
caption into something like ``Do you mean this resulted from a
quantum fluctuation?''
\eq

Once again, I will defer to whatever decision you two come to.

Thanks once again for all your attentive help.  Now I spend the next
two days preparing for my Amherst visit.  I will indeed say hello to
Bob Hallock for you.

\section{06 February 2000, \ ``Can of Worms''}

The only thing my (computer) dictionary says is:
\bq
\noindent {\bf can of worms}, Informal. a source of many unpredictable or
unexpected problems: Buying a company we know nothing about would be
opening up a whole new can of worms.  [1965--70]
\eq

As you see, they don't trace it back to a biblical origin.  But who
knows how accurate they are.

\section{08 February 2000, \ ``Copyright Sent''}

I just mailed off the copyright agreement for our PT article.  Also,
yesterday (by accident) I talked to Steve {\Benka} on the phone.  (I
say by accident, because I returned a call of his \ldots\ but
apparently he had called last week and after our email transactions
his questions were no longer relevant.)  In any case, he informed me
that he changed the caption of the cartoon to be in accord with your
suggestion.

Also I received your fax this morning (and your latest email
instructions yesterday).

\section{02 May 2000, \ ``Physics Today and Way Behind''}

\bap
Shouldn't we have received from {\sl Physics Today\/} the comments on
our paper, so as to prepare a reply?
\eap

Yes we should have.  Do you remember the name of the person who
contacted us about this?  Maybe one of us should send her a note
asking what's up.

I too have been having computer troubles.  I got my new laptop last
week and it has taken me over five days to set it up properly.  (And
I've only met partial success.)

Yesterday, I gave my second talk in my series of lectures.  It went
much too slowly:  I tried to pay attention to the ones in the
audience who needed more introduction, but that was a mistake.  The
one high point for everyone, however, was when I exhibited
{\Wigner}'s 1961 paper where he just missed noticing the no-cloning
theorem (and in fact got it wrong).  This brought a great smile to
Wojciech.

I do still plan to comment on your cats!  (I just hope they don't
die before I get there.)

\section{10 May 2000, 9:19 AM, \ ``Alive and Well''}

\bap
[9 May 2000:] I was distressed to learn of your forest fire. We
occasionally have such problems here. In 1979 the fire was at the
bottom of our garden.
\eap
\bap
[10 May 2000:] Please keep me posted that everything is under control
with forest fires, etc. Today (independence day) Aviva and I climbed
on our roof to clean it and add some white paint to increase its
albedo.
\eap
\bap
[10 May 2000:] Please write, just to reassure me that you and your
family are safe. Richard {\Jozsa} told me that he had seen our paper,
and that he disagreed with us.
\eap

Thanks for the concern.  Yes, we are still safe \ldots\ though the
winds have started to howl again.  I can hear the slurry bombers
flying outside, so I have a feeling they may be bombing very close
to us. Unfortunately, I missed the news this morning:  I had some
very minor surgery yesterday and it has made it difficult for me to
sit up.  Also {\Kiki} hasn't been able to connect to the emergency
update number.  We keep our fingers crossed that things won't get
worse because of the winds.

I knew that Richard would disagree with us.  He has some rather
silly notions running around in the back of his mind.

\section{11 May 2000, 11:02 AM, and to others, \ ``Alive but Depressed''}

A small note to let you know that {\Emma}, {\Kiki} and I are all OK.
We're fairly sure our house and all its contents have been
destroyed:  we only managed to escape with two suitcases worth.  (I
even left all my old calculations behind, which struck me as very
painful once I realized it.)  The whole town is ablaze; I suspect
that nothing will be left.

Thank you all for your concern.

\pagebreak

\subsection{ {\Asher}'s Reply: Displaced Persons}

\noindent Dear Chris and {\Kiki},\medskip

It must be a terrible feeling to be ``displaced persons'' but you
will overcome that drama. When I was 6 years old, my mother and I
were ``displaced'' from Paris by the advance of German troops (my
father was at the front). It took three days of train zigzags to
reach the little village where we spent the rest of the war. Five
years later, when we returned to Paris, nothing was left of our
furniture and other belongings.

In 1956, Aviva and her family were expelled from Egypt (only two
suitcases allowed). They had to restart their lives in a new country,
speaking an unknown language and with many strange customs. Aviva's
parents never fully adjusted to their new life, but she and her
brothers have nothing to regret about Egypt.

As painful as things are for you now, you are in your country, there
is no war to fear, and you are still young and strong enough to
recover and enjoy life. In my office there is a poster with ``IF'' by
Rudyard Kipling. Here are some excerpts:
\bv
If you can keep your head when all around you are losing theirs
\medskip
\\
If you can lose \ldots\ and start again at your beginnings, \\
And never breathe a word about your loss \medskip
\\
Yours is the Earth and everything that's in it
\ev

\noindent Be strong! We are thinking of you,\medskip\\
\noindent Aviva and  {\Asher}

\section{18 June 2000, \ ``Absence''}

I apologize for my long absence from email.  I will be able to look
at all you've written (and the original letters) starting Tuesday
morning.  Thanks for already putting so much time into the project.

\section{11 July 2000, \ ``It's a Boy!''}

Many congratulations on the new grandchild!  All children are
impatient and authoritarian; take it as a compliment that you are so
young!

As you should be able to see from this note, I am finally in
cyberspace again.  The Italian and Greek phone systems were
deplorable!  I arrived in Munich Sunday, but haven't had a chance to
connect until now.  Unfortunately the only way I've found a simple
means to work completely out of my laptop (given the Compaq and LANL
firewalls), is to dial directly to the United States for a
connection.  So I will likely only be checking and sending mail once
a day.

Later this afternoon, I will start the painful task of working on
our {\sl Physics Today\/} replies.

\section{11 July 2000, \ ``End o' Day''}

It's getting near the end of the day here, and {\Kiki} and her mom
will be returning from Austria soon, at which point I will once again
have social demands placed on me.

Today I finally had the courage to look at all the replies to our
article and all your related writings---I have procrastinated and
procrastinated on this, mostly because I think the fire has made me
mentally weak and wanting to shun any criticism.  If these letters
are the best of the lot sent to Physics Today, then that's sad.

I'm quite pleased with much of your drafted reply to the replies.
Thank you so for being a better person than me these past weeks. But
as you can guess, I disagree with some points (e.g.\ ``physical
states'' and ``it is a set of rules giving the probabilities of
macroscopic detection events, FOLLOWING SPECIFIED PREPARATION
PROCEDURES.'') We'll have to work again to come to some common
ground, while still keeping the text snappy and pithy.  Tomorrow I
will send you a revised draft to see what you think.

Did {\Ptolemy} and {\Euclid} both really reside in Alexandria?
Wasn't there quite an amount of historical time between them?  And
{\Euclid} was Greek, wasn't he?

\section{12 July 2000, \ ``What Muck!''}

I think these letters are horrible.  What's to be gained from being
pitted against this lot?  We pored over every word in that article,
just to have every deeper idea summarily ignored.  I wonder which is
more to blame here, the ``rigorous in-house review cycle,'' or
instead that every submission was of this quality?

This whole affair, upon reading and thinking more, has turned out to
be more complicated than I had wished.  What on earth is {\Sobottka}
trying to say?  I don't think your reply hits the mark, because it
seems that ultimately he does {\it not\/} want to ascribe an
objective reality to the wavefunction.

And {\Holladay} \ldots\ yuck.  Why was he not worried with trying to
make some sense of our statements about quantum theory also giving an
``effective reality'' for some very gross aspects of the world?
I.e., those aspects that we can still talk about even without
detailed knowledge of any microphysics?  If he had done this, maybe
he would have answered his own questions.  I think we should
reinsert some discussion of this into your reply, and make it clear
that we are not really instating yet another consideration that we
had not in our original article.

{\Brun} and {\Griffiths}:  who could have guessed that they wouldn't
take this opportunity to advertise their interpretation of quantum
mechanics yet again?  (I.e., the one Murray {\GellMann} calls the
``modern interpretation of quantum mechanics.'')

When they talk about ``coarse-grainings'' what the heck do they think
determines such things in a completely objective description of the
universe?  They can only be expressions of {\it ignorance\/} if the
theory does not specify them (and it doesn't), and then they are
exactly back to the point we started from:  saying that the best one
can say is about a few degrees of freedom of Nature that do not
include the observer.  There can be nothing in Nature itself that
says, ``Thou shalt coarse grain \ldots''

Maybe what bugs me about their attempt to inject a free-standing
reality into the theory is the impotence of the whole affair.  Why
can they not appreciate how liberating an idea it is?

Even their reporting of the efficacy of experiment is naive:
``Experimental results guide the development of theory, and are the
ultimate arbiters of correctness or incorrectness.''

Here's an idea attributed to {\Einstein} by {\Heisenberg} that I
love. It captures an extremely important point (in this fuller
version that is not quoted so often):
\bq
It is quite wrong to try founding a theory on observable magnitudes
alone.  In reality the very opposite happens.  It is the theory
which decides what we can observe.  You must appreciate that
observation is a very complicated process.  The phenomenon under
observation produces certain events in our measuring apparatus.  As
a result, further processes take place in the apparatus, which
eventually and by complicated paths produce sense impressions and
help us to fix the effects in our consciousness.  Along this whole
path---from the phenomenon to its fixation in our consciousness---we
must be able to tell how nature functions, must know the natural
laws at least in practical terms, before we can claim to have
observed anything at all.  Only theory, that is, knowledge of
natural laws, enables us to deduce the underlying phenomena from our
sense impressions.  When we claim that we can observe something new,
we ought really to be saying that, although we are about to
formulate new natural laws that do not agree with the old ones, we
nevertheless assume that the existing laws---covering the whole path
from the phenomenon to our consciousness---function in such a way
that we can rely upon them and hence speak of ``observation.''
\eq

I am not pushed to the rejection of a free-standing reality in the
quantum world out of a predilection for positivism.  I am pushed
there because that is the overwhelming message quantum THEORY is
trying to tell me.

I guess now that the end of the day has arrived again, I will be
sending some more concrete changes to the article tomorrow.  I hope
you will forgive me:  this idle time of thinking about things (and
running through the whole gamut of angers) seems necessary to
congeal my thoughts.

\section{13 July 2000, \ ``Sunny Day Apology''}

Today for the first time in about two weeks, {\Kiki} tells me, the
sun is shining in Munich.  The forecast says that it will disappear
again by tomorrow.  She has asked if we could go into town, enjoy
the sun, window shop, and have lunch at an outside cafe.  Despite my
obligations, I just couldn't turn her down:  I apologize.  I want to
get these replies off our back too, but it does seem a shame to pass
up this opportunity to make {\Kiki} and {\Emma} happy.

I should be able to return to our project this evening or tomorrow
with a clean palette.  I don't think we'll have any real problem
getting things in for a September issue appearance.

\section{14 July 2000, \ ``Lunchtime Report''}

I'm working hard on hammering out my changes to our reply:  I
thought I would give you a quick report before I join {\Kiki}
downstairs for lunch.  The only {\it real\/} trouble I'm having is
with {\Holladay}, which I think requires some special care.  I hope
beyond hope to have you a completed manuscript before the end of the
day.

\section{14 July 2000, \ ``Here It Comes''}

The end of another day again arrives, but at least this time I am
sending a draft of the manuscript with it.  There are two versions,
one with explanatory footnotes, and one without.  My word processor
gives the word count in the unfootnoted version as 1520 \ldots\ well
below the 1737 that Ms {\Hanna} suggested.

As you will see, the main changes I made in your draft were to the
{\Holladay} discussion.  I didn't feel that your earlier discussion
of him was really getting at the right point.  I hope that you find
you can agree with some of what I've written.

Then there are further smaller changes throughout.  Some of them
were put in place to lead up to my other main change:  I tried to
build a more positive and far-reaching statement into our
conclusion.  We said exactly that idea throughout (both in our
original Opinion and in various places of both this draft and your
earlier one), but I thought it should have the ending emphasis.  The
main reason is that people continually accuse us with pure
positivism or pure idealism because of their very sloppy reading of
us, and because of their preconceptions (probably based on some
childhood trauma).  I want to do my best to stop that.

I will be back online tomorrow morning, but there is some chance
that {\Kiki} and I will spend much of the day in Munich.  In any
case, I'm hoping to get the chance to do one more reiteration (if
need be) before we fly out early Monday morning.

Being Friday night, I presume you will see Lydia tonight.  Please
give my warm regards to her and her family, and especially the new
little one.

\subsection{Parsing the Paper}

\noindent {\bf Quote from Draft:}
We do not claim that contemporary quantum theory is the last word
for a description of Nature; it would be hard to imagine how we
could have written the last words in our ``Opinion'' if we did.
\smallskip\\
{\bf Comment:} A little slap on the wrist for the empty attention
span of {\Harris} and others.
\bigskip\\
\noindent {\bf Quote from Draft:} Unfortunately, as he
hints, \ldots
\smallskip\\
{\bf Comment:} I added this since he does seem to turn around by the
end of his letter.
\bigskip\\
\noindent {\bf Quote from Draft:}
Some people may deplore this situation, but we
should make it clear that we were not pushed to the rejection of a
free-standing reality in the quantum world out of a predilection for
positivism. We were pushed there because that is the overwhelming
message quantum theory is trying to tell us.
\smallskip\\
{\bf Comment:}  These sentences were placed here to make it clear
that I, personally, walked into the quantum foundations question as
a hard-core realist.  I came out of it a new man.  That is to say,
the cement of my philosophy wasn't set before I got interested in
quantum foundations. By the way, the last sentence of this paragraph
is throwing a bone to a title of one of David {\Mermin}'s recent
``Ithaca Interpretation'' papers.
\bigskip\\
\noindent {\bf Quote from Draft:}
We do not find any merit in the various alternatives that were
proposed to the straight\-forward interpretation of quantum theory:
It is a set of rules giving the probabilities of macroscopic
detection events based on any previous experimental information we
may have gathered.
\smallskip\\
{\bf Comment:} Changed this because, as you know, I am reluctant to
say that the only relevant information is preparation information.
\bigskip\\
\noindent {\bf Quote from Draft:}
If that provides a ``straitjacket'' to our thought, it is only in
that it keeps us from contemplating an endless stream of conundrums
that come about solely from shunning quantum theory's greatest
lesson---that the experimenter cannot be detached from the world he
is attempting to describe.
\smallskip\\
{\bf Comment:} I really would like to end the paper with a statement
that makes the whole affair look positive and hopeful.  To say that
the experimenter in quantum mechanics is not ``detached'' is a
phrase I like very much.  It's one that {\Pauli} used, and now
{\Anton} {\Zeilinger} uses a lot.

\section{16 July 2000, \ ``Getting Slow Start''}

I'm getting a slow start today:  how I can't wait to have my own
office at my own home again, with NO visitors to waste my time.  (My
in-laws are having several house guests this weekend.)

Thanks for sending the revised draft.  And MANY thanks for removing
that horrible paragraph I wrote on diamonds:  immediately upon
seeing your comments, I realized that even I didn't agree with what
I had written!  Still though, I think we have some rough edges to
work out on {\Holladay}.  I think it is quite important that we
tackle his rhetorical question in a head-on way:  it (i.e., if the
theory is not about reality, then what is it about?) and the
misunderstanding of {\Brun}/{\Griffiths} (i.e., so if there were no
people, then there would be no world!) are the same comments we've
gotten over and over and over.  Beside the written comments we've
seen, several have expressed these concerns in person to me
({\Jozsa}, {\Gisin}, {\Omnes}, and {\Wootters}).  And we should have
a pithy, good answer:  in a way, this is a good opportunity for us to
flesh that out.  There must be a way of expressing it that will make
it easy for people to see what we are talking about.  Why is it so
difficult for them to understand?

I don't think you understood the point of my gambling-house
paragraph.  I think I will try again on that, because I do think
it's an important point, and relevant to the reply on {\Holladay}.
Please give me at least one more shot at it.  The point---which I
will try to reexpress more eloquently in the next draft---is that a
theory need make NO direct referral to reality in order to be
successful.  Probability theory is the prime example of that.  That
is because it is a theory of how to reason in light of the
information we have, regardless of the origin of that information.
Its success is gauged by the fact of our reasoning as well as we can
(by various criteria) in the light of our unpredictable world.
Quantum theory is more like that than any other physical theory
heretofore.  Its formal structure is more a ``law of thought'' than a
``law of physics'':  the physics is in the information we feed into
that formal structure (Hamiltonians, initial conditions, known
symmetries, etc.).  Nature and our reasoning about it should not be
confused.

You took out ``straitjacket'' because it is a torture device \ldots\
but it was not I who put it there in the first place:  it was {\Brun}
and {\Griffiths}, describing our ideas as working within a
straitjacket.  I think we should replace it (perhaps making it more
explicit that {\it they\/} said it first), to help make a frontal
attack on their silliness.

I will send you the next iteration by the end of the day today.
(With some luck for my sanity, all the house guests \ldots\ and
perhaps {\Kiki} and her mother \ldots\ will all take off for a day
trip to somewhere soon!)

\section{16 July 2000, \ ``Blue in the Face''}

A couple of more thoughts about {\Deutsch}.  [See note to David
{\Mermin}, dated 23 July 2000.]  Seeing his vitriol does reinforce
my resolve for us to be as absolutely clear and as positive as we
can be in our reply to critics.  But on the other hand, it also
reinforces my opinion that to some people we could talk until we are
blue in the face, and never make a crack in their preconceived
notions.  I strongly doubt that {\Deutsch} even read our paper:  but
{\it yet\/} he has an opinion of it.  Which of us are the cranks,
and which of us are the physicists.

More of substance later.

\section{16 July 2000, \ ``A Little Progress''}

I think I will go ahead and send a partial draft now.  The part in
italics I haven't yet touched:  it is all the stuff dealing with
{\Holladay}'s rhetorical question, and, I think, should really be the
core of our reply.  However, I think I have come to closure on the
rest:  it seems to me that that it is an independent unit, and you
can start evaluating your thoughts on it if you wish.

I apologize for not sending you a full draft, but this whole family
thing is really getting in my way.  If I get a chance to do more
this evening (probably not), it will be solely confined to the
italicized part of what I'm sending you presently.  I really want to
shoot for having this whole lot finished before I depart Munich.

In any case, I'll be back tomorrow morning.

\subsection{Parsing the Paper}

\noindent {\bf Quote from Draft:}
We do not claim that contemporary quantum theory is the final word
for a description of Nature; this should be clear from the last
paragraph of our ``Opinion'' essay.
\smallskip\\
{\bf Comment:} How's this version?  I didn't think your version was
forceful enough (starting with a dependent clause): the guy didn't
take more than a moment to read us---probably didn't even read to
the end of the article---and he ought to be punished for that.
\bigskip\\
{\bf Quote from Draft:} It is these gratuitous additions to quantum
theory that should have been the analogs of {\Ptolemy}'s epicycles in
{\Harris}'s story.
\smallskip\\
{\bf Comment:} Again, here, I am going for a slightly stronger tone.
\bigskip\\
{\bf Quote from Draft:} The mistake {\Holladay} makes in his
presumptive question is a common one for those not yet accustomed to
quantum lines of thought.
\smallskip\\
{\bf Comment:} Presumptive: Founded on presumption.  Presumption:
behavior or language that is boldly arrogant or offensive.
\bigskip\\
{\bf Quote from Draft:} He fails to see that we have accepted a
distinction that he does not: There is Nature, and then there is our
best description of our experimental interventions into it.
\smallskip\\
{\bf Comment:} You had changed the wording of this before, but that
completely deleted the effect the sentence was supposed to have.
\bigskip\\
{\bf Quote from Draft:}  {\Brun} and {\Griffiths} call this a
``straitjacket,'' but if it is, it is only in that it keeps us from
getting mired in the endless conundrums that arise solely from
shunning quantum theory's greatest lesson---that the notion of
experiment plays an irreducible role in the world we are trying to
describe.
\smallskip\\
{\bf Comment:} OK, so now I myself have removed the word
``detached'' that I like so much.  But I'm still shooting for
conveying the idea, and would like the paper to end on this note.
Does it sound a little better now?

\section{17 July 2000, \ ``A Little Pride''}

In the next two notes I will again send two versions of the draft:
with and without footnotes.

I must say, I think I am quite proud of the version now!  I think it
sounds really good, and most importantly conveys some essential
thoughts that complement our original ``Opinion'' exposition.  Given
so many people's questions, I am quite confident that this was
needed.

I hope you will agree with the points \ldots\ as I am quite settled
on them in my mind now.  The discussion of {\Holladay} is now a
hybrid of your original and my later versions, and (I think!) greatly
improved at getting at the essential points.

We leave very early tomorrow morning for the airport.  However, I
hope to download email one last time just before then.  In any case,
I will almost surely check it one last time before dinner this
evening (about 8:00 PM Munich time).

It'd be great if we could send our final draft off to Ms {\Hanna} by
Wednesday.

\section{17 July 2000, \ ``Great''}

I think we've done a fine job.  In the next email, I'll send you my
latest (only {\it with\/} footnotes this time).  My changes were only
very, very minor changes of English.

Sorry I don't have time to write more:  we are starting the long
process of a European dinner now.

I will await your last word when I arrive in America; I think we can
plan to send everything off to Ms {\Hanna} Wednesday morning.

\section{19 July 2000, \ ``Bigger Change''}

I just sent off the final draft to Ms {\Hanna}.  You'll note that I
modified the final sentence even more than you had suggested:  I
completely removed the phrase ``our getting mired in.''  I think the
sentence flows even better now.

I hope you don't mind me doing that last minute.  If you disagree,
we can always make a final change at the proof stage.

Now we do a small amount of shopping and then take off for Los
Alamos.  We had the most horrible experience flying home:  delays
and delays, so much so that we had to spend the night in Albuquerque
last night.  And all our checked luggage was lost, one suitcase is
still missing (the one with my quantum de {\Finetti} presentation
slides!).

\section{21 July 2001, \ ``God, etc.''}

\bap
Thank you for sending to {\Petra} a list of interesting problems. I'm
afraid some of them are a bit difficult for her at this stage of her
studies.
\eap
You can see I've had an undue influence from John {\Wheeler}!  I once
saw him give an undergraduate a summer research project of ``derive
quantum mechanics from principles of distinguishability''!!  The
fellow was floored.

\bap
I must also say that I am worried by your frequently invoking God in
your text. Also, in our last paper, you wrote several times
``Nature'' which is just a euphemism for the same purpose. As you
know, I am a devout atheist, but since we were under the pressure of
a deadline, I did not argue with you on this point.
\eap

Yes, you bring up a point that even I have been a little worried
about.  I would say that I also am a devout atheist.  Sometimes, I
bring up the imagery of a god creating the world not to be taken
seriously, but to emphasize that I am not a pure positivist.  I.e.,
that I believe there is ``something'' to the universe that is there
independently of us.  Whether that ``something''---I am averse to
using the word ``reality'' because that has no interactive aspect to
it, and does not convey the idea that the ``something'' itself might
be malleable and changeable---can be probed with the methods of
science, I don't know.  But that's not so important to me right now.

I would say that I use the word ``nature'' for the same reason---to
label that which is there independent of us.  It is true that
capitalizing the word ``Nature'' denotes it as an object of worship,
and in that sense makes it serve as a euphemism for God.  In fact I
did feel a little uncomfortable in capitalizing it throughout our
paper.  Looking back on it, I know that I only did that out of peer
pressure, i.e., because I have been noticing so many other people do
it in print lately.

Shall we decapitalize all the Natures in our paper?  I think I would
like to do that now myself.

Did you get the fax Ms {\Hanna} sent us?

\section{21 July 2000, \ ``24 Changes''}

Well, I count 24 changes to our manuscript.  Luckily, only two of
them annoy me.  Also, they themselves started to decapitalize our
Natures:  so I'm happy about that.  But they missed three, so we'll
have to point that out to them.

Now the question is how best can I get these things to you.  I did
not get a fax of the her changes, so I can't fax them to you
either.  Maybe I will re\TeX\ the changes and send them to you that
way.

\chapter{Letters to John {\Preskill}}

\section{18 November 1997, \ ``The Radical Eye''}

Let me send you some prep work \ldots\ while I'm procrastinating from
things I really need to be doing.  I'll convey the information in my
favorite way:  quoting/forwarding old emails!

The first, an old---kind of poorly written---note ``Dreams of a More
Ethereal Quantum,'' concerns one aspect of the below-the-water stuff:
the idea about distilling the MaxEnt-looking part out of quantum
theory.  It doesn't say much more than I said in the proposal \ldots\
but you can't expect too much, it's still just a vague program for a
program.  That stuff is at the bottom of the note; the early part of
the note says a few---even more sketchy---things about {\Mermin} and
Quantum Probabilities \ldots\ so I thought I would include it for
that reason.

The only other readable thing (I think) I've written on this aspect
of the below-the-water stuff is on pages 22--26 of my paper with
{\Caves}, ``Quantum Information: How Much Information in a State
Vector'' (it's on {\tt quant-ph}).  These same pages also contain an
extended discussion on probability.

The second note, ``Rainy Mondays and Consistent Histories,'' is a
note I wrote Bob {\Griffiths} concerning what I think of his
Consistent History stuff.  I think it says quite plainly and clearly
what I think about probability in quantum mechanics.  It's pretty
self-contained (i.e., not referring to old conversations, etc.), so
you might find it useful reading.

The third, ``A Little More Reality,'' is an example of the excesses
that scanners and character-recognition software can lead to!
Anyway, it says a bit more about what ontology (or rather what lack
of an ontology) I think quantum theory gives us.  If you want to see
how wacky your postdoc can be in private---I told you I was coming
out!---look at the discussion after Asherism \ref{OldIsm7}.

The fourth, ``Tyche or Moira?''  is again about quantum ontology---it
clears up something about the discussion of Asherism \ref{OldIsm6} in
the previous note.

In general, as you'll see if you look at any of this gunk, I'm rarely
willing to carry quantum theory so far as to make ontological
statements with it---I am more inclined to think that it is a theory
about our knowledge.  For instance, you might have noticed that I
opened up my proposal with ``The world we live in is well-described
by quantum mechanics.''  I didn't say, ``The world we live in is a
quantum mechanical one.''  (I caught myself in that particular
instance \ldots\ I came oh so close to compromising my morals!) There
must be some ontology---some theory of the ``real''---out there, it
just doesn't seem to me to be quantum theory.  Trying to fit quantum
theory to that task, seems more like trying to stuff a foot into a
glove.

That's about it.  Maybe I've said some of these things more clearly
in other notes, but these are the most recent \ldots\ and easiest to
find.  I should be around tomorrow (if I can get to sleep tonight)
\ldots\ as is my way though, you'll probably find my writings more
clear than my speech.

\section{21 November 1997, \ ``Closing the Door''}

By the way, you know, I believe in reality too.  (It's pretty
difficult to gulp a world that is completely dependent upon our
dreams for its existence.)  It's just that the reality I'm groping
for is significantly more interactive---or evolutionary---than yours:
you have to remember, I grew up in the day of the Graphical Users
Interface.  You really shouldn't lump all non-Everett'istas with the
people at {\it Social Text}.  [[I am saying all this with a smile.]]

\section{02 December 1997, \ ``Boojumating''}

Thinking about boojumating for the next couple of days has brought my
thoughts back to your many-worlds lecture.  And I ended up reading a
lot of my old correspondence on the subject:  in slightly more detail
it covers most of the points we discussed the other day, maybe a
couple more.  Also it shows that I guess in the end I had never
really come to grips with accepting the coarse-graining stuff after
all.

Anyway, I know that your interest in this has likely long since
vanished away.  But, just in case there's a little interest left,
I'll forward on those old e-mails.  Also, I have a hard copy of some
of Don Page's objections (in a note sent to {\Hartle})---if you want
to see it, I can bring it around some time.
\bq
\noindent
But oh, beamish nephew, beware of the day, \\
If your Snark be a Boojum!  For then \\
You will softly and suddenly vanish away, \\
And never be met with again!
\eq

\section{02 December 1997, \ ``Flying Equations''}

I couldn't help but think of the anecdote about John {\Wheeler}'s
(non-)flying equations you told the other day when I came across the
following little passage (presumably Biblical in origin):
\bq
\noindent ``I forbade any simulacrum in the temples because the
divinity that breathes life into nature cannot be represented.''
\eq

\section{07 August 1998, \ ``Impressions''}

Another cute thing I thought you might like is a quote that ended a
letter Adrian {\Kent} sent me this morning.  See the connection to
one thing we talked about yesterday?

\bq
\noindent In fragment D 125 \ldots\ [{\Democritus}] introduces the intellect
in a contest with the senses.  The former says ``Ostensibly there is
colour, ostensibly sweetness, ostensibly bitterness, actually only
atoms and the void''; to which the senses retort: ``Poor intellect,
do you hope to defeat us while from us you borrow your evidence?
Your victory is your defeat.''  You simply can not put it more
briefly and clearly.
\begin{flushright}
 --- Erwin {\Schroedinger}, ``Nature and the Greeks''
\end{flushright}
\eq

\section{06 April 1999, \ ``Round 1''}

I'm finally getting off of my lazy duff and taking care of some of my
external duties.  In this note I'll fulfill your request that I look
at your paper, ``Quantum Information and Physics: Some Future
Directions.''

You wanted citations to {\Wootters} and me for our ``visions.''  For
{\Wootters} the best place I know of is his PhD thesis:  W. K.
{\Wootters}, {\em The Acquisition of Information from Quantum
Measurements}, PhD thesis, The University of Texas at Austin,
Austin, TX, 1980.  For me, you can cite either my research proposal
C. A. Fuchs, ``The Structure of Quantum Information,'' or my
collected emails on the subject C. A. Fuchs, ``Collecting My
Thoughts:  Some Historical, Foundational, and Forward-Looking
Thoughts on the Quantum.''

Random comments:

3)  Hilbert space is a big place.  I think {\Carl} and I were the
first people to say it (in print at least, but also even in word)
and give the idea a quantitative meaning/analysis:  C. M. {\Caves}
and C. A. Fuchs, ``Quantum Information: How Much Information in a
State Vector?,'' in {\sl The Dilemma of {\Einstein}, {\Podolsky} and
{\Rosen} - 60 Years Later}, edited by A. {\Mann} and M. {\Revzen},
(Annals of The Israel Physical Society {\bf 12}), 226--257 (1996).
In 1994 after a dinner-time discussion with us, Jeff {\Kimble}
popularized it a bit in his quantum-state preparation talks
(1-900-SUM-FOCK). One can also find the following phrase quoted at
the beginning of Chapter 2 of Volume 1 of Reed and {\SimonB}:
``Gentlemen: there's lots of room left in Hilbert space.'' --- S.
MacLane

5)  Holographic universes and other parts of Section III.  The other
night {\Charlie} {\Bennett}, {\Herb} {\Bernstein}, and I were
discussing the foundations of QM, and {\Charlie} was really
revealing his distaste for the word ``man.''  (At the root of it, he
seems not to realize that the only way {\Shannon} information theory
has to define {\it information\/} is with respect to ignorance,
i.e., probability distributions.  Information in universes without
ignorant agents \ldots\ what could it mean?  More importantly, why
would one go to the trouble of defining it?  Why not stick with a
more objective sounding language---like dynamical laws, etc.---and
just forget about information?)  Anyway, in a moment of
exasperation, I exclaimed, ``Why is it that you can trust my
observations and thoughts on quantum channels, but not on quantum
mechanics itself!?! Why is it that you refuse to give me some
benefit of the doubt here?''  I suppose I should be happy:  his
reply was to compare me to {\Newton} and his follies with alchemy.
So, OK, I pass on the good will:  In my eyes you are like {\Newton}.

\section{01 September 1999, \ ``Flushing Fuchses''}

\bjp
But furthermore, as sympathetic as I am with your remarks about
{\Gleason}'s theorem, I'm not sure I know how to motivate
noncontextuality either.  Indeed, as you know, advocates of hidden
variables ({\Bell}, {\Bohm}, ..) have argued that noncontextuality is
not well motivated.
\ejp

That's quite a good point.  Noncontextuality of the {\it
probabilities\/} bothers me too.  (See notes below for whatever
they're worth.)  This almost always goes as an unemphasized
assumption behind {\Gleason}: people just say that a frame function
is assumed and let that be that.  (At least I haven't seen the term
``noncontextuality'' used in this way before; the comment by David
{\Mermin} indicates that he may not have either.)  What you're
thinking of when you refer to {\Bohm} and {\Bell} is that they didn't
think the notion of a noncontextual {\it hidden-variable theory\/}
was well motivated.  That is, neither of them liked the kind of
hidden variable theory that the {\Kochen}-{\Specker} results seem to
disallow (but see the {\Meyer}-{\Kent}-{\Clifton} articles of this
summer). For a hidden-variable theory to be noncontextual, the truth
values of all potential measurement results have to be set before the
interaction. For a probability assignment to be noncontextual, the
probability of a Hilbert-space projector must be independent of
which basis it's associated with (i.e., which measurement
interaction it's associated with).  Who ordered that?  I don't know
(but I list my present leaning in the note to Howard below).

\bjp
Therefore, {\Deutsch}'s framework (suitably sharpened) is of
potential interest if it enables us to derive the quantum
probability rule without directly invoking noncontextuality (as
{\Gleason}'s argument does).  [\ldots]  So maybe there is something
of merit that we can sift from the wreckage of {\Deutsch}'s paper
\ldots\ Do you agree?
\ejp

It could be, though I'm inclined NOT to keep BOTH the state space and
the observable space as primitive, when we already know that we can
get by without one of those---clearly my personal taste.  But if one
does accept Insufficient Reason in this context---i.e. that equal
amplitudes give equal probabilities---one would still have to be very
careful with the remainder of {\Deutsch}'s {\it particular\/}
argument. We were never able to make the rest of it sufficiently
rigorous to examine before getting so frustrated as to give up
\ldots\ and just write up what we had.  (Recall our slight
discussion of the principle of substitutability.)  Since then,
Howard and Jerry have continued to think about it a bit, but I don't
think they've been able to come to a sufficient conclusion either.

Safer ground can probably be found in the stuff that {\Cleve} and
{\Braunstein} (and {\Nielsen}?) started to write up but never posted.
(I talked to Richard about it this summer:  he said he believed the
result rigorous but not overly compelling and didn't have plans to
complete the project.)  They explicitly took Insufficient Reason as
their starting point.  {\Zurek} also has a similar argument that I
saw him present in Baltimore (but God knows if there's anything
rigorously proved there without loads of side assumptions \ldots\
{\Cleve} sure seemed to need them, and I find {\Cleve} very
trustworthy). The reference to {\Zurek} is: Phil.\ Trans.\ Roy.\
Soc.\ London {\bf A356} (1998) 1793.

All that said, I have my doubts that your invariance property doesn't
already build in noncontextuality almost immediately.  The
assumptions that {\Deutsch} does use (without invariance) already
gets to the fact that the value functions can be expressed as
expected utilities for some probability distribution.  And then from
there is it that hard? \ldots\ OK, I don't see it immediately:  let
me think a little harder about this.

\section{07 September 1999, \ ``Two Rabbis in a Bar''}

Thanks for the latest note.  {\Carl} and I thought a little about
your point.  We came across nothing great or deep, but let me record
some of the equations and thoughts that came out of that discussion
(mostly so WE won't forget them).

Basically it looks like there are are loads and loads of value
functions that satisfy the zero-sum, additivity, and unitary
invariance properties.  Unitary invariance gives that the value
function can only depend upon the amplitudes and utilities (not the
basis used in the expansions) and also gives permutation symmetry
thereafter:
\bea
{\cal V}\!\left(\sum_i \lambda_i|\phi_i\rangle;\sum_i
x_i|\phi_i\rangle\langle\phi_i|\right)
&=&
{\cal V}(\lambda_1,\ldots,\lambda_n;x_1,\ldots,x_n)\\
&=&
{\cal V}\Big(\lambda_{\sigma(1)},\ldots,\lambda_{\sigma(n)};
x_{\sigma(1)},\ldots,x_{\sigma(n)}\Big)\;,
\eea
where $\sigma$ is any permutation.  Adding to that {\Deutsch}'s
additivity or {\Savage}'s complete transitive ordering we have,
\bea
{\cal V}(\lambda_1,\ldots,\lambda_n;x_1,\ldots,x_n)
&=&
\sum_i f_i(\lambda_1,\ldots,\lambda_n)x_i \\
&=&
\sum_i
f_i(\lambda_{\sigma(1)},\ldots,\lambda_{\sigma(n)})x_{\sigma(i)}\;.
\eea
If this is going to hold true for all sets of $\{x_i\}$, then we must
also have,
\be
f_i(\lambda_1,\ldots,\lambda_n)=
f_{\sigma(i)}\Big(\lambda_{\sigma(1)},\ldots,\lambda_{\sigma(n)}\Big)\;.
\ee
Presumably, noncontextuality would force us down to
\be
f_i(\lambda_1,\ldots,\lambda_n)=f(\lambda_i)\;.
\ee

Anyway, within this context some of the probability laws we've
discussed take the following form:
\bea
& \mbox{standard QM rule:} &
f_i(\lambda_1,\ldots,\lambda_n)=|\lambda_i|^2\\
& \mbox{BCF$^2$S rule:} &
f_i(\lambda_1,\ldots,\lambda_n)=\frac{1}{\mbox{\# of nonzero
$\lambda$'s}}\\
& \mbox{JP rule (slightly generalized):} &
f_i(\lambda_1,\ldots,\lambda_n)=\frac{g(|\lambda_i|)}{\sum_i
g(|\lambda_i|)}\quad\mbox{for any function $g$.}
\eea
Notice that all of these rules reduce to the standard one when the
$\lambda$'s are equal.  The point to be made here is that
{\Deutsch}'s development up to the ``pivotal result'' does not make
a cut between any of these proposed rules.  At least up to that
stage in the argument, unitary invariance is still pretty weak. So,
if unitary invariance does add some extra power it is going to have
to be at later stages in his argument, perhaps where he invokes the
mysterious principle of substitutability.  In particular, we
probably ought to retract the statement we wrote you last week,
``Indeed you have a point that unitary invariance will pretty much
give the whole shebang.''

As a point of philosophy, it is quite interesting to see that the
unitary invariance of the value functions is distinct from the
{\Gleason} assumption of noncontextuality---this may be important
even for the strict Bayesian.  (You know, we do have significant
sympathies with the idea of deriving as much of the quantum
probability rule from decision theory as we can.)  But, upon thinking
about it, I don't understand why you find noncontextuality so
unmotivated from the {\Everett} point of view.  Can you articulate
why you believe that?

I suppose part of my problem in seeing the difficulties you're seeing
is that the {\Everett} structure continues to look completely
arbitrary to me.  If one is going to go already that far in
disconnecting the formalism from the world of our experience, why
bother with such relatively minor details?  Despite your words,
\bjp
{}From the {\Everett} viewpoint, then, probability is about the
expectations of an agent who is himself part of the system,
expectations that are manifested by the behavior of the agent.
[\ldots] So if we were able to argue that an observer who is part of
the quantum system will play a game according to a certain strategy,
then perhaps we will have succeeded in ``deriving'' a probability
rule.
\ejp
I continue to see no place for probability within the {\Everett}
point of view.  I just don't know how to say it more forcefully than
this: the notion of a decision-making agent is a contradiction in
terms within that model of the world.  I know you likely read what I
wrote Howard the other day even if you didn't comment, but I want
the words to gnaw at you:
\bq
\noindent In the {\Everett} interpretation, we can never ARRANGE to do any
experiment, to play any game.  There is no sense in which this phrase
is meaningful in that interpretation. The {\Everett} world just IS\@.
Experimenters have NO choice to arrange anything in an {\Everett}
multiverse; their actions are at best a grand conspiracy set by the
initial condition of the universe.  They have no will, no possibility
to do anything that was not preordained.  If an experimenter in the
{\Everett} theory perceives himself choosing one game over another,
it can be but ILLUSION that he had any choice in the matter.
\eq

{\Charlie} {\Bennett} likes to tell the story of two rabbis.  One
walks into a congregation, throws himself before the ark and
exclaims, ``Lord, before you I am nothing!''  Seeing that, a second
rabbi walks in, throws himself violently to the ground; groveling,
he screams, ``LORD, BEFORE YOU I AM NOTHING!!!''  Carried away in
the excitement, a common member of the congregation goes to the ark
and does the same thing.  In shock, the first rabbi looks to the
second and says, ``Who is HE to think HE's nothing?!?!''

So too, I find myself chuckling, ``Who is HE to think that
noncontextuality is unmotivated for HIS interpretation of quantum
mechanics?!?!'' From the point of view I'm trying to construct---that
quantum mechanics is about the tradeoff between INFORMATION gain and
INFORMATION disturbance---it is truly a problem.  For in that model
of the world, measurements are invasive creatures and the theory of
quantum mechanics is our best attempt to say something in spite of
that fact (that truth, reality, noumenon, ding an sich, \ldots\
whatever you wish to call it). My question is why are measurements
invasive but not too invasive?  Why are noncontextual hidden
variables ruled out, but noncontextual probability assignments
endorsed?  From this point of view noncontextuality screams out for a
deeper explanation. What for the Everettista makes it scream the same
way?

\section{08 September 1999, \ ``Smug Note''}

\bjp
``Decision-making agent is a contradiction of terms \ldots'': \
I'm not sure I'm ready and willing to debate the issue of free
will.  The illusion of being able to make decisions (if it is an
illusion) is a property of the conscious mind, and can be explained
only as such. To the Everettista, this would require a theory of how
a conscious observer functions, an observer who is part of a quantum
system \ldots\
\ejp

This is a new tack to me.  Since when has not having an explicit
``theory of conscious observers'' ever slowed down the {\Everett}
movement before??  What else should we make of it when {\Everett}
says [{\DeWitt} and {\Graham}, p.\ 65]:
\bq
\noindent We note that there is no longer any independent system state
or observer state, although the two have become correlated in a
one-one manner.  However, in each {\it element\/} of the
superposition \ldots\ the object system state is a particular
eigenstate of the observer, and {\it furthermore the observer-system
state describes the observer as definitely perceiving that particular
system state.}
\eq
How else could we take seriously Figure 2, ``A Reversible Simple
Mind,'' in Cooper and van Vechten paper's [D \& G, pp.\ 223--223].
How else could {\Deutsch} in his misconceived thought-experiment
[IJTP, 1985] say,
\bq
\noindent After the completion of the measurement, the observer
records (in his memory, or in his notebook if necessary)---not the
value ``$N$\,'' or ``$S$\,'' of the spin, but only {\it whether or
not he knows this value}.  He may write ``I, Professor X, F.R.S.,
hereby certify that at time $t'''$ I have determined whether the
value of the North component of the spin of atom-1 is
$+\frac{1}{2}\hbar$ or $-\frac{1}{2}\hbar$.  At this moment I am
contemplating in my own mind one, and only one of those two values.
\ldots''
\eq
How could John {\Preskill} in his much lauded lecture notes even be
able to write his Eq.~(3.178)??

As advertised, I say all this with complete smugness.  The point is,
how can all these things be said without a tacit (but nevertheless
grand) theory of consciousness working in the background?

How does the Everettista know {\it a priori\/} that conscious states
are always associated with ORTHOGONAL quantum states?  The brain's a
pretty complex system:  if we know so little about it, how on earth
can we be so confident of capturing its essence in such a glib way?
[[{\Kiki} asks me what I'm hungry for.  I say, ``I don't know, I'm
not even hungry.''  Within a couple of minutes my stomach is
roaring. Her information gathering caused a disturbance to my
conscious state, you might say.  In this respect, my conscious
states act a little like nonorthogonal quantum states.  Who's to say
how far that analogy can or cannot be pushed?  I say that tongue in
cheek, but my point is firm:  I don't think there's a better
justification for the standard quantum measurement analysis (in
terms of orthogonal observer states) than that the pace was set in
the last few pages of von Neumann's book.]]  How does the
Everettista know {\it a priori\/} that conscious states should be
associated with PURE states?  [[My body usually runs about 99
degrees F; people who run significantly colder, are usually
significantly deader---and I've never caught a dead man thinking. Or
to make it sound more relevant, think of the poor girl who was
locked in a closet for 13 years while her brain should have been
developing speech skills.  She was never able to recover: should she
be considered as conscious as I?  How does the Everettista know {\it
a priori\/} that conscious entities do not have to be significantly
ENTANGLED with other conscious entities to be conscious in the first
place?  Again I say that tongue in cheek, but how do you know?]]

Or here's a better one.  I feel strongly---you might even say
``know''---that I love {\Kiki}.  I also feel strongly---you might
even say ``know''---that I observed a single ion in one of Herbert
{\Walther}'s ion traps last spring.  In the first case, most of us
wouldn't presume that a single pure quantum state could be associated
with my state of consciousness:  ``love'' is just too loosey-goosey
of a notion to imagine being captured in such a simple-minded way.
How does the second case differ except out of expedience for the
problem at hand?

So the point again:  It seems hard to deny that the {\Everett}
analysis of measurement doesn't dip into concepts well beyond the
domain of standard physical theory.  As such I view it as little
more than a comforting religion.  The point (among others I could
give) which the Everettistas have yet to analyze is precisely the
point I CONSCIOUSLY leave UNanalyzed: how do experimental outcomes
come about and how do they become known?

That is, I take ``experiment'' as the basic atom out of which we
build scientific theories and models of the world.  If a model so
constructed does NOT cough back up the very possibility of
experiment, then I would view it with a leery eye \ldots\ and I
suppose that is what I am attempting to get you to do also.

I am a little uncomfortable with the phrase ``free will'' as you used
it above.  What I am trying to get at is perhaps better captured by
what Hans Primas had to say:
\bq\noindent
[I]t is a tacit assumption of all engineering and experimental
sciences that nature can be manipulated and that the initial
conditions required by experiments can be created by interventions
using means external to the object under investigation. That is, {\it
we take it for granted that the experimenter has a certain freedom of
action which is not accounted for by first principles of physics.}
Man's free will implies the ability to carry out actions, it
constitutes his essence as an actor. Without this freedom of choice,
experiments would be impossible. {\it The framework of experimental
science requires this freedom of action as a constitutive though
tacit presupposition.}
\eq
The words ``free will'' are indeed used there, but the more central
issue is the one captured in the first sentence.  That is about the
notion of science and hardly more.  If with each succeeding
experiment we become more and more confident that quantum mechanics
is a good description of the world, how can we at the same time
become ever more confident that those same knob-twiddling actions
were preordained in the initial condition of the universe?

I keep thinking that if I hit this from enough angles something will
eventually shake loose in you.  {\it I want to wake you from your
dogmatic slumber \ldots\ }

So I'll end with one last act of smugness.  You say,
\bq\noindent
To the Everettista, this would require a theory of how a conscious
observer functions, an observer who is part of a quantum system
\ldots\
\eq
I'll go you one better.  In the view I am promoting, I am not only
part of a quantum system, I AM a quantum system.  You hint that you
don't think a Quantum Bayesian can have that.  But I do have it: what
it means for me to be a quantum system is that I can be analyzed by
you or anyone else profitably with the methods of quantum mechanics.
If you are lucky enough to have maximal knowledge of me, you would
describe what you know via some pure state.

What fear is it that drives you and so many of our colleagues to the
excesses (and emptiness) of the {\Everett} worldview?

You want to do quantum cosmology, and you think you can't get by
without it?  Is that it?   Here's how it's done; I'll show you how to
do it.  You write down a wave function for the universe:
$$
|\,\Psi_{\rm \scriptscriptstyle universe}\,\rangle\;.
$$
(It {\it is\/} that easy.)  For some aspects of the universe, this
might be an excellent approximation TO WHAT YOU KNOW ABOUT IT. As far
as I can see---and I don't believe I am working with blinders
on---there are no conceptual difficulties with doing quantum
cosmology in just this way.

To think that it would first require a theory of consciousness before
we could do quantum cosmology (or any physics at all) in the Bayesian
style is just a bunch of malarkey.  That's no more true than saying
we would need a theory of consciousness before we could open a
gambling house in the state of New Mexico.  The theory of gambles and
decisions seems to work perfectly well in spite of psychology.

You say in your lecture notes:
\bjp
To those who hold the contrary view (that, even if there is an
underlying reality, the state vector only encodes a state of
knowledge rather than an underlying reality) tend to believe that the
current formulation of quantum theory is not fully satisfactory, that
there is a deeper description still awaiting discovery.  To me it
seems more economical to assume that the wavefunction does describe
reality, unless and until you can dissuade me.
\ejp
I've now put my best emotion into it, and hopefully enough reasoning
to pique your interest.  I'm not likely to come back to pester you
again if I haven't made a dent by now.  But do know this:  I do not
fit into the category of those who ``tend to believe that the current
formulation of quantum theory is not fully satisfactory.''  The only
thing that I think is not fully satisfactory is our understanding of
why we are compelled to use this theory.  When we understand why we
cannot obtain more information about a system's behavior than can be
captured by a symbol $|\psi\rangle$, we will have learned something
profound.

\section{10 September 1999, \ ``Meeker Note''}

Thanks for the discussion last night, and thanks for bringing the
weakness of my linearity argument to the surface.  I got a few things
out of the meeting.  The biggest one might be this.  I was struck
more than usual how the Bayesian view seems a bit like beer: that is,
in being an acquired taste, or, more accurately, tasting good only
after a concerted effort to swallow it more than once. Talking with
you helps emphasize how the mode of language {\Carl} and I have
gotten ourselves into is so very distinct from the usual language of
theoretical physics.  This is something we really need to be careful
about in our presentations.

Our discussion reminded me of something I once read in an article of
Ed {\Jaynes}:
\bq\noindent
The main suggestion we wish to make is that how we look at basic
probability theory has deep implications for the {\Bohr}--{\Einstein}
positions.  \ldots\  {\Einstein}'s thinking is always on the
ontological level traditional in physics, trying to describe the
realities of Nature.  {\Bohr}'s thinking is always on the
epistemological level, describing not reality but only our
information about reality.  The peculiar flavor of his language
arises from the absence of words with any ontological import; the
notion of a ``real physical situation'' was just not present and he
gave evasive answers to questions of the form: ``What is really
happening?''
\eq
In some of your formulations of the ways in which we agree, I felt
that precisely the same distinction was coming out. The use of the
phrase ``dynamical laws'' is probably a good example. Whenever I used
the phrase, I meant something that tells us how to update our
predictions. [[A symptom of that is that I find myself questioning
whether the dynamical laws (both classical and quantum) are linear
simply because Bayes rule is linear.]]  But then I wished to stop
there. You, on the other hand, seemed always to use ``dynamical
laws'' in a more ontologized way:  those dynamical laws exist in some
sense independent of our theory.

I don't find the compulsion to do that.  For though I consider myself
a realist at heart (for precisely the reason you express in your
lecture notes), I don't see physics as an expression of our great
knowledge of the world \ldots\ but more accurately as an expression
of our great ignorance.  I see it as an expression of the best stab
we can make at things.

Our greatest difference at this stage seems to be a sociological one,
or one of temperament or emotion (as you said).  I---just stealing
the method from {\Socrates}---think we will find our greatest
strength by more fully realizing our ignorance and digging for its
roots. This is the direction I've been turning toward ever more
forcefully since reading the {\Jozsa}, {\Robb}, {\Wootters} paper in
1994 (and choosing my thesis topic soon after that).

When I think of running with the theory (another of your phrases), I
mostly think of things like these words of John {\Wheeler}'s:
\bq
I want you \ldots\ to jolt the world of physics into an understanding
of the quantum because the quantum surely contains---when
unraveled---the most wonderful insight we could ever hope to have on
how this world operates, something equivalent in scope and power to
the greatest discovery that science has ever yet yielded up: Darwin's
Evolution.

You know how {\Einstein} wrote to his friend in 1908, ``This quantum
business is so incredibly important and difficult that everyone
should busy himself with it.'' \ldots\ Expecting something great when
two great minds meet who have different outlooks, all of us in this
Princeton community expected something great to come out from {\Bohr}
and {\Einstein} arguing the great question day after day---the
central purpose of {\Bohr}'s four-month, spring 1939 visit to
Princeton---I, now, looking back on those days, have a terrible
conscience because the day-after-day arguing of {\Bohr} was not with
{\Einstein} about the quantum but with me about the fission of
uranium. How recover, I ask myself over and over, the pent up
promise of those long-past days? Today, the physics community is
bigger and knows more than it did in 1939, but it lacks the same
feeling of {\bf desperate} puzzlement. I want to recapture that
feeling for us all, even if it is my last act on Earth.
\eq
I cannot shake the feeling that we are lucky to be in a unique time,
finally amassing the technical tools to get at what John is asking us
to do.  Quantum mechanics is about information (in the deepest of
senses): run with it, I say --- that's the sense.

Another thing that came out of yesterday's discussion is that I
became more confident in the force of something I wrote you the other
day:
\bq\noindent
If with each succeeding experiment we become more and more confident
that quantum mechanics is a good description of the world, how can we
at the same time become ever more confident that those same
knob-twiddling actions were preordained in the initial condition of
the universe?
\eq
Let me try to say it slightly differently.  I find it kind of
lopsided to accept that experiments help us amass evidence for this
or that dynamical theory,\footnote{Don't forget, I'm likely to be
stating the phrase ``dynamical theory'' in a more epistemology-soaked
way than you are used to.} and not at the same time accept that they
have helped us amass evidence that we are getting ever finer-tuned
control of nature---that finer-tuned control being expressed in our
setting the initial conditions of the experiments in any way we
please.  The force of these statements, I think, finds expression in
the general structure we use for all our endeavors in physics:  there
are dynamical laws, and there are initial conditions.  As far as I
understand, no one has ever been able to tie the two distinct
structures together, as they say, with a ``theory of initial
conditions.''  (I suppose I'm willing to bet they never will \ldots\
but that is certainly a statement of temperament and emotion.)

Where should quantum mechanics go in my eyes?  Finding better ways to
describe and quantify the ``ticklishness'' of the world that has been
given a technical expression by quantum cryptography and its
subsidiaries (for instance I'm thinking of ``nonlocality without
entanglement'' here). To see if there's a good sense in which that
same ticklishness can be thought of (in a useful way) as the power
behind quantum computing. To get things tidied up from this point of
view (or see if it falls) so that the next generation can do
something truly astounding.  But that's just a feeling in my gut (and
probably the spirit of John {\Wheeler}).

\section{12 December 1999, \ ``Freedom''}

\bjp
Free will usually means the ability of conscious beings to influence
their own future behavior. Its existence would seem to imply that
different physical laws govern conscious systems and inanimate
systems.  I know of no persuasive evidence to support this viewpoint,
and so I am inclined to reject it.
\ejp

Can't agree with your second sentence.  When you think of physical
law, you should say the chant epistemic, epistemic, epistemic.  Then
you can chime in with MLK:  free at last, free at last, thank god,
I'm free at last!

Had fun talking to you the other day.

\section{10 October 2000, \ ``Sneak Previews''}

\bjp
In the course of thinking about what to tell my class about
{\Gleason}'s theorem, I looked again at your paper {\tt
quant-ph/9907024}, and pondered again your statement about the
noncontextuality assumption:
\bq
\noindent\rm
    This important assumption, which might be called
    the ``nocontextuality'' of probabilities, means that
    the probabilities are consistent with the Hilbert
    space structure of the observables.
\eq
And I realized \ldots\ I have no idea what is meant by ``consistent
with the Hilbert space structure of the observables.''

What {\em does} it mean?
\ejp

That's a good question.  \ldots\ Because I don't completely know the
answer yet.  But I think I'm on the tail of the tiger.  You'll see
what I mean in the next email.  That document is a problem set I was
writing up for a little meeting {\Gilles} and I organized in
Montr\'eal on ``Quantum Foundations in the Light of Quantum
Information.'' [See letter to {\Gilles} {\Brassard}, titled ``Problem
Set Based on Information-Disturbance Foundation Quest,'' dated 15
May 2000.] It looks a little incomplete because it is. Unfortunately
the fire hit just as I was starting to write Problem \#2, so I never
completely finished the document properly.

The part that's most relevant to your question is Problems \#1 and
\#3 and Proto-Problems \#5 and \#6.  In particular, Problem \#1.
Have a look at those and then come back to what I'm about to write
below.

Since then, we've made progress on a few fronts.  With regard to
Problem \#1, Nurit {\Baytch} and I have now shown that Conjecture 1
holds for at least a set of measure 1 in the parameters.  There are
a few stubborn holdouts (a discrete set's worth), but I think
ultimately we'll fix that.  I hope that gives you a better feeling
for what the noncontextuality assumption is about:  it only meshes
well with the orthogonality structure of Hilbert space, not with the
linear structure per se or even any other ``fixed-angles''
structure.  Nurit has \TeX'd up some of this, but I haven't had a
thorough look at her document yet.  If you'd like, I can send the
details by the end of the weekend.

A more constructive bit of progress has come from Problem \#3.  I
ultimately scrapped the precise wording there because a more natural
question is this.  Let us think of the most basic elements in
quantum measurement theory as the POVMs, not simply the ODOPs
(one-dimensional orthogonal projectors).  Then a natural
generalization of {\Gleason}'s problem is to suppose that the
probabilities for measurement outcomes are given by a function $f$
from positive operators to the real interval $[0,1]$ such that
$\sum_i f(E_i) = 1$ whenever the $E_i$ form a POVM.  This assumption
embodies an even stronger kind of ``noncontextuality'' for quantum
probabilities than before.  What's neat is that if one makes these
assumptions, then one can prove the standard quantum law---i.e. that
there must exist a density operator $\rho$ such that $f(E_i)= {\rm
tr} \rho E_i$---in a way that mere mortals can understand.  The
proof is quite simple and memorable (by that I really mean
memorizable). This is joint work with {\Caves}, {\Renes}, and another
student of {\Carl}'s (don't know his last name or even the spelling
of his first) \ldots\ with actually {\Renes} doing the brunt of the
details. A further payoff of this effort is that the proof works in
2-D where the standard {\Gleason} theorem falters.  Actually I think
that's my favorite feature of the whole thing.  (That that might
work was my first scientific thought after the fire, so I guess I
hold it particularly dear.)  {\Renes} has started working on a draft
of this: we can send it to you once it's in reasonable form.

Finally, let me say a smidgen about Proto-Problem \#5 in conjunction
with Proto-Problem \#3.  I've gathered a very small amount of
evidence that the project there might possibly work out.  Namely it
dawned on me that in all known {\Kochen}-{\Specker} examples all the
rays have algebraic-number coordinates.  This is a prerequisite for
the possibility of a {\Gleason} theorem on that structure.  You might
sneer that this is an overly technical question, but I think its
interpretational import is quite relevant (though I didn't express
that so well in my shoddy notes).  Let me try to do that idea a
little more justice here.  The main question is whether most of the
structure of quantum mechanics is about what ``is'' in the world, or
instead how it goes ding to our interactions with it.  (You know I'm
more inclined to the latter.)  Well a number of decent scientists
(for instance Birkhoff and von Neumann) have thought so much that
quantum mechanics is about what ``is'' that they were willing to
rethink what it means for a system to have properties.  Out of this
arose the idea of a quantum logic:  algebraic operations AND, OR,
and NOT that satisfy the postulates of orthomodular lattices rather
than the postulates of Boolean algebra.  Interestingly, B and vN
noticed that AND, OR, and NOT can be identified with INTERSECTION,
SPAN, and ORTHOGONAL COMPLEMENT for subspaces in a vector space if
and only if the (skew-) field of the vector space is one of the
three:  real numbers, complex numbers, or quaternionic numbers.
Rationals won't work; algebraic won't work.  So my question to the
quantum logicians (like Jeff {\Bub}) is, ``Do you hold {\Gleason}'s
theorem dear?''  ``If so, then what do you make of the fact that it
might work for a structure that is not an orthomodular lattice?''
You see, I have no compulsion to believe that QM should only be a
tight structure for a theory of ``properties'' (i.e., a quantum
logic).  In fact, I would consider it evidence in my favor if
{\Gleason} still worked for such a non-lattice-theoretic entity.

If you want updates on any of the other problems in the Montr\'eal
problem set, I can send them, but I don't think they're completely
relevant to your question.

Hope that helps.  I'll try to answer your more logistical email a
little later in the week.

\chapter{Letters to Joseph {\Renes}}

\section{22 June 2000, \ ``References for {\Gleason}''}

Some relevant references.  The important ones are {\Gleason}'s paper,
Cooke et al's paper, and {\Pitowsky}'s paper.

\begin{enumerate}

\item
R.~Cooke, M.~Keane, and W.~Moran, ``An Elementary Proof of
{\Gleason}'s Theorem,'' Math.\ Proc.\ Camb.\ Phil.\ Soc.\ {\bf 98},
117--128 (1981).

\item
A.~M. {\Gleason}, ``Measures on the Closed Subspaces of a Hilbert
Space,'' J. Math.\ Mech.\ {\bf 6}, 885--894 (1957).

\item
I.~{\Pitowsky}, ``George {\Boole}'s `Conditions of Possible
Experience' and the Quantum Puzzle,'' Brit.\ J.\ Phil.\ Sci.\ {\bf
45}, 95--125 (1994).

\item
I.~{\Pitowsky}, ``Infinite and Finite {\Gleason}'s Theorems and the
Logic of Indeterminacy,'' J. Math.\ Phys.\ {\bf 39}, 218--228 (1998).

\item
F.~Richman and D.~Bridges, ``A Constructive Proof of {\Gleason}'s
Theorem,'' J. Func.\ Anal.\ {\bf 162}, 287--312 (1999).

\end{enumerate}

\section{11 December 2000, \ ``New Ideas and Old Men''}

Thanks for the new ideas.  Let me try to make a few comments.  I'm
in a smoky hotel bar right now; I wonder how this is affecting my
laptop's health?  I know it's wreaking havoc on mine.  I'm still in
Vienna, by the way.  I leave Tuesday morning.  Unfortunately I
haven't gotten an ounce written of what I had hoped to.

\bjr
this got me thinking about this stuff, and it seems that in some
sense what we're gathering information about, when we make
measurements, is the ability to predict future measurements.
\ejr

You haven't read my paper with Kurt {\Jacobs} yet, have you?  [See
{\tt quant-ph/0009101}.]  Come on, at least read the Introduction and
Footnote 27. Also it might be useful to have a look at the section
where we formulate what the notion of information ought to be! (Come
on, I didn't write the paper for nothin'.)

\bjr
with this notion it's simple to define the information gained (or
lost!) about a hypothetical measurement $X$ when actually measuring
$Y$. it seems first that the structure of measurements makes things
such that $I_Y(X)$ can be negative, whereas classically this can't be
the case (right? knowing more of anything classically means we know
more of everything).
\ejr

It feels like you're kind of on the right track here, but I think
there is a definite problem in grounding this exercise in one
observer.  That is, if one observer makes a measurement of $X$, he
can always rig the measurement interaction so that he ends up with
perfect predictability over $Y$.  The interaction just has to have a
conditional unitary so that $X$-eigenstates are taken to
$Y$-eigenstates.

This is why I think it might be better to ground the problem with
respect to two people.  Suppose Alice and Bob start with a third
system for which they both ascribe a state $\rho$.  Now imagine
Alice interacts with it in such a way as to increase her
predictability of an $X$ measurement on it, while Bob interacts with
it in such a way as to increase his predictability of a $Y$
measurement on it.  How high can they make their mutual
predictability as a function of $\rho$, $X$, and $Y$?

Why am I so obsessed with always having two players in the game?
Because I want to connect all the concerns in quantum mechanics with
Bayesianism as much as I can.  The issue is:  Under what conditions
can the various agents in a situation come into better agreement
even though they may start with differing or incomplete states of
knowledge?  That sort of thing is the bread and butter of {\Bernardo}
and {\Smith}'s book, for instance.  But quantum mechanics further
adds the issue of noncommutivity to this

So, there will be some limitation to how high the mutual
predictability can be.  You might find some earlier work on the
subject in the papers of Michael {\Hall}.  (M.~J.~W. {\Hall})  I'm
thinking in particular of a PRL of his from about 1994 or 1995. You
might ask {\Carl} about it.  But I don't think the two player
problem is precisely the one Michael addressed.

\bjr
because i don't know what quantum information is -- to me information
should be classical, or we should build up the information theory of
quantum systems to the point where we make use of von Neumann entropy
rather than making quantum information theory an analogous theory to
classical information theory
\ejr

I certainly like this statement.  In fact I just made a slide for my
talk here last week that said:
\bq
\noindent
QM is about Information.  $H(X)$. Plain old information in
{\Shannon}'s sense:  uncertainty, lack of predictability.
\eq

\bjr
listening to bill wootters's talk yesterday i had a slightly
different idea about approaching this -- the relationship to
entanglement.
\ejr

I think that's a good idea too.  One gets the feeling that the more
entangled two systems are, the more one can predict about
measurements on the other system, given only the identity of the
measurement.  In the ideal case of an EPR pair, one just performs
the identical measurement on the system in one's possession.

Unfortunately I can't think of anything more intelligent to say than
that.

Let me end by coming back to the point above:  noncommutivity.  What
I really want to know is, ``Why noncommutivity?''  (That's another
way of saying, ``Why Hilbert space?'')  The greatest missing link in
my philosophical program is now the answer to that question.  How
can we make an information-disturbance tradeoff IMPLY the existence
of noncommuting states of knowledge?  I don't have a clue how to
make that happen, but it's got to happen.

It's got to happen.  And I've got to go to bed.

\section{05 January 2001, \ ``In Defense of Interactionism''}

I don't know what {\Carl} has said, but I like the sound of your
``research direction.''  I doubt any of the references below will be
overly useful, but I met a guy at the quantum information tutorials
in Edinburgh who has spent a lot of time generalizing quantum logics
to effect algebras.  His name is Roberto {\Giuntini}.  I just looked
up the following articles on {\sl Web of Science}.

I say I doubt it will be overly useful, because I gathered that, in
his heart, he too was looking for the holy grail of most quantum
logicians:  To find a way of pinning observer-independent properties
on the world via algebraic properties of the theory's surface terms
(i.e., states, observables, Hamiltonians).  In opposition to
{\Carl}---I won't say it's heresy---I think that's the wrong way to
proceed.  I think we'll ultimately find A QUANTUM REALITY, but we'll
have to dig deeper than that.

All that said, the reason I like your research direction is because
I think understanding the purely algebraic properties of POVMs is an
important link nevertheless.  I would like to think that quantum
mechanics is more about our interface with the world than the world
itself \ldots\ and that it's at that interface that we'll find our
glimpse of a ``quantum reality.''  What is it about the interface
that makes our best Bayesian predictions of the quantum mechanical
form? That can be an algebraic question just as much as trying to
pin naive properties on the world (regardless of how contrived the
logic).  In that sense, there is some chance of maybe learning
something from the {\Giuntini}s of the world.

\section{30 March 2001, \ ``Web Page Snooping''}

Great quote!
\begin{center}
\parbox{2.7in}{
As far as we can discern, the sole purpose of human existence is to
kindle a light in the darkness of mere being. \\
\hspace*{\fill} --- {\it Carl {\Jung}} }
\end{center}
What's the precise source?

\section{11 April 2001, \ ``Piggybacking Philosophy on Physics''}

\bjr
answering physics questions relative to a static background
philosophy will not, i believe, yield what one would want it to ---
the grand questions are necessarily partly philosophical and thus
must be confronted head on.
\ejr

I do appreciate the spirit, and I am part of the choir.  John
{\Wheeler} once said something that made a big impression on me:
``Philosophy is too important to be left to the philosophers!''

You can see the implication.  If we're going to hope to make real
progress we've got to have our feet firmly planted in the practice
of physics.  Really good physics is the very best of philosophy.

In the words of George Castanza, ``Do me a solid buddy.''

\chapter{Letters to Mary Beth {\Ruskai}}

\section{17 February 2000, \ ``The Quantum View''}

What a really nice article!  I read it once this evening with dinner
(at a place that served grilled chicken---not so easy to scroll
through a computer screen under those conditions!).  Now, with my
jetlag in distinct control---its 3:30 AM for me---let me read it
again and make a few comments here and there.  (I just arrived in
Haifa yesterday for a visit with  {\Asher} {\Peres}; I'll be here a
little over a week.)

$\diamondsuit$  Why don't you write the following for your requested
reference: C.~A. Fuchs, ``The Structure of Quantum Information,''
available at [...]  [Of course, it is no longer available.  Instead,
see note to {\Carl} {\Caves}, dated 21 December 1997.]

$\diamondsuit$  I like this view of physics as being like a jig-saw
puzzle. Note a typo in this paragraph:  you misspelled de {\Broglie}.

\bbr
As far as I am aware, there is no evidence that either {\Heisenberg}
or {\Schroedinger} was motivated by political or social factors.
\ebr
Actually there is a pretty strong case that these two guys were
indeed quite motivated by cultural factors to find an
indeterministic element in physics.  And that was long before
quantum mechanics took a final form.  Two important articles to read
in this regard are:
\begin{enumerate}
\item
P.~{\Forman}, ``Weimar Culture, Causality, and Quantum Theory,
1918--1927: Adaptation by German Physicists and Mathematicians to a
Hostile Intellectual Environment,'' in {\it Historical Studies in the
Physical Sciences, Vol.~3}, edited by R.~McCormmach (U. Pennsylvania
Press, Philadelphia, 1971), pp.~1--115.
\item
P.~A. {\Hanle}, ``Indeterminacy Before {\Heisenberg}: The Case of
Franz {\Exner} and Erwin Schr\"o\-dinger,'' in {\it Historical
Studies in the Physical Sciences, Vol.~10}, edited by R.~McCormmach,
L.~Pyenson, and R.~S. Turner (Johns Hopkins U. Press, Baltimore,
1979), pp.~225--269.
\end{enumerate}
Both articles are quite good and do get to the nub of the matter.
However similar to the point you made in the
``suppose-for-the-sake-of-argument'' sentence, none of that desire
played out in the particular forms for quantum dynamics that these
gentlemen proposed.  It's probably worth reading and citing these
articles, if for no other reason to show that you are aware of all
the issues.

\bbr
It is often said that mathematics is the language of science. But it
would be more accurate to say that mathematics is a family of
related languages.  The languages of algebra, analysis, geometry,
etc. often give very different insights into a problem.
\ebr
I like that sentence!  Later when you write about gender:
\bbr
In these situations, I believe that the insights of gender can play
a role analogous to that of a mathematical language such as geometry.
\ebr
I was reminded that I tend to think similar things about the issue
of quantum interpretations.  Maybe the different interpretations
make no difference in what is and is not possible in quantum
phenomena, but they do play a role in directing thought.  It seems
to me that some interpretations are just dead ends for the
inquisitive spirit (I'm thinking of Bohmianism in particular here).
Whereas some interpretations---even some that I don't like, like
many-worlds---have led to the contemplation of whole new phenomena,
and, based on that criterion, are useful regardless of their ultimate
efficacy.

$\diamondsuit$  In your quote of {\SimonB}, ``I speak probability
with a marked functional analysis accent: lecturing in Zurich, I
couldn't help feeling that I was speaking not hoch probability but
only a kind of Schweitzer probability,'' is there a typo?

\bbr
Similar remarks apply to the recent attempts by {\Goldstein}, et al
to develop `Bohmian mechanics', although its advocates assert that it
is consistent with standard quantum theory because the
{\Schroedinger} equation still holds.  However, there is more to
quantum theory than the {\Schroedinger} equation, and I am not
convinced that no experimental distinction is possible.
\ebr
Here, here!!!!  Have you read {\Ghose}'s paper?  Is there anything to
it?

\bbr
Delicate experiments, such as those of done by the laboratory groups
associated with Aspect, Leggett, and {\Zeilinger} are useful in
clarifying some critical aspects of quantum theory.  However, to my
mind, the most convincing verification of quantum theory is not the
individual microscopic experiments, but the fact that no other
theory can even come close to explaining so many diverse phenomena
(including macroscopic as well as microscopic phenomena); literally,
``from atoms to stars.''  Few jigsaw puzzles fit together so neatly.
We are forced to overcome the biases arising from our experience
with the familiar macroscopic world of classical mechanics despite
the challenge of resolving all questions about the foundations of
quantum theory.  In the end, quantum theory remains a human
construct subject, in principle, to social forces.  But it is a
theory so remarkable, so different from ordinary experience, that it
transcends social and cultural forces.
\ebr
Excellent paragraph!  We have quite similar views about this.  Note
one small typo:  Leggett should have two t's.  More importantly
though, Leggett is a theorist:  so what laboratory group are you
referring to there?

That's about it.  Thanks for giving me the opportunity to read
through this.

\section{20 February 2000, \ ``Like It Even Better''}

Despite your reservations, I enjoyed the added discussion about
{\Bohm}.  In November, I attended a conference in Naples organized by
{\Duerr} and {\Goldstein}, and I was completely turned off to their
movement.  In my talk, I posed as a challenge for them to draw the
Bohmian trajectories associated with any quantum teleportation
experiment.  And furthermore to tabulate how much work it took to do
that versus the extra insight it gives into the phenomenon.  Does it
give any extra insight?  My challenge was summarily dismissed by
almost all the participants.  The reaction was uniform:  ``You're
missing the point Chris. Bohmians DON'T NEED TO draw the
trajectories.''  Those people got under my skin.  When in my second
talk, I expressed the point of view that  {\Asher} {\Peres} and I
take to quantum mechanics (I'll place our {\sl Physics Today\/}
article on the subject in the next mail), they labeled ME as being
unscientific!

Anyway, good job.  I'll try to dig up those references for you.
(It's a little harder with my being in Israel.)

\section{18 April 2000, \ ``Extremal Maps''}

I was cleaning up my office today, trying to file some old things,
and I came across a paper that might interest you:
\bq
\noindent S.-H. Kye, ``On the convex set of completely positive linear
maps in matrix algebra,'' Math.\ Proc.\ Camb.\ Phil.\ Soc.\ {\bf
122}, 45--54 (1997).
\eq
That, however, also prodded my memory that I had once somewhere seen
something special about CPMs that leave two points touching the
surface of the Bloch sphere.  You should probably have a look at:
\bq
\noindent
C.-S. {\Niu} and R.~B. {\Griffiths}, ``Two Qubit Copying Machine for
Quantum Eavesdropping,'' {\tt quant-ph/9810008}.  (I'm sure the
paper's long since appeared; most likely in PRA.)
\eq
and see if it is indeed relevant to you or has some connection to
your work.

\section{19 December 2000, \ ``Myopic Readers''}

\noindent [NOTE: Beth and I then exchanged some correspondence
about the responses she had already received on her article mentioned
above. One of those responses referred to my article with Asher
{\Peres} in {\sl Physics Today}.]

\bbr
You are only a rather innocuous footnote.
\bq
\noindent
\rm See in this connection the return to the attitude of ``quantum
mechanics works'' in the article by Fuchs and {\Peres} \ldots\ ``What
[it] does'', these authors write, ``is provide an algorithm for
computing the consequences of our experimental predictions.''
\eq
\ebr
But that is a misquote, and it is exactly the sort of thing that
makes me want to keep my ignorant bliss.  The actual words were:
\bq
\noindent
Contrary to those desires, quantum theory does {\it not\/} describe
physical reality.  What it does is provide an algorithm for computing
{\it probabilities\/} for the macroscopic events (``detector
clicks'') that are the consequences of our experimental
interventions.
\eq
The word ``interventions'' is prominent there.  It is there {\sl
precisely\/} to refer to the previous paragraph where we write:
\bq
\noindent
To begin, let us examine the role of experiment in science. An
experiment is an active intervention into the course of nature: We
set up this or that experiment to see how nature reacts.
\eq
The word ``nature'' is a stand-in for all the ``stuff'' of the world.
Why would we use such a word if we didn't believe deep inside that
the existence of humankind is only a contingent sort of thing?  The
main point we were trying to get at is captured by the next few
sentences:
\bq
\noindent
We have learned something new when we can distill from the
accumulated data a compact description of all that was seen and an
indication of which further experiments will corroborate that
description. This is what science is about. If, from such a
description, we can {\it further\/} distill a model of a
free-standing ``reality'' independent of our interventions, then so
much the better. Classical physics is the ultimate example of such a
model. However, there is no logical necessity for a realistic
worldview to always be obtainable. If the world is such that we can
never identify a reality independent of our experimental activity,
then we must be prepared for that, too.
\eq
To paraphrase that for the myopic readers who couldn't see past the
large print of our title:  the point is, quantum mechanics is much
more about our {\sl interface\/} with the world than the world in and
of itself.

The really exciting question to ask, I think, is:  What is it about
the quantum world that blocks us from going that {\it further\/}
step?  (I.e., the further step of distilling a free-standing
reality.)  Can we identify the aspect of the theory that
demonstrates that in the crispest way?  It's here that I think the
tools of quantum information theory and cryptography will help us
the most.  Optimistically, I see us as carrying out that project in
our lifetimes.  Like I told Jeffrey {\Bub} the other day:  [See note
to Jeff {\Bub}, dated 10 December 2000.]

\bbr
On the plus side, I learned that {\Heisenberg} also raised my
objection that Bohmian mechanics destroys the symmetry between
position and momentum representation.
\ebr
I'm sorry, I guess I forgot to tell you that.  I knew it too.  If
she didn't give you the reference, I think I can dig it up.

\section{24 March 2001, \ ``For Other Myopic Readers''}

And your note prompted me to reread the rebuttal  {\Asher} and I
wrote after the appearance of our original {\sl Physics Today\/}
thing. It's pasted below.

It's haughty of me, but I do think that myopia is running rampant in
our community \ldots\ not only in the reading of our little article,
but in the much bigger picture---our hopes for a physical theory.
Bohmism is an example of such a dull point of view:  If we can just
return to the womb of classical physics, everything will be oh so
much more warm and comfortable.  Yuck!

\chapter{Letters to {\Ruediger} {\Schack}}

\section{28 February 1996, \ ``QM is a Law of Thought''}

By the way, did you ever read the paper {\Carl} and I wrote titled
``Quantum Information: How Much Information in a State Vector?''  In
it we talk at some length about this idea of ``Quantum Mechanics as
a Law of Thought'' and other Bayesian sounding things.  (We even
quote Gertrude Stein?!?!)  Of course, all the ideas are loose now,
but one day they'll be tight.

\section{13 September 1996, \ ``Silly Latin''}

Allow me to continue my tradition of writing you only when I need
something translated!!

I've decided on an epithet for myself.  Either:
\begin{center}
{\large Quantum Theory is a Law of Thought} \\
or \\
{\large Quantum Mechanics is a Law of Thought}
\end{center}
What do you think?  (\ldots and don't tell me I'm plagiarizing
{\Boole}!) Anyway, the problem is that I would like to have these
translated as closely as possible into Latin.  Do you think you're
up to the task? I would appreciate it so much.

I can't wait to see you in Japan; I really miss the old gang.

\subsection{{\Ruediger}'s First Reply:  19 September 1996}

\bq
I like ``Quantum Mechanics is a Law of Thought'' better, but I
can't give a good reason for that.  I asked my father to translate
your epithet into Latin, and he has come up with
\begin{center}
\large Mechanica quantica ex mente orta lex est.
\end{center}

``orta'' means something like ``emerging from''.

I have a friend who is a Latin specialist. If you want, I can ask him
for a second opinion. I am sure my father's translation is correct,
but there might be a more elegant and more idiomatic
version.

Looking forward to seeing you in Japan,
\eq

\subsection{{\Ruediger}'s Second Reply: 3 May 2001}

\bq
\noindent [Looking over a preprint version of this samizdat, {\Ruediger}
wrote:]\medskip

Concerning your addition at the beginning of the {\Schack} chapter, I
told you that ``Mechanica quantica ex mente orta lex est'' mostly
means something trivial like ``Quantum mechanics is a law created by
mind'' or something like that, didn't I? All the experts (three) I
asked chickened out in the end. I don't know if you still want to
keep the translation, but you should add a footnote that the Latin
phrase does NOT capture what you mean. In any case, I am going to
make a last attempt to find a better Latin translation.
\eq

\subsection{{\Ruediger}'s Third Reply: 4 May 2001}

\bq
The big problem for all the Latin specialists I ask (four of them
now) is that they misinterpret the English phrase.
\begin{center}
Mechanica quantica ex mente orta lex est \ $=$ \ qm is a law that has
arisen from mind
\end{center}
is not what you want.

\begin{center}
Mechanica quantica lex cogitationis est \\ $=$ \\ qm is a law of
thought
\end{center}
[That is:]
\bv
lex $=$ law \\
cogitatio $=$ [thinking , conception, reflection, reasoning];\\
\ \ \ \ \ \ \ sometimes a particular [thought, idea or intention]. \\
cogitationis $=$ genitive of cogitatio \\
lex cogitationis $=$ law of thought
\ev

You see this is a literal translation, and therefore avoids any
narrow interpretation of the English original.\footnote{After much
further discussion between {\Ruediger}, David {\Mermin}, and {\it
the Vatican\,}(!)---well, actually a professor in Rome who had
organized a conference in the Vatican once---it was finally decided
that this {\it really\/} is the appropriate translation of ``Quantum
Mechanics is a Law of Thought.''}
\eq

\subsection{{\Ruediger}'s Fourth Reply: 4 May 2001}

Either you capitalize everything or nothing:

\begin{center}
\large Mechanica Quantica Lex Cogitationis Est \medskip
\\
MECHANICA QUANTICA LEX COGITATIONIS EST \medskip
\\
mechanica quantica lex cogitationis est
\end{center}
\bigskip

\section{15 October 1996, \ ``Mechanica Quantica''}

I'm kind of confused about what you're trying to get at.  See my
notes below, and try it out on me one more time please.  (I realize
my notes are pretty muddled in comparison to your one-sentence
formulation \ldots\ so I hope you have a little patience with me.)

\brs
I have read {\Hartle}'s 1968 paper and was quite confused at first.
After some thinking, however, I have come to the conclusion that the
finite-$N$ part of {\Hartle}'s analysis supports my understanding of
what makes quantum probabilities special. Here it is in one sentence:
\bq
\rm
\noindent The knowledge of how to prepare the pure state $|\psi\rangle$
implies the knowledge of how to prepare the product state
$|\psi\rangle\otimes|\psi\rangle\otimes\cdots\otimes|\psi\rangle$.
\eq
Or:
\bq
\rm
\noindent If one can prepare a pure state, one can prepare any number
of independent copies of it.
\eq
\ers

How is this any different than my being able to prepare $N$
independent (but identical) probability distributions?  For instance,
what is wrong with saying, ``The knowledge of how to prepare a
probability distribution $p(x)$ implies the knowledge of how to
prepare the product distribution $p(x_1)p(x_2)p(x_3)\cdots$.''  There
are certainly cases where I can do that, and they are purely
classical by nature:  I could produce 10,000 identical 75/25-weighted
coins, flip each of them independently and place them in sealed
envelopes (without looking at the outcomes).  That creates for me the
situation you describe.\footnote{I didn't understand it at the time,
but {\Ruediger} had hit upon one of the key distinctions between the
classical and quantum cases.  The two situations are not the same.
In the classical case, one is using a nontrivial probability
distribution precisely because one is admitting that maximal
information is not at hand.  Therefore, for just this reason, one
would never assign an i.i.d.\ distribution to the multi-trial
space.} (Though, granted, it's not doing it for states of maximal
knowledge.) Alternatively, in the quantum mechanical setting, I
could prepare 10,000 identical mixed states by repeatedly preparing
a pure state on a bipartite system and throwing one half of it away.

What is it that makes {\it pure\/} quantum states unique in your
formulation?  And what makes it particularly quantum mechanical in
the first place?  I'm probably just missing something.

I understand that there are classical situations where I cannot
conjure up the situation you describe.  For instance, I can give a
probability assignment to tomorrow's weather, but I cannot repeat the
preparation. But isn't that beside the point?

It sort of seems to me that {\Carl} hit on the crucial point: any
``scientific'' theory must be capable of dealing with repeatable
phenomena.  Thus any ``scientific'' enterprise must be, of it's
nature, capable of extrapolating from ``knowledge of how to prepare''
the individual instance to ``knowledge of how to prepare'' any number
of instances.

\brs
I know you have a strong dislike for the word ``objective''
(Yuck!). Yet it may not be completely misleading to call a quantum
pure-state preparation procedure ``objective.'' The preparation of a
classical system, however, should be called ``objective'' only if it
leads to a completely known state. Otherwise such a preparation
always relies on some environment being in an unknown state.
\ers

I don't know that I have an opinion on this comment.  Does it connect
to the previous part of the note?

As a side note, though.  What about a universal computer built on
Toffoli's billiard ball model (for instance).  I might start it in a
precise configuration that will eventually lead to a proof one way or
the other of Riemann's conjecture.  I don't know what the answer will
be, so---even without environments---I can be in a situation of
making probability assignments for the configurations of classical
systems. Does this have any impact?

\section{09 August 1997, \ ``Correspondentia''}

\noindent
\underline{\bf NOTE}: This letter was written with the intent of
ultimately being sent to {\Herb} {\Bernstein}.  However, the note was
{\it never finished}.  During a visit to England soon after the date
listed above, I gave {\Ruediger} {\Schack} a copy of the unfinished
note. This explains my reason for placing the text in this slot.
\medskip

\begin{flushright}
\baselineskip=13pt
\parbox{2.4in}{\baselineskip=13pt
``Just as the sexual drive frequently transforms man into a monster,
so the elementary category of causality can assume the character of a
need, an insatiable craving which overruns everything, and which
people will even sacrifice their lives to gratify.  It is an
indefatigable longing which inflames us \ldots''}\medskip\\
---{\it Carl {\Jung}}\\
The Zofingia Lectures
\end{flushright}
\medskip

\begin{flushright}
\baselineskip=13pt
\parbox{2.4in}{\baselineskip=13pt
``As philosophers we may well find the concept of objective chance
troublesome, but that is no excuse to deny its existence, its
legitimacy, or its indispensability.  If we can't understand it, so
much the worse for us.''}\medskip\\
---{\it David Lewis}\\
A Subjectivist's Guide to \\ Objective Chance
\end{flushright}
\bigskip

\noindent Hey {\Herb},
\bigskip

It's been a good two weeks since I've talked to you \ldots\ how are
things?  Actually, I was looking for a little philosophical
companionship this afternoon; I hope you don't mind my picking on
you!  For a while I've been wanting to expand on some {\it silly\/}
thoughts I've been thinking.  Maybe you'll be the one to have a
sympathetic ear to this stuff.  Let me break you in gently, by first
repeating two very, very lengthy quotations.  I'm sure you'll be able
to ferret out the common theme.  The first comes from a letter I
wrote {\Greg} {\Comer} in 1992 (itself containing lengthy quotations
of {\Pauli}).  The second is a passage from one of Arthur {\Fine}'s
articles. (By the way, the latter appears before you with the help
of modern technology:  a good scanner and a character recognition
program.)
\bigskip

\noindent -----------------------------------------
\bigskip

\noindent 28 April 1992 (to {\Greg} {\Comer})
\bigskip

\ldots\ \ As per your request, I'll use this as an opportunity to
repeat the quote I wrote you the other day.  On 29 December 1947,
{\Pauli} wrote {\Fierz}:

\begin{quotation}
I'm more and more expecting a further revolutionizing of the basic
concepts in physics.  In connection with this particularly the manner
in which the space-time continuum is currently introduced into it
appears to me to be increasingly unsatisfactory.  \ldots\ Something
only really happens when an observation is being made, and in
conjunction with which, as {\Bohr} and Stern have finally convinced
me, entropy necessarily increases.  Between the observations nothing
at all happens, only time has, ``in the interval'', irreversibly
progressed on the mathematical papers.
\end{quotation}

Now I think even more important to me personally is another long
quote of {\Pauli}'s from a letter to {\Fierz} dated 26 November 1949.
For typing practice (ha!), I'll relate this to you in a second.
First let me tell you a little about the origin of my interest in
this randomness business.  In 1986, while writing a paper for a
philosophy course I was taking, I used a phrase something like: each
individual quantum mechanical measurement outcome is completely
undetermined and this indeterminism is NOT due to our ignorance of
the true picture of things (as would be the case if a hidden
variable theory existed). Well, after the paper was turned in, that
simple phrase (or something like it) led me to start wondering:
``What could it possibly mean to be `completely undetermined'?'' and
``Why is it that if each individual outcome is completely
undetermined, nevertheless, in the long run, the outcome statistics
eventually settle down to a predictable mean, etc.?''  You see I
could have easily answered the latter question if there were a
hidden variable theory; for then the quantum mechanical
probabilities would just represent our ignorance of the actual
situation---a situation that would explain one set of frequencies
over another.  But in the absence of such an external controlling
factor how is it that a ``completely undetermined'' individual
outcome also knows something about all the other outcomes that could
come about in repeated trials of the same experiment? Doesn't this
question hint at a contradiction in terms?  It was, in fact, exactly
these sort of questions that eventually led me to identify ``the
wave function''  with a random string rather than sticking with the
standard thought of identifying ``the wave function'' with the
``state of the system.''  Well, anyway, strangely enough, until
yesterday I had never seen these questions I raised appear in
print.  And this is where {\Pauli}'s quote comes into the picture.

\begin{quotation}
{\Bohr}'s expression ``correspondence'' served as an aid to me when,
then, I was trying to give a name to the positive principle which
lies at the basis of quantum mechanics.  ({\it After\/} putting
forward wave mechanics he continues to speak of a ``correspondence
argument'' see Naturw. {\bf 21}, 245-250, 1933, particularly the
passage on page 246, top of the second column.)  The statistical
behavior of many similar individual systems which have no contact
whatsoever with one another (``windowless monads''), without, on the
other hand, being causally determined, has, of course, in quantum
mechanics been interpreted as the {\it last\/} law-governed fact
which cannot be further reduced (approximately as was the case for
Galileo with respect to uniformly accelerated falling bodies).  In my
lecture on complementarity, originally published in the journal {\it
Experientia\/} and now available as an offprint, I thus tried to use
the expression ``correspondence'' in a more general sense than
{\Bohr} had, in a way which would specifically characterize the
positive side of a quantum mechanical description of nature.  It is
certainly this {\it statistical correspondence\/} which mediates
between continuum (wave image) and discontinuum (particle image).
(This in a somewhat more general way than the mediation between
``quantum theory'' and ``classical theory'' in {\Bohr}'s writings).
There I did {\it not\/} explicitly state that for me the
intellectual derivative of the ``correspondentia'' of the Middle
Ages (``correlations'') clearly seems to glimmer through in the term
``correspondence.''  In both cases, however, we are concerned with a
form of describing nature in terms of laws which transcends normal
causality and which is based on some kind of analogy.  (This is also
the case with {\Leibniz}'s prestabilized harmony.)

The single systems of quantum mechanics are ``windowless monads''
and, nevertheless, the correct fraction can always be found which
reacts as calculated (apart, naturally, from the expected statistical
fluctuations).

The tertium comparationis of the quantum mechanical case with that of
the synchronistic phenomenon is the {\it mutually tuned behavior of
different events\/} (not bound in a deterministic-causal sense).  (It
is on this, of course, that the concept of ``physical situation'' in
quantum mechanics rests, not on a direct mutual influence of the
objects in question.)

The quantum mechanical situation is naturally not only a degeneration
of the more general ``synchronicity'' (this to be understood as a
working hypothesis suggested here), but {\it also\/} a ``rational
generalization'' of normal deterministic causality ({\Bohr}).  When
the fraction is one (instead of between zero and one), that is, of
course, a special borderline case, as, indeed, the old deterministic
causality seems when observed from the standpoint of quantum
mechanics.  I have no doubts that the quantum mechanical
``statistical correspondence'' lies much closer on the side of old
determinism than on the side of the synchronicity phenomenon.
Observed from the standpoint of this phenomenon quantum mechanics
must appear to be a {\it very weak\/} generalization of the old
causality. And, nevertheless, quantum mechanics seems to me also to
have that road sign towards the other direction, towards the one
where it is no longer possible to speak of arbitrary reproducibility
at all.  To me quantum mechanics seems to occupy a kind of
intermediate place.

This is my momentary view \ldots.  The {\it success\/} of the
``reasonable belief''---and with it also the possibility of laws of
nature---appears to me {\it always\/} to rest on an archetypically
conditioned coincidence of our expectation (psychologically) with an
external natural occurrence (physically).  For the abstract arranger
there is just {\it not\/} any actual difference between
``physical--psychological.''
\end{quotation}

Read this about five times and you'll start to see the connections.
Long before randomness came around, I studied {\Leibniz} and his
monads for just this same reason.  The words ``synchronicity'' and
``archetype'' in the quote above refer to the concepts used by
{\Jung}. You are probably more familiar with this set of concepts
than I am---seeing as Joseph Campbell is a fan of {\Jung}.
\bigskip

\noindent -----------------------------------------
\bigskip

\noindent From Arthur {\Fine}, ``Do Correlations Need To Be
Explained,'' in {\it Philosophical Consequences of Quantum Theory:
Reflections on {\Bell}'s Theorems}, edited by James T. Sherman and
Ernan McMullin (University of Notre Dame Press, Notre Dame, Indiana,
1989).
\bigskip

\begin{quotation}
If we adopt an indeterminist attitude to the outcomes of a single,
repeated measurement, we see each outcome as undetermined by any
factors whatsoever.  Nevertheless we are comfortable with the idea
that, as the measurements go on, the outcomes will satisfy a strict
probabilistic law.  For instance, they may be half positive and half
negative.  How does this happen?  What makes a long run of positives,
for example, get balanced off by the accumulation of nearly the very
same number of negatives?  If each outcome is really undetermined,
how can we get any strict probabilistic order?  Such questions can
seem acute, deriving their urgency from the apparent necessity to
provide an explanation for the strict order of the pattern, and the
background indeterminist premise according to which there seems to be
nothing available on which to base an explanation.  If one accepts
the explanationist challenge, then one might be inclined to talk of a
``hidden hand'' that guides the outcome pattern, or its modern
reincarnation as objective, probability-fixing ``propensities.''

This talk lets us off the hook, and it is instructive to note just
how easily this is accomplished.  For if propensities were regular
explanatory entities, we would be inclined not just to investigate
their formal features and conceptual links, but we would make them
the object of physical theorizing and experimental investigation as
well.  However, even among the devotees of propensities, few have
been willing to go that far.  The reason, I would suggest, is this.
Once we accept the premise of indeterminism, we open up the idea that
sequences of individually undetermined events can nevertheless
display strict probabilistic patterns.  When we go on to wed
indeterminism to a rich probabilistic theory, like the quantum
theory, we expect the theory to fill in the details of under what
circumstances particular probabilistic patterns will arise. The
state/observable formalism of the quantum theory, as is well known,
discharges this expectation admirably.  Thus indeterminism opens up a
space of possibilities.  It makes room for the quantum theory to
work.  The theory specifies the circumstances under which patterns of
outcomes will arise and which particular ones to expect. It simply
bypasses the question of how any patterns could arise out of
undetermined events, in effect presupposing that this possibility
just is among the natural order of things.  In this regard, the
quantum theory functions exactly like any other, embodying and taking
for granted what Stephen Toulmin (1961) has nicely called ``ideals of
natural order.''  What then of correlations?

Correlations are just probabilistic patterns between two sequences of
events.  If we treat the individual events as undetermined and
withdraw the burden of explaining why a pattern arises for each of
the two sequences, why not adopt the same attitude toward the
emerging pattern between the pairs of outcomes, the pattern that
constitutes the correlation?  Why, from an indeterminist perspective,
should the fact that there is a pattern between random sequences
require any more explaining than the fact that there is a pattern
internal to the sequences themselves?

We have learned that it is not necessary to see a connection linking
the random events in a sequence, some influence from one event to
another that sustains the overall pattern.  Why require a connection
linking the pairs of events between the sequences, perhaps some
influence that travels from one event in a pair to another (maybe
even faster than the speed of light) and sustains the correlation? We
have explored part of the answer above.  Our experience with
correlations that arise in a context in which there generally are
outcome-fixing circumstances has led us to expect that where
correlations are not coincidental, we will be able to understand how
they were generated either via causal influences from one variable to
another or by means of a network of common background causal factors.
The tangled correlations of the quantum theory, however, cannot be so
explained.

The search for ``influences'' or for common causes is an enterprise
external to the quantum theory.  It is a project that stands on the
outside and asks whether we can supplement the theory in such a way
as to satisfy certain a priori demands on explanatory adequacy. Among
these demands is that stable correlations require explaining, that
there must be some detailed account for how they are built up, or
sustained, over time and space.  In the face of this demand, the
tangled correlations of the quantum theory can seem anomalous, even
mysterious.  But this demand represents an explanatory ideal rooted
outside the quantum theory, one learned and taught in the context of
a different kind of physical thinking.  It is like the ideal that was
passed on in the dynamical tradition from {\Aristotle} to {\Newton},
that motion as such requires explanation.  As in the passing of that
ideal, we can learn from successful practice that progress in
physical thinking may occur precisely when we give up the demand for
explanation, and shift to a new conception of the natural order. This
is never an easy operation, and it is always accompanied by
resistance and some sense of a lost paradise of reason.  If we are to
be serious about the science that we now have, however, we should
step inside and see what ideals it embodies.

The quantum theory takes for granted not only that sequences of
individually undetermined events may show strict overall patterns, it
also takes for granted that such patterns may arise between the
matched events in two such sequences.  From the perspective of the
quantum theory, this is neither surprising nor puzzling.  It is the
normal and ordinary state of affairs.  This ideal is integral to the
indeterminism that one accepts, if one accepts the theory.  There was
a time when we did not know this, when the question of whether the
theory was truly indeterminist at all was alive and subject to real
conjecture.  Foundational work over the past fifty years, however,
has pretty much settled that issue (although, of course, never beyond
any doubt).  The more recent work related to EPR and the {\Bell}
theorem has shown, specifically (although again, not beyond all
doubt), that the correlations too are fundamental and irreducible,
so that the indeterminist ideal extends to them as well.  It is
time, I think, to accept the ideals of order required by the
theory.  It is time to see patterns between sequences as part of the
same natural order as patterns internal to the sequences themselves.

A nonessentialist attitude toward explanation can help us make this
transition, for it leads us to accept that what requires explanation
is a function of the context of inquiry.  So when we take quantum
theory and its practice as our context, then we expect to look to it
to see what must be explained.  This leads us to the indeterminist
ideal discussed above, and to the ``naturalness'' of (even distant)
correlations.  There is a small bonus to reap if we shift our
thinking in this direction.  For the shift amounts to taking the
correlations of the theory as givens not in further need of
explanation and using them as the background resources for doing
other scientific work.  One thing they can do is to help us
understand why the theory has correlational gaps.  From the very
beginning, one wondered about the incompatible observables and why
one could not even in principle imagine joint measurements for them.
After all, as {\Schroedinger} (I believe) first pointed out, in the
EPR situation, one could measure position in one wing and momentum
in the other and, via the conservation laws, attribute simultaneous
position and momentum in both wings.

The conventional response here has been to point out that only the
direct measurements yield values that are predictively useful. (See
note 4) Not everyone has been happy with the positivism that seems
built into this response.  But if we recall the discussion in section
1, then we see that (at least in part) there is a better response at
hand.  For we have seen how the correlations that the theory does
provide actually exclude the possibility that there could be any
stable joint distributions for incompatible observables in those
states where the correlations are tangled.  This shows us that there
is no way of augmenting the theory with values for incompatible
observables, and distributions for those, that would follow the same
lawlike patterns as do the distributions of the theory itself.  To
put it dramatically, the shadow of the given correlations for
compatible observables makes it impossible to grow stable
correlations for the incompatibles.  There is a sense, then, in which
there would be no point in trying to introduce more for incompatible
observables than what the theory already provides.

This way of thinking turns the {\Bell} theorem around.  Instead of
aiming to demonstrate some limitation or anomaly about the theory,
this way proceeds in the other direction and helps us understand why
the probability structure of the theory is what it is.  That
understanding comes about when we take a nonessentialist attitude
toward explanation, letting the indeterminist ideals of the theory
set the explanatory agenda.  Such an attitude means taking the
theoretical givens seriously, and trusting that they will do good
explanatory work.  Thus, in the {\Bell} situation, we shift our
perspective and use the given quantum correlations (and the simple
sort of counting argument rehearsed in section 1) to explain why,
even in principle, correlations forbidden by the theory cannot arise.
Nonessentialism leads us to engage with our theories seriously, and
in detail.  In the end, that is how better understanding comes about.

What then of nonlocality, influences, dependencies, passions, and the
like, all diagnosed from correlational data?  As one good
statistician remarked about the similar move from linear regression
to causal connection, and as we have seen demonstrated above, ``Much
less is true.''
\end{quotation}

\noindent -----------------------------------------
\bigskip

What do you think of {\Fine}'s writing?  Isn't it nice and clear?  In
any case, what I'd like to do in this note is focus mostly on the
common thread between the two passages.  Namely I want to talk about
how quantum phenomena are like ``windowless monads.''  I may come
back to {\Fine}'s stuff on comparing correlations in EPR-type
phenomena to the fundamental assumption of indeterminism, but that
won't be in the main of this discussion.

Did you understand my point in the letter to {\Greg} about how one
{\it might\/} find the existence of a limiting frequency for
measurement outcomes in quantum mechanics mysterious?  I'm not so
sure I said it as clearly as I would have liked to.  So perhaps I'll
try it one more time \ldots\ just to see if I can explain it to {\it
my\/} satisfaction. (I realize that {\Fine} may have already
clarified this to {\it your\/} satisfaction with his discussion, but
indulge me:  I like to beat my potatoes until they're really mushy.)

The point is this.  Despite my long-running obsession with and love
of quantum mechanics and indeterminism, I have never been completely
at ease with the idea of ``probabilistic causality.''  It's there
\ldots\ and I know that \ldots\ and I {\it like\/} that \ldots\ but
I've never felt completely at grips with it.  That is to say, I've
never been completely at ease with the ``hidden hand'' that {\Fine}
speaks of. The problem seems to be that my mind always tries to slip
into the ``point of view'' of the individual event.

Imagine the following.  I have a friend {\Hideo} who is singularly
obsessed with preparing his favorite quantum system in the state
$|\psi\rangle$ as cleanly and as quickly as he can. His purpose is to
make a huge number of copies of $|\psi\rangle$ so that our mutual
friend Quentin can test various quantum mechanical statements.  In
particular, suppose Quentin worries about what he should expect of a
measurement corresponding to the projector
$|\phi\rangle\langle\phi|$.  (Quentin, being just as conscientious as
{\Hideo}, is careful to perform his measurements just as cleanly as
he can.)  So what is it that Quentin can expect?  The standard dogma
of quantum mechanics gives us two things:
\begin{enumerate}
\item
The probability of a ``yes'' answer on a single trial of the
measurement is $|\langle\phi|\psi\rangle|^2$.
\item
Beyond that probability, the outcome of the measurement is completely
unpredictable and, in fact, undetermined.
\end{enumerate}

From a ``global'' point of view, one might say that there's nothing
out of synch about these two statements.  This should be especially
reasonable given the phrasing I used in item 2 above---namely,
``Beyond that probability \ldots\ blah, blah, blah.''

\section{17 August 1997, \ ``New Babies of Quantum Information''}

Very many congratulations on the new baby!  I'm glad to hear that
both mother and daughter (and presumably the whole family) are doing
well. This is quite a family you and {\SchackD} are building.

It would be nice if there were a generally-agreed-upon definition of
``quantum information,'' but I don't think there is one presently.
The problem is the following.  A large portion of the community,
like the Schumacherites,  might say, ``Quantum information theory is
the subject of quantifying the resources required to transmit intact
nonorthogonal quantum states and entanglement through noisy
channels.''  Another portion of the community, like the Holevoites,
might say, ``Quantum information theory is the subject of
quantifying the maximal rate at which classical bits can be sent
over quantum mechanical channels.'' (These, by the way, are the same
people that worry about how distinguishable nonorthogonal states
are.)  Still another portion of the community, like the
{\Caves}-Schackites, might say, ``Quantum information concerns itself
with the various ways in which Hilbert space is big in comparison to
classical state spaces.''  And, I guess I shouldn't forget about the
part of the community that worries about the efficient processing of
classically defined questions---the quantum computing clique, the
Ekertites.  They would probably have their own way of saying things.

The truth of the matter is that somehow one needs a definition that
encompasses all these ideas, i.e., a reason behind why we all see
each other at the same conferences.  It can be done \ldots\ you just
string all the things above together \ldots\ but I guess it wouldn't
come out too pithily.  Good luck.  I don't know about a standard
reference that emphasizes all the aspects---one should be written.
Maybe a good overall reference would be {\Bennett}'s {\sl Physics
Today\/} article (last year or the year before last).  Other than
that I really can't say.  {\Carl} and I tried to say something of a
summary of ``what quantum information is'' in our {\Rosen}
festschrift, but the things we said weren't nearly all-purpose
enough.

\section{29 September 1997, \ ``The Hidden Hand''}

Anyway, on a connected subject, I've spent a little time this week
reacquainting myself with what has been written on propensities.  I'm
not yet finished with the project, but maybe I'll go ahead and record
some things.

First off, I think it would be worthwhile for you to take a look at
the {\Giere} (1973) paper cited in Howson and Urbach (and in the
paper {\Carl} and I wrote).  If the guy should be damned for not
being a subjectivist, he shouldn't be damned for anything too
extreme.  He is certainly no lover of frequentism.  In fact, even
now---upon another reading---I think he only differs from us
substantially in the argument {\Carl} and I noted (and rejected):
\bq
\ldots\ Having withdrawn earlier objections to {\Laplace}'s view, he
[Mill] concludes: ``Every event is in itself certain, not probable;
if we knew all, we should either know positively that it will happen
or positively that it will not. But its probability to us means the
degree of expectation of its occurrence which we are warranted in
entertaining by our present evidence.''  The role of the assumption
of determinism in this argument is clear enough.
   One will not find the above argument in the writings of {\Laplace}'s
contemporary heirs, the subjective Bayesians, but they are tacitly
committed to a similar position.  The most prominent personalists,
e.g., {\Savage} and de {\Finetti}, insist that there is only one
legitimate concept of probability, that which identifies probability
with subjective uncertainty.  Once this identification is made,
however, one lacks the conceptual apparatus to distinguish
uncertainty due to lack of information from uncertainty which no
physically possible increase in present knowledge could eliminate.
But this is just the distinction between physical determinism and
physical indeterminism. Not being able to make the distinction,
Bayesians are forced to assume that all uncertainty is due to lack
of information, i.e., to assume determinism. Indeed, to admit the
possibility of uncertainty not due to lack of information would be
to admit the possibility of physical, i.e., nonsubjective,
probabilities---an admission personalists refuse to make.
\eq

You'll probably remember that Howson and Urbach were pretty rough on
the guy.  They wrote:  ``The second objection is more fundamental and
seems to be unanswerable.  Von {\Mises}'s theory may seem stubbornly
deficient in empirical content, but the present account is, if
anything, even worse.  For {\Giere}'s single-case propensity theory
conveys no information of any sort about observable phenomena, not
even, we are told, about relative frequencies in the limit. \ldots''
Well, if that's the view they take of him, then it'll most certainly
be the view they take of us too.  I say this because {\Giere} takes a
stand pretty close to us:
\bq
A preliminary point worth emphasizing is that a single-case
propensity interpretation provides an extremely natural understanding
of the standard formalism for dealing with series of trials.  \ldots\
Thus, for example, a series of trials is conventionally represented
by a sequence of random variables [i.i.d., presumably] \ldots\  On my
interpretation, the density function gives the propensity
distribution on the $i$th trial.  It is as simple as that.  There is
no need for vague talk about ``virtual'' infinite sequences, etc.
Independence of trials just means that there is no causal connection
between the outcome of one trial and the propensity distribution of
any other trial.  [We would at this point say that maximal
information has allowed us identical preparations instead of the no
causal connection thing.] \ldots\
    A central question concerning the relation between single-case
propensities and frequencies is whether it is possible to {\it
deduce\/} values of one from values of the other.  The answer, as one
would expect since propensities are theoretical, is negative.
Consider a series of Bernoullian trials in which the propensity for
success is $r$ on each trial.  Let $f_n$ be the relative frequency of
successes after $n$ trials.  In this case the strongest connection
one can establish between $r$ and $f_n$ is given by the {\Bernoulli}
theorem, namely: [he writes the standard WLLN], where $P$ measures
the propensity of the compound trial which consists of $n$ trials of
the original chance setup.  It does {\it not\/} follow that the
sequence of values of $f_n$ has limit $r$ in the ordinary sense [of
calculus, that is].  Thus the sequence need not be a [collective] in
the sense of von {\Mises} or {\Reichenbach}.  Indeed, it is {\it
logically\/} possible, for example, that $r=1/2$ and that $f_n=1$ for
any $n$, although of course the propensity for this outcome
approaches zero.  The limiting frequency interpretation rules out all
such possibilities by convention.
\eq

Note his emphasis on ``deduce'' and ``logically''.  If Howson and
Urbach read his paper carefully---and it doesn't look like they
did---that should be the only lack of connection between propensities
and frequencies that they are criticizing.

Anyway, taking a look at all this I am again struck by how similar it
all is.  We say ``maximal information cannot be completed'' where
{\Giere} says ``propensity'' and vice versa (I think).  Our main
point of departure from {\Giere} is most likely in viewing a
distinction between ignorance and objective chance as dangerous
(when it comes to mixed-state ensembles for instance).  So I think
we need to flesh that out more carefully.

I'd like to get hold of {\Giere}'s 1976 paper, but it's not in the
library here.  So it'll probably have to wait until I'm at UCLA
Friday---maybe it's in their library.

Let's see, let me say a couple of more things.  I also read a paper
``Single-Case Probabilities'' by David Miller ({\it Found.\ Phys.}
{\bf 21} (1991) 1501--1516.  It's not really worth reading, don't
worry about it.  But I did find one thing interesting in it about the
``hidden hand.'' Miller writes:  ``A second but not less urgent
problem confronted by the propensity interpretation is {\it the
fundamental problem of the theory of chance}---the explanation of
objective statistical stability. I shall not discuss this problem
here, except to remark that it seems to be well beyond the resources
of the subjectivist theory  of probability.''  He cites {\Popper}'s
Logic of Scientific Discovery (Sec. 49) for the ``fundamental
problem.''  Thus the ``hidden hand'' goes back at least that far!
(By the way, it might be fun and/or useful to look at {\Popper} in
those pages in a little more detail---I haven't yet---if for no other
reason because {\Popper} has had such an inordinate influence on so
many.)

The other thing is that I invested a little time in van Fraassen (his
book The Scientific Image).  One thing that I did learn is that the
frequentist conception of probability is more distinct from
{\Kolmogorov} probability than in just it's interpretation.  van
Fraassen goes to some length to explain that the domain of the
``relative frequency function'' (that thing which will eventually be
labeled ``probability'' by the frequentists) is not a Borel field.
Then he cites de {\Finetti} and {\Suppes} (independently) as having a
more complicated argument that shows that it is not even a field.  I
don't know what all this is worth, but it did seem interesting. Then
van Fraassen went on to give his interpretation---``the modal
frequency interpretation''---of probability, but I got pretty lost.
His main point seems to be that, though a frequency interpretation
is useless empirically and can't stand on its own, one can construct
a {\it model\/} for the world that takes ideas from the old
frequentist camp.  And that model, he thinks, has some desirable
properties.  I doubt there's anything to this \ldots\ but his name
is huge in the philosophy of science since way back in the 70s  and
he does have a major book on quantum mechanics.  So maybe it's at
least worth understanding the flavor of his interpretation.

\section{02 October 1997, \ ``Wake Up Call''}

I haven't heard from you in a while.  You did get my last long
letter, right?  This morning I think I'll send you a little gift: a
passage from J. L. {\Heilbron}'s {\sl The Dilemmas of an Upright Man:
Max {\Planck} as Spokesman for German Science\/} (pp.\ 127--128).  I
very much enjoyed this because (I think) it led me to the path of
thinking about the connections---whether real or not---between
``maximal information not being complete'' and ``introspection.''

\bq
Although {\Planck}'s reconciliation of free will and determinism had
been anticipated by others, notably William {\James}, and so was
familiar to philosophers, it turns on an argument perhaps too
compressed for easy understanding.  {\Planck} later set it out
clearly and distinctly in his correspondence with Theodor {\Haering},
professor of philosophy at the University of Tubingen, whose
enlightenment demanded six long letters. First {\Planck} explained
that determinism makes sense only when the data on which prediction
rests can be obtained without influencing or changing the system
under study. ``This is the basic presupposition of any sort of
scientific knowledge.''  It does not hold for self-analysis.  Hence
one's will cannot be a subject of science for one's self, and the
question of the will's determination makes no sense to the willing
individual. {\Haering} replied that he could not understand why, if
scientific determinism holds generally, it does not apply also to
``the I''; nor did he perceive why the self must be disturbed in
self-examination.

{\Planck} insisted that a disturbance must occur: self-knowledge is a
conscious experience; every such experience implies a change of
mental state; but self-knowledge has as its object the mental state;
therefore self-knowing of the knowing-self is impossible.  This was
not enough to complete {\Haering}'s understanding.  It appeared that
the philosopher had confused the ``phenomenological I (`Me')'' with
the ``actual I (`I')''; the latter can know the former, but not
itself. ``Therefore,'' {\Planck} concluded, having reduced his
argument, he thought, to the level of his grandchildren,
``determinism can never invalidate freedom of the will.''  The
philosopher still had his difficulties: he did not comprehend, he
said, how observing an object could change it.  {\Planck} replied
that observing does not change objects, but subjects; it happens
that in the case under investigation, the subject, the knowing I, is
identical with the object, the willing I.  That did the trick.
{\Haering} grasped the argument and {\Planck} the difficulty of
discourse with philosophers.
\eq

Oh, another paper you may wish to look at: Adrian {\Kent}, ``Against
Many-Worlds'' on the LANL server, {\tt gr-qc/9703089}.  Andrew
{\Whitaker} brought it to my attention.

\section{06 October 1997, \ ``Germinal Jewel''}

Pay no attention to [\ldots]; I read the draft tonight while sitting
at a coffee shop and am about as pleased as I was two weeks ago when
we had our nice conversations.  I think there is something of real
substance to this ``principle of quantum indeterminism'' (PQI)---it
somehow just pinpoints the right thing.  The fact that it gives such
a simple, believable justification to the Law of Large Numbers
already sets it apart \ldots\ as you show.  However, the draft does
make me feel like maybe we're closer to the beginning of the road
than the end---it'll take me a few days to put my finger on it all.
Mainly, don't lose faith in the mean time. It may be just a case of
saying things in a slightly different order, and with a little more
chutzpah.  For instance, I still don't feel that the PQI was stated
with sufficient force and in such a way that its necessity will
become as pleasing to the reader as we see it.  This'll get sorted
out; this is the germ of something good.

Let me leave you with a quote from {\Mermin}'s 1983 book review of
{\Popper}'s ``Postscript.''  In it, you can see the germ of {\it
his\/} present views:
\bq
Physicists who have come to take their subject for granted would do
well to suppress their aversion to rubbish and read these volumes,
both to refresh their awareness of how bizarre their subject has
become, and to test their own grasp of its foundations against
{\Popper}'s view that what is most marvelously intricate and subtle
in the behavior of the atomic world is just a mystery and horror to
be dispelled by some clear thinking about probability.
\eq

\section{07 October 1997, \ ``Two Forgotten Questions''}

I forgot to ask you two things last night.  The first is, when do you
start lecturing?  The second is: you cite Percival in the second
paragraph for criticism of the assumption that ``there exists, in
principle, a completely specified preparation procedure for an
arbitrary (pure or mixed) quantum state.''  What is that paper?  Can
you give me its coordinates?

Last night I did a little more reading on propensity theories---I've
been trying to get a handle on what the heck such things are supposed
to be.  Not much luck so far; with some good fortune, perhaps I can
give you a report on my understanding this weekend (along with many
further comments on the draft)  One thing of note, though, I did
reread Feynman's old paper ``The Concept of Probability in Quantum
Mechanics'' (1951).  It's not of much use to us, but there is maybe
one passage that summarizes his views (as far as they were
developed):
\bq
I should say, that in spite of the implication of the title of this
talk the concept of probability is not altered in quantum mechanics.
When I say the probability of a certain outcome of an experiment is
$p$, I mean the conventional thing, that is, if the experiment is
repeated many times one expects that the fraction of those which give
the outcome in question is roughly $p$.  I will not be at all
concerned with analyzing or defining this concept in more detail, for
no departure from the concept used in classical statistics is
required.

What is changed, and changed radically, is the method of calculating
probabilities. \ldots\
\eq

\section{12 October 1997, \ ``The Deep End''}

I feel like I've fallen off the deep end:  the weekend has now passed
and I still have no detailed comments for you.  How many apologies
will you be able to take before they start seeming meaningless???

Anyway, I'm finding myself doing exactly what I warn the students
around me not to do!  I can't stop reading and bring myself to
action!!!  I keep reading about propensities, trying to see what the
(careful) propensitists think different about their concept than you
with your principle of indeterminism.  The differences keep seeming
smaller to me.  Also I've taken some time to read {\Jaynes} a little
better, for instance his Chapter 18 on the $A_p$ distribution.

I guess what I'm trying to do is find a way of expressing your
principle of quantum indeterminism in a way that I'm happy with
\ldots\ so that I might take my own shot at pith.  (At least as far
as that detail is concerned.)  I hope that doesn't take much longer.

On a different note, reading {\Jaynes} has made me think of the
following thing.  I wonder if we might make use of the example
{\Jaynes} does in his two subsections on ``{\Laplace}'s rule of
succession'' (pp.\ 1807--1811). In particular, it would be nice to
derive a quantum analog of Eq.\ (18-22).  Namely, where {\Jaynes}'
assumes that ``the underlying `causal mechanism' is assumed the same
at every trial,'' we might assume that the same (unknown) quantum
state is prepared and the same measurement is performed.  Then,
putting a uniform (unitarily invariant) distribution on the set of
(pure) quantum states, we could go through the same steps as he.
Now, the wonderful thing is \ldots\ I think \ldots\ that nothing
whatsoever changes, though it certainly had every right to.  This is
because of Sykora's old result that uniform distribution on quantum
states corresponds to an induced uniform distribution on the
probability simplex.  That is, as long as the Hilbert space is
complex and not real or quaternionic.

\section{21 November 1997, \ ``Deadbeats for Quantum Reality''}

May I ask a favor of you?  Would you read the statements below, and
tell me whether you think they express the issue fairly accurately.
So that someone who hasn't seen MaxEnt before won't be misled, but on
the other hand won't be bowled over with details either.

I'll send you the finished product (i.e., the full proposal) in a
couple days for your enjoyment.

\noindent ------------------

The year 1957 is significant in physical thought because it marks the
penetration of Information Theory into physics in a systematic
way---into statistical mechanics in particular. This refers to the
{\it Maximum Entropy\/} or ``MaxEnt'' program for statistical
mechanics set into motion by E.~T. {\Jaynes}. With the tools of
Information Theory, one was able for the first time to make a clean
separation between the purely {\it statistical\/} and the purely
{\it physical\/} aspects of the subject matter.

Perhaps it would be good to present a mild example of this.  Because
of MaxEnt, a standard statistical mechanical ensemble, like the
canonical en\-sem\-ble, can finally be seen for what it really is:
an expression of the physicist's {\it state of knowledge} (specified,
of course, by the experimental parameters under his control).  Though
this reveals a subjective aspect for statistical mechanics, the
ensemble is not arbitrary:  Two physicists working on a single
experiment, if true to their states of knowledge, will derive the
same distributions for the system's variables. The {\it structure\/}
of the canonical distribution, with its exponential form, is due
purely to the kind of knowledge the experimenter possesses---in this
particular case, the expectation value of some observable and
nothing else.  That is to say, the canonical distribution's form is
a theorem of the laws of inference, {\it not physics}.  The physics
of the system rests solely in its Hamiltonian and boundary
conditions.  Thinking that the canonical distribution somehow comes
out of physics alone, without reference to what one knows, is to
greatly misunderstand the content of statistical mechanics.

\section{14 January 1998, \ ``Practice and Perfection''}

Anyway, not that it'll help my standing---it'll probably only
hurt---but I offer a little bathroom reading for your contemplation.
It takes the form of a letter I wrote David {\Mermin}. It carries the
beginning of my new slogan ``{\Bohr} was a Bayesian.'' (In fact a
Bayesian good enough to know that Bayes' rule for conditionalizing
isn't always valid \ldots\ that is, especially when one is
confronted with the situation where one's information gathering
measurements {\it necessarily\/} disturb someone else's
predictability.)  I hope you enjoy, especially the phraseology that
starts up right around Merminition \ref{OldAsm10}.

\section{13 March 1998, \ ``Maximal Information and {\Schroedinger}''}

You'll be happy to know that I am jumping into the project today.
Wish me sweet luck \ldots\ I may need it.

In complement to that, I wonder if you would be interested in looking
at some old archival material.  I've come across what I think is a
remarkable exchange of letters between {\Pauli}, {\Heisenberg}, and
{\Schroedinger} in the wake of the EPR paper (June--July 1935).
(Steven van {\Enk} has done enough of a translation for me to come
to this opinion.)  In particular, {\Schroedinger} goes to pains to
say things about how in standard quantum theory, pure states
correspond to ``maximal knowledge.''   He also seems to stress that,
in transferring from one state to another, one gains knowledge but
also loses knowledge:  the level of knowledge stays maximal, but the
particulars of it are about different ``things.''

Also---though not too connected with our present project---there seem
to be some nice remarks by {\Pauli} on how ``the separation and
combination of systems'' should be better understood as a
``foundation of quantum theory.''  And a small (unpublished and
never-translated) paper by {\Heisenberg} is included that makes a
little more of those remarks.

Would you willing to read over these and write a summary about their
essential content \ldots\ especially as it relates to our paper?
It'd be even better if you could put a translation onto tape as you
did once before for me, but I know that that's asking for a lot of
work.  Any enthusiasm?  The letters are number 412, 413, 414, and
415 in {\sl Wolfgang {\Pauli}:\  Scientific Correspondence with
{\Bohr}, {\Einstein}, {\Heisenberg} a.o.\ Vol.\ 2}, edited by Karl
von Meyenn (Springer, 1985). If you can't find the book in your
library, I could fax you the pages.

\subsection{{\Ruediger}'s First Reply}

\bq
The letter exchange is indeed very interesting. I made quite
extensive translations, which are a little rough in places. In case
you want to cite anything, I'd like to go over it again.

{\Heisenberg}'s paper doesn't seem to be relevant for what we are
doing: it's all about the impossibility of hidden variables.

{\Schroedinger} is very interesting. He is very hard to translate: he
is obviously struggling with the language, trying to formulate deep
thoughts in an inadequate language. {\Pauli}, by contrast, is easy to
read and easy to translate.  That does not mean he is less deep, of
course.

I greatly enjoyed {\Pauli}'s insults in the first letter.

The most interesting passage is in letter 413. I am not certain
though that I do understand it well.

YOU SHOULD LEARN GERMAN! \medskip

\noindent ------------------

\noindent {\bf Letter 412 [{\Pauli} to {\Heisenberg}]}

\noindent p402

\bq
1. {\Einstein} has once more made public remarks on quantum
mechanics, in Phys Rev of May 15 (with {\Podolsky} and {\Rosen} -- no
good company btw). As is well known, this always amounts to a
catastrophe when it happens. ``His conclusion, razor-sharp -- that
cannot be which should not be'' (Morgenstern).

Admittedly I would like to grant him that, should an undergraduate
advance objections of this kind, I would consider him quite
intelligent and promising. -- Because this publication causes a
certain danger of public confusion -- especially in America -- it
might be good to send a comment to Phys Rev, and I'd like to
convince YOU to do it.
\eq

Now follows a description of the EPR setup and an {\Einstein} quote.
Then, bottom of p403, he says that a pedagogical reply should
clarify the difference between
\bq
a) The systems 1 and 2 do not interact ($=$ absence of interaction
energy)
\eq
and
\bq
b) The total system is in a state in which subsystems 1 and 2 are
independent. (The eigenfunction can be written as a product.)

Definition: This is the case if after a measurement on 2 of an
arbitrary quantity $F_2$ with known result $F_2=(F_2)_0$ (number) the
expectation values of the quantities $F_1$ stay the same as without
the measurement on 2.

Independently of {\Einstein} it seems to me that in a systematic
foundation of quantum mechanics one should, to a higher degree, {\it
begin} with the composition and separation of systems, than has been
done so far (e.g.\ by {\Dirac}). \ldots

One must distinguish different layers of reality: One $R$,
containing all information which can be obtained from measurements
on 1 and 2, and one (deducible from $R$) $r$, which contains only
the information which can be obtained from measurements on 1 alone.
One must then show how announcing a measurement result on 2 implies
discontinuous changes of $r$ ($r\rightarrow r_A$ or $r\rightarrow
r_B$ etc.) (unless the subsystems [Teilchensysteme must be a typo:
this means ``particle systems'', whatever that means. I am sure it
should read Teilsysteme, meaning ``subsystems''] are independent).
And that one is led to contradictions if one tries to explain these
changes, without reference to 2, classically or half-classically --
e.g.\ as ``hidden properties'' of 1.

Elderly Herren like Laue and {\Einstein} are haunted by the idea that
qm is correct but incomplete. They think qm can be completed by
statements which are not part of qm, without changing  the
statements which are part of qm. (A theory with such a property I
denote -- in the logical sense -- incomplete. Example: the kinetic
gas theory). Maybe you could -- in the reply to {\Einstein} --
clarify with authority that such a completion of qm is impossible
without changing its content.

NB. It is probably only because I recently got an invitation to
Princeton for the next winter semester that I have gone to such
trouble on these things, which for us are just trivialities. It will
be a lot of fun to go there. In any case, I want then to make the
Morgenstern motto popular.
\eq

The rest of the letter are comments on the magnetic moment of the
proton.\medskip

\noindent {\bf Letter 413 [{\Schroedinger} to {\Pauli}]}

\noindent 3rd paragraph:

\bq
I'd like to know very much if you agree with the following version
of the case, behind which I can certainly not go back.

It is well known that there are preparative measurement methods
through which one can make a system maximally known, i.e., through
which one can transform it into a ``pure case''. I claim: there are
preparative methods for which, after the final isolation of the
system being prepared, the experimenter is still free to continue
the method either in such a way that a pure case of type A results
with certainty, or in such a way that a pure case of type B results
with certainty.

So far you agree? Yes? You permit me also to call the pure case a
``state'' and to say: the experimenter is therefore free, after the
system has been isolated, to cause or to prevent the system to end
up in a state of type A?

Now, ``state'' is a word which everybody uses, even Saint PAM [Paul
Adrien Maurice], but that doesn't add to its content. One can,
however, easily convince oneself that, and in which way, a system
really changes when its psi function changes. Any psi function
conveys {\it the same\/} amount of knowledge. When psi changes,
there is never only an increase but always also a loss of knowledge.
But knowledge can only be acquired, never lost (except when one goes
gaga, which is disregarded here). A loss of knowledge can only occur
when the state of affairs has changed. In his sense I say: different
psi functions certainly correspond to different states of affairs --
or ``states''. I do not regard this as an illegitimate invoking of
the reality dogma. But I'd really like to know what you think about
it. And whether you really think that the {\Einstein} case -- let's
call it thus -- doesn't give anything to think about, but is
completely clear and simple and self-evident. (This is what
everybody said with whom I talked about it for the first time,
because they had well learned their Copenhagen credo in unum
sanctum. Three days later usually there came the statement: what I
said earlier was of course wrong, much too complicated. \ldots
\eq

\noindent {\bf Letter 414 [{\Heisenberg} to {\Pauli}]}

Mentions {\Bohr}'s reply to EPR in PR, which claims that EPR is the
same as the double slit experiment and therefore nothing new. Then
says he is nevertheless tempted to write a paper about the ``cut''
between system and measurement apparatus. And mentions a paper (or
book) by Grete {\Hermann} on the questions, which he likes
\medskip overall.

\noindent {\bf Appendix to letter 414:}

``Is a deterministic completion of qm possible'' (by {\Heisenberg},
unpublished)

The basic argument is that qm predictions are independent of the
location of the ``cut'', but that any such ``completion'' would not
be able to make the same prediction.

Curious, because refuted in a very simple way by Bohmian
mechanics.\medskip

\noindent {\bf Letter 415 [{\Pauli} to {\Schroedinger}]}

Agrees generally with {\Schroedinger}'s view of the {\Einstein} case.

Says that {\Bohr}'s reply to EPR contains nothing new, but that he
agrees with {\Bohr} that the {\Einstein} case contains nothing but
very elementary and direct consequences of the indeterminacy
relation.

Then summarizes {\Bohr}'s argument. \medskip

\noindent p420, second paragraph

\bq
Now, if one should denote ``pure case'' by ``state''?  Already
Kramers in his ``Lehrbuch der Wellenmechanik'' always puts ``physical
situation'' instead of ``state''. A pure case of A is a total
situation in which the results of certain measurements on A are (to
a maximal extent) predictable with certainty. If you call that a
``state'', I don't object -- but it is then indeed so that a change
in the state of A -- i.e., what can be predicted about A -- also in a
way different from a direct perturbation of A -- i.e. also {\it
after\/} isolation of A -- is within the {\it free choice\/} of the
experimenter.
\eq

\noindent 3rd para:
\bq
\noindent in my opinion {\it there is simply  no problem} --
and we know this state of affairs even without the {\Einstein}
example.
\eq

\noindent 4th para:

\bq
I would now like to comment on the general question raised by your
handwritten addition -- independent of {\Einstein}'s example. To what
extent is the additional assumption possible that there exist
properties, unknown to us today, that distinguish the single
realizations of the qm collectives?

I would like to raise this question from the level of the ``thou
shalt not'' to that of a statement that can be decided logically
within qm. {\it Which type of additional assumptions is possible
without changing the statistical consequences of qm itself?}

This question is very important, because in the heads of the
conservative older gentlemen ({\Einstein}, Laue, etc) there is an
unexterminable wrong analogy with kinetic gas theory. \ldots
\eq

Then he argues that it is impossible to make additional
micro-assumptions without changing expectation values -- a different
approach from {\Heisenberg}'s, but refuted by {\Bohm} just the same.
\eq

\subsection{{\Ruediger}'s Second Reply}

\bq
I missed one interesting part of letter 413 [{\Schroedinger} to
{\Pauli}]: the ``handwritten addition'' on p407:

\bq
I realize that ``pure case'' (or known $\psi$ function) is a
collective as well; and that this collective is created each time
through {\it selection\/} resulting from reading the (already
detached) measurement apparatus. And in both cases because of
different choice principles. Thus everything seems intelligible.

But I think that every collective in qm is constituted of completely
identical single cases. If one grants me differences between the
single cases that constitute the collective, then one admits
automatically the incompleteness of the quantum-mechanical
[``mechanical] description''.\footnote{Note: Concerning the funny
construction  [``mechanical] description'' {\Ruediger} later wrote:
``It's not a typo. It's exactly like that in the original. I don't
know what to make of it.''} Incomplete because I am ordered to regard
cases as equal although I know that they are not; because I am
forbidden to continue asking although I know that there is something
left to ask.
\eq

This seems quite relevant to our enterprise.
\eq

\section{17 March 1998, \ ``Schnitt''}

\brs
Curious, because refuted in a very simple way by Bohmian mechanics.
\ers

Do you think maybe he is relying implicitly in some form or another
on locality in his argument?  Say, via no interactions between the
two sides of the schnitt?

By the way, thanks for the summaries!!!

\section{18 March 1998, \ ``Hi''}

Thanks again for doing all the translation work.  Yes, I very much
wish that I knew German.  I can see it happening in the future, but,
unfortunately, not immediately.

I was intrigued by {\Pauli}'s phrase ``layers of reality.''  What do
you think he really meant by that?  Did he give any further
indication? Perhaps in the no-hidden-variable argument he gave
{\Schroedinger}?

\subsection{{\Ruediger}'s Reply}
There is nothing more on this question in the parts I didn't
translate. I am intrigued, too. Maybe it is just not a very clear
statement. If you replaced ``reality'' by ``description'', it would
become rather standard.

\section{23 March 1998, \ ``Frightening Depths''}

Would you remind me again what you see as the difference between
Lewis's ``principal principle'' and your ``principle of quantum
indeterminism''? That is, beside the trivial sort of thing that the
quantum case applies to an infinite number of random variables (i.e.,
one for each measurement).  It seems to me presently that the
statement ``maximal information cannot be completed for quantum
phenomena'' provides the principal principle for the ``principal
principle.''

By the way, did you ever hear from your friend, the Latin expert?
One day last week, {\Hideo} snuck into my office, placed a banana on
my laptop computer, and changed the screensaver to ``Mechanica
bananica ex mente orta lex est!''  So far I've left it without
further modification.

\section{23 March 1998, \ ``{\Bohr}'s Palate''}

\begin{flushleft}
\parbox{2.9in}{
``Between us, we cover all knowledge; he knows all that can be known
and I know the rest.''\\
\hspace*{\fill} --- {\it Mark Twain}\\
\hspace*{\fill} (ruminating on entanglement)}
\end{flushleft}
\brs
Doesn't the principal principle involve two kinds of probability
and is therefore unpalatable?
\ers
Indeed it does, but I'm thinking about what similarities and
distinctions we can draw between the PP (principal principle) and
the PQI (principle of quantum indeterminism) \ldots\ don't worry by
the way, I don't intend to use these acronyms in the paper:  I was
just trying to save some writing now \ldots\ though I see that I'm
defeating that purpose.

The thing I'm thinking is that a Lewisian might see $|\psi\rangle$ as
a compact description of all probabilities (for a given physical
system) for which the PP is operant.

\brs
``Maximal information cannot be completed for quantum phenomena''
entails (does it?) that two copies of a quantum system can be placed
into exactly the same state \ldots
\ers
Not completely, I don't think. One needs something slightly more,
perhaps the principle that ``knowledge is power'' over nature.  (Do
you know who first said the thing between the quotes?  The earliest
thing I can find in my quote book is George Eliot.)  That is to say,
maximal knowledge gives an in-principle reproducibility in
experiment (I guess it need not have been so).  That maximal
information cannot be completed is the quantum mechanical piece of
it all.

\brs
Didn't we conclude that statements like the principal principle
can be replaced by statements about joint probability distributions?
\ers
True enough.  But still the question above.

\section{22 April 1998, \ ``Infinite Patience''}

I thought about opening up this note with the sentence, ``If there be
saints, then by virtue of your infinite patience you are surely
one.'' But I knew that was too sappy and, so, thought better of
writing it.

In any case let me do thank you.  I have spent the whole day
navigating the space of your mind, or so it seems.  I finally read
again, from start to finish (and pretty carefully I might add), your
original manuscript, {\Carl}'s commentary on it, and finally your
commentary on his commentary.  I had been avoiding this because I
wanted to make some thoughts of my own before doing that---I wanted
to sketch an outline for the paper that had a reasonable chance of
being independent.  I'm surely glad that I did that; but also now
I'm surely glad that I read the old stuff again.  It again injected
me with a warm fuzzy feeling that has so far lasted all day.

Part of that is also surely due to my first real encounter with de
{\Finetti}.  I've read about half of his 1931 essay ``Probabilism''
and find that I can hardly put it down (though I had to \ldots\ so
that I could write this note).  It's wonderful.  I'm not sure why I
wasn't drawn to him so much before \ldots\ maybe the time just
wasn't ripe. But I'm certainly ready for it now.

Unfortunately I have to close shop in 24 minutes so that we can take
Scott Parkins to dinner.  So let me just ramble for a couple of
minutes.

First, I found the de {\Finetti} article in a special issue of the
journal {\sl Erkenntnis\/} (vol.\ 31, nos.\ 2--3, 1989) titled,
``Bruno de {\Finetti}'s Philosophy of Probability.''  There look to
be several interesting articles in it.  So if you haven't encountered
this yet, you might interested in having a look.

Second, I came across a passage that intrigues me in connection to
your Principle of Quantum Indeterminism.  It is in a book review of
Donald Gillies' book {\sl An Objective Theory of Probability}.  It
states,
\bq
His thesis that {\Kolmogorov}'s Axioms should be regarded as
formulating the general laws of an explanatory theory of random
phenomena leads naturally to the realisation that as such they are
incomplete and require to be augmented with what Gillies calls an
Axiom of Independent Repetitions (p.~90ff).  His criticism
(pp.~108--118) of the popular thesis that the various limit theorems
of mathematical probability supply the connection between theory and
experience is to my mind convincing \ldots\
\eq

I wonder what that's about?  Unfortunately, we don't have a copy of
this book here and it'll take me a few days to get it by interlibrary
loan (though I'll order it tomorrow).  If you get a chance, could you
check whether you guys have a copy there?  Maybe you're more likely
to; I see from his web page that he (Gillies) is at King's College.

\section{23 April 1998, \ ``It Sinks Deeper''}

The master de {\Finetti} declares that we should not speak of
``unknown'' probabilities.  For probabilities are not objective
entities in their own right.  Probability theory is a law of thought.
\ldots\ \ldots\ \ldots\ But quantum mechanics is a law of thought.
What does it mean to speak of attempting to clone an unknown quantum
state?

I sink deeper into myself.

\section{24 April 1998, \ ``Groping and Moping''}

\brs
What does it mean that quantum mechanics is a law of thought?
\ers

That's a good question, the answer of which I keep groping for.  I
think it means that QM's formal structure or mathematics corresponds
to ``nothing more'' than Bayesian-style reasoning applied to any
situation where maximal information is necessarily incomplete. This
could be a world in which we must come to intersubjective agreement
but yet has the mischievous quality that MY information-gathering
questions disturb YOUR predictability. (This appears to be the case
in the physical world described by QM.)  Or it might just as well be
a human mind when it tries to reach some understanding of its future
thoughts: the process of introspection.

\brs
Don't take him [de {\Finetti}] too literally here. He understood
nothing about quantum mechanics.
\ers

Not to worry, I wasn't.  I was just extrapolating based on the
thoughts above.  This issue of being an ``unknown'' quantum state has
been something that's been bothering more for quite a while. I've
tried to invent a better term, but I haven't found one that I am
completely satisfied with.  If one fails to remember that when one is
speaking of an ``unknown'' quantum state, there is a secret
preparator {\it somewhere\/} in the background---who in principle
could be watching all actions as they unfold---it can lead to a load
of trouble.

\brs
If you had a clone of an unknown quantum state, you could not
test if it was indeed a clone, could you?
\ers

Indeed, never with infinite precision:  just as the flipping of a
coin that continually comes up heads can never give---on it's
own---infinite security that there is actually a bias.  Our friend
David {\Mermin} keeps asking, ``What is the structure of quantum
mechanics trying to tell us?''  I think it's trying to tell us that
it is a law of thought.

\section{27 April 1998, \ ``Fenomeno Aleatorio''}

So I ended up not being philosophical at all yesterday---sorry for my
hollow promises.  I ended up thinking more about making my life more
efficient on the computer.

\brs
I don't believe {\Mermin} read it a second time. But I did (not the
edited version, though) --- and it made me think. {\em Mechanica
quantica ex mente orta lex est.} A bold idea, and I don't think I
have ever understood it as clearly as after my last rereading of your
notes.
\ers
Thank you for the pleasant compliments.  More than is
healthy I'm sure, my self-esteem has lately been thriving on such
simple personal things. (Rejection letter Number Aleph came in this
morning.)  If you get a chance, do take a look at the last subsection
of the edited version. You'll now find not only Asherisms and
Merminitions, but \ldots\

\brs
Could you remind me of {\Wheeler}'s twenty questions?
\ers

The quickest place to look is the notes I wrote on one of  {\Asher}
{\Peres}'s review articles, the discussion after Asherism
\ref{OldIsm7} in particular. (Somewhat connected also is the
discussion following Asherism \ref{OldIsm10}.)  I believe I sent you
those notes, but in case you've discarded them I'll send them on
again.  If you really wish, I can look for the clearest statement of
it in {\Wheeler}'s own writings and give you that reference a little
later.

\brs
Your {\Heisenberg} quote ``What we learn about is not nature itself,
but nature exposed to our methods of questioning'' does apply just as
well to pre-quantum science. It is a pithy summary of {\Kant}, which
gives one the choice of imagining that reality exists inpendently of
any observer.
\ers
True enough.  I tried to make some bow to this in my note to
{\Mermin}---take a look at the discussion following Merminition
\ref{OldAsm11}. (Alternatively do a word search on ``noumena.'')
More to the point though, I do think that {\Heisenberg} (who
certainly new of {\Kant}'s ideas) was trying to go further than this
with his statement \ldots\ though perhaps not as far as I would like
him to go.  I now think the natural---most conservative,
actually---follow-up or rewrite to {\Kant} is in this thing that
{\Herb} likes to call ``realit{\it t\/}y'' (i.e., the
game-of-twenty-questions ontology).  But that hasn't yet been
defined well enough to be of great use (in a philosophy of science).

\brs
The existence of a world out there is not a matter of taste any
more, not ``a mere metaphysical question.''
\ers
I don't know, it seems to me to depend upon what lengths one is
willing to go to preserve an ``independent reality.''  They tell me
that the {\Bohm} theory does it just fine.  The thing that is most
important to me---in my public behavior at least---is really which
point of view will be most productive for going the next step beyond
quantum theory.

\brs
How is {\Plato}nism related to your ideas about physical reality?
Are mathematical statements like Fermat's last theorem true in an
absolute, {\Plato}nist sense?
\ers
I've never thought much about the philosophy of mathematics.  It
seems to me presently though that there's not a lot of difference
between Fermat's Last Theorem and my toaster.  (See the discussion
following Asherism \ref{OldIsm7} again.)  I think I'd likely even
say the same thing for mathematical logic (as encoded in Boolean
algebra, for instance).

\brs
The Latin phrase may be rather inaccurate --- not bad Latin, but
not expressing the right thing. I'll make another effort at finding a
better translation.
\ers
Oh now this is important!  Please do find the best possible
translation! (\ldots\ So we'll know for sure that we're in the best
of all possible worlds.)

\brs
``Wait, wait, one more last thing: I defer discussion of objective
probability once again.''  Did you ever discuss it with {\Mermin}?
\ers
Indeedy, but you now know everything that I know.  I do
think what he's trying to do is get at something different than
simple frequentism or propensitism.  His ``objective probability'' is
founded upon {\it joint\/} probability distributions only.  So, at
least in that simple sense, it must be different from the standard
tacks on ``objective probability.''

\brs
Wouldn't it be marvelous if one could derive Hilbert space from
something like the principle of quantum indeterminism?
\ers
Yes, yes, yes, I say!  But beyond that, I'm sort of empty of thought
right now.

OK, to work finally on THE project.  I want to finish reading the de
{\Finetti} volume today.

\section{30 April 1998, \ ``One Stone Unturned''}

The point of the last note?  Nothing really, I just like lists
\ldots\ you'll see what I mean.  The next note I'm sending contains
(much of) the bibliography for the paper.  Please check it over when
you get a chance and tell me what I've missed \ldots\ certainly no
hurry on this. One thing I know that I have missed is de {\Finetti}'s
books. Unfortunately I haven't been able to get hold of them yet:
it's a shame, because I'm (mentally) ready for them now.  Maybe I
should buy them \ldots\ they're just so darned expensive.

Believe it or not, I do believe that I will cite each and every one
of the items you're about to see.  There's a lot of emphasis on the
objectivist/propensity stuff.  Reason?  Basically because
subjectivists---the few that there are in this game---will feel
pretty happy with the argument.  However each and every
objectivist---take Ian Percival as an example---will be thinking
something like, ``Well, you clearly haven't taken the time to
understand my version of the story!''  Oh yeah?

Now the challenge:  To make the paper longer than the reference
list!!!

\noindent ---------------

\begin{enumerate}\itemsep -2pt
\footnotesize

\item
L.~E. {\Ballentine}, ``Can the Statistical Postulate of Quantum
Theory Be Derived?---A Critique of the Many-Uni\-verses
Interpretation,'' Found.\ Phys.\ {\bf 3}, 229--240 (1973).

\item
P.~{\Benioff}, ``Possible Strengthening of the Interpretative Rules
of Quantum Mechanics,'' Phys.\ Rev.\ D {\bf 7}, 3603--3609 (1973).

\item
P.~A. {\Benioff}, ``Finite and Infinite Measurement Sequences in
Quantum Mechanics and Randomness: The {\Everett} Interpretation,''
J.\ Math. Phys. {\bf 18}, 2289--2295 (1977).

\item
P.~{\Benioff}, ``A Note on the {\Everett} Interpretation of Quantum
Mechanics,'' Found. Phys. {\bf 8}, 709--720 (1978).

\item
P.~{\Benioff}, ``On the Correct Definition of Randomness,'' in {\sl
PSA 1978: Proceedings of the 1978 Biennial Meeting of the Philosophy
of Science Association, Vol.~2}, edited by P.~D. Asquith and
I.~{\Hacking} (Philosophy of Science Association, East Lansing,
Michigan, 1981), pp.~63--78.

\item
F.~C. Benenson, ``Randomness and the Frequency Definition of
Probability,'' Synthese {\bf 36}, 207--233 (1977).

\item
G.~{\Boole}, {\sl An Investigation of the Laws of Thought}, (Dover,
New York, 1958).

\item
R.~B. Braithwaite, ``On Unknown Probabilities,'' in {\sl Observation
and Interpretation: A Symposium of Philosophers and Physicists},
edited by S.~K\"orner (Academic Press, New York, 1957), pp.~3--11.

\item
S.~L. {\Braunstein} and C.~M. {\Caves}, ``Wringing Out Better {\Bell}
Inequalities,'' Ann.\ Phys.\ {\bf 202}, 22--56 (1990).

\item
J.~{\Bub}, ``{\Popper}'s Propensity Interpretation of Probability and
Quantum Mechanics,'' in {\sl Induction, Probability, and
Confirmation}, Minnesota Studies in the Philosophy of Science, Vol.\
VI, edited by G.~Maxwell and R.~M. Anderson, Jr. (U. of Minnesota
Press, Minneapolis, 1975), pp.~416--429.

\item
M.~Bunge, ``Possibility and Probability,'' in {\sl Foundations of
Probability Theory, Statistical Inference, and Statistical Theories
of Science, Vol.~III}, edited by W.~L. Harper and C.~A. Hooker
(D.~Reidel, Dordrecht, 1976), pp.~17--33.

\item
A.~Cassinello and J.~L. S\'anchez-G\'omez, ``On the Probabilisitic
Postulate of Quantum Mechanics,'' Found.\ Phys.\ {\bf 26},
1357--1374 (1996).

\item
C.~M. {\Caves}, ``Information and Entropy,'' Phys.\ Rev.\ E {\bf 47},
4010--4017 (1993).

\item
C.~M. {\Caves} and C.~A. Fuchs, ``Quantum Information:~How Much
Information in a State Vector?,'' in {\sl The Dilemma of {\Einstein},
{\Podolsky} and {\Rosen} -- 60 Years Later}, edited by A.~{\Mann} and
M.~{\Revzen}, Ann.\ Israel Phys.\ Soc.\ {\bf 12}, 226--257 (1996).

\item
R.~Cooke, M.~Keane, and W.~Moran, ``An Elementary Proof of
{\Gleason}'s Theorem,'' Math.\ Proc.\ Camb.\ Phil.\ Soc.\ {\bf 98},
117--128 (1981).

\item
B.~de {\Finetti}, ``Probabilism,'' Erkenntnis {\bf 31}, 169--223
(1989).

\item
D.~{\Deutsch}, ``Quantum Theory as a Universal Physical Theory,''
Int.\ J. Theor.\ Phys.\ {\bf 24}, 1--41 (1985).

\item
B.~S. {\DeWitt}, ``The Many-Universes Interpretation of Quantum
Mechanics,'' in {\sl Proceedings of the International School of
Physics ``Enrico {\Fermi}'' Course IL: Foundations of Quantum
Mechanics}, edited by B.~d'Espagnat (Academic Press, New York,
1971), pp.~211--262.

\item
J.~Earman, {\sl Bayes or Bust?: A Critical Examination of Bayesian
Confirmation Theory}, (MIT Press, Cambridge, MA, 1992).

\item
H.~{\Everett}, III, ``\,`Relative State' Formulation of Quantum
Mechanics,'' Rev.\ Mod.\ Phys.\ {\bf 29}, 454--462 (1957).

\item
H.~{\Everett}, III, ``The Theory of the Universal Wave Function,'' in
{\sl The Many-Worlds Interpretation of Quantum Mechanics}, edited by
B.~S. {\DeWitt} and N.~{\Graham} (Princeton U. Press, Princeton,
1973), pp.~3--140.

\item
E.~{\Farhi}, J.~{\Goldstone}, and S.~{\Gutmann}, ``How Probability
Arises in Quantum Mechanics,'' Ann.\ Phys.\ {\bf 192}, 368--382
(1989).

\item
J.~H. Fetzer, ``Statistical Probabilities: Single Case Propensities
vs.\ Long-Run Frequencies,'' in {\sl Developments in the Methodology
of Social Science}, edited by W.~Leinfellner and E.~K\"ohler
(D.~Reidel, Dordrecht, 1974), pp.~387--397.

\item
J.~H. Fetzer, ``Probability and Objectivity in Deterministic and
Indeterministic Situations,'' Synthese {\bf 57}, 367--386 (1983).

\item
R.~P. Feynman, ``The Concept of Probability in Quantum Mechanics,''
in {\sl  Proceedings of the Second Berkeley Symposium on Mathematical
Statistics and Probability}, edited by J.~Neyman (U. of California
Press, Berkeley, 1951), pp.~533--541.

\item
A.~{\Fine}, ``Do Correlations Need To Be Explained?,'' in {\sl
Philosophical Consequences of Quantum Theory: Reflections on
{\Bell}'s Theorem}, edited by J.~T. Cushing and E.~McMullin (U. of
Notre Dame Press, Notre Dame, 1989), pp.~175--194.

\item
D.~Finkelstein, ``The Logic of Quantum Mechanics,'' Trans.\ N.Y.\
Acad.\ Sci.\ {\bf 25}, 621--635 (1963)

\item
M.-C.~{\Galavotti}, ``Comments on Patrick {\Suppes} `Propensity
Interpretations of Probability'\,'' Erkenntnis {\bf 26}, 359--368
(1987).

\item
M.-C.~{\Galavotti}, ``Anti-Realism in the Philosophy of Probability:
Bruno de {\Finetti}'s Subjectivism'' Erkenntnis {\bf 31}, 239--261
(1989).

\item
A.~J.~M. {\Garrett}, ``Making Sense of Quantum Mechanics: Why You
Should Believe in Hidden Variables,'' in {\sl Maximum Entropy and
Bayesian Methods (Paris, France, 1992)}, edited by
A.~Mohammed-Djafari and G.~Demoment (Kluwer, Dordrecht, 1993),
pp.~79--83.

\item
R.~N. {\Giere}, ``Objective Single-Case Probabilities and the
Foundations of Statistics,'' in {\sl Logic, Methodology and
Philosophy of Science IV}, edited by P.~{\Suppes}, L.~Henkin,
A.~Jojo, and Gr.~C.~Moisil (North-Holland, Amsterdam, 1973),
pp.~467--483.

\item
R.~N. {\Giere}, ``Review of D.~H. Mellor, {\sl The Matter of
Chance},'' Ratio {\bf 15}, 149--155 (1973).

\item
R.~N. {\Giere}, ``A Laplacean Formal Semantics for Single-Case
Propensities,'' J.\ Phil.\ Logic {\bf 5}, 321--353 (1976).

\item
R.~N. {\Giere}, ``Propensity and Necessity,'' Synthese {\bf 40},
439--451 (1979).

\item
D.~A. Gillies, {\sl An Objective Theory of Probability}, (Methuen,
London, 1973), Chaps.~4--5.

\item
N.~{\Gisin}, ``Propensities and the State-Property Structure of
Classical and Quantum Systems,'' J.\ Math.\ Phys.\ {\bf 25},
2260--2265 (1984).

\item
A.~M. {\Gleason}, ``Measures on the Closed Subspaces of a Hilbert
Space,'' J. Math.\ Mech.\ {\bf 6}, 885--894 (1957).

\item
N.~{\Graham}, ``The {\Everett} Interpretation of Quantum Mechanics,''
Ph.D. thesis, University of North Carolina, Chapel Hill, NC (1970).

\item
N.~{\Graham}, ``The Measurement of Relative Frequency,'' in {\sl The
Many-Worlds Interpretation of Quantum Mechanics}, edited by B.~S.
{\DeWitt} and N.~{\Graham} (Princeton U. Press, Princeton, 1973),
pp.~229--253.

\item
S.~{\Gutmann}, ``Using Classical Probability to Guarantee Properties
of Infinite Quantum Sequences,'' Phys.\ Rev.\ A {\bf 52}, 3560--3562
(1995).

\item
J.~B. {\Hartle}, ``Quantum Mechanics of Individual Systems,'' Am.\ J.
Phys. {\bf 36}, 704--712 (1968).

\item
G.~{\Hellman}, ``Randomness and Reality,'' in {\sl PSA 1978:
Proceedings of the 1978 Biennial Meeting of the Philosophy of Science
Association, Vol.~2}, edited by P.~D. Asquith and I.~{\Hacking}
(Philosophy of Science Association, East Lansing, Michigan, 1981),
pp.~79--97.

\item
R.~I.~G. Hughes, {\sl The Structure and Interpretation of Quantum
Mechanics}, (Harvard U. Press, Cambridge, MA, 1989), Chap.~8.

\item
P.~Humphreys, ``Why Propensities Cannot Be Probabilities,'' Phil.\
Rev.\ {\bf 94}, 557--570 (1985).

\item
E.~T. {\Jaynes}, ``Clearing Up Mysteries -- The Original Goal,'' in
{\sl Maximum Entropy and Bayesian Methods (Cambridge, England,
1988)}, edited by J.~Skilling (Kluwer, Dordrecht, 1989), pp.\ 1--27.

\item
E.~T. {\Jaynes}, ``Probability in Quantum Theory,'' in {\sl
Complexity, Entropy and the Physics of Information}, edited by W.~H.
{\Zurek} (Addison-Wesley, Redwood City, CA, 1990), pp.~381--403.

\item
E.~T. {\Jaynes}, {\sl Probability Theory:~The Logic of Science}, book
in preparation. Incomplete preliminary draft available at {\tt
ftp://bayes.wustl.edu/}.

\item
R.~C. {\Jeffrey}, {\sl The Logic of Decision}, (McGraw-Hill, New
York, 1965).

\item
R.~C. {\Jeffrey}, ``{\Mises} Redux,'' in {\sl Basic Problems in
Methodology and Linguistics}, edited by R.~E. Butts and J.~Hintikka
(D.~Reidel, Dordrecht, 1977), pp.~213--222.

\item
R.~{\Jeffrey}, ``Reading {\it Probabilismo},'' Erkenntnis {\bf 31},
225--237 (1989).

\item
A.~{\Kent}, ``Against Many-Worlds Interpretations,'' Int.\ J.\ Mod.\
Phys.\ {\bf A5}, 1745--1762 (1990).

\item
H.~E. {\Kyburg}, ``Propensities and Probabilities,'' Brit.\ J.\
Phil.\ Sci.\ {\bf 25}, 358--375 (1974).

\item
H.~E. {\Kyburg}, Jr., ``Chance,'' J.\ Phil.\ Logic {\bf 5}, 355--393
(1976).

\item
I.~Levi, ``Probability Exists (but just barely)!''\ in {\sl Logic,
Methodology and Philosophy of Science VII}, edited by R.
Barcan~Marcus, G.~J.~W. Dorn, and P.~Weingartner (North-Holland,
Amsterdam, 1986), pp.~367--385.

\item
D.~Lewis, ``A Subjectivist's Guide to Objective Chance,'' in {\sl
Studies in Inductive Logic and Probability, Vol.~II}, edited by
R.~C. {\Jeffrey} (U. of California Press, Berkeley, 1980),
pp.~263--293.

\item
T.~J. Loredo, ``From {\Laplace} to Supernova SN\,1987A: Bayesian
Inference in Astrophysics'' in {\sl Maximum Entropy and Bayesian
Methods (Dartmouth, U.S.A., 1989)}, edited by P.~F. Foug\`ere
(Kluwer, Dordrecht, 1990), pp.~81--142.

\item
D.~H. Mellor, {\sl The Matter of Chance}, (Cambridge U. Press,
Cambridge, 1971).

\item
N.~D. {\Mermin}, ``The Ithaca Interpretation of Quantum Mechanics,''
Pramana {\bf 51}, 549--565 (1998).

\item
N.~D. {\Mermin}, ``What Is Quantum Mechanics Trying To Tell Us?,''
Am.\ J. Phys.\ {\bf 66}, 753--767 (1998).

\item
D.~Miller, ``Single-Case Probabilities,'' Found.\ Phys.\ {\bf 21},
1501--1516 (1991).

\item
P.~{\Mittelstaedt}, ``The Objectification in the Measuring Process
and the Many Worlds Interpretation,'' in {\sl Symposium on the
Foundations of Modern Physics}, edited by P.~{\Lahti} and
P.~{\Mittelstaedt} (World Scientific, Singapore, 1991), pp.\
261--279.

\item
P.~{\Mittelstaedt}, ``Is Quantum Mechanics a Probabilistic Theory?,''
in {\sl Potentiality, Entanglement and Passion-at-a-Distance},
edited by R.~S. Cohen, M.~Horne, and J.~Stachel (Kluwer, Dordrecht,
1997), pp.~159--175.

\item
M.~Mugur-Schachter, ``The Quantum Mechanical Hilbert Space Formalism
and the Quantum Mechanical Probability Space of the Outcomes of
Measurement,'' in {\sl Foundations of Quantum Mechanics and Ordered
Linear Spaces}, edited by A.~Hartk\"amper and H.~Neumann
(Springer-Verlag, Berlin, 1974), Lecture Notes in Physics, No.~29,
pp.\ 288--308.

\item
W.~{\Ochs}, ``On the Strong Law of Large Numbers in Quantum
Probability Theory,'' J. Phil.\ Logic {\bf 6}, 473--480 (1977).

\item
W.~{\Ochs}, ``Concepts of Convergence for a Quantum Law of Large
Numbers,'' Rep.\ Math.\ Phys.\ {\bf 17}, 127--143 (1980).

\item
D.~N. Page, private communication to J.~B. {\Hartle}, 28 October
1996.

\item
A.~{\Peres}, ``What is a State Vector?,'' Am.\ J.\ Phys.\ {\bf 52},
644--650 (1984).

\item
A.~{\Peres}, {\sl Quantum Theory: Concepts and Methods}, (Kluwer,
Dordrecht, 1993).

\item
I.~{\Pitowsky}, ``From George {\Boole} to John {\Bell}---The Origins
of {\Bell}'s Inequality,'' in {\sl {\Bell}'s Theorem, Quantum Theory
and Conceptions of the Universe}, edited by M.~Kafatos (Kluwer,
Dordrecht, 1989), pp.~37--49.

\item
I.~{\Pitowsky}, ``George {\Boole}'s `Conditions of Possible
Experience' and the Quantum Puzzle,'' Brit.\ J.\ Phil.\ Sci.\ {\bf
45}, 95--125 (1994).

\item
I.~{\Pitowsky}, ``Infinite and Finite {\Gleason}'s Theorems and the
Logic of Indeterminacy,'' to appear in J. Math.\ Phys.\ (1998).

\item
K.~R. {\Popper}, ``The Propensity Interpretation of the Calculus of
Probability and the Quantum Theory,'' in {\sl Observation and
Interpretation: A Symposium of Philosophers and Physicists}, edited
by S.~K\"orner (Academic Press, New York, 1957), pp.~65--70.

\item
K.~R. {\Popper}, {\sl Quantum Theory and the Schism in Physics},
(Hutchinson, London, 1982).

\item
J.~{\Preskill}, ``Chapter 3.\ Foundations of Quantum Theory II:
Measurement and Evolution,'' in {\sl Lecture Notes for Physics 229:
Advanced Mathematical Methods of Physics}, available on the World
Wide Web at \\
{\tt http://www.theory.caltech.edu/people/preskill/ph229/}.

\item
H.~{\Reichenbach}, {\sl The Theory of Probability: An Inquiry into
the Logical and Mathematical Foundations of the Calculus of
Probability}, (U. of California Press, Berkeley, 1949).

\item
C.~{\Rovelli}, ``Relational Quantum Mechanics,'' Int.\ J.\ Theor.\
Phys.\ {\bf 35}, 1637--1678 (1996).

\item
W.~C. Salmon, ``Propensities: A Discussion Review,'' Erkenntnis {\bf
14}, 183--216.

\item
L.~J. {\Savage}, {\sl The Foundations of Statistics}, (Dover, New
York, 1972).

\item
C.~Schneider, ``Two Interpretations of Objective Probabilities,''
Philosph.\ Naturalis {\bf 31}, 107--131 (1994).

\item
B.~Skyrms, {\sl Causal Necessity: A Pragmatic Investigation of the
Necessity of Laws}, (Yale U. Press, New Haven, 1980), Chap.~IA.

\item
B.~Skyrms, ``Statistical Laws and Personal Propensities,'' in {\sl
PSA 1978: Proceedings of the 1978 Biennial Meeting of the Philosophy
of Science Association, Vol.~2}, edited by P.~D. Asquith and
I.~{\Hacking} (Philosophy of Science Association, East Lansing,
Michigan, 1981), pp.~551--562.

\item
B.~Skyrms, {\sl Pragmatics and Empiricism}, (Yale U. Press, New
Haven, 1984).

\item
B.~Skyrms, ``Probability and Causation,'' J.\ Econometrics {\bf 39},
53--68 (1988).

\item
S.~Spielman, ``Physical Probability and Bayesian Statistics,''
Synthese {\bf 36}, 235--269 (1977).

\item
W.~Spohn, ``A Brief Remark on the Problem of Interpreting
Probability Objectively,'' Erkenntnis {\bf 26}, 329--334 (1987).

\item
E.~J. Squires, ``On an Alleged `Proof' of the Quantum Probability
Law,'' Phys.\ Lett.\ A {\bf 145}, 67--68 (1990).

\item
M.~Strauss ``Two concepts of probability in physics,'' in {\sl Logic,
Methodology and Philosophy of Science IV}, edited by P.~{\Suppes},
L.~Henkin, A.~Jojo, and Gr.~C.~Moisil (North-Holland, Amsterdam,
1973), pp.~603--615.

\item
P.~{\Suppes}, ``New Foundations for Objective Probability: Axioms for
Propensities,'' in {\sl Logic, Methodology and Philosophy of Science
IV}, edited by P.~{\Suppes}, L.~Henkin, A.~Jojo, and Gr.~C.~Moisil
(North-Holland, Amsterdam, 1973), pp.~515--529.

\item
P.~{\Suppes}, ``Some Further Remarks on Propensity: Reply to
Maria-Carla {\Galavotti},'' Erkenntnis {\bf 26}, 369--376 (1987).

\item
P.~{\Suppes}, ``The Transcendental Character of Determinism,'' in
{\sl Midwest Studies in Philosophy, Volume XVII: Philosophy of
Science}, edited by P.~A. French, T.~E. Uehling, Jr., and H.~K.
Wettstein (U. of Notre Dame Press, Notre Dame, 1993), pp.~242--257.

\item
S.~S\'ykora, ``Quantum Theory and the Bayesian Inference Problems,''
J.\ Stat.\ Phys.\ {\bf 11}, 17--27 (1974).

\item
B.~C. van Fraassen, {\sl Laws and Symmetry}, (Clarendon Press,
Oxford, 1989).

\item
B.~C. van Fraassen, {\sl Quantum Mechanics: An Empiricist View},
(Clarendon Press, Oxford, 1991).

\item
R.~von {\Mises}, {\sl Probability, Statistics and Truth}, (Dover, New
York, 1981).

\item
J.~von Neumann, {\sl Mathematical Foundations of Quantum Mechanics},
translated by R.~T. Beyer, (Princeton U. Press, Princeton, 1955),
p.~298, fn.~156.

\item
J.~von Neumann, ``Quantum Logics (Strict- and
Prob\-a\-bil\-ity-Logics),'' in {\sl John von Neumann: Collected
Works, Vol. IV}, edited by A.~H. Taub (Macmillan, New York, 1962),
pp.~195--197.

\item
J.~von Plato, ``De {\Finetti}'s Earliest Works on the Foundations of
Probability,'' Erkenntnis {\bf 31}, 263--282 (1989).

\item
C.~F. von Weizs\"acker, ``Probability and Quantum Mechanics,''
Brit.\ J.\ Phil.\ Sci.\ {\bf 24}, 321--337 (1994).

\item
E.~P. {\Wigner}, ``On Hidden Variables and Quantum Mechanical
Probabilities,'' Am.\ J.\ Phys.\ {\bf 38}, 1005--1009 (1970).
\end{enumerate}

\section{24 May 1998, \ ``Nature Says No''}

Well, nature apparently doesn't like our tinkering with these
anti-realist ideas so much.  Despite my attempt at optimism the
other day, I've hardly had a productive moment since the last time I
wrote you.  I had to stay at home both Thursday and Friday.  Friday
I became feverish.  This morning my ears started filling with fluid,
feeling as if they were going to burst.  So I finally went to see a
doctor:  now my system is stuffed full of clarithromycin, codeine,
acetominophen, phenylephrine HCl, and God only knows what else.  I
hope these drugs will have some effect soon---I will soon go crazy
otherwise.

In some of my delusional moments I have been thinking about your
mathematical Platonism question.  The more I think about it, the
more I really dislike the idea.  A world with such a stable,
untouchable substrate strikes me as a horrible, still-born world.

\section{31 August 1998, \ ``Comfort in Numbers''}

I found something today that I think you might be interested in:  an
essay by {\Heisenberg} titled ``Quantum Mechanics and {\Kant}ian
Philosophy'' in his book {\sl Physics and Beyond}.  He makes some
points about {\Kant}'s a priori that I heard you making last week.
Take a look if you get a chance.

\section{21 January 1999, \ ``Stupendous Literature Man''}

Apparently, the phrase ``The world is presented to us only once'' and
slight variants thereof are popular ones.  For I have now seen them
used to mean three different things.  The quotes below speak for
themselves.  In the original outline of Fuji, however, I had only
meant the version that Richard {\Jeffrey} (and Ernst Mach) are
talking about.\medskip

\noindent
Mara {\Beller}, ``The Sokal Hoax:~At Whom are We Laughing?,'' {\sl
Physics Today}, September 1998, pp.~29--34:
\bq
\noindent
In fact, {\Einstein} was no ``naive realist,'' despite such
caricaturing of his stand by the Copenhagen orthodoxy.  He ridiculed
the ``correspondence'' view of reality that many scientists accept
uncritically.  {\Einstein} fully realized that the world is not
presented to us twice---first as it is, and second, as it is
theoretically described---so we can compare our theoretical ``copy''
with the ``real thing.''  The world is given to us only
once---through our best scientific theories.
\eq
Erwin {\Schroedinger}, {\sl Mind and Matter\/} (1958), p.~51:
\bq
\noindent
It is the same elements that go to compose my mind and the world.
This situation is the same for every mind and its world, in spite of
the unfathomable abundance of `cross-references' between them.  The
world is give to me only once, not one existing and one perceived.
Subject and object are only one.  The barrier between them cannot be
said to have broken down as a result of recent experience in the
physical sciences, for this barrier does not exist.
\eq
Erwin {\Schroedinger}, {\sl Mind and Matter\/} (1958), p.~63:
\bq
\noindent
Nay, may we call a world that nobody contemplates even that?
[\ldots] But a world, existing for many millions of years without
any mind being aware of it, contemplating it, is it anything at
all?  Has it existed? For do not let us forget:  to say, as we did,
that the becoming of the world is reflected in a conscious mind is
but a clich\'e, a phrase, a metaphor that has become familiar to
us.  The world is given but once. Nothing is reflected.  The
original and the mirror-image are identical. The world extended in
space and time is but our representation.
\eq
Erwin {\Schroedinger}, {\sl Nature and the Greeks\/} (1954):
\bq
\noindent
{\Plato}, {\Aristotle}, and {\Epicurus} emphasize the import of being
astonished. And this is not trivial when it refers to general
questions about the world as whole; for, indeed, it is given us only
once, we have no other one to compare it with.
\eq
Albert {\Einstein}, ``Elementare \"Uberlegungen zur Interpretation
der Grundlagen der Quan\-ten\-Me\-chanik,'' in M. {\Born}, Scientific
Papers Presented to Max {\Born} (Hafner Publishing, New York, 1953),
pp.\ 33--40:
\bq
\noindent
Nature as a whole can only be viewed as an individual system,
existing only once, and not as a collection of systems.
\eq

{\Carl}'s further attribution of the translation of this to
H.~Freistadt is apparently incorrect; or at least I couldn't find
this phrase in that article---it is just a review article of
Bohmianism.  {\Carl}, where did you actually see this line quoted?
Was it in Cushing's book or something?

Now for the {\it pi\`ece de resistance}.  I'm not sure I agree with
all of the larger context that {\Jeffrey} adds to the quote---I
think it still imagines that the observer is ``detached'' in the way
of classical physics (and you guys know I hate that)---but here it
is, the thing I was originally thinking of.\medskip

\noindent Richard {\Jeffrey}, ``de {\Finetti}'s
Radical Probabilism,'' in {\sl Probabilit\`a e Induzione ---
Induction and Probability}, edited  by P.~Monari and D.~Cocchi,
Biblioteca di Statistica, CLUEB, Bologna, pp.~263--275:
\bq
De {\Finetti}'s probabilism is ``radical'' in the sense of going all
the way down to the roots: he sees probabilities as ultimate forms of
judgement which need not be based on deeper all-or-none knowledge.
\dots

In science de {\Finetti} expected determinism to give way to
probabilism. \dots

[Physical arguments are not alone decisive.]  Still less are
philosophical arguments alone decisive.  But de {\Finetti}'s
philosophical view does see determinism as a state of mind
masquerading as a state of nature, and sees causality as a fancied
magical projection into nature of our own patterns of expectation.
Beneath the mask of determinism is a state of mind---certainty---that
is intelligible enough, but (de {\Finetti} argues) generally
inappropriate. This state of mind needs to be replaced by a more
appropriate, probabilistic one.  That's one prong in his attack on
determinism.

The other corresponds to Mach's aphorism\footnote{``In der Natur gibt
es keine Ursache und keine Wirkung.  Die Natur ist nur einmal da''
{\sl Die Mechanik in ihrer Entwicklung}, Leipzig, 1883, ch.~4, \S 4
(Die Oekonomie der Wissenschaft), par.~3; p.~455.}, ``Nature is only
there once.''  In indecision or ignorance we envision alternative
courses of events, each of which is governed by the laws of nature,
i.e.\ each is logically consistent with certain sentences in
universal form which (let us suppose) correctly describe and predict
the one actual course of events.  Causal laws are magical projections
into nature of such confident descriptions and expectations.  But it
is over the whole mental array of {\it possibili\/} courses of nature
that we distribute judgmental probabilities, giving no special
treatment to the unknown course that corresponds to reality --- much
as we might like to.
\eq

\section{25 January 1999, \ ``Oh Hibernia''}

I think my most accurate way of describing this time after {\Emma}'s
birth can be summed up with the word ``hibernation.''  On the one
hand, it feels like we get no sleep at all; while on the other, it
feels like we do nothing but sleep!  I hope that sympathy you
offered me last week is still holding steady.  I've hardly been
working at all, I must say.

I have read all the {\Galavotti} papers again.  They are marvelous.
Have you read all of them?  I'll append the whole list below.

Another paper that I think should be of {\it great\/} interest to us
is {\Schroedinger}'s ``The Present Situation in Quantum Mechanics,''
translated by J.~D. Trimmer, Proc.\ Am.\ Philos.\ Soc.\ {\bf 124},
323--338 (1980). You can also find it in the {\Wheeler}-{\Zurek}
volume. The important piece in it is his emphasis (and eloquent
explanation) that a pure state corresponds to ``maximum
information'' even though that information is incomplete.  Sound
familiar?  What he sees as the real cutting point between classical
and quantum, however, is that maximal information about the whole
generally does not determine maximal information about the parts.
This, he tries to state, is the essence of entanglement. Note, in
particular, that he does not draw the line where we usually do in
our discussions: namely, at the fact that maximal information is
complete classically but not complete quantum mechanically.  I think
this may be an important point.  And I think it is perfectly in
keeping with his pre-quantum prejudices: {\Schroedinger} believed in
physical {\it indeterminism\/} well before quantum mechanics ever
came upon the scene.  (See for instance Paul {\Forman}'s discussion
of {\Schroedinger}'s and {\Exner}'s indeterminism in his huge
article.) Thus---I think---for him, {\it classical\/} physical
description already included cases where maximal information could
not be made complete \ldots\ so of course that concept alone
couldn't be used to make the cut between classical and quantum.  I
wish he had said these things outright in his article (and not leave
me to so much detective work), but unfortunately he didn't.

By the way, I doubt Hibernia (i.e., Ireland) has anything to do with
the word hibernate, but I thought it would help make a good title.

\begin{enumerate}
\item
M.~C. {\Galavotti}, ``Comments on Patrick {\Suppes} `Pro\-pen\-sity
Interpretations of Probability'\,'' Erkenntnis {\bf 26}, 359--368
(1987).

\item
M.~C. {\Galavotti}, ``Anti-Realism in the Philosophy of Probability:
Bruno de {\Finetti}'s Subjectivism,'' Erkenntnis {\bf 31}, 239--261
(1989).

\item
M.~C. {\Galavotti}, ``The Notion of Subjective Probability in the
Work of Ramsey and de {\Finetti},'' Theoria {\bf LVII}, 239--259
(1991).

\item
M.~C. {\Galavotti}, ``Operationism, Probability and Quantum
Mechanics,'' Found.\ Sci.\ {\bf 1}, 99--118 (1995).

\item
M.~C. {\Galavotti}, ``F.~P. Ramsey and the Notion of `Chance','' in
{\sl The British Tradition in 20th Century Phi\-losophy: Proceedings
of the 17th International Witt\-gen\-stein Symposium, 14th to 21th
August 1994, Kirchberg am Wechsel (Austria)}, edited J. Hintikka and
K. Puhl (H\"older-Pichler-Tempsky, Vienna, 1995), pp.~330--340.

\item
M.~C. {\Galavotti}, ``Probabilism and Beyond,'' Erkenntnis {\bf 45},
253--265 (1996).
\end{enumerate}

\section{25 January 1999, and to {\Carl} {\Caves}, \ ``Oh Hernia''}

Just when you think things are getting better and better ... then
you stay up most all night.  Last night was not a good night.
{\Carl} sometimes says I should thank him for the hernia I got back
in '94 and the good life it brought.  Thank you {\Carl}.

\brs
I reread that section of {\Forman}'s article.  Do you have the
complete reference?  My copy of it does not give the source.
\ers
The reference is:
\bq
\noindent
P.~{\Forman}, ``Weimar Culture, Causality, and Quantum Theory,
1918--1927: Adaptation by German Physicists and Mathematicians to a
Hostile Intellectual Environment,'' in {\sl Historical Studies in
the Physical Sciences}, Vol.~3, edited by R.~McCormmach (U. of
Pennsylvania Press, Philadelphia, 1971), pp.~1--115.
\eq
A more focused study on {\Schroedinger}'s thoughts on indeterminism
is:
\bq
\noindent P.~A. {\Hanle}, ``Indeterminacy before {\Heisenberg}: The Case of
Franz {\Exner} and Erwin {\Schroedinger},'' in {\sl Historical
Studies in the Physical Sciences}, Vol.~10, edited by R.~McCormmach,
L.~Pyenson, and R.~S. Turner, (Johns Hopkins U. Press, Baltimore,
MD, 1979), pp.~225--269.
\eq
This reference has little to do with {\Schroedinger}, but while we're
on the {\Forman} thesis and I have the paper out in front of me:
\bq
\noindent
P.~{\Forman}, ``{\it Kausalit\"at}, {\it Anschaulichkeit} and {\it
Individualit\"at}, or How Cultural Values Prescribed the Character
and the Lessons Ascribed to Quantum Mechanics,'' \ldots oops I don't
have the full reference on this one.  I'll have to get back to you on
that one.
\eq

\section{01 August 1999, \ and to {\Carl} {\Caves}, ``Recording the
Vague''}

I've got a few moments (hours?) to kill while waiting in
Heathrow---we just discovered that {\Kiki}'s passport expired a month
ago---so I thought I'd use this time to record some rather vague
thoughts.

The first of my troubles has to do with the usual way of justifying
that quantum time evolutions correspond to trace-preserving
completely positive maps.  Preserving positivity and trace, I have
no problem with: we can justify this straight out with {\Gleason}'s
theorem.  But where does the strong assumption of linearity come
from?  Where does the strong assumption of {\it complete\/}
positivity come from?  The question I want to ask is whether we can
give Bayesian motivations for either of these assumptions?

Think first about linearity.  The motivation for it given by {\Kraus}
(and the other operationalists) is that quantum states correspond to
equivalence classes of preparations.  So, if we don't know the
preparation completely, we will know just as little about the time
evolved version of it as we did about the original:  the time
evolved state will be a convex combination of the individual time
evolved states.  Another way of saying this is that the set of
states forms a convex structure, and the time evolutions complement
that convex structure quite nicely.  This is their motivation for
linearity.

But from our point of view, a quantum state need not necessarily
correspond to a preparation---to use {\Carl}'s phrase, it is a
subjective entity.  Quantum states are states of knowledge
regardless as to how that knowledge actually came about.  So the
first thing we might ask is, why should the class of subjective
entities possess a convex structure?  I think we've started to find
a partial answer to that in the quantum de {\Finetti} theorem.  Can
we, by twist or turn, answer the question of linearity in a similar
fashion?

The point here is that it is only the de {\Finetti} theorem (both
classically and quantumly) that allows us to make sense of
probability distributions over subjective entities.  If linearity is
going to come into quantum mechanics via purely Bayesian means, then
it seems to me that it is going to have to come in at just this
point.

But---today at least---I'm having a hard time making this idea
concrete.  Perhaps we could play with the following sort of ideas.
Assume outright that time evolutions must map density operators to
density operators, but make no a priori assumptions about linearity,
etc.  Say that a time evolution is individual to a system (and
identical on identical systems), if whenever we have an exchangeable
sequence of density operators, the time evolved versions also form
an exchangeable sequence---that seems like a perfectly good notion
of individual time evolution to me.  Now use the quantum de
{\Finetti} theorem for both the initial and final states.  Is there
some argument we can use to ensure that probabilities in both de
{\Finetti} representations should be the same?  Perhaps by invoking
some principle about the difficulty of learning?  But then, even if
we could get that far, where do we take the argument from there?
Just something to think about.  Clearly there's a long way to go,
but I have a hard time shaking the feeling that there might just be
something here that a little clear thought can conquer.

Perhaps one more way to say it all is that maybe just maybe
linearity comes straight out of Bayes rule as applied to learning.

Next point: {\it complete\/} positivity, who ever said that entangled
states had to exist and that they must be preserved?  My trouble is
mainly this.  The usual argument for complete positivity relies on
extending the system (to an arbitrarily large one) and making the
assumption that a ``superagent'' trying to describe the extended
system must also have his valid states of knowledge (density
operators) evolve to valid states of knowledge.  But why should we
have to invoke such a superagent within our point of view?  Why
should we ever have to look outside the given system?  Why would we,
who are concerned with a fixed set of $n$-level quantum systems, ever
worry about the point of view of someone outside our little world?
(For all we know, there is no one outside our little world?) Clearly
we {\it should\/} care if we believe in the existence of the Church
of the Larger Hilbert Space \ldots\ if we were lovers of the
many-worlds point of view.  But we're not, so why should {\it we\/}
bother?

What I'm trying to get at is this.  Surely there is a
characterization of the completely positive maps that is intrinsic
to the given system.  The question is, what in our program will get
at that characterization in a natural way?  This may be a little
lame, but the only thing I can think of right now is something to do
with a Dutch book.  If there were a bookie who didn't believe in the
complete positivity of quantum evolutions, could we find a series of
bets that would lead him to a sure loss?  Maybe.  Any ideas?

OK, that's my two vague thoughts about time evolution.  I look
forward to any reactions you might have.

Finally, let me---for the benefit of {\Carl}---turn to something
that I discussed very briefly with {\Ruediger} while still in
Cambridge. What is the firm ground behind the Bayesian point of view
of QM? What is the real stuff underneath it all?  {\Carl} and I both
have had tendencies to think that it is the Hamiltonians and
interactions between systems.  But that may be troublesome.  Just as
we can ask ``What is an unknown quantum state?'' we should also ask
``What is an unknown Hamiltonian or interaction?''  The {\Kraus}
representation of a completely positive map is not unique, nor is a
unitary representation in the Church of the Larger Hilbert Space. If
{\Carl} finds the nonuniqueness of the density operator decomposition
as the most damning thing for an objective interpretation of quantum
probabilities, then it seems that he should find the same pretty
damning for the objective interpretation of quantum time
evolutions.  Where we go now with this, I'm not sure \ldots\ but as
above, it may be worth thinking about.

That's it for now.

\section{18 August 1999, \ and to {\Carl} {\Caves}, ``A Hated Phrase''}

I'm coming ever more to hating phrases like the following:
\bq
\noindent However, if we assume that macroscopic objects can be
described by quantum theory, then the orthodox view predicts that
under certain conditions, objects such as the pointers on the dials
of an apparatus do not have the property of being at a well defined
location, or more dramatically, that a cat does not have the property
of being dead or alive. This is clearly in conflict with our
observations.
\eq

Since when has such a property as ``live'' versus ``dead'' been such
a clear cut thing?  The breaking point between live and dead in the
years before cardiopulmonary resuscitation was certainly very
different from what it was afterward.  With better and better
technology, one finds that death is not so irreversible.  So, what is
it that makes ``live'' versus ``dead'' such an all-fired objective
property?  And maybe most importantly, what point of principle
distinguishes this situation from our ability to manipulate
superpositions?

I swiped the phrase above from Robert {\Spekkens}' webpage; he's one
of the modal interpretation guys I met last week.  (He and his
advisor {\Sipe}, by the way, along with a student of Leslie
{\Ballentine} were quite interested in the Q de {\Finetti} theorem.
They were the only ones beside {\DAriano} who spent a significant
amount of time trying to understand its statement thoroughly and the
method of proving it.)

Just thinking out loud.

\section{29 August 1999, \ and to {\Carl} {\Caves}, ``{\Penrose} Tiles''}

\begin{flushright}
\baselineskip=12pt
\parbox{2.8in}{\baselineskip=12pt
\small I want you \ldots\ to jolt the world of physics into an
understanding of the quantum because the quantum surely
contains---when unraveled---the most wonderful insight we could ever
hope to have on how this world operates, something equivalent in
scope and power to the greatest discovery that science has ever yet
yielded up: Darwin's Evolution.
}\medskip\\
\small --- {\it John Archibald {\Wheeler}}\\ 13 March 1998
\end{flushright}

Quantum states are states of knowledge, we say, not states of nature.
Embracing this is surely the surest path to John's vision. But, Roger
{\Penrose} and Richard {\Jozsa} counter with the opposite opinion.
Here's their argument.  From {\sl The Emperor's New Mind}, pages
268--269:
\bq
Despite the fact that we are normally only provided with
probabilities for the outcome of an experiment, there seems to be
something {\it objective\/} about a quantum-mechanical state. It is
often asserted that the state-vector is merely a convenient
description of `our knowledge' concerning a physical system---or,
perhaps, that the state-vector does not really describe a single
system but merely provides probability information about an
`ensemble' of a large number of similarly prepared systems. Such
sentiments strike me as unreasonably timid concerning what quantum
mechanics has to tell us about the {\it actuality\/} of the physical
world.  [Actually, in Richard {\Jozsa}'s version of this sentence he
substituted the word ``cowardly'' for ``timid.'']

Some of this caution, or doubt, concerning the `physical reality' of
state-vectors appears to spring from the fact that what is physically
measurable is strictly limited, according to theory. Let us consider
an electron's state of spin, as described above. Suppose that the
spin-state happens to be $|\alpha\rangle$, but we do not know this;
that is, we do not know the {\it direction\/} $\alpha$ in which the
electron is supposed to be spinning. Can we determine this direction
by measurement? No, we cannot. The best that we can do is extract
`one bit' of information---that is, the answer to a single yes/no
question. We may select some direction $\beta$ in space and measure
the electron's spin in that direction. We get either the answer {\bf
YES} or {\bf NO}, but thereafter, we have lost the information about
the original direction of spin. With a {\bf YES} answer we know that
the state is {\it now\/} proportional to $|\beta\rangle$, and with a
{\bf NO} answer we know that the state is {\it now\/} in the
direction opposite to $\beta$. In neither case does this tell us the
direction $\alpha$ of the state {\it before\/} measurement, but
merely gives some probability information about $\alpha$.

On the other hand, there would seem to be something completely {\it
objective\/} about the direction $\alpha$ itself, in which the
electron `happened to be spinning' before the measurement was made.
[Footnote: This objectivity is a feature of our taking the standard
quantum-mechanical formalism seriously. In a {\it non}-standard
viewpoint, the system might actually `know', ahead of time, the
result that it would give to {\it any\/} measurement. This could
leave us with a {\it different}, apparently objective, picture of
physical reality.]  For we {\it might\/} have chosen to measure the
electron's spin in the direction or $\alpha$---and the electron has
to be prepared to give the answer {\bf YES} with {\it certainty}, if
we happened to have guessed right in this way! Somehow, the
`information' that the electron must actually give this answer is
stored in the electron's state of spin.

It seems to me that we must make a distinction between what is
`objective' and what is `measurable' in discussing the question of
physical reality, according to quantum mechanics. The state-vector of
a system is, indeed, {\it not measurable}, in the sense that one
cannot ascertain, by experiments performed on the system, precisely
(up to proportionality) what that state is; but the state-vector {\it
does\/} seem to be (again up to proportionality) a completely {\it
objective\/} property of the system, being completely characterized
by the results that it must give to experiments that one {\it
might\/} perform. In the case of a single spin one-half particle,
such as an electron, this objectivity is not unreasonable because it
merely asserts that there is {\it some\/} direction in which the
electron's spin is precisely defined, even though we may not know
what that direction is. (However, we shall be seeing later that this
`objective' picture is much stranger with more complicated
systems---even for a system which consists merely of a {\it pair\/}
of spin one-half particles.)

But need the electron's spin have any physically defined state {\it
at all\/} before it is measured? In many cases it will {\it not\/}
have, because it cannot be considered as a quantum system on its own;
instead, the quantum state must generally be taken as describing an
electron inextricably entangled with a large number of other
particles. In particular circumstances, however, the electron (at
least as regards its spin) {\it can\/} be considered on its own. In
such circumstances, such as when its spin has actually previously
been measured in some (perhaps unknown) direction and then the
electron has remained undisturbed for a while, the electron {\it
does\/} have a perfectly objectively defined direction of spin,
according to standard quantum theory.
\eq
And, almost identically, from {\sl Shadows of the Mind}, pages
312--315:
\bq
There are many versions of the viewpoint according to which the state
vector $|\psi\rangle$ is {\it not\/} regarded as providing an actual
picture of a quantum-level physical reality.  Instead, $|\psi\rangle$
would be taken to serve only as a calculational device, useful merely
for calculating probabilities, or as an expression of the
experimenter's `state of knowledge' concerning a physical system.
\ldots\ Sometimes, it is even argued that $|\psi\rangle$ {\it
cannot\/} describe a quantum-level reality since it makes no sense
at all to talk of a `reality' for our world at that level, reality
consisting only of the results of `measurements'. \ldots\

Yet, if we accept that there must be a reality of some kind that
holds at the quantum level, we may still have doubts that this
reality can be accurately described by a state vector $|\psi\rangle$.
There are various arguments that people raise as objections to the
`reality' of $|\psi\rangle$. \ldots\ [He lists a few arguments here.]

All this is true, yet it remains hard to take the opposite position
either:  that the state vector $|\psi\rangle$ is somehow physically
`unreal', perhaps encapsulating merely the sum total of `our
knowledge' about a physical system. This, I find very hard to accept,
particularly since there appears to be something very subjective
about such a role of `knowledge'. {\it Whose\/} knowledge, after all,
is being referred to here? Certainly not mine. I have very little
actual knowledge of the individual state vectors that are relevant to
the detailed behaviour of all the objects that surround me. Yet they
carry on with their precisely organized actions, totally oblivious to
whatever might be `known' about the state vector, or to whomever
might know it. Do different experimenters, with different knowledge
about a physical system, use different state vectors to describe that
system?  Not in any significant way; they might only do so if these
differences were about features of the experiment that would be
inessential to the outcome.

One of the most powerful reasons for rejecting such a subjective
viewpoint concerning the reality of $|\psi\rangle$ comes from the
fact that whatever $|\psi\rangle$ might be, there is always---in
principle, at least---a {\it primitive measurement\/} whose {\bf YES}
space consists of the Hilbert-space ray determined by $|\psi\rangle$.
The point is that the physical state $|\psi\rangle$ (determined by
the ray of complex multiples of $|\psi\rangle$) is {\it uniquely\/}
determined by the fact that the outcome {\bf YES}, for this state, is
{\it certain}.  No other physical state has this property.  For any
other state, there would merely be some probability, short of
certainty, that the outcome will be {\bf YES}, and an outcome of {\bf
NO} might occur. Thus, although there is no measurement which will
tell us what $|\psi\rangle$ actually {\it is}, the physical state
$|\psi\rangle$ is uniquely determined by what it asserts must be the
result of a measurement that {\it might\/} be performed on it. This
is a matter of counterfactuals again, but we have seen how important
counterfactual issues actually are to the expectations of quantum
theory.

To put the point a little more forcefully, imagine that a quantum
system has been set up in a known state, say $|\phi\rangle$, and it
is computed that after a time $t$ the state will have evolved, under
the action of {\bf U}, into another state $|\psi\rangle$. For
example, $|\phi\rangle$ might represent the state `spin up'
($|\phi\rangle=|\!\uparrow\rangle$) of an atom of spin $\frac{1}{2}$,
and we can suppose that it has been put in that state by the action
of some previous measurement. Let us assume that our atom has a
magnetic moment aligned with its spin (i.e.~it is a little magnet
pointing to the spin direction). When the atom is placed in a
magnetic field, the spin direction will precess in a well-defined
way, that can be accurately computed as the action of {\bf U}, to
give some new state, say $|\psi\rangle=|\!\rightarrow\rangle$, after
a time $t$. Is this computed state to be taken seriously as part of
physical reality? It is hard to see how this can be denied. For
$|\psi\rangle$ has to be prepared for the possibility that we {\it
might\/} choose to measure it with the primitive measurement referred
to above, namely that whose {\bf YES} space consists precisely of the
multiples of $|\psi\rangle$.  Here, this is the spin measurement in
the direction $\rightarrow$. The system has to know to give the
answer {\bf YES}, with {\it certainty\/} for that measurement,
whereas {\it no\/} spin state of the atom {\it other\/} than
$|\phi\rangle=|\!\rightarrow\rangle$ could guarantee this.
\eq

How do we move past this?  I am coming more and more to the opinion
that the solution lies in discarding the term ``measurement'' for
labeling what it is that occurs in the ``quantum measurement
process.'' The problem is that the word ``measurement'' conveys too
much of a confusing sense that the purpose of the measurement outcome
is to reveal something that {\it is\/} (or at least {\it was\/})
``there'' independently of our having looked.  I think a much better
terminology---one that may start to crack this problem---can be built
from the terminology used in L.~J. {\Savage}'s book.

Decisions, Acts, and Consequences.  Let me introduce this in a rather
long-winded way by quoting some long passages that have had a
significant influence on me.  The first is a piece from Doug
{\Bilodeau}'s paper ``Why Quantum Mechanics is Hard to Understand,''
(not yet published and apparently not going to be, {\tt
quant-ph/9812050}): [See note to  {\Asher} Peres, titled ``The Deep
Intervention,'' dated 26 December 1998.]

The second selection is an old article that I will quote in full.  It
is by Markus {\Fierz}: ``Does a physical theory comprehend an
`objective, real, single process'?'' from {\sl Observation and
Interpretation in the Philosophy of Physics}, edited by S. K\"orner
(Dover, NY, 1957), pp.\ 93--96.

\bq
\begin{center}
Does a physical theory comprehend an `objective, real, single
process'?\\ by M. {\Fierz} \medskip
\end{center}

\small The so-called `orthodox' interpretation of quantum theory has
been criticized in recent years from different sides, by {\Einstein},
{\Schroedinger}, and de {\Broglie}. These critics think that the
ideas about quantum theory put forward by {\Bohr} and others are
much too far away from those points of view that have been
successful and productive in the development of our science for the
last three hundred years.  They claim, against the `orthodox' or
`Copenhagen' school, that wave-mechanics is an incomplete theory,
even inside its field of application, because this theory does not
comprehend really an `objective and real single physical process'.

Now everybody acknowledges the great success of wave-mechanics in
explaining quantitatively a multitude of physical phenomena. And
nobody denies that our theory, as it stands, is incomplete, as there
is no mathematically irreproachable relativistic generalization. Just
because of this the question arises, in what direction we should
search for a better relativistic theory. It seems to us that our
`heterodox' colleagues think it might be helpful to look for a theory
which can be interpreted more akin to classical mechanics or field
theory. Surely it is not their aim to translate only the physical
contents of quantum theory as it stands, into a language different
from the one of {\Bohr}.

I think, indeed, that the ideas developed by {\Bohr} during the
growth of quantum theory will not lose their leading character. I
further expect that the new features characterizing {\Bohr}'s way of
thinking, will even be more dominant in a new and better
relativistic quantum theory. Such a theory will lead to physical
ideas even more different from those of old.

Only the future can decide if I am right or wrong. At the moment
there is no possibility of proof, as no such a theory has been
discovered. But we can try to justify our point of view by plausible
arguments. To do so, we may ask: does classical theory describe
really an `objective and real single process'? Our critics would
answer this question in the positive---at least I think so, as they
are in quest of a `realistic' theory.

If one takes `real' as opposed to `ideal'---against this one could
argue with good reasons indeed---one might say: a physical theory
never describes real events. It treats always idealized systems, as
only these are capable of mathematical treatment. One may however,
claim the idealization to be a simplification only, not referring to
any essential features of the system under consideration. But the
distinction between the essential and the inessential already
contains a theoretical element; and theory itself seems to me
something ideal. I don't want to pursue further the questions
following from this consideration. This would lead us to rather
general problems, not directly linked to our special question. This
question is: does classical physics describe or comprehend single
physical events? It does indeed, but only in so far as these can be
looked at as an example, contained in a class or ensemble of similar
events. The judgement that such events are similar and form a class
is naturally always a theoretical one. And by this judgement some
features of a given event must be taken as inessential. To these
so-called inessential features now belong those which make our event
a single and unique one, showing up here and now. So, in some sense,
physical theory never comprehends a real single process.

To make this statement clearer, I refer to {\Kepler}'s theory of the
solar system, a theory belonging to a pre-physical state of science.
{\Kepler} was convinced that the sun is the centre of the universe.
He held the universe to be a singular harmonic structure, a cosmos,
contained in the sphere of the fixed stars. The infinite space of
Giordano Bruno, containing an infinity of worlds, seemed to him to be
a horrible exile. His own theory was meant to explain the one
harmonic universe. He wanted to understand why there are just 6
planets, circling at well defined distances around the sun. By
constructing, in a most ingenious way, the spheres of the planets
with the help of the 5 regular or `platonic' bodies, he thought he
had reached his goal. The order in which he had to arrange the bodies
seemed to him to be mathematically unique, and as there are just five
of them, he claimed to have explained the uniqueness of our world.
This still remarkable theory is very beautiful, but we no longer
believe in it.

In {\Newton}'s theory, which we think to be right, the solar system
is just one example taken from the innumerable systems of planets
created by God in infinite space. {\Newton} had no ability, nor did
he feel any need, to explain the number 6 of our planets---and there
are really more than 6.

So {\Newton}'s theory just does not achieve what {\Kepler} held to
be of utmost importance for a theory of the solar system. It
foretells neither the number of the planets nor their distances from
the sun. There is also no longer any essential difference between
planets and comets, a difference which was very marked in all old
theories of the solar system.

The strength of {\Newton}'s theory is not to explain the unique
structure of this system, but to comprehend in general terms the
movement of any system of planets, independently of the number and
the bulk of the celestial bodies contained in it.

Classical mechanics does not give equations for a definite movement
of the masses in a system, but for all their possible movements. The
real movement taking place depends on initial conditions,
corresponding to the constants of integration of differential
equations. These initial conditions have to be known experimentally.
From the theoretical point of view they are arbitrary. If there were
a law stating only one initial condition to be possible, the theory
would lose much of its sense, as `almost all' of its statements would
refer to something impossible. There would be no reason either to
assume the only path possible to be embedded in a family of curves,
fulfilling certain differential equations of second order. I think
the theoretical arbitrariness of the initial conditions, the fact
that they have to be given experimentally, is an essential feature of
classical mechanics. It shows this theory to be one belonging to an
experimental science, where the experimenter has the freedom to
interfere with a system and may form an initial state corresponding
to his aims.

Against this, one may point out that there are systems, as the system
of planets, where we cannot change the initial state. That's true.
But the laws ruling over the movements of the planets are the same as
those ruling over the fall of a stone on earth. These we can explore
experimentally, and we can verify them with different examples. One
is the solar system, in which Jupiter with his moons can be looked at
as a second small and independent one. Similarly we are able to
understand the processes in the interior of the sun, where we can't
do any experiments. They can be explored with the help of terrestrial
experiments, because these processes are not peculiar to the sun. By
this again the sun becomes but an example of a system, where such
processes take place. It seems to me, that all this was clear to
{\Newton}, when he wrote his second {\it regula philosophandi}:
\begin{quote}
Ideoque effectuum naturalium eiusdem generis eaedem sunt causae:
descensus lap\-i\-dum in Europa et America, lucis in igne culinari et
in Sole, reflexionis lucis in terra et in planetis.
\end{quote}

From our point of view, the idea of {\Laplace} that the whole world
is nothing but a huge mechanical machine seems rather queer. In this
picture there is no room for an experimenter as all initial
conditions are given for ever. This is quite contrary to {\Newton}'s
outlook, who even in his cosmological speculations assumed God to be
the great experimenter, who changes from time to time, according to
his purpose, the state of the world. In {\Laplace}'s view, we cannot
take the whole world as an example of a possible world, as by
definition there doesn't exist any other one. The situation becomes
quite different indeed if we base our cosmology on the assumption
that the world is homogeneous in the large scale. If this is done,
every region big enough can be looked at as representative of the
whole world. A given region, for example the one we are able to
survey, can be taken as an example for any region. In such a theory
everything distinguishing one region from another must be looked at
as accidental, and so in some sense as inessential. Thus such a
theory needs to be a statistical one.

We may sum up all this with the statement that physical laws refer to
reproducible phenomena only.  Correspondingly, an experiment is
meaningful only if it is reproducible. Although physics is not bound
to experimentally reproducible events, it can only treat phenomena
where every single event can be taken as representative of an
ensemble or class of similar events.

Now in some sense every real event is something single and unique and
happens never again. But this feature of reality is outside physical
theories. As this is true for classical theory, we should not be
astonished, if it is more so in quantum theory. This is a statistical
theory, and as such very well fitted to treat reality, in so far it
consists of reproducible phenomena. Therefore it is a logical and
natural development. The dominant role the experimenter or observer
plays here, if we interpret the theory as {\Bohr} does, clarifies the
general feature of physical theories. To me this seems definite
progress.

If somebody wants to construct a theory, comprehending the
unreproducible reality of a single event, and this seems to me the
only other consequent alternative, he should not only discard
{\Bohr}'s ideas on what a physical theory can be, but even those of
{\Newton}. He should look for a theory which in its whole character
is more akin to {\Kepler}'s ideas---and even {\Kepler}'s theory, as
he himself understood it, is highly platonic!

I do not believe that in such a way the actual problems of
relativistic quantum theory come nearer to a solution.
\eq

The point of both stories is that it is our ``acts'' and their
``consequences'' that are the ground upon which scientific theories
are built.  Only with hindsight, it seems to me, can one {\it in
certain circumstances\/} distill the idea of a ``measurement'' of a
property intrinsic to the system acted upon.  Our acts and their
consequences are the iron posts of scientific reality; the intrinsic
properties they ``reveal'' is just so much papier mach\'e we place in
between (when we can).

The wonderful lesson of quantum mechanics is that it brings these
points to life.  When confronted with a quantum system, we can {\it
decide\/} to act upon it in one way or the other in any of a number
of nonequivalent ways.  The set of acts available to us is the set of
positive operator valued measures:  Each POVM corresponds to an {\it
act\/} we might perform on the system.  The set of possible {\it
consequences\/} associated with an act corresponds to the separate
operators within the act's affiliated POVM.

{\Savage}, in his development, goes further to tip his hat to
``states of the world.''  These, when combined with the acts, lead
to the consequences.  But it doesn't seem to me that this notion is
overly needed to get at a system much like the one he develops:
What role do the ``actual'' states of the world play in our
decisions? None, it seems.  ({\Ruediger}, by the way, has been
wanting to push this point of view for quite some time in the
context of many worlds.  But there it seems doomed to failure to
me:  The multiverse is not only a world void of states; it is a
world void of decisions, void of acts, and certainly void of
consequences.)

So we are left with the point of view that quantum measurement is
about decisions, acts, and consequences.  What is a {\it quantum
system}?  It is nothing more than a conduit from acts to
consequences. It takes the place of the ``states of the world'' in
the {\Savage} system, but it is a black box that need not be cracked
(and appears not to be crackable).  A {\it quantum state\/} is an
``if-then.''  It is the compendium of probabilities that summarizes
our beliefs about the consequences of our acts.

Let me return to {\Penrose}.  The point is that from this view the
quantum state is specifically {\it not\/} a property intrinsic to the
quantum system.  It is a statement about what we can expect of the
consequences to our actions.  If you want to call it a ``property,''
then it seems to me at best it is a {\it relational\/} property.

I don't think it is unsafe to liken this situation to the
cryptographic problem of remote coin tossing.  Alice places either
bit of her choosing, 0 or 1, in a locked safe and sends it to Bob.
Bob, after receiving the safe, publicly announces whatever bit he has
in mind. Alice then relinquishes the key to her safe. The outcome of
the official ``coin toss'' is the XOR of the two previous bits.  How
would {\Penrose} interpret this process?  Is the outcome of the coin
toss an {\it objective\/} property of the bit in Alice's safe? Would
he answer {\bf YES}\@? I don't think he would, and I am left
wondering why he would do otherwise for the quantum state.

Decisions.  Acts.  Consequences.

\section{22 September 1999, \ ``Andrei''}

My present opinion---the one that has really evolved since we last
met---is that there is probably no choice in the classical equations
of motion.  That is, once one fixes that the subject of discourse is
about the evolution of probability distributions over a phase space.
Then to say that one has maximal knowledge about the evolution is to
say that the distinguishability of any ``unknown probabilities''
neither increases nor decreases.  One then has that the overlaps
between any two phase space distributions must be constant.  And
presumably---but this is a research program, not a definitive
statement---an analogue of {\Wigner}'s (quantum) theorem on
symmetries will step in to save the day.  The only ``flows'' with
that property (connected to the identity) are ``unitary'' flows. And
these correspond to the standard Hamiltonian dynamics that we all
know. That's the idea anyway.  (This, by the way, has evolved from
my starting to think of classical physics as well as quantum physics
as a ``law of thought.''  We have become accustomed to thinking of
classical physics as more ``ontological'' than its quantum
counterpart \ldots\ which we have started think of as more
``epistemological.''  But I now think that was just wrong-headed.
{\Kant} was a pretty smart man.)

\section{17 June 2000, \ ``Quantum Pedophiles''}

I'm reading this article of {\Jaynes} (``Predictive Statistical
Mechanics'') that I wrote you about and I'm realizing that actually
I had never seen it before.  This is the first time I'm reading it,
and it (or at least the first seven pages of it) is wonderful!

The astonishing thing is that the program he pushes for in this
paper is significantly similar to the one I've been pushing for:  an
information--disturbance foundation for quantum mechanics.  The main
difference is that he hasn't yet consoled himself to the idea that
``maximal information cannot be made complete (even conceptually)
because of the strength of the disturbance.''  (The part of the thing
in quotes following the parenthesis is the little spin that
distinguishes my hopes and dreams from {\Carl}'s.)

Remember what {\Bennett} wrote me:
\bq
\noindent
Probably to one uncorrupted by many worlds, your idea seems
perfectly sensible and beautiful.  My second thought is the
realization is that you encountered {\Jaynes} in your formative
years, at around the mental age I was when I encountered whatever
makes me loyal to my Church.  In other words, in both cases, we
encountered what might be described as a pedophilic idea, an idea so
seductive as to be dangerous to our youthful selves and make us
dangerous to others when we grew up, or were thought by others to
have grown up.
\eq
Until today I hadn't realized how true that might be.

\section{03 September 2000, \ ``One Sentence?''}

What could I do to entice you to carefully translate just one
sentence for me?  It's below, you can't miss it.  The passage is
from: C.~P. {\Enz}, ``Quantum Theory in the Light of Modern
Experiments,'' in {\sl Advances in Scientific Philosophy:~Essays in
Honour of Paul Weingartner on the Occassion of the 60th Anniversary
of his Birthday}, edited by G.~Schurz and G.~J.~W. Dorn (Rodopi,
Amsterdam, 1991), pp.~191--201.

\bq
\indent
I hope to have conveyed with the above discussion of experiments the
{\Schopenhauer}\-esque feeling for the {\it freedom\/} of the quantum
objects for potentially {\it being\/} both, particle and field,
located and moving, in phase and numbered, \ldots, before facing the
{\it necessity\/} of being determined by the {\it action\/} of the
measurement.  Indeed, one of the new ideas {\Schopenhauer} proclaimed
is this:  ``Ich hingegen sage: jedes Wesen, ohne Ausnahme, {\it
wirkt\/} mit strenger {\it Nothwendigkeit}, dasselbe aber {\it
existiert\/} und ist was es ist, verm\"oge seiner {\it Freiheit}.''
But this new thinking became possible only through an ever finer
subdivision of the objects of science or, to let {\Pauli} speak once
more: ``Science is a systematic refinement of the concepts of
everyday life revealing a deeper and [\ldots] not directly visible
reality behind the everyday reality of the colored, noisy things.''
\eq

\subsection{{\Ruediger}'s Reply}

My pleasure:
\bq
\noindent I, however, say: every being, without exception, {\it acts\/}
with strict {\it necessity}, but {\it exists\/} and is what it is, by
virtue of its {\it freedom}.
\eq

\section{05 April 2001, \ ``Waking in the Hundred Acre Wood''}

Thanks for the vote of confidence.  I needed that first thing this
morning.  (But I guess insecure people always need that, no matter
what the timing.)

I'll fax you the figures as soon as I get in to the office this
morning.

Thinking of von Neumann measurements as more special than other
POVMs carries a certain anti-Bayesian danger.  Not least of all by
laying more credence on the Church of the Larger Hilbert Space point
of view than I prefer to do:  it says there is something special
about extending the Hilbert space; that we have a necessary reason
to do it, even if only to understand measurement.  (One can write
incomplete sentences in email without my wrath.)  But also, I am
reluctant to take the popular expo of POVMS as a necessary thing
because of this way of viewing quantum-state change as a
generalization of Bayes rule (that I've been promoting).  You can
read about it in my NATO paper when it's finished (hopefully by the
end of the month), or we can talk about it in Sweden.

From still another front, I try to make the point sharp in my
talks:  I put up a slide that has two columns.  On the left-hand
side there is a list of various properties for von Neumann
measurements.  On the right-hand side, there is an almost identical
list of properties for POVMs.  The only difference is that it is
MINUS the orthonormality condition required of a von Neumann
measurement.  Then I have an arrow pointing up to the orthonormality
condition on the von Neumann side with this caption:  ``Does the
addition of this one extra assumption really make the process of
measurement any less mysterious?''  I follow that by saying, ``I
imagine myself teaching quantum mechanics for the first time and
taking a vote with the unconditioned students.  Which set of
postulates for measurement would you prefer?  [Pause.]  They'd only
look at me with blank faces.  And that's the point!  It'd make no
difference to them, and it should make no difference to us.''

Damn, now I spoiled my talk in Sweden!

\chapter{Letters to Robert {\Schumann}}

\section{22 February 2000, \ ``Your Project''}

(Please remind me of our conversation in Italy.  I remember your
name, but I can't quite picture you.)

Your proposal sounds interesting:  it wasn't too bold.  Why don't we
carry on just as you suggested.  I'd be happy to supervise a project
of yours (or perhaps collaborate) \ldots\ as long as the work
concerns something that interests me.  Presently most of my interest
is in quantifying different aspects of the trade-off between
information and disturbance in quantum mechanics.  This includes
deriving various properties for some information functions that have
little coverage in the existing literature so far.  If we could find
something that pleases us both in that general area, I think it'd be
great.  (Presently I'm not too interested in tackling entanglement
questions:  I can't see the forest for the trees in that subject.)

About collaborations, I only have one hard and fast rule at the
outset:  that is that author ordering be alphabetical.  I place my
form letter concerning this issue below.  If you feel that this
would hurt you, we can agree to keep MY role in the project in a
strictly supervisory capacity.

Let me know what you think:  I'm always within email contact.  (I'm
in Haifa, Israel for the week.  Next week I will be back in New
Mexico.  Then a week following that I will be in Japan.)

\section{23 February 2000, \ ``OK''}

Now that you've prodded my memory, I remember you quite well.  I'm
sorry that I had forgotten.

I'm glad to hear that you wouldn't mind thinking about information
vs.\ disturbance issues.  My ultimate hope is to use them for a
better grounding of quantum foundations, but there are plenty enough
practical things (like quantum cryptography) that they are connected
to that the wider community will still have respect for the work.

Two possible projects come to my mind presently.  One has to do with
extending our knowledge of a quantity that {\Bill} {\Wootters} calls
the subentropy.  This is because I want to use it as a
quantification of ``how much'' one knows, when one possesses a
density operator.  From there I plan to use it to explore how much
``information'' {\it I\/} lose about a system when {\it you\/} learn
something about it (via a measurement).  I already have preliminary
results showing that this will be a quite interesting
investigation.  The practical implication of this question is that
it can also tell us something about controlling quantum systems (a
kind of quantum control theory).

The other project has to do with quantifying how ``quantum'' a set of
quantum states are with respect to each other.  In particular I
quantify it by a game related to quantum cryptography:  how good is
the best intercept-resend strategy.  (The is closely connected to
something I call the ``minimal disturbance at maximal information.'')
What is interesting about the measure I propose is that it {\it
appears\/} that for a Hilbert space of dimension $d$, one can always
find a set of $d^2$ states that is just as quantum (with respect to
each other) as they can possibly be.  I.e., one does not have to
make use of the full Hilbert space of quantum states to see maximal
quantumness.  I have just a load of open questions about this new
information function.

In the case of the first project, the two things to read are:
\begin{enumerate}
\item
R.~{\Jozsa}, D.~{\Robb}, and W.~K. {\Wootters}, ``Lower bound for
accessible information in quantum mechanics,'' {\em Physical Review
A}, vol.~49, pp.~668--677, 1994.

\item
W.~K. {\Wootters}, ``Random quantum states,'' {\em Foundations of
Physics}, vol.~20(11), pp.~1365--1378, 1990.
\end{enumerate}
Especially the first of these papers.  What I mean by ``reading'' a
paper is working through every step (with pencil and paper), making
sure that you understand exactly what they are doing.  I estimate
that it will take about a week (or possibly two) of real work to get
through these papers in that kind of way.  As I recall, that is how
long it took me, and I was at the same academic level as you.

In the case of the second project, the only thing I have in print so
far is:  C. A. Fuchs, ``Just Two Nonorthogonal Quantum States,''
{\tt quant-ph/9810032}.  This unfortunately is a sort of review paper
with no detailed calculations.  Also the direction I'm going in now
is to generalize that concept to $n$-state ensembles.  I do, however,
have 23 pages of notes and calculations prepared already on this
project.  I think they're pretty readable already.  If you will send
me mailing address I will have them shipped express to you as soon
as I get back to Los Alamos early next week.

I have picked both projects so that only a minimal amount of reading
is required.  (That is because in both cases almost no literature
exists.)  So it is just a case of understanding a minimal amount of
stuff, and then jumping into the research waters yourself.  Once you
understand both these areas, I'll send you a list of open questions,
ideas and suggestions.  Then we can sift through that and see where
to go next.  I'd be more than happy to roll up my sleeves and
contribute to either project as you develop it.

Also could you send me a fax number?  I may be able to fax you a
subset of the notes from here.  Finally, can you read and write
\LaTeX?  When I collaborate long distance, I most always write my
notes in \LaTeX\ so that formulas can be displayed easily.

I'm happy to hear that you've gotten {\Bill} {\Wootters} thesis.  In
my opinion there's no deeper thinker in our field than {\Bill}. (And
I think that thesis was really on the right track:  it would be
wonderful to get quantum mechanics out of a maximization principle.
I think the only thing {\Bill} was missing is that it will have to do
with both information AND disturbance.)  By the way, to get to know
me a little better philosophically---and the direction I'll be
sending you, but in a more technical way---you might {\it skim\/}
some of the stuff I've posted at my website.

\section{20 March 2000, \ ``Ontology''}

I see you found my letter ``Fuchsian Genesis'' but somehow you
didn't absorb the main point of it.  I do want ontology, just as
much as the next guy.  I just don't want a cheap ontology.  I think
quantum mechanics is giving us the grand opportunity to lift the
veil a bit, to see how alive and changeable the world really is.
Maybe look over that letter once again, but also have a look at:
\bq
\noindent
Letter to {\Landauer}, 6 July 1998 \\
Letter to {\Landauer}, 28 July 1998 \\
Pages [$X$]--[$X+1$], starting at Merminition \ref{OldAsm10} \\
Letter to {\Bill} {\Wootters}, 2 July 1999.
\eq

The project I intend to have you working on, as I see it, is one of
quantum ontology:  refining the tools (in your case the subentropy
function) that will help us lift the veil even further.  I want to
understand the subentropy's properties better to see what they can
better tell us about the tradeoff between information gain and
information loss by various observers in a measurement situation.
It is an epistemological tool that will help us say something about
an ontological property of the world:  the world is sensitive to our
touch.

You were right that I do want a principle.  But I don't think that
precludes ``inferential power.''  Just the opposite.  The way I see
it is that once we pin down the principle behind quantum mechanics,
then the floodgates to new science will open before us.  I like to
liken our situation with quantum mechanics to that of Lorentzian
mechanics five or so years before {\Einstein}.  The Lorentz
transformations (as listed on the back cover of Taylor and
{\Wheeler}'s book, for instance) existed---that's why they're called
the Lorentz transformations.  But the principle they captured was a
mystery. When {\Einstein} finally saw a principle---that spacetime
is a manifold---everything changed.  This was not because it
immediately gave much more inferential power, but because it laid
the groundwork for the much greater realization to come, the general
theory of relativity.  {\Einstein} needed to see that spacetime was a
manifold first before he could even imagine asking whether it was a
curved manifold.  We too need to see that quantum mechanics hints
that the world is a \underline{\phantom{substrate}} before we can
contemplate asking whether it is a \underline{\phantom{malleable}}
\underline{\phantom{substrate}}. Whatever this great principle is,
it's likely to be right here in front of us just waiting that we fix
our attention upon it. When we break the mental barrier and see it,
the world will change.

By the end of the week, I will send you a concrete project that I
think you should go through the motions of.  It'll be based on a
paper that Kurt {\Jacobs} and I are writing presently.  It is to
explicitly work out a relation for the gain and loss of subentropy
for a two-level system.  It is the sort of thing that I know can be
done---I could do it myself if I had sufficient stamina---and so, at
the very least, will make you feel that you've done something real.
(I once had a quite satisfying summer digging post holes for a gas
pipeline company.  There was no question that at the end of the day
I had done something.)  Also, I will collect the book by Aczel and
Daroczy before departing for Scotland and start to think more deeply
while I am there whether we can give some classical sense to the
subentropy.

See further poetry below.  [See note to {\Greg} {\Comer}, titled
``Rabbits on the Moon,'' dated 20 February 2000.]

\section{20 February 2000, \ ``Interventions''}

You might also have a look at the article  {\Asher} {\Peres} and I
wrote for {\sl Physics Today\/} (March issue).  It comes out sounding
slightly more negative than I had hoped for---between fighting with
 {\Asher} and fighting with the editors over the appropriate choice of
words, I finally just gave up.  But the message is there between the
lines if you look closely.  I say, give it up if what you're looking
for is a free-standing reality independent of our interventions:
we're part of the world and we're learning that we have to take that
into account in a deep way.

By the way, I never adequately said anything about {\Jaynes}.
{\Charlie} was right, he was quite an influence on me.  I'm pretty
Bayesian to the core now when it comes to probability theory.  Where
{\Jaynes} and I differ though is in what we think the ignorance can
be about when one writes down a probability function.  {\Jaynes}
explicitly thought that probability could only concern the truth
value of a hypothesis.  This led him to think that there must be
some kind of hidden variables underneath quantum mechanics.  I, on
the other hand, am willing to attach probabilities to the
consequences of our interventions into the course of Nature.  Those
consequences have no truth value before the intervention (and maybe
not even after it). Just because one has ignorance, it does not
follow that it can always be relieved (as {\Jaynes} had hoped).

\section{20 February 2000, \ ``The Blank''}

\brschu
Essentially, talking about disturbance reminds me of {\Bohr}'s
rhetoric, and indeed your position sounds like his because you don't
have the \underline{\phantom{substrate}} in which quantum theory
takes place.
\erschu

OK, I'll fill the blank:
\bq
\noindent {\bf nou$\cdot$me$\cdot$non} (`n\"ume,n\"an) {\it noun} \\
{\it plural\/} {\bf nou${\cdot}$me${\cdot}$na} (-ne) \\
{\it Philosophy}. \\
{\bf 1.} An object that can be intuited only by the intellect and not
perceived by the senses. \\
{\bf 2.} An object independent of intellectual intuition of it or of
sensuous perception of it. Also called thing-in-itself. \\
{\bf 3.} In the philosophy of {\Kant}, an object, such as the soul,
that cannot be known through perception, although its existence can
be demonstrated.
\eq

The noumenon, whatever it is, is malleable, changeable, utterly
sensitive to the touch.  So much so, that the only grasp we can get
on it necessarily involves information, unadulterated agent-centered
information.  (Information is, by definition, not an ontological
entity.)  That's the direction I want to go anyway, at least until
someone shakes some better sense into my head.

It's indeed probably better to focus on a little mathematics today.
Do you completely understand the derivation of Eq.\ 3.273 in my
thesis?  Also the manipulations between Eqs.\ 3.284 and 3.288? You'll
have to be able to do manipulations like that for the project I'll
propose to you by the end of this week.

\section{25 August 2000, \ ``Start of the Day''}

\brschu
My general feeling is to agree with the Bayesian view because of its
consistency -- although it leaves a hollow feeling in someone who,
deep down, is a realist. (Can we use a subjective interpretation in
understanding the physical universe?)
\erschu

Deep down, I am a realist, though one of a strange kind of sort
(more like {\Schopenhauer}).  The thing is not to confuse probability
and reality; the two things are different.  Bayesians are pretty
good about this.  But most of the time they carry that forward in
too naive a fashion.

Anyway, I say this more eloquently below.  [See note to David
{\Mermin}, titled ``Great {\Garrett} Quote,'' dated 23 July 2000.]
Today, I spend the whole day with your thesis.

\chapter{Letters to Abner {\Shimony}}

\section{16 December 1999, \ ``Evolutionary Laws Paper''}

I obtained this email address for you from  {\Asher} {\Peres}.  I
have seen in an advertisement that you have written an article titled
``Can the Fundamental Laws of Nature Be the Results of Evolution?''
I'm quite interested in that subject.  If possible, I would like to
obtain a copy of your paper.  You can either send a hard copy to my
address below, or simply email a file of it (I can read \TeX, \LaTeX,
PostScript, MS Word). \medskip

\noindent [NOTE:  The updated reference to that paper is  A.~{\Shimony},
``Can the Fundamental Laws of Nature Be the Results of Evolution?,''
in {\sl From Physics to Philosophy}, edited by J.~{\Butterfield} and
C.~Pagonis (Cambridge U. Press, Cambridge, 2000), pp.~208--223.]

\section{20 January 2001, \ ``Nonontology''}

I'm writing to invite you to a meeting and to remind you of a
challenge you once presented (in the hope that the latter might help
entice you to come).  All the information about the meeting is
further below.  That's the easy part.

About the challenge, I refer to something you wrote in 1978:  [See
note to {\Jon} {\Waskan}, dated 11 May 1999.]

Well, opposing you, I do find a nonontological interpretation of
quantum mechanics quite appealing.  On the other hand, I think you
were absolutely on the mark with your statement, ``I do not believe
that a fully worked out and coherent formulation of a nonontological
interpretation ... exists in the literature.''  Moreover, I don't
think much has been done to fill the gaps in the 22 years since you
wrote that.  BUT, as David {\Mermin} will tell you, I am a mad
optimist, and I believe in my heart that we're just on the verge of
getting there.  The missing link all this time has been that we
didn't have the proper tools for EVEN asking the proper questions.
That, however, I believe is changing with the advent of quantum
information theory and quantum computing.  And thus my involvement
in this meeting and the particular list of invitees you'll find
below.  (Also in a separate note, I'll forward the invitation I
wrote to Henry {\Folse}.  I think this will give you a better feeling
about what I'm up to.)

I do hope you can join us.  I want to crack this problem you posed
so long ago, and I think your attendance would be just wonderful in
that regard.

\section{04 April 2001, \ ``Using {\Shannon}
to Build on {\Bohr}''}

\bas
I am puzzled by your remark that you would read the paper that I sent
you. What paper was that?
\eas

I was referring to the paper you recommended in this passage:
\bas
You have a good memory and an even better temperament, to continue
to be optimistic about your nonontological program. I hope to discuss
the matter with you, in Sweden or elsewhere, but I suggest strongly
that before we meet you read something relevant that I've written
since 1978:  ``Reality, Causality, and Closing the Circle'', in vol.\
1 of my collection of essays  entitled {\sl Search for a Naturalistic
World View\/} (Cambridge U. Press, 1993).
\eas

You later sent it to me in the mail.  I have absorbed most of it,
but will certainly give it another read before June.

\bas
I have no plans to write a paper especially for the conference at
present, though I should be happy to send old papers that might be of
interest to you and other participants.
\eas

I had only meant that we would try to find time to discuss the paper
above during the meeting.  If you have any other papers that are
appropriate, please send them to me and I'll see to it that they
make it to onto the display table with the other
preprints/reprints.  That's what I really meant by being a virtual
participant.

The main thing I want to explore the prospects for is captured in
the original note I wrote you concerning V\"axj"o.  I'll place that
note below so you can read it again.  I was especially intrigued by
your remarks about a {\it similarity\/} between the programs of
{\Kant} and {\Bohr}.  In my own mind, I'm even more intrigued by the
possibility of a {\it similarity\/} between the programs of
{\Schopenhauer} and {\Bohr} \ldots\ and that's what I want to flesh
out. It seems to me that quantum mechanics does give us a little
knowledge of the thing-in-itself. But that thing-in-itself is not
the state vector, nor is it some set of hidden variables:  Rather it
is that {\it property\/} of quantum systems which keeps us from ever
knowing more of how they will react to measurement situations than
can be captured by a state vector. It is the thing that shunts us
away from the potential to even theorize a reasonable
hidden-variable theory. That property I view as something active,
interesting, and good. Giving some mathematical substance to it is
what I'm about, and precisely where I think the tools of quantum
information will be most handy.  As I see it, most of quantum
mechanics---but not all---is about the optimal processing of
subjective information (not so unlike Bayesian probability) in light
of that fundamental situation (imposed by the world).

I hope that helps you in the selection of papers you might send.

\chapter{Letters to {\Jon} {\Waskan}}

\section{24 April 1999, \ ``Thinking {\Kant}ian Thoughts?''}

The other day I wrote the little essay below [cf.\ ``Fuchsian
Genesis'' addressed to {\Greg} {\Comer}, dated 22 April 1999] for
myself, and then searched up a reason to send it to my friend {\Greg}
{\Comer}.  It would have been better to send it to you I think.  It
strikes me as sounding a little {\Kant}ian, though with an extra
twist: namely, that we are even further removed from the ``ding an
sich'' than he had imagined.  On my view, we know that the quantum
formalism CANNOT correspond to reality (whatever it may be). My
understanding of {\Kant} is that he would have said that {\it
classical\/} physics may or may not be a reflection of the real
thing, but we can never know.  It does form part of the categories
of understanding and is therefore necessary to make sense of the
world; but correspondence to reality is a different issue---in fact,
one that cannot be decided.  In the case of the quantum world, I
think it is (reasonably) safe to assume that the terms in quantum
theory cannot be mapped to any sort of ``ding an sich.''  That's why
it strikes me as a still stronger form of {\Kant}ianism.

Am I way off track here?  Any thoughts?

\subsection{{\Jon}'s Reply}

\bq
\noindent\vspace{-12pt}
\bq
\noindent
Chris said: My understanding of {\Kant} is that he would have said
that {\it classical\/} physics may or may not be a reflection of the
real thing, but we can never know.
\eq

This is what {\Kant} {\it should\/} have said.  In fact, he said that
the noumenal realm does not have any of the properties (time, space,
causation) attributable to the known world of objects and events.  I
have often wondered about this claim myself, and I have never been
able to come up with a convincing reason to prefer {\Kant}'s view to
yours.

\bq
\noindent
Chris said:  It does form part of the categories of understanding and
is therefore necessary to make sense of the world; but correspondence
to reality is a different issue---in fact, one that cannot be
decided. In the case of the quantum world, I think it is (reasonably)
safe to assume that the terms in quantum theory cannot be mapped to
any sort of ``ding an sich.''  That's why it strikes me as a still
stronger form of {\Kant}ianism.
\eq

This part I'm not so sure about.  With the disclaimer that {\Kant}'s
project was to give a sense of necessity to {\Newton}'s laws (they
were not just inductive generalizations a la {\Hume}): When we do
science, according to {\Kant}, we are examining a world of objects
and events conditioned by the faculties of understanding.  This does
not always require immediate perception (Critique A225/B272).  It
only requires that these objects have some causal influence on us
(again, we're in the phenomenal realm).

{\Kant} was a realist about theoretical entities.  E.g., though we
cannot see it, he thought there were good reasons (e.g., hot springs
and volcanoes at various points around the planet) for believing in a
`central fire'.  He also wrote, ``From the perception of the
attracted iron filings we know of the existence of a magnetic matter
\ldots\ Our knowledge of the existence of things reaches, then, as
far as perception and its advance according to empirical laws can
extend'' (A226/B273).  While there remains the possibility that the
noumenal realm conforms to the same laws as those governing thought,
science has no bearing on this issue according to {\Kant}. The big
question is this:  What would {\Kant} say about the world of very
small theoretical objects?  Even subatomic objects should conform to
the rules of the understanding that he outlined.  When mere
observation starts altering the object of inquiry something seems to
be amiss.

Maybe he has three options:
\begin{enumerate}
\item
{\Kant} can say that we just don't know the mechanisms, so it all
looks mysterious (And how many times has {\it that\/} happened in the
history of science?).
\item
{\Kant} can revise his system and say that there were more rules of
thought conditioning heaven and earth than were dreamt of in his
philosophy (highly unlikely that {\Kant} would admit this since his
analysis of the faculties was both exhaustive and without possibility
of error; this may be close to what you're after though).
\item
He can admit he was wrong and that when we do science we really are
investigating the noumenal realm.  At the macro level, the laws of
thought and the laws of nature are in conformity (it sure makes good
evolutionary sense that this would be the case). In the case of the
really big and really small, however, all bets are off.  The best we
can do is gain some predictive leverage within the constraints
imposed by our own representational system.  That would explain the
proliferation of metaphors rooted in Newclidian (hey, I like that)
spacetime.  Space is curved, malleable, and what-not.
\end{enumerate}
{\Kant}, I imagine, would have to go with (1).
\eq

\section{11 May 1999, \ ``Nails and Heads''}

You hit the nail on the head with the second of your three options:
\bjw
{\Kant} can revise his system and say that there were more rules of
thought conditioning heaven and earth than were dreamt of in his
philosophy (highly unlikely that {\Kant} would admit this since his
analysis of the faculties was both exhaustive and without possibility
of error; this may be close to what you're after though).
\ejw
That is to say, I think this is indeed the sort of thing I'm after.
(But---warning!---I don't care so much about what {\Kant} himself
would have thought of the program.  I'm just encouraged to know that
it is along the same lines as the one he set out with respect to
classical physics.)

I read a passage in an old paper by Abner {\Shimony} the other day
[A. {\Shimony}, ``Metaphysical Problems in the Foundations of Quantum
Mechanics,'' International Philosophical Quarterly {\bf 18}, 3--17
(1978)], and it struck me that it was a partial elaboration of (and
answer to) the point you made.  I'll scan in the passage for both our
sakes:
\bq
I shall discuss the latter of these two options first---the
abandonment of a realistic interpretation of quantum mechanics of
physical systems---since it is the more familiar in the literature
and indeed seems to have been the view of Niels {\Bohr}, and with
some variations, of various other advocates of the Copenhagen
interpretation of quantum mechanics.  {\Bohr}'s former assistant,
{\Petersen}, quotes {\Bohr} as saying, ``There is no quantum world.
There is only an abstract quantum physical description.  It is wrong
to think that the task of physics is to find out how nature {\it is}.
Physics concerns what we can say about nature.'' {\Petersen} says
elsewhere that the radical way in which {\Bohr} broke with tradition
was to abandon an {\it ontological\/} mode of thinking.  I am not
fully confident that this is the most accurate exegesis of {\Bohr}'s
rather cryptic philosophical statements, but it does seem to me to
be a way in which one can deny the strong version of EPR's reality
criterion. It is possible, in my opinion, that we shall be forced to
accept this nonontological philosophy as a last resort, but for
several reasons I find it very unappealing.

(3) Most important, I do not believe that a fully worked out and
coherent formulation of a nonontological interpretation of quantum
mechanics exists in the literature, in writings of {\Bohr} or of any
one else.  We have a standard of comparison, for there is one great
and quite fully worked out nonontological philosophy which antedates
quantum mechanics:  namely, the transcendental philosophy of
{\Kant}. It is, of course, central to {\Kant}'s philosophy that we
have no knowledge of things in themselves, but only knowledge of
phenomena. But then {\Kant} undertakes the obligation of explaining
why our knowledge has the structure that it does have, if that
structure cannot be attributed to the things-in-themselves:  the
structure is imposed by the knowing subject, the faculties of
intuition supplying the forms of space and time and the faculty of
understanding supplying the categories.  I am completely unconvinced
by {\Kant}'s explanation of the structure of our knowledge, but I do
admire him for his sense of responsibility in undertaking to work
out a detailed epistemology, instead of elliptically stating a
program. What, according to {\Bohr}, plays the role of {\Kant}'s
transcendental self in establishing the structure of our physical
knowledge?  It seems not to be the knowing self since {\Bohr}
explicitly denies that he is any kind of idealist. If it is the
character of experimental apparatus, then a nest of troublesome
problems is opened: is the apparatus to be considered as ``real'' in
an unequivocal sense, so that the position is not nonontological at
all, but rather is a kind of ontological commitment to
macrophysicalism?  And is it not strange that the macrophysical
apparatus has to be described in microscopic terms, and specifically
quantum mechanically, in order to understand how it works
which---would be a peculiar kind of macrophysicalism? In short, I am
gladly willing to say that {\Bohr} is a great phenomenologist of
scientific research---that he has said some profound things about
experimental arrangements---but I do not see that he has been able
{\it successfully\/} to bypass the question of characterizing the
intrinsic states of physical systems.
\eq

In opposition to Mr.\ {\Shimony}, I do find a nonontological
interpretation of quantum theory quite appealing.  However, I have to
agree with him that no one has worked out the details of such a thing
yet.  (But then he gets way off track again by talking about
experimental apparatuses.)

There's so much great work to be done!

\section{23 August 1999, \ ``The Point''}

I'm sitting here reading another article about the similarities
between {\Bohr}'s thought and {\Kant}ianism [C. Chevalley,
``Philosophy and the Birth of Quantum Theory''].  And I found a
couple of sentences that seemed relevant to my discussions with
you---ones that I think should be kept in mind whenever we tackle
this subject again.  Let me record them so that I'll have them in my
files.

\bq
\noindent
If we change our standpoint and look back at the genesis of quantum
theory as if it belonged to the history of philosophy, we see that
the destructuration of the theoretical language of classical theories
had as its counterpart a destructuration of the philosophical
language of {\Kant}'s theory of knowledge. This internal breakdown
took support from a tradition that is little visible now but can be
identified, and it left almost nothing intact except {\Kant}'s
fundamental stance that we are finite beings and hence cannot ever
compare our representations to essences.
\eq

\subsection{{\Jon}'s Reply}

\bq
I see your Chevalley and raise you a {\Hegel}:
\bq
\noindent
{\Kant}'s so-called critique of the cognitive faculties, {\Fichte}'s
[doctrine that] consciousness cannot be transcended nor become
transcendent, Jacobi's refusal to undertake anything impossible for
Reason, all amount to nothing but the absolute restriction of Reason
to the form of finitude \ldots\\
\hspace*{\fill} --- ({\Hegel}, {\sl Faith and Knowledge}, pp.~64)
\eq
The philosophical tradition ``that is little visible now'' is
probably that of Jacobi and {\Fichte}.
\eq

\chapter{Letters to {\Bill} {\Wootters}}

\section{31 October 1997, \ and to  {\Asher} {\Peres}, ``Premonition?''}

I was awakened this morning just in the middle of a complicated
dream.  I only remember very fuzzy fragments, but there is one in
particular that I thought you two would enjoy hearing about.  Someone
walked up and handed me a newspaper clipping.  The headline read,
``{\Bill} {\Wootters} and World Renowned Teleporter  {\Asher}
{\Peres} Attain New Heights.''  I didn't see the article, but the
headline was accompanied by a picture of you two both wearing
lederhosen and Alpine hats, smiling at the camera!!

\section{21 July 1998, \ ``Quantum Giggles''}

I'm sitting at a little sidewalk cafe in Benasque, Spain thinking of
you and all your efforts to find a compelling structure underneath
quantum mechanics.

It's strange, but for me this has turned out to be a very
foundation-oriented conference.  {\Carl} {\Caves} and I have been
working to construct a quantum version of de {\Finetti}'s theorem in
classical probability theory.  De {\Finetti}'s theorem gives the
``subjectivist'' a way to interpret an ``unknown probability''
within his framework: probability distributions over the probability
simplex are nothing more than shorthand for a certain kind of
probability assignment (called ``exchangeable'') over a large
multi-trial space.  What is an unknown quantum state?  If you take
quantum states to be states of knowledge, as we do, then an unknown
quantum state is a troublesome concept.  We think that it can be
fixed up in roughly the same way---one just has to identify the
right notion of an ``exchangeable density operator.''  We've got a
weak result in that respect, but are still searching for a stronger
one.

On top of all that, Adrian {\Kent} showed up and so I've been having
more philosophical fun than I should.  Even Richard {\Jozsa} has
jumped into the debate.  Richard thinks quantum states are much
realer than I think is safe.  So the discussions have been lively at
the least \ldots\ and scathing at the most!

I don't know that I've ever heard you take a strong stance on what
you think a quantum state represents.  What is a quantum state?  I
know you once mentioned that you think the many worlds interpretation
is a clean solution, but that you like a mystery.  What's the best
indication your mystery gives to the meaning of a quantum state?

On another strand, I reread your little paper ``Is Time Asymmetry
Logically Prior to Quantum Mechanics?'' yesterday.  That was fun: I
do like to think that time asymmetry is prior to it all---for quite a
while I've believed that the flow of time is one of those iron posts
that John {\Wheeler} speaks of.  The reversible equations of motion
that we use in our predictions are just the papier-m\^{a}ch\'e of
theory we put in between.  I don't suppose you've made an progress
in the direction defined by that paper?  If you've ever had any more
thoughts on it, I'd love to hear them.

Let me now give you a bit of an introduction to the silly stuff I'm
gonna send you.  I think it is pretty darned hard to uphold the idea
that a quantum state somehow corresponds to an objective state of
affairs out there in nature.  That is not to say, however, that I
think quantum mechanics is completely divorced from some kind of
ontological statement.  There are lots of structures within quantum
mechanics:  POVMs, tensor products, entanglement, unitary evolutions,
etc., etc.  To fixate on taking the quantum state itself as the sole
possible referent to reality in the theory and to struggle hard to
interpret that weird reality, strikes me as really misguided.  That's
what these scattered notes are about.

%
%

\section{30 August 1998, \ ``Oh So Close''}

For historical interest, let me tell you about something I ran across
the other the other day in case you haven't seen it.  It's an essay
by Eugene {\Wigner} titled ``The Probability of the Existence of a
Self-Reproducing Unit'' (found in {\sl The Logic of Personal
Knowledge: Essays in Honor of Michael Polanyi}, Routledge and Kegan
Paul, London, 1961).  What's so interesting about it is that it is
concerned with the possibility of ``cloning'' in the biological sense
and it identifies that question precisely with ``cloning'' in the
quantum mechanical sense!!  Somehow though---amazingly!---he never
quite got to a crisp statement of the no-cloning theorem.  Instead he
takes as given an interaction $S$ between the ``organism'' and the
``nutrient'' and asks if there are any solutions $|v\rangle$ (state
of organism), $|w\rangle$ (initial state of nutrient), and
$|r\rangle$ (final state of nutrient) to the cloning equation
$$
S|v\rangle|w\rangle=|v\rangle|v\rangle|r\rangle.
$$
He concludes that the for a random $S$, the probability of there
being a solution is vanishingly small.

What's even more surprising is that he really did get to the brink of
the problem with cloning. He just didn't see the conclusion.  For he
says,  ``There must be many states $|v\rangle$ all of which represent
a living organism.  \ldots\ Let us denote the $n$ vectors which
represent living organisms by $|v^{k}\rangle$ \ldots\ .  Then every
linear combination of the $|v^{k}\rangle$ will also represent a
living state.''  He apparently didn't see that this just can't be
done at all for any $S$!!!  Instead, he says that only $S$'s
``tailored so as to permit reproduction'' will work.

In light of this, I find it now even more impressive that you fellows
were able to get to the nub of what unitarity is:  no cloning, no
information without disturbance.  {\Wigner} had just the right sort
of question in front of himself for ferreting this out and somehow
missed it.

On another matter, I took your story of your conversation with your
son about moving his hand with his will as a good excuse to read and
think about some of {\Schroedinger}'s old writings again.  (I always
get a little overly philosophical on Sundays.)  They're just
marvelous.

\section{16 September 1998, \ ``The Aphorism''}

Last night, I saw another episode of the show on PBS that I told you
about:  the name is ``Oliver Sacks: The Mind Traveler.''  This one
was nice, I think, in that it really brought even closer to home your
aphorism about the dog and its understanding of the universe.  The
topic was a genetic disease called William's Syndrome (due to part of
the 7th chromosome being missing).  What's really intriguing about
this is that this ``defect'' seems to enhance people's verbal and
social skills \ldots\ making them extraordinarily pleasant to be
around and extraordinarily receptive to other people's feelings,
facial expressions, etc.  Moreover, it seems even to enhance natural
musical ability!  Many of the afflicted have perfect pitch.  In the
tradeoff however, one looses a strong sense of spatial relations and
even very basic mathematical skills (as in counting accurately).  The
example in the show that most clearly exhibited your aphorism was in
the asking of a six year old, socially very adept girl---almost
adult-like actually---to rebuild a pattern in blocks that was just
built in front of her and left for her examination:  it was just a
simple cross, like this $+$.  After something of a struggle, the best
she could come up with was a T \ldots\ and then she didn't even
perceive that there was a difference between the two patterns. Really
very intriguing.

\section{14 November 1998, \ ``Magic Eight Ball''}

By the way, have I shared the imagery of the Magic Eight Ball with
you before?  It has its roots in a debate/discussion with Micha{\l}
{\HorodeckiM} about what it is that is really sent through quantum
channels anyway. We were talking about teleportation in particular.
Micha{\l} said, ``Something certainly travels from Alice to Bob.''  I
said, ``If you ask me, the only thing that is teleported is what a
Victor has the right to say about what's in Bob's hands.  Quantum
systems are more like empty boxes out of which answers pop when we
ask questions; they don't actually contain the state vectors we
ascribe them.''  Anyway, I returned from Spain and told {\Kiki} about
this business---I caught her in a rare moment when she actually
wanted to talk about this(!)---and she said, ``Oh you mean quantum
systems are really Magic Eight Balls?''  I loved it!  Especially
since the classical point of view is that the world is made of little
billiard balls.  The last vestige of {\Democritus}'s atomism is the
eight ball \ldots\ the magic eight ball!

\section{17 November 1998, \ ``Supplement on Undergrads:  In Case
It's Useful''}

\bq
\begin{center}
Undergraduate Research and Quantum Information
\end{center}
\medskip

When I was an undergraduate at The University of Texas I had the
opportunity to be associated briefly with the research group of John
Archibald {\Wheeler}.  Two things about {\Wheeler}'s style made a
great impression on me.  First, he viewed research for both
graduates and undergraduates in precisely the same light:  It's a
frying pan and you've just got to jump into it! It got him results,
and it trained a generation of excellent theoretical physicists. The
second thing came in a question-answer session at the end of a
seminar.  Someone in the audience asked, ``Do you see a difference
between the students at Princeton and the students at UT?''  His
answer was just as clean and as simple as it should have been, ``Yes
I do; the students at Princeton {\it know\/} they're smart.  Next
question.''

If you want to know my philosophy of how to advise research, then you
need go no further than the paragraph above.  Great discoveries are
waiting to be made at all levels of science.  And if there ever was a
frying pan to jump into for the undergraduate, it is quantum
information.  Some of the greatest discoveries of our field have been
very literally at the level of a third-year undergraduate quantum
mechanics course.  There is not a student who has studied Vol.~III of
{\sl The Feynman Lectures on Physics\/} who could not have discovered
quantum teleportation for himself.  There is not a student at that
level who could not have discovered the idea behind quantum
cryptography.  Wonderfully, these are not isolated incidents: there
is a sense in which they define what the field is about.  The field
is about looking at quantum mechanics in a new way and wringing
everything we can from it.  The only tool a student really needs for
a start in quantum information is to {\it know\/} that he's smart.

The bulk of present-day research in quantum information is truly an
interdisciplinary effort.  Take quantum mechanics, computer science,
information theory, and linear algebra, put them in a bowl and mix.
Because the field is in its infancy, the use of undergraduate-level
ingredients from each of those disciplines is far from exhausted.
There is just so much fun work to be done; one cannot help but be
thankful for the army of eager, questioning undergraduates that will
teach their professors so much.
\eq

\section{02 January 1999, \ ``The Holidays''}

A little while ago I opened up a nice holiday greeting from  {\Asher}
that contained a photo of you, Peter {\Shor}, Richard {\Jozsa}, and
me.  I wrote  {\Asher}, ``Who would guess that these giants would
get so close to me as to be captured in a single photograph?!?''
Anyway, I hope you are doing well and are happy for the holidays.

My romantic side tends to think that there's something quite
important about 1999 and indeed about 2000.  I keep coming back to
the talk John {\Wheeler} gave in Santa Fe in 1994.  He showed a
slide of a coin with {\Planck}'s face on it and said, ``In 1900
{\Planck} discovered the quantum. The end of the century is drawing
near; we only have six years left to understand why the quantum is
here. Wouldn't that be a tribute?!'' Maybe I'm foolish, but I really
don't think it's out of our reach.

\section{24 January 1999, \ ``Throwing Away Old Scraps''}

I was just cleaning up my old office (i.e., {\Emma}'s new room) and
ran across a note I had made about a paper of Araki's.  The paper is
H.~Araki, ``On a Characterization of the State Space of Quantum
Mechanics,'' Comm.\ Math.\ Phys {\bf 75}, 1--24 (1980).  In it (near
the end as I recall), he makes the same observation that you do in
your paper ``Local Accessibility of Quantum Information.''  The reals
are too big, the quaternions are too small; complex spaces are just
right. There, now that that information is part of the community of
communicators, I can safely throw away my old scrap.

\section{24 May 1999, \ ``The Consequence of Indeterminism''}

Yesterday, I made the decision to go with LANL over Amherst this
year. Today, with the decision set in stone, I find myself feeling
rather empty.  The wonderful thing about indeterminism is the
openness it gives our futures.  But the awful thing about
indeterminism is the openness it gives our futures.

Thanks a million for the thoughts you sent.  The second set was
especially inspiring.  Connected to one comment you made, I couldn't
help but think of two small stories about Max {\Planck} that I had
scanned into my computer a few months ago.  I'll place them below;
you'll see the connection.
\bq
{\Planck} traced the discovery of his vocation to the teaching of an
instructor at the gymnasium, Hermann {\Muller}, who awakened an
interest, which became a passion, to `investigate the harmony that
reigns between the strictness of mathematics and the multitude of
natural laws.'  In 1878, at the age of twenty, {\Planck} chose
thermodynamics as the subject of his doctoral dissertation, which he
wrote in four months. He recalled that his professor at the
University of Munich, Philipp von {\Jolly}, had counseled against a
career in physics on the ground that the discovery of the principles
of thermodynamics had completed the structure of theoretical physics.
That had not dissuaded {\Planck}, who had his compulsion and also an
objective far removed from the principal ambition of today's
physicists.  He had no wish to make discoveries, he told {\Jolly},
but only to understand and perhaps to deepen the foundations already
set. {\it --- J. L. {\Heilbron}}
\eq
\bq
Many kinds of men devote themselves to Science, and not all for the
sake of Science herself.  There are some who come into her temple
because it offers them the opportunity to display their particular
talents.  To this class of men science is a kind of sport in the
practice of which they exult, just as an athlete exults in the
exercise of his muscular prowess.  There is another class of men who
come into the temple to make an offering of their brain pulp in the
hope of securing a profitable return.  These men are scientists only
by the chance of some circumstance which offered itself when making a
choice of career.  If the attending circumstance had been different
they might have become politicians or captains of business.  Should
an angel of God descend and drive from the Temple of Science all
those who belong to the categories I have mentioned, I fear the
temple would be nearly emptied. But a few worshippers would still
remain---some from former times and some from ours.  To these latter
belongs our {\Planck}.  And that is why we love him. {\it --- A.
{\Einstein}}
\eq

\section{02 July 1999, \ ``Small Piece of WV''}

Last night after our conversation, on my 40-minute walk home, I kept
thinking about ``what's missing?'' from my two little pieces below. I
think I can put it in a slogan:  ``The world can be moved.''

There is a reason we are stuck with a physics that is the ability to
win a bet:  if the world can be moved, you can't ask for much more.

That sentence will make more sense if you read the stuff below. [See
note to Paul {\Benioff} dated 10 June 1999.]

\section{13 July 1999, \ ``Weyling Away My Time''}

Here's that {\Weyl} quote concerning indeterminism that I was talking
about tonight.  I snatched it from Paul {\Forman}'s article ``Weimar
Culture, Causality, and Quantum Theory, 1918-1927:  Adaptation by
German Physicists and Mathematicians to a Hostile Intellectual
Environment.''

\bq
Finally and above all, it is the essence of the continuum that it
cannot be grasped as a rigid existing thing, but only as something
which is in the act of an inwardly directed unending process of
becoming \ldots\ .  In a given continuum, of course, this process of
becoming can have reached only a certain point, i.e. the quantitative
relations in an intuitively given piece $\cal S$ of the world
[regarded as a four-dimensional continuum of events] are merely
approximate, determinable only with a certain latitude, not merely in
consequence of the limited precision of my sense organs and measuring
instruments, but because they are in themselves afflicted with a sort
of vagueness \ldots\ .  And only ``at the end of all time,'' so to
speak, \ldots\ would the unending process of becoming $\cal S$ be
completed, and $\cal S$ sustain in itself that degree of definiteness
which mathematical physics postulates as its ideal \ldots\ .  Thus
the rigid pressure of natural causality relaxes, and there remains,
without prejudice to the validity of natural laws, room for
autonomous decisions, causally absolutely independent of one another,
whose locus I consider to be the elementary quanta of matter.  These
``decisions'' are what is actually real in the world.
\eq

\section{23 July 1999, \ ``Autumn Leaves''}

Boy you really took me back in time tonight---both to old thoughts
about the generation of reality and to memories of a sad time.  Below
is an old attempt to capture some of those thoughts \ldots\ strangely
enough, both the kinds you brought back tonight.  (The source is from
a letter to my friend {\Greg} {\Comer} dated 19 September 1992.)
Looking back on it, I wonder why I stopped thinking about the ideas
there.

I truly had a good time tonight.  Thanks for sharing all those
thoughts with me!  There is just so much to this world, isn't there?

\bq
Greetings from Chapel Hill---once the seat of some of the deepest
theoretical investigations around revealing the mutual underlying
structure of spacetime and quantum theory.  Or should I have said
``underlying structurelessness?''  So how goes it buddy?  Nothing new
on this end---just as the greeting hints.  Just a lonely old man
living out his days in the most mindless fashion possible.  Remember
the good old days when my notes were filled with nothing but
philosophy---good old metaphysics.  Remember the days when there was
nothing else on my mind?

Waking up the other morning, I discovered part of the origin of the
dream I wrote you about [9 July 1992] where you had entered five
paintings in an art show \ldots\ one painting of which wispily
revealed the words ``Where Randomness Persists.''  For some reason I
woke up, put on my glasses, but remained in bed---just staring at the
ceiling and walls. On one patch of wall next to my chest-of-drawers I
noted some thumbtacks still remaining from Mary's move.  The five
thumbtacks were in just the formation I wrote you about!  They used
to hold up five small impressionistic paintings (postcards) all by
Monet I think.  This contrasts with the dream; for there all the
paintings were surreal. It's funny, though, in my first days here at
Chapel Hill---when I was first really focusing on this randomness
business---I was seriously starting to consider an impressionistic
picture for the world's structure.  I would walk down the ``Quantum
Gravity Trail'' and look at the falling autumn leaves.  Here or there
I would see a pattern that would faintly remind me of
something---maybe a pot or a shovel or a kite.  And I would think
that that sort of process was the origin of structure just as in
quantum mechanics.  But then later it dawned on me that, for
instance, there was nothing intrinsically kite-like in the set of
leaves that I was picturing as a kite.  With another glance I might
see that set of leaves as depicting something more akin to an arrow.
The structure emerging from the leaves was imposed by me \ldots\
within bounds, of course, but nevertheless imposed by me.  More of
essence here I realized was the fact that I need never impose a
structure when looking at some random pattern, but if I ever did,
that very action would bring into existence another clear-cut
requirement.  If I were to ever say ``Oh, that's an arrow pointing
that way,'' I could never simultaneously say ``But it's also an arrow
pointing the other way.'' Everything I could ever say of the leaves
was necessarily constrained to be consistent.  This is where the old
idea of ``randomness with consistency'' comes from.  And where this
idea that ``consistency'' is an adequate (and perhaps preferred)
replacement for ``realism'' as a foundation for metaphysics.  And
hence this is how I started to think of the world as an
impressionistic painting. Up close one can see whatever one wants in
the painting, but from a view far away one is almost assuredly
constrained to seeing one unique image.  (I remember a note from
you---after your first visit to Paris---saying that you've had
similar thoughts.)  Now for some reason, in the dream, impressionism
was replaced with surrealism.  I can only wonder if the phrase
``Where Randomness Persists'' dually references the five-painting
pattern that used to house a set of impressionistic works and my
decaying but unsettled relationship with Mary.  And I am led to
ponder more deeply the reason for surrealism over impressionism.  You
know, I think I eventually discarded part of the autumn leaves
analogy---it just didn't take into account the crucial aspect of
quantum theory that the leaves are falling because somebody's shaking
the tree.  The surrealist painter creates his own world---not one
completely constrained by something already given. Whether this fits
into anything, I'm not sure.
\eq

\subsection{{\Bill}'s Reply}

\bq
I too really enjoyed our conversation that night.

Your 1992 letter, with its invocation of the consistency constraint,
is wonderful and also happened to remind me of something I once
wrote in a journal (though I think what you said makes more sense
than my journal entry).  I just now dug it up.  It was pretty far
out stuff, mostly about how there is only one consciousness, which
manifests itself as different people in pretty much the same way as
I can focus my attention at one moment on one thing and at another
moment on something else and yet still be a single conscious being.
``Now suppose that there is in fact only one consciousness, the
results of whose focussing its attention are consciousnesses as we
know them.  When it focusses on a certain thing, it is me; when it
focusses on something else, it is someone else, or it is me at a
later time, etc.  Thus, my own focussing of my attention is not only
an analogy but also a special case of this.''  At some point in this
metaphysical prose I suggested that, though inconsistency was somehow
necessary, the world we experience might be a world of minimum
inconsistency:  ``[One could] make a mathematical model of focussing
[in the above sense], and define `degree of inconsistency' and try to
minimize it. Local minima will correspond to stable
consciousnesses.''   That's from January, 1973.
\eq

\section{02 August 1999, \ ``Consistent Inconsistencies''}

What a wonderful idea that the world in totality may not be
consistent, but that local minima in its inconsistency might be
interesting little islands.  January 1973!  Don't you think it's
high time we started doing something scientific with these ideas!?!?
(Ok, ok, so your PhD was an example in that direction. But there's
still so much further to go!)

\section{20 August 1999, \ ``Echoes''}

Once upon a time, you wrote me this:
\bbw
What I believe about the meaning of a quantum state is mostly
negative: I don't think it can be taken literally as an element of
reality; neither is it merely an expression of what we know.  I think
it somehow straddles the fence between objective and subjective, just
as information does.  Information is embodied physically, as
{\Landauer} points out, but at the same time, information is not
information unless it is information for somebody.  The same
physical object can hold different amounts of information for
different people.  (E.g., for someone who already knows the state of
the object, it holds no information.)  I think a quantum state
likewise, though maybe in a more profound way, lies in the region
between objective and subjective, or between ontology and
epistemology.
\ebw
In going through some {\Pauli} writings again yesterday, I thought I
heard a faint echo.  From a letter to Markus {\Fierz}, 12 August
1948:
\bq
\noindent The layman usually thinks that when he says `reality' he is
speaking of something obviously known; while it seems to me just to
be the most important and exceedingly difficult task of our time to
work for the establishment of a new idea of reality. This is also
what I mean when I always emphasize that science and religion {\it
must\/} have something in common. (I do {\it not\/} mean `religion
within physics' nor also `physics within religion'---since both would
be one-sided---but inclusion of both into a whole.) What appears to
me to contain the new idea of reality I would like to tentatively
call: the {\it idea of the reality of the symbol}. A symbol is, on
the one hand, a product of the efforts of man and, on the other hand,
a sign for an objective order in the universe of which man is only
part. It owns something of the old notion of God and also something
of the old notion of object.
\eq
And again, from a letter to {\Fierz} dated 26 October 1954:
\bq
\noindent
{\it Sought\/}:  an {\it ``intermediate realm''\/} (between matter
and spirit, mental or respectively ``neutral'')
\eq

If you have a little time lying about, I would really appreciate
it---to the level of a beer, if not a whole meal or an evening of
child sitting---if you would tackle (at least in a meager way) the
question I posed after you wrote what you did above.

\bq
\noindent
{\bf Challenge:} What do you see as the cleanest, simplest argument
indicating that the quantum state has some objective (ontological)
content?  Can you articulate that?  What aspect of the quantum state
lies on the objective side of the straddle?
\eq

Is your argument different than what Roger {\Penrose} writes in {\sl
The Emperor's New Mind\/} (pages 268 and 269) or, almost
identically, in {\it Shadows of the Mind\/} (pages 312 through
315)?  (If you don't have a copy of that, I have some of the
passages scanned in and I can send them to you.)  This issue is a
bit important to me for the paper that {\Carl}, {\Ruediger} and I
are writing on the quantum de {\Finetti} theorem; but it's also quite
important for the large scheme of things.  If your reasoning is
different from {\Penrose}'s, I would like to understand it.

\subsection{{\Bill}'s Reply}

\bq
I confess that I haven't taken the time to think as hard about this
as I should have, or if I did think hard about it when you first
posed the challenge, the thoughts I had then are not immediately
accessible to me now.  But here is a response, for what it's worth.

Consider {\Penrose}'s spin-1/2 particle (of pp.\ 268--9 in ENM) in a
pure state.  Now, not everyone will necessarily regard it as being in
a pure state.  It may be in a mixed state for you and in a pure state
for me.  But if it is in a pure state for me, then I will find that
everyone I talk with for whom the particle is in a pure state will
agree on {\it which\/} pure state it's in.  In that sense the pure
state is an objective property.  It's just as objective as the
orientation of the desk that I'm sitting at.  Now, whether or not
that makes it part of {\it reality\/} is something one can still
argue about.  (In the many-worlds theory neither the direction of
spin nor the orientation of the desk is an element of reality.)  I
don't want to claim that the direction of spin is ``pure reality.''
Like everything, its existence requires perceivers.  But the fact
that everyone I talk to agrees on the state is the aspect that lies
on the ``objective side.''
\eq

\section{30 September 1999, \ ``Your Dog''}

\parbox{2.75in}{
Physics is mathematical not because we know so much about the
physical world, but because we know so little:  it is only the
mathematical properties we can discover. \\ \hspace*{\fill}
---~{\it Bertrand Russell}}

\section{12 February 2000, \ ``Feels Right''}

Still another thing to come out of the process is that I've been
reading Arthur {\Zajonc}'s book {\sl Catching the Light:\  The
Entwined History of Light and Mind}.  It's great!  I think you'd
enjoy it too.  Especially part of the discussion on some of Rudolf
{\Steiner}'s ideas:
\bq
[T]he physical world is the fruit of the moral world.  Pure hearts
truly will illumine future worlds.  Or, if we harbor darkness
within, then a dark world will be its lawful consequence in the
distant future.  We are cocreators of the world not only through the
deeds of our hands but, in even greater measure, through the
spiritual impulses we foster inwardly.
\eq

\section{13 February 2000, \ ``Zteiner''}

\bbw
I like the {\Steiner} quote.
\ebw
Actually its a {\Zajonc} quote, from the section on {\Steiner}.
(Sorry to bug you with another email:  I just couldn't leave a
misimpression.)

\chapter{Letters to {\Anton} {\Zeilinger}}

\section{06 January 1999, \ ``The (Un-)Detached
Observer''}

Yesterday I ran across your paper ``On the Interpretation and
Philosophical Foundation of Quantum Mechanics'' posted on the web,
and was very much impressed with its depth.  Most importantly,
however, I am sad to say, it reminded me that I still owe you some
papers.

I have remedied that now by sending you a pile of things that I have
written on the relation between information gain and quantum state
disturbance.  These are relevant to your request above, and, I think,
even more relevant to the thoughts in your paper.  I am in great
agreement with the idea that there may be a deeper foundation to
quantum mechanics, one along the lines that {\Pauli} and {\Wheeler}
attempted to sketch.  Quantum mechanics, it seems to me, is an
expression of the best we can say in a world where the observer is
not detached (to use {\Pauli}'s phrase).  Recognizing and
appreciating that lack of detachedness is the most exacting tool we
have for peering into the greater future of quantum theory.  This I
think is the ultimate role of quantum information theory.  (For the
opposite point of view, however, you may be interested in having a
look at John {\Preskill}'s paper ``Quantum Information: Its Future
Impact on Physics'' at {\tt
http://theory.caltech.edu/$\,\tilde{\;\;}\!$preskill/}.)

I hope you will enjoy the things I have sent you; it is always my
attempt to be as clear as I possibly can be when I write.  If you
have written anything more on the point of view you espouse in the
paper above, I would very much appreciate learning its coordinates.

\subsection{{\Anton}'s Reply}

\bq
Thank you very much for your e-mail. I look forward to receiving the
package of your papers. I fully agree with the basic spirit of what
you said.  We have to learn a little more about the role of us in the
Universe, which certainly cannot be the role of a ``detached
observer''.

Since I wrote the paper you mentioned I have kept thinking about
these problems and I feel that I made a significant step forward (one
always loves one's own children best!)\ in having been able to
identify a foundational principle for quantum mechanics.  A copy is
in the mail to you.

The basic point in all these considerations is that the old idea in
physics of a world ``out there'' which we just observe and the
consequent role of physics to just describe what already ``is'' out
there is too naive.  In essence, what can be said about Nature has a
constitutive contribution on what can be ``real''.  I would be very
interested to hear your responses on my paper.
\eq

\section{11 July 2000, \ ``Need Reference''}

It was great to finally meet you the other day.  Could you do me a
favor and give me the proper page numbers (both beginning and end)
for the article below?  I'm preparing a little article titled
``Resource Material for a Paulian/{\Wheeler}ish Conception of
Nature'' and I want it to be as complete as possible.
\bq
A.~{\Zeilinger}, ``On the Interpretation and Philosophical Foundation
of Quantum Mechanics,'' in {\sl  Vasta\-koh\-tien todellisuus,
Juhlakirja Professori K.~V. Lau\-rikaisen 80-vuotisp\"aiv\"an\"a},
edited by U.~Ketvel, A.~Airola, R.~H\"am\"a\-l\"ai\-nen, S.~Laurema,
S.~Liukkonen, K.~Rainio and J.~Rastas (Helsinki University Press,
Helsinki, 1996), pp.~167--178. Also archived at \\
{\tt
http://www.ap.univie.ac.at/users/\-%
An\-ton.\-{\Zeilinger}/philosop.html}.
\eq

\section{20 July 2000, \ ``Our Mutual Interest''}

I have just informed the ESI secretary that I would like to come to
Vienna for the dates November 27 through December 17.  I would
mostly like to have that time to think, and reason, and have some
time to interact with you on our mutual interest (namely, getting
down to the essence of the undetached observer in QM).

I have had a chance to look at your FP paper that you suggested, and
also a paper by you and {\Brukner}.  I sympathize with the direction
of your thought, but I am reluctant to think that that specific
course goes deep enough for my tastes.  So we should have plenty to
discuss.

In the mean time, if you would like to become more familiar with my
ideas, please go to my website listed below.  The long file titled
``Notes on a Paulian Idea'' should be particularly interesting for
you.  (If you're worried that it might be a waste of your time, get
a referee report from David {\Mermin} first.)  I plan to go public
with a longer version of that file soon, on {\tt quant-ph}, just
before my move.

Which brings me to some other good news:  I have now obtained a
permanent research position at Bell Labs.  We will be building a
strong group in quantum information, and my personal bent will be in
turning it toward foundational studies in quantum cryptography. I'm
quite serious about getting at this ``undetached observer'' business
in a more technical way---that provides my real motivation.  I should
be starting there in September or October.

\chapter{Other Letters}

\section{11 April 1996, \ to Michiel van {\Lambalgen}, \ ``Thesis''}

About myself (in case you are wondering) I am a physicist.  My
research specialty is ``quantum information theory,'' a field that
concerns a hodgepodge of things including quantum cryptography,
quantum computing, and statistical inference problems having to do
with quantum mechanical systems.  You may have heard of some of these
things through Paul {\Vitanyi} or Andr\'e {\Berthiaume} ({\Vitanyi}'s
postdoc). I am presently a postdoc working for {\Gilles} {\Brassard}
and Claude {\Crepeau} in Montreal; starting October, I will have a
three year position at Caltech.

I was once interested in the mathematics of randomness because,
though I am Bayesian through and through for all other uses of
probability, I believed that probabilities for quantum mechanical
measurement outcomes were something different \ldots\ in fact
something more akin to the frequentist conception.  Thus I put a lot
of effort into studying von {\Mises}, {\Church}, {\Kolmogorov},
{\MartinLof}, {\Chaitin}, etc.  (I didn't find your papers until I
had pretty much abandoned this belief, though I'm not sure that I am
completely over it!)  I had hoped that there might be some
mathematical connection between the structures found in quantum
theory (vector spaces, positive-operator valued measures, etc.) and
the structures required to formalize the notion of random sequences
\ldots\ at least that was my motivation.

I thought about you again the other day because---while in search of
another paper---I ran across a citation of your thesis.  (The paper
is ``A Limiting Frequency Approach to Probability Based on the Weak
Law of Large Numbers'' by R.~E. {\Neapolitan}, {\sl Philosophy of
Science\/} {\bf 59} (1992) 389.)  Also there have been some claims
lately by Sidney {\Coleman}---I'm sure you've never heard of him, but
he's a pretty well-known physicist---that he has {\it derived\/} that
strings of repeated quantum mechanical measurements on identical
preparations give rise to strings random in the sense of
{\MartinLof}.  I'm pretty confident that this is humbug \ldots\
because it is based on some earlier work by {\Hartle} claiming to do
such a thing for frequencies. And I know that that stuff, in the end,
turned out to be a misinterpretation of something that was nothing
more than a Law of Large Numbers argument (with a bit of a twist).

(By the way, there were some efforts in the early 1970's by Paul
{\Benioff} to explicitly introduce ``randomness'' as a postulate of
quantum mechanics.  Are you aware of these?  The papers are pretty
steeped in the jargon of symbolic logic and should be right up your
alley.  However, I don't think anything useful ever came of them.)

In any case, I have now started thinking about these things again. I
look forward to seeing your thesis and latest things \ldots\ and, if
you are interested, continuing this conversation.

\section{15 April 1996, \ to Michiel van {\Lambalgen}, \ ``Getting
Started''}

Just a few replies with my coffee to get the morning started.

You say,
\bq
No work on randomness that I am aware of differentiates between
classical and quantum mechanics; but if you have an idea how this
could be done, I would be most grateful for a few hints \ldots
\eq

I wish I had an idea!  As I say, I am quite Bayes- (i.e.\ subjective
probability-) oriented in all matters except possibly quantum
phenomena.  I've never been able to buy it that an infinite
repetition of tosses of a classical coin cannot give rise to all
heads, HHHHH\ldots \ {\it However} I am sometimes---though now {\it
very\/} rarely---inclined to believe that cannot be the case for
quantum coins \ldots\ for just the sorts of reasons von {\Mises} used
mistakenly for classical ones.  Thus, in my mind, all the work on
defining randomness was tacitly about truly indeterministic
phenomena---of which quantum measurement outcomes are the
paradigmatic (and ONLY as far as I am concerned) example---whether
the authors intended it that way or not.  This is why I would have
said that ``no work on randomness [\ldots] differentiates between
classical and quantum mechanics.''  It is a tautology.  And this is
why I had hoped to see some similarity in the structure of the two
theories.  However, who knows, perhaps after reading your thesis
I'll come away with a different opinion.

Of course, compounding the problems of such an idea (i.e. relating
quantum and randomness) is the fact that there is no unique notion of
randomness within mathematics (as your papers were the first to
really teach me).  But as I've said, I'm already disposed to leaving
these ideas behind.

Now I much more strongly believe that even the probabilities in
quantum theory are of the Bayesian cast, quantifying degrees of
belief based on prior information.  One must dig deeper for the
indeterminism \ldots\ which is most certainly there whether one
adopts a subjective view of probability or not.  I think the point
worth hanging on is this.  In going from classical physics to quantum
physics, it is the nature of the alternatives in the probability
assignments that changes, not the notion of probability itself.  In
this respect, maybe I do have something for you to think about,
because I find myself using language somewhat similar to the things
I see in your writings.

(Oops, my coffee's already run out; so much for a short note!)  There
is a pretty clear sense in which the outcomes of measurements on
classical systems can be said to be ``already there'' before any
measurement is performed; any surprisal on the experimenter's part
upon seeing the outcome can be taken to be due to his ignorance of
the ``actual'' situation.  However, for quantum systems, the
measurement outcomes ``are not already there'' independent of the
measurement (much as I've seen you contend for random sequences).
The surprisal on the part of the experimenter is again due to
ignorance, but not an ignorance that could have been removed
beforehand as in the classical situation.  I like to paraphrase this
situation with the following words.  ``Maximal'' information in
classical physics is complete in the sense that, once it is fixed,
all possible measurement outcomes can be predicted perfectly; maximal
information in quantum physics is {\it not\/} complete (in this
sense) and cannot be completed.  Because, again, the outcomes ``are
not already there'' before measurement.  Is there some real
resemblance between these two uses of ``not already there''??

There is a new idea that intrigues me: that perhaps, with very few
side conditions, quantum theory is the unique expression for how to
use probabilities when ``maximal information is not complete.''  That
is to say, this situation so restricts the use of probability
assignments that, in the end, what is left can be identified with
the standard structure of a quantum theory (i.e.\ probabilities being
generated by vectors in a linear space and positive-operator valued
measures).  However, I don't have much of a clue about how to tackle
this (still too vague) idea.

If any of this has piqued your interest, I can send you a paper by
{\Carl} {\Caves} and myself where we lay these ideas out in a little
more detail.

\section{19 July 1996, \ to Sam {\Braunstein}, \ ``The Prior''}

While in Torino, you really got me interested in the old Cox Box
question again.  I noticed in this version of the book that {\Jaynes}
makes some points about how there are still quite a few questions
about how to set priors when you don't even know how many outcomes
there are to a given experiment, i.e., you don't even know the
cardinality of your sample space.  That, it seems to me, has
something of the flavor of quantum mechanics \ldots\ where you have
an extra freedom not even imagined in classical probability.  The
states of knowledge are now quantum states instead of probability
distributions; and one reason for this is that the sample space is
not fixed---any POVM corresponds to a valid question of the system.
The number of outcomes of the experiment can be as small as two or,
instead, as large as you want.

However I don't think there's anything interesting to be gained from
{\it simply\/} trying to redo the Coxian ``plausibility'' argument
but with complex numbers. It seems to me that it'll more necessarily
be something along the lines of: ``When you ask me, ``Where do all
the quantum mechanical outcomes come from?''  I must reply, ``There
is no where there.''  (with apologies to Stein again!)  That is to
say, my favorite ``happy'' thought is that when we know how to
properly take into account the piece of prior information that
``there is no where there'' concerning the origin of quantum
mechanical measurement outcomes, then we will be left with
``plausibility spaces'' that are so restricted as to be isomorphic to
Hilbert spaces.  But that's just thinking my fantasies out loud.

\section{22 January 1997, \ to Harald {\Weinfurter}, \ ``Random Recall''}

I see that I had a somewhat poor recall of what von Neumann actually
said about the generation of random numbers.  Nevertheless, they're
still kind of interesting.  The reference is an article titled
``Various Techniques Used in Connection with Random Digits,'' and it
is found in The Collected Works of von Neumann, p.\ 768 (I don't have
the particular volume written down).

The first quote is the following:  ``\ldots\ Two quantitatively
different methods for the production of random digits have been
discussed: physical processes and arithmetical processes.  \ldots\
There are nuclear accidents, for example, which are {\it the ideal of
randomness\/} [my emphasis], and up to a certain accuracy you can
count on them.''   Then he goes on to discuss technical problems
about collecting random numbers in this way.

A little later he comes to pseudo-random numbers, i.e., ones
generated by an algorithm, about which he says the following:  ``Any
one who considers arithmetical methods of producing random digits is,
of course, in a state of sin.  For, as has been pointed out several
times, there is no such thing as a random number---there are only
methods to produce random numbers, and a strict arithmetic procedure
of course is not such a method.''

I hope you can find a use for these nice quotes.

\section{29 July 1997, \ to Ken Alder, \ ``Rosenthal Thesis''}

Actually I have read the books by {\Kuhn} and {\Heilbron} that you
suggest.  My interest is much more in Paul {\Forman}'s ideas; I've
read, I believe, three of his papers \ldots\ the main one being the
one you cite.  The thing that caught my eye about the Rosenthal
thesis is that it apparently provides some commentary on {\Forman}.
So, yes, I would still like to obtain a copy of it:  if nothing else,
perhaps the secondary references will be useful to me. Also there's
the basic curiosity of seeing what someone else thinks of
{\Forman}---I haven't seen too much commentary on his ideas up to
now.

My main interest along these lines concerns scientists' thoughts on
``indeterminism'' before the birth of quantum mechanics.  Of course,
I am also interested in this issue of their social and cultural
conditioning, but that is somewhat secondary for me.  On top of the
things mentioned so far, I've taken some time to study C.~S.
{\Peirce}, {\Pais}'s comments on the air about Copenhagen surrounding
the {\Bohr}-Kramers-Slater paper, and also a long paper on
{\Schroedinger} and {\Exner} (I forget the author's name).  If you
have any further suggestions on this subject, I would certainly
appreciate it---my literature search, I'm sure, has been far from
exhaustive.  I profess nothing more than dilettantism in this!

Also, if you can find it, I would like the reference to Garber---I
don't believe that I've heard of her before.  And, finally, thanks
for the tip about Kojernikov; I will look him up soon \ldots\ I
didn't realize that we had any historians around here!

PS.  If I can find it, I'll dig up a letter I wrote a few years ago
that may amuse you:  the subject is a (playful) application of
{\Forman}'s ideas to my research in the early 1990's.  (It'll be
written in \LaTeX, but it should still be readable with a little
effort if you don't have the means to compile it.)  Maybe, it'll
also help you see what interests me.

\bq
\noindent 18 February 1992 (from a letter to Greg {\Comer})
\bigskip

So anyway, as promised, I intend to make this note cheerier than the
last.  In fact I think I'll do this by describing something that
delighted me to no end the other day.  (And in the process maybe
we'll both learn a little more about the ``proper'' way to implement
the ``Law without Law program.''  At any rate, I'll get some quotes
into my computer that I've been wanting to get there.)  You see,
lately I've become quite interested in the way the LWL idea has
started to infiltrate popular culture.  Examples of it can now be
found everywhere.  I'll show you a few in just a couple of seconds.
Why am I thinking about this?  I'm not really sure, but probably
because I don't think I'm up to any real physics right now.
Nevertheless, I should say that, to some extent, I suppose that I've
always considered myself a ``child of the times'' ({\em
??zeitkinder??\/}). For instance, whenever predicting social
behaviors, my usual justification is simply, ``I'm an average guy
and I figure he'd do what I'd do.''  (The word ``do'' here not
necessarily to be taken in Charlie's{\index{Rasco, B. Charles}} usual
sense!) But this of course leads me to wonder whether my randomness
obsession is nothing more than a product of my cultural milieu
(that's a \$5.00 word) or, alternatively, evidence that I'm part of
the fuel for this fire.  As of right now I'm not sure which of these
is the proper point of view to take or whether it really matters at
all.  What these considerations do lead to, though, is the following
story and set of thoughts.

I suppose I should start with the examples I alluded to.  At the most
superficial level recall the postcard in my office that I found in
the Haight-Ashbury district of San Francisco---a scene of Einstein
observing the hand of God rolling a pair of dice with the Crab Nebula
in the background.  Or similarly the graffito I saw on a restroom
wall at the Hole-In-The-Wall in Austin signed by ``The Random Man.''
(Speaking of which, I came very close to writing below this the
formal expression for a random string and signing it ``The REAL
Random Man.'')  Stretching it---I know---but consider some of the
lines from the movie {\sl Slacker\/} (a film about the fringe culture
in Austin).  For instance, the line of the Dostoyeveski Wannabe,
``Who's ever written a great work about the immense effort required
in order not to create?''  Or that of the Old Anarchist, ``And
remember the passion for destruction is also a creative passion.''
Or the Disgruntled Grad Student, ``Every action is a positive action
even if it has a negative result.''  And, in particular, the
soliloquy of Having A Breakthrough Day:
\begin{quote}
It's like I've had a total recalibration of my mind, you know.  I
mean it's like I've been banging my head against this nineteenth
century thought type ... ah, what ... thought mode, constructs, human
construct.  Well the wall doesn't exist; it's not there, you know.  I
mean they tell you to look to the light at the end of the
tunnel---well there is no tunnel.  There is no structure.  The
underlying order is chaos.  Man.  I mean everything's in one big ball
of fluctuating matter---constant state of change.  You know.  I mean
it's like across that great quantum divide is this new consciousness.
And you know I don't know what that's gonna be like, but I know that
we're all part of it.  I mean it's new physics.  You can't look at
something without changing it.  You know, anything.  I mean man
that's like almost beyond my imagination.  Just like that butterfly
flappin it's wings in Galveston and somewhere down the road a pace
it's gonna create a monsoon in China.  You know.
\end{quote}
A mouth full, huh?  But I should also add that the same character
followed this with the later remark, ``That's all right, time doesn't
exist.''  (By the way, the definition of a slacker is a member of ``a
new generation of young people, primarily centered around college
campuses, that rejects the values of the generation before them.'')
Finally I'd like to cap all this off with the words of the new Rush
song ``Roll The Bones'' from the album of the same name:
\begin{quote}
Well, you can stake that claim\\
Good work is the key to good fortune\\
Winners take that praise\\
Losers seldom take that blame\\
If they don't take that game\\
And sometimes the winner takes nothing\\
We draw our own designs\\
But fortune has to make that frame\\
We go out in the world and take our chances\\
Fate is just the weight of circumstances\\
That's the way that lady luck dances\\
Roll the bones, Roll the bones\\
Why are we here?\\
Because we're here.\\
Roll the bones, Roll the bones\\
Why does it happen?\\
Because it happens.\\
Roll the bones, Roll the bones\\
Faith is cold as ice\\
Why are little ones born only to suffer\\
For the want of immunity\\
Or a bowl of rice?\\
Well, who would hold a price\\
On the heads of the innocent children\\
If there's some immortal power\\
To control the dice?\\
We come into the world and take our chances\\
Fate is just the weight of circumstances\\
That's the way that lady luck dances\\
Roll the bones, Roll the bones
\end{quote}
(Maybe, as evidence for the earlier worry, I should add that I grew
up with the music of Rush and even at one time considered them my
favorite band.)

So what do you think?  A valiant attempt of the literate layman? I
don't know; I don't know.  I do know, though, that something about
all this (or at least the more extensive quotes) gives me the wrong
feel.  I wish I could articulate it a little better at this point,
but I don't think I can.  It's just something about the turning of
phrase in these passages (and the countless others that I haven't
been lucky enough to get down) that leads me to believe their authors
see nature's indeterminism as happening within some ``background.''
But it seems that the true indeterminism---quantum mechanical
indeterminism---is the farthest thing it could possibly be from that.
The quantum mechanical indeterminism doesn't come about from an
indiscriminate swerve in the path of an atom; it comes from the point
of contact between the theory and the world---the measurement.  And
the laws of nature aren't the accumulated effect of a multitude of
lawless fireflies flashing against a black screen.  The laws of
nature arise out of some sort of G\"odelian web:  only
pattern/law/guess confirming bodies (like people) can perform quantum
mechanical measurements even though they themselves can in turn be
treated quantum mechanically by a similar body.  Furthermore, you've
already seen my guess that the ``throw of a die'' isn't quite the
correct model for quantum mechanical indeterminism:  the repeated
throw of a die can create a nonrandom string.

``On with the story!'' you say.  ``Tell me how this delighted you to
no end.''  Well \ldots\ actually all that didn't, but what it
reminded me of did.  And that was {\Jammer}'s contention that,
``\ldots certain philosophical ideas of the late nineteenth century
not only prepared the intellectual climate for, but contributed
decisively to, the formation of the new conceptions of the modern
quantum theory.''  The excitement started when this memory led me to
two papers that I had long ago copied but never bothered to read.
Their subject---``indeterminism in German physics before quantum
theory.'' Let me just start with a list of some
physicists/philosophers who were indeterminists (or at least toyed
with the notion) well before quantum mechanics was around:  Maxwell
(!!), C.~S.~{\Peirce}, Larmor, Boltzmann, {\Exner} ({\Schroedinger}'s
graduate advisor), {\Schroedinger} himself, von {\Mises}, Schottky,
Nernst, {\Reichenbach}, and {\Weyl}.  What's very interesting is that
almost exactly across the board when these people spoke of
indeterminism they were speaking of an indeterminism for ``little''
objective events (or for the collisions of bodies) within
spacetime!  (Perhaps not so different from the pop culture we see
now.)  The one notable exception was Weyl.  And this is exactly
where the delight comes in.  Recall that once upon a time a certain
brash young man wrote to you that:
\begin{quote}
If you turn to \ldots\ Wheeler's contribution ``Law without Law''
[you will find] an illustration of a 3-D letter {\bf R} with the
following caption:  ``What we call reality, symbolized by the letter
R in the diagram, consists of an elaborate papier-mach\'e
construction of imagination and theory fitted in between a few iron
posts of observation.''  From various considerations \ldots\ I see no
alternative but to view quantum mechanical measurements as Wheeler's
``few iron posts of observation'' and spacetime as at least part of
his ``elaborate papier-mach\'e construction of imagination and theory
fitted in between.''
\end{quote}

Compare and contrast this (and the bits and pieces of our discussions
that you can recall) with the wonderful set of words by {\Weyl} that
I found in the paper, ``Weimar Culture, Causality, and Quantum
Theory, 1918-1927:  Adaptation by German Physicists and
Mathematicians to a Hostile Intellectual Environment'' by Paul
Forman.
\begin{quote}
Finally and above all, it is the essence of the continuum that it
cannot be grasped as a rigid existing thing, but only as something
which is in the act of an inwardly directed unending process of
becoming \ldots .  In a given continuum, of course, this process of
becoming can have reached only a certain point, i.e. the quantitative
relations in an intuitively given piece ${\cal S}$ of the world
[regarded as a four-dimensional continuum of events] are merely
approximate, determinable only with a certain latitude, not merely in
consequence of the limited precision of my sense organs and measuring
instruments, but because they are in themselves afflicted with a sort
of vagueness \ldots .  And only ``at the end of all time,'' so to
speak, .... would the unending process of becoming ${\cal S}$ be
completed, and ${\cal S}$ sustain in itself that degree of
definiteness which mathematical physics postulates as its ideal
\ldots . Thus the rigid pressure of natural causality relaxes, and
there remains, without prejudice to the validity of natural laws,
room for autonomous decisions, causally absolutely independent of one
another, whose locus I consider to be the elementary quanta of
matter.  These ``decisions'' are what is actually real in the world.
\end{quote}

Perhaps there's not all that much to compare---{\Weyl}, I think, sees
the continuum as being in a state of formation by a physical process
rather than as a thought construct and he certainly views the
fundamental ``decisions'' in a more objective light than allowed by
quantum mechanics.  But nevertheless there is a certain similarity
between these ideas.  I guess it's just kind of nice to know that
even if you can't keep good company with your contemporary
colleagues, there's at least somebody from the past with which you
can.
\eq

\section{16 October 1997, \ to Lucien {\Hardy}, \ ``Wispy Words''}

I was rummaging through the library today, and I found this little
thing that reminded me of you.  It is a transcript of a panel
discussion at the Symposium on the Foundations of Modern Physics 1994
in Helsinki.  The point of interest is something that {\Zeilinger}
said:

\bq
\ldots\ we don't know why events happen, as expressed by {\Bell}.
Let me explain a little bit what I mean by that.  By quantum
phenomenon we mean the whole unity from preparation via evolution and
propagation to detection.  Then there is an uncontrollable element
somewhere in this chain.  It can be called the reduction of the wave
packet.  Or it can be in the many worlds interpretation the
unexplainability of the fact that I find my consciousness in one
given universe and not in the others.  In a {\Bohm} interpretation it
can be the fact that I cannot control the initial condition.  As an
experimentalist I would say that there is some uncontrollable element
from the following point of view. When doing a Stern-Gerlach
experiment, for example, with an x-polarized spin, I cannot predict
that this spin will go up, this one will go down, etc.  There is
something beyond my control. My personal feeling is that we have
found for the first time in physics that {\it there are things which
happen without sufficient reason}.  This, I think, is a very profound
discovery.  I don't know whether there is a way to understand this or
not.  I feel there might be a way to understand why the world is so
strange but we have not understood that yet.  In my opinion this so,
because we really don't know what information is.  We don't know what
it means to collect information about the world.  There is some world
out there.  In the words of Professor {\Laurikainen}, in a very
specific sense we have created the whole universe.  But in some
sense it exists without us. We have to understand therefore what it
means to collect information about something which is not as much
structured as we think.
\eq

Is that roughly what you remember?  They are wispy words indeed!

\section{10 October 1997, to Eli {\Yablonovitch}, ``Entanglement and
Correlation''}

Thanks for being such a good participant at my talk last week.  I
really enjoyed having you there \ldots\ I must have because I woke
up the other day thinking about one of the points you brought up.  I
just want to address it quickly while it's still on my mind.

If I can paraphrase, I think you said at some point something like
``entanglement is another word for correlation.''  And then we had a
slight discussion of things to do with {\Bell} inequalities.  Anyway,
what I'd like to do is emphasize again the distinction between
``classical correlation'' and ``entanglement''---this is something
that people have only started to focus on in an applied way in the
last five years. (One thing that prompts me to this is something I
overheard at a meeting in England three weeks ago:  this annoying
fellow named [\ldots]---have you ever met him?---said far more
graphically than you, ``As far as I'm concerned, `entanglement' is
just baby talk for `correlation'.''  So I'm taking my frustrations
with him out on you!! Aren't you lucky?)

The main point is that ``classical correlation'' ultimately boils
down to \ldots\ or can be thought of as \ldots\ simple probabilistic
correlation between things objectively existent on the two player's
sides.  ``Alice and Bob previously entered a lottery for which they
were the only two players.  The lottery commission sent them an
announcement of the winnings.  They haven't opened their envelopes
yet, but the message in one envelope says that one is the winner and
the message in the other says that one is the loser.''  Entanglement,
on the other hand, expresses the {\it potential\/} for correlation.
``Alice and Bob will eventually perform some measurements on their
EPR pairs and the outcomes of their measurements {\it will be\/}
correlated.  However, before their measurements are performed, there
are no objectively existent variables that are correlated in the
sense of the last example. Different measurements can and will lead
to different correlations.'' The distinction is important
conceptually and also useful for technical applications.  In a
certain sense entanglement is a kind of ``all-purpose correlation''
just waiting to be baked into something real. (You should think of it
as a modern version of the ``Martha White's All-Purpose Flour'' that
Tennessee Ernie Ford used to advertise.)

One of the tasks of quantum information theory is to invent ways that
such all-purpose correlation can be more useful than plain old
classical correlation.  There are now several:  {\Bennett} and
Weisner's ``superdense coding'', quantum teleportation, quantum
cryptography, entanglement-enhanced classical communication, Richard
{\Cleve}'s communication complexity games, better control of
frequency standards, etc., etc.  That's the motivation.  Because of
this, one of the hot research topics in this field to actually
quantify the amount of classical correlation {\it and\/}
entanglement in a bipartite density operator.  One would like to
know the exact ways the quantities trade off each other, etc.

\section{16 October 1997, \ to Glenn {\Starkman}, \ ``Michelson Lecture
Proposal''}

A few weeks ago I wrote a wispy, philosophical-sounding little
letter to an old friend.  I think I can give you no better
introduction to what I would like to say in more detail in the
Michelson Lectures than to quote it directly.

\begin{quotation}
I'm nearing the end of the flight and feeling a little philosophical.
I hope you'll let me entertain you for a while.  Lately I've been
thinking about the airy nothings of quantum mechanics again.  It's
been a long time since I've done that to any extent---it's sort of
refreshing.

Indeterminism and entanglement.  The first is an old friend, that you
know.  The second, though, every day takes my heart a little more.
In a certain way, indeterminism couldn't live without entanglement:
the EPR argument would have triumphed over indeterminism if
entanglement hadn't {\it also\/} led to a necessary violation of
{\Bell} inequalities.  I believe in the ultimate indeterminism of
quantum mechanical measurement outcomes just because of the
experimental confirmation of {\Bell} inequality violations and the
experimental confirmation of Special Relativity.  I've said this to
you before (probably three years ago), but now it's starting to weigh
on my mind more heavily.  If I want to understand quantum
indeterminism, then I must also understand entanglement: the argument
goes in just that order.

Luckily for me, I think, the field of Quantum Information is
especially suited to that purpose.  Viewing entanglement as a new
resource is the main thing on everybody's mind.  In fact, I'm
starting to feel that the situation we're in can be likened to the
beginning of thermodynamics.  What is heat, energy, work?  No one
knew at the outset; some thought them fluids, some thought them vital
forces much like the soul, and so on.  However, one thing did become
clear eventually: no informed judgment on that fundamental question
stood a chance until there existed a quantitative theory of
thermodynamics.  Without that, one could have never come across the
mechanical theory of heat and the corollary of atomism that it led
to.

So what is this thing called entanglement?  What is its use?  That
we're just starting to understand.  If I had to put it in a phrase
right now, I would say it is ``all-purpose correlation.''  Alice and
Bob come to me and say, ``Give us a little correlation, something
that we can both have and no one else can possess.  We think we're
going to need it pretty badly tomorrow.''  I say, ``Sure, no problem,
just tell me which variables you'll be needing correlated and I'll do
the trick for you.''  They say, ``Sorry, we don't know which ones
we'll need correlated yet.  A lot of that will depend upon what we
actually encounter tomorrow.''

In the classical world, Alice and Bob would have been out of luck.
But because the world is quantum, I actually can do something for
them.  I can give them a little ``all-purpose correlation.''  And it
turns out that that stuff can be really useful for several tasks. (In
fact, we're finding ever more uses all the time.)

Thus, in a certain way, I'm starting to be impressed that
``entanglement'' shares a strong analogy to ``energy.''  Both
fulfill similar roles in our engineering endeavors:  they are
``all-purpose'' essences that can be used for various beneficial
tasks.  Once we understand that in real depth, I think we'll finally
put a dent in this question of ``How come the quantum.''
\end{quotation}

The ``tasks'' this letter refers to include quantum cryptography,
quantum state teleportation, error correction for quantum computers,
entanglement-enhanced classical communication over noisy channels,
better control of frequency standards, and the list goes on.  All
these things sum together to form the new and growing fields of
Quantum Information Theory and Quantum Computation.\footnote{For your
reference, two recent introductory/expository articles on the subject
can be found in the pages of {\it Physics Today}: pages 19--21 of the
October 1997 issue, and pages 24--30 of the October 1995 issue.  Web
links to several introductory expositions and resources on the
subject can be found in John {\Preskill}'s ``Physics 229'' homepage:
{\tt http://www.theory.caltech.edu/people/preskill/ph229/}.}

I propose for the {\it colloquium\/} associated with the Michelson
Lectures to give an introductory talk on some of these new
applications of quantum entanglement.  Chief in the list will be
quantum state teleportation and a clean, pretty example from
Communication Complexity Theory due to {\Cleve} and
{\Buhrman}.\footnote{See ``Substituting Quantum Entanglement for
Communication'' by Richard {\Cleve} and Harry {\Buhrman} at {\tt
http://xxx.lanl.gov/abs/quant-ph/9704026}.}  These examples and the
technical details associated therewith will form the backbone of the
talk.  However, along the way, I plan to weave connections all the
way from the very fundamental (the {\Einstein}-{\Podolsky}-{\Rosen}
argument and the {\Bell} inequalities) to the very applied (the power
of quantum computing to factor large integers).  The hope is that it
will come out as a fairly entertaining and useful mix. The level of
presentation should be accessible to anyone who has had a basic
undergraduate quantum mechanics course and has some knowledge of
simple {\Dirac} notation.  The title of the talk could be, ``What Can
You Do With Quantum Entanglement?''

I propose to explain, for the {\it three technical talks\/} of the
Michelson Lectures, detailed aspects of Quantum Information Theory
more closely associated with my own research.  The three talks will
build on each other, each expanding upon one of the ``tasks''
mentioned above.
\bigskip

{\bf Lecture 1:} {\em Optimal Quantum Measurements and the
Distinguishability of Quantum States}.

\noindent
Arbitrarily large amounts of ``classical'' information can be dumped
into a single finite quantum system---for instance, the spin of a
simple spin--1/2 particle---just by preparing it one of many
nonorthogonal quantum states.  The amount of information so dumped
corresponds to the logarithm of the number of possible
preparations.  On the other hand, in general very little of that
information can be retrieved reliably.  This is because there are no
physical means for distinguishing nonorthogonal quantum states.
This principle, closely related to the ``no-cloning theorem'' of
quantum mechanics, is the engine that powers quantum cryptography.
In this lecture I will build the background required for quantifying
just how much information can be retrieved in a situation like
this.  Topics will include generalized quantum measurements, the
distinguishability of mixed-state density operators, and various
information theoretic measures associated with that question.
\bigskip

{\bf Lecture 2:} {\em Sending Classical Information on Noisy Quantum
Mechanical Channels}.

\noindent
People like to put ``classical'' information---like the stories in
today's newspaper---into quantum systems for a very simple reason:
to get it from one place to another.  Since the world is quantum
mechanical, this, in the last analysis, is exactly what one {\em
always\/} does in transmitting information.  So, what is the highest
rate at which one can hope to transmit information on a quantum
system?  That is, what is the largest number of bits per
transmission that one can hope to obtain? The answer to this question
is not yet completely known, but great progress toward a general
conclusion has been found recently.  This lecture will focus on
reporting the most general results known so far; this will require
integral use of the measures introduced in the first lecture.  With
that, one can start to explore at last two other questions that
reveal rather surprising aspects of the world.  Can a channel's
noise ever be evaded more effectively by first entangling the
separate transmissions before sending them on their way?  Can it ever
help to encode classical information in nonorthogonal quantum
states?  Simple examples of both effects will be demonstrated.
\bigskip

{\bf Lecture 3:} {\em Quantifying Entanglement}.

\noindent
The name of the game in Classical Information Theory is to make the
correlation between sender and receiver as high as possible.  This
is what communication is.  A corresponding game in Quantum
Information Theory is to try to make the amount of ``all-purpose
correlation'' or entanglement between sender and receiver as high as
possible.  Such a thing is useful, for instance, when one needs to
port the program on one quantum computer onto another. In the
classical case, {\Shannon}'s ``mutual information'' quantifies the
question.  What is the corresponding quantity in Quantum Information
Theory?  How does one quantify entanglement? This is the subject of
this lecture, which focuses on the two most interesting ways of
carrying this project out.  The first has to do with how much
``pure'' entanglement one can hope to squeeze out of a noisy
transmission.  The second has to do with reversing the process. Both
of these quantities are interesting in their own right, and go quite
some way toward showing how different ``classical information'' is
from ``quantum information.''
\bigskip

All three lectures will assume the same level of expertise on the
part of the audience that the colloquium does, though they will
likely be presented in a more intensive way.  Audience participation
will be encouraged (and even begged for if it's not forthcoming on
its own!).  In all cases, the goal will be on getting the essential
ideas across to the audience---to draw pictures in their minds that
won't vanish upon leaving the auditorium.

My most personal stake in giving these lectures is to try to convey
some of the excitement in this new field, to spread the word.  I
firmly believe the field holds forth great potential not only in its
technical application but for our deeper understanding of the world
in its entirety.  It is just a question of planting the seed in
enough fresh young minds.  I deem that the Michelson Lectures will
provide a good vehicle for just this purpose and I thank you for
your consideration.

\section{19 January 1998, \ to Daniel {\Kevles}, \ ``{\Herzfeld} History''}

I'm wondering if you can help me track down the identity of my
academic great-great-great-grandfather.  My interest in this came
originally from my winning the Michelson award at Case Western
Reserve U.  Since John {\Wheeler} is my academic great-grandfather
and had once described his PhD advisor as ``the great optical
physicist \ldots'' in front of me, I was hoping there was some
chance that his advisor's advisor was Michelson. (I would have liked
to use this to start off my lectures at CWRU.)  Anyway that turned
out not to be the case:  {\Wheeler}'s advisor, I have since found
out, was Karl~F. {\Herzfeld} (whom you write a bit about in your book
{\sl The Physicists}). I have been able to find out that {\Herzfeld}
received his doctorate in Vienna in 1914 (both from {\sl Sources for
History of Quantum Physics\/} and {\sl American Men of Science\/}),
but I have not been able to trace his advisor.  Could it have
possibly been Boltzmann??  Do you know of any other source by which
I might ferret out this information?

PS.  One tidbit that I find really interesting is that Mehra and
Rechenberg have traced the usage of the phrase ``quantum mechanics''
back to at least a 1921 paper by {\Herzfeld}!

\section{19 January 1998, \ to Gary {\Herling}, \ ``{\Herzfeld}''}

Thanks for sending on your friend's suggestions.  It would have
indeed been great if {\Herzfeld} had been Boltzmann's student!  I
have been able to ascertain that {\Herzfeld} got his doctorate in
Vienna in 1914---this, I think, rules out Boltzmann in the lineage.
Boltzmann committed suicide in 1906.  {\Herzfeld} focused mostly on
spectroscopy and statistical mechanics.  He had at least {\Wheeler}
and {\Heitler} as students.  However I still don't know who
{\Herzfeld}'s advisor was.

In any case, my search hasn't been fruitless!  The most interesting
tidbit I've picked up is this:  {\Herzfeld} may actually have been
the first person to introduce the term ``quantum mechanics'' into the
literature!!  {\Born} thought that he had done it himself, but his
first usage of it was in a paper of 1924.  Lorentz preceded that in
1923 with ``the mechanics of quanta.''  However, {\Herzfeld} had a
paper of 1921 titled, ``Quantum Mechanics of Atomic Models.''  (All
this according to a footnote in Mehra and Rechenberg, vol 4.)

\section{22 January 1998, \ to Alan Hajek, \ ``{\Bohr} was a Bayesian?''}

I was quite intrigued the other day to find (from Sam {\Braunstein})
that you have some interest in the subject of ``objective chance.''
I do too, in connection with quantum mechanics \ldots\ though in a
somewhat negative way.  I would very much like to talk to you in
detail about these things.  A small part of my research program in
quantum information theory is to give some substance to the slogan
``{\Bohr} was a Bayesian'' (\ldots\ though a Bayesian good enough to
know that Bayes' rule for conditionalizing isn't always valid
\ldots\ especially when one is confronted with a world where one's
information gathering measurements {\it necessarily\/} disturb
someone else's predictability.)  Or more importantly, historical
facts aside, to give some substance to the idea!

\section{14 March 1998, \ to Ryszard {\HorodeckiR}, \ ``{\Jaynes}' Principle''}

A warm thank you to you and Micha\l\ for sending me a copy of your
upcoming paper on {\Jaynes}' Principle and quantum-state
compression.  I gave it a quick read with much interest this
morning.  Indeed I am very sympathetic with efforts such as this.
For a long time now it has been my pet ``idea for an idea'' to use
quantum information theory in aid of finding a foundation for
quantum theory itself that has something of the flavor of {\Jaynes}'
efforts in classical statistical mechanics.  As I wrote  {\Asher}
{\Peres} once:
\bq
  My own pet ``idea for an idea'' (as {\Wheeler} would say) is that the
  quantum theoretical description---along with its Hilbert-space
  structure---is forced upon us when the information we have about a
  physical situation/system is of a certain variety.  In this way,
  quantum theory is not so different from Ed {\Jaynes}' molding of
  classical statistical mechanics and the maximum entropy
  principle.  It just so happens in the latter case that the prior
  information is in the form of an expectation value for some physical
  quantity that can be assumed objective and independent of
  measurement: this prior information along with a small few
  desiderata lead uniquely to the canonical probability assignment.
  In the case of uniquely quantum problems, we start with a
  different kind of prior information.  The {\it hope\/} is that, upon
  pinpointing the nature of that information (and again a few small
  desiderata), one would see that a ``wave-function assignment'' is the
  natural (and unique) way of summarizing what we know and what we can
  predict.
\eq
The extra piece of knowledge that we seem to have concerning quantum
phenomena is that the probability assignments we make will be
ultimate \ldots\ and, in particular, non-updateable.  I.e., if we
make a formal assumption of ``no hidden variables'' and use a decent
principle of unbiased inference---as {\Jaynes}' principle seems to be
for the case where one is given an expectation value and nothing
more---, will we end up with a structure formally identical to
quantum mechanics?  I wish I knew.  Of course, the principle,
whatever it may turn out to be, will not be the MaxEnt principle
itself:  the emphasis is rather on the ``flavor'' of the MaxEnt
principle.  Anyway, I say slightly more about this idea on pages
23--26 of my paper {\tt quant-ph/9601025} with {\Carl} {\Caves}. And
I hope to say still a little more in an upcoming paper with {\Caves}
and {\Schack} about Bayesianism and the quantum probability law.

\section{05 May 1998, \ to Ronald {\Giere}, \ ``Quantum Propensity \&
Quote''}

I am a physicist trying to make sense of whether ``the'' notion of
propensity is of any use for understanding quantum theory.  (In fact,
some colleagues and I are preparing a manuscript on the subject
presently.)  Thus I have come across several of your old papers from
the 1970s on propensity and objective chance.  (I might add that I
think these are the {\it very best\/} that I've seen on the subject
so far.)

Perhaps you can help me out in some of my bibliographic troubles for
the project?  Most importantly, in your paper ``A Laplacean Formal
Semantics for Single-Case Propensities'' (J.\ Phil.\ Logic, 1976),
you attribute a nice little quote to Albert {\Einstein}:
\bq
\noindent
I can, if worst comes to worst, still realize that God may have
created a world in which there are no natural laws.  In short, a
chaos.  But that there should be statistical laws with definite
solutions, i.e., laws which compel God to throw the dice in each
individual case, I find highly disagreeable.
\eq
But then you say, ``I have been unable to find the exact reference
for this one.''

Have you been able to find the source of the quote in the intervening
years?  If not, what was your original source in the first place?  Is
this ``quote'' actually your own paraphrase of a conversation that
you had with someone?  If so, who?  I would like to quote it myself
and want to make the reference as complete as possible.

Besides that, though---but less importantly for my immediate
manuscript---can you point me to any significant
critiques/endorsements of your theory of propensity?

\section{05 May 1998, \ to Ronald {\Giere}, \ ``Propensity Stuff''}

Many thanks for trying to rethink your steps on the {\Einstein}
quote. As far as I recall, there's no mention of any such thing in
{\Fine}'s book.  (I read it pretty carefully a couple of years ago,
and, in any case, a quote like this would have caught my attention
\ldots\ given my interests.  Also there's one technical point in
that the oldest essay in the volume didn't appear until 1976, the
same year as your Laplacean-semantics paper---so it's not very
likely you could have read the full book then.)

The closest that I've found to it is from a letter, {\Einstein} to
Solovine, dated 30 March 1952:
\bq
You find it strange that I consider the comprehensibility of the
world (to the extent that we are authorized to speak of such a
comprehensibility) as a miracle or an eternal mystery.  Well, a
priori one should expect a chaotic world which cannot be grasped by
the mind in any way.  One could (yes one should) expect the world to
be subjected to law only to the extent that we order it through our
intelligence. Ordering of this kind would be like the alphabetical
ordering of the words of a language.  By contrast, the kind of order
created by {\Newton}'s theory of gravitation, for instance, is wholly
different.  Even if the axioms of the theory are proposed by man, the
success of such a project presupposes a high degree of ordering of
the objective world, and this could not be expected a priori.  That
is the ``miracle'' which is being constantly reinforced as our
knowledge expands.
\eq
But this really doesn't have the same intent as your quote.  If you
ultimately think of where you found it, please let me know at any
time.

Thanks also for the remarks on the old propensity theory.  I was
aware of Humphreys' critique. In case you would like the appropriate
references, at least two of them are:
\begin{enumerate}
\item
P.~W. Humphreys, ``Is `Physical Randomness' Just Indeterminism in
Disguise?,'' in {\sl PSA 1978: Proceedings of the 1978 Biennial
Meeting of the Philosophy of Science Association}, Vol.~2, edited by
P.~D. Asquith and I.~{\Hacking} (Philosophy of Science Association,
East Lansing, Michigan, 1981), pp.~63--78.
\item
P.~Humphreys, ``Why Propensities Cannot Be Probabilities,'' Phil.\
Rev.\ {\bf 94}, 557--570 (1985).
\end{enumerate}
A related argument was proposed by Peter Milne:
\begin{enumerate}
\setcounter{enumi}{2}
\item
P.~Milne, ``Can There Be a Realist Single-Case Interpretation of
Probability?,'' Erkenntnis {\bf 25}, 129--132 (1986).
\end{enumerate}

I myself am striving to formulate a relatively personalist account of
quantum probabilities---that is one with a notion of ``objective
chance'' that is no more reified than Ramsey's 1928 version of the
concept.  In searching through the propensity/chance literature, I've
mostly been trying to see that all my bases are covered.

\section{10 August 1998, \ to Richard {\Beyler}, \ ``Your Thesis''}

Last night I read your article ``Targeting the Organism'' (Isis,
1996) and enjoyed it very much.  Lately I've taken a dilettante
interest in the particular thoughts of {\Pauli} and {\Jordan} on the
foundations of quantum mechanics, and your article has been helpful
in that regard.  (I am a physicist at Caltech, by the way, not a
historian.)  My interest mostly comes from a ``fixation on the issue
of determinism'' (as you describe {\Jordan}), which lays at the
foundation of my line of research (quantum information theory /
quantum cryptography) more than any other aspect quantum mechanics.
Anyway, in this regard, I am quite interested in reading your Ph.D.
thesis.  I'm writing this note, mostly, to ask how I might obtain a
copy.  If you have an offprint of it lying around that you could
mail to me that would be great!  Alternative to that, though, if
have a Microsoft Word, \TeX /\LaTeX, or PostScript file of it, then
you could email it here and I could print it myself.

\section{16 August 1998, \ to Adan {\Cabello}, \ ``{\Mermin} and His
Correlata''}

Last week, I was fortunate enough to receive a sneak preview of your
paper ``Quantum Correlations as Local Elements of Reality'' from my
(visiting) office-mate  {\Asher} {\Peres}.  It is a nice paper and
very well written.  In particular, I was very pleased to discover the
existence of another level-headed Copenhagenist out there!  Given
the number of physicists trying to turn away from the Copenhagen
interpretation, we are in grave need of as many eloquent supporters
for our point of view as possible.

However, unfortunately, I think your article will make no affect on
David {\Mermin}.  I know because I have tried exactly the same
argument as yours on him before (once in person last December, then
once in some email correspondence).  He rejected it then; I suspect
he'll reject it now.  In a separate mail, I'll send you an excerpt
from my ``Quantum Correspondence File'' with the appropriate
passages:  the most relevant piece of it is in the section titled
Merminition
\ref{OldAsm9}.

The problem hinges on what David means by the word ``correlation.''
His definition is fairly strict:  from his perspective, the
correlation between particles 1 and 4 does {\it not\/} change as a
result of the situation you describe.  For what he means by
``correlation'' is the set of all joint probability distributions for
``measurement outcomes'' (on systems 1 and 4 separately) that can be
derived from the {\it reduced density operator\/} of that joint
system.  Because the reduced density operator on the 1-4 system does
not depend on whether Experiment 1 or 2 is performed---in fact, it
doesn't depend at all on actions taken on the 2-3 system---the
``correlation'' between systems 1 and 4 does not change either.  For
David, it doesn't matter that the experiments you describe allow the
experimenter to update {\it his\/} quantum state for the 1-4 system.
Moreover, it doesn't matter that the experimenter can toggle between
different amounts of {\it entanglement\/} for those systems.  It
doesn't seem to matter to him that with the receipt of the classical
information gathered from your experimenter, Alice and Bob at systems
1 and 4 can ``activate'' different amounts of useful entanglement.
(For the whole spectrum of things that can be done in between these
extremes for a situation more general than the one you describe, you
might be interested in taking a look at our paper ``Entanglement of
Assistance'' {\tt quant-ph/9803033}.)

You may not like it that he is ultimately inconsistent in his use of
the term quantum state---i.e., not specifying with what/whom it is
respect to---but that is his usage, and there's not much we can do
about that. You also may not like his ignoring entanglement as a
reasonable notion of (potential) correlation---I didn't---but, again,
it's his prerogative to define his terms as he wishes.  In essence,
David says this himself in Section VIII of his paper ``What Is
Quantum Mechanics Trying to Tell Us?''.

In summary, though I liked your exposition, much of the force of your
argument is lost explicitly because in your Assumption (b) David is
only talking about the reduced density operator \ldots\ not the
correlations due to the quantum states determined from measurements
on other systems.

Thank you again for the opportunity---unbeknownst to you---for a
little thought!

\section{19 August 1998, \ to Richard {\Cleve}, \ ``The Word
Entanglement''}

The two best sources I know of on the history of the original EPR
argument and the subsequent development of the idea of entanglement
are: the article, Max {\Jammer}, ``The EPR Problem In Its Historical
Development,'' in {\sl Symposium on the Foundations of Modern
Physics: 50 Years of the {\Einstein}-{\Podolsky}-{\Rosen}
Gedankenexperiment}, edited by P. {\Lahti} and P. {\Mittelstaedt}
(World Scientific, Singapore, 1985), pp.\ 129--149, and the book,
{\sl The Shaky Game: {\Einstein} Realism and the Quantum Theory}, by
Arthur {\Fine} (U. Chicago Press, 1986).

Indeed {\Schroedinger} was the first person to use the word
``entanglement'' \ldots\ and he meant just exactly what we mean in
the modern sense.  The first appearance in print is with the words,
\bq
\noindent
When two systems, of which we know the states by their respective
representatives, enter into temporary physical interaction due to
known forces between them, and when after a time of mutual influence
the systems separate again, then they can no longer be described in
the same way as before, viz.\ by endowing each of them with a
representative of its own.  I would not call that {\it one\/} but
rather {\it the\/} characteristic trait of quantum mechanics, the one
that enforces its entire departure from classical lines of thought.
By the interaction the two representatives (or $\psi$-functions) have
become entangled.
\eq
This comes from his article ``Discussion of Probability Relations
Between Separated Systems,'' Proceedings of the Cambridge
Philosophical Society, vol. 32, pp. 555--563, 1935 (submitted 14
August, read 28 October).  He makes it clear that he is thinking of
the modern concept of entanglement in Section 2 of the article, where
he writes,
\bq
Let $x$ and $y$ stand for all the coordinates of the first and second
systems respectively and $\Psi(x,y)$ for the normalized
representative of the state of the composed system, when the two have
separated again, after the interaction has taken place.  What
constitutes the entanglement is that $\Psi$ is not a product of a
function of $x$ and a function of $y$.
\eq

According to {\Fine}, {\Schroedinger} was already saying these
things to {\Einstein} in a June 7 letter,
\bq
{\Schroedinger} continues by focusing on certain mathematical aspects
of the EPR example.  These have to do with the expansion of the state
function of a composite system into a bilinear series of functions
defined only on the component systems.  He points out that the
composite EPR case, after the interaction has effectively ceased, is
a very exceptional one in this regard, for there all the coefficients
of the bilinear expansion are identical. \ldots\
\eq
The original EPR paper appeared May 15 in Physical Review.

{\Schroedinger} continues with a much more detailed set of thoughts
on entanglement in his paper ``The Present Situation in Quantum
Mechanics,'' reprinted (and translated from German) in the book
edited by {\Wheeler} and {\Zurek} ``Quantum Theory and Measurement.''
It made its first appearance in Die Naturwissenschaften, vol.\ 23,
pp.\ 807--812, 824--828, 844--849, 1935.  It's worth reading if
you've got the time.

What is intriguing about the {\Jammer} article is that it makes it
clear that {\Einstein} had a rudimentary notion of entanglement in
his head as early as 9 July 1931.  This is exhibited in a letter from
{\Ehrenfest} to {\Bohr} on that date explaining a conversation he had
had with {\Einstein}.  {\Jammer} wrote:
\bq
\ldots\ what {\Einstein} had in mind is confirmed by a letter which
{\Ehrenfest} wrote to {\Bohr} on July 9, 1931.  As {\Ehrenfest}
reports, {\Einstein} uses the photon-box no longer to disprove the
uncertainty relation but ``for a totally different purpose.''  For
the machine, which {\Einstein} constructs, emits a projectile; well
after this projectile has left, a questioner can ask the machinist,
by free choice, to predict by examining the machine alone {\it
either\/} what value a quantity A {\it or\/} what value an even
conjugate quantity B would have if measured on the projectile. ``The
interesting point,'' continued {\Ehrenfest}, ``is that the
projectile, while flying around isolated on its own, must be able of
satisfying totally different non-commutative predictions without
knowing as yet which of these predictions will be made \ldots''
\eq

\section{01 October 1998, \ to Richard {\Beyler}, \ ``More {\Pauli}''}

It would be wonderful if you could convert your dissertation into MS
Word format.  Your effort will certainly benefit more than just me:
we write these words so that people will read them.  If you are
going to do the conversion, I will hold off contacting University
Microfilms for a while---just let me know when/if you've given up
and what I should do.

Quantum cryptography concerns the business of disseminating ``secret
key'' among communicators who may want to communicate privately at
some point in the future.  Once the secret key is in their hands,
and they have verified that no one else knows it, then they can rest
assured that their encryption scheme will be completely secure.  The
engine that powers it is the particular form of indeterminism
provided by quantum mechanics:  it gives a method by which the
complete secrecy of the key can be verified.  This is something that
cannot be done with classical physics.  I've dug up for you an old
list of popular articles that {\Gilles} {\Brassard} put together in
1994; I'll paste it below.

Often in these articles, you will see it expressed that the source of
quantum cryptography's efficacy is the ``{\Heisenberg} uncertainty
principle.''  But, mostly I think that comes from a lack of a very
deep understanding of quantum mechanics on these writers' parts.
Also it comes from taking the easy way out---i.e., attaching a quick
explanation in terms the writers think the readers will feel
comfortable with.  If you ask me, the real source of the efficacy is
what {\Pauli} so often called ``the lack of the detached observer''
in the quantum world.  Thus a little of my interest in {\Pauli}.  If
you take a real interest in this, I'll provide you with some
technical references later.

About {\Pauli}'s thought, my main sources are the following:
\begin{enumerate}
\item
W. {\Pauli}, {\sl Writings on Physics and Philosophy\/}
(Springer-Verlag, 1994).
\item
K. V. {\Laurikainen}, {\sl Beyond the Atom:~The Philosophical
Thought of Wolfgang {\Pauli}\/} (Spring\-er-Verlag, 1988).
\item
K. V. {\Laurikainen} and C. Montonen, eds., {\sl Symposia on the
Foundations of Modern Physics 1992: The Copenhagen Interpretation and
Wolfgang {\Pauli}\/} (World Scientific, 1993)
\end{enumerate}

{\Laurikainen} can be a bit annoying in his repetition and in his
hero worship, but still his articles are good (if meager) sources of
{\Pauli}'s more private thoughts.

I haven't run across any comparative studies of {\Pauli} and
{\Jordan} outside of the references I found in your paper.  If you
know of any newer material on these fellows' thoughts, please let me
know.

\section{13 October 1998, \ to Itamar {\Pitowsky}, \ ``Rational
{\Gleason}''}

Well I'm finally giving your paper ``Infinite and Finite {\Gleason}'s
Theorems \ldots'' the frontal attack it deserves.  I've worked my way
up to the top of page 224---where the proof of the regular version of
{\Gleason}'s theorem ends---and have been having a great time. (That
in spite of the myriad typos infused throughout the manuscript,
AARGH!)  In fact I'm giving a talk on it at this week's ``quantum
information journal club'' that we have here.  If you'd ask me for
one thing that {\it I\/} wish you had included, I would have to say a
motivational sermon or two for the graphs $G_1$ and $G_2$.  Who
ordered them?  What line of reasoning led to considering those
graphs in particular?  You haven't written a longer, more
expository, version of the paper where you do this, have you?

Let me run one speculative idea past you; tell me what you think.
One of the great things I'm going to learn---when I get all the way
through your paper(!)---is the interesting restrictions placed on
states even when they're defined on finite sets of rays.  My
question, then, is this:  How far do you think one has to go before
recovering the whole standard probability rule?  For instance, how
much is the continuum really needed for this problem?  Might one be
able to derive a full version of {\Gleason}'s theorem for Hilbert
spaces over the {\it rational\/} complex field?

\subsection{Itamar's Reply}

\bq
I don't think the theorem will remain true if you stick with a dense
set such as the rationals. Years ago I've shown a very bizarre
example to the contrary. I ``constructed'' an example of a three
dimensional function $f$ with values 0 and 1 with the following
property: If $x$ is ANY direction then  $f(x) + f(y) + f(z) = 1$ for
all directions $y$, $z$ orthogonal to $x$ (and to each other) {\it
except countably many}. The proof is based on the continuum
hypothesis (!!!) and is not very difficult. The set on which $f(x)+
f(y) + f(z) = 1$ is nonmeasurable and big and dense. The proof
appears in {\sl Phil.\ of Science\/} around 1985.  (I'm in Canada
right now so I don't have the reference, I can send you a copy when
I'm back in a couple of weeks.) There may be dense subsets on which
{\Gleason}'s theorem is true but they have to have a special
structure.
\eq

\section{19 October 1998, \ to John B.~{\Kennedy}, \ ``Tensor Products''}

Actually my question is pretty simple \ldots\ nothing to get too
intrigued about.  I read your paper ``On the empirical foundations of
the quantum no-signalling proofs'' (Phil.\ Sci., 1995) and enjoyed it
quite a bit. I think you're really on the mark with it!

At the end of the Introduction you say, ``One direction for future
research would be to canvas such systems [i.e., nonstandard axiom
systems] for clean justifications of the tensor product formalism.''
To what extent have you carried that out?  Have you found any
interesting tidbits (or even uninteresting ones, for that matter) in
that direction? Alternatively, or even better, have you written any
follow-up papers to this one that I might get hold of?  Has there
been any discussion in the literature about your paper?  Those are my
main questions.

One little point about some of the language you use.  In your summary
you write, ``Tensor Product Space: assumes the impossibility of
wholly new states in the combined system \ldots\ in short, no gain or
loss of states.''  I think that's a little bit of a dangerous
expression.  In a certain sense, we get loads more states when we
combine systems.  There are all the product states (the ones isolated
to each system alone), but then on top of that we have all the
entangled states!  The product states form a set of measure zero!
Moreover just look at the dimensionality change.  When one combines
systems of dimensions $x$ and $y$, the result is a system with
dimension $xy$.  These remarks, I'm sure, are trivial to you \ldots\
and that leads me to think you mean something else.  But I'm not
completely sure what that is.

One thing that might interest you in relation to the tensor product
combo rule, is that there is a certain sense in which it is only well
behaved for complex Hilbert spaces.  It is not such a nice thing for
real or quaternionic spaces.  Suppose that one means by a ``possible
state'' any density operator (Hermitian, positive semi-definite,
trace 1) on the Hilbert spaces of a given dimension.  Then, for
instance, if we take the tensor product of two real Hilbert spaces,
one of dimension $x$ and one of dimension $y$, the set of operators
constructible from (real) superpositions of tensor product operators
will not turn out to be equivalent to the set of ``possible states.''
The set of possible states is strictly larger:  for it consists of
all density operators constructible on a space of $xy$ dimensions.
The two things are not the same for real Hilbert spaces (though they
are for complex spaces). Another way to say it is the following.  For
separate isolated systems described by complex Hilbert space quantum
mechanics, we can completely reconstruct the overall state (if we
have many copies) from local measurements only.  We need only know
the measurement outcomes and the correlations between them.  For real
Hilbert space quantum mechanics, local measurements and the
correlations between the outcomes is not enough for state
reconstruction.  You can read about this in:
\bq
\noindent W. K. {\Wootters}, ``Local Accessibility of Quantum States,''
in {\sl Complexity, Entropy and the Physics of Information}, edited
by W.~H. {\Zurek} (Addison-Wesley, Redwood City, California, 1990),
pp.\ 29--46.
\eq

In general, I'm just interested sorting out the physical essence of
the tensor producting assumption and how that fits into a
Copenhagenistic framework of understanding quantum mechanics.

\section{28 October 1998, \ to Charles {\Seife}, \ ``Yeeks!''}

Let me first tell you my view of what defines ``quantum
teleportation.''  It's an affair that involves at least three
participants: Alice and Bob, who run the business, and Victor, their
customer.  Victor has some quantum system that he knows something
about; the description of what he knows is called a ``quantum
state.''  It's that description that he want's transferred from the
physical system he has, onto a system in the possession of Bob.
Perhaps Victor is sitting in New York, and he needs a physical system
in Paris to be describable in some particular way:  for instance, he
needs to know that there is at least one photon in Paris for which he
can ascribe a polarization angle equal to a binary encoding of his
{\it secret\/} middle name, Rumpelstiltskin.  When teleportation is
complete, it is that description that has been teleported.  People
often say that it is an unknown quantum state that is teleported, but
what that means is that it is unknown to Alice and Bob.  There always
has to be a Victor in the background, or there wouldn't be a hell of
a lot of use in running this business.  It's much like AT{\&}T:
they're more than willing to transport secret messages for you,
without ever snooping, as long as you pay your bill.  No customers,
no business.  This is what teleportation strives for. But there's a
little more to it.  Alice and Bob can't just be a courier service
like Federal Express; they're not allowed to physically transport the
actual system Victor gives them to its destination.  They are only
allowed to physically transport some ``side information'' that they
generate themselves.  I present this as if it's a restriction, but
actually it's an advantage---they advertise it.  This method of
getting the quantum state from one place to the other is what makes
their business special.  It means that they don't have to worry about
stormy weather screwing up Victor's package while in transit.  And on
top of that, Bob might even be subcontracting for Victor in that he
promises not to reveal his location to Alice.  These things can be
done as long as Alice and Bob have previously gotten together and
shared a little ``entanglement.''  This is the physical resource
their company relies on to stay in business.  Intel would be out of
business without silicon for their chip; AT{\&}T would be out of
business without copper for their wire.  Alice and Bob would be
(should be!) out of business without entanglement.

\section{03 November 1998, \ to Daniel {\Gottesman}, \ ``The
{\Bill} {\Wootters} Fan-Club Page''}

Well, I've finally gotten off my duff and written the reviews I
promised you:  you're right it is a pretty enjoyable process---I hope
it won't be too long before I do some more!   As you'll see below, I
can be a pretty wordy fellow.

\begin{center}
Review of:\\ William K. {\Wootters}, {\sl The Acquisition of
Information from Quantum Measurements}, \\ Ph.D. Thesis, The
University of Texas at Austin, 1980.
\end{center}

Lately I've been finding in my work in quantum information that I'm
starting to feel a little more like an engineer than a physicist.
{\sl Transactions on Information Theory\/} is after all published by
the IEEE.  I find it becomes easier and easier to forget the original
goal of my education---to understand nature itself \ldots\ regardless
of how we might make use of that knowledge.  I fall asleep thinking,
``What can I do that will impress IBM or AT\&T?''  There's got to be
more to our work than this!

In that regard, this little book is a bright star of hope.  For his
Ph.D. work, {\Wootters} had the courage to ask where quantum
mechanics itself might come from.  And more interesting than that,
he saw that the answer might have something to do with information
theory! This work sets out, for the most part, to derive the
standard quantum probability rule from a variational principle, one
that involves information.  For instance, could it be the case that
Nature is so constructed that the following game tells us something
deep?

Someone hands you many, many identical copies of an otherwise
completely unknown (pure) quantum state; on each shot, you perform
the same quantum measurement and tabulate the outcome.  The goal of
the game is to attempt to identify the unknown state from the
statistical information that so accumulates.  Since the information
is statistical, only some of it is useful for the task.  But, how
much?  Could it be the case that if the probability rule for the
measurement outcomes were anything other than the one we know and
love, then the rate at which useful information accumulates would not
be optimal?

Well, it is \ldots\ ALMOST\@.  I say, ``almost'' because it is true
for quantum mechanics with real Hilbert spaces, not for quantum
mechanics with complex Hilbert spaces.  Aarrgh!  Unfortunately,
things turn out to be not so clean for complex spaces, though perhaps
not all hope is lost ({\Wootters} has a chapter devoted to this
issue). Some crucial connection between quantum mechanics and
information theory is still missing.  There's still work to be done!

This book is a delight to read, and surely inspirational for getting
as far as it did.  I recommend it to all in our field for a little
reading just before sleep:  we should all dare to dream.

\begin{center}
Review of:\\ Eugene P. {\Wigner}, ``The Probability of the Existence
of a Self-Reproducing Unit,'' \\ in {\sl The Logic of Personal
Knowledge: Essays Presented to Michael Polanyi on his \\ Seventieth
Birthday\/} (Routledge \& Kegan Paul, London, 1961), pp.\ 231--238.
\end{center}

The no-cloning theorem first discussed by {\Wootters} and {\Zurek}
and (independently) by Dieks is now understood to be a significant
part of the foundation of quantum information theory.  But have you
ever explained it to another physicist and received a reaction of the
form, ``Is that it?  That's the big deal everyone is talking about?''
I have.  And it's no wonder:  the issue of no-cloning boils down to
almost an immediate consequence of unitarity---inner products cannot
decrease.  In fact, {\Wigner}'s famous theorem on symmetries even
shows that the group of time-continuous, inner-product preserving
maps on Hilbert space is strictly equivalent to the unitary group.
Therefore, it comes as quite a shock to see that {\Wigner} himself
just missed the no-cloning theorem!  In this paper, {\Wigner} tackles
the question, ``How probable is life?'' He does this by identifying
the issue of self-reproduction with the existence of the types of
maps required for the cloning of quantum states.  He doesn't tackle
the question of cloning for a completely {\it unknown\/} quantum
state head on, but instead analyses the ``fraction'' of unitary
operators on a tensor-product Hilbert space that can lead to a
cloning transformation for at least some states.  Nevertheless, he
states quite clearly that an arbitrary linear superposition of
clonable states ought also to be clonable. But this, of course,
cannot be.

I think this paper is quite interesting from the historical point of
view of our field:  it gives us an appreciation of the beauty and
simplicity of that little theorem in a way that simply learning about
it cannot provide.  It gets at the heart of something deep in very
present physical terms, terms that even a great mind like {\Wigner}'s
missed.

\begin{center}
Review of:\\ William K. {\Wootters}, ``Entanglement of Formation of
an Arbitrary State of Two Qubits''\\ Physical Review Letters, Vol.
80, No. 10, pp.\ 2245--2248 (1998).
\end{center}

This paper is a linear-algebraic tour de force.  It proves the
correctness of a surprisingly weird-looking closed expression for the
entanglement of formation of a general mixed state on two qubits. The
expression itself had been conjectured in an earlier paper by Hill
and {\Wootters}.  For those of you who hope to find a surprise or
two in the ``information functions'' of quantum
information---something, say, beside the same old, same old von
Neumann  entropy---this paper is a winner.  More importantly for our
field, however, are the nontrivial mathematical techniques this
paper introduces for issues just as this.  These techniques should
have a wider range of application, for instance, in obtaining exact
expressions for other entanglement measures.

\begin{center}
Review of:\\ Lane P. {\Hughston}, Richard {\Jozsa}, and William K.
{\Wootters},\\ ``A Complete Classification of Quantum Ensembles
Having a Given Density Matrix,'' \\ Physics Letters A, Vol.\ 183,
pp.\ 14--18, 1993.
\end{center}

Abner {\Shimony} likes to say that entanglement gives rise to
``passion at a distance.''  He does this because when Alice performs
a measurement on A of an entangled system AB, {\it something\/}
changes for B, BUT that change cannot be used for the purpose of
communication with a Bob at B.  If you ask me, this is language that
is just asking for trouble; it is language that is poised to confuse
a generation of new physicists.  Something indeed does change for B,
what Alice can {\it say\/} of it.  But it is nothing more than that;
to think that it is truly a physical change with respect to B
alone---especially one that is so contrived as to not lead to
communicability---is to open up a sink hole.  We are dealing here
with changes of states of knowledge.

Within this context, it is quite reasonable and quite interesting to
ask how many different ways Alice's knowledge can change.  Depending
upon which measurement Alice wishes to perform on A, there will be
any of a number of different state assignments for B that follow from
that.  What are they, and what are their probabilities?  This is the
main question addressed in this little paper.  It has a very clean
answer:  a state assignment can be created for B from a measurement
on A if and only if that state falls within a pure-state
decomposition of the marginal density operator of B.  Moreover, the
probabilities of Alice's measurement outcomes correspond precisely to
the probabilities of the ultimate state assignments.

This result, it turns out, is not of purely academic interest. It has
had a wide range of application in several more applied problems in
quantum information:  it is crucial for the proof that no quantum bit
commitment schemes exist, it plays a crucial role in proving the
exact expression for the entanglement of formation for two qubits,
and it is crucial for defining the notion of the entanglement of
assistance.  This theorem is one that all quantum information
theorists should have incorporated into their tool bag.

Can more be said?  Actually, it turns out that this result even has
some historical significance.  For, unknown to the authors above, the
same question was raised and even partially answered by Erwin
{\Schroedinger} in a 1936 paper!  [E. {\Schroedinger}, ``Probability
Relations between Separated Systems,'' Proc.\ Cam.\ Phil.\ Soc.\
{\bf 32}, 446--452 (1936).]

\begin{center}
Review of:\\ Richard {\Jozsa}, Daniel {\Robb}, and William K. {\Wootters},\\
``Lower Bound for Accessible Information in Quantum Mechanics,''\\
Physical Review A {\bf 49}, 668--677 (1994).
\end{center}

There is a sense in which much of quantum information theory reduces
to only a small subset of classical information theory.  For
ultimately there is always classical information going into a problem
(in terms of quantum-state assignments) and classical information
coming out of a problem (in terms of the results of measurement
outcomes).  In between there is just a very restricted class of
transition probabilities between the input and output alphabets---a
class that is absolutely miniscule in terms of the ones contemplated
by classical information theory.

However, there is a much more important sense in which quantum
information is something new, something sui generis.  Within it, we
can free ourselves from explicitly having to consider the ultimate
input and output:  transmitting quantum states for the sake of simply
transmitting quantum states has a certain beauty of its own.  For, at
the very least, it gives rise to some miraculous information
quantities that stand out in their own right.  It is hard to shake
the feeling that these quantities will not ultimately be just as
important as {\Shannon}'s original quantification of information.

The quantity of information that arises in this paper, the
subentropy, though it is ultimately motivated by a problem of
classical information transmission, is exactly of this flavor.  Who
would imagine that by replacing a trace with a determinant in a
rather abstruse expression of the {\Shannon} entropy one would find
something of any interest?  Moreover that that quantity would be the
answer to a well-defined, physically motivated problem?  No one I
know, except maybe these three guys.

The subentropy is a strange beast; here is how it arises.  Consider
the problem of extracting the most information possible about the
identity of an unknown (pure) quantum state.  What ``unknown'' means
in this context is that a set of states along with a listing of their
a priori probabilities---an ensemble---is given; it is just the
actual identity of the state in front of the observer that is not
known.  The maximal information that can be extracted via a
measurement is called the accessible information.  The subentropy of
the density operator of the given ensemble turns out to be a lower
bound on that information.  Moreover, the subentropy is the best
lower bound that depends only on the density operator---for a given
density operator, there is always an ensemble for which the
accessible information equals the subentropy.  What is really most
interesting is how the subentropy can be so similar to the von
Neumann entropy at the same time as being so different from it.  For
instance, the von Neumann entropy is bounded above by the logarithm
of the dimension of the Hilbert space; the subentropy, on the other
hand, is bounded above by a more complicated function of the
dimension with in turn is ultimately bounded above by one minus
Euler's constant.

What other uses will we find for this intriguing little quantity?
That remains to be seen.  But this paper is one to keep in the back
one's mind.

\section{27 November 1998, \ to Tim {\Ralph}, \ ``Quantum
Teleportation:~A QND View?''}

We have read with interest your comment ``Quantum Teleportation:~An
Operational View.''  Thank you for giving us that opportunity. But
our learning from you in this case is a double-edged sword. This is
because your writings show that you pretty much missed the boat by
analyzing teleportation through the language of conventional QND\@.
No doubt you did this because you were most familiar with that
language, but that really only gets in the way for this problem. The
problem is that the questions for which QND was invented are too
remote from the question at hand to be an efficient tool for its
analysis.  Quantum information is something new.  As such, its
operational analysis requires tools of a different sort than have
arisen in the old classical-communication and parameter-estimation
scenarios you are familiar with.  We think your comment proves this
point.

In this note, we plan to make this just as crystal clear as we can.

In quantum teleportation, the task set before Alice and Bob is to get
a {\it quantum state\/}---one that is {\it unknown\/} to them---from
one side of the laboratory bench to the other.  It is assumed at the
outset that the only resources available for this transmission are a
classical channel and some previously shared entanglement.  (By the
term ``classical channel'' we mean, strictly speaking, a channel that
allows only the transmission of some set of orthogonal quantum
states, not arbitrary superpositions of them.) The question of the
operational verification of teleportation becomes this:  how can we
know when Alice and Bob had to have used some of their entanglement
in the aid of their transmission?  If there are certain transmission
characteristics that they could not possibly have met without the use
of their previously shared entanglement, then that will be the sure
signature.  Meeting the minimal such signature is what we call
``unconditional quantum teleportation.''  On these last three
sentences  it seems we all can agree.

Where our differences appear to hinge is in the italicized words that
start the last paragraph: {\it quantum state\/} and {\it unknown}.
Let us go at them one at a time.

What does it mean to get a quantum state from one side of the bench
to the other with minimal distortion?  We chose to gauge this by the
``fidelity'' between the input and output.  To be precise, this is
defined in the following way.  If the input state---as known by some
third party Victor---is $|\psi_{\rm in}\rangle$ and the output is
(generally a mixed-state density operator) $\rho_{\rm out}$, then the
fidelity is given by
\be
F=\langle\psi_{\rm in}|\rho_{\rm out}|\psi_{\rm in}\rangle\;.
\label{CoffeeBean}
\ee
This measure has the nice property that it equals 1 if and only if
$\rho_{\rm out}=|\psi_{\rm in}\rangle\langle\psi_{\rm in}|$. Moreover
it equals 0 if and only if the input and output states can be
distinguished with certainty by {\it some\/} quantum measurement. The
thing that is really important about our particular measure of
``fidelity'' is these last two properties.  It captures in a single,
simple and convenient package the extent to which {\it all
possible\/} measurement statistics that can be produced by the output
state---if a measurement were performed on it(!)---will match what
the input would have specified.

For instance, take {\it any\/} observable (generally a positive
operator-valued measure or POVM) $\{E_\alpha\}$ with measurement
outcomes $\alpha$. (What you call ``symmetric detection'' is an
example of this where
$E_\alpha=\frac{1}{\pi}|\alpha\rangle\langle\alpha|$ and the
$|\alpha\rangle$ are coherent states, but it could be any observable
whatsoever:  photon counting, measuring a single quadrature, or what
have you.)  If that observable were performed on the input system, it
would give a probability density for the outcomes $\alpha$ given by
\be
P_{\rm in}(\alpha)=\langle\psi_{\rm in}|E_\alpha |\psi_{\rm
in}\rangle\;.
\ee
If the same observable were performed on the output system, it would
have instead given a probability density of
\be
P_{\rm out}(\alpha)={\rm tr}(\rho_{\rm out} E_\alpha)\;.
\ee
One can gauge the similarity of these two densities by their
statistical overlap:
\be
{\rm overlap}=\int \sqrt{P_{\rm in}(\alpha)P_{\rm out}(\alpha)} \,
d\alpha\;.
\label{TeaTime}
\ee
It turns out that regardless of what observable we are talking about,
\be
{\rm overlap}^2\ge \langle\psi_{\rm in}|\rho_{\rm out}|\psi_{\rm
in}\rangle\;.
\ee
Moreover there exists an observable that will actually give precise
equality in this expression.\footnote{For the original proof of this
inequality and an expression of the observable that gives it
equality, see Fuchs and {\Caves}, ``Mathematical Techniques for
Quantum Communication Theory,'' Open Sys.\ and Info.\ Dyn.\ {\bf 3},
345--356 (1995).  Or you can find it in the more readily available
paper: {\Barnum}, {\Caves}, Fuchs, {\Jozsa}, and {\Schumacher},
``Noncommuting Mixed States Cannot Be Broadcast,'' Phys. Rev. Lett.
{\bf 76}, 2818--2821 (1996).}  So, it is in this sense that the
fidelity captures a fact about all possible measurements.

What now about the measures of transmission quality that you would
have your readers consider?  Well they are certainly good measures of
something that has been of great interest in QND, but they hardly
capture the essence of good teleportation.  In fact there is a sense
in which they only capture an infinitesimal bit of it:  the upshot is
that you are gauging teleportation quality by the performance of {\it
just two\/} of the possible observables in Eq.~(\ref{TeaTime}).  From
our point of view that is an immediate handicap for your criterion.
And this remains, regardless of how it may be used in combination
with any other ingredients.  For true-quality teleportation, one
should transfer {\it every\/} aspect of the unknown quantum state
from one side of the bench to the other (at least to within some
tolerance).  With this point, it is easy to see the trouble spot in
your criterion. The problem is that two state vectors can be
completely orthogonal---and therefore just as different as they can
possibly be---and still give rise to the same $x$ statistics {\it
and\/} the same $p$ statistics.

For an easy example of this, just consider the two state vectors
$|\psi_+\rangle$ and $|\psi_-\rangle$ whose representations in
$x$-space are
\begin{equation}
\psi_\pm(x)= \left(\frac{2a}{\pi}\right)^{\! 1/4}\exp\!\left((-a\pm i
b)x^2\right),
\end{equation}
for $a,b\ge0$.\footnote{This nice example was provided by Jason
McKeever, one of our new graduate students at Caltech.}  In $k$-space
representation, these state vectors look like
\begin{equation}
\tilde\psi_\pm(k)=\left(\frac{a}{2\pi}\right)^{\!
1/4}\sqrt{\frac{a\pm ib}{a^2+b^2}}\, \exp\!\left(\frac{-a\mp
ib}{4(a^2+b^2)}k^2\right).
\end{equation}
Clearly neither $x$ measurements nor $p$ measurements can distinguish
these two states; for in both observables both wave functions differ
from each other only by a local phase function. However, if we look
at the overlap between the two states we find:
\begin{equation}
\langle\psi_-|\psi_+\rangle=\sqrt{\frac{a(a+ib)}{a^2+b^2}}\;.
\end{equation}
By taking $b\rightarrow\infty$, we can make these two states just as
orthogonal as we please. Suppose now that $|\psi_+\rangle$ were
Victor's input into the teleportation process, and---by whatever
means---$|\psi_-\rangle$ turned out to be the output.  By your
criterion, this would be perfect teleportation.  But it certainly
isn't so!

Of it course it is clear that you're trying to get at something else.
But the example above, contrived though it is, already shows a
certain weakness in your criterion. It is true, as you point out,
that a {\it cheating\/} Alice cannot simultaneously get hold of two
conjugate observables.  This is just as true for these two state
vectors as for any others---the uncertainty principle is a fact of
life.  But so what of it?  What does it necessarily have to do with
the ability to detect a cheating Alice and therefore define an
operational criterion for experimental teleportation?

Just to give you a far-fetched example, let us make use of the
(almost orthogonal) states $|\psi_+\rangle$ and $|\psi_-\rangle$ once
again.  Consider the case where Alice and Bob are privy to the fact
that Victor wishes to teleport these and only these states.  At any
shot, they know that Alice will be given one of these states; they
just don't know which one.  Then, clearly, they need use no
entanglement whatsoever to get the quantum states across the bench.
A cheating Alice need only perform some measurement $\cal O$ with
eigenstates very close to $|\psi_+\rangle$ and $|\psi_-\rangle$ and
send that classical information to Bob over the classical channel.
Bob simply uses that information to resynthesize the appropriate
state at his end.  No entanglement has been used, and yet your
criterion doesn't catch them in the cheat.

Well, of course, our criterion of fidelity in Eq.~(\ref{CoffeeBean})
doesn't catch them either.  But this example will help us define the
second problem in your comment much more sharply.  The issue is this:
What does it mean to say that Alice and Bob are given an {\it
unknown\/} quantum state?  If you dig deep, you'll see that there's
about only one useful meaning that can be given to this phrase.  It
means that Alice and Bob know that Victor will be drawing his states
$|\psi_{\rm in}\rangle$ from some fixed, given set $\cal S$.  Alice
and Bob's lack of knowledge about which state Victor will actually
draw is necessarily quantified by some probability ascription
$P(|\psi_{\rm in}\rangle)$. That's what an unknown quantum state
means in the most general setting.

The last example above forces the following statement upon us:
\begin{quote}
\it
All useful criteria for ``unconditional quantum teleportation'' must
be anchored in whatever $\cal S$ and $P(|\psi_{\rm in}\rangle)$ are
given.  A criterion is senseless if the states to which it is to be
applied are not mentioned explicitly.
\end{quote}
If $\cal S$ consists of orthogonal states, then no criterion
whatsoever (short of watching Alice and Bob's every move) will ever
be able to draw a distinction between true teleportation and the sole
use of the classical side channel.  Things only become interesting
when the set $\cal S$ consists of two or more {\it nonorthogonal\/}
quantum states.

With that caveat, one can start to make a little sense out of why you
might have been led to the particular criterion that you and Lam
proposed.  From our first days as a physics majors we are taught a
mantra:  Measurement causes disturbance.  But then almost immediately
all our instructors botch it by saying things like, ``Measuring $x$
necessarily disturbs $p$ and vice versa.''  What could it possibly
mean to say this when we know from the observed {\Bell} inequality
violations that not both $x$ and $p$ can exist with simultaneously
definite values?  How can something be disturbed when it's not even
there?  Fortunately quantum information and quantum cryptography in
particular provide us a way out of that historical conundrum.  When
one is given an unknown state drawn from a set of nonorthogonal
states, it is the gathering of information about the state's identity
that causes a necessary disturbance.  But that disturbance has
nothing {\it a priori\/} to do with conjugate {\it classical\/}
variables.  Instead the disturbance is to the state
itself.\footnote{If you would like a lengthier discussion of this
issue, have a look at:  C. Fuchs and A. {\Peres}, ``Quantum State
Disturbance vs.\ Information Gain:~Uncertainty Relations for Quantum
Information,'' Phys.\ Rev.\ A {\bf 53}, 2038--2045 (1996) or C.
Fuchs, ``Information Gain vs.\ State Disturbance in Quantum Theory,''
Fort.\ der Physik {\bf 46}, 535--565 (1998).}  Herein lies the key to
a good teleportation criterion.

We, in particular, used Eq.~(1) to define a notion of average
fidelity between input and output:
\be
F_{\rm av} = \int_{\cal S}P(|\psi_{\rm in}\rangle)\, F\Big(|\psi_{\rm
in}\rangle,\rho_{\rm out}\Big)\, d|\psi_{\rm in}\rangle\;.
\ee
Depending upon the precise states in $\cal S$ and the given form of
$P(|\psi_{\rm in}\rangle)$ there will be any number of different
maximum values that $F_{\rm av}$ can take on in the presence of a
cheating Alice and Bob (i.e., an Alice and Bob who only use their
classical channel and make \underline{no} use of their entanglement).
Only one thing is for sure:  if $\cal S$ contains some distinct
nonorthogonal states, a cheating Alice and Bob can never achieve
$F_{\rm av}=1$.

Let us drive this home before returning to a comment about your
criterion.  Suppose ${\cal S}=\{|\psi_0\rangle\}$ consists of just
one state.  Then a cheating Alice and Bob can certainly achieve
$F_{\rm av}=1$.  Alice needs perform no measurement nor send any
classical information in order to do this:  Bob only has to
synthesize the single fixed state for his output at the correct
times.  One way of saying this is that our teleportation criterion is
so ridiculously nonstringent that it is trivial to achieve.

So let us consider a more stringent criterion.  Take ${\cal S}=
\{|\psi_0\rangle,|\psi_1\rangle\}$ to consist of just two
nonorthogonal states (with inner product $x$).  Suppose the two
states are deemed to occur with equal probability.  Then it can be
shown\footnote{Fuchs and {\Peres}, op. cit.} that the best thing for
the cheating duo to do is this.  Alice should measure an operator
whose orthogonal eigenvectors symmetrically bestride
$|\psi_0\rangle$ and $|\psi_1\rangle$.  She then sends that
information to Bob.  He synthesizes two states
$|\tilde\psi_0\rangle$ and $|\tilde\psi_1\rangle$ depending upon the
outcomes, being careful to tweak them so that the overlap between
input and output is as high as it can possibly be on average. (In
this case that means that Bob's states will have a slightly larger
inner product than the original two states.)  When all that is said
and done, the best fidelity that can be achieved in this way is
\be
F_{\rm av}=\frac{1}{2}\!\left(1+\sqrt{1-x^2+x^4}\right)\,.
\label{herpolhode}
\ee
Even in worst case (when $x=1/\sqrt{2}$), this fidelity is always
relatively high---it is always above 0.933.

What does this mean?  Again it means that the criterion of
unconditional teleportation is awfully weak.  It is so easy for Alice
and Bob to cheat when they know that their input is only one of two
possibilities that the average fidelity can just about be made to
approach one with hardly any effort.

As we pile more and more states into $\cal S$ it becomes ever harder
for Alice and Bob to cheat.  But even then things can still be
nontrivial for finite-dimensional Hilbert spaces.  Consider the case
where $\cal S$ consists of every normalized vector in a Hilbert space
of dimension $d$ and assume $\cal S$ equipped with the uniform
probability distribution.  Then it turns out that the maximum value
$F_{\rm av}$ can take on is\footnote{H. {\Barnum}, {\sl Quantum
Information}, Ph.D. thesis, University of New Mexico, 1998.}
\be
F_{\rm av}=\frac{2}{d+1}\,.
\ee
For the case of a single qubit, as in the original teleportation
scheme of {\Bennett} and company, Alice and Bob would {\it only\/}
have to achieve a fidelity of $2/3$ before they could claim
unconditional quantum teleportation for Victor.  BUT, again, that is
only if Victor can be sure that Alice and Bob know absolutely
nothing about which state he will input other than the dimension of
the Hilbert space it lives in.

This last case tells us something very nice about the teleportation
of continuous-variable quantum states.  For this essentially
corresponds to the limit $d\rightarrow\infty$ above.  If Victor can
be sure that Alice and Bob know nothing whatsoever about the quantum
states he intends to teleport, then on average the best fidelity they
can achieve in cheating is strictly zero!  In this case, seeing any
nonzero fidelity whatsoever in the laboratory would signify that
unconditional quantum teleportation had been achieved.

To use this last criterion for our experiment here at Caltech,
however, would have been a little ridiculous.  For any reasonable
Alice and Bob that had wanted to cheat would know that it is pretty
difficult for the Victor using their services to create nonclassical
states for his light.  Therefore we explicitly made the assumption
for our case that $\cal S$ contains the coherent states
$|\alpha\rangle$ with a gaussian distribution describing the
probability density on that set (it matters not where the mean of the
gaussian is placed).  It turns out that in the limit that the
variance of the gaussian approaches infinity, i.e., the distribution
of states becomes ever more uniform, the {\it very best possible\/}
average fidelity that can be achieved by a cheating Alice and Bob is
\be
F_{\rm av}=\frac{1}{2}.
\label{polhode}
\ee
Or to say it one more time, if Alice and Bob know that Victor will be
handing off a coherent state but they have no clue which one it will
be, then the best fidelity they can get upon cheating is $1/2$.
``Very best possible'' means that one must consider all measurements
allowed within quantum mechanics for Alice's part and all possible
translations of that classical information into quantum-state
synthesis for Bob's part.

\section{17 April 1999, \ to Salman Habib, \ ``Tightivity / Quantum
Sensitivity''}

Well it's a little tight, but I believe I said everything I've ever
wanted to say \ldots\ on a single page!  (Shows how little I have to
say.)

\bq
\begin{center}
Quantum Sensitivity and Control
\end{center}
\medskip

Our world is a quantum mechanical one.  What should we make of that?
In a way, the answer to this question was once less positive than it
is today.  For though quantum theory is a tool of unprecedented
success in understanding the phenomena about us, the usual
intellectual lesson derived from it is one of limitations. This is
seen in almost any introductory lecture on the {\Heisenberg}
uncertainty principle: it is presented as if the world had
short-changed us. ``The task of physics is to sober up to this and
make the best of it.''

But that is a lesson of the past.  Recent years have seen the start
of a significantly more positive, almost intoxicating, attitude about
the basic role of quantum mechanics. This is evidenced no more
abundantly than in the field of quantum information and computation.
Its point of departure is not to ask what limits quantum mechanics
places upon us, but instead what novel, productive things we can do
in the quantum world that we could not have done otherwise. In what
ways is the quantum world fantastically better than the classical
one?

Perhaps the two most striking examples are quantum key distribution
and {\Shor}'s quantum factoring algorithm. In the first, quantum
mechanics allows communicators to transmit cryptographic keys in a
way that (hidden) eavesdropping on them can be excluded out of hand.
This is impossible in the classical world because no connection
exists between the information gatherable about a physical system and
the disturbances induced upon it in the process. In the second, one
finds that algorithms designed for computers built of unabashedly
quantum components can factor large integers exponentially faster
than anything designed for classical computers. To give a quick
instance, consider a 600-digit number that is known to be the product
of two (secret) primes.  The number of computational steps required
of a classical computer to crack it into its two components is
roughly $10^{34}$. In contrast, the corresponding number of steps for
a quantum computer is only $10^{11}$.  Quantum computing gives 23
orders of magnitude greater efficiency here!

These two examples are the most outstanding of the class, and there
is well-founded hope that they are the tip of a technological
iceberg.  There is no doubt, however, that they are the tip of a
physical iceberg.  What is being uncovered in a {\it rigorous\/} way
for the first time since quantum theory's inception is a sense in
which the quantum world is wildly more ``sensitive to the touch''
than the classical world: irritate it in the right way and the result
is a pearl. One has to wonder if this itself is not the very essence
of quantum mechanics.

Exploring three well-defined aspects of this sensitivity is the
underpinning of my research program.  Information, disturbance,
entanglement.  Let me say a word about each.

{\it Information}.  People encode information in the states of
quantum systems for a reason: to get it from one place to another.
Since the world is quantum mechanical, this is, in the last analysis,
what is always done for information transfer.  Once this is
recognized forthright, hosts of issues arise that have no analog in
classical information theory. Key among these are ways of coding
information into {\it nonorthogonal\/} quantum states so that it {\it
cannot\/} be retrieved with complete reliability. What are the
ultimate limits on that retrievability? This question defines a field
of research in itself.  Why bother? It turns out that for noisy
channels it is better to take a hit in retrievability at the outset
before transmission---the information becomes more resilient to noise
in the long run. Understanding the ``capacity'' of noisy channels is
the greatest of open problems. Solving it would put quantum
information theory where the classical theory was in 1948.

{\it Disturbance}.  Nonorthogonal states are not only remarkable in
that {\it some\/} of the information they encode is lost forever, but
in that they are magical canaries for the theft of {\it any\/} of the
remaining. Measurements on nonorthogonal states cause detectable
disturbances. This is the engine that powers quantum
cryptography---an engine only now beginning to be understood. What is
the ultimate tradeoff between information gain and state disturbance
for nonorthogonal states? What is the ultimate secrecy capacity---the
fraction of secret bits per transmission---that can be achieved with
such alphabets?  These are technical questions that will find solid
answers in the near term.  They are also the beginning of a deep set
of foundational questions. Is there a sense in which the structure of
quantum mechanics is optimal for these numbers? Much work remains to
be done.

{\it Entanglement}.  In 1935 {\Schroedinger} coined the term and said
of it, ``I would not call that {\it one\/} but rather {\it the\/}
characteristic trait of quantum mechanics.'' It is the stuff
{\Einstein}-{\Podolsky}-{\Rosen} pairs and {\Bell} inequalities are
made of. On the one hand, entanglement is the very embodiment of the
information-disturbance principle---the quantum sensitivity to touch.
On the other, it is a phenomenon {\it sui generis}. Most distinctly,
it can be viewed as a physical resource in its own right, lending
itself to information applications of every sort.  The greatest issue
is how to quantify entanglement and, further, to distill how such
measures connect with the previous two categories. Progress along
these lines is inevitable.

The T-8 group at LANL offers a unique set of resources for the
continuance and expansion of the research just described.
Information, disturbance, and entanglement all enter in crucial ways
into the theory of quantum control systems.  Environmental
perturbations become entangled with the system to be controlled. To
stave that off, control systems must gather information about the
perturbation.  But then the system is perturbed further in the
process. Ultimately one must understand the interplay and
interrelation of all these factors to make the best of them---to
irritate the quantum world in just the right way.  Working on this
project would be a grand adventure.  Thank you for your
consideration.
\eq

\section{08 June 1999, \ to {\Todd} {\Brun}, \ ``Information Theoretic
Entanglement''}

\btb
As you might have guessed, I have a question.  I've recently been
toying with ideas of quantum information in consistent histories,
with the vague intent of trying to understand what it is.  Most
recently, I've been looking at entanglement.

A definition of entanglement that occurs fairly naturally in CH is
the following (translated into standard QM terms):

``Entanglement is quantified by the difference between the minimum
  missing information in a general measurement and the minimum missing
  information in a set of local measurements (suitably defined).''
\etb

What a coincidence that you should ask that just now!  Within the
last few days I (along with a student here, Bob {\Gingrich}, and
Michael {\Nielsen}) have been readdressing just that question.
Unfortunately things don't look so well.  As you know, ``the minimum
missing information in a general measurement'' is identical to the
von Neumann entropy of the density operator under consideration (so
long as the general measurements are restricted to be rank-one
POVMs).  So one might consider comparing $S(\rho_{AB})$ to
$S(\rho_{A})$ and/or $S(\rho_{B})$ to get at a measure of
entanglement.  Indeed a good start is to think about
$S(\rho_{AB})-S(\rho_{A})$.  If the state $\rho_{AB}$ is separable
(i.e., violates {\it no\/} {\Bell} inequalities, or equivalently can
be written as a convex combination of product states), then that
quantity is always positive.  However, for a lot of entangled states
it goes negative. Unfortunately it does not go negative for all
entangled states.

Nicolas {\Cerf} and Chris {\Adami} have put in a lot of time talking
about the quantity
$$
S(\rho_{A}\otimes\rho_{B})-S(\rho_{AB}) =
S(\rho_{A})+S(\rho_{B})-S(\rho_{AB}).
$$
and calling that a ``measure of entanglement'' (at least in their
early papers).  But almost all of that work is casuistry as far as I
am concerned:  this quantity too makes no unique break between
separable and nonseparable.  It ranges all over the map for both
kinds of states.

You gave yourself some good leeway by saying ``the minimum missing
information in a set of local measurements (suitably defined) \ldots\
(Only one party?  Two uncorrelated?  Two correlated?  Any product
state?).'' The point of the remark above is that two-party
uncorrelated seems no good.  Two-party correlated may be better, but
how to get a clean characterization of all such things?  The
difficulty with this is what our nine-state eight-author paper is
about ({\tt quant-ph/9804053}).  POVMs consisting of product states?
The problem with this is that I don't know what meaning can be given
to such a construction (again because of the considerations in the
nine-state paper).

The reason I've become interested in this question again is mostly
because I'd like to give another forward thrust to an old idea of
mine. It is that ``entanglement'' itself is a shorthand for a more
fundamental situation:  namely that information gathering (about what
someone else knows) and disturbance (to what they can predict) go
hand in hand in the quantum world.  The theory of quantum mechanics
is our best attempt at synthesizing what can be said in light of that
situation.  ``I know something about the system in front of you, how
it will react (statistically) when I prod it this way or that.  And I
know something about you---how you will react statistically when I
prod you this way or that---and I even know that you will try to
learn something about what I know of the system.  BUT after you have
carried that project through, I know that I can say less of the
system than I could have before.  And indeed I can say less of you
than I could have before.''  That is what entanglement IS ABOUT
\ldots\ or at least this is what is at trial in my mind.  (I put a
little more meat on these bones in the letters to David {\Mermin}
dated 15 February 1998, 04 September 1998, and 08 September 1998
\ldots\ just in case you're not immediately seized with the desire
to call this rubbish.)

In this connection, it is my present hope that one can use this as a
background to give some substance to (and, more importantly, extend)
{\Schroedinger}'s 1935 remark:
\begin{quote}
This is the reason that knowledge of the individual systems can {\it
decline\/} to the scantiest, even to zero, while that of the combined
remains continually maximal.  Best possible knowledge of a whole does
{\it not\/} include best possible knowledge of its parts---and that
is what keeps coming back to haunt us.
\end{quote}

Let me give you a couple of pointers for where to do a little reading
about entanglement (from an information theoretic point of view):
\begin{enumerate}
\item
R.~{\HorodeckiR} and M.~{\HorodeckiM}, ``Information-Theoretic
Aspects of Inseparability of Mixed States,'' Phys.\ Rev.\ A {\bf
54}, 1838--1843 (1996).
\item
M.~{\HorodeckiM}, P.~{\HorodeckiP}, and R.~{\HorodeckiR},
``Separability of Mixed States: Necessary and Sufficient
Conditions,'' Phys.\ Lett.\ A {\bf 223}, 1--8 (1996).
\end{enumerate}
The second of these papers is also on {\tt quant-ph}; the first
appears not to be.

I hope that's a little helpful.  If you have any good ideas or
thoughts, please keep me informed!

\section{03 August 1999, \ to Alan {\Fowler}, \ ``{\Landauer} Memorial''}

I apologize for not having replied to your invitation to Rolf's
memorial meeting until now, but as it turns out today was my first
day back at the office after being away for six weeks.

Very unfortunately, I must decline your warm invitation; my schedule
is already booked for the days of Sept.\ 13 and 14.  However, I would
still like to make a contribution to the meeting if in some way the
opportunity arises.  Below, I paste a part of one of my email
exchanges with Rolf.  It stands out in my mind because of the sheer
sweetness with which he won me over \ldots\ just before telling me
that I didn't understand him at all.  In return, I spanked him a bit
by pointing out that he was fighting a battle with someone who was
not me.  Rolf's final response was short but gracious and, I think,
indicative of his personality as a whole.  [See correspondence with
Rolf {\Landauer} collected herein.]

He was a good man, and I will miss him.

\section{18 August 1999, \ to Rob {\Spekkens}, \ ``Modal Stuff''}

Please do send me a copy of your paper as soon as you get it
written.  I want to devote more attention to understanding what I
like/dislike about the modal point of view.  I haven't thought about
it that much before.

A few things crop up in my mind that I want to record now.

1)  Are the probabilities in the decomposition objective or
subjective?  If subjective, from whose point of view?  If objective,
then define explicitly what an objective probability might mean.

2)  Why is it that the universal wave function in your theory plays
the role of a compendium of possible worlds (along with their
associated probabilities), but the other wave functions---the ones
cropping up in your preferred decomposition---correspond to actual
properties?  Why would two things with such different ontological
status be described by mathematical objects with indistinguishable
mathematical structures?  (They're both given by vectors in a vector
space.)

3)  Why a decomposition into $d$ orthogonal components?  What
physical criterion singles out that number?  I know you do it
because you want an easy rule to get to the probabilities.  But what
dismisses, for instance, always decomposing the state into $d^2$
components (all of which when turned into projectors are linearly
independent in the operator sense or some other such rule) and using
a more contrived method for distilling probabilities?  What compels
you to making the assumptions you do?  Why such a blatantly minor
variation on {\Everett}?

4)  Why minimize the {\Shannon} entropy {\it per se}, other than the
herd effect?  Why not any of the Renyi entropies?  Why not the
Bayesian probability of error for inferring the initial state via
knowledge of the final state?  (This of course would require you to
look at the universal wave function at two different times when
ferreting out your decompositions \ldots\ and would thus be harder.
But if you're trying to  get at a certain stability property --- the
world sort of hops about the least --- this to me seems to be better
motivated than raw {\Shannon} entropy.)

If you write me back, I'm sure to think about and mull over your
answers.  But there's no guarantee that I'll write back until I get
the itch again.

\section{23 August 1999, \ to Sylvan {\Schweber}, \ ``Living with Law
without Law''}

I have just had the most wonderful time reading your two articles:
\begin{enumerate}
\item S.~S. {\Schweber}, ``Physics, Community and the Crisis in Physical
Theory,'' in {\sl Physics, Philosophy, and the Scientific Community},
edited by K.~Gavroglu, J.~Stachel, and M.~W. Wartofsky (Kluwer,
Dordrecht, 1995), pages 125--152.

\item S.~S. {\Schweber}, ``The Metaphysics of Science at the End of a
Heroic Age,'' in {\sl Experimental Metaphysics: Quantum Mechanical
Studies for Abner {\Shimony}, Vol.~1}, edited by R.~S. Cohen,
M.~Horne, and J.~Stachel (Kluwer, Dordrecht, 1997), pages 171--198.
\end{enumerate}
and starting a search on many of your references cited therein.

What I would like to know is if you have written anything further
expanding on the theme in these two articles?  If you have, I would
appreciate learning of their coordinates or perhaps receiving
reprints/preprints.

I am a physicist at Caltech working in quantum information theory
and quantum cryptography.  I have had a long-standing interest in
John {\Wheeler}'s ideas on ``law without law''---that is how your
articles attracted my attention.  (I'm giving a plenary talk on
quantum information at the New England Section Fall APS meeting this
November; perhaps I'll meet you there and we can talk about all this
much more?)

\section{24 August 1999, \ to Pawel {\Wocjan}, \ ``Quantum Theory from
QKD''}

I've finally found a few moments to think about some of the things I
discussed with you.  Namely, that of all theories with a trade-off
between information and disturbance, quantum theory might just be the
optimal one.  I did a search through my files and found three papers
that might be of relevance for such a project:
\begin{enumerate}
\item
I.~Csisz\'ar and J.~K\"orner, ``Broadcast Channels with Confidential
Messages,'' IEEE Trans.\ Inf.\ Theory {\bf IT-24} (1978) 339--348.
\item
R.~Ahlswede and I.~Csisz\'ar, ``Common Randomness in Information
Theory and Crypto\-graphy---Part I: Secret Sharing'' IEEE Trans.\
Inf.\ Theory {\bf 39} (1993) 1121--1132.

and, maybe most importantly,

\item
U.~Maurer, ``Secret Key Agreement by Public Discussion from Common
Information'' IEEE Trans.\ Inf.\ Theory {\bf 39} (1993) 733--742.
\end{enumerate}

If these papers exist, then probably even more relevant things can be
found in the literature.  But these may be a good starting point for
some thought.  (And they may form a good base for a literature search
for some more up-to-date things.)

If you have any ideas or thoughts, I would love to hear them.

\section{24 August 1999, \ to Dominik {\Janzing}, \ ``{\Bopp} Stuff''}

About {\Bopp}, if it is Friedrich (Fritz) Arnold {\Bopp} that you are
speaking of, then I have found a little bit about him.  Max
{\Jammer}, in his book {\sl The Philosophy of Quantum Mechanics},
has a few pages devoted to him.  He writes, ``{\Bopp}'s theory of
which only a very brief outline could be presented seems to have
found, in spite of its logical and philosophical consistency, only a
very limited amount of interest among physicists and
philosophers.''  Almost all the references are in German except
one:  F.~{\Bopp}, ``The Principles of the Statistical Equations of
Motion in Quantum Theory,'' in {\sl Observation and Interpretation
in the Philosophy of Physics}, edited by S. K\"orner (Dover, NY,
1957), pp.\ 189--196.  As it turns out, I have that volume at home;
so I will have a look at the paper. {\Jammer} also cites the
following paper: W.~M. Machado and W.~Sch\"utzer, ``{\Bopp}'s
Formulation of Quantum Mechanics and the
{\Einstein}-{\Podolsky}-{\Rosen} Paradox,'' Anais da Academia
Brasileira de Ciencias {\bf 35} (1963) 27--35.  I would like to get
my hands on that article, but it may be pretty hard to.

\section{07 January 2000, \ to Jeremy {\Butterfield}, \ ``Page Numbers
for {\Shimony}''}

Could you do me a favor?  {\Shimony} sent me a copy of his paper
``Can the Fundamental Laws of Nature Be the Results of Evolution?''
appearing in your book {\sl From Physics to Philosophy\/}:  I would
like to enter the paper into my database, but I don't have the page
numbers.  Could you please give me both the beginning and ending
page numbers for the article?

In case the book hasn't made it to press yet, here's one typo that
you may have overlooked.  {\Shimony}'s spell checker must not have
known the word ``observership,'' for the word ``ownership''
consistently appears in its place (there are at least two
instances).  For instance, {\Wheeler}(1977) is listed as ``Genesis
and Ownership'' instead of ``Genesis and Observership.''

\section{07 February 2000, \ to Eugen {\Merzbacher}, \ ``Missing
Copenhagen''}

Aha!  You ask me a new question, but you don't answer my old one?
Perhaps I should try to extract some leverage with this \ldots!  Oh,
I suppose I'll just place it below to prod your memory and leave it
to your conscience.

You said,
\bq
Fortunately, your piece leaves a number of issues untouched or
unresolved. For a long time I have thought about writing an essay on
what is meant by the term ``physical system'', including some history
of that slippery concept.  I'll probably never do it, but your
opinion piece has brought this matter back to my mind.
\eq
and I replied with
\bq
I would be interested in hearing your thoughts about the ``slippery
concept of physical system.''  One of the things that fascinates me
now is our freedom to decompose a large Hilbert space into any tensor
product decomposition we wish.  We label the components in that
decomposition to be ``systems,'' but that is imposed by us, not
Nature.  In what way do you mean ``slippery?''
\eq

Thanks for sending me the information about the play {\sl
Copenhagen}. I had tried to get to London to see it this summer when
I was in Cambridge for six weeks, but things never panned out.  My
colleague {\Ruediger} {\Schack} saw it and said that it's marvelous.

\section{27 July 2000, \ to Robert {\Garisto}, \ ``The Zing of Empty
Space''}

Thanks for the note.  I'm still thinking about your question, which
now strikes me as a truly important one.

\bq
\noindent Q:  What is real about a quantum system?

\noindent A:  That ``sensitivity to the touch,'' that ``zing!'' that keeps
us from ever having more information about it than can be encoded in
a quantum state.
\eq

That's where I left it.  But you asked, ``Where are these systems?
Where do they live?''  Could this say anything about spacetime that
we may have missed before?

Good question!

\section{14 November 2000, \ to Steven van {\Enk}, \ ``Always One Theory
Behind''}

If (by a miracle) you end up reading any of the thick document I
gave you yesterday {\it and\/} (by chance) you find any typos in it,
I'd appreciate it if you would note them and let me know.

I've been thinking about your ``philosophical'' question all
morning.  (I'm in the Chicago airport right now.)  And, it dawned on
me that maybe the following is a better language than {\Bohr}'s.  As
you might remember, I'm already quite inclined to reexpress the
process of quantum ``measurement'' in a language borrowed from the
economists.  Let me review that briefly before getting back to some
Bohrish things.  (I'll place an older, long account far below.)  The
keywords are DECISIONS, ACTS, and CONSEQUENCES.  The experimentalist
DECIDES to perform this or that ACTION upon a quantum system.  Such
actions will generally have some unpredictable CONSEQUENCES and it
is those consequences that we more standardly call a measurement
outcome.

Within this wording, your question becomes this:  Must our acts and
their consequences always be expressed in the language of classical
physics?  Here's where I think I've made some headway.  A better
way, I think, to express the situation is that we have no choice but
to use a language that is ``one theory behind.''  In this, I have
been greatly influenced by a passage (attributed to {\Einstein}) from
one of {\Heisenberg}'s books.  You can find the passage in ``The
Activating Observer,'' but let me repeat it here.

\bq
\noindent W.~{\Heisenberg}, {\sl Physics and Beyond:~Encounters and
Conversations}, translated by A.~J. Pomerans (Harper \& Row, New
York, 1971), pp.~63--64.  {\Heisenberg} reports {\Einstein} as having
once said the following to him:
\bq
It is quite wrong to try founding a theory on observable magnitudes
alone.  In reality the very opposite happens.  It is the theory
which decides what we can observe.  You must appreciate that
observation is a very complicated process.  The phenomenon under
observation produces certain events in our measuring apparatus.  As
a result, further processes take place in the apparatus, which
eventually and by complicated paths produce sense impressions and
help us to fix the effects in our consciousness.  Along this whole
path---from the phenomenon to its fixation in our consciousness---we
must be able to tell how nature functions, must know the natural
laws at least in practical terms, before we can claim to have
observed anything at all.  Only theory, that is, knowledge of
natural laws, enables us to deduce the underlying phenomena from our
sense impressions.  When we claim that we can observe something new,
we ought really to be saying that, although we are about to
formulate new natural laws that do not agree with the old ones, we
nevertheless assume that the existing laws---covering the whole path
from the phenomenon to our consciousness---function in such a way
that we can rely upon them and hence speak of ``observation.''
\eq
\eq

Now, {\Einstein} probably thought the situation of having to express
``observations'' in a language ``one theory behind'' was to be a
provisional thing:  At the completion of the new theory's
construction, one would just do away with it.  But maybe that was
never the case, and {\Bohr} was onto that train of thought.  (Or
maybe he wasn't, and it's just me instead.)

Anyway, that's my little thought for the morning.  I'll mail this to
you when I get to Caltech.

\section{14 January 2001, \ to Rob {\Pike}, \ ``Wispy Sunday''}

I'm listening to Dinah {\Washington} this Sunday morning.  And that
always makes me feel wispy.  But part of it is coming from this
session I'm organizing for the Sweden meeting.

I was impressed with some things you said at cookie time Friday.  I
think they show we have some overlap in our thoughts about quantum
information.  In particular, your remark about how computer
scientists work, perhaps, in an overly restrictive framework by
thinking quantum computing is captured (exclusively?) by the idea of
an exponentiation of computational paths struck me as on the mark.
In that connection, I think you might enjoy reading Andy {\Steane}'s
paper, ``A Quantum Computer Needs Only One Universe,'' {\tt
quant-ph/0003084}.  ({\Steane}, if you don't know him, independently
invented the idea of quantum error correction at the same time as
Peter {\Shor}.)

I too keep thinking that entanglement and exponentiation are more a
surface phenomenon of a deeper kind of ``zing!'' that quantum systems
have.  The two old notes below put that more in the context of our
business.  And they seem suitably wispy to fit my Sunday mood:  so I
thought I'd forward them to you.  When I rediscovered the remark I
made to {\Mermin} about Bell Labs at the end of one of them, I
thought it was so apropos!  [See note to {\Gilles} {\Brassard},
titled ``Memories of Montr\'eal,'' dated 17 June 2000, and note to
David {\Mermin}, titled ``Zing!,'' dated 20 July 2000.]

\section{18 January 2001, \ to Max {\Tegmark}, \ ``100 Years of the
Quantum''}

[AUTHOR'S NOTE:  Given the influence of John {\Wheeler} on my passion
for the Paulian idea, it seems only appropriate that this volume
should include a bibliography of John's post-{\Everett}-enthusiasm
quantum writings. (By the adjective post-{\Everett}-enthusiasm, I
refer to the period in {\Wheeler}'s published writings from roughly
1971 through to the end of 2000.) Max {\Tegmark}'s posting of {\tt
quant-ph/0101077} on 17 January 2001 presented the right
circumstances for me to place my personal bibliography from that
period into the medium of email. Five references and two quotes have
been added since the email's original delivery.]

\begin{enumerate}\itemsep -1pt
\item\label{Wheeler71a}
J.~A. {\Wheeler}, ``Transcending the Law of Conservation of
Leptons,'' in {\sl Atti del Convegno Internazionale sul Tema:\ The
Astrophysical Aspects of the Weak Interaction\/} (Cortona ``Il
Palazzone,'' 10-12 Giugno 1970), Accademia Nationale die Lincei,
Quaderno N.~{\bf 157} (1971), pp.~133--164.

\item\label{Wheeler73a}
J.~A. {\Wheeler}, ``From Relativity to Mutability,'' {\sl The
Physicist's Conception of Nature}, edited by J.~Mehra (D.~Reidel,
Dordrecht, 1973), pp.~202--247.

\item\label{Misner73}
C.~W. Misner, K.~S. Thorne, and J.~A. {\Wheeler}, ``Beyond the End of
Time,'' Chap.~44 in {\sl Gravitation}, (W.~H. Freeman, San
Francisco, CA, 1973), pp.~1196--1217.

\item\label{Wheeler74a}
J.~A. {\Wheeler}, ``From Mendel\'{e}ev's Atom to the Collapsing
Star,'' in {\it Philosophical Foundations of Science}, edited by
R.~J. Seeger and R.~S. Cohen (Reidel, Dordrecht, 1974), pp.~275--301.

\item\label{Wheeler74b}
J.~A. {\Wheeler}, ``The Universe as Home for Man,'' Am.\ Sci.\ {\bf
62}, 683--691 (1974).  This is an abridged, early version of
Ref.~[\ref{Wheeler75c}].

\item\label{Wheeler75a}
J.~A. {\Wheeler}, ``From Magnetic Collapse to Gravitational
Collapse:~Levels of Understanding Magnetism,'' in {\sl Role of
Magnetic Fields in Physics and Astrophysics}, Ann.\ NY Acad.\ Sci.\
{\bf 257}, edited by V.~Canuto (NY Academy of Sciences, NY, 1975),
pp.~189--221.

\item\label{Wheeler75b}
J.~A. {\Wheeler}, ``Another Big Bang?'' Am.\ Sci.\ {\bf 63}, 138
(1975). This is a letter in reply to a reader's comments on
Ref.~[\ref{Wheeler74b}].

\item\label{Wheeler75c}
J.~A. {\Wheeler}, ``The Universe as Home for Man,'' {\sl The Nature
of Scientific Discovery:\ A Symposium Commemorating the 500th
Anniversary of the Birth of Nicolaus Copernicus}, edited by
O.~Gingerich (Smithsonian Institution Press, City of Washington,
1975), pp.~261--296, discusion pp.~575--587.

\item\label{PattonWheeler75}
C.~M. Patton and J.~A. {\Wheeler}, ``Is Physics Legislated by
Cosmogony?,'' in {\sl Quantum Gravity:~An Oxford Symposium}, edited
by C.~J. {\Isham}, R.~{\Penrose}, and D.~W. {\Sciama} (Clarendon
Press, Oxford, 1975), pp.~538--605.

\item\label{Wheeler76a}
J.~A. {\Wheeler}, ``Include the Observer in the Wave Function?,''
Fundamenta Scientiae:\ Seminaire sur les Fondements des Sciences
(Strasbourg) {\bf 25}, 9--35 (1976).

\item\label{Wheeler77a}
J.~A. {\Wheeler}, ``Genesis and Observership,'' in {\sl Foundational
Problems in the Special Sciences:  Part Two of the Proceedings of the
Fifth International Congress of Logic, Methodology and Philosophy of
Science, London, Canada -- 1975}, edited by R.~E. Butts and
J.~Hintikka (D.~Riedel, Dordrecht, 1977), pp.~3--33.

\item\label{Wheeler77b}
J.~A. {\Wheeler}, ``Include the Observer in the Wave Function?,''
{\sl Quantum Mechanics, a Half Century Later:~Papers of a Colloquium
on Fifty Years of Quantum Mechanics, Held at the University Louis
Pasteur, Strasbourg, May 2--4, 1974}, edited by J.~Leite Lopes and
M.~Paty (D.~Reidel, Dordrecht, 1977), pp.~1--18.  This is a reprint
of Ref.~[\ref{Wheeler76a}].

\item\label{PattonWheeler77}
C.~M. Patton and J.~A. {\Wheeler}, ``Is Physics Legislated by
Cosmogony?,'' in {\sl Encyclopedia of Ignorance:\ Everything You
Ever Wanted to Know about the Unknown}, edited by R.~Duncan and
M.~Weston-Smith (Pergamon, Oxford, 1977), pp.~19--35. This is an
abridged version of Ref.~[\ref{PattonWheeler75}].

\item\label{Wheeler78a}
J.~A. {\Wheeler}, ``The `Past' and the `Delayed-Choice' Double-Slit
Experiment,'' {\sl Mathematical Foundations of Quantum Theory},
edited by A.~R. Marlow (Academic Press, New York, 1978), pp.~9--48.

\item\label{Wheeler79a}
J.~A. {\Wheeler}, ``Parapsychology---A Correction,'' Science {\bf
205}, 144 (1979).  This correction refers to something stated in
Wheeler's talk at ?? (reported later in corrected form in
Ref.~[\ref{Wheeler81a}).

\item\label{Wheeler79b}
J.~A. {\Wheeler}, ``Frontiers of Time,'' in {\sl Problems in the
Foundations of Physics, Proceedings of the International School of
Physics ``Enrico {\Fermi},'' Course LXXII}, edited by G.~Toraldo de
Francia (North-Holland, Amsterdam, 1979), pp.~395--492.

\item\label{Wheeler79c}
J.~A. {\Wheeler}, ``The Quantum and the Universe,'' in {\sl
Relativity, Quanta, and Cosmology in the Development of the
Scientific Thought of Albert {\Einstein}, Vol.~II}, edited by
F.~de~Finis (Johnson Reprint Corp., New York, 1979), pp.~807--825.

\item\label{Wheeler79d}
J.~A. {\Wheeler}, ``The Superluminal,'' New York Review of Books, 27
September 1979, p.~68.

\item\label{Wheeler79e}
J.~A. {\Wheeler}, ``Collapse and Quantum as Lock and Key,'' Bull.\
Am.\ Phys.\ Soc., Series II {\bf 24}, 652--653 (1979).

\item\label{Wheeler80a}
J.~A. {\Wheeler}, ``Beyond the Black Hole,'' in {\sl Some
Strangeness in the Proportion:~A Centennial Symposium to Celebrate
the Achievements of Albert {\Einstein}}, edited by H.~Woolf
(Addison-Wesley, Reading, MA, 1980), pp.~341--375, discussion
pp.~381--386.

\item\label{Wheeler80b}
J.~A. {\Wheeler}, ``Pregeometry:~Motivations and Prospects,'' in {\sl
Quantum Theory and Gravitation:  Proceedings of a Symposium Held at
Loyola University, New Orleans, May 23-26, 1979}, edited by A.~R.
Marlow (Academic Press, New York, 1980), pp.~1--11.

\item\label{Wheeler80c}
J.~A. {\Wheeler}, ``Law without Law,'' {\sl Structure in Science and
Art}, edited by P.~Medawar and J.~Shelley (Elsevier, Amsterdam,
1980), pp.~132--154.

\item\label{Wheeler80d}
J.~A. {\Wheeler}, ``Delayed-Choice Experiments and the
{\Bohr}-{\Einstein} Dialogue,'' in {\sl American Philosophical
Society and the Royal Society:~Papers Read at a Meeting, June 5,
1980}, (American Philosophical Society, Philadelphia, 1980),
pp.~9--40.

\item\label{Wheeler81a}
J.~A. {\Wheeler}, ``Not Consciousness but the Distinction Between the
Probe and the Probed as Central to the Elemental Quantum Act of
Observation,'' in {\sl The Role of Consciousness in the Physical
World}, edited by R.~G. Jahn (Westview Press, Boulder, CO, 1981),
pp.~87--111.

\item\label{Wheeler81b}
J.~A. {\Wheeler}, ``The Participatory Universe,'' Science81 {\bf
2}(5), 66--67 (1981).

\item\label{Wheeler81c}
J.~A. {\Wheeler}, ``The Elementary Quantum Act as Higgledy-Piggledy
Building Mechanism,'' in {\sl Quantum Theory and the Structures of
Time and Space:~Papers Presented at a Conference Held in Tutzing,
July, 1980}, edited by L.~Castell and C.~F. von Weizs\"acker (Carl
Hanser, Munich, 1981), pp.~??--??.

\item\label{Wheeler82a}
J.~A. {\Wheeler}, ``The Computer and the Universe,'' Int.\ J. Theo.\
Phys.\ {\bf 21}, 557--572 (1982).

\item\label{Wheeler82b}
J.~A. {\Wheeler}, ``Particles and Geometry,'' in {\sl Unified
Theories of Elementary Particles:~Critical Assessment and
Prospects}, Lecture Notes in Physics {\bf 160}, edited by
P.~Breitenlohner and H.~P. D\"urr (Springer-Verlag, Berlin, 1982),
pp.~189--217.

\item\label{Wheeler82c}
J.~A. {\Wheeler}, ``{\Bohr}, {\Einstein}, and the Strange Lesson of
the Quantum,'' in {\sl Mind in Nature:~Nobel Conference XVII,
Gustavus Adolphus College, St.~Peter, Minnesota}, edited by R.~Q.
Elvee (Harper \& Row, San Francisco, CA, 1982), pp.~1--23, discussion
pp.~23--30, 88--89, 112--113, and 148--149.

\item\label{Wheeler82d}
J.~A. {\Wheeler}, ``Physics and Austerity:~Law without Law,''
University of Texas preprint, 1--87 (1982).

\item\label{Wheeler82e}
J.~A. {\Wheeler}, ``Black Holes and New Physics,''
Discovery:~Research and Scholarship at the University of Texas at
Austin {\bf 7}(2), 4--7 (Winter 1982).

\item\label{Wheeler83a}
J.~A. {\Wheeler}, ``On Recognizing `Law without Law': Oersted Medal
Response at the Joint APS-AAPT Meeting, New York, 25 January 1983,''
Am.\ J. Phys.\ {\bf 51}, 398--404 (1983).

\item\label{Wheeler83b}
J.~A. {\Wheeler}, ``Elementary Quantum Phenomenon as Building Unit,''
in {\sl Quantum Optics, Experimental Gravity, and Measurement
Theory}, edited by P.~Meystre and M.~O. Scully (Plenum Press, New
York, 1983), pp.~141--143.

\item\label{Wheeler83c}
J.~A. {\Wheeler}, ``Law without Law,'' in {\sl Quantum Theory and
Measurement}, edited by J.~A. {\Wheeler} and W.~H. {\Zurek}
(Princeton University Press, Princeton, 1983), pp.~182--213.

\item\label{Wheeler83d}
J.~A. {\Wheeler}, ``Guest Editorial:\ The Universe as Home for
Man,'' in {\sl The Dynamic Universe:~An Introduction to Astronomy},
edited by T.~P. Snow (West Pub.\ Co., St.~Paul, Minnesota, 1983),
pp.~108--109. This is an excerpt from Ref.~[\ref{Wheeler75c}].

\item\label{Wheeler83e}
J.~A. {\Wheeler}, ``Physics and Austerity,'' in {\sl Krisis}, Vol.~1,
No.~2, edited by I.~Masculescu (Klinckscieck, Paris, 1983),
pp.~671--675.

\item\label{Wheeler83f}
J.~A. {\Wheeler}, ``Jenseits aller Zeitlichkeit,'' in {\sl Die Zeit:
Schrifter der Carl Friedrich von Seiemens-Stiftung}, Vol.~{\bf 6},
edited by A.~Peisl and A.~Mohler (Oldenbourg, Munchen, 1983),
pp.~17--34.

\item\label{ArmourWheeler83}
R.~S. Armour, Jr.\ and J.~A. {\Wheeler}, ``Physicist's Version of
Traveling Salesman Problem:~Statistical Analysis,'' Am.\ J. Phys.\
{\bf 51}, 405--406 (1983).

\item\label{Wheeler84a}
J.~A. {\Wheeler}, ``Quantum Gravity:~The Question of Measurement,''
in {\sl Quantum Theory of Gravity:~Essays in Honor of the 60th
Birthday of Bryce S. {\DeWitt}}, edited by S.~M. Christensen (Adam
Hilger, Bristol, 1984), pp.~224--233.

\item\label{Wheeler84b}
J.~A. {\Wheeler}, ``Die Experimente der verz\"ogerten Entscheidung
und der Dialog zwischen {\Bohr} und {\Einstein},'' in {\sl Moderne
Naturphilosophie}, edited by B.~Kanitscheider (K\"onigshausen and
Neumann, W\"urzburg, 1984), pp.~203--222.  This is an slightly
abridged, German version of Ref.~[\ref{Wheeler80d}].

\item\label{Wheeler84c}
J.~A. {\Wheeler}, ``Bits, Quanta, Meaning,'' in {\sl Problems in
Theoretical Physics}, edited by A.~Giovannini, F.~Mancini, and
M.~Marinaro (University of Salerno Press, Salerno, 1984),
pp.~121--141.

\item\label{Wheeler84d}
J.~A. {\Wheeler}, ``Bits, Quanta, Meaning,'' in {\sl Theoretical
Physics Meeting:~Atti del Convegno, Amalfi, 6-7 Maggio 1983},
(Edizioni Scientifiche Italiene, Naples, 1984), pp.~121--134. This
should be a reprint of Ref.~[\ref{Wheeler84c}].

\item\label{Miller84}
W.~A. Miller and J.~A. {\Wheeler}, ``Delayed-Choice Ex\-periments and
{\Bohr}'s Elementary Quantum Phenomenon,'' in {\sl Proceedings of the
International Symposium, Foundations of Quantum Mechanics in the
Light of New Technology}, edited by S.~Kamefuchi, H.~Ezawa,
Y.~Murayama, M.~Namiki, S.~Nomura, Y.~Ohnuki, and T.~Yajima
(Physical Society of Japan, Tokyo, 1984), pp.~140--152.

\item\label{Wheeler85a}
J.~A. {\Wheeler}, ``{\Bohr}'s `Phenomenon' and `Law without Law',''
in {\sl Chaotic Behavior in Quantum Systems:~Theory and
Applications}, edited by G.~Casati (Plenum Press, New York, 1985),
pp.~363--378.

\item\label{Wheeler85b}
J.~A. {\Wheeler}, ``Niels {\Bohr}, the Man,'' Phys.\ Today {\bf
38}(10), 66--72 (1985).

\item\label{Wheeler85c}
J.~A. {\Wheeler}, ``Delayed-Choice Experiments and the
{\Bohr}-{\Einstein} Dialogue,'' in {\sl Niels {\Bohr}:~A Profile},
edited by A.~N. Mitra, L.~S. Kothari, V.~Singh, and S.~K. Trehan
(Indian National Science Academy, New Delhi, 1985), pp.~139--168.
This is a reprint of Ref.~[\ref{Wheeler80d}].

\item\label{Wheeler86a}
J.~A. {\Wheeler}, ``Hermann {\Weyl} and the Unity of Knowledge,''
Am.\ Sci.\ {\bf 74}, 366--375 (1986).

\item\label{Wheeler86b}
J.~A. {\Wheeler}, ``Niels {\Bohr}:~The Man and his Legacy,'' in {\sl
The Lesson of Quantum Theory}, edited by J.~de~Boer, E.~Dal, and
O.~Ulfbeck (Elsevier, Amsterdam, 1986), pp.~355--367.

\item\label{Wheeler86c}
J.~A. {\Wheeler}, ``\,`Physics as Meaning Circuit':~Three Problems,''
in {\sl Frontiers of Nonequilibrium Statistical Physics}, edited by
G.~T. Moore and M.~O. Scully (Plenum Press, New York, 1986),
pp.~25--32.

\item\label{Wheeler86d}
J.~A. {\Wheeler}, ``Foreword,'' in J.~D. Barrow and F.~J. Tipler,
{\sl The Anthropic Cosmological Principle}, (Oxford University Press,
Oxford, 1986), pp.~vii--ix.

\item\label{Wheeler87a}
J.~A. {\Wheeler}, ``How Come the Quantum'' in {\it New Techniques and
Ideas in Quantum Measurement Theory}, edited by D.~M. Greenberger,
Ann.\ New York Acad.\ Sci.\ {\bf 480}, 304--316 (1987).

\item\label{Wheeler87b}
J.~A. {\Wheeler}, ``Foreword'' in H.~{\Weyl}, {\sl The Continuum:~A
Critical Examination of the Foundation of Analysis}, translated by
S.~Pollard and T.~Bole (Thomas Jefferson University Press,
Kirksville, MO, 1987), pp.~ix--xiii.  This is an excerpt from
Ref.~[\ref{Wheeler86a}].

\item\label{Wheeler88a}
J.~A. {\Wheeler}, ``World as System Self-Synthesized by Quantum
Networking,'' IBM J. Res.\ Develop.\ {\bf 32}, 4--15 (1988).

\item\label{Wheeler88b}
J.~A. {\Wheeler}, ``World as System Self-Synthesized by Quantum
Networking,'' in {\sl Probability in the Sciences}, edited by
E.~Agazzi (Kluwer, Dordrecht, 1988), pp.~103--129.  This is a reprint
of Ref.~[\ref{Wheeler88a}].

\item\label{Wheeler88c}
J.~A. {\Wheeler}, ``Hermann {\Weyl} and the Unity of Knowledge,'' in
{\sl Exact Sciences and their Philosophical Foundations}, edited by
W.~Deppert {\it et al}.~(Lang, Frankfurt am Main, 1988),
pp.~366--375. This is an expanded version of Ref.~[\ref{Wheeler86a}].

\item\label{Wheeler89a}
J.~A. {\Wheeler}, ``Bits, Quanta, Meaning,'' in {\sl Festschrift in
Honour of Eduardo R. Caianiello}, edited by A.~Giovannini,
F.~Mancini, M.~Marinaro, and A.~{\Rimini} (World Scientific,
Singapore, 1989), pp.~133--154.  This should be a reprint of
Ref.~[\ref{Wheeler84c}].

\item\label{Wheeler90a}
J.~A. {\Wheeler}, ``Information, Physics, Quantum:\ the Search for
Links,'' in {\sl Proceedings of the 3rd International Symposium on
Foundations of Quantum Mechanics in the Light of New Technology},
edited by S.~Kobayashi, H.~Ezawa, Y.~Murayama, and S.~Nomura
(Physical Society of Japan, Tokyo, 1990), pp.~354--368.

\item\label{Wheeler92a}
J.~A. {\Wheeler}, ``Recent Thinking about the Nature of the Physical
World:~It from Bit,'' in {\sl Frontiers in Cosmic Physics:~Symposium
in Memory of Serge Alexander Korff}, Ann.\ NY Acad.\ Sci.\ {\bf 655},
edited by R.~B. Mendell and A.~I. Mincer (NY Academy of Sciences, NY,
1992), pp.~349--364.

\item\label{Wheeler92b}
J.~A. {\Wheeler}, {\sl At Home in the Universe}, (American Institute
of Physics Publishing, New York, 1992).

\item\label{Wheeler94a}
J.~A. {\Wheeler}, ``Time Today,'' in {\sl Physical Origins of Time
Asymmetry}, edited by J.~J. Halliwell, J.~P\'erez-Mercader, and W.~H.
{\Zurek} (Cambridge University Press, Cambridge, 1994), pp.~1--29.

\item\label{Wheeler98a}
J.~A. {\Wheeler} (with K.~W. Ford), {\sl  Geons, Black Holes, and
Quantum Foam:~A Life in Physics}, (W.~W. Norton, New York, 1998).

\item\label{Wheeler98b}
J.~A. {\Wheeler}, letter to Carrol {\Alley}, 13 March 1998.

\bq
\indent
I want you and {\Einstein} to jolt the world of physics into an
understanding of the quantum because the quantum surely
contains---when unraveled---the most wonderful insight we could ever
hope to have on how this world operates, something equivalent in
scope and power to the greatest discovery that science has ever yet
yielded up: Darwin's Evolution.

You know how {\Einstein} wrote to his friend in 1908, ``This quantum
business is so incredibly important and difficult that everyone
should busy himself with it.'' \ldots\ Expecting something great
when two great minds meet who have different outlooks, all of us in
this Princeton community expected something great to come out from
{\Bohr} and {\Einstein} arguing the great question day after
day---the central purpose of {\Bohr}'s four-month, spring 1939 visit
to Princeton---I, now, looking back on those days, have a terrible
conscience because the day-after-day arguing of {\Bohr} was not with
{\Einstein} about the quantum but with me about the fission of
uranium. How recover, I ask myself over and over, the pent up
promise of those long-past days? Today, the physics community is
bigger and knows more than it did in 1939, but it lacks the same
feeling of {\bf desperate} puzzlement. I want to recapture that
feeling for us all, even if it is my last act on Earth.
\eq

\item\label{Wheeler98c}
J.~A. {\Wheeler}, ``The Eye and the U,'' letter dated 19 March 1998
to Abner {\Shimony}, with carbon copies to Arthur Wightman, Peter
Mayer, Demetrios Cristodoulou, Larry Thomas, Elliot {\Lieb}, Charles
Misner, Frank Wilczek, Kip Thorne, Ben {\Schumacher}, William
{\Wootters}, Jim {\Hartle}, Edwin Taylor, Ken Ford, Freeman {\Dyson},
Peter Cziffra, and Arkady {\Plotnitsky}.

\bq
\indent
You have known me long enough and well enough to believe me when I
say that I would willingly give up an arm or a leg to understand
``How Come the Quantum?'' or ``How Come Existence?'' -- or give up
both to understand both.  As you know, nobody is taking a question
seriously unless he educates himself enough about it to hypothesize
an answer to it.  My tentative answer is stated nowhere more
compactly than in Act IV, Scene 1, of Shakespeare's {\sl The
Tempest}, ``We are such stuff as dreams are made on'' -- and
symbolized in the Eye and the U at the right:
\eq

\item\label{Wheeler99a}
J.~A. {\Wheeler}, ``Information, Physics, Quantum:~The Search for
Links,'' in {\sl Feynman and Computation:~Exploring the Limits of
Computers}, edited by A.~J.~G. Hey (Perseus Books, Reading, MA,
1999), pp.~309--336.

\item\label{Wheeler00}
J.~A. {\Wheeler}, ``\,`A Practical Tool,' But Puzzling, Too,'' New
York Times, 12 December 2000.

\bq
\indent
What is the greatest mystery in physics today? Different physicists
have different answers. My candidate for greatest mystery is a
question now a century old, ``How come the quantum?''

What is this thing, the ``quantum''? It's a bundle of energy, an
indivisible unit that can be sliced no more. Max {\Planck} showed us
a hundred years ago that light is emitted not in a smooth, steady
flow, but in quanta. Then physicists found quantum jumps of energy,
the quantum of electric charge and more. In the small-scale world,
everything is lumpy.

And more than just lumpy. When events are examined closely enough,
uncertainty prevails; cause and effect become disconnected. Change
occurs in little explosions in which matter is created and destroyed,
in which chance guides what happens, in which waves are particles and
particles are waves.

Despite all this uncertainty, quantum physics is both a practical
tool and the basis of our understanding of much of the physical
world. It has explained the structure of atoms and molecules, the
thermonuclear burning that lights the stars, the behavior of
semiconductors and superconductors, the radioactivity that heats the
earth, and the comings and goings of particles from neutrinos to
quarks.

Successful, yes, but mysterious, too. Balancing the glory of quantum
achievements, we have the shame of not knowing ``how come.'' Why does
the quantum exist?

My mentor, the Danish physicist Niels {\Bohr}, made his peace with
the quantum. His ``Copenhagen interpretation'' promulgated in 1927
bridged the gap between the strangeness of the quantum world and the
ordinariness of the world around us. It is the act of measurement,
said {\Bohr}, that transforms the indefiniteness of quantum events
into the definiteness of everyday experience. And what one can
measure, he said, is necessarily limited. According to his principle
of complementarity, you can look at something in one way or in
another way, but not in both ways at once. It may be, as one French
physicist put it, ``the fog from the north,'' but the Copenhagen
interpretation remains the best interpretation of the quantum that
we have.
\eq
\end{enumerate}

\section{20 January 2001, \ to John {\Smolin}, \ ``{\Shannon} Meets {\Bohr}''}

As I told you on the phone, it's happening to me again:  I've gotten
another opportunity to gather my friends in an exotic place to
thrash out the idea of ``Quantum Foundations in the Light of Quantum
Information.''  This time I've been asked to organize a session at a
larger conference in V\"axj\"o, Sweden titled ``Quantum Theory:
Reconsideration of Foundations.''  The main organizer is Andrei
{\Khrennikov}, and he is planning to have 40-50 people attending,
with further sessions on Bohmian mechanics, GRW mechanics, and other
issues in quantum foundations ({\Bell} inequalities,
{\Kochen}-{\Specker} theorems, etc.).

In flavor and constitution, I plan to make our session much like the
meeting {\Gilles} {\Brassard} and I held in Montreal last spring.
There the theme was organized around the strong feeling that we'll
find the greatest things and technologies to come out of quantum
mechanics when we finally grasp the parts of it that make us feel
the most uncomfortable.  The theory is begging us to ask something
new and profound of nature.

This time the list of confirmed attendees (all invitees so far)
includes: {\Gilles} {\Brassard}, Jeffrey {\Bub}, {\Carl} {\Caves},
Lucien {\Hardy}, Richard {\Jozsa}, Pekka {\Lahti}, David {\Mermin},
{\Asher} {\Peres}, Itamar {\Pitowsky}, John {\Preskill}, {\Ruediger}
{\Schack}, and Ben {\Schumacher}. Depending upon how the money holds
out, I might be able to call up even a few more in the community.

I think we'll have a double-edged opportunity at this meeting, so
I'm looking quite forward to it.  First, we'll get a chance to
continue the lines several of us started in Montreal---at that time,
the attendants were {\Bennett}, {\Bernstein}, {\Brassard}, {\Hayden},
{\Jozsa}, {\Mermin}, {\Schack}, {\Schumacher}, and {\Wootters}.  But
second, we may be able to play the role of educators to a larger
community interested in dabbling in these same issues.  That is, we
will have an opportunity to get them to think of quantum information
as a tool for exploring quantum foundations.

The dates for the meeting are June 17 to 22.  We think this choice
will help make for quite a pleasant time:  Midsummer's eve is June
21, and {\Mermin} has pointed out that picking wild strawberries in
the midnight twilight can be huge fun.  Furthermore, there are
significant local festivities planned around that time of year.

The information I'd like to get from you (as soon as possible!) is:
\bv
1)  Can you confirm positively that you would like to come? \\
and \\
2) What kind of financial resources will it take for us to get you
there?\\
\hspace*{.15in} a) Full expenses paid? \\
\hspace*{.15in} b) Local expenses paid? \\
\hspace*{.15in} c) Nothing? \\
\ev

Concerning question 2), you can count this letter as an invitation,
so feel free to answer any of the three options a), b), or c).
However, the more money I can get the invitees to throw into the pot
(if they have it available), the more interesting people I can get
to Sweden to keep us entertained.  I've been told that I can have
3-4 invitees all expenses paid and 5 invitees with local expenses
paid.  You can see that I'm trying to stretch my limits!  The way
I've been able to do this is that, indeed, a significant number of
attendees have {\it volunteered\/} to pay their nonlocal expenses,
i.e. their travel expenses.  (In this group is {\Caves}, {\Hardy},
{\Mermin}, and {\Preskill}.  {\Jozsa} and {\Schack} have also
committed to contribute in the case that it will make for a more
interesting meeting.  I think your presence would do that!)

I really hope you can attend; I think we'll have a lot of fun.
Please let me know your thoughts as soon as you can!  (Like, today
or tomorrow maybe?)  The sooner {\Khrennikov} and I can get this
preliminary information (dates in particular), the sooner we can
start throwing stones at all the other problems.

\section{26 January 2001, \ to Jeremy {\Butterfield}, \ ``Worlds in
the {\Everett} Interpretation''}

Anyone who knows me knows that I am rather down on attempts to
interpret quantum mechanics along {\Everett}-like lines.  I think the
most funny and telling statement of this in the present context is
that, whereas Mr.\ {\Wallace} speaks of ``Everettians,'' I often
speak of ``Everettistas.''  Thus, I am almost surprised that you
sent me this paper to referee.

My difficulties come not so much from thinking that an
{\Everett}-like interpretation is inherently inconsistent or that
parallel worlds tax the imagination too much.  It's more that this
line of thought strikes me, at best, as a complete dead end in the
physical sense. At worst, I fear it requires us to tack on even more
ad hoc structures to quantum theory than we already have.  (Here, I'm
thinking of a preferred basis for the Hilbert space and a preferred
tensor-producting of it into various factors.)  For these reasons,
among umpteen others, I have always been inclined to an epistemic
interpretation of the quantum state.  Doing this has helped me
(personally) to focus the issue to asking, ``What is this {\it
property\/} of the quantum world---i.e., reality---that keeps us from
ever knowing more of it than can be captured by the quantum state?''
To that extent, I consider myself something of a realist who---just
as David {\Deutsch}---takes the wavefunction absolutely seriously.
BUT absolutely seriously as a state of knowledge, not a state of
nature.  I do well believe we will one day shake a notion of reality
from the existing theory (without adding hidden variables, etc.),
but that reality won't be the most naive surface term floating to
the top (i.e., the quantum state).  When we have it, we'll really
have something; there'll be no turning back. Physics won't be at an
end, but at a beginning.  For then, and only then, will we be able to
recognize how we might extend the theory to something bigger and
better than quantum mechanics itself.

All that said, you're going to be surprised by my evaluation of this
paper.  Without hesitation, I recommend you publish it!  Of all the
things I've read on {\Everett}-like interpretations over the years,
this paper has struck me as the most reasonable of the lot.  This is
a nice paper.  It changed none of my views, but it caused me deep
pause for thought.  What better honor can one have from a sparring
partner?

I am not in a position to judge whether the paper is a significant
step forward over the many {\it recent\/} papers it cites---I've
never read any of them---but I do know this much:  I have walked
away from several talks of the ``most modern'' treatment of
{\Everett} and not had a clue what was being talked about. So, even
if my evaluation (as an outsider) of the paper's technical merit
leaves something to be desired, I think Mr.\ {\Wallace}'s paper
clearly serves a purpose within our community. The analogies between
the {\Everett}-like structure he proposes and the spacetime of
general relativity are indeed intriguing and worthy of thought.  For
me personally---whatever their ultimate merit---they give a new stone
on which to hone the arguments for an epistemic interpretation of
the quantum state.

Let me just make a few minor comments to round out this report.

I very much enjoyed the discussion:
\bq
\noindent So there are no theory-neutral observations:  rather, there
is an existing theory in terms of which our observations are
automatically interpreted, and which we must take as our starting
point when interpreting the theories of physics.
\eq
I've seen a discussion like this before, and it probably should be
cited.  Here are two references from my personal archive:
\begin{enumerate}
\item
W.~{\Heisenberg}, {\sl Physics and Beyond:~Encounters and
Conversations}, translated by A.~J. Pomerans (Harper \& Row, New
York, 1971), pp.~63--64.
\item
M.~{\Jammer}, ``The Experiment in Classical and in Quantum
Physics,'' in {\sl Proceedings of the International Symposium,
Foundations of Quantum Mechanics in the Light of New Technology},
edited by S.~Kamefuchi, H.~Ezawa, Y.~Murayama, M.~Namiki, S.~Nomura,
Y.~Ohnuki, and T.~Yajima (Physical Society of Japan, Tokyo, 1984),
pp.~265--276.
\end{enumerate}

Section 10 (the big table).  This may have been my favorite part of
the paper.  But I would do this:  Put the relativity column on the
left and the {\Everett} column on the right.  Somehow, I found it
much easier to read in that manner.  It just seemed more natural to
recognize a feature in relativity first, and then look for the
analogous feature in {\Everett}.

\section{27 January 2001, \ to David {\Wallace}, \ ``The Test
Particle''}

I presume by now you've seen my thoughts on your {\Everett}
article.  I meant it all; it's a very nice paper.  I'd like to
encourage you to place the article on the {\tt quant-ph} archive.
I'd like to refer some of my friends to it to the get a debate
going.  Having the paper easily accessible will help get that off
the ground.

One piece of analogy (or disanalogy) that you didn't explore very
much is the geodesic.  Is there an analog to the geodesic in your
Everettian system?  One thing that strikes me is that you might
start seeing a conceptual divergence here.  It's not clear to me
that one can concoct a good notion of ``test particle'' for this
game: even adding the smallest system possible to an existing
multiverse (I hate that term) doubles its Hilbert space dimension
(acting as if the Hilbert space is finite to begin with).  There is
no such thing as a small system maintaining a negligible perturbation
to the larger system.

\section{27 February 2001, \ to V\"axj\"o Meeting participants, \
``Claude Elwood {\Shannon}''}

\noindent To those friends I contacted for the ``{\Shannon} meets {\Bohr}''
session in V\"axj\"o: \medskip

We at Bell Labs were all deeply moved yesterday to learn of Claude
{\Shannon}'s death over the weekend.  Like for many Alzheimer's
patients, however, we knew that it may have been a blessing in
disguise.  You can read more about {\Shannon}'s life in the New York
Times obituary section today:\\
{\tt http://www.nytimes.com/pages/national/text/index.html\#obits}

Some weeks ago, I found myself using the following words for a
recommendation letter I was asked to write:
\bq
\noindent
As it turns out, today January 16, we had a dedication ceremony at
Bell Labs for a bronze bust of Claude {\Shannon}, the founder of
classical information theory.  That struck me as symbolic.  Since
joining the laboratory, I have been asking myself over and over what
role I might play in furthering the legacy of {\Shannon}.
\eq
Those words hold just as true for me for this midsummer's meeting.
Let us use information theory as a sword and finally defeat this
mystery of the quantum.

\chapter*{Postpartum}
\addcontentsline{toc}{chapter}{Postpartum}

\section{16 May 2000, \ ``History Reduced to Eight Pages''}



\begin{flushright}
The past exists only insofar as it is recorded in the present.\\
 --- {\it John Archibald {\Wheeler}}
\end{flushright}


\noindent \parbox{3.0in}{Dear friends and family who have written or called
     with concern over the Los Alamos fires,}
     \bigskip

I apologize at the outset for sending a mass mailing like this.  All
of you hold a unique place in my mind, but time is scarce at the
moment.

Sometime Wednesday or Thursday last week, the Cerro Grande fire
invaded our home in Los Alamos and took essentially all that was in
it.  The only exceptions were our car, a couple of suitcases of
clothes and memories, and most importantly, our lives.  {\Kiki},
{\Emma}, and I are all physically well.  So are our ``boys,'' the
golden retrievers {\Albert} and {\Wizzy}.  The last few days we have
received shelter and comfort in Albuquerque in the home of {\Carl}
{\Caves} and Karen {\Kahn}.  The boys have been in the foster care of
Chris {\Hains} (where they likely view themselves as having a resort
vacation).

We saw the remains of our home Sunday:  all that was left solid was
the chimney.  The thought of it makes us quite sad at times.  I
count as my greatest loss all the ``intellectual property'' that
never made it or could not make it into my computer---my thousands of
pages of meticulous calculation, my sad poems from sad times and my
pre-electronic correspondence, my collection of over 700 physics,
math, and philosophy books.  The evidence of my academic life has
been reduced to the eight pages of my vita.  {\Kiki} counts as her
greatest loss her many family heirlooms, her red and her yellow
``biker'' jackets, her antique cookbooks, the quilts and cookie
cutters of her grandmother, some forgotten jewelry of great
sentiment, and the large library of children's books she had built
up for {\Emma}.

In the grand scheme of things, all of this will of course fade.  Our
daughter is a great source of strength and purpose.  The
responsibility of rebuilding a good life for her sake, if for
nothing else, is at the top of our minds.

With this loss, as with most events in my life, I have tried to look
for a lesson of deeper significance.  It won't come as a surprise to
most of you that my thoughts have turned toward quantum mechanics.
What I have always found wonderful in the theory is its indication
that our world is more malleable than was thought in classical
times.  With our experimental interventions into Nature, we have the
opportunity to shape it in unforeseen and perhaps wonderful ways.
How, now that so many of my records are gone, I hope there is a
grain of truth in John {\Wheeler}'s words.  For then, the past would
be every bit as open as the future.  I could see that as a kind of
strange recompense in this time of trouble.  But in this, just as
for all things, only time will tell. \medskip

\noindent Thoughts of all,\medskip

\noindent Chris

\pagebreak

\addcontentsline{toc}{chapter}{Index of Names}

\small



\begin{theindex}

  \item Adami, Chris, 75, 460
  \item Adler, Steven, 125, 158, 246, 283
  \item Aharonov, Yakir, 327
  \item Albert, David Z., 279, 317
  \item Alley, Carol, 84, 470
  \item Archimedes, 7, 159, 316
  \item Aristotle, 377, 397

  \indexspace

  \item Baker, David B. L., 1, 144
  \item Baker, Howard, 9
  \item Ballentine, Leslie E., 109, 110, 393, 402
  \item Banville, John, 50, 53, 161
  \item Barad, Karen, 163, 164
  \item Barnum, Howard, ii, 17, 76, 92, 103, 121, 125, 126, 140, 183,
        231, 232, 235, 298, 301, 318, 322, 325, 454, 457
  \item Baudrillard, Jean, 45, 46, 248
  \item Baytch, Nurit E., 361
  \item Bell, John S., 10, 15, 21, 59, 92, 97, 117, 118, 121--123, 147,
        174, 175, 212, 216, 220, 228, 241, 258, 259, 272, 279,
        281, 294, 295, 297, 305--307, 309, 311, 333, 338, 353,
        375, 377, 378, 393--395, 438--441, 456, 459, 460, 471
  \item Beller, Mara, 44, 123, 302, 397
  \item Benioff, Paul, ii, 24, 26--29, 108, 110, 112, 207, 210, 331,
        393, 425, 432
  \item Benka, Stephen G., 298, 299, 301, 303, 304, 314, 327, 330, 331,
        336, 339--341
  \item Bennett, Charles H., ii, 12, 17, 33--36, 38--40, 43, 53, 65--69,
        92--95, 120--122, 143, 153, 155, 161, 176, 177, 190,
        212, 243, 249, 258, 274, 283, 295, 301, 317, 331, 353,
        356, 379, 408, 413, 439, 457, 471
  \item Bennett, Theodora, 17
  \item Bennett, Tony, 144
  \item Bernardo, Jos{\'e} M., 30, 364
  \item Bernoulli, James, 131, 380
  \item Bernstein, Herbert J., ii, 5, 6, 20--22, 39, 47, 70, 76, 84,
        93--95, 127, 151, 155, 161, 163, 275, 292, 301, 353,
        373, 392, 471
  \item Berthiaume, Andr\'e, 432
  \item Beyler, Richard H., 445, 447
  \item Bhattacharya, Tanmoy, 336
  \item Bilodeau, Doug, ii, 53, 74, 80, 277, 405
  \item Bohm, David, 31, 45, 49, 50, 59, 98, 102, 103, 125, 184, 225,
        242, 258, 259, 303, 309, 311, 353, 367, 388, 392, 438
  \item Bohr, Niels, 13, 44, 45, 47--50, 72, 79, 82, 116, 123, 124, 130,
        132--134, 139, 141, 146, 147, 150, 151, 154, 155,
        162--165, 173--177, 191, 198, 206, 207, 211, 213,
        215--219, 221--224, 229--231, 234, 236, 237, 253, 254,
        257, 259--263, 265--269, 271, 272, 288, 290, 300, 301,
        307, 320, 359, 360, 374, 375, 384, 385, 387, 389, 405,
        407, 414, 416, 419, 420, 435, 443, 447, 464, 465,
        467--471, 474
  \item Boole, George, 119, 122, 123, 247, 363, 370, 393, 395
  \item Bopp, Friedrich (Fritz), 463
  \item Born, Max, vi, 86, 87, 183, 202, 203, 207, 225--227, 270, 281,
        285, 293, 295, 303, 310, 315, 319, 332, 397, 443
  \item Brassard, Gilles, ii, 82, 85, 93, 95--97, 99, 120--122, 139,
        161, 190, 232, 234, 237, 295, 331, 361, 432, 448, 465,
        471
  \item Braunstein, Samuel L., 97, 120, 143, 198, 258, 354, 393, 434,
        443
  \item Bretthorst, G. Larry, 152
  \item Broglie, Louis, 309, 366, 405
  \item Brukner, \v {C}aslav, 323, 431
  \item Brun, Todd A., v, 92, 128, 291, 297, 344, 347, 348, 460
  \item Bu\v {z}ek, Vladimir, 121, 157
  \item Bub, Jeffrey, ii, v, 18, 91, 92, 94, 95, 99, 234, 243, 257--260,
        263, 264, 266--268, 300, 301, 362, 369, 393, 471
  \item Buck, Joseph R., 125, 156
  \item Buhrman, Harry, 122, 441
  \item Busch, Paul, 105, 243, 289
  \item Bush, George W., 134
  \item Butterfield, Jeremy, 93, 317, 415, 463, 472

  \indexspace

  \item Cabello, Adan, 92, 225, 282, 445
  \item Cardano, Girolamo, 71
  \item Carvalho, Alvaro, 239, 255
  \item Caves, Carlton M., ii, v, 11, 15, 24, 25, 75, 77, 79, 105, 106,
        114, 116, 121--123, 125, 129, 133, 134, 153, 154,
        157, 158, 162, 166, 181, 183, 190, 193, 195, 210, 229,
        231, 232, 243, 253, 258, 273, 276, 277, 287, 298, 301,
        303, 305, 307, 312, 336, 351, 353, 354, 359, 361,
        364--366, 370, 372, 379, 380, 390, 393, 398--402, 408,
        421, 428, 434, 444, 454, 471, 472, 475
  \item Cerf, Nicolas J., 75, 312, 460
  \item Chaitin, Gregory J., 25, 26, 109, 432
  \item Chau, H. F., 82
  \item Chuang, Isaac L., 120
  \item Church, Alonzo, 32, 432
  \item Cirac, Juan I., 121, 122, 135
  \item Cleve, Richard, 122, 354, 439, 441, 446
  \item Clifton, Rob, 283, 353
  \item Coleman, Sidney R., 106--110, 112, 113, 210, 212, 432
  \item Colgate, Stirling A., 291
  \item Comer, Gregory L., ii, 28, 48, 50, 53, 58, 83, 127, 136, 145,
        150, 180, 193, 207, 230, 265, 284, 373, 374, 378, 413,
        417, 426, 435
  \item Cr{\'e}peau, Claude, 121, 295, 432

  \indexspace

  \item D'Ariano, G. Mauro, 97, 402
  \item D\"urr, Detlef, 319, 367
  \item de Finetti, Bruno, 70, 77, 78, 80, 158, 166, 167, 232, 242,
        273, 274, 298, 312, 317, 349, 380, 381, 390--394, 396,
        398--402, 421, 428
  \item De\nobreakspace  {}Martini, Francesco, 315, 317
  \item Democritus, 186, 187, 193, 352, 423
  \item Derrida, Jacques, 5, 44, 45, 146, 196, 198
  \item Deutsch, David, 37, 62, 63, 121, 179, 207, 239, 253, 347,
        354--356, 393, 472
  \item Dewdney, Chris, 317
  \item DeWitt, Bryce S., 292, 356, 393, 394, 468
  \item Dirac, Paul A. M., 151, 211, 266, 271, 386, 441
  \item DiVincenzo, David P., 120--122, 318, 323
  \item Dyson, Freeman J., 120, 470

  \indexspace

  \item Ehrenfest, Paul, 13, 447
  \item Einstein, Albert, 3, 10, 13, 42, 44, 51--53, 79, 82, 98, 100,
        102, 103, 116, 117, 119, 122--124, 130, 132, 137, 138,
        142, 149, 150, 153, 163, 168, 175, 191, 211, 216, 217,
        219, 223--227, 241, 242, 253, 254, 257, 264--266,
        270, 271, 275, 285, 344, 353, 359, 360, 385--388, 393,
        397, 405, 413, 425, 441, 444, 446, 447, 459, 463--465,
        467--470
  \item Ekert, Artur, 37, 120--122
  \item Enk, Steven J. van, v, 276, 286, 287, 384, 464
  \item Enz, Charles P., 71, 171, 172, 290, 409
  \item Epicurus, 397
  \item Euclid, 295, 296, 344
  \item Everett, Hugh, 2, 19--22, 24, 29, 31, 93, 108, 112, 183, 187,
        196, 199, 200, 207, 210, 300, 303, 355--358, 393, 394,
        462, 466, 472, 473
  \item Exner, Franz, 230, 297, 366, 399, 400, 435, 437
  \item Ezekiel, 301

  \indexspace

  \item Farhi, Edward, 110, 111, 113, 394
  \item Faye, Jan, 44, 146, 175, 198
  \item Feinstein, Amiel, 9
  \item Fermi, Enrico, 317, 393, 467
  \item Feuer, Lewis S., 44
  \item Fichte, J. G., 420
  \item Fierz, Markus, 20, 71, 74, 172, 235, 272, 282, 374, 405, 428
  \item Fine, Arthur, 142, 174, 175, 242, 373, 375, 378, 394, 444,
        446, 447
  \item Flatt, Lester, 121, 148, 193
  \item Fleck, Ludwik, 53--55, 58, 65, 71--73, 233
  \item Folse, Henry J., 44, 146, 164, 173--176, 198, 230, 237, 238,
        415
  \item Forman, Paul, 366, 399, 400, 425, 434, 435
  \item Forney, G. D., 13
  \item Fortun, Mike, 47, 49
  \item Fowler, Alan, 461
  \item Frost, David, 150
  \item Fuchs, Albert Winston, 3, 475
  \item Fuchs, Emma Jane, 36, 37, 243, 298, 300, 306, 318, 321, 323,
        327, 331, 342, 345, 398, 424, 475
  \item Fuchs, Kristen M. (Kiki), v, 3, 4, 36, 45, 47, 53, 65, 123, 138,
        140, 149, 193, 194, 248, 250, 298, 306, 318, 321, 323,
        336, 340, 342, 343, 345--347, 357, 400, 423, 475
  \item Fuchs, Wizzy, 475
  \item Furusawa, Akira, 120

  \indexspace

  \item G\"odel, Kurt, 44, 51, 257
  \item Galavotti, Maria Carla, 125, 394, 396, 398, 399
  \item Garcia-Alcaine, Guillermo, 283
  \item Gardner, Martin, 45
  \item Garfunkel, Art, 144
  \item Garisto, Robert, 70, 464
  \item Garrett, Anthony J. M., 154, 155, 239, 240, 242, 394, 414
  \item Gell-Mann, Murray, 178, 262, 302, 344
  \item Gershefeld, Neil A., 120
  \item Ghirardi, Giancarlo, 31, 309
  \item Ghose, Partha, 367
  \item Giacometti, Alberto, 151
  \item Giere, Ronald N., 379--381, 394, 444
  \item Gingrich, Robert M., 460
  \item Ginsberg, Allen, 41, 42
  \item Gisin, Nicolas, 120, 121, 171, 195, 290, 301, 312, 314, 316,
        347, 394
  \item Giuntini, Roberto, 365
  \item Gleason, Andrew M., 17, 18, 86--90, 92, 94, 101, 104, 126,
        202, 203, 215, 217, 222, 235, 242, 243, 293, 295,
        310, 311, 314, 319, 328, 329, 353--355, 361--363,
        393--395, 400, 448, 449
  \item Goldstein, Sheldon, 123, 242, 302, 319, 327, 367
  \item Goldstone, Jeffrey, 110, 113, 394
  \item Good, Irving John, 292
  \item Gordon, J. P., 13
  \item Gottesman, Daniel, 122, 176, 289, 323, 331, 450
  \item Gottfried, Kurt, 197, 227, 228
  \item Graham, Neill, 207, 356, 394
  \item Griffiths, Robert B., ii, 40, 121, 178, 191, 197, 205, 207, 284,
        290, 297, 302, 303, 344, 347, 348, 351, 368
  \item Gutmann, Sam, 110, 113, 210, 394

  \indexspace

  \item Hacking, Ian, 26, 29, 169, 170, 393, 394, 396, 445
  \item Haering, Theodor, 382
  \item Hagar, Sammy, 250
  \item Hains, Chris, 250, 475
  \item Hall, Michael J. W., 121, 364
  \item Hanle, Paul A., 230, 297, 366, 400
  \item Hanna, Marty, 345, 348, 349
  \item Hardy, Lucien, 43, 70, 96, 128, 438, 471, 472
  \item Harris, Paul, 295, 296, 339, 346, 348
  \item Hartle, James B., 25, 74, 106, 107, 109--113, 178, 183, 210,
        212, 262, 302, 352, 372, 394, 395, 432, 470
  \item Hausladen, Paul, 9, 13
  \item Hawking, Stephen W., 6, 51
  \item Hayden, Patrick, 94, 96, 471
  \item Hegel, G. W. F., 420
  \item Heilbron, J. L., 123, 381, 425, 434
  \item Heisenberg, Werner, 10, 50, 51, 79, 94, 114, 116, 129, 150, 151,
        182, 185, 202, 211, 218, 224, 230, 236, 263, 266, 268,
        271, 280, 282, 285, 296, 297, 344, 366, 369, 384, 385,
        387, 388, 392, 397, 400, 448, 458, 464, 465, 473
  \item Heitler, Walter, 270, 443
  \item Hellman, Geoffrey, 26, 27, 30, 394
  \item Hepburn, Audrey, 4
  \item Herling, Gary, 442
  \item Hermann, Grete, 387
  \item Herzfeld, Karl F., 79, 270, 442, 443
  \item Hillery, Mark, 121
  \item Hirota, Osamu, 97, 121, 151
  \item Hitler, Adolf, 4
  \item Holevo, Alexander S., 9, 11, 13, 14, 49, 115, 120, 121, 277,
        288, 331
  \item Holladay, Wendall, 296, 344, 345, 347, 348
  \item Holton, Gerald, 44
  \item Honner, John, 44, 146, 198, 230
  \item Hood, Christina J., 120
  \item Horodecki, Micha{\l }, 423, 461
  \item Horodecki, Pawe{\l }, 461
  \item Horodecki, Ryszard, 443, 461
  \item Hughes, Richard J., 120
  \item Hughston, Lane P., 203, 215, 452
  \item Hume, David, 418

  \indexspace

  \item Isham, Chris J., 143, 183, 252, 466

  \indexspace

  \item Jacobs, Kurt, 97, 104, 194, 236, 336, 363, 413
  \item James, Henry, 1
  \item James, Henry Sr., 1
  \item James, William, ii, 1, 2, 67, 188, 381
  \item Jammer, Max, 13, 242, 324, 437, 446, 447, 463, 473
  \item Janzing, Dominik, 463
  \item Jaynes, Edwin T., 32, 34, 36, 118, 122, 129, 131, 132, 134, 152,
        154, 155, 215, 216, 238, 242, 257, 273, 287, 359,
        383, 384, 394, 408, 413, 414, 434, 443
  \item Jeffrey, Richard C., 125, 395, 397, 398
  \item Jolly, Phillip von, 424, 425
  \item Jordan, Pascual, 445, 448
  \item Jozsa, Richard, 9, 13, 14, 37, 91, 94, 96, 99, 120, 121, 140,
        143, 195, 203, 215, 232, 234, 237, 279, 295, 329, 342,
        347, 359, 402, 412, 421, 424, 452, 454, 471, 472
  \item Jung, Carl G., 6, 144, 145, 365, 373, 375

  \indexspace

  \item Kahn, Karen L., 475
  \item Kalckar, J{\o }rgen, 44
  \item Kampen, N. G. van, 301
  \item Kant, Immanuel, 2, 57, 75, 98, 99, 140, 174, 186, 187, 206, 221,
        229, 233, 250, 332, 392, 397, 408, 414, 416--420
  \item Kennedy, John B., 276, 449
  \item Kennedy, John F., 4
  \item Kent, Adrian, ii, 82, 184, 217, 283, 311, 317, 352, 353, 382,
        395, 421
  \item Kepler, Johannes, 20, 44, 52, 278, 405--407
  \item Kermode, Frank, 52
  \item Kevles, Daniel J., 442
  \item Khrennikov, Andrei, 471, 472
  \item Kierkegaard, S{\o }ren, 2
  \item Kimble, H. Jeffrey, 120, 155, 161, 353
  \item Kochen, Simon, 17, 103, 118, 209, 235, 241, 245, 249, 282, 287,
        353, 362, 471
  \item Kolmogorov, Andrei N., 381, 390, 432
  \item Koresh, David, 60
  \item Kraus, Karl, 217, 218, 221, 277, 280, 400, 401
  \item Kuhn, Thomas S., 54, 170, 206, 434
  \item Kyburg, Henry E., 30, 273, 312, 395

  \indexspace

  \item Laflamme, Raymond, 286, 287
  \item Lahti, Pekka J., 172, 174, 237, 395, 446, 471
  \item Lambalgen, Michiel van, 25, 110, 210, 432, 433
  \item Landau, Lev D., 210--212, 307
  \item Landauer, Rolf, ii, 79, 97, 152, 153, 155, 189, 220, 281, 413,
        427, 461
  \item Lange, Wolfgang, 120
  \item Laplace, Pierre Simon, 131, 132, 162, 380, 383, 395, 406
  \item Laurikainen, Kalervo V., 71, 172, 175, 439, 448
  \item Leibniz, Gottfried W., 51, 211, 375
  \item Levin, Leonid, 26, 109
  \item Lieb, Elliott H., 115, 120, 252, 317, 470
  \item Lifshitz, Evgenii M., 210--212, 307
  \item Lindblad, G\"oran, 115, 116, 120, 121, 232, 289
  \item Lloyd, Seth, 53
  \item Lo, Hoi-Kwong, 82, 323

  \indexspace

  \item M\"uller, Hermann, 424
  \item Mabuchi, Hideo, ii, 12, 47, 113, 120, 122, 150, 193, 378, 379,
        389
  \item Macchiavello, Chiara, 121, 122
  \item MacKinnon, Edward, 44
  \item Mann, Ady, 122, 195, 284, 353, 393
  \item Martin-L\"of, Per, 26, 108, 109, 432
  \item Mayers, Dominic, 82, 250, 251
  \item McCartney, Paul, 150
  \item Mermin, N. David, ii, v, 40, 43, 44, 46--48, 76, 94, 96, 97,
        105, 121--123, 145, 146, 149, 150, 162, 173, 177, 191,
        195, 198, 209, 225, 227, 240, 243, 269, 270, 281, 290,
        293, 305, 318, 331, 346, 347, 351, 353, 371, 382, 384,
        391, 392, 395, 414, 415, 431, 445, 460, 465, 471, 472
  \item Merzbacher, Eugen, 298, 463
  \item Meyer, David A., ii, 17, 92, 126, 235, 245, 282, 283, 286, 287,
        314, 353
  \item Minnelli, Liza, 3
  \item Mises, Richard von, 108, 110, 111, 212, 380, 395, 396, 432, 433,
        437
  \item Mittelstaedt, Peter, 110, 112, 113, 172, 174, 237, 395, 446
  \item Mor, Tal, 122, 288
  \item Murdoch, Dugald, 45

  \indexspace

  \item Neapolitan, R. E., 113, 432
  \item Newton, Isaac, 51, 97, 133, 157, 163, 164, 167, 278, 284, 353,
        377, 406, 407, 417, 444
  \item Nicholson, Jeffrey W., 161, 247, 250
  \item Nielsen, Michael A., ii, 239, 249, 323, 354, 460
  \item Niewood, Gerry, 144
  \item Niu, Chi-Sheng, 121, 178, 179, 368

  \indexspace

  \item Ochs, Wolfgang, 112, 113, 395
  \item Omn\`es, Roland, 129, 284, 290, 302, 304, 347
  \item Oppenheimer, J. Robert, 1, 3, 4, 340
  \item Orwell, George, 42, 43, 61, 111, 482

  \indexspace

  \item Pagels, Heinz R., 141
  \item Pais, Abraham, 45, 219, 224, 435
  \item Pauli, Wolfgang, vii, 71, 73, 79, 96, 98, 116, 130, 136, 137,
        145, 153, 154, 171, 172, 175, 206, 210, 211, 216, 219,
        221--223, 230, 231, 236, 237, 244, 264, 266, 272, 282,
        290, 346, 373, 374, 384--388, 409, 428, 430, 445,
        447, 448
  \item Pearle, Philip, 301
  \item Peierls, Rudolf, 93, 94, 232, 233, 235, 236, 284, 306, 307,
        309, 310, 329
  \item Peirce, Charles Sanders, 1, 2, 46, 189, 197, 254, 435, 437
  \item Penrose, Roger, 156, 233, 235, 329, 402, 407, 408, 428, 466
  \item Peres, Asher, ii, 3, 11, 15, 29--31, 43, 53, 77, 79, 88, 92,
        101, 102, 104, 105, 120, 121, 123, 133, 142, 159, 167,
        191, 193, 197, 209, 214, 222, 224, 225, 232, 233,
        235, 236, 239, 242, 250, 255, 257, 267, 269, 279, 285,
        292, 293, 295, 298, 310, 320, 337--339, 341, 343, 366,
        368, 369, 391, 395, 405, 413, 415, 421, 424, 443, 445,
        456, 457, 471
  \item Petersen, Aage, 142, 266, 419
  \item Piaget, Jean, 2
  \item Pierce, John R., 80
  \item Pike, Rob, 465
  \item Pitowsky, Itamar, 18, 92, 123, 126, 222, 314, 363, 395, 448,
        471
  \item Plaga, Rainer, 306
  \item Planck, Max, 10, 84, 123, 381, 382, 424, 425, 470
  \item Plath, Sylvia, 148, 156
  \item Plato, 169, 170, 174, 392, 397
  \item Plenio, Martin B., 122
  \item Plotnitsky, Arkady, 44, 45, 162, 219, 237, 290, 470
  \item Podolsky, Boris, 3, 10, 13, 116, 117, 122, 217, 353, 385, 393,
        441, 446, 459, 463
  \item Poincar\'e, Henri, 168, 295, 312, 316, 320, 326, 330
  \item Polzik, Eugene S., 120
  \item Popescu, Sandu, 121, 122
  \item Popper, Karl R., 43, 212, 300, 381, 382, 393, 395
  \item Preskill, John, ii, vi, 76, 78, 114, 149, 161, 176, 205, 231,
        243, 269, 301, 318, 351, 357, 395, 430, 441, 471, 472
  \item Ptolemy, 296, 344, 348

  \indexspace

  \item Quine, Willard V., 32

  \indexspace

  \item Ralph, Timothy C., 453
  \item Rasco, B. Charles, 141, 435
  \item Reichenbach, Hans, 108, 380, 395, 437
  \item Renes, Joseph, ii, 105, 243, 361, 363
  \item Revzen, M., 122, 195, 353, 393
  \item Rimini, Alberto, 31, 309, 469
  \item Robb, Daniel, 359, 412, 452
  \item Rorty, Richard, 169
  \item Rosen, Nathan, 3, 10, 13, 79, 116, 117, 122, 217, 276, 325, 326,
        353, 379, 385, 393, 441, 446, 459, 463
  \item Rosenfeld, L\'eon, 75, 76, 82, 152, 213, 214, 284, 320
  \item Rovelli, Carlo, 123, 242, 395
  \item Ruskai, Mary Beth, ii, 235, 317, 366

  \indexspace

  \item Savage, Leonard J., 21, 22, 354, 380, 395, 404, 407
  \item Schack, Dorothee, 379
  \item Schack, R\"udiger, ii, 74, 75, 77, 80, 93, 94, 116, 121, 122,
        125, 141, 154, 158, 166, 168, 210, 216, 229, 235, 253,
        258, 273, 277, 312, 370--373, 385, 388, 389, 401, 407,
        409, 428, 444, 464, 471, 472
  \item Schilpp, P. A., 224, 266
  \item Schopenhauer, Arthur, 221, 409, 414, 416
  \item Schr\"odinger, Erwin, 13, 73, 92, 125, 128, 132, 133, 158,
        186, 187, 227, 229, 230, 234, 238, 239, 244, 248, 260,
        262, 271, 279, 296, 307, 352, 366, 367, 377, 384--388,
        397--400, 405, 422, 435, 437, 446, 447, 452, 459, 461
  \item Schumacher, Benjamin W., 9, 13, 14, 80, 84, 93, 96, 120--122,
        140, 157, 166, 177, 196, 232, 331, 454, 470, 471
  \item Schumann, Robert H., ii, 411
  \item Schweber, Sylvan S., 46, 47, 151, 189, 253, 462
  \item Sciama, Dennis W., 466
  \item Scruggs, Earl, 121, 148, 193
  \item Scudo, Petra F., 287, 349
  \item Seife, Charles, 450
  \item Shannon, Claude E., 9, 10, 20, 35, 89, 109, 112, 116, 118, 122,
        353, 364, 416, 442, 453, 462, 471, 474
  \item Shenker, Orly R., 317
  \item Shimony, Abner, ii, 224, 253, 268, 415, 418, 419, 452, 462, 463,
        470
  \item Shor, Peter W., 10, 16, 84, 114, 120--122, 161, 237, 424, 458,
        465
  \item Simon, Barry, 353, 367
  \item Simon, Paul, 138, 144, 201, 252, 253, 255
  \item Simone, Nina, 3
  \item Sipe, John E., 18, 402
  \item Smith, Adrian F. M., 30, 364
  \item Smolin, John A., 121, 122, 143, 177, 258, 317, 471
  \item Smolin, Lee, 189
  \item Sobottka, Stanley, 296, 344
  \item Socrates, 214, 270, 359
  \item Sommerfeld, Karl, 270
  \item Specker, Ernst, 17, 103, 118, 209, 235, 241, 245, 249, 282, 353,
        362, 471
  \item Spekkens, Robert W., 18, 402, 461
  \item Spinoza, Baruch, 234, 254
  \item Sponzo, Amy, 250
  \item Stapp, Henry P., 67, 68, 215, 273, 274, 282
  \item Starkman, Glenn D., 440
  \item Steane, Andrew M., 120, 465
  \item Steiner, Rudolf, 429
  \item Stevens, Wallace, 51, 52
  \item Styer, Daniel, 296
  \item Suppes, Patrick, 381, 394, 396, 399
  \item S{\o }rensen, Jens L., 120

  \indexspace

  \item Tegmark, Max, 68, 93, 466
  \item Terhal, Barbara M., 121
  \item Thacker, William D., 158
  \item Thapliyal, Ashish V., 122
  \item Turchette, Quentin A., 12, 120
  \item Turing, Alan, 32, 37, 108

  \indexspace

  \item Uffink, Jos, 317
  \item Uhlhorn, Ulrich, 168
  \item Uhlmann, Armin, 89, 115, 120, 122

  \indexspace

  \item Vaidman, Lev, 318, 324--327
  \item Vidal, Guifre, 135
  \item Vitanyi, Paul, 432

  \indexspace

  \item Wallace, David, 472, 473
  \item Wallach, Nolan, 17, 18, 331
  \item Walther, Herbert, 357
  \item Washington, Dinah, 465
  \item Waskan, Jonathan A., 415, 417, 420
  \item Weaver, Leslie, 336
  \item Weber, Tullio, 31, 309
  \item Weinfurter, Harald, 122, 434
  \item Westmoreland, Michael D., 9, 13, 14, 120, 121
  \item Weyl, Hermann, 425, 437, 438, 469
  \item Wheeler, John Archibald, 2, 5, 7, 10, 31, 46, 47, 50, 61--63,
        72, 73, 75, 79, 84, 125, 126, 137, 138, 144, 145,
        149--151, 154, 158, 182, 189, 191, 192, 196, 200,
        229, 230, 250, 251, 257, 264--266, 270, 271, 278, 323,
        349, 352, 359, 360, 365, 391, 398, 402, 413, 421,
        423, 424, 430, 431, 442, 443, 447, 462, 463, 466--470,
        475
  \item Whitaker, Andrew, 382
  \item Wiesner, Stephen J., 122, 161
  \item Wigner, Eugene P., 120, 139, 167, 168, 276, 284, 286, 289, 342,
        396, 408, 422, 451
  \item Wocjan, Pawel, 462
  \item Wolf, Stefan, 290
  \item Wolpert, David H., 61
  \item Wootters, William K., ii, 11--13, 15, 68, 84, 88, 89, 93, 96,
        121, 122, 125, 126, 138, 158, 161, 179, 195, 196, 198,
        202, 203, 211, 213--215, 232, 236, 237, 270, 283, 295,
        331, 347, 353, 359, 411--413, 421, 427, 428, 449--452,
        470, 471

  \indexspace

  \item Yablonovitch, Eli, 439

  \indexspace

  \item Zajonc, Arthur, 96, 340, 429
  \item Zbinden, Hugo, 120
  \item Zeilinger, Anton, ii, 122, 228, 229, 298, 300, 302, 304, 323,
        346, 367, 430, 431, 438
  \item Zurek, Wojciech H., 36, 121, 123, 126, 129, 178, 195, 230, 242,
        257, 262, 279, 290, 298, 354, 394, 398, 447, 449, 451,
        468, 469

\end{theindex}


\pagebreak

\thispagestyle{empty}

\vspace*{6in}

\begin{flushright}
\parbox{2.9in}{
All writers are vain, selfish and lazy, and at the very bottom of
their motives lies a mystery. Writing a book is a long, exhausting
struggle, like a long bout of some painful illness. One would never
undertake such a thing if one were not driven by some demon whom one
can neither resist nor \medskip understand. \hspace*{\fill} --- {\it
George {\Orwell}}}\hspace*{0.5in}
\end{flushright}

\end{document}